\documentclass[12pt,a4paper]{report}
\usepackage[utf8]{inputenc}

\usepackage{amsmath,amssymb,bm,epsf,epsfig,graphicx,dsfont,url}
\usepackage{dsfont,caption,subcaption}
\usepackage[section]{placeins}
\usepackage{tensor}
\usepackage{cancel,accents,mathdots,multirow}

\usepackage{amsfonts}       
\usepackage{amsthm}         
\usepackage{bbding}         
\usepackage{fancyvrb}       
\usepackage{dcolumn}        
\usepackage{booktabs}       
\usepackage{paralist}       
\usepackage[usenames]{xcolor}  
\usepackage{breakurl}

\input epsf.sty

\newcommand{\Tr}{\mathop{\rm Tr}\nolimits}
\newcommand{\re}{\mathop{\rm Re}\nolimits}
\newcommand{\im}{\mathop{\rm Im}\nolimits}
\newcommand{\bpz}{\mathop{\rm bpz}\nolimits}

\let\eps = \varepsilon

\DeclareMathOperator*{\res}{res}

\def \del {\partial}
\def \delb {\bar \partial}
\def \rar {\rightarrow}

\def \lrar {\leftrightarrow}
\def \la {\langle}
\def \ra {\rangle}
\def \rra {\rangle\!\rangle}
\def \lla {\langle\!\langle}
\def \ww {|\hspace{-1pt}|}
\def \ps {\phantom{+}}
\def \zb {\bar z}
\def \nn {\nonumber}
\def \Id {\mathds{1}}
\def \inf {\infty}
\def \mr {\multirow}

\def \vv {{\mathcal V}}
\def \aa {{\mathcal A}}
\def \mm {{\mathcal M}}
\def \hh {{\mathcal H}}
\def \cc {{\mathcal C}}
\def \kk {{\mathcal K}}
\def \tt {{\mathcal T}}
\def \nnn {{\mathcal N}}
\def \ff {{\mathcal F}}
\def \ee {{\mathcal E}}

\def\lS{\lambda_{\rm S}}
\def\lB{\lambda_{\rm B}}
\def\Pade{Pad\'{e}\ }
\def\PadeBorel{Pad\'{e}-Borel\ }

\def \dexp#1{\times 10^{#1}}
\def \rowh#1{\rule{0pt}{#1}\hspace{-3.5pt}}

\def \normo { \mathrel{\raisebox{.07ex}{$\colon$}}\!\! }
\def \normc { \mathrel{\raisebox{.69ex}{\tiny$\circ$}}\hspace{-0.74ex} \mathrel{\raisebox{-.11ex}{\tiny$\circ$}} }
\def \norms { \mathrel{\raisebox{.69ex}{\tiny$\star$}}\hspace{-0.74ex} \mathrel{\raisebox{-.11ex}{\tiny$\star$}} }
\def \normsf { \mathrel{\raisebox{.69ex}{\tiny$\star$}}\hspace{-0.86ex} \mathrel{\raisebox{-.11ex}{\tiny$\star$}} }

\newcommand{\Cbulk}[3] {C\indices{_#1_#2^#3}}

\newcommand{\Cbound}[6] {C\indices*{^{(#1#2#3)#6}_{\ #4#5}}}
\newcommand{\CboundL}[6] {C\indices*{^{(#1#2#3)}_{\ #4#5#6}}}
\newcommand{\Cbb}[3] {\tensor[^{(#1)}]{B}{_#2^#3}}
\newcommand{\CbbL}[3] {\tensor[^{(#1)}]{B}{_#2_#3}}

\newcommand{\Fmatrix}[6] {
F_{#1#2}\!
\begin{bmatrix}
 #4 & #5 \\
 #3 & #6 \\
\end{bmatrix}}

\newcommand{\BRmatrix}[6] {
B_{#1#2}\!
\begin{bmatrix}
 #4 & #5 \\
 #3 & #6 \\
\end{bmatrix}}

\begin{document}

\begin{center}
 {\LARGE \bf $\,$\\
\vskip2cm
Level Truncation Approach to Open String Field Theory}
\vskip.5cm

\vskip 1.1cm

{\large  Mat\v{e}j Kudrna}\footnote{Email:
kudrnam at fzu.cz}$^{(a)}$
\vskip 1 cm

$^{(a)}${\it {Institute of Physics of the ASCR, v.v.i.} \\
{Na Slovance 2, 182 21 Prague 8, Czech Republic}}
\vspace{.5cm}

\end{center}

\vspace*{6.0ex}

\centerline{\bf Abstract}
\bigskip
Given a D-brane background in string theory (or equivalently boundary conditions in a two-dimensional conformal field theory), classical solutions of open string field theory equations of motion are conjectured to describe new D-brane backgrounds (boundary conditions). In this thesis, we study these solutions in bosonic open string field theory using the level truncation approach, which is a numerical approach where the string field is truncated to a finite number of degrees of freedom.

We start with a review of the theoretical background and numerical methods which are needed in the level truncation approach and then we discuss solutions on several different backgrounds. First, we discuss universal solutions, which do not depend on the open string background, then we analyze solutions of the free boson theory compactified on a circle or on a torus, then marginal solutions in three different approaches and finally solutions in theories which include the A-series of Virasoro minimal models. In addition to known D-branes, we find so-called exotic solutions which potentially describe yet unknown boundary states.

This paper is based on my doctoral thesis submitted to the Faculty of Mathematics and Physics at Charles University in Prague \cite{KudrnaThesis}.
\newpage

\tableofcontents

\chapter*{Introduction}
\addcontentsline{toc}{chapter}{Introduction}
Open string field theory (OSFT) was introduced by Edward Witten \cite{WittenSFT} as a non-pertubative formulation of string theory. The theory can be written both for bosonic string and for superstring \cite{BerkovitsSuperstring}, but we will consider only the original bosonic theory, which is much simpler. It has a very simple cubic action:
\begin{equation}\label{action intr}
S[\Psi] =-\frac{1}{g_o^2} \int \left( \frac{1}{2}\Psi \ast Q\Psi + \frac{1}{3} \Psi\ast\Psi\ast\Psi \right).
\end{equation}
This action is formally very similar to the Chern-Simmons theory, the one-form is analogous to the string field $\Psi$, the wedge product to the star product $\ast$ and the exterior derivative to the BRST charge $Q$.

String field theory provides a second quantization framework for string theory, which is usually developed from a first quantized point of view. Their relation is somewhat similar to the relation between quantum mechanics and quantum field theory. Quantum mechanics is formulated using path integrals over all possible particle trajectories, while quantum field theory is formulated using path integrals over all field configurations. Similarly, path integrals in string theory go over all string worldsheets, while in string field theory, we integrate over string field configurations. Simple scattering amplitudes of strings are usually computed in ordinary string theory, but string field theory can provide additional insight into more complicated amplitudes. It has also other applications, most notably, it can describe D-brane dynamics.

The most famous process described by OSFT is tachyon condensation. Ta\-chy\-ons, which live on all D-branes in bosonic string theory and on certain D-brane systems in superstring theory, cause instability of D-branes and lead to D-brane decay.
In \cite{SenUniversality}, Ashoke Sen published three conjectures regarding tachyon condensation\footnote{See also \cite{TaylorZwiebach}, where Sen's conjectures are recapitulated (and given this somewhat nonintuitive order).}. The first conjecture states that the tachyon potential has a locally stable minimum, which is known as the tachyon vacuum, and that the energy density at this minimum cancels the energy density of the original D-brane. The third conjecture states that the tachyon vacuum solution describes the closed string vacuum, which means that the original D-brane is absent and there are no open string excitations around this solution. These conjectures have been proven thanks to Schnabl's analytic solution \cite{AnalyticSolutionSchnabl}\cite{EllwoodTVCohomology}. Sen's second conjecture concerns a somewhat different process, it states that lower dimensional D-branes are described by solitonic solutions on a D25-brane.

A generalization of these conjectures is known as background independence of string field theory\footnote{See for example \cite{SenBackground1}\cite{SenBackground2}\cite{SenBackground3}\cite{SenBackground4}, although these references do not discuss this topic using the current D-brane point of view. A modern formulation of the conjecture is given in \cite{ErlerMaccaferri3}.}. This conjecture states that if we are given a string field theory formulated on an arbitrary D-brane system, there should be classical solutions describing all other possible D-branes systems in the given string theory. Furthermore, string field theories on the two different D-brane systems should be related by field redefinition. If we write the string field as $\Psi=\Psi_0+\Phi$, where $\Psi_0$ is a classical solution of the equations of motion, then the OSFT action reads
\begin{equation}
S[\Psi_0+\Phi]=S[\Psi_0]-\frac{1}{g_o^2} \int \left( \frac{1}{2}\Phi \ast Q_{\Psi_0}\Phi + \frac{1}{3} \Phi\ast\Phi\ast\Phi \right).
\end{equation}
The action for $\Phi$ is very similar to (\ref{action intr}), which allows us to identify it with the OSFT action around the new D-brane system. The action $S[\Psi_0]$ is conjectured to be equal to the difference between energies of the two D-brane systems. Note that the string field $\Phi$ consists of fields that live on the reference D-brane system, so, to fully match the action on the new D-brane system, we also need to find a map between the two Hilbert spaces. The modified BRST operator $Q_{\Psi_0}$ is by definition nilpotent, which means that it is possible to compute its cohomology, which can help with identification of the new D-brane system and with finding the field redefinition.

Another possibility how to characterize a D-brane system is using its boundary state. Ian Ellwood conjectured \cite{EllwoodInvariants} that the boundary state of the new D-brane system is described by a set of linear gauge invariant observables of the form
\begin{equation}
\la I | \vv(i,-i)| \Psi\ra,
\end{equation}
where $\vv$ is an on-shell bulk operator. A generalization of this conjecture was introduced in \cite{KMS}.

Evidence for these conjectures are provided by many solutions, which can be found using two different strategies: analytically and numerically.
Analytic solutions are based on the subalgebra of wedge states with insertions \cite{WedgeSchnabl}\cite{AnalyticSolutionSchnabl}\cite{AnalyticSolutionOkawa}. A great success of the analytic approach is a solution found by Erler and Maccaferri \cite{ErlerMaccaferri}\cite{ErlerMaccaferri2}\cite{ErlerMaccaferri3}, which can describe an arbitrary background. Therefore it allows verification of the aforementioned conjectures on abstract level. However, a certain limitation of the solution is that if we want to use it beyond formal level, we need a good understanding of underlying CFT. In particular, we need to understand boundary condition changing operators between the initial and the final background.

In this thesis, we focus on the other approach. We use a numerical method called the level truncation scheme.
Since it is a numerical approach, we can get only results with limited precision, but they are usually good enough to make unique identification of solutions. The advantage of this approach is that it is quite universal and it can be used even on many nontrivial backgrounds as long as we understand at least one boundary CFT. Using this approach, we have two main goals:

First, we want to gather independent evidence about background independence of OSFT and the related conjectures. Therefore we compute energies of OSFT solutions and the corresponding boundary states and we compare them with the expected results. Unfortunately, it is not known how to reliably compute spectra of excitations around high level numerical solutions, so we cannot verify this part of the conjecture.

Second, we want to introduce a new application for open string field theory, which is search for new boundary states in irrational conformal field theories\footnote{In this context, we mean theories which are irrational with respect to the full energy-momentum tensor. Formulation of OSFT requires understanding of at least one boundary theory, for which we usually need some stronger symmetry.}. By solving the level-truncated equations, we find so-called exotic solutions which cannot be identified as conventional D-branes. Therefore we conclude that they describe boundary states which break the full symmetry of the theory (for example U(1)$^n$ symmetry in some free boson theory) and preserve only the conformal symmetry. So far, it is not known how to search for such solutions analytically. As we mentioned above, numerical calculations cannot give us exact results, but we get strong evidence about existence of these boundary states and our results can be compared with boundary conformal field theory perturbation techniques or used as hints for finding such boundary states analytically.

This thesis is organized as follows. In chapter \ref{sec:CFT}, we provide a brief review of conformal field theory, which serves as a tool for string field theory. We describe general properties of bulk and boundary CFTs and then we take a closer look at several models that will appear in this thesis. In chapter \ref{sec:SFT}, we focus on description of string field theory. First, we provide a short review, then we introduce the backgrounds we are interested in and we define observables on these backgrounds. Finally, we derive conservation laws for the cubic vertex and for Ellwood invariants, which play a crucial role in our calculations. In chapter \ref{sec:Numerics}, we describe our numerical algorithms for computing OSFT solutions and their observables.
In chapters \ref{sec:universal} to \ref{sec:MM}, we summarize our results in the individual settings. In chapter \ref{sec:universal}, we discuss universal solutions. We focus on the tachyon vacuum solution in Siegel gauge and in Schnabl gauge, but we also mention other solution in these gauges and universal solutions without any gauge fixing condition. In chapter \ref{sec:FB circle}, we discuss free boson solutions on a circle. We focus on single and double lump solutions, but we also add few Wilson line solutions. In chapter \ref{sec:marginal}, we present marginal solutions in three different approaches (marginal, tachyon and perturbative). As the background, we choose the free boson theory on the self-dual circle. In chapter \ref{sec:FB D2}, we discuss solutions in the free boson theory on a 2D torus. We show few regular solutions and introduce the concept of exotic solutions, which describe non-conventional boundary states. In chapter \ref{sec:MM}, we analyze solutions in theories including the Virasoro minimal models. We consider the following settings: Ising model, Lee-Yang model, double Ising model and Ising$\otimes$tricritial Ising model. In chapter \ref{sec:Discussion}, we summarize generic properties of numerical solutions and offer some possible future directions for numerical OSFT. In appendix \ref{sec:MM Fmatrix}, we show some properties of F-matrices in minimal models. In appendix \ref{sec:characters}, we provide characters for the models that appear in this thesis. In appendix \ref{sec:time}, we discuss time and memory requirements of numerical calculations in OSFT.

This paper is based on the doctoral thesis \cite{KudrnaThesis}. We have done only small changes, which mostly include stylistic changes and correction of errors. In few places, we also extended discussion of some topics (mainly in the introduction and summary) and added more references.

\chapter{Conformal field theory in two dimensions}\label{sec:CFT}
In this chapter, we provide a brief review of two-dimensional conformal field theory (CFT), which is crucial tool for formulation of string theory. Writing a self-contained text is beyond the scope of this thesis, so we focus on topics which are useful in the context of string field theory. We also describe in more detail some particular CFTs that will play a role in our calculations.

Most of the information provided in this chapter is well known and it can be found is many books and articles. In section \ref{sec:CFT:Bulk}, which describes bulk CFT, we mostly follow \cite{DiFrancesco}\cite{Blumenhagen} and in section \ref{sec:CFT:Boundary}, which describes boundary CFT, we follow \cite{RecknagelSchomerus}. The books by Polchinski \cite{Polchinski}\cite{Polchinski2} also discuss many aspects of conformal field theory and we use his conventions for the free boson theory, the ghost theory and for bosonic string theory. Another string theory book with a CFT review is \cite{BlumenhagenStringTheory}.

\section{Bulk conformal field theory}\label{sec:CFT:Bulk}
A conformal field theory is a quantum field theory which is invariant with respect to so-called conformal transformations. A conformal transformation is defined as an invertible coordinate transformation $x^\mu\rar x'^\mu$ which leaves the metric tensor $g_{\mu\nu}$ invariant up to a scale factor:
\begin{equation}\label{coordinates plane}
g'_{\mu\nu}(x')=\Lambda(x)g_{\mu\nu}(x).
\end{equation}
In more than two dimensions, conformal transformations around the standard Euclidean metric $g_{\mu\nu}=\rm{diag}(1,1,\dots,1)$\footnote{In the Minkowski space, we can obtain the Euclidean metric by the Wick rotation $t\rar i\tau.$} are given by
\begin{eqnarray}
x'^\mu&=&x^\mu+a^\mu \quad \rm({translation)}, \\
x'^\mu&=&M^{\ \mu}_\nu x^\nu \quad \rm{(rotation)}, \\
x'^\mu&=&\lambda\, x^\mu\quad \rm{(dilation)}, \\
x'^\mu&=&\frac{x^\mu-b^\mu x^2}{1-2b_\mu x^\mu+b^2 x^2} \quad \rm{(special\ conformal\ transformation)}.
\end{eqnarray}
Therefore the conformal group for $d>2$ has a finite number of generators and it is isomorphic to the group SO$(d+1,1)$. However, two dimensions are exceptional and the conformal group there is much bigger.

We usually formulate a two-dimensional conformal field theory on a complex plane (or more precisely on the Riemann sphere), where we define complex coordinates as
\begin{equation}
z=x^1+i x^2,\quad \zb=x^1-i x^2.
\end{equation}
The derivatives with respect to $z$ and $\zb$ are given by
\begin{equation}
\del\equiv \del_z=\frac{1}{2}(\del_0-i\del_1),\quad \delb\equiv \del_{\zb}=\frac{1}{2}(\del_0+i\del_1)
\end{equation}
and the metric tensor in these coordinates is
\begin{equation}
g_{\mu\nu}=\left(\begin{array}{cc}
0 & \frac{1}{2} \\
\frac{1}{2} & 0 \\
\end{array}\right).
\end{equation}

Using the complex coordinates, we can easily see that all conformal transformations are given by holomorphic maps
\begin{equation}\label{conformal map}
z'=f(z) 
\end{equation}
and that the metric transforms as
\begin{equation}
dz'd\zb'=\del f(z) \delb\bar f(\zb)dz d\zb.
\end{equation}
These transformations are however defined only locally. Global conformal transformations must be well-defined and invertible in the whole complex plane. All functions with such properties can by written as
\begin{equation}\label{conformal map glob}
f(z)=\frac{az+b}{cz+d},
\end{equation}
where $(a, b, c, d)$ are complex numbers that satisfy $ad-bc=1$. They are also invariant under $(a,b,c,d)\rar (-a,-b,-c,-d)$, therefore global conformal transformations form the M\"{o}bius group SL(2,$\mathbb{C}$)/$\mathbb{Z}_2$.
Other conformal transformations map the complex plane on a different surface.

\begin{figure}
\centering
\includegraphics[width=12cm]{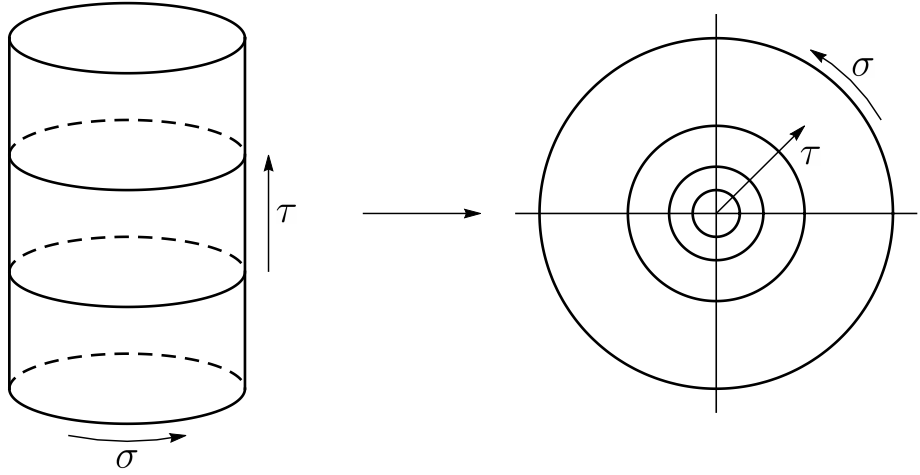}
\caption{Conformal transformation from the cylinder to the complex plane.}
\label{fig:cylinder map}
\end{figure}

There is no clear distinction between time and space in the Euclidean spacetime and therefore the usual definition of the time direction is inspired by string theory. A closed string is described by its worldsheet, which forms a cylinder parameterized by Euclidean coordinates $\tau$ and $\sigma$. Using these two coordinates, we define a new complex coordinate
\begin{equation}
w=\tau+i\sigma,
\end{equation}
which has the identification $w\sim w+2\pi i$. The cylinder can be mapped to the complex plane using the following coordinate transformation
\begin{equation}
z=e^w=e^{\tau+i\sigma}.
\end{equation}
This map changes meaning of the cylindrical coordinates as follows: $\tau$ becomes a radial coordinate, with past infinity mapped to $z=0$ and future infinity to $z=\inf$, and $\sigma$ becomes an angular coordinate running counterclockwise, see figure \ref{fig:cylinder map} for illustration.

\subsection{Conformal fields}\label{sec:CFT:Bulk:fields}
Local fields in a conformal field theory can by classified according to their behavior under conformal transformations. First, we consider a field which transforms under the scaling $z\rar \lambda z$ as
\begin{equation}
\tilde\phi(z,\zb)=\lambda^h \bar \lambda^{\bar h}\phi(\lambda z,\bar\lambda\zb).
\end{equation}
Such field is said to have conformal dimension (or weight) $(h,\bar h)$. The sum of these numbers is called the total dimension of the field, $\Delta=h+\bar h$, and their difference the spin of the field, $s=|h-\bar h|$.

A field that transforms under a general conformal map $z\rar f(z)$ as
\begin{equation}\label{primary transformation}
\tilde\phi(z,\zb)=\left(\frac{df}{dz}\right)^h \left(\frac{d\bar f}{d\zb}\right)^{\bar h}\phi(f(z),\bar f(\zb))
\end{equation}
is called primary. Fields that are not primary are called secondary or descendant fields. We can also define quasi-primary fields, for which (\ref{primary transformation}) holds only for global conformal transformations.

A field with definite conformal dimension $(h,\bar h)$ can be expanded as
\begin{equation}\label{primary expansion 2}
\phi(z,\zb)=\sum_{m,n=-\inf}^\inf\phi_{m,n} z^{-m-h}\zb^{-n-\bar h}.
\end{equation}
However, many conformal fields depend only on one of the two coordinates. A field that depends only on $z$ is called chiral (or holomorphic) and a field that depends only on $\zb$ is called anti-chiral (antiholomorphic). A chiral field can be expanded as
\begin{equation}\label{primary expansion 1}
\phi(z)=\sum_{m=-\inf}^\inf\phi_{m} z^{-m-h}
\end{equation}
and its modes $\phi_{m}$ can be extracted using contour integrals
\begin{equation}
\phi_{m}=\oint \frac{dz}{2\pi i} z^{m+h-1}\phi (z).
\end{equation}
Sometimes it is useful to expand a field around a point different from the origin, in that case we define
\begin{equation}
\phi_{m}(w)=\oint \frac{dz}{2\pi i} (z-w)^{m+h-1}\phi (z).
\end{equation}

Conformal field theories have one special property which is not present in other quantum field theories. There is an isomorphism between the state space and local operators at a given point,
\begin{equation}
\phi(z,\zb)\leftrightarrow |\phi\ra.
\end{equation}
This correspondence can be motivated by the transformation from the cylinder to the complex plane. An asymptotic state at $\tau=-\inf$ on the cylinder is mapped to $z=0$ on the complex plane. Therefore we can write
\begin{equation}
|\phi\ra=\lim_{z,\zb\rar 0}\phi(z,\zb)|0\ra,
\end{equation}
where $|0\ra$ is the conformal field theory vacuum state. The local operator associated to a state is called vertex operator. Using the expansion (\ref{primary expansion 2}), we find
\begin{equation}
\phi_{m,n}|0\ra=0, \quad m>-h\ {\rm or} \ n>-\bar h
\end{equation}
and
\begin{equation}
|\phi\ra=\phi_{-h,-\bar h}|0\ra.
\end{equation}

\subsection{Operator product expansion}\label{sec:CFT:Bulk:OPE}
Another special property of conformal field theories is that a product of two local fields can be replaced by a new local field, which can be expressed as a Laurent series in the distance of the two operators:
\begin{equation}\label{OPE general}
\phi_i(z_1,\bar z_1)\phi_j(z_2,\bar z_2)=\sum_{k} \Cbulk ijk \phi_k(z_2,\bar z_2) z_{12}^{h_{ijk}}\zb_{12}^{\bar h_{ijk}},
\end{equation}
where $z_{12}=z_1-z_2$, $h_{ijk}=h_k-h_i-h_j$ and similarly for the antiholomorphic quantities. The numbers $\Cbulk ijk$ are called structure constants. This type of expression is called the operator product expansion (OPE). The sum on the right hand side goes in principle over all local fields in the theory, but there are usually some selection rules that restrict which operators may appear in the sum. They are called fusion rules, see (\ref{fusion rules}).

The OPE is closely connected to commutators between modes of local fields. Consider two chiral fields, $\phi$ and $\chi$. A commutator of two modes of these fields $[\phi_m,\chi_n]$ can be derived using contour manipulations as
\begin{eqnarray}\label{commutators contour}
[\phi_m,\chi_n] &=& \oint \frac{dz}{2\pi i}\oint\frac{dw}{2\pi i} z^{m+h_\phi-1} w^{n+h_\chi-1} [\phi(z),\chi(w)] \nn\\
   &=& \oint\frac{dw}{2\pi i} \res_{z\rar w} z^{m+h_\phi-1} w^{n+h_\chi-1} \phi(z)\chi(w).
\end{eqnarray}
The residuum picks only the singular part of the OPE and therefore fields with regular OPE have trivial commutators.

When two fields approach each other, they often have singularities. Therefore, in order to define a product of two operators at a single point, we have to use a prescription called normal ordering, which is denoted by $\normo\normo$. The normal ordering is defined in terms of the OPE as
\begin{equation}
\phi(z)\chi(w)={\rm singular\ part}+\sum_{n=0}^\inf \frac{(z-w)^n}{n!}\normo\! \del^n \phi\,  \chi \!\! \normo \! (w).
\end{equation}
Therefore we can express the normal ordering in terms of a contour integral as
\begin{equation}
\normo\! \phi\,  \chi \!\!\normo \! (w)=\oint \frac{dz}{2\pi i} \frac{\phi(z)\chi(w)}{z-w}.
\end{equation}
This type of normal ordering is usually called the conformal normal ordering. When working with states instead of operators, it is more convenient to introduce the so-called creation-annihilation normal ordering, denoted by $\normc\ \normc$, which places all annihilation operators to the right of creation operators. For the $n$-th mode of a product of two operators, we get
\begin{equation}
\normc \! \phi\, \chi \! \normc_n\,\equiv \sum_{m>-h_\phi}\chi_{n-m}\phi_m+\sum_{m\leq-h_\phi}\phi_m\chi_{m-n}.
\end{equation}
There are other possible prescriptions for normal orderings and the definitions are sometimes not equivalent. Fortunately, these ambiguities will not play a big role in this thesis. All definitions are equivalent in the free boson theory, so the only issue we encounter lies in the ghost theory, where we can choose different ground states. We will fix this ambiguity in section \ref{sec:CFT:ghost}.

\subsection{The energy-momentum tensor}\label{sec:CFT:Bulk:EM tensor}
The energy-momentum tensor of a two-dimensional CFT has some special properties, which are not present in other theories. The conformal symmetry makes it traceless, which in the complex coordinates translates to
\begin{equation}
T_{z\zb}=T_{\zb z}=0.
\end{equation}
The conservation of the energy-momentum tensor, $\del_\mu T^{\mu\nu}=0$, implies that the remaining two components depend only on one of the two coordinates,
\begin{equation}
\delb T_{zz}=0,\quad \del T_{\zb\zb}=0.
\end{equation}
The usual notation for the nonzero components is
\begin{eqnarray}
T(z)&\equiv & T_{zz}(z), \\
\bar T(\zb)&\equiv& T_{\zb\zb}(\zb).
\end{eqnarray}

The OPE of the energy-momentum tensor with itself is fixed by the conformal symmetry to
\begin{equation}\label{TT OPE}
T(z)T(w)\sim\frac{c/2}{(z-w)^4}+\frac{2T(w)}{(z-w)^2}+\frac{\del T(w)}{z-w}
\end{equation}
and similarly for $\bar T$. The constant $c$ that appears in this OPE is called the central charge and it represents a trace anomaly of the energy-momentum tensor in a curved background.

The energy-momentum tensor is not a primary field and under a conformal transformation, it transforms as
\begin{equation}\label{T transformation}
\tilde T(z)=\left(\frac{\del f(z)}{\del z}\right)^2 T(f(z))+\frac{c}{12}S(f(z),z),
\end{equation}
where the function $S(f(z),z)$ is called the Schwarzian derivative,
\begin{equation}\label{Schwarzian derivative}
S(w,z)=\frac{(\del_z w)(\del_z^3 w)-\frac{3}{2}(\del_z^2 w)^2}{(\del_z w)^2}.
\end{equation}
The Schwarzian derivative vanishes for global conformal transformations, which makes $T(z)$ a quasi-primary operator.

The modes of the energy-momentum tensor
\begin{equation}
L_{n}=\oint \frac{dz}{2\pi i} z^{n+1}T(z)
\end{equation}
are called Virasoro generators. Their commutators can be computed using (\ref{commutators contour}) and they satisfy the well known Virasoro algebra:
\begin{equation}\label{Virasoro algebra}
[L_m,L_n]=(m-n)L_{m+n}+\frac{c}{12}(m^3-m)\delta_{m,-n}.
\end{equation}
The operators $L_n$ and $\bar L_n$ generate local conformal transformations. Global conformal transformations are generated by $L_{-1}$, $L_0$ and $L_1$ (and their antiholomorphic counterparts), which form a closed subalgebra of the Virasoro algebra.

The conformal symmetry also fixes the OPE between the energy-momentum tensor and primary fields. For a primary field $\phi(w)$ of weight $h$, we find
\begin{equation}
T(z)\phi(w)\sim\frac{h \phi(w)}{(z-w)^2}+\frac{\del \phi(w)}{z-w}.
\end{equation}
This OPE also gives us commutators between modes of the primary field and $L_m$:
\begin{equation}
[L_m,\phi_n]=((h-1)m-n)\phi_{m+n}.
\end{equation}

\subsection{State space}\label{sec:CFT:Bulk:states}
The states space of a unitary conformal field theory can be decomposed into irreducible representations of the Virasoro algebra $\hh_i$. Since the two copies of the Virasoro algebra commute, the total Hilbert space consists of a sum of products of holomorphic and antiholomorphic representations:
\begin{equation}
\mathcal{H}=\bigoplus_{i,\bar i}\mm_{i\bar i}\ \hh_i\otimes \bar{\hh}_{\bar i},
\end{equation}
where $\mm_{i\bar i}$ are multiplicities of the representations.

Representations of the Virasoro algebra are highest weight representations called Verma modules. No pair of generators commutes in the Virasoro algebra, so it is possible to diagonalize only one generator, conventionally $L_0$. Therefore highest weight states are distinguished only by their conformal weights. The highest weight state in a given representation satisfies
\begin{eqnarray}
L_0|h\ra &=& h|h\ra, \\
L_n |h\ra &=& 0,\quad n>0.
\end{eqnarray}
All other states in this representation, which are called descendant states, can be written as
\begin{equation}\label{Verma module}
L_{-k_1}L_{-k_2}\dots L_{-k_n}|h\ra, \quad k_1\geq\ldots\geq k_n\geq 1.
\end{equation}
The $L_0$ eigenvalue of such state is $h+k_1+\dots+k_n$. The sum $k_1+\dots+k_n$ is called level of the state\footnote{This is the conformal field theory definition of level. Later, we will also define level in string field theory using a different prescription.}. For some purposes, it is useful to introduce the notation
\begin{equation}
L_{-K}\equiv L_{-k_1}\dots L_{-k_n},
\end{equation}
where $K=\{k_1,\dots,k_n\}$ is called multiindex. We also define $|K|=k_1+\dots +k_n$.

In the operator-state correspondence, a highest weight state $|h\ra$ corresponds to a primary fields $\phi(0)$. We also find that
\begin{equation}
(L_{-1})^n|h\ra \leftrightarrow \del^n \phi (0).
\end{equation}
Descendant states which contain higher Virasoro modes correspond to normal ordered products of derivatives of $\phi$ and the energy-momentum tensor. There is no simple formula for these fields, so it is usually more convenient to represent the Virasoro generators using contour integrals.

Some Verma modules are not irreducible. They may contain submodules, which are also representations of the Virasoro algebra. However, it can be shown that these submodules are orthogonal the whole Verma module,
\begin{equation}
\la \psi |\chi\ra=0,
\end{equation}
where $\la \psi |$ is an arbitrary state and $|\chi\ra$ is a state from a submodule. These states are therefore called null states. Null states do not contribute to any observable quantity. To construct an irreducible representation of the Virasoro algebra, we take a quotient of the full Verma module by its submodules.

The structure of an irreducible representation is well captured in the character of the representation:
\begin{eqnarray}\label{character representation}
\chi_{\hh}(\tau)&=&\Tr_{\hh}q^{L_0-c/24}\\
&=&q^{h-c/24}\sum_{n=0}^{\inf} P(n)q^n, \nn
\end{eqnarray}
where $q=e^{2\pi i \tau}$ and $P(n)$ is the number of independent states at level $n$.

Many conformal field theories have a larger symmetry algebra than just the Virasoro algebra. We can decompose the state space of such theoris into irreducible representations of the extended symmetry algebra. Let us denote generators of this symmetry as $W^a$ and $\bar W^a$. Then we can essentially repeat the construction above. The holomorphic part of the state space is spanned by states
\begin{equation}
W_{-k_1}^{a_1}\dots W_{-k_n}^{a_n}|\phi_i\ra,
\end{equation}
where $|\phi_i\ra$ are highest weight states with respect to the extended symmetry. These highest weight states are always primary with respect to the Virasoro algebra, but the opposite is not true. We will encounter these representations in the free boson theory, which has the U(1) symmetry, and in some decompositions of the state space of the ghost theory.

A conformal field theory is called rational if it has finite number of primaries and irrational otherwise.

\subsection{BPZ product and Hermitian product}\label{sec:CFT:Bulk:BPZ}
In a conformal field theory, we can define two different products between states: the BPZ product and the Hermitian product. Let us begin with the BPZ product, which is more natural in CFT and which plays a role in string field theory. The BPZ product of two states $|\phi_1\ra$ and $|\phi_2\ra$ is defined in terms of a two-point function as
\begin{equation}\label{BPZ definition}
\la \phi_1 |\phi_2\ra_{BPZ}=\la \mathcal{I}\circ\phi_1(0)\phi_2(0)\ra,
\end{equation}
where $\mathcal{I}\equiv-\frac{1}{z}$ is the inversion transformation, which maps $0$ to $\inf$. The BPZ product is symmetric and linear in both arguments.
For two primary states with the same conformal weights, we find
\begin{equation}\label{BPZ product primary}
\la \phi_1 |\phi_2\ra_{BPZ}=C_{12},
\end{equation}
where $C_{12}$ is the two-point structure constant defined in (\ref{two-point}).

Evaluation of BPZ products of descendant states is this way would be more complicated. Therefore we introduce the so-called BPZ conjugation, which maps ket states to bra states. In order to match (\ref{BPZ definition}), we define
\begin{equation}\label{BPZ reverse}
{\rm bpz}(W_{-n_1}\dots W_{-n_1}|\phi\ra)=\la \phi|{\rm bpz}(W_{-n_1})\dots{\rm bpz}( W_{-n_1}),
\end{equation}
where $W_n$ stands either for Virasoro or other symmetry generators. The BPZ conjugation of these modes is
\begin{equation}\label{BPZ state}
{\rm bpz}(W_n)\equiv\oint \frac{dz}{2\pi i}z^{n+h_W-1}\mathcal{I}\circ W(z)=(-1)^{n+h_W} W_{-n}
\end{equation}
and
\begin{equation}
\la \phi|=\lim_{z\rar 0} z^{-2h} \la 0|\phi (-1/z).
\end{equation}
If some of the operators are fermionic, the formula (\ref{BPZ reverse}) receives an additional sign $(-1)^{g(g-1)/2}$, where $g$ is the number of anticommuting operators. Now we can easily compute the BPZ product of two descendant states as follows: We remove all $W_n$ operators one by one using their commutators until we get a product of two primary fields, which is given by (\ref{BPZ product primary}).
\\

The definition of the Hermitian product in a conformal field theory is a bit tricky because there is no unique concept of time in the Euclidean space. Therefore it is useful to return to the Lorentzian theory on the cylinder, where the mode expansion of a chiral primary field is
\begin{equation}
W(t,\sigma)=\sum_{n=-\inf}^\inf W_{n}\, e^{-i n (t+\sigma)}.
\end{equation}
A Hermitian field satisfies the condition $W(t,\sigma)^\dagger=W(t,\sigma)$, which gives us Hermitian conjugation of its modes, $W_{n}^\dagger=W_{-n}$.
If we consider a more generic field given by a complex function of Hermitian fields (for example $J^{\pm}=J^1\pm i J^2$ or $e^{ikX}$), we have to take into account complex conjugation of this function as well and we get
\begin{equation}\label{HC state}
W_{n}^\dagger=W^\ast_{-n}.
\end{equation}
Once we know how to conjugate $W_{n}$, we can compute the Hermitian product similarly to the BPZ product.

The main difference between the two conjugations is that the BPZ conjugation is a linear operation, while the Hermitian conjugation is antilinear. This leads to differences in some products of primary states, for example $\la e^{i k_1 X}|e^{i k_2 X}\ra_{BPZ}=\delta_{k_1,-k_2}$, while $\la e^{i k_1 X}|e^{i k_2 X}\ra_{HC}=\delta_{k_1,k_2}$. Otherwise, the two conjugations usually differ only by a sign, which comes from (\ref{BPZ state}).

The BPZ product appears much more often in this thesis and therefore we will drop its subscript and keep the subscript only for the Hermitian product.

\subsection{The Kac determinant}\label{sec:CFT:Bulk:Kac}
A representation of the Virasoro algebra is called unitary if it contains no negative-norm states. Unitarity of Virasoro representations can be analyzed using the so-called Kac determinant. The Kac determinant can be also used to detect the presence of null states, which is, for our purposes, more important than unitarity of representations.

First, we define the so-called Gram matrix with respect to the Hermitian product\footnote{The Gram matrix of BPZ products can be used for identification of null states as well. This matrix is actually more convenient because it has other uses in string field theory. If we consider a Virasoro Verma module, both matrices differ only by an overall sign in blocks at odd levels, so we can easily relate the two matrices.} as
\begin{equation}
G_{ij}=\la i| j\ra_{HC},
\end{equation}
where $|i\ra$ are states in the Verma module (\ref{Verma module}). For simplicity, we assume that the product of the two highest weight states is normalized as $\la h|h\ra_{HC}=1$.

The Gram matrix is obviously Hermitian, therefore it can be diagonalized and it has real eigenvalues. Its matrix elements can be nonzero only if both states have the same level and therefore the matrix is block diagonal. We denote a block at level $L$ as $G^{(L)}$.

The determinant of the Gram matrix, which was found by Kac, is given by
\begin{equation}\label{Kac determinat}
\det G^{(L)}=\alpha_L\prod_{\substack{r,s\geq1 \\ rs\leq L}}(h-h_{r,s}(c))^{P(L-rs)},
\end{equation}
where $P(k)$ is the number of partitions of the integer $k$ and $\alpha_L$ is a positive integer given by
\begin{equation}
\alpha_L=\prod_{\substack{r,s\geq1 \\ rs\leq L}}\left((2r)^s s!\right)^{P(L-rs)-P(L-r(s+1))}.
\end{equation}
We are usually interested only in the functions $h_{r,s}(c)$, which determine zeros of the Kac determinant. If the conformal weight $h$ matches $h_{r,s}(c)$ for some $r$ and $s$, then we know that this Verma module is reducible and that there are null states at levels $L\geq rs$. The number of null states at a given level may be more difficult to determine because there can be more overlapping submodules.

The functions $h_{r,s}(c)$ can be expressed in several different forms. In terms of the central charge, they are given by
\begin{eqnarray}\label{Kac roots 1}
h_{r,s}(c)&=&\frac{1}{24}(c-1)+\frac{1}{4}(r\alpha_+ + s\alpha_-)^2,\\
\nn \alpha_\pm&=&\frac{\sqrt{1-c}\pm\sqrt{25-c}}{\sqrt{24}}.
\end{eqnarray}
A different parametrization is
\begin{eqnarray}\label{Kac roots 2}
h_{r,s}(m)&=&\frac{((m+1)r-ms)^2-1}{4m(m+1)},\\
c&=&1-\frac{6}{m(m+1)}. \nn
\end{eqnarray}
This parametrization is useful for description of the unitary Virasoro minimal models, where $m$ in an integer greater than 2.

Yet another expressions for the roots of the Kac determinant is
\begin{eqnarray}\label{Kac roots 3}
h_{r,s}(p,q)&=&\frac{(qr-ps)^2-(p-q)^2}{4pq},\\
c&=&1-\frac{6(p-q)^2}{pq} \nn.
\end{eqnarray}
This parametrization is convenient for the nonunitary Virasoro minimal models, where $(p,q)$ are coprime integers. We can choose $p<q$ without loss of generality. When $|q-p|=1$, this parametrization reduces to (\ref{Kac roots 2}).

\subsection{Correlators}\label{sec:CFT:Bulk:correlators}
Correlation functions in a conformal field theory are strongly restricted by the conformal symmetry. They must be invariant under the global conformal maps (\ref{conformal map glob}). This fully fixes two-point functions of (quasi-)primary operators to
\begin{equation}\label{two-point}
\left\la \phi_1(z_1,\zb_1)\phi_2(z_2,\zb_2) \right\ra=\delta_{h_1h_2}\delta_{\bar h_1\bar h_2}\frac{C_{12}}{z_{12}^{2h_1}\zb_{12}^{2\bar h_1}},
\end{equation}
where $C_{12}$ are two-point structure constants and we use the usual notation $z_{ij}=z_i-z_j$.

Three-point functions of primary operators are also fully fixed to
\begin{equation}\label{three-point}
\left\la \phi_1(z_1,\zb_1)\phi_2(z_2,\zb_2)\phi_3(z_3,\zb_3) \right\ra=C_{123}\prod_{\stackrel{i,j=1}{i<j}}^3
z_{ij}^{h-2h_i-2h_j} \bar z_{ij}^{\bar h-2\bar h_i-2\bar h_j},
\end{equation}
where $C_{123}$ are three-point structure constants and $h=h_1+h_2+h_3$. If one of the three operators equals to the identity, we get $C_{ij \Id}=C_{ij}$.

A four-point correlator of primary fields is the first one which not fully fixed by the conformal symmetry. Using the 4 points, it is possible construct an invariant cross-ratio $\eta=\frac{(z_1-z_2)(z_3-z_4)}{(z_1-z_3)(z_2-z_4)}$ and the correlator includes an arbitrary function of this variable,
\begin{equation}\label{four-point}
\left\la \phi_1(z_1,\zb_1)\phi_2(z_2,\zb_2)\phi_3(z_3,\zb_3) \phi_4(z_4,\zb_4)\right\ra=f(\eta,\bar\eta)\prod_{\stackrel{i,j=1}{i<j}}^4
z_{ij}^{h/3-h_i-h_j}\bar z_{ij}^{\bar h/3-\bar h_i-\bar h_j},
\end{equation}
where $h=\sum_{i=1}^4 h_i$. A similar pattern holds for correlators with more insertions. An $n$-point correlator is a function of $n-3$ independent cross-ratios.

The formulas above give us correlation functions only for primary and quasi-primary operators, but we often need correlators of descendant fields. These are usually computed using contour deformations. Consider the following correlator:
\begin{equation}
\la (L_{-k}\chi)(z,\zb)\ \phi_1(z_1,\zb_1)\dots \phi_n(z_n,\zb_n)\ra.
\end{equation}
We can express $L_{-k}$ as $\oint_z \frac{dw}{2\pi i} v_k(w)T(w)$, where $v_k(w)=(w-z)^{-k+1}$, and deform the contour around $z$ to contours around the other insertions (and possibly around infinity, which however does not contribute in this case)
\begin{eqnarray}\label{contour manipulation}
&&\la (L_{-k}\chi)(z,\zb)\ \phi_1(z_1,\zb_1)\dots \phi_n(z_n,\zb_n)\ra \nn\\
&&\qquad=\oint_{z} \frac{dw}{2\pi i} v_k(w)\la T(w)\  \chi(z,\zb)\ \phi_1(z_1,\zb_1)\dots \phi_n(z_n,\zb_n)\ra \\
&&\qquad=-\sum_{i=1}^n \oint_{z_i} \frac{dw}{2\pi i} v_k(w) \la T(w)\  \chi(z,\zb)\ \phi_1(z_1,\zb_1)\dots \phi_n(z_n,\zb_n)\ra.\nn
\end{eqnarray}
If the operators $\phi_i$ are primary, we can use the OPE between $T(w)$ and $\phi_i(z_i,\zb_i)$ to derive
\begin{eqnarray}\label{Ward identity}
&&\la (L_{-k}\chi)(z,\zb)\ \phi_1(z_1,\zb_1)\dots \phi_n(z_n,\zb_n)\ra \\
&&\qquad=\sum_{i=1}^n\left(\frac{(k-1)h_i}{(z_i-z)^k}-\frac{1}{(z_i-z)^{k-1}}\del_{z_i} \right)\la \chi(z,\zb)\ \phi_1(z_1,\zb_1)\dots \phi_n(z_n,\zb_n)\ra\nn.
\end{eqnarray}
For $\chi=\Id$ and $k=2$, this expression reproduces the famous conformal Ward identity.

Similar expressions can be derived for other chiral operators, particularly for the symmetry generators $W^a$.
\\

Correlation functions of primary fields are closely related to the OPE. We can write the OPE (\ref{OPE general}) for two primary operators as
\begin{equation}\label{OPE primary}
\phi_i(z_1,\zb_1)\phi_j(z_2,\zb_2)=\sum_{k} \Cbulk ijk \phi_k(z_2,\zb_2) z_{12}^{h_{ijk}}\zb_{12}^{\bar h_{ijk}}+{\rm descendants},
\end{equation}
where $k$ now runs only over primary operators. The structure constants for primary operators $\Cbulk ijk$ fully determine the OPE. Including the descendant fields, the full OPE can be written as
\begin{eqnarray}\label{OPE beta}
\phi_i(z_1,\zb_1)\phi_j(z_2,\zb_2)&=&\sum_{k,N,M} \Cbulk ijk \beta_{ij}^{k,N}\bar\beta_{ij}^{k,M}(L_{-N}\bar L_{-M} \phi_k)(z_2,\zb_2) \nn  \\
&\times & z_{12}^{h_k+|N|-h_i-h_j}\zb_{12}^{\bar h_k+|M|-\bar h_i-\bar h_j},
\end{eqnarray}
where $\beta_{ij}^{k,N}$ and $\bar \beta_{ij}^{k,M}$ are constants that depend only on the cental charge and the three conformal weights. They can be computed independently of the structure constants or other details of the particular CFT.

The conditions that determine which conformal families appear in (\ref{OPE primary}) are called fusion rules. They can be schematically written as
\begin{equation}\label{fusion rules}
\phi_i\times\phi_j=\sum_k N_{ij}^{\ \ k} \phi_k,
\end{equation}
where $N_{ij}^{\ \ k}$ are multiplicities of the representations labeled by $k$.

By comparing (\ref{OPE primary}) and the three-point function (\ref{three-point}), we find a relation between the structure constants:
\begin{equation}
C_{ijk}=\sum_l \Cbulk ijl C_{lk}.
\end{equation}
We can often choose an orthonormal basis of primary fields with $C_{ij}=\delta_{ij}$, for which we find $C_{ijk}=\Cbulk ijk$.

Next, we take a closer look at four-point functions. Using global conformal transformations, we can always fix positions of three insertions to $0,1$ and $\inf$. Then we define
\begin{equation}
G_{14}^{23}(\eta,\bar \eta)=\lim_{z_1,\zb_1\rar\inf}z_1^{2h_1}\zb_1^{2\bar h_1}\la\phi_1(z_1,\zb_1)\phi_2(1,1)\phi_3(\eta,\bar \eta)\phi_4(0,0)\ra.
\end{equation}
This correlator can be computed by replacing $\phi_1$ and $\phi_2$ by their OPE. Then we can write the function $G_{14}^{23}$ as\footnote{For simplicity, we assume that the constants $C_{ij}$ are diagonal.}
\begin{equation}\label{four point expansion}
G_{14}^{23}(\eta,\bar \eta)=\sum_p \Cbulk 12p \Cbulk 34p C_{pp}\ \ff_{\!\;14}^{23}(p|\eta)  \mathcal{\bar F}_{14}^{23}(p|\bar\eta),
\end{equation}
where $\ff$ and $\mathcal{\bar F}$ are called conformal blocks. They can be expressed as
\begin{equation}
\ff_{\;\!14}^{23}(p|\eta)=\eta^{h_p-h_3-h_4}\sum_{K,M} \beta_{34}^{p,K}\beta_{12}^{p,M}\eta^{|K|}G_{KM},
\end{equation}
where $G_{KM}$ is the Gram matrix of BPZ products. A similar expression holds for $\mathcal{\bar F}$. Like the $\beta$ coefficients, the conformal blocks depend only on the conformal weights and the central charge.

A four-point function of bosonic operators does not depend on their ordering. Therefore we can permute the operators and then return their arguments to the canonical form using conformal transformations. In this way, we can derive several conditions for $G_{14}^{23}$, for example
\begin{eqnarray}\label{four point crossing sym}
G_{14}^{23}(\eta,\bar \eta) &=&\eta^{h_1+h_2-h_3-h_4}\bar \eta^{\bar h_1+\bar h_2-\bar h_3-\bar h_4} G_{43}^{12}(1-\eta,1-\bar \eta), \\
G_{14}^{23}(\eta,\bar \eta) &=&\eta^{h_1-h_2-h_3-h_4}\bar \eta^{\bar h_1-\bar h_2-\bar h_3-\bar h_4} G_{14}^{32}(1/\eta,1/\bar \eta).
\end{eqnarray}

When expanded in terms of conformal blocks, the left hand side and the right hand side of (\ref{four point crossing sym}) describe the same correlator computed using different order of the OPEs. The conformal blocks for $\eta$ and $1-\eta$ are related by so-called F-matrices:
\begin{equation}\label{Fmatrix definition}
\ff_{\; il}^{jk}(p|\eta)=\sum_q \eta^{h_i+h_j-h_k-h_l}\Fmatrix pqijkl \ff_{\, lk}^{ij}(q|1-\eta).
\end{equation}

By combining (\ref{four point crossing sym}), (\ref{four point expansion}) and (\ref{Fmatrix definition}), we obtain a consistency relation for bulk structure constants
\begin{equation}\label{sewing constr bulk 4}
\Cbulk ijp \Cbulk klp C_{pp} \Fmatrix pqijkl = \Cbulk jkq \Cbulk ilq C_{qq} \Fmatrix qpilkj.
\end{equation}
This condition imposes nontrivial constraints on the structure constants and in some cases (for example for the Virasoro minimal models), it is strong enough to find a solution for the structure constants in terms of the F-matrices.

Similarly to F-matrices, we also define braiding matrices by
\begin{equation}
\ff_{\; il}^{jk}(p|\eta)=\sum_q \eta^{h_i-h_j-h_k-h_l}\BRmatrix pqijkl \ff_{\,il}^{k\!j}\left(q\left|\frac{1}{\eta}\right)\right.
\end{equation}
These are useful for deriving similar conditions in a boundary conformal field theory.

\subsection{Modular invariance}\label{sec:CFT:Bulk:modular}
In subsection \ref{sec:CFT:Bulk:states}, we mentioned that the Hilbert space of a CFT can be decomposed into irreducible representations of the Virasoro algebra,
\begin{equation}
\mathcal{H}=\bigoplus_{i,\bar i} \mathcal{M}_{i\bar i}\ \hh_i\otimes \bar{\hh}_{\bar i},
\end{equation}
but we have not set any rules for the multiplicities $\mathcal{M}_{i\bar i}$. To do so, we have to define CFT on a torus instead of the complex plane.

A two-dimensional torus is specified by two independent vectors, which can be described by two complex numbers $\omega_1$ and $\omega_2$. In conformal field theory, the overall scale or orientation of the torus does not matter, so the only relevant parameter, which is called the modular parameter, is $\tau=\omega_2/\omega_1$. The same torus can be described by many different parameters. It turns out that equivalent tori are related by SL(2,$\mathbb{Z}$)/$\mathbb{Z}_2$ transformations
\begin{equation}\label{modular transformation}
\tau\rar \frac{a\tau+b}{c\tau+d},
\end{equation}
where $a,b,c,d\in \mathbb{Z}$ and $ad-bc=1$. The group is generated by two transformations, which are usually denoted as $\mathcal{T}$ and $\mathcal{S}$:
\begin{eqnarray}
\mathcal{T}&:&\tau\rar \tau+1,\\
\mathcal{S}&:&\tau\rar -\frac{1}{\tau}.
\end{eqnarray}

The basic quantity to study on a torus is called the partition function and it is defined as
\begin{equation}
Z(\tau)=\Tr_\mathcal{H} \left(q^{L_0-c/24} \bar{q}^{\bar L_0-c/24}\right),
\end{equation}
where
\begin{equation}
q=e^{2\pi i \tau},\quad \bar q=e^{-2\pi i \bar\tau}.
\end{equation}
The partition function can be expressed in terms of irreducible characters of Virasoro representations (\ref{character representation}),
\begin{equation}
Z(\tau)=\sum_{i,\bar i} \mathcal{M}_{i\bar i}\ \chi_i(\tau)\bar\chi_{\bar i}(\bar \tau).
\end{equation}

The partition function must be invariants with respect to the modular transformations (\ref{modular transformation}). If we write the action of the basic modular transformations on characters as
\begin{equation}
\chi_i(\tau+1)=\sum_j T_i^{\ j}\chi_j(\tau)
\end{equation}
and
\begin{equation}\label{S matrix def}
\chi_i(-1/\tau)=\sum_j S_i^{\ j}\chi_j(\tau),
\end{equation}
the modular invariance is equivalent to
\begin{eqnarray}
\mathcal{M}_{i\bar i}&=&\sum_{j,\bar j}T_i^{\ j} \bar T_{\bar i}^{\ \bar j} \mathcal{M}_{j\bar j},\\
\mathcal{M}_{i\bar i}&=&\sum_{j,\bar j}S_i^{\ j} \bar S_{\bar i}^{\ \bar j} \mathcal{M}_{j\bar j}.
\end{eqnarray}
The matrix $T$ is always diagonal for Virasoro characters,
\begin{equation}
T_i^{\ j}=\delta_i^{\ j} e^{2\pi i(h_i-c/24)},
\end{equation}
so the first condition is easy to solve. It determines that all primary operators must have integer spin, i.e. $h-\bar h\in \mathbb{Z}$. The second equation gives us more complicated restrictions on the spectrum. In rational CFTs, which have finite number of primaries, these two conditions allow us to classify all modular invariant spectra.

The modular invariance also allows us to derive the Verlinde formula, which connects the $S$-matrix and the fusion numbers $N_{ij}^{\ \ k}$:
\begin{equation}\label{Verlinde}
\sum_k S_k^{\ p} N_{ij}^{\ \ k}=\frac{S_i^{\ p} S_j^{\ p}}{S_0^{\ p}}.
\end{equation}

\section{Boundary CFT}\label{sec:CFT:Boundary}
In the previous section, we have discussed the bulk conformal field theory, which, in the context of string theory, describes closed strings. However, this thesis is about open string field theory and open strings are described by the boundary conformal field theory (BCFT), which we review in this section.

\subsection{CFT on the upper half-plane}\label{sec:CFT:Boundary:UHP}
A boundary conformal field theory is a CFT defined on a surface with a boundary. For simplicity, we consider theories on the upper half-plane, which has the real axis as its boundary. A boundary conformal field theory is usually constructed by restriction of a parent bulk CFT.

The introduction of a boundary clearly breaks some symmetries of the theory, most notably translations perpendicular to the boundary. Out of the full conformal group, only transformations that respect the boundary survive. These transformations must satisfy
\begin{equation}\label{boundary preservation}
f(x)\in \mathbb{R}\quad {\rm for} \quad x\in \mathbb{R}.
\end{equation}
Globally defined transformations with this property form the group SL(2,$\mathbb{R}$)/$\mathbb{Z}_2$.

The condition (\ref{boundary preservation}) also implies that there is no energy flow across the boundary. In terms of the energy-momentum tensor, we get
\begin{equation}\label{gluing T}
T(z)=\bar T(\zb)|_{z=\zb}.
\end{equation}
This relation is known as the gluing condition for the energy-momentum tensor. It means that the holomorphic and anti-holomorphic parts of the energy-momentum tensor are no longer independent and therefore we can define only one set of Virasoro generators:
\begin{equation}
L_n=\int_{\mathcal{C}_U}\frac{dz}{2\pi i} z^{n+1}T(z)-\int_{\mathcal{C}_L}\frac{d\zb}{2\pi i} \zb^{n+1}\bar T(\zb),
\end{equation}
where $\mathcal{C}_U$ and $\mathcal{C}_L$ are half-circles in the upper and lower half-plane respectively.

This expression is cumbersome to work with, so we use the so-called doubling trick and extend the holomorphic part of the energy-momentum tensor to the whole complex plane using the prescription
\begin{equation}\label{doubling trick T}
T(z)=\bar T(\bar z'),\quad \mbox{Im}(z)\leq 0,\quad z'=\zb.
\end{equation}
This trick allows us to trade a theory of a chiral and anti-chiral field on the half-plane for a theory of a chiral field on the full complex plane. There we can define Virasoro generators using the usual formula
\begin{equation}
L_n=\int\frac{dz}{2\pi i} z^{n+1}T(z).
\end{equation}

The OPE between the energy-momentum tensor and a bulk primary operator $\phi(w,\bar w)$ is given by
\begin{eqnarray}
T(z)\phi(w,\bar w)&\sim&\frac{h}{(z-w)^2}\phi(w,\bar w)+\frac{1}{z-w}\del\phi(w,\bar w) \nn\\
&+&\frac{\bar h}{(z-\bar w)^2}\phi(w,\bar w)+\frac{1}{z-\bar w}\delb\phi(w,\bar w).
\end{eqnarray}
That means that the bulk primary $\phi(w,\bar w)$ essentially splits into a product of two chiral operators $\phi_L(w)\phi_R(\bar w)$ with weights $h$ and $\bar h$.

If the theory contains other chiral symmetry generators $W^a(z)$ and $\bar W^a(\zb)$, we may impose gluing conditions for them as well,
\begin{equation}\label{gluing W}
W^a(z)=\Omega(\bar W^b(\zb))|_{z=\zb}.
\end{equation}
These gluing conditions admit a local automorphism $\Omega$, which leaves the energy-momentum tensor invariant, $\Omega(T)=T$. Following (\ref{doubling trick T}), we can extend the fields $W^a(z)$ to the whole complex plane using the doubling trick as
\begin{equation}\label{doubling trick W}
W^a(z)=\Omega(\bar W^b(\bar z')),\quad \mbox{Im}(z)\leq 0,\quad z'=\zb.
\end{equation}
The gluing automorphism $\Omega$ has to be taken into account also when we use the doubling trick on bulk operators, so we have $\phi(w,\bar w)\rar \phi_L(w)\Omega (\phi_R(\bar w))$.

We emphasize that while imposing gluing condition (\ref{gluing T}) is necessary in any BCFT, imposing (\ref{gluing W}) is not. Conformal boundary conditions that (at least partially) break the extended symmetry are difficult to find because the corresponding BCFT is usually irrational with respect to the remaining symmetry. However, we will encounter string field theory solutions that correspond to such boundary conditions.

\subsection{Boundary operators}\label{sec:CFT:Boundary:operators}
Boundary conformal field theory contains a new type of fields, which are called boundary fields. They are localized on the boundary and, in general, they cannot be moved into the interior of the upper half-pane. Since the boundary is parameterized only by one coordinate $x$, we denote them as $\psi(x)$.

We distinguish two types of boundary fields. Ordinary boundary fields appear when a bulk operator approaches the boundary (see the OPE (\ref{Bulk-boundary OPE})). Such operators have the same boundary conditions on both sides and, in the context of string theory, they correspond to open strings living on a single D-brane.

The second type of boundary fields corresponds to strings stretched between two different D-branes. When we map a strip with two different boundary conditions $\alpha$ and $\beta$ to the UHP, we find a discontinuity: The boundary condition $\alpha$ applies for $x<0$ and the boundary condition $\beta$ for $x>0$. The discontinuity is caused by a so-called boundary condition changing operator, which we denote $\psi^{(\alpha\beta)}(x)$. Of course, when the boundary conditions are identical, $\alpha=\beta$, a boundary condition changing operator reduces to a normal boundary operator.

The state space of a boundary conformal field theory decomposes into irreducible representations $\hh_i$ of the remaining symmetry, which is either a single copy of the Virasoro algebra or some extended algebra. In general, for two different boundary conditions $\alpha,\ \beta$, the state space $\hh_{\alpha\beta}$ is given by
\begin{equation}
\hh_{\alpha\beta}=\bigoplus_i n_{\alpha\beta}^{\ \ i}\hh_i,
\end{equation}
where $n_{\alpha\beta}^{\ \ i}$ are multiplicities of the irreducible representations.

\subsection{OPE and correlators}\label{sec:CFT:Boundary:OPE}
Boundary conformal field theories include three types of OPE. The OPE of bulk operators is the same as (\ref{OPE general}). Next, we have bulk-boundary OPE, which indicates singularities between the chiral and anti-chiral part of a bulk operator and which produces boundary operators:
\begin{equation}\label{Bulk-boundary OPE}
\phi_i(z,\zb)\sim\sum_k \Cbb \alpha ik\ \psi_k^{(\alpha\alpha)}(x)\ (2y)^{h_k-h_i-\bar h_i},
\end{equation}
where we write the complex coordinate as $z=x+iy$ and the numbers $\Cbb \alpha ik$ are called bulk-boundary structure constants.

Finally, there can be OPE between two boundary operators,
\begin{equation}\label{Boundary OPE}
\psi_i^{(\alpha\beta)}(x_1)\psi_j^{(\beta\gamma)}(x_2)\sim \sum_k \Cbound \alpha\beta\gamma ijk \ \psi_k^{(\alpha\gamma)}(x_2)\ (x_1-x_2)^{h_k-h_i-h_j},
\end{equation}
where $\Cbound \alpha\beta\gamma ijk$ are boundary structure constants. This OPE makes sense only if the right boundary condition of the first operator matches the left boundary condition of the second operator. The boundary OPE has similar properties as the bulk OPE, including fusion rules and expansion to conformal families. However, boundary operators are often not mutually local, that is
\begin{equation}
\psi_1(x_1)\psi_2(x_2)\neq\psi_2(x_2)\psi_1(x_1).
\end{equation}
We can see that from the $(x_1-x_2)$ term in (\ref{Boundary OPE}), which has generally non-integer power and therefore it produces a complex phase when we exchange the arguments. Therefore we typically consider this OPE only for $x_1>x_2$. We do not encounter such problems in the bulk OPE because the phases from $z_{12}$ and $\zb_{12}$ usually cancel each other.

Correlators in a boundary conformal field theory can include both bulk and boundary operators:
\begin{equation}
\la \psi_1(x_1)\dots \psi_n(x_n)\phi_1(z_1,\zb_1)\dots \phi_m(z_m,\zb_m)\ra^{(\alpha)}.
\end{equation}
The conformal symmetry restricts the possible coordinate dependence of correlators of (quasi-)primary operators similarly to (\ref{three-point}) and (\ref{four-point}). Correlators with descendant fields can be simplified using contour manipulations as well.

We will be mostly interested in two types of correlators, in three-point functions of boundary primary fields,
\begin{equation}\label{three-point boundary}
\la\psi_i^{(\alpha\beta)}(x_1)\psi_j^{(\beta\gamma)}(x_2)\psi_k^{(\gamma\alpha)}(x_3)\ra^{(\alpha)}=\frac{\CboundL \alpha\beta\gamma ijk}{x_{12}^{h_i+h_j-h_k} x_{23}^{h_j+h_k-h_i}x_{13}^{h_i+h_k-h_j}},
\end{equation}
and in two-point functions of one bulk and one boundary primary fields,
\begin{equation}\label{two-point bulk boundary}
\la\phi_i(z,\zb)\psi_j^{(\alpha\alpha)}(x)\ra^{(\alpha)}=\frac{\CbbL \alpha ij}{|z-\zb|^{h_i+\bar h_i-h_j} (z-x)^{h_j+h_i-\bar h_i}(\zb-x)^{h_j+\bar h_i-h_i}}.
\end{equation}
The relations between structure constants in the OPEs and in the correlation functions are given by
\begin{equation}
\CboundL \alpha\beta\gamma ijk=\Cbound \alpha\beta\gamma ijk \Cbound \alpha\gamma\alpha kk\Id \la\Id\ra^{(\alpha)}
\end{equation}
and by
\begin{equation}
\CbbL \alpha ij=\Cbb \alpha ij \Cbound \alpha\alpha\alpha jj\Id \la\Id\ra^{(\alpha)},
\end{equation}
where $\la\Id\ra^{(\alpha)}$ is the normalization of the empty correlator, which, in unitary theories, coincides with the $g$-function of the corresponding boundary state.

\subsection{Boundary states}\label{sec:CFT:Boundary:BS}

The information about boundary conditions in a given BCFT can be captured in a so-called boundary state. A boundary state, denoted as $\ww \alpha\rra$, is an element of the state space of the bulk CFT. Boundary states are usually defined in terms of 1-point functions of spinless bulk primaries on the UHP:
\begin{equation}\label{boundary state def}
|z-\zb|^{2h_i}\la \phi_i(z,\zb)\ra^{(\alpha)}=\la \phi_i\ww \alpha \rra.
\end{equation}
Boundary states must incorporate gluing conditions on the boundary. For the gluing condition $T(z)=\bar T(\zb)|_{z=\zb}$, we consider the following correlator on the UHP:
\begin{equation}
\la T(z)-\bar T(\zb)\ra^{(\alpha)}.
\end{equation}
Using the conformal transformation $\xi=\frac{1+iz}{1-iz}$, we can map the UHP to the unit disk. The new correlator can be interpreted as a closed string amplitude
\begin{equation}
\la 0|\left[ \left(\frac{d\xi}{dz}\right)^2 T(\xi)-\left(\frac{d\bar\xi}{d\zb}\right)^2 \bar T(\bar \xi)\right]\ww\alpha\rra.
\end{equation}
We find that $\xi=\bar \xi^{-1}$ on the unit circle and, by expanding $T$ and $\bar T$ in powers of $\xi$, we get
\begin{equation}\label{Ishibashi conditions}
(L_n-\bar L_{-n})\ww\alpha\rra=0.
\end{equation}
Similar relations can be derived also for the more general gluing conditions (\ref{gluing W}):
\begin{equation}\label{Ishibashi conditions W}
(W_n^a-(-1)^{h_W}\Omega(\bar W_{-n}^b))\ww\alpha\rra=0.
\end{equation}
These equations are called Ishibashi conditions.

Solutions of the Ishibashi conditions, denoted as $|i\rra$, are called Ishibashi states and they are labeled by spinless bulk primary operators. Ishibashi states form a basis for boundary states, therefore we can write
\begin{equation}
\ww\alpha\rra=\sum_i B_\alpha^{i}|i\rra.
\end{equation}
Assuming that bulk primaries are normalized as $\la \phi_i|\phi_j\ra=\delta_{ij}$, we get
\begin{equation}
|z-\zb|^{2h_i}\la \phi_i(z,\zb)\ra^{(\alpha)}=B_\alpha^i.
\end{equation}
By comparing this equation with (\ref{two-point bulk boundary}), we find
\begin{equation}
B_\alpha^i=\CbbL \alpha i\Id.
\end{equation}

The Ishibashi conditions (\ref{Ishibashi conditions}) are easy to solve and the solution for Ishibashi states is given by
\begin{equation}
|i\rra=\sum_{I,J}G^{IJ}(-1)^{|I|}L_{-I}\bar L_{-J}|\phi_i\ra
\end{equation}
where $G^{IJ}$ is the inverse Gram matrix in the Verma module with the given central charge and conformal weight. A similar solution can be found for the conditions (\ref{gluing W}), but it is more complicated because it includes the gluing automorphism $\Omega$.

\subsection{Nonlinear constraints}\label{sec:CFT:Boundary:Constraints}
Boundary conformal field theories are subject to several consistency conditions, which, in some rational BCFTs, make it possible to find explicit solutions for boundary states and structure constants.

We start with the so-called Cardy condition, which involves only boundary states. Consider a cylinder of length $L$ and circumference $T$ with boundary conditions $\alpha$ and $\beta$. We define a partition function on the cylinder as
\begin{equation}
Z_{\alpha\beta}(q)=\Tr_{\hh_{\alpha\beta}} q^{L_0-c/24}=\sum_i n\indices{_\alpha_\beta^i}\chi_i(q),
\end{equation}
where $q=e^{2\pi i \tau}$ and $\tau=\frac{iT}{2L}$. In string theory, this formula can be interpreted as a loop diagram of an open string. But the same partition function can be also interpreted as a closed string amplitude with the Hamiltonian $H=L_0+\bar L_0-\frac{c}{12}$:
\begin{equation}
Z_{\alpha\beta}(\tilde q)=\lla \alpha\ww \tilde q^{\frac{1}{2}\left(L_0+\bar L_0-\frac{c}{12}\right)}\ww \beta\rra=\sum_i B_\alpha^i B_\beta^i \chi_i(\tilde q),
\end{equation}
where $\tilde q=e^{-\frac{2\pi i }{\tau}}$.

The two partition functions can be compared using the modular transformation (\ref{S matrix def}). Assuming linear independence of the characters, we get
\begin{equation}\label{Cardy condition}
\sum_j n\indices{_\alpha_\beta^j} S_j^{\ i}=B_\alpha^i B_\beta^i.
\end{equation}
This equation is called the Cardy condition. In theories with diagonal spectrum, $\mathcal{M}_{i\bar i}=\delta_{i\bar i}$, this condition admits a simple solution. We consider a set of boundary states labeled by Virasoro representations, for which $n\indices{_\alpha_\beta^i}$ equal to the fusion numbers $N\indices{_\alpha_\beta^i}$. Then the Verlinde formula (\ref{Verlinde}) leads to
\begin{equation}
\frac{S_\alpha^{\ i} S_\beta^{\ i}}{S_\Id^{\ i}}=B_\alpha^i B_\beta^i.
\end{equation}
This equation is easily solved by so-called Cardy boundary states:
\begin{equation}\label{Cardy BS}
B_\alpha^{\ i}=\frac{S_\alpha^{\ i}}{\sqrt{S_\Id^{\ i}}}.
\end{equation}

Other constraints, which involve various combinations of structure constants, arise from correlators on the half-plane. These conditions are called sewing constraints and they were originally derived by Lewellen \cite{LewellenSewing} (see also \cite{MooreSeiberg}\cite{BCRational}\cite{Runkel2}\cite{Runkel3} for more detailed discussion and generalizations).
These sewing constraints can be derived similarly to (\ref{sewing constr bulk 4}). We take a given correlator of bulk and boundary fields and express it in terms of conformal blocks. By taking the OPE in two different orders, we find two different expressions for the correlator, which can be related using the F-matrices or the braiding matrices. Their comparison then leads to the desired sewing relations.

A correlator of four boundary fields $\la \psi_i^{(\alpha\beta)}(x_1)\; \psi_j^{(\beta\gamma)}(x_2)\; \psi_k^{(\gamma\delta)}(x_3)\; \psi_l^{(\delta\alpha)}(x_4)\ra$ allows us to derive a sewing constraint for boundary structure constants:
\begin{equation}\label{sewing constr bound 4}
\Cbound \beta\gamma\delta jkq \; \Cbound \alpha\beta\delta iql \; \Cbound \alpha\delta\alpha ll\Id
=\sum_p \Cbound \alpha\beta\gamma ijp \; \Cbound \gamma\delta\alpha klp \; \Cbound \alpha\gamma\alpha pp\Id \; \Fmatrix pqijkl.
\end{equation}
A correlator of one bulk field and two boundary fields $\la \phi_i(z) \; \psi_p^{(\alpha\beta)}(x_1) \; \psi_q^{(\beta\alpha)}(x_2)\ra$ leads to a constraint for boundary and bulk-boundary structure constants:
\begin{eqnarray}\label{sewing constr BB 12}
\Cbb \beta il \; \Cbound \alpha\beta\beta plq \; \Cbound \alpha\beta\alpha qq\Id &=&
\sum_{k,m} \Cbb \alpha ik \; \Cbound \alpha\beta\alpha pqk \; \Cbound \alpha\alpha\alpha kk\Id  \\
&\times & e^{i\pi(2h_m+\frac{1}{2}h_k-h_p-h_q-2h_i+\frac{1}{2}h_l)} \; \Fmatrix kmpqii \Fmatrix mlpiiq . \nn
\end{eqnarray}
Similarly, from a correlator involving two bulk fields and one boundary field $\la \phi_i(z_1) \; \phi_j(z_2) \; \psi_k^{(\alpha\alpha)}(x)\ra$, we find
\begin{eqnarray}\label{sewing constr BB 21}
\Cbb \alpha iq \; \Cbb \alpha jt \; \Cbound \alpha\alpha\alpha qtk &=&
e^{i\frac{\pi}{2}(h_t-h_k-h_q-2h_j)} \sum_{p,r} \Cbulk ijp \; \Cbb \alpha pk \; e^{i\pi h_r} \nn\\
&&\times  \Fmatrix prpkji \Fmatrix pqijri \Fmatrix rtqjjk.
\end{eqnarray}
By choosing $k=\Id$, this equation simplifies and we find
\begin{equation}\label{sewing constr BB 20}
\Cbb \alpha iq \; \Cbb \alpha jq \; \Cbound \alpha\alpha\alpha qq\Id=\sum_p \Cbulk ijp \; \Cbb \alpha p\Id \; \Fmatrix pqjiij.
\end{equation}

\section{Free boson theory}\label{sec:CFT:FB}
In this and the following sections, we move from a general description of conformal field theory to description of particular theories that appear in this thesis. We start with the free boson theory. Our conventions in this theory follows the book by Polchinski \cite{Polchinski} with $\alpha'=1$.

\subsection{Bulk theory}\label{sec:CFT:FB:bulk}
We start with description of a single free boson $X$, which has the action
\begin{equation}\label{free boson action}
S=\frac{1}{2\pi}\int d^2z\ \del X \delb X.
\end{equation}
The corresponding equation of motion is
\begin{equation}
\del \delb X (z,\zb)=0.
\end{equation}
This equation is easily solved by writing $X(z,\zb)$ as a sum of holomorphic and antiholomorphic part
\begin{equation}
X(z,\zb)=X_L(z)+X_R(\zb).
\end{equation}

The OPE of the $X$ field with itself is given by
\begin{equation}
X(z,\zb)X(w,\bar w)=-\frac{1}{2}\ln |z-w|^2.
\end{equation}
This indicates that $X(z,\zb)$ is not a proper scalar field of zero weight. Nevertheless, its derivative $\del X(z)$ is a weight one current with the OPE
\begin{equation}
\del X(z)\del X(w)=-\frac{1}{2}\frac{1}{(z-w)^2}
\end{equation}
and similarly for the antiholomorphic derivative $\delb X(\zb)$.

In order to write down the Laurent expansion of the $X$ field, we have to first specify its boundary conditions. In the bulk theory, we usually consider periodic conditions
\begin{equation}\label{periodicity circle}
X(\tau,\sigma+2\pi)=X(\tau,\sigma)+2 \pi R w,\quad w \in \mathbb{Z},
\end{equation}
where the integer $w$ is called the winding number. This condition means that the free boson lives in a compact space, namely on a circle of radius $R$. This periodic condition fixes the mode expansions of $\del X(z)$ and $\delb X(\zb)$ to
\begin{eqnarray}
\del X(z) &=& -\frac{i}{\sqrt2} \sum_{m=-\inf}^{\inf}\frac{\alpha_m}{z^{m+1}},\\
\delb X(\zb) &=& -\frac{i}{\sqrt2} \sum_{m=-\inf}^{\inf}\frac{\bar\alpha_m}{\zb^{m+1}}.
\end{eqnarray}
The mode expansion of the whole field can be obtained by integrating these equations:
\begin{equation}\label{X expansion}
X(z,\zb)=x_0-\frac{i}{\sqrt{2}} \left(\alpha_0 \ln z+\bar\alpha_0 \ln \zb\right)+
\frac{i}{\sqrt{2}}\sum_{\substack{m=-\inf\\m\neq 0}}^{\inf} \frac{1}{m}\left(\frac{\alpha_m}{z^m} + \frac{\bar\alpha_m}{\zb^m}\right).
\end{equation}
The oscillators $\alpha_m,\bar \alpha_m$ satisfy the following canonical commutation relations:
\begin{eqnarray}
\left[\alpha_m,\alpha_n\right]&=&\left[\bar\alpha_m,\bar\alpha_n\right]=m\delta_{m,-n}, \\
\left[\alpha_m, \bar \alpha_n \right]&=&0, \\
\left[x_0,\alpha_0\right]&=&\left[x_0,\bar \alpha_0\right]=i\sqrt{2}.
\end{eqnarray}
The spacetime momentum of the free boson is proportional to the sum of the zero modes,
\begin{equation}
p=\frac{1}{2\pi}\oint(dz\ \del X-d\zb\ \delb X)=\frac{1}{\sqrt2}( \alpha_0+\bar\alpha_0).
\end{equation}
Momentum eigenvalues must be quantized because we want the theory to be invariant under the action of the translational operator $e^{i2\pi R p}$. Therefore the only allowed momenta are
\begin{equation}
k=\frac{n}{R}, \quad n\in \mathbb{Z}.
\end{equation}
The other combination of zero modes is proportional to the winding number as
\begin{equation}
wR=\frac{1}{2\pi}\oint(dz\ \del X+d\zb\ \delb X)=\frac{1}{\sqrt2}( \alpha_0-\bar\alpha_0).
\end{equation}

Instead of the momentum and the winding number, it is sometimes more convenient to work with so-called left and right momenta
\begin{eqnarray}
p_L=\sqrt{2}\alpha_0,\quad p_R=\sqrt{2}\bar\alpha_0,
\end{eqnarray}
which have eigenvalues
\begin{eqnarray}\label{momentum winding}
k_L=\frac{n}{R}+wR,\\
k_R=\frac{n}{R}-wR.\nn
\end{eqnarray}

The energy-momentum tensor of the free boson theory is given by
\begin{eqnarray}
T(z)&=&-\del X\del X, \\
\bar T(\zb)&=&-\delb X\delb X.
\end{eqnarray}
The $TT$ OPE implies that the theory has central charges $c=\bar c=1$. The Virasoro generators can be written in terms of the $\alpha$ oscillators as
\begin{equation}
L_n=\frac{1}{2}\sum_{k=-\inf}^\inf\normc\! \alpha_{n-k}\alpha_k\! \normc.
\end{equation}
Concretely, for the $L_0$ operator, we find
\begin{equation}
L_0=\frac{p^2}{4}+\sum_{k=1}^\inf \alpha_{-k}\alpha_k.
\end{equation}

The state space of the free boson theory is spanned by states of the form
\begin{equation}\label{FB state space}
\alpha_{-i_1}\alpha_{-i_2}\dots\alpha_{-i_m}\bar\alpha_{-j_1}\bar\alpha_{-j_2}\dots\bar\alpha_{-j_n}|k_L,k_R\ra,
\end{equation}
where $|k_L,k_R\ra$ are primary states with respect to the U(1)$\times$U(1) algebra. They are eigenstates of $p_L$, $p_R$ (with the eigenvalues given by (\ref{momentum winding})) and they are annihilated by $\alpha_m$, $\bar\alpha_m$ with $m>0$. The basis states (\ref{FB state space}) have conformal weights $\left(\frac{k_L^2}{4}+\sum_{k=1}^m i_k,\frac{k_R^2}{4}+\sum_{k=1}^n j_k\right)$.

Unlike in a generic CFT, vertex operators corresponding to these states are actually very simple to write down. Momentum primary operators correspond to
\begin{equation}
|k_L,k_R\ra \lrar  e^{ik_L X_L(0)+ik_R X_R(0)}
\end{equation}
and $\alpha$ oscillators can be replaced by
\begin{equation}
\alpha_{-n}\lrar \frac{\sqrt{2}i}{(n-1)!}\del^n X(0)
\end{equation}
and similarly for $\bar\alpha$. These replacements work independently for each oscillator because the theory is free. Precise definition of these vertex operators requires additional factors, which are called cocycles (see for example \cite{Polchinski}). These factors however affect only certain bulk correlators, which do not appear in this thesis, so we will not discuss them.

Sometimes it is useful to know how the U(1) representations decompose under the Virasoro algebra. Roots of the Kac determinant (\ref{Kac roots 1}) for $c=1$ have a very simple form:
\begin{equation}
h_{r,s}=\frac{1}{4} (r-s)^2.
\end{equation}
Therefore Verma modules of weights $j^2$, where $j$ is a half-integer, are reducible and they decompose into infinite sums of irreducible Virasoro representations
\begin{equation}
\hh^{U(1)}_{j}=\bigoplus_{k=0}^\inf \hh^{Vir}_{(j+k)^2}.
\end{equation}
The Virasoro primary states that appear in this decomposition can be classified according to irreducible representations of the SU(2) algebra. They are labeled by numbers $(j,m)$ and they have conformal weight $j^2$ and momentum $2m$. We will encounter mostly the zero momentum primaries, which have weights given by squares of integers. We label these primaries as $P_{j^2}$. The first three primary states (with unit norm) are
\begin{eqnarray}\label{primary states boson}
|P_1\ra &=& \alpha_{-1} |0\ra, \nn\\
|P_4\ra &=&\sqrt{\frac{2}{27}} \left(\alpha_{-3}\alpha_{-1}-\frac{3}{4}(\alpha_{-2})^2-\frac{1}{2}(\alpha_{-1})^4\right) |0\ra,\\
|P_9\ra &=& \sqrt{\frac{2}{1125}} \left(\alpha_{-5}\alpha_{-3}\alpha_{-1}-\frac{3}{4}\alpha_{-5}(\alpha_{-2})^2-\frac{15}{16}(\alpha_{-4})^2\alpha_{-1}+\frac{5}{4}\alpha_{-4}\alpha_{-3}\alpha_{-2}\right.\nn\\
&-&\frac{5}{9}(\alpha_{-3})^3-\frac{1}{2}\alpha_{-5}(\alpha_{-1})^4+\frac{5}{4}\alpha_{-4}\alpha_{-2}(\alpha_{-1})^3-\frac{5}{4}\alpha_{-3}(\alpha_{-2})^2(\alpha_{-1})^2 \nn\\
&+&\left.\frac{15}{32}(\alpha_{-2})^4\alpha_{-1}+\frac{1}{6}\alpha_{-3}(\alpha_{-1})^6-\frac{1}{8}(\alpha_{-2})^2(\alpha_{-1})^5-\frac{1}{72}(\alpha_{-1})^9\right) |0\ra.\nn
\end{eqnarray}

\subsection{Boundary theory}\label{sec:CFT:FB:boundary}
The free boson theory admits two basic boundary conditions
\begin{eqnarray}
\del X(z)&=&\delb X(\zb)|_{z=\zb}, \\
\del X(z)&=&-\delb X(\zb)|_{z=\zb}.
\end{eqnarray}
The first boundary conditions are called Neumann boundary conditions and they can be also written as $\del_\sigma X|_{\sigma=0,\pi}=0$. The second conditions are called Dirichlet boundary conditions and they are equivalent to $\del_\tau X|_{\sigma=0,\pi}=0$ or $X|_{\sigma=0,\pi}=x_0$. In string theory, these boundary conditions are interpreted as follows: The ends of a string with Neumann boundary conditions are free to move along the circle, while the ends of a string with Dirichlet boundary conditions are firmly attached to the position $x_0$. The two boundary conditions are very similar (in string theory, they are related by the T-duality), so we will focus on description of the Neumann boundary conditions.

The mode expansion of $X(z,\zb)$ with Neumann boundary conditions is given by
\begin{equation}
X(z,\zb)=-\frac{i}{\sqrt{2}} \alpha_0 \ln z\zb+
\frac{i}{\sqrt{2}}\sum_{\substack{m=-\inf\\m\neq 0}}^{\inf} \frac{\alpha_m}{m}\left(\frac{1}{z^m} + \frac{1}{\zb^m}\right).
\end{equation}
The zero mode has a slightly different meaning than in the bulk, now it is related to the momentum as $p=\sqrt{2} \alpha_0$. That means that the definition of the $L_0$ generator changes to
\begin{equation}
L_0=p^2+\sum_{k=1}^\inf \alpha_{-k}\alpha_k.
\end{equation}

The boundary spectrum is obtained by limits of bulk operators approaching the boundary. There are usually singularities between holomorphic and antiholomorphic parts of bulk operators, so we define the so-called boundary normal ordering, denoted as $\norms\ \norms$, which cancels these singularities:
\begin{equation}
\norms\!\phi(x)\! \norms=\lim_{z\rar \zb} (\phi(z,\zb)-{\rm singular\ part}).
\end{equation}
If we consider a generic exponential operator $e^{i(k_L X_L(z)+k_R X_R(\zb))}$, we find that the winding part of the operator disappears in this limit. Therefore we define a boundary momentum operator in terms of the free field as
\begin{equation}
\norms\! e^{i \frac{n}{R} X}(x)\! \norms=\lim_{z\rar \zb}  e^{i \frac{n}{R} X}(z,\zb)\, (z-\zb)^{-\frac{n^2}{2R^2}}.
\end{equation}
This operator has conformal weight $\frac{n^2}{R^2}$, which is four times as much as the parent bulk operator.

Boundary states corresponding to the Neumann and Dirichlet boundary conditions can be written in terms of U(1) Ishibashi states, which are
\begin{eqnarray}
|k,-k\rra_N &=& \exp\left(-\sum_{n=-1}^\inf\frac{1}{n}\ \alpha_{-n}\bar\alpha_{-n}\right)|k,-k\ra,\\
|k,k\rra_D &=& \exp\left(\ \sum_{n=-1}^\inf\frac{1}{n}\ \alpha_{-n}\bar\alpha_{-n}\right)|k,k\ra.
\end{eqnarray}
The precise form of these boundary states can be derived analogously to the Cardy condition \cite{GaberdielBoundary}. Including the normalization, the Neumann boundary state reads
\begin{equation}\label{boundary state N}
\ww N(\tilde x_0)\rra=\frac{\sqrt{R}}{2^{1/4}}\sum_{n\in \mathbb{Z}} e^{i nR\tilde x_0} |n R,-n R \rra_N.
\end{equation}
The constant $\tilde x_0$ can be interpreted in string theory in terms of the Wilson line on the T-dual circle. We will encounter it only in few rare occasions because we usually use the Neumann boundary state as the initial configuration in string field theory, where we set $\tilde x_0$ to zero for simplicity. This constant affects only phases of some bulk-boundary correlators.

The Dirichlet boundary state equals to
\begin{equation}\label{boundary state D}
\ww D(x_0)\rra=\frac{1}{2^{1/4}\sqrt{R}}\sum_{n\in \mathbb{Z}} e^{i\frac{n}{R} x_0} |\frac{n}{R},\frac{n}{R} \rra_D,
\end{equation}
where $x_0$ is interpreted in string theory as the position of the D-brane the strings are attached to.

\subsection{Correlators}\label{sec:CFT:FB:correlators}
The free boson theory is fully solvable and it is possible to explicitly compute all correlators using the Green's function \cite{Polchinski}.

Bulk correlators of generic momentum operators are equal to\footnote{If the operators include both nonzero momentum and winding number, this formula additionally needs cocycles to ensure mutual locality of the operators.}
\begin{equation}\label{FB correlator bulk}
\left\la \prod_{i=1}^n e^{i(k_{Li} X_L(z_i)+k_{Ri} X_R(\zb_i))} \right\ra=\nnn_{bulk}\, \delta_{\sum_i k_{Li}}\delta_{\sum_i k_{Ri}}
\prod_{\substack{i,j=1\\ i<j}}^{n} z_{ij}^{k_{Li}k_{Lj}/2}\;\zb_{ij}^{k_{Ri}k_{Rj}/2}.
\end{equation}
The normalization is usually chosen to be $\nnn_{bulk}=1$, so that the BPZ product of momentum states is $\la k_1|k_2\ra=\delta_{k_1,-k_2}$. See also section \ref{sec:SFT:observables:energy} for discussion of our normalization of correlators in the context of OSFT.

More general correlators include derivatives of the $X$ field
\begin{equation}\label{FB correlator bulk2}
\left\la \prod_{i=1}^n e^{i(k_{Li} X_L(z_i)+k_{Ri} X_R(\zb_i))} \prod_{j=1}^p \del^{m_j} X(z_j) \prod_{k=1}^q \delb^{n_k} X(\zb_k)\right\ra.
\end{equation}
This type of correlator can be computed by removing the $\del^{m} X$ operators one by one using contractions. We pick one $\del^{m} X$ operator and contract it with all $e^{i k X}$ and $\del^{n} X$ operators at other points, where the contractions work as replacements
\begin{eqnarray}\label{FB correlator contractions}
\left\la \dots \del^{m_i} X(z_i)\dots \del^{m_j} X(z_j)\dots \right\ra &\rar& (-1)^{m_i} \frac{(m_i+m_j-1)!}{2(z_i-z_j)^{m_i+m_j}}\la  \dots \ra, \\
\left\la \dots \del^{m_i} X(z_i)\dots e^{ik_{Lj} X_L(z_j)}\dots \right\ra &\rar&(-1)^{m_i} \frac{i k_{Lj}(m_i)!}{2(z_i-z_j)^{m_i+1}}\la  \dots e^{i k_{Lj} X_L(z_j)} \dots \ra \nn.
\end{eqnarray}
We repeat this process until we remove all $\del^{m} X$ and $\delb^{n} X$ operators. In this way, we can reduce (\ref{FB correlator bulk2}) to (\ref{FB correlator bulk}).

Boundary correlators with Neumann boundary conditions work in the same way. A correlator with just momentum primaries is very similar to (\ref{FB correlator bulk}):
\begin{equation}\label{FB correlator bound}
\left\la \prod_{i=1}^n \norms e^{i k_i X(x_i)} \norms \right\ra^{(N)}=\nnn_{bound}\, \delta_{\sum_i k_i}
\prod_{\substack{i,j=1\\ i<j}}^{n} x_{ij}^{2k_i k_j}.
\end{equation}
where the normalization is fixed by the boundary state (\ref{boundary state N}) to $\nnn_{bound}=\frac{\sqrt{R}}{2^{1/4}}$.
If we want to compute a correlator with additional $\del^m X$ operators,
\begin{equation}
\left\la \prod_{i=1}^n \norms e^{i k_i X(x_i)} \norms \prod_{j=1}^p \del^{m_j} X(x_j) \right\ra^{(N)},
\end{equation}
we can use the contractions (\ref{FB correlator contractions}), but we have to double the momenta in contractions of $\del^m X$ with $e^{i k X}$ to account for the boundary normal ordering.

Finally, we can consider a completely generic correlator of bulk and boundary operators
\begin{equation}\footnotesize
\left\la \prod_{i=1}^{m} e^{i(k_{Li} X_L(z_i)+k_{Ri} X_R(\zb_{i}))}\prod_{j=1}^{n} \normsf e^{i k_j X(x_j)} \normsf \prod_{k_1=1}^{p} \del^{m_{k_1}} X(z_{k_1}) \prod_{k_2=1}^{q} \delb^{m_{k_2}} X(\zb_{k_2})  \prod_{l=1}^r \del^{m_l} X(x_l)\right\ra^{(N)}.
\end{equation}
This type correlator is more complicated, so, instead of considering all possible types of contractions, it is more convenient to compute it using the doubling trick. We can replace all bulk and boundary operators by chiral operators as
\begin{eqnarray}
\delb^{m} X(\zb) &\rar& \del^{m} X(\zb), \nn \\
e^{i(k_L X_L(z)+k_R X_R(\zb))} &\rar& e^{ik_L X(z)}e^{i k_R X(\zb)}, \\
\norms e^{i k X(x)} \norms &\rar& e^{i 2 k X(x)}. \nn
\end{eqnarray}
Then we can treat all operators equally and compute the correlator using the same contractions as for the holomorphic part of (\ref{FB correlator bulk2}). Similar formulas can be used for correlators with Dirichlet boundary conditions. We just have an additional replacement $X_R(\zb)\rar -X_R(\zb)$ in antiholomorphic parts of bulk operators.

\subsection{Free boson on a torus}\label{sec:CFT:FB:torus}
At the end of this section, let us discuss free boson theory compactified on a torus. A $D$-dimensional torus is characterized by linearly independent vectors $v_i^\mu$, $i=1,\dots,D$, which determine the lengths of its sides and the angles between them.
If we consider the bulk theory on such torus, the periodicity conditions (\ref{periodicity circle}) generalize to
\begin{equation}
X^\mu(\tau,\sigma+2\pi)=X^\mu(\tau,\sigma)+\sum_iw_i v_i^\mu,\quad w_i \in \mathbb{Z}.
\end{equation}
This affects the quantization of momentum and winding eigenvalues. The winding now belongs to the lattice spanned by the vectors $v_i^\mu$, while the momentum belongs to the dual lattice.

\begin{figure}
\centering
   \includegraphics[width=8cm]{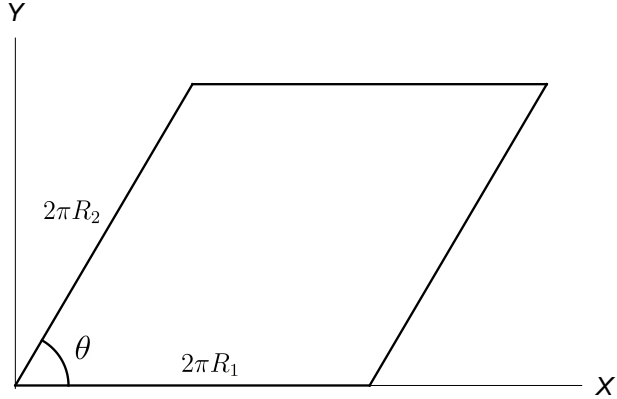}
\caption{A two-dimensional torus parameterized by two radii $R_1$ and $R_2$ and by an angle $\theta$.}
\label{fig:torus}
\end{figure}

We will focus on a two-dimensional torus. It can be characterized by two radii $R_1$, $R_2$ and by an angle $\theta$, see figure \ref{fig:torus} for illustration. The identification of the two free fields, which we call $X$ and $Y$, is
\begin{eqnarray}
(X,Y)&\sim& (X+2\pi R_1,Y),  \\
(X,Y)&\sim& (X+2\pi R_2\cos\theta,Y+2\pi R_2\sin\theta).
\end{eqnarray}
The quantization of momentum and winding eigenvalues gets more complicated than in a single dimension:
\begin{equation}\label{FB momentum torus}
k^\mu=\left(\frac{n_1}{R_1},\frac{n_2}{R_2\sin\theta}-\frac{n_1\cos\theta}{R_1\sin\theta}\right),
\end{equation}
\begin{equation}\label{FB winding torus}
\tilde k^\mu=\left(w_1 R_1+w_2 R_2 \cos\theta,w_2 R_2\sin\theta\right).
\end{equation}
Notice that the quantization in the two directions is not independent.

The bulk theory on a torus is not much different from the theory on a circle, but the boundary theory offers much richer spectrum of boundary conditions. Classification of all possible boundary conditions remains unknown, so we will describe just boundary conditions that respect the U(1)$\times$U(1) symmetry. The gluing conditions for the $\del X^\mu$ currents have the form
\begin{equation}
\del X^\mu=\Omega^{\mu}_{\ \nu}\delb X^\nu|_{z=\zb},
\end{equation}
where $\Omega$ must be an O(2) matrix in order to preserve the gluing conditions for the energy-momentum tensor. A generic SO(2) matrix describes free bosons in the background of a constant magnetic field $B_{\mu\nu}$, the special values $\Omega=\Id$ and $\Omega=-\Id$ describe the Neumann and Dirichlet boundary conditions in both directions respectively. The disconnected component of the O(2) group describes mixed Neumann-Dirichlet boundary conditions, which, in string theory, correspond to a D1-brane wrapped around the torus.

In this thesis, we restrict our attention to the simplest background, the Neumann boundary conditions in both directions, which describes a D2-brane. The boundary theory in this case factorizes into the $X$ and $Y$ directions with the exception of quantization of the momentum modes, which is given by (\ref{FB momentum torus}).

\section{Minimal models}\label{sec:CFT:MM}
In this section, we describe the Virasoro minimal models. These models are rational conformal field theories characterized by two integers $(p,q)$ and by their modular invariant partition function, which follows the A-D-E classification. We will describe only the simplest case, the A-series, which has a diagonal partition function.

\subsection{Chiral theory}\label{sec:CFT:MM:chiral}
It can be shown that for a generic value of the central charge, the OPE of two primary operators generates infinite number of new primaries. Conformal field theories which are rational with respect to the Virasoro algebra can be therefore constructed only for special values of the central charge
\begin{equation}
c=1-\frac{6(p-q)^2}{pq},
\end{equation}
where $p$ and $q$ are coprime integers. We can choose $p<q$ without loss of generality. Conformal dimensions of primary operators in these theories are
\begin{eqnarray}
h_{r,s}(p,q)=\frac{(qr-ps)^2-(p-q)^2}{4pq},\quad  1\leq r<p,\ 1\leq s<q.
\end{eqnarray}
These weights exactly match the roots of the Kac determinant (\ref{Kac roots 3}), so the associated Verma modules in are reducible. Notice that the weights satisfy $h_{r,s}=h_{p-r,q-s}$, which means that we can identify the corresponding primary operators, $\phi_{(r,s)}=\phi_{(p-r,q-s)}$. That leaves us with $\frac{(p-1)(q-1)}{2}$ independent primaries.

In principle, we can label a given primary both by $(r,s)$ and $(p-r,q-s)$, but, in order to be consistent with the formulas for F-matrices \cite{Runkel3}, which we show in appendix \ref{sec:MM Fmatrix}, we choose the following convention: If $p$ is odd, then we have to pick the expression which makes $r$ odd as well, otherwise the second parameter $s$ must be made odd.

Primary operators in this theory satisfy fusion rules
\begin{equation}\label{MM fusion rules}
\phi_{(r,s)}\times \phi_{(m,n)}=\sum_{\substack{k=1+|r-m| \\ k+r+m\ {\rm odd}}}^{k_{max}}\sum_{\substack{l=1+|s-n| \\ l+s+n\ {\rm odd}}}^{l_{max}}\phi_{(k,l)},
\end{equation}
where $k_{max}=\min(r+m-1,2p-r-m-1)$ and $l_{max}=\min(s+n-1,2q-s-n-1)$.

A generic minimal model contains operators with negative conformal weights and therefore it is nonunitary. Only models with $|p-q|=1$ are unitary, in that case we can write $(p,q)$ as $(m,m+1)$ and the central charge and conformal weights reduce to (\ref{Kac roots 2}).

\subsection{Bulk and boundary theory}\label{sec:CFT:MM:bulk boundary}
The full theory is specified by its partition function. In this thesis, we consider only the A-series of minimal models, which have diagonal spectrum,
\begin{equation}
\hh=\bigoplus_{r,s} \hh_{(r,s)}\otimes \bar\hh_{(r,s)}.
\end{equation}
Bulk operators therefore carry exactly the same labels $(r,s)$ as chiral operators.

The conformal boundary conditions are labeled by $(r,s)$ as well. The spectrum of primary operators for a given boundary condition $(r,s)$ is determined by fusion of the corresponding operator with itself, $\phi\in \phi_{(r,s)}\times \phi_{(r,s)}$.

Boundary states of unitary minimal models are given the Cardy solution (\ref{Cardy BS}):
\begin{equation}\label{boundary state minmod}
B_\alpha^i=\frac{S_\alpha^{\ i}}{\sqrt{S_\Id^{\ i}}},
\end{equation}
where the $S$ matrix equals to
\begin{equation}
S_{(r,s)}^{\ \ \ (m,n)}=2\sqrt{\frac{2}{pq}}(-1)^{1+rn+sm}\sin(\pi \frac{q}{p}r m)\sin(\pi\frac{p}{q}sn).
\end{equation}

The formulas for nonunitary minimal models are a bit more complicated. Some primary operators have negative two-point functions, which cannot be made real by normalization (assuming we want to avoid complex structure constants).
In that case, there may be differences between coefficients of boundary states and one-point functions,
\begin{equation}
\la \phi_i\ra^{(\alpha)}=B_\alpha^i \la i|i\ra,
\end{equation}
where $\la \phi_i\ra^{(\alpha)}=|z-\zb|^{2h_i}\la \phi_i(z,\zb)\ra^{(\alpha)}$ and $\la i|i\ra$ is the product of bulk operators. The Cardy condition for the nonunitary models is discussed in detail in \cite{Runkel1}. The theory allows a change of normalization
\begin{eqnarray}
\la 0|0\ra &\rar& \mu^2 \la 0|0\ra,  \\
\la \Id \ra^{(\alpha)} &\rar& \mu \la \Id \ra^{(\alpha)}, \\
\phi_i &\rar& \alpha_i \phi_i.
\end{eqnarray}
By setting $\mu=\frac{1}{\sqrt{|S_\Id^{\ \Id}|}}$ and $\alpha_i=\sqrt{\left|\frac{S_\Id^{\ \Id}}{S_\Id^{\ i}}\right|}$ in the formulas from \cite{Runkel1}, we get
\begin{equation}
\Cbulk ii\Id= {\rm sgn} \frac{S_\Id^{\ i}}{S_\Id^{\ \Id}}
\end{equation}
and
\begin{eqnarray}
\la \phi_i\ra^{(\alpha)}&=& \frac{S_\alpha^{\ i}}{\sqrt{|S_\Id^{\ i}|}},\\
B_\alpha^i &=& \frac{S_\alpha^{\ i}}{\sqrt{|S_\Id^{\ i}|}}\ {\rm sgn} S_{\Id}^{\ i}.
\end{eqnarray}
Curiously, in this normalization, it is possible that even the product of two bulk vacuum states is negative,
\begin{equation}
\la 0|0\ra={\rm sgn} S_\Id^{\ \Id}.
\end{equation}

Finally, we will write down the explicit solution for structure constants, which was found in \cite{Runkel1} by solving the sewing relations (\ref{sewing constr bound 4})-(\ref{sewing constr BB 20}). Bulk structure constants are given by
\begin{equation}\label{MM bulk constants}
\Cbulk ijk = \left(\Fmatrix k\Id jiij \right)^{-1},
\end{equation}
boundary structure constants by
\begin{equation}\label{MM bound constants}
\Cbound \alpha\beta\gamma ijk = \Fmatrix \beta ki\alpha \gamma j
\end{equation}
and finally bulk-boundary structure constants by
\begin{equation}\label{MM BB constants}
\Cbb \alpha ij=\frac{S_\Id^{\ i}}{S_\Id^{\ \Id}}\sum_m e^{i\pi (2h_m-2h_\alpha-2h_i+\frac{1}{2}h_j)} \Fmatrix \Id m\alpha\alpha ii \Fmatrix mj\alpha ii\alpha.
\end{equation}
Explicit formulas for the F-matrices and some of their basic properties can be found in appendix \ref{sec:MM Fmatrix}. It can be shown that all structure constants (even (\ref{MM BB constants}), which include complex phases) are real.

The structure constants are unique up to normalization of primary operators. If we scale the operators as $\phi_i\rar \alpha_i \phi_i$ and $\psi_i^{(\alpha\beta)}\rar \lambda_i^{(\alpha\beta)}\psi_i^{(\alpha\beta)}$, the structure constants change as
\begin{eqnarray}
\Cbulk ijk &\rar& \frac{\alpha_i \alpha_j}{\alpha_k}\Cbulk ijk, \\
\Cbound \alpha\beta\gamma ijk &\rar& \frac{ \lambda_i^{(\alpha\beta)} \lambda_j^{(\beta\gamma)}}{\lambda_k^{(\gamma\alpha)}}\Cbound \alpha\beta\gamma ijk, \\
\Cbb \alpha ij &\rar& \frac{\alpha_i}{\lambda_j^{(\alpha\alpha)}} \Cbb \alpha ij.
\end{eqnarray}
We use this normalization to set bulk and boundary two-point structure constants to $\pm 1$.

\section{The ghost system}\label{sec:CFT:ghost}
The last theory that we are going to describe is the system of $b$ and $c$ ghosts. The ghosts appear in string theory as a result of gauge fixing of the worldsheet metric \cite{Polchinski} and they play an important role in string field theory.

\subsection{Chiral theory}\label{sec:CFT:ghost:chiral}
The $b$ and $c$ ghosts are anticommuting primary fields with conformal weights $h_b=2$ and $h_c=-1$. The action of the ghost theory reads
\begin{equation}
S=\frac{1}{2\pi}\int d^2z\ b\delb c
\end{equation}
and the corresponding equations of motion imply that both fields are holomorphic,
\begin{equation}
\bar\del b(z)=\bar\del c(z)=0.
\end{equation}

The OPE between the $b$ and $c$ fields is very simple
\begin{equation}
b(z_1)c(z_2) \sim \frac{1}{z_1-z_2}.
\end{equation}
The OPEs of $b$ and $c$ with themselves are nonsingular because of anticommutativity of the ghosts. Using the usual mode expansions
\begin{equation}
b(z)=\sum_{n=-\inf}^\inf \frac{b_n}{z^{n+2}},\quad  c(z)=\sum_{n=-\inf}^\inf \frac{c_n}{z^{n-1}},
\end{equation}
we find that the modes $b_m$, $c_m$ satisfy anticommutators
\begin{eqnarray}
\left\{b_m,c_n \right\}&=&\delta_{m,-n}, \\
\left\{b_m,b_n \right\}&=&\left\{c_m,c_n \right\}=0. \nn
\end{eqnarray}

The energy-momentum tensor of the ghost theory is given by
\begin{equation}
T(z)=(\del b)c-2\del(bc)
\end{equation}
and the corresponding Virasoro generators are equal to\footnote{We have decided to split creation and annihilation operators with respect to the ghost number one vacuum $c_1|0\ra$, i.e. $b_n$ with $n\leq -1$ and $c_n$ with $n\leq 0$ are considered creation operators. This normal ordering is not equivalent to the conformal normal ordering, see for example \cite{Polchinski}, however it is useful for practical purposes because the state $c_1|0\ra$ is the ground state in string field theory.}
\begin{equation}
L_m^{gh}=\sum_{n=-\inf}^\inf(2m-n)\normc b_n c_{m-n} \normc-\delta_{m0}.
\end{equation}
The $TT$ OPE implies that the ghost central charge is $c^{gh}=-26$.

The theory has an additional ghost number symmetry $\delta b=-i\epsilon b$, $\delta c=i\epsilon c$, which is generated by the ghost current
\begin{equation}
j^{gh}(z)=-bc(z).
\end{equation}
The modes of the ghost current are given by
\begin{equation}
j^{gh}_m=-\sum_{n=-\inf}^\inf\normc b_n c_{m-n} \normc+\delta_{m0}.
\end{equation}
The ghost current allows us to define ghost number of a state or operator as the eigenvalue of the $j_0^{gh}$ operator. The ghost number of a given state equals to the number of $c$ ghosts minus the number of $b$ ghosts.

The ghost current is not a proper weight 1 primary field because its OPE with the energy-momentum tensor is
\begin{equation}
T(z)j^{gh}(0)\sim -\frac{3}{z^3}+\frac{1}{z^2} j^{gh}(0)+\frac{1}{z}\del j^{gh}(0).
\end{equation}
This implies that the ghost current transforms as
\begin{equation}
\tilde j^{gh}(w)=\frac{dz}{dw}j^{gh}(z)-\frac{3}{2}\frac{d^2z}{dw^2}\left(\frac{dz}{dw}\right)^{-1}.
\end{equation}
This anomalous transformation law is related to an unusual property of ghost correlators. Imagine that we insert $j_0^{gh}$ into a correlator in form of a contour integral that encircles all insertions. By deforming the contour around infinity, we find that the correlator is nonzero only if the total ghost number of all insertions is equal to 3. Therefore the simplest nonzero correlator in this theory is
\begin{equation}\label{ghost correlator}
\la c(z_1) c(z_2) c(z_3) \ra=(z_1-z_2)(z_1-z_3)(z_2-z_3).
\end{equation}
More complicated correlators with $b$ insertions can be reduced to (\ref{ghost correlator}) by removing all $b$ operators by contour manipulations. The basic BPZ product, which is equivalent to (\ref{ghost correlator}), is given by
\begin{equation}
\la 0| c_{-1}c_0c_1|0\ra=1.
\end{equation}

The state space of the ghost theory is spanned by states
\begin{equation}\label{ghost state space}
b_{-k_1}\dots b_{-k_m}c_{-l_1}\dots c_{-l_n}|0\ra, \quad k_i\geq2,\ l_i\geq-1.
\end{equation}
We will discuss the structure of the ghost state space in more detail in section \ref{sec:SFT:string field:ghost}.

\subsection{Bulk and boundary theory}\label{sec:CFT:ghost:full}
Once we understand the chiral theory, the construction of bulk and boundary theories is easy. The bulk theory is given by the product of a holomorphic and antiholomorphic copy of the chiral theory. The action reads
\begin{equation}
S=\frac{1}{2\pi}\int d^2z(b\delb c+\bar b\del \bar c),
\end{equation}
the Hilbert space is given by $\hh\otimes \bar\hh$ and correlators factorize into holomorphic and antiholomorphic parts (up to possible signs from anticommutators).

When it comes to the boundary theory, there is only one consistent set of gluing conditions
\begin{eqnarray}
c(z)=\bar c(\zb)|_{z=\zb},\\
b(z)=\bar b(\zb)|_{z=\zb}.
\end{eqnarray}
We can use these gluing conditions for the doubling trick and the resulting boundary theory exactly matches the chiral theory.

The boundary state in the ghost theory has ghost number 3, so we have to slightly modify the definition (\ref{boundary state def}) to
\begin{equation}
|z-\zb|^{2h_i}\la c_0^-\phi_i(z,\zb)\ra=\la \phi_i|c_0^-\ww B \rra,
\end{equation}
where $c_0^-=c_0-\bar c_0$ and we assume that the bulk state $\la \phi_i|$ has ghost number 2. The boundary state must obey (\ref{Ishibashi conditions}) and also
\begin{eqnarray}
(b_n-\bar b_{-n})\ww B \rra &=& 0, \\
(c_n+\bar c_{-n})\ww B \rra &=& 0, \\
(j_n^{gh}+\bar j^{gh}_{-n}-3\delta_{n0})\ww B \rra &=& 0.
\end{eqnarray}
These conditions are satisfied by
\begin{equation}
\ww B_{gh}\rra=(c_0+\bar c_0)\exp\left(-\sum_{n=1}^\inf (\bar b_{-n}c_{-n}+b_{-n}\bar c_{-n})\right)c_1\bar c_1|0\ra.
\end{equation}

\subsection{Commutators}\label{sec:CFT:ghost:commutators}
For later convenience, we list all commutators between various operators that appear in the ghost theory. In addition to the ghost Virasoros and modes of the ghost current, we introduce 'twisted' ghost Virasoros
\begin{equation}\label{twisted Virasoro 1}
L'^{gh}_n=L^{gh}_n + n j^{gh}_n + \delta_{n0} = \sum_{m=-\infty}^{\infty}(n-m)\normc b_m c_{n-m}\normc,
\end{equation}
which will play a role later in string field theory in Siegel gauge.

Commutators of the ghost Virasoros are
\begin{eqnarray}
\left[L^{gh}_m,L^{gh}_n\right]&=&(m-n)L^{gh}_{m+n}-\frac{13}{6}(m^3-m)\delta_{m,-n}, \\
\left[L^{gh}_m,b_n\right]&=&(m-n)b_{m+n}, \\
\left[L^{gh}_m,c_n\right]&=&-(2m+n)c_{m+n}.
\end{eqnarray}
Next, commutators involving modes of the ghost current are
\begin{eqnarray}
\left[j^{gh}_m,j^{gh}_n\right]&=&m\delta_{m,-n},\\
\left[L^{gh}_m,j^{gh}_n\right]&=&-n j^{gh}_{m+n}-\frac{3}{2}(m^2+m)\delta_{m,-n}, \\
\left[j^{gh}_m,b_n\right]&=&-b_{m+n},  \\
\left[j^{gh}_m,c_n\right]&=&c_{m+n}.
\end{eqnarray}
Finally, we list commutators of the 'twisted' ghost Virasoros:
\begin{eqnarray}
\left[L'^{gh}_m,L'^{gh}_n\right]&=&(m-n)L'^{gh}_{m+n}-\frac{1}{6}(m^3-m)\delta_{m,-n}, \\
\left[L'^{gh}_m,L^{gh}_n\right]&=&(m-n)L'^{gh}_{m+n}+n^2 j^{gh}_{m+n}-\frac{1}{6}(4 m^3+9 m^2-m)\delta_{m,-n},\qquad\qquad \\
\left[L'^{gh}_m,j^{gh}_n\right]&=&-n j^{gh}_{m+n}-\frac{1}{2}(m^2+3m)\delta_{m,-n}, \label{commutator Lp j} \\
\left[L'^{gh}_m,b_n\right]&=&-n b_{m+n}, \\
\left[L'^{gh}_m,c_n\right]&=&-(m+n)c_{m+n}.
\end{eqnarray}

\section{Bosonic string theory}\label{sec:CFT:strings}
At the end of this chapter, we briefly review of some aspects of bosonic string theory following \cite{Polchinski}\footnote{Other classic textbooks for string theory are \cite{ZwiebachStrings}\cite{GreenSchwarzWitten1}\cite{GreenSchwarzWitten2}\cite{BeckerBeckerSchwarz}\cite{BlumenhagenStringTheory}.}. We focus on topics which are relevant for string field theory, that is the relation between string theory and conformal field theory and the BRST quantization.

\subsection{The Polyakov action}\label{sec:CFT:strings:action}
A string in $D$ dimension is described by $D$ scalar fields, which represent its position in the spacetime. The Polyakov action for the bosonic string in Minkowski spacetime is\footnote{We set the constant $\alpha'$ to 1 for simplicity.}\footnote{The full action also includes some topological terms, which are however not important for our purposes.}
\begin{equation}\label{Polyakov action}
S[X,g]=-\frac{1}{4\pi}\int d^2\sigma \sqrt{g}\ g^{ab}\eta^{\mu\nu} \del_a X_\mu \del_b X_\nu,
\end{equation}
where $g^{ab}$ is the worldsheet metric (with Euclidean signature), $g=\det g_{ab}$ and $\eta^{\mu\nu}$ is the spacetime Minkowski metric. The action is invariant under worldsheet diffeomorphisms, $\sigma^a\rar \sigma'^a$, and under two-dimensional Weyl transformations, $g_{ab}\rar \omega(\sigma) g_{ab}$.

The Euclidean path integral in this theory,
\begin{equation}
\int [dX\; dg]\ e^{-S[X,g]},
\end{equation}
includes an enormous overcounting given by the diffeomorphisms and by the Weyl symmetry. If we want to have a well-defined part integral, we have to remove the overcounting. This is usually done by the Faddeev-Popov method. This procedure allows us to fix the metric to some specific functional form $\hat g_{ab}(\sigma)$. For simplicity, we choose the flat metric $\hat g_{ab}(\sigma)=\delta_{ab}$, although such choice is in general possible only locally. After the gauge fixing, the path integral changes to
\begin{equation}
\int [dX\; db\; dc]\ e^{-S[X,b,c]},
\end{equation}
where the new action reads
\begin{equation}\label{action gauge fix}
S[X,b,c]=\frac{1}{2\pi}\int d^2z\ \del X_\mu \delb X^\mu+\frac{1}{2\pi}\int d^2z\ (b\delb c+\bar b\del \bar c).
\end{equation}
We can see that the bosonic string is now described by a two-dimensional conformal field theory, which includes $D$ free bosons and the $bc$ ghost system, which are described in sections \ref{sec:CFT:FB} and \ref{sec:CFT:ghost} respectively.

This action can be used for both open and closed strings. Closed strings are described by a bulk CFT, while open strings are described by a boundary CFT with some boundary conditions (usually Neumann or Dirichlet) at the ends of strings.

\subsection{BRST quantization}\label{sec:CFT:strings:BRST}
The state space of a bosonic string theory is given by the product of the free boson state space and the ghost state space. The theory is not unitary because the Hermitian product in this space is not positive definite. Negative-norm states come from the $X^0$ boson, which has a wrong sign kinetic term, and from the ghosts. There are three common methods to obtain the physical spectrum: the light-cone quantization, the old covariant quantization and the BRST quantization. We will describe the BRST quantization, which is relevant for string field theory. For simplicity, we describe only the open string spectrum.

The BRST quantization allows us to fix the gauge in a covariant way. Following a procedure similar to the Faddeev-Popov method, we arrive to the action (\ref{action gauge fix}) plus a gauge-fixing term. This action is invariant under the BRST symmetry and the corresponding Noether's theorem gives us the conserved current
\begin{equation}
j_B(z)=cT^m(z)+\frac{1}{2}\!\normo\! c T^{gh}\! \!\normo\!(z)+\frac{3}{2}\del^2 c(z).
\end{equation}
The zero mode of this current, which plays a crucial role in string field theory, is called the BRST charge and denoted as $Q_B$,
\begin{equation}\label{BRST charge}
Q_B=\oint\frac{dz}{2\pi i}j_B(z)=\sum_{n=-\infty}^\infty c_{-n} L_n^m+\frac{1}{2}\sum_{n=-\infty}^\infty \normc c_{-n} L_n^{gh}\normc-\frac{1}{2} c_0.
\end{equation}
For consistency, the BRST charge must be nilpotent,
\begin{equation}
Q_B^2=0.
\end{equation}
This condition is satisfied only when the matter central charge is $c^m=26$. This condition therefore fixes the dimension of bosonic string theory to $D=26$.

Physical spectrum of the bosonic string is determined by the cohomology of the BRST charge, which means that on-shell states must be annihilated by the BRST charge,
\begin{equation}\label{BRST condition}
Q_B|\psi\ra=0,
\end{equation}
and there is an equivalence relation
\begin{equation}\label{BRST equivalence}
|\psi\ra\cong|\psi\ra+ Q_B|\chi\ra,
\end{equation}
where $|\chi\ra$ is an arbitrary state. Physical states must also satisfy one additional condition:
\begin{equation}
b_0|\psi\ra=0.
\end{equation}
Using the anticommutator $\{Q_B,b_0\}=L_0^{tot}$, we find the usual on-shell condition
\begin{equation}
L_0^{tot}|\phi\ra=(p^2+m^2)|\phi\ra=0.
\end{equation}
The $L_0^{tot}$ eigenvalue consists of momentum squared and mass squared, which equals to level of all creation operators in $|\phi\ra$. If some of the spacetime dimensions are compact, the momentum in these dimensions is conventionally also added to the mass squared.

As an example, let's see what these conditions imply for states at the lowest levels. The only on-shell state at level 0 is
\begin{equation}
c_1|k_\mu\ra,
\end{equation}
where the momentum satisfies $k^2-1=0$, which means that this state is tachyonic. There is no equivalence relation at this level.

The most general state at level 1 is
\begin{equation}
(e_\nu\alpha^\nu_{-1}+\beta b_{-1}+\gamma c_{-1})c_1|k_\mu\ra.
\end{equation}
The on-shell conditions are $k^2=0$, which implies that these states are massless, and $k^\mu e_\mu=\beta=0$. The equivalence relation leads to identifications $e_\mu \cong e_\mu+a_1 k_\mu$ and $\gamma \cong \gamma +a_2$ for any constants $a_{1,2}$. Therefore we can set $\beta=\gamma=0$ and we are left with $D-2$ polarizations of $e_\mu$, which is the expected result for a massless vector particle.

The analysis can be carried on to higher levels and it is possible to show that the cohomology contains no negative-norm states. Physical states can be represented only by free boson excitations in $D-2$ dimensions.

\subsection{D-branes and T-duality}\label{sec:CFT:strings:D-branes}
Boundary conditions in string theory have a somewhat different interpretation than in a generic conformal field theory. Consider an open string theory with Neumann boundary conditions in $p+1$ directions and Dirichlet boundary conditions in the remaining $25-p$ directions. These conditions determine a $p$-dimensional hyperplane. Ends of open strings with these boundary conditions are attached to this hyperplane and they can move along it. This hyperplane is called a D$p$-brane and it turns out to be a non-perturbative object in string theory\footnote{See for example \cite{PolchinskiDbranes}\cite{BachasDbranes}.}. It has its own mass, charges and its own dynamics. An open string attached to a D-brane can be understood precisely as a small perturbation of this D-brane. Large deformations of D-branes can be described by string field theory, which is the subject of this thesis. D-branes have many uses in string theory, they help us for example to understand various dualities or they give us an insight into black hole physics.

String theory contains a duality which relates different compactifications, which is called T-duality. In the context of open strings, it exchanges Neumann and Dirichlet boundary conditions and therefore it changes D-brane dimensions. Consider a simple example, free boson on a circle. We can easily match the spectra of closed strings at radii $R$ and $\frac{1}{R}$, the identification just exchanges momentum and winding modes. If we want to match the boundary theory, we find that it is necessary to exchange Neumann and Dirichlet boundary conditions. Therefore a D1-brane\footnote{When referring to D-branes, we usually consider only their dimensionality with respect to the compact space of interest. Therefore we can imagine that they have Dirichlet boundary condition in all other spatial directions.} at radius $R$ is T-dual to a D0-brane at radius $\frac{1}{R}$. The T-duality can be extended to more general toroidal compactifications. There it can be performed in different directions and these transformations form the group O($D,D,\mathbb{Z}$). It can also involve more generic types of D-branes, for example D-branes with flux.

\subsection{Introduction of other CFTs}\label{sec:CFT:strings:other}
When we construct a string theory, we can replace some of the 26 free bosons by another conformal field theory of interest. This construction is analogous to Gepner models in superstring theory \cite{Gepner}\cite{GepnerDbranes}. The new theory can be pretty much arbitrary (for some purposes even nonunitary), we just require that the total matter central charge remains 26. We assume that the full theory takes the form CFT$^X\otimes$CFT'$\otimes$CFT$^{gh}$, where CFT$^X$ is the theory of interest, CFT' is the rest of the matter theory and CFT$^{gh}$ is the ghost theory. The central charge of CFT$^X$ does not have to be an integer, so CFT' may also have to include CFTs different from the free boson theory, but the exact content of CFT' is not important for our purposes. In string field theory, we will consider only the universal part of the Hilbert space of CFT', which includes only Virasoro descendants of the vacuum state.

Physically, we can understand these theories as follows. We can imagine that the new CFT comes from a compactification of several free bosons on some nontrivial manifold. Such theories are generally unsolvable, but there are special manifolds which can be described by products of rational CFT, one of which becomes the theory of interest. A simple example is described in section \ref{sec:MM:Ising2}, there is a duality between the double Ising model and the free boson theory on $\mathbb{Z}_2$ orbifold of a circle with $R=\sqrt{2}$. This duality allows us to introduce the Ising model to string theory.

To use the same terminology as in the free boson theory, we will also call boundary states in these models D-branes (for example, the Ising model includes $\Id,\eps,\sigma$-branes), although they usually do not have a simple geometric interpretation.

In this thesis, we consider string theories which include the Virasoro minimal models, another interesting option would be to explore some WZW models. We need to understand boundary theory for at least one boundary condition to construct string field theory, so we are in general restricted to some product of rational or free CFTs.

\chapter{String field theory}\label{sec:SFT}
In this chapter, we focus on the main subject of this thesis, the bosonic open string field theory. As in the previous chapter, the whole theory it too broad to be fully reviewed here. Therefore we focus on topics which are relevant for our numerical approach. In this regard, we attempt to be self-contained, so the reader should be able to find all formulas necessary for reproducing our results. We also discuss some possible extensions of our calculations to more complicated string backgrounds. On the other hand, we only briefly touch the $KBc$ algebra and modern analytic methods.

In section \ref{sec:SFT:basic}, we provide a brief review of the bosonic open string field theory. We introduce basic elements of OSFT, show its action and discuss its symmetries. Then we define the level truncation scheme and sketch our methods to solve the equations of motion.  In section \ref{sec:SFT:string field}, we introduce several backgrounds we are interested in and we discuss structure of state spaces in the corresponding BCFTs. In section \ref{sec:SFT:observables}, we define various gauge invariant observables in OSFT, which are used to identify solutions, and we discuss their specifics in the chosen backgrounds. Finally, in sections \ref{sec:SFT:cubic cons} and \ref{sec:SFT:Elw cons}, we derive conservation laws for the cubic vertex and for Ellwood invariants, which are needed for  construction of recursive numerical algorithms in the next chapter.

\section{Introduction to open string field theory}\label{sec:SFT:basic}
Open string field theory was first introduced by Edward Witten in \cite{WittenSFT}. This original formulation of string field theory, which includes path integrals over length of the string, is not well suited for practical calculations, so this review follows mainly classical references \cite{TaylorZwiebach}\cite{RastelliZwiebach}\cite{OhmoriReview}\footnote{Some other useful references regarding OSFT are \cite{ThornSFT}\cite{ZwiebachTensor}\cite{OkawaLectures}\cite{ErlerLectures}.}, where the string field theory is formulated using a more convenient CFT formalism.

The central object of string field theory is a so-called string field $\Psi$. The string field belongs to the Hilbert space of a first-quantized open string theory formulated on some D-brane background. For now, we will consider a single space-filling D25-brane for simplicity. However, the string field is an \emph{off-shell} element of the Hilbert space, which means that it is generally not annihilated by the BRST charge,
\begin{equation}
Q_B |\Psi\ra\neq 0.
\end{equation}
It is possible consider string field of any ghost number, but the physical string field, which enters the OSFT action, inherits ghost number 1 from on-shell string states. We can write such string field as an expansion in open string states,
\begin{equation}\label{string field expansion}
|\Psi \ra=\int\! d^{26}k\left(T(k)+A_\mu(k)\alpha_{-1}^\mu+\beta(k)c_0b_{-1}+B_{\mu\nu}(k)\alpha_{-1}^\mu\alpha_{-1}^\nu+\dots\right)e^{ik.X}(0)c_1|0\ra,
\end{equation}
where the integral over $k$ is not restricted by the mass-shell condition. In this expansion, we recognize various spacetime fields like the tachyon $T(k)$, massless vector field $A_\mu(k)$, etc.

There are several operations that we can perform with string fields. First, we can multiply two string fields. The multiplication is called the star product and denoted by the symbol $\ast$. It maps two string fields back to a string field. It is associative,
\begin{equation}
(\Psi_1\ast\Psi_2)\ast\Psi_3=\Psi_1\ast(\Psi_2\ast\Psi_3),
\end{equation}
but, in general, it is neither commutative nor anticommutative,
\begin{equation}
\Psi_1\ast\Psi_2\neq\pm \Psi_2\ast\Psi_1.
\end{equation}

Next, there is an odd derivative $Q$. It is defined using the string theory BRST charge (\ref{BRST charge}), which acts on a string field in the usual operator sense. Therefore the derivative is nilpotent, $Q^2=0$, and it satisfies a "Leibnitz rule" when acting on a star product of two string fields:
\begin{equation}
Q(\Psi\ast\Phi)=Q(\Psi)\ast\Phi+(-1)^\Psi\Psi \ast Q(\Phi),
\end{equation}
where $(-1)^\Psi$ is a $\mathbb{Z}_2$ degree of the string field. Unless there are some fermionic matter fields, the degree in the bosonic theory is given by minus one to the ghost number. The degree of the derivative $Q$ is therefore equal to $-1$.

Finally, we introduce an integration operation, which maps a string field to a complex number, $\int \Psi\in \mathbb{C}$. The integration is linear, $\int(\Psi+\Phi)=\int \Psi+\int \Phi$, and it satisfies the cyclic property
\begin{equation}\label{vertex cyclicity}
\int \Psi\ast\Phi=(-1)^{\Psi\Phi}\int \Phi\ast\Psi.
\end{equation}
The integration can be nonzero only if the total ghost number of the argument is equal to 3 (see for example the definition (\ref{vertices CFT def})) and therefore the factor $(-1)^{\Psi\Phi}$ is always equal to $+1$ in the bosonic theory. Another important property of the integration is that an integral of a derivative always equals to zero,
\begin{equation}
\int Q\Psi=0.
\end{equation}

Using the star product and the integration operation, we define an $n$-vertex, which maps $n$ string fields to a complex number,
\begin{equation}
\int\Psi_1\ast \Psi_2\ast \dots \ast \Psi_n \in \mathbb{C}.
\end{equation}
It follows from (\ref{vertex cyclicity}) that the $n$-vertex also has a cyclic symmetry
\begin{equation}
\int\Psi_1\ast \Psi_2\ast \dots \ast \Psi_n=\int\Psi_2\ast \dots \ast \Psi_n\ast \Psi_1 .
\end{equation}

Using these elements, we can write a cubic action for the bosonic open string field theory as
\begin{equation}\label{SFT action}
S =-\frac{1}{g_o^2} \int \left( \frac{1}{2}\Psi \ast Q\Psi + \frac{1}{3} \Psi\ast\Psi\ast\Psi \right),
\end{equation}
where $g_o$ is the open string coupling constant. The equations of motion derived from this action are
\begin{equation}\label{SFT equations}
Q\Psi+\Psi\ast\Psi=0.
\end{equation}
If we neglect the nonlinear term, this equation reproduces the physical state condition (\ref{BRST condition}).

The action (\ref{SFT action}) has a large amount of gauge symmetry, which is given by an infinitesimal transformation of the string field of the form
\begin{equation}\label{SFT symmetry}
\delta \Psi=Q\Lambda+(\Psi\ast\Lambda-\Lambda\ast\Psi),
\end{equation}
where $\Lambda$ is an arbitrary string field with ghost number 0. If we neglect the nonlinear terms, we can once again see a relation to the BRST quantization of string theory; this equation reproduces the equivalence relation (\ref{BRST equivalence}).

In this thesis, we consider mostly OSFT formulated only on a single D-brane background, but, for some purposes, it is also useful to understand how the theory looks on multiple D-branes. In that case, the string field gains a Chan-Patton-like matrix structure,
\begin{equation}
\Psi=\left(\begin{array}{ccc}
\Psi_{11} & \Psi_{12} & \dots \\
\Psi_{21} & \Psi_{22} &   \\
\vdots &  & \ddots
\end{array}\right),
\end{equation}
where $\Psi_{ij}$ belongs to the Hilbert space characterized by mixed boundary conditions $i$ and $j$. Therefore the off-diagonal part of the string field includes nontrivial boundary condition changing operators. The corresponding action acquires a trace over the matrix indices,
\begin{equation}\label{SFT action Tr}
S =-\frac{1}{g^2_o} \Tr \int \left( \frac{1}{2}\Psi \ast Q\Psi + \frac{1}{3} \Psi\ast\Psi\ast\Psi \right).
\end{equation}
After expanding the trace, we find that the action consists of a sum of terms of the form ${\int \Psi_{ij} \ast Q\Psi_{j\,i}}$ and $\int \Psi_{ij} \ast \Psi_{jk}\ast\Psi_{ki}$.

\subsection{Evaluation of the action}\label{sec:SFT:basic:action}
So far, we have defined the star product and the integration only using their algebraic properties, but we need an explicit realization of these operations for actual calculations. First, let us have a look at geometric interpretation of the star product.

\begin{figure}
\centering
\includegraphics[height=8cm]{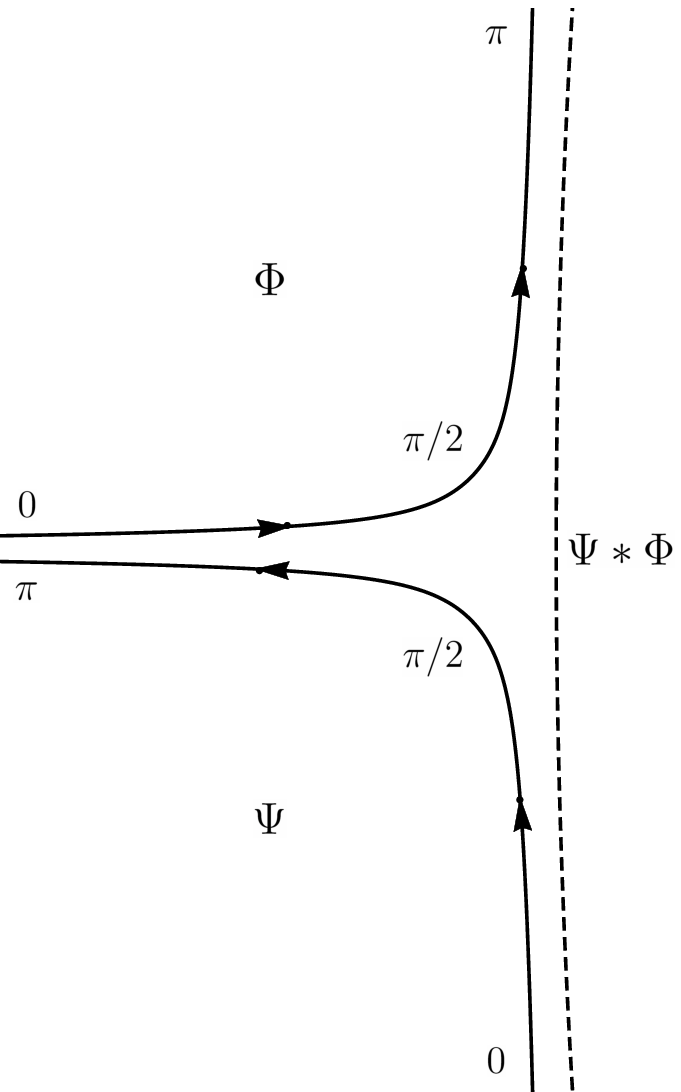}\hspace{4cm}
\includegraphics[height=8cm]{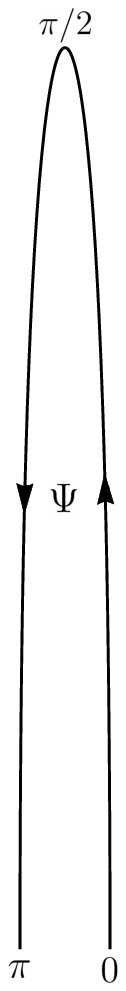}\hspace{3cm}
\caption{On the left: Schematic depiction of a star product of two string fields $\Psi\ast \Phi$. On the right: Schematic depiction of the integration operation  $\int\Psi$.}
\label{fig:star product}
\end{figure}

We can imagine that a string field represents worldsheet of a string propagating from time $\tau=-\inf$ to $\tau=0$, when it interacts with other strings. The star product, which represents the interaction, is interpreted as gluing two strings together. We take the right part ($\pi/2\leq\sigma\leq\pi$) of the first string and glue it together with the left part ($0\leq\sigma\leq\pi/2$) of the second string. Such operation can be done repeatedly and it is clearly associative. Similarly, the integration is represented by gluing the left and the right part of a string together. Finally, the $n$-vertex is constructed by repeated multiplication followed by gluing of the two remaining half-strings together. See figure \ref{fig:star product} for illustration of these operations.

To formalize these ideas, we use tools from conformal field theory. First, we are going formulate string field theory using conformal maps and correlation functions and later, we will discuss an alternative operator formulation.

\begin{figure}
\centering
\includegraphics[width=9cm]{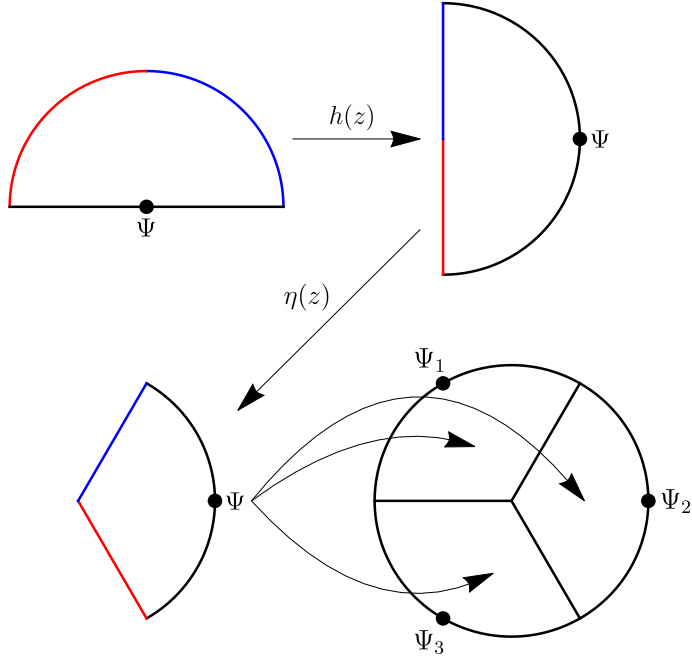}
\caption{Map of three half-disks on a full disk. We denote the left and the right edge of the string worldsheet by blue and red colors respectively.}
\label{fig:threevertex1}
\end{figure}

\begin{figure}
\centering
\includegraphics[width=12cm]{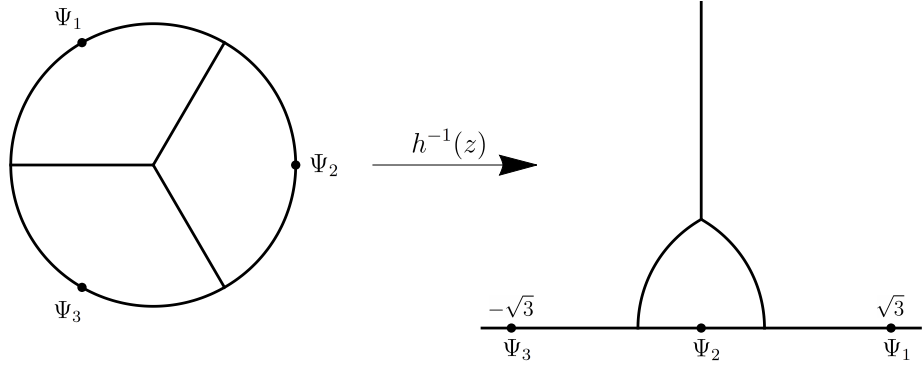}
\caption{Map of a unit disk with three insertions on the upper half-plane.}
\label{fig:threevertex2}
\end{figure}

A string field will be represented by a unit half-disk with a vertex operator inserted at the origin. In order to define the $n$-vertex, we need to map all the half-disks to the UHP in a way which respects the cyclic symmetry of the vertex. The first step is to map the half-disks on a full disk. First, we consider the transformation
\begin{equation}
h(z)=\frac{1+iz}{1-iz},
\end{equation}
which maps a half-disk back on a half-disk and which exchanges the straight and the curved boundary (see figure \ref{fig:threevertex1}). It also maps the common interaction point $z=i$ to the origin of the new coordinates. Next, we have to shrink the half-disks so that they can fit into the full disk. This can be done by the transformation
\begin{equation}
\eta(z)=z^{2/n}.
\end{equation}
Finally, we have to rotate the surfaces to avoid overlap. By composition of all three transformations, we obtain maps\footnote{There are two conventions for these conformal transformations, which differ by sign of the complex phase. We follow the conventions of \cite{RastelliZwiebach}, where the insertions are positioned clockwise around the unit circle. Distinguishing between the two conventions is important if the boundary structure constants are not fully symmetric, then one has to make sure that the order of insertions is the same as in the definition of boundary OPE (\ref{Boundary OPE}).}
\begin{equation}
g_k(z_k)=e^{-\frac{2(k-1)\pi i}{n}}\left(\frac{1+iz_k}{1-iz_k}\right)^{2/n}.
\end{equation}
Now we can write the $n$-vertex as a correlator on the unit disk:
\begin{equation}
\int\Psi_1\ast \Psi_2\ast \dots \ast \Psi_n=\la g_1\circ \Psi_1(0) g_2\circ\Psi_2(0) \dots g_n\circ\Psi_n(0)\ra_{disk}.
\end{equation}
This representation makes the cyclic symmetry of the vertex manifest, but, for practical computations, it is more convenient to work with correlators on the UHP. The disk can be mapped back on the UHP using the inverse of the $h$ transformation (see figure \ref{fig:threevertex2}),
\begin{equation}
h^{-1}(z)=-i\frac{z-1}{z+1}.
\end{equation}
Then we find
\begin{equation}\label{vertices CFT def}
\int\Psi_1\ast \Psi_2\ast \dots \ast \Psi_n=\la f_1\circ \Psi_1(0) f_2\circ\Psi_2(0) \dots f_n\circ\Psi_n(0)\ra_{UHP},
\end{equation}
where $f_k(z_k)=h^{-1}(g_k(z_k))$. There is an explicit formula for the functions $f_k$, which reads
\begin{equation}\label{vertices general map}
f_k(z_k)=\tan\left(\frac{2}{n}\arctan z+ \frac{(k-1)\pi}{n}\right).
\end{equation}
It can be derived by mapping the vertex on a cylinder instead of the disk in the intermediate step.

In the cubic theory, we usually encounter only vertices with one to three entries. The 1-vertex is essentially just an alternative definition of the integration operation. It is defined by the conformal map
\begin{equation}\label{identity map}
f_1(z)\equiv f_I(z)=\frac{2z}{1-z^2}
\end{equation}
and, using this function, the integration can be written as
\begin{equation}
\int \Psi=\la f_I\circ \Psi(0)\ra.
\end{equation}
The function $f_I(z)$ also defines the so-called identity string field, which we are going to introduce in subsection \ref{sec:SFT:basic:wedge}. Using this string field, we can alternatively write the integration as
\begin{equation}
\int \Psi=\la I|\Psi\ra,
\end{equation}
where we use the BPZ product instead of a correlation function. We never need to use the 1-vertex explicitly in this thesis, but it will play a role in the construction of Ellwood invariants in section \ref{sec:SFT:observables:Ellwood}.

The 2-vertex mainly appears in the kinetic term of the OSFT action. The conformal maps for $n=2$ are
\begin{eqnarray}
f_1(z_1) &=& z_1, \nn \\
f_2(z_2) &=& -\frac{1}{z_2}.
\end{eqnarray}
Therefore we observe that the 2-vertex reduces just to the usual BPZ product and we do not have to introduce any new formalism to compute it,
\begin{equation}
\int\Psi_1\ast \Psi_2=\la \Psi_1|\Psi_2\ra.
\end{equation}

Finally, we get to the cubic vertex, which defines the interaction term of the OSFT action. The three maps that define the vertex are given by\footnote{We use a cyclic rotation of the functions (\ref{vertices general map}) to match the conventions from \cite{RastelliZwiebach}.}
\begin{eqnarray}\label{vertex map3}
f_1(z_1)&=& \tan\left(\frac{2}{3} \arctan z_1 +\frac{\pi}{3}\right),\nn \\
f_2(z_2)&=& \tan\left(\frac{2}{3} \arctan z_2\right), \\
f_3(z_3)&=& \tan\left(\frac{2}{3} \arctan z_3-\frac{\pi}{3}\right)\nn.
\end{eqnarray}
Assuming that all three string fields are just primary operators, which transform as $f_k\circ\Psi_k(0)=(f_k'(0))^{h_k}\Psi_k(f_k(0))$, we can compute the cubic vertex directly
\begin{eqnarray}
\int\Psi_1\ast \Psi_2\ast \Psi_3&=&\la f_1\circ \Psi_1(0) f_2\circ\Psi_2(0) f_3\circ\Psi_3(0)\ra\nn\\
&=&\left(\frac{8}{3}\right)^{h_1}\left(\frac{2}{3}\right)^{h_2}\left(\frac{8}{3}\right)^{h_3} \la \Psi_1(\sqrt{3}) \Psi_2(0) \Psi_3(-\sqrt{3})\ra \nn\\
&=& C_{123}K^{-h_1-h_2-h_3}, \label{vertex primaries}
\end{eqnarray}
where $C_{123}$ is the boundary 3-point structure constant from (\ref{three-point boundary}) (with suppressed boundary indices) and $K$ is a constant defined as
\begin{equation}
K\equiv \frac{3\sqrt{3}}{4}.
\end{equation}
Unfortunately, this method of evaluation of the cubic vertex is not very practical for secondary fields, which usually have complicated transformation laws.

Another option how to compute string field theory vertices is the operator approach. The interaction of $n$ string fields in this approach is expressed as
\begin{equation}
\int\Psi_1\ast \dots \ast \Psi_n=\la V_n|\Psi_1\ra \dots |\Psi_n\ra,
\end{equation}
where the representative of the $n$-vertex $\la V_n|$ belongs to the $n$-fold tensor product of the dual Hilbert space, $\la V_n|\in \underbrace{\hh^\ast\otimes \dots \otimes \hh^\ast}_{n}$.

In this formalism, the 1-vertex is equal to the aforementioned identity string field, $\la V_1|\equiv\la I|$ and the 2-vertex is just an alternative representation the BPZ product. Therefore we will use this formalism mainly for the cubic vertex. This representation of the cubic vertex has the advantage that it allows us to write an explicit formula for the star product:
\begin{equation}
\Psi_1 \ast \Psi_2=\bpz(\la V_3|\Psi_1\ra|\Psi_2\ra).
\end{equation}

In the standard theory on a space filling D-brane, the cubic vertex can be explicitly written in terms of oscillators. The zero momentum part of the vertex is given by
\begin{eqnarray}
\la V_3|_{k_i=0} &=& \nnn (\la 0|c_{-1}c_0)^{(1)}\otimes(\la 0|c_{-1}c_0)^{(2)}\otimes(\la 0|c_{-1}c_0)^{(3)}  \\
 &\times& \exp\left(\frac{1}{2} \sum_{r,s=1}^3\sum_{m,n=0}^\inf \alpha_m^{\mu(r)} N_{mn}^{rs} \alpha_n^{\nu(s)}\eta_{\mu\nu}+
 \sum_{r,s=1}^3\sum_{\substack{m=0, \\n=1}}^\inf b_m^{(r)} X_{mn}^{rs} c_n^{(s)}\right),  \nn
\end{eqnarray}
where the constants $N_{mn}^{rs}$ and $X_{mn}^{rs}$ are called Neumann coefficients. They can be explicitly computed using the conformal maps (\ref{vertices general map}), see for example \cite{TaylorZwiebach}.

However, this representation of the cubic vertex is not well suited for practical computations either. The main reason is that the cubic vertex cannot be easily transformed to other bases which we have to use in various BCFTs, most notably to the Virasoro basis. Therefore we will borrow the notation from this approach, because it is more practical to treat the string field as a state than as a field, but in the actual computations, we will use the conservation laws for the cubic vertex, which we are going to introduce in section \ref{sec:SFT:cubic cons}.

\subsection{Level truncation}\label{sec:SFT:basic:level}
The string field has infinite number of degrees of freedom, so if we want to solve the equations of motion (\ref{SFT equations}), we have to reduce the number of degrees of freedom to a manageable amount. There are two very different approaches to this problem. The first is a so-called level truncation approach, in which we truncate the string field (\ref{string field expansion}) to a finite number of terms and solve the corresponding equations numerically. The other possibility is to solve the equations of motion analytically using a set of coherent states that form a closed algebra under the star product. In this thesis, we focus on the level truncation approach, analytic methods will be mentioned only briefly in subsection \ref{sec:SFT:basic:wedge}.

In the level truncation approach, we expand the string field in eigenstates of the total Virasoro operator $L_0^{tot}$. Following \cite{MSZ lump}, we define the level $L$ of a state $|\phi\ra$ as the $L_0^{tot}$ eigenvalue plus one:
\begin{equation}
L|\phi\ra=(L_0^{tot}+1)|\phi\ra.
\end{equation}
The shift by one is introduced to compensate for the $L_0^{tot}$ eigenvalue of the ghost number one ground state $c_1|0\ra$, which we want to have level 0. In a similar way, we also define level in the constituent BCFTs; the level in matter theories is given just by the $L_0^m$ eigenvalue, the shift by one appears only in the ghost BCFT. The level is additive under tensor product of the individual BCFTs.

The definition of level in string field theory differs from the CFT definition of level, which we mentioned in section \ref{sec:CFT:Bulk:states}. In the context of string theory, the CFT level is usually called number operator and denoted as $N$. The two quantities differ by the weight of the primary operator included in the state $|\phi\ra$, $L=N+h$. The number operator also plays a role in string field theory, for example in the twist symmetry.

We can also define level of an $n$-vertex contracted with string fields of definite level as the sum of levels of all the involved string fields. This applies mainly to the cubic vertex because the 2-vertex is nontrivial only if both states have the same level.

Truncation of a string field to level $L$ means that we reduce the infinite sum over fields only to terms which have level lower or equal to $L$. Truncation of the OSFT action is characterized by two numbers $(L,M)$, where $L$ is the maximal allowed level for the string field and $M$ is the maximal allowed level for the vertices. It is reasonable to choose $M$ only from the interval $2L\leq M\leq3L$. If we choose $M<2L$, we would miss part of the kinetic term and there no terms with level higher than $3L$ in the cubic action.

In the literature, we usually encounter either $(L,2L)$ or $(L,3L)$ approximations. The $(L,2L)$ approximation scheme was mostly used in earlier works, for example \cite{SenZwiebachTV}\cite{MSZ lump}\cite{MoellerTaylorTV}, because it is computationally less demanding. However, the $(L,3L)$ scheme offers a better precision of results \cite{GaiottoRastelli}\cite{KMS}, so, in this thesis, we show results exclusively in this scheme. Considering the structure of our algorithms, the $(L,2L)$ scheme would not lead to a dramatic acceleration of our calculations anyway and it would make our algorithms more complicated.

\subsection{Twist symmetry}\label{sec:SFT:basic:twist}
In addition to the gauge symmetry, the string field theory has a second symmetry of different nature. It is a $\mathbb{Z}_2$ symmetry called the twist symmetry and denoted by $\Omega$. It corresponds to reversal of orientation of all strings, which is generated by the worldsheet transformation $\sigma\rar \pi-\sigma$. Therefore it relates a star product of two string fields to the star product in the opposite order
\begin{equation}
\Omega(\Psi_1\ast\Psi_2)\sim(\Omega\Psi_2)\ast(\Omega\Psi_1).
\end{equation}
The action of the twist symmetry depends on properties of the particular string field theory, so we will describe only the most common form of the symmetry, which nevertheless covers all backgrounds that we encounter in this thesis. First, we restrict the background only to a single D-brane to avoid boundary condition changing operators. Then we assume that modes of all operator are labeled by integers and that the boundary structure constants satisfy\footnote{This property is not satisfied in theories with cocycles, where we find
\begin{equation}
C_{123}=(-1)^{\omega(\phi_1,\phi_2)}C_{321},\nn
\end{equation}
where $(-1)^{\omega(\phi_1,\phi_2)}$ is a non-factorizable complex phase. An example of such theory is the free boson theory on a D-brane with magnetic flux. In theories of this type, we cannot split the string field into twist even and twist odd part in the usual sense.}
\begin{equation}\label{twist 3-point}
C_{123}=(-1)^{\phi_1}(-1)^{\phi_2}(-1)^{\phi_3}C_{321},
\end{equation}
where $(-1)^{\phi_i}=\pm 1$ is the twist parity of primary operators. Then we find
\begin{equation}\label{twist star product}
\Omega(\Psi_1\ast\Psi_2)=(-1)^{\Psi_1\Psi_2+1}(\Omega\Psi_2)\ast(\Omega\Psi_1),
\end{equation}
where the action of the twist operator on a string field is given by
\begin{equation}\label{twist state}
\Omega\Psi=(-1)^{N+\phi}\Psi,
\end{equation}
where $N$ is the number operator.

In the matter theories that we consider in this thesis, which are the free boson theory with Neumann boundary conditions and the Virasoro minimal models, the factor $(-1)^{\phi}$ equals to $+1$ for all fundamental primary states, so it will not affect our calculations. The ground state in the ghost theory, $c_1|0\ra$, must be twist even, which makes the ghost SL$(2,\mathbb{R})$ vacuum state twist odd.

As a consequence of the twist symmetry, we find that the 2-vertex and the 3-vertex have the following properties:
\begin{equation}
\la \Psi_1,\Psi_2\ra=\la \Omega\Psi_2,\Omega\Psi_1\ra,
\end{equation}
\begin{equation}
\la \Psi_1,\Psi_2,\Psi_3\ra=(-1)^{\Psi_1\Psi_2+1}\la \Omega\Psi_3,\Omega\Psi_2,\Omega\Psi_1\ra.
\end{equation}
By substitution of (\ref{twist state}) into the second equation, we find
\begin{equation}\label{twist vertex}
\la \Psi_1,\Psi_2,\Psi_3\ra=(-1)^{N_1+N_2+N_3+\phi_1+\phi_2+\phi_3+\Psi_1\Psi_2+1}\la\Psi_3,\Psi_2,\Psi_1\ra.
\end{equation}
We will prove this formula in section \ref{sec:SFT:cubic cons} using properties of the cubic vertex conservation laws.

The twist symmetry naturally introduces decomposition of the string field into twist even part and twist odd part,
\begin{equation}
\Psi=\Psi_e+\Psi_o,
\end{equation}
which satisfy
\begin{equation}
\Omega\Psi_e=\Psi_e,\qquad \Omega\Psi_o=-\Psi_o.
\end{equation}
By combining the twist and the cyclic symmetry of the vertices, we find that twist odd states must always appear in the action in pairs, otherwise their contribution disappears. Concretely, the cubic vertex satisfies
\begin{equation}
\la \Psi_e,\Psi_e,\Psi_o\ra=\la \Psi_o,\Psi_o,\Psi_o\ra=0.
\end{equation}
This property implies that we can consistently set the twist odd part of the string field to zero. We can show this using a more general argument.

Let's assume that the string field can be split as $\Psi=\Psi_1+\Psi_2$, where $\Psi_2$ does not appear in the action linearly. The variation of the action with respect to $\Psi_2$ gives us an equation of the form
\begin{equation}
Q\Psi_2+\{\Psi_2,\Psi_1\}_\ast+\Psi_2\ast\Psi_2=0.
\end{equation}
This equation is proportional to $\Psi_2$ and therefore it is identically satisfied if we set $\Psi_2=0$.

Most of known numerical solutions are twist even, so we usually work in the twist even ansatz, which simplifies the calculations. However, some theories include important states which do not survive the twist even projection, for example the state $\alpha_{-1}c_1|0\ra$ which is needed for $\del X$ marginal deformations. In such cases, we can define a deformed twist symmetry by combining the twist symmetry with some other $\mathbb{Z}_2$ symmetry of the theory $S$. In this particular example of $\del X$ marginal deformations, $S$ is given by the spacetime parity (see section \ref{sec:SFT:string field:FB}). The new twist symmetry is given by $\Omega'=S\Omega$. The action of the new symmetry is different from (\ref{twist state}), but it still satisfies the basic property (\ref{twist star product}). Therefore we can look for solutions which are even with respect to the new symmetry.

\subsection{Reality conditions}\label{sec:SFT:basic:reality}
The string field contains an infinite tower of fields, which, in general, have complex coefficients. However, physical solutions of the OSFT equations of motion must have a real action. In order to identify such solutions, we introduce a complex conjugation. We define it as the combination of the Hermitian and the BPZ conjugation:
\begin{equation}\label{complex conjugation SFT}
\Psi^\ast={\rm bpz}({\rm hc}(\Psi))={\rm hc}({\rm bpz}(\Psi)).
\end{equation}
The reality condition then reads $\Psi^\ast=\Psi$ and we can use it to find reality conditions for individual components of the string field.

In many simple cases, which essentially cover all setting that we are going to introduce in section \ref{sec:SFT:string field}, the complex conjugation acts as
\begin{equation}\label{complex conjugation univ}
\Psi^\ast=(-1)^N\Psi.
\end{equation}
Assuming that $(-1)^\phi=1$ for all primary fields, this equation is very similar to (\ref{twist state}). That means that the complex conjugation act is the same way as twist conjugation, $\Psi^\ast=\Omega\Psi$. Therefore a real string field has real coefficients in the twist even part and purely imaginary coefficients in the twist odd part. Moreover, since we usually work with twist even string fields, the reality condition often reduces just to reality of all string field coefficients.

However, while it is guaranteed that a real string field has real action, the opposite is not true. We are going to see several solutions that have real action although they do not satisfy the reality condition. Such string fields usually satisfy a reality condition modified by some $\mathbb{Z}_2$ symmetry, $\Psi^\ast=S\Psi$. We denote such string fields as pseudo-real. As an example, we can consider a twist non-even string field with real coefficients. The twist odd part of this string field is purely imaginary with respect to the usual reality condition, but it is easy to check that it has a real contribution to the action because it appears only quadratically.

The difference between real and pseudo-real solutions often shows in Ellwood invariants, which are defined in section \ref{sec:SFT:observables:Ellwood}. These observables are linear functions of the string field and therefore they are more sensitive to its imaginary part. Determining reality of the string field from observables is also a preferable option when direct check of the reality condition is nontrivial, for example when we work with complex symmetry generators (like $J^{\pm}$ in SU(2) WZW model) or when the string field is real only up to a null state.

\subsection{Wedge states and the $KBc$ algebra}\label{sec:SFT:basic:wedge}
In this subsection, we briefly describe some analytical aspects of string field theory. We introduce the so-called wedge states and the $KBc$ algebra, which is used to construct most of analytic solutions. For more information about this subject, see for example \cite{WedgeSchnabl}\cite{AnalyticSolutionSchnabl}\cite{AnalyticSolutionOkawa}\cite{FuchsKroyterL0}\!\! \cite{ErlerKBC1}\cite{ErlerKBC2}\cite{FuchsKroyterAnalytic}\cite{SchnablLightningReview}\cite{OkawaLectures}\cite{ErlerLectures}.

Recall that a conformal transformation $f(z)$ acts on a primary field $\phi$ as
\begin{equation}
f\circ \phi (z)=\left(\frac{df}{dz}\right)^h \phi(f(z)).
\end{equation}
We can rewrite this equation as
\begin{equation}
f\circ \phi (z)=U_f\phi(z) U_f^{-1},
\end{equation}
where the operator $U_f$ is given by an exponential of the Virasoro generators,
\begin{equation}
U_f=\exp\left(\sum_{n\geq 0} v_n L_n\right).
\end{equation}
The coefficients $v_n$ can be uniquely determined from the map $f$.

To every conformal map $f$, we can associate a state $\la f|$, which is defined by
\begin{equation}
\la f|\phi\ra=\la f\circ\phi(0) \ra.
\end{equation}
Such states are known as surface states and they are related to the previously defined operators $U_f$ as
\begin{equation}
\la f|=\la 0| U_f.
\end{equation}

A subset of surface states related to transformations\footnote{Notice a shift of $r$ by 1 compared to (\ref{vertices general map}) and \cite{WedgeSchnabl}. Using this notation, $r$ matches the geometrical width of a wedge state.}
\begin{equation}
f_r(z)=\tan\left(\frac{2}{r+1}\arctan(z)\right)
\end{equation}
is called wedge states. These functions coincide with (\ref{vertices general map}) for integer $r$, but we can take $r$ to be an arbitrary nonnegative number. We denote the wedge states as
\begin{equation}
\la r|_W\equiv\la f_{r}|=\la 0|U_{f_{r}},\quad r\geq 0.
\end{equation}
The ket form of wedge states is
\begin{equation}
|r\ra_W=U_{f_{r}}^\ast |0\ra,
\end{equation}
where $U_{f_{r}}^\ast$ denotes the BPZ conjugation of $U_{f_{r}}$. The notation of these operators is usually simplified as $U_r\equiv U_{f_{r}}$.

The star product of wedge states can be explicitly computed and it has a very simple form
\begin{equation}
|r\ra_W \ast |s\ra_W=|r+s\ra_W.
\end{equation}
From this equations, it follows that
\begin{equation}
|0\ra_W \ast |r\ra_W=|r\ra_W.
\end{equation}
Therefore the state $|0\ra_W$ acts as the identity element in the wedge state algebra. In fact, it acts as the identity element in the whole star product algebra. Therefore it is called the identity string field and denoted as
\begin{equation}\label{identity definition}
|I\ra\equiv |0\ra_W.
\end{equation}
The conformal transformation associated with the identity string field is given by (\ref{identity map}).

The wedge state $|1\ra_W$ equals to the vacuum state $|0\ra$. Therefore we can write wedge states with integer $r$ as $r$-th power of the vacuum state,
\begin{equation}
|r\ra_W=\underbrace{|0\ra\ast\dots \ast|0\ra}_{r}.
\end{equation}

An important property of wedge states is that they are annihilated by the BRST operator,
\begin{equation}\label{wedge and Q}
Q|r\ra_W=0.
\end{equation}
This follows from the fact that wedge states are made out of only the total Virasoro operators, which commute with $Q$.

Wedge states can be generalized by insertion of primary operators,
\begin{equation}
U_r^\ast \phi(x)|0\ra.
\end{equation}
Wedge states with insertions also have a relatively simple star product
\begin{equation}
U_r^\ast \phi_1(x)|0\ra\ast U_s^\ast \phi_2(y)|0\ra=U_{r+s}^\ast\, g_1\circ\phi_1(x)\,g_2\circ \phi_2(y)|0\ra,
\end{equation}
where
\begin{eqnarray}
g_1(x)=\cot \left(\frac{r}{r+s-1}\left(-\arctan x+\frac{\pi}{2}\right)\right),\nn \\
g_2(y)=\cot \left(\frac{s}{r+s-1}\left(-\arctan y-\frac{\pi}{2}\right)\right).
\end{eqnarray}

\begin{figure}
\centering
\includegraphics[width=8cm]{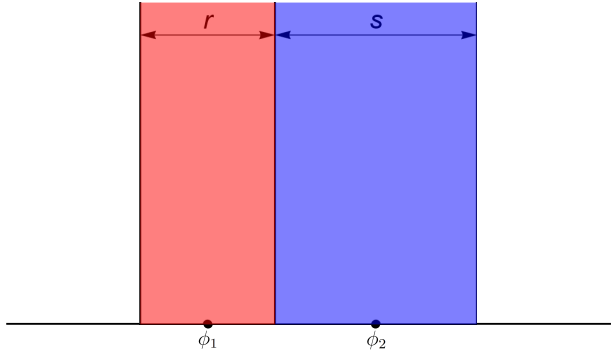}
\caption{Star product of wedge states with insertions in the sliver frame, $U_r^\ast \phi_1(x)|0\ra\ast U_s^\ast \phi_2(y)|0\ra$.}
\label{fig:wedge}
\end{figure}

The wedge state algebra can be nicely visualized in the so-called sliver frame, which is given by the conformal map
\begin{equation}\label{sliver frame map}
f_S(z)=\frac{2}{\pi}\arctan z.
\end{equation}
In this frame, wedge states are represented by strips of width $r$ and the star product by gluing of these strips (see figure \ref{fig:wedge}). The integration operation is implemented by gluing the edges of the strip together to form a cylinder, which can be then scaled to a canonical circumference and mapped to the UHP.

Wedge states can be expressed using a string field $K$ as
\begin{equation}
|r\ra_W=\exp(-r K),
\end{equation}
where the exponential is defined using the star product. The string field $K$ is defined as a line integral in the sliver frame:
\begin{equation}
K=\int_{-i\inf}^{i\inf}\frac{d\tilde z}{2\pi i}\tilde T(\tilde z)|I\ra,
\end{equation}
where $\tilde z$ is the sliver frame coordinate and $\tilde T(\tilde z)$ is the full energy-momentum tensor in this frame. By mapping this expression back to the UHP, we get
\begin{equation}
K=\frac{\pi}{2}\int_{\mathcal{C}_L}\frac{dz}{2\pi i}(1+z^2)T(z)|I\ra,
\end{equation}
where $\mathcal{C}_L$ is the left half of the unit circle.

Similarly to $K$, we define a string field $B$
\begin{equation}
B=\int_{-i\inf}^{i\inf}\frac{d\tilde z}{2\pi i}\tilde b(\tilde z)|I\ra
\end{equation}
and also a string field $c$
\begin{equation}
c=\tilde c\left(\frac{1}{2}\right)|I\ra,
\end{equation}
where tildes denote tensor fields in the sliver frame.

These three string fields have the following (anti)commutators under the star multiplication:
\begin{eqnarray}
[K,B]&=&0, \nn\\
\{B,c\}&=&1, \\
B^2=c^2&=&0\nn
\end{eqnarray}
and they also satisfy
\begin{eqnarray}
QK&=&0,\nn \\
QB&=&K, \\
Qc&=&cKc.\nn
\end{eqnarray}
These algebraic relations are known as the $KBc$ algebra, which is the fundamental tool for construction of analytic solutions.

\subsection{Solving the equations of motion}\label{sec:SFT:basic:eom}
As we mentioned before, there are two vastly different approaches to string field theory, which use different techniques to solve the equations of motion. In the analytic approach, solutions are usually expressed in terms of the $K$, $B$ and $c$ fields, which we introduced in the previous subsection, and they can be found just by algebraic manipulations with these fields. However, in this thesis, we focus on the numerical approach. We use the level truncation scheme, which reduces the exact equations of motion (\ref{SFT equations}) to a finite system of quadratic equations. These equations can be then solved by various numerical techniques. The solutions of the truncated equations obviously do not solve the full equations, but it is believed that stable numerical solutions approach exact solutions in the infinite level limit. In this subsection, we briefly describe our general strategy for solving the equations of motion, detailed numerical algorithms can be found in sections \ref{sec:Numerics:Newton} and \ref{sec:Numerics:homotopy}.

The first issue we have to deal with is the gauge fixing. As we mentioned earlier, the string field theory action has the gauge symmetry
\begin{equation}
\delta \Psi=Q\Lambda+(\Psi\ast\Lambda-\Lambda\ast\Psi).
\end{equation}
These gauge transformations do not commute with $L_0^{tot}$. As a result, the level truncated action loses the exact gauge symmetry. Naively, it looks like an advantage because the truncated equations have only a discreet set of solutions, but the opposite is true. There are still approximate remnants of the gauge symmetry, which cause instabilities in the level truncation scheme and prevent us from finding solutions at higher levels (see section \ref{sec:universal:no gauge} for illustration). Therefore it is necessary to impose some gauge fixing conditions, we consider linear conditions of the form
\begin{equation}
\mathcal{G}\Psi=0.
\end{equation}
The gauge fixing reduces the number of degrees of freedom of the string field. Together with the broken gauge symmetry, it means that we end up with an overdetermined system of equations, which is generally unsolvable. The usual approach to this problem is to solve only a subset of the full equations of motion. The remaining equations are left unsolved, but, for consistent solutions, they must asymptotically approach zero in the infinite level limit.
An alternative approach, which has not been attempted so far, would be to find a string field that minimizes violation of the full equations of motion.

The most convenient gauge for numerical calculations is called Siegel gauge and its gauge condition reads
\begin{equation}\label{Siegel gauge}
b_0|\Psi\ra=0.
\end{equation}
This gauge is particularly useful because the gauge condition commutes with $L_0^{tot}$ and therefore it can be implemented independently level by level. The implementation is very simple, it can be realized just by removing states including $c_0$ from the spectrum or by using only SU(1,1) singlets in the ghost sector (see section \ref{sec:SFT:string field:ghost}).

It can be shown that Siegel gauge fully fixes the linearized gauge transformations $\delta \Psi=Q\Psi$. However, there is no guarantee that every string field can be brought to Siegel gauge by the full nonlinear transformation (see \cite{EllwoodTaylorGauge}). Therefore it is possible that some D-brane configurations do not have a corresponding solution in Siegel gauge. There is also a possibility that some D-brane configurations can have multiple representatives in Siegel gauge\footnote{There seem to be such solutions in OSFT formulated on a D2-brane with flux \cite{KudrnaVosmera}.}.

Analytic solutions use different types of gauges, for example Schnabl gauge \cite{AnalyticSolutionSchnabl}
\begin{equation}\label{Schnabl gauge}
\mathcal{B}_0\Psi=\left(b_0+\sum_{k=1}^\inf\frac{2(-1)^{k+1}}{4k^2-1}\right)\Psi=0.
\end{equation}
Some analytic solutions can be even constructed without an explicit specification of gauge conditions.

After we deal with implementation of the gauge fixing, we need to find solutions that are relevant for the given physical problem. The string field theory equations form a system of coupled quadratic equations, which have generically $2^N$ solutions for $N$ equations. Therefore finding and analyzing all solutions of thousands of equations is simply out of question\footnote{For illustration, let us estimate how many equations could be solved using all the world's computer resources. We can make a simple order estimate using the memory required to store the solutions. Let's say that the Earth's population is 8 billion and that there is 1 TB of memory for every person. That gives us approximately $8\dexp{21}$ B of memory. Storing one complex solution requires $16N$ B of memory and therefore all solutions require $16N2^N$ B. By comparison, we get that it would be possible to solve approximately 63 equations.}. Instead, we focus only on one or few interesting solutions and ignore the rest. A very efficient tool for that is the well known Newton's method. It is an iterative algorithm, which allows us to find a solution of nonlinear equations given a seed which is close to the solution.

This leads us to a problem how to find correct seeds for Newton's method. It is believed that physically relevant solutions in string field theory look similar when truncated to different levels and therefore a solution from a lower level is usually a good seed for a solution at higher level. However, every solution has obviously a minimal level below which it does not exists, so we still need a tool to find seeds at the minimal level.

In some simple cases, for example for the tachyon vacuum solution, the seed can be guessed, but in more complicated theories, where we expect many solutions corresponding to various boundary states, finding all interesting solutions by a guess is not possible. Therefore we use a different numerical method, which is called the homotopy continuation method, to find all solutions at some low initial level. Then we analyze these solutions and we pick the interesting ones as seeds for Newton's method. The initial level cannot be very high. In the simplest possible case, which is the twist even universal string field in Siegel gauge, we have managed to solve the equations of motion up to level 6, but in more complicated theories, we usually work with the initial level between 1 and 3.

To summarize, our setup to find and identify solutions is
\begin{enumerate}
   \item We choose a low initial level and find all solutions at this level using the homotopy continuation method.
   \item We select interesting solutions based on criteria relevant for the given physical problem (gauge invariant observables, symmetries, etc.).
   \item We improve the chosen solutions to higher levels using Newton's method.
   \item We make final identification of these solutions using various observables, which we usually extrapolate to infinite level for improved precision.
\end{enumerate}

\section{String field in various backgrounds}\label{sec:SFT:string field}
In the previous section, we have reviewed basic properties of open string field theory, but we have been vague about the actual content of the string field. In this section, we discuss structure of the string field in more detail and we introduce several backgrounds that will appear in our calculations. See also appendix \ref{sec:characters}, where we write down characters for the individual theories.

\subsection{Structure of the ghost theory}\label{sec:SFT:string field:ghost}
First, let us take a better look at the structure of the ghost states space, which is common to all string field theories. We follow the reference \cite{KudrnaUniversal}. As we already mentioned in section \ref{sec:CFT:ghost}, a generic element of the ghost state space is spanned by modes of the $b$ and $c$ ghosts:
\begin{equation}
b_{-k_1}\dots b_{-k_m}c_{-l_1}\dots c_{-l_n}|0\ra, \quad k_i\geq2,\ l_i\geq-1.
\end{equation}
We denote this basis as the $bc$ basis. This basis is quite convenient for some purposes. The modes $b_k$ and $c_k$ have simple (anti)commutators between themselves and with other ghost operators, which simplifies many calculations. We can also easily implement the Siegel gauge condition just by removing the $c_0$ mode from the spectrum, which allows us to avoid use of the projector that we will introduce in subsection \ref{sec:Numerics:Newton:gauge}.

However, it also has one disadvantage. Manipulations with the modes $b_k$ and $c_k$ (for example when using cubic vertex conservation laws) change ghost numbers of states. The physical string field has ghost number 1, so when we work with this basis, we have to introduce various auxiliary objects with other ghost numbers. Therefore, for some purposes, it is more convenient to use other bases.

The character (\ref{character ghost 1}) suggests that the ghost state space at a fixed ghost number $g$ can be reinterpreted as a Virasoro Verma module over a primary of weight $\frac{g^2-3g}{2}$. These primaries are given by
\begin{equation}
|g\ra=\begin{cases}
   c_{-g+2}\dots c_1|0\ra\quad\ \ \, {\rm for}\ g>0, \\
   |0\ra   \qquad\qquad\qquad\ \,{\rm for}\  g=0, \\
   b_{g-1}\dots b_{-2}|0\ra\quad\ \ \,{\rm for}\  g<0.
\end{cases}
\end{equation}
However, by analyzing the Kac determinant (\ref{Kac determinat}) for the central charge $c^{gh}=-26$, we find that these Verma modules are not irreducible at ghost numbers ${g=0}$ mod 3. For ghost numbers $g=0,-3,-6,\dots$,  null states disappear when expressed it terms of the $b$ and $c$ ghosts. The states that appear instead of the null states are neither primaries nor descendants with respect to the ghost Virasoro algebra. An example of such state is $b_{-2}c_1|0\ra$, which appears instead of $L^{gh}_{-1}|0\ra$. This state is clearly not a Virasoro descendant and, since it is not annihilated by $L^{gh}_{1}$, it is not a primary either. The 'null states' at ghost numbers $g=3,6,9,\dots$ have unusual properties. They vanish when contracted with Virasoro primaries or descendants, but they are nonzero when expressed it terms of the $b$ and $c$ modes. Such strange structure is possible because the ghost CFT is not unitary.

These properties make the Virasoro basis impractical except for ghost number 1. Fortunately, there is a similar basis which does suffer from these problems. It is spanned by modes of the ghost current:
\begin{equation}
j^{gh}_{-k_1}j^{gh}_{-k_2}\dots j_{-k_n}^{gh}|g\ra, \quad k_i\geq1,
\end{equation}
where the primary states $|g\ra$ are the same as for the Virasoro basis. This state space has the correct character and one can easily check that there are no null states because the corresponding Gram matrix is diagonal.

In summary, there are three possible bases in the ghost sector: the $bc$ basis, the Virasoro basis and the ghost current basis. The $bc$ basis allows a simple implementation of Siegel gauge (and somewhat simpler implementation of Schnabl gauge), but calculations in this basis it require various auxiliary objects. The Virasoro basis and the ghost current basis allow simpler computation of vertices, but they are not compatible with Siegel gauge. Out of these two, the ghost current basis is more universal because it works well at any ghost number.

Now, let us focus on Siegel gauge, which is used in most of OSFT calculations. The $bc$ basis works without any problems in this gauge, however the ghost theory possesses an additional structure, which we can use to our advantage. In \cite{ZwiebachSU11}, it was found that the string field in Siegel gauge has a continuous SU(1,1) symmetry generated by
\begin{equation}
J_3 = \frac{1}{2}\sum_{n=1}^\infty \left(c_{-n} b_{n}-b_{-n} c_{n}\right),
\quad J_{+} = \sum_{n=1}^\infty n c_{-n} c_{n},
\quad J_{-} = \sum_{n=1}^\infty \frac{1}{n} b_{-n} b_{n}.
\end{equation}
These three generators commute both with $b_0$ and $L_0$ and therefore all Siegel gauge states can be decomposed into irreducible representations of the SU(1,1) group. The states in these representations are labeled by a half-integer spin $j$ and by the $J_3$ eigenvalue $m$, which belongs to the interval $-j\leq m\leq j$. The $J_3$ generator is closely related to the zero mode of the ghost current as
\begin{equation}
j_0^{gh}=2J_3+c_0 b_0+1.
\end{equation}
Therefore we find that the ghost number in Siegel gauge is given by $g=2m+1$. Notice that odd ghost numbers correspond to integer spins and even ghost numbers to half-integer spins.

There is a very convenient basis in the ghost theory which respects this symmetry. States in this basis are generated by 'twisted' Virasoro generators \cite{GaiottoRastelliSU11}, which act on the corresponding primaries. The 'twisted' Virasoro generators are given by
\begin{equation}\label{twisted Virasoro}
L'^{gh}_n=L^{gh}_n + n j^{gh}_n + \delta_{n0} = \sum_{m=-\infty}^{\infty}(n-m)\normc b_m c_{n-m}\normc.
\end{equation}
The operators $L'^{gh}_{n}$ satisfy the Virasoro algebra with central charge $c'^{gh}=-2$. They commute both with $b_0$ and with the generators of the SU(1,1) group and therefore their action does not change the irreducible representations.

Primary operators with respect to the 'twisted' Virasoros are also labeled by $(j,m)$ according to the SU(1,1) representations. They can be constructed from the SU(1,1) highest weight states
\begin{equation}
|j,j\ra\equiv c_{-2j}\dots c_{-1}c_1|0\ra
\end{equation}
by repeated action of $J_-$:
\begin{equation}\label{spin primary}
|j,m\ra\equiv N_{j,m} (J_-)^{j-m}|j,j\ra,
\end{equation}
where $N_{j,m}=\prod_{k=m+1}^j \left(j(j+1)-k(k-1)\right)^{-\frac{1}{2}}$. These states obey the standard SU(2) recursion relations
\begin{equation}
J^{\pm} |j,m\ra = \sqrt{(j \mp m)(j \pm m +1)} |j,m\pm 1\ra,
\end{equation}
but they have a somewhat unusual normalization with respect to the BPZ pro-duct,
\begin{equation}
\la j,m|c_0|j,-m\ra=(-1)^{j-m}.
\end{equation}
The primaries obey $b_0|j,m\ra=0$, which means that the Siegel gauge state space is spanned by states
\begin{equation}
L'^{gh}_{-k_1}L'^{gh}_{-k_2}\dots L'^{gh}_{-k_n}|j,m\ra, \quad k_i\geq1.
\end{equation}
In appendix \ref{sec:characters:ghost}, we use characters to show the equivalence of this basis with the usual $bc$ basis.

It was noticed in \cite{ZwiebachSU11} that the generators $J_{\pm,3}$ obey
\begin{equation}
\la V_3|\left(J_{\pm,3}^{(1)}+J_{\pm,3}^{(2)}+J_{\pm,3}^{(3)}\right)|\Psi_1\ra|\Psi_2\ra|\Psi_3\ra=0
\end{equation}
assuming that all $\Psi_i$ are in Siegel gauge. Therefore we can derive the Wigner-Eckart theorem for the cubic vertex following the usual procedure from quantum mechanics,
\begin{equation}\label{ghost Wigner-Eckart}
\la V_3|L'^{gh}_{-I_1}|j_1,m_1\ra L'^{gh}_{-I_2}|j_2,m_2\ra L'^{gh}_{-I_3}|j_3,m_3\ra=
\begin{pmatrix} j_1& j_2& j_3 \\ m_1& m_2& m_3 \end{pmatrix}
C(j_1,j_2,j_3,I_1,I_2,I_3),
\end{equation}
where $\begin{pmatrix} j_1& j_2& j_3 \\ m_1& m_2& m_3 \end{pmatrix}$ is the usual SU(2) 3-j symbol and $C(j_1,j_2,j_3,I_1,I_2,I_3)$ is a reduced vertex, which does not depend on $m_i$. Using the symmetry of the 3-j symbols
\begin{equation}
\begin{pmatrix} j_1& j_2& j_3 \\ -m_1& -m_2& -m_3 \end{pmatrix}=
(-1)^{j_1+j_2+j_3}\begin{pmatrix} j_1& j_2& j_3 \\ m_1& m_2& m_3 \end{pmatrix}
\end{equation}
we find that the cubic vertex contracted with ghost number 1 ($m_i=0$) states can be nonzero only if the sum of the three spins $j_1+j_2+j_3$ is an even integer. This implies a $\mathbb{Z}_2$ symmetry of the cubic vertex
\begin{equation}
(-1)^J:\quad L'^{gh}_{-I}|j,m\ra\rar (-1)^j L'^{gh}_{-I}|j,m\ra,
\end{equation}
which plays a role in the structure of some universal solutions \cite{KudrnaUniversal}.

The full SU(1,1) basis is not very convenient for OSFT calculations. The reason is that the cubic vertex conservation laws for $L'^{gh}_{n}$ (see sections \ref{sec:SFT:cubic cons} and \ref{sec:Numerics:V3:ghost SU11}) are not compatible with the SU(1,1) symmetry. They produce additional operators (in our conventions $j^{gh}_n$), which do not annihilate the primaries $|j,m\ra$.

However, it is possible to truncate the Siegel gauge state space only to SU(1,1) singlets. The equation (\ref{ghost Wigner-Eckart}) implies that the cubic vertex is nonzero only if the insertions obey the SU(1,1) fusion rules, $|j_1-j_2|\leq j_3\leq j_1+j_2$, which means that the cubic vertex contracted with two singlet states and one non-singlet state vanishes. Therefore we can use the argument from subsection \ref{sec:SFT:basic:twist} to show consistency of the truncation to SU(1,1) singlets. There is an efficient algorithm to evaluate the cubic vertex contracted with SU(1,1) singlets \cite{KudrnaUniversal} and it seems that SU(1,1) singlet solutions describe most of expected boundary states, so we will impose this condition in all of our Siegel gauge calculations with the exception of some parts of chapter \ref{sec:universal}\footnote{In fact, it is actually quite difficult to find solutions which are not SU(1,1) singlets. Newton's method preserves the symmetry, so we need non-singlet seeds. First twist even non-singlet states appear at level 4 and first twist non-even non-singlet states at level 3. In most settings, we are not able to use the homotopy continuation method at such levels because there are too many states.}.

\subsection{Universality of the string field}\label{sec:SFT:string field:universal}
Next, we focus on general structure of the full state space of string field theory. The underlying BCFT is given by a product of the matter and ghost part, BCFT$^{tot}=\rm{BCFT}^m\otimes$BCFT$^{gh}$. The matter part usually consists of 26 free bosons that live in some spacetime. Even if we ignore the issue of continuous momentum in noncompact dimensions, the full Hilbert space of this theory is far too large to be dealt with. Therefore we need to impose some restrictions on the string field to reduce the number of its degrees of freedom.

We assume that the matter BCFT can be split as BCFT$^m=\rm{BCFT}'\otimes$BCFT$^X$, where BCFT$^X$ is some relatively simple theory of interest and BCFT$'$ is an arbitrary theory with the correct central charge, see section \ref{sec:CFT:strings:other}. In BCFT$'$, we allow only states with the maximal possible symmetry, which are the Virasoro descendants of the vacuum state. Therefore any specific properties of BCFT$'$ do not matter and the theory can be characterized only by its central charge. This restriction vastly reduces the number of components the string field and it makes  numerical calculations possible.

We can show the consistency of this truncation following \cite{SenUniversality}. We assume that BCFT$'$ consists of several free bosons with Neumann or Dirichlet boundary conditions plus possibly some other unitary BCFT. First, we consider states with nonzero momentum $k_\mu$. Any boundary correlator with only one such state always disappears thanks to the momentum conservation. Therefore we can consistently set these states to zero using the argument from subsection \ref{sec:SFT:basic:twist} and focus on the zero momentum part of the spectrum. The remaining states can be always decomposed into primary states and their descendants\footnote{This decomposition cannot be done with the whole spectrum because the theory includes the nonunitary timelike boson $X^0$. Among states which include momentum in the timelike direction, there are some which are neither primaries nor descendants.}. One-point function of a nontrivial primary operator is always zero,
\begin{equation}
\la \phi(z)\ra=0,\qquad \phi\neq\Id,
\end{equation}
so nontrivial primary operators never appear linearly in the action and we can remove them as well. The same goes for their descendants. Therefore we are left only with descendants of the identity operator.

This universal structure of OSFT leads us the first setting, where we allow only the universal states in the matter theory, BCFT$^m$=BCFT$'$. This universal theory is the simplest possible string field theory and it contains the least amount of states. This allows us to perform calculations to very high levels and makes it an ideal setting for various experiments with gauge choice and other conditions imposed on the string field. In this thesis, we consider three different gauge choices: Siegel gauge (\ref{Siegel gauge}), Schnabl gauge (\ref{Schnabl gauge}) and no gauge imposed. In each gauge, we can choose whether we apply the twist even condition or not and, when in Siegel gauge, we can additionally apply the SU(1,1) singlet condition in the ghost sector.

The precise form of the string field depends on these choices and on the choice of basis in the ghost theory. The most general form is
\begin{equation}
|\Psi\ra =\sum_{I,J,K} t_{IJK} L'_{-I}b_{-J}c_{-K}c_1|0\ra,
\end{equation}
where the multiindex $I$ does not contain $-1$ (because $L'_{-1}|0\ra=0$) and $K$ may or may not include 0 depending on the gauge. The SU(1,1) singlet string field takes the form
\begin{equation}\label{string field univ singlet}
|\Psi\ra =\sum_{I,J} t_{IJ} L'_{-I}L'^{gh}_{-J}c_1|0\ra.
\end{equation}
Both multiindices $I,J$ exclude $-1$ and therefore this string field has the same structure in the matter and in the ghost sector. If one decides to use the Virasoro or the ghost current basis, the string field is similar to (\ref{string field univ singlet}), but $L'^{gh}_{-J}$ are replaced by the appropriate operators and the multiindex $J$ may include $-1$.

The complex conjugation acts on the universal string field as in (\ref{complex conjugation univ}), $|\Psi\ra^\ast=(-1)^N |\Psi\ra$. This is easiest to see using the Virasoro basis in the ghost sector. In other bases, one has to be more careful about signs when deriving the formula, but the final result is the same. Therefore a real string field has real coefficients in the twist even part and purely imaginary coefficients in the twist odd part.

Solutions of this theory also appear in any other string field theory in the same gauge because all theories contain the universal subsector. The solutions may differ by an overall scaling of gauge invariant observables and they may be written in different bases, but we can easily recognize them for example by the value of the tachyon coefficient. Unfortunately, the only known nontrivial universal solution free of any problems it the tachyon vacuum solution.

\subsection{Free boson theory}\label{sec:SFT:string field:FB}
Now we can turn our attention towards more interesting backgrounds. First, we consider the most natural setting for string theory: the free boson theory, where we choose toroidal compactification for simplicity. This theory allows us to observe transitions between D-branes of various dimensions and we can also discover some non-conventional boundary states.

In one dimension, a torus reduces to a circle, which is characterized by its radius $R$. Although the free boson CFT is not rational, it is well understood. At a generic radius, there are two possible boundary conditions, Neumann and Dirichlet. These two are connected by the T-duality, so we can choose the Neumann boundary conditions, which correspond to a D1-brane\footnote{As we mentioned earlier, we label D-branes by their dimension with respect to the compact space we are interested in. We do not care about their properties in the remaining part of the matter theory, so we can assume that all free bosons in BCFT' have Dirichlet boundary conditions.}, as our initial setting without loss of generality. At special radii of the form $\frac{M}{N}$, $M,N\in\mathbb{N}$, there are also boundary states which interpolate between Neumann and Dirichlet boundary conditions \cite{GaberdielBoson}, but these can be usually found only using the marginal setting, see later.

In two dimensions, there are far more possible initial configurations. The torus has three independent parameters and boundary conditions preserving the U(1)$\otimes$U(1) symmetry correspond to several different types of D-branes. Some of them (D2-branes with magnetic flux, tilted D1-branes) have more complicated BCFT, so we stick with the simplest choice: a D2-brane with no magnetic flux, which has the Neumann boundary conditions. There are also symmetry-breaking boundary states. These are not classified and we cannot use them as the initial background, but they may appear among our solutions.

We have not done any calculations in more than two dimensions. The OSFT construction can be easily generalized to higher dimensions, but each additional dimension significantly increases volume of the state space, which reduces the maximal accessible level.

The free boson theory is almost factorizable, so, in many calculations, we can treat each dimension separately. The Hilbert space of a single free boson with Neumann boundary conditions is spanned by states
\begin{equation}\label{basis FB}
\alpha_{-i_1}\alpha_{-i_2}\dots\alpha_{-i_n}|k\ra.
\end{equation}
The momentum in a single dimension is quantized simply as $k=n/R$, but the quantization in more dimensions depends on all parameters of the torus, see (\ref{FB momentum torus}). However, the full state space cannot be used for numerical calculations. The reason is that the string field has a continuous translation symmetry
\begin{equation}
|\Psi\ra\rar e^{i \Delta x \alpha_0}|\Psi\ra
\end{equation}
because the zero more $\alpha_0$ is a derivation of the star product. This symmetry makes the Jacobian matrix in Newton's method non-invertible (see section \ref{sec:Numerics:Newton}) and therefore it prevents us from finding discrete numerical solutions. To solve this problem, we introduce the spacetime parity operator $P$, which acts on the free scalars as
\begin{equation}
P(X^\mu)=-X^\mu.
\end{equation}
The parity generates a $\mathbb{Z}_2$ symmetry of the free boson CFT. The string field splits into the even and odd part with respect to this symmetry and, following subsection \ref{sec:SFT:basic:twist}, we can consistently truncate the string field to the parity even part. As we defined it, the parity acts in all dimensions at once. In principle, we could impose the parity even condition independently in each dimension, but by doing so, we would loose some important solutions.

Momentum states are not invariant under the action of the parity operator, $P|k\ra=|\!-\! k\ra$, so it is more convenient to use their (anti)symmetric combinations
\begin{equation}\label{parity even odd}
|k\ra_\pm=\frac{1}{2}(|k\ra\pm |\!-\! k\ra),
\end{equation}
which are eigenstates of the parity operator
\begin{equation}
P|k\ra_\pm=\pm|k\ra_\pm.
\end{equation}
Notice that odd states are in a certain sense imaginary, $|k\ra_-=i\sin (k X)(0)|0\ra$. This definition is somewhat unusual, but it leads to nicer reality properties of the string field. Since $|k\ra_\pm=\pm|\!-\! k\ra_\pm$, we restrict the momentum to $k \geq 0$ to avoid doubling of the Hilbert space.

The parity acts on the $\alpha$ oscillators as $P(\alpha_k)=-\alpha_k$, so parity even states take form
\begin{equation}\label{states FB even}
\alpha_{-i_1}\alpha_{-i_2}\dots\alpha_{-i_n}|k\ra_\epsilon,
\end{equation}
where $\epsilon$ is given by the number of $\alpha$ oscillators, $\epsilon=(-)^{\#\alpha}$. At zero momentum, the parity even condition selects states with even number of $\alpha$ oscillators. For other momenta, one only one of the two states (\ref{parity even odd}) is allowed and therefore this part of the spectrum is halved.

Notice that in two (or more) dimensions, the states (\ref{parity even odd}) are not in a factorized form,
\begin{equation}
|k_1,k_2\ra_{\pm}\neq |k_1\ra_\pm\otimes |k_2\ra_\pm,
\end{equation}
so we cannot apply the parity projection in each dimension separately. Therefore we work with states of definite momentum in the individual dimensions and we introduce effects of the parity projection only when we consider the full string field. We will discuss this issue in more detail in chapter \ref{sec:Numerics}.

In the free boson theory, we always impose Siegel gauge and the SU(1,1) singlet condition. Therefore the full parity even string field takes form
\begin{equation}
|\Psi\ra=\sum_{k\geq 0}\sum_{I,J,K} t_{IJK,k}\, L'_{-I}L'^{gh}_{-J}\alpha_{-K} c_1|k\ra_\epsilon.
\end{equation}
If we consider more than one dimension, the $\alpha$ oscillators and the momentum gain vector indices.

The complex conjugation acts on this string field in the same way as in the universal case, $|\Psi\ra^\ast=(-1)^N|\Psi\ra$, but the derivation of this formula is a bit more complicated. The BPZ conjugation of $\alpha$ oscillators generates the factor $(-1)^{N+\#\alpha}$, but part of it is canceled by the sign $(-1)^{\#\alpha}$, which comes from the Hermitian conjugation of (\ref{parity even odd}). Therefore we get the standard result, but only for a parity even string field.

We always consider parity even string field and we usually independently impose the twist even condition (with the exception of one solution described in subsection \ref{sec:FB circle:double:twist}). However, there is an alternative condition which can be imposed on the string field. Instead of the parity and twist symmetries, we can define a 'twisted parity' $P\Omega$ and keep only states which are even with respect to this symmetry. The advantage of this condition is that it does not project out the state $\alpha_{-1}c_1|0\ra$, which is needed for example for $\del X$ marginal deformations.

\subsection{Minimal models}\label{sec:SFT:string field:MM}
Minimal models are another simple possibility for a string field theory background, we consider the A-series of Virasoro minimal models. The central charge of these models is not an integer, so BCFT' must be a more complicated theory than just the usual free boson theory, but its details are not reflected in OSFT calculations. As the initial setting, we choose one of the Cardy boundary states (\ref{Cardy BS}) and we expect to see solutions describing other Cardy boundary states or their linear combinations.

We can work even with nonunitary minimal models. In this case, there can be problems with physical interpretation of some of observables (for instance, some solutions have negative energy), but our numerical methods work without any technical problems.

The string field in this theory is given by
\begin{equation}
|\Psi\ra=\sum_{(r,s)}\sum_{I,J,K} t_{IJK,(r,s)}\, L'_{-I}L'^{gh}_{-J}L_{-K} c_1|\phi_{(r,s)}\ra,
\end{equation}
where the spectrum of primary operators $\phi_{(r,s)}$ is determined by the initial boundary conditions. There are no complications like in the free boson case, but we have to remove null states to avoid problems in Newton's method, see section \ref{sec:Numerics:V2:MM}. This string field also has the usual reality condition (\ref{complex conjugation univ}) because minimal model primaries are invariant under the complex conjugation.

A more interesting setting is a product of two minimal models. Such theories are irrational with respect to the $c^{(1)}+c^{(2)}$ Virasoro algebra. Their boundary states are generally not classified, but we can easily construct string field theory on a background given by a product of two Cardy boundary states. Then we have a chance to find some non-conventional boundary states among the solutions.

\subsection{Marginal deformations}\label{sec:SFT:string field:marginal}
String field theory can also describe marginal deformations. They are not specific to one concrete BCFT, but they are allowed in any theory which has at least one weight 1 current with the OPE
\begin{equation}\label{current OPE}
j(z)j(w)\sim \frac{k}{(z-w)^2}.
\end{equation}
If nonzero, the constant $k$ can be always set to a desired value by rescaling the current.

The deformed BCFT is described by deformation of the boundary action by the current,
\begin{equation}
S(\lB)=S_0+\lB\int_{\del M} j(x)dx,
\end{equation}
where $\lB$ parameterizes the deformation in BCFT. These theories have been studied in \cite{RecknagelSchomerusMarginal} and many of them are exactly solvable. In particular, when the current belongs to the chiral symmetry algebra of the theory, the deformation changes the gluing conditions (\ref{gluing W}) to
\begin{equation}
W^a(z)=\Omega\circ \gamma_{\bar j}(\bar W^b(\zb))|_{z=\zb},
\end{equation}
where $\gamma_{j}$ is an automorphism defined by
\begin{equation}
\gamma_{j}(W^a)=e^{-i\lB j_0}W^a e^{i\lB j_0}.
\end{equation}
The boundary state of the deformed theory is given by
\begin{equation}
\ww \alpha\rra_{\lB}=e^{i\lB j_0}\ww \alpha\rra.
\end{equation}

Marginal deformations should be possible to describe by OSFT, which means that we expect a one-parametric family of solutions $\Psi(\lS)$, which correspond to  deformed boundary states. These solutions are usually parameterized by the coefficient $\lS$ in front of the marginal field,
\begin{equation}
|\Psi(\lS)\ra=\lS j_{-1}c_1|0\ra+\dots.
\end{equation}
However, it turns out that the two parameterizations are not the same, $\lB\neq\lS$, and the two parameters are related by some nonlinear function $\lB=\lB(\lS)$. Furthermore, this function depends of the OSFT gauge condition. The study of this relation is one of our main goals in chapter \ref{sec:marginal}.

A solution for marginal deformations can be constructed perturbatively \cite{MarginalSchnablPert}. We expand the string field as a power series in $\lS$
\begin{equation}\label{marginal pert expansion}
\Psi(\lS)=\sum_{n=1}^\inf \lS^n \Psi_n,
\end{equation}
where $\Psi_1=j_{-1}c_1|0\ra$. Using this ansatz, the equations of motion can we solved order by order and we find
\begin{equation}\label{marginal pert solution}
\Psi_n=-Q^{-1}\sum_{k=1}^{n-1}\Psi_k\ast \Psi_{n-k}.
\end{equation}
This form of the marginal solution implies that $\lB$ is an odd function of $\lS$, ${\lB(-\lS)=-\lB(\lS)}$.

However, evaluation of the perturbative solution is quite difficult both numerically and analytically. There are problems with inversion of the BRST charge and the solution gets more and more complex with increasing order. It can be also shown that it has only a finite radius of convergence. Therefore the perturbative solution gives us precise results only for small $\lB$ and solutions describing large $\lB$ must be obtained using different methods.

In this thesis, we consider $j=\cos X$ marginal deformations in the free boson theory on the self-dual radius $R=1$. This theory is also dual to the SU(2)$_1$ WZW model. We have chosen this setting because it allows us to use the code developed for the free boson theory and because we can make connections with the some of lump solutions from chapter \ref{sec:FB circle}.

\section{Observables and consistency checks}\label{sec:SFT:observables}
On of the main goals of this thesis is to check validity of Sen's conjectures, background independence of OSFT and Ellwood's conjecture. Solutions of the equations of motion depend on gauge choice, so we need gauge invariant observables that can be compared to BCFT quantities.
Therefore the goal of this section is to define several string field theory observables, which will help us to identify solutions, and also some pseudo-observables, which are not gauge invariant, but which serve as consistency checks. Usually, three types of observables are considered in OSFT\footnote{Recently, the article \cite{MasudaAmplitudes} introduced a new type of observables, which seem to describe on-shell scattering amplitudes around the corresponding D-brane configuration. However, these observables do not seem suitable for numerical solutions.}:

The first observable quantity is the energy, which, according to Sen's conjectures \cite{SenUniversality}, should be proportional to the mass of the D-brane described by a solution (or to the boundary entropy from the BCFT point of view). The energy can be easily computed from the action, but there are many cases when it is not enough to fully identify a solution.

The second set of observables is given by so-called Ellwood invariants. It is conjectured \cite{EllwoodInvariants} that the Ellwood invariants reproduce the boundary state corresponding to a given solution. For some analytic solutions, it is possible to compute all invariants and verify the conjecture. When it comes to numerical solutions, only some invariants are convergent in the level truncation approximation, but they are still the best tool we have to identify solutions.

Finally, it is possible analyze the spectrum of excitations around a solution. Consider a string field of the form $\Psi+\Phi$, where $\Psi$ is a solution of the equations of motion. The action of this string field reads
\begin{equation}
S[\Psi+\Phi]=S[\Psi]-\frac{1}{g_o^2} \int \left( \frac{1}{2}\Phi \ast Q_\Psi\Phi + \frac{1}{3} \Phi\ast\Phi\ast\Phi \right),
\end{equation}
where
\begin{equation}\label{Q Psi}
Q_\Psi\equiv Q+[\Psi,\,\cdot\,]_\ast
\end{equation}
is a modified BRST operator and $[\,\cdot\,,\,\cdot\,]_\ast$ is graded star commutator. This operator is by definition nilpotent, $Q_\Psi^2=0$. The cohomology of $Q_\Psi$ is conjectured \cite{SenBackground1} to give us the spectrum of on-shell excitations around the background described by the solution.

This cohomology can be computed for some analytic solutions, but computing it for numerical solutions is difficult. There have been some attempts \cite{EllwoodTaylorCohomology}\cite{GiustoCohomology}\cite{ImbimboCohomology}\cite{OhmoriLumps}, which focus mainly on the tachyon vacuum. It is not clear whether these techniques would work on other types of solutions and whether they are usable at high enough levels. We will not attempt to do these calculations in this thesis.

\subsection{Energy}\label{sec:SFT:observables:energy}
In order to define the energy of a solution, we interpret the OSFT action as an integral over a Lagrangian density, which is a function of the spacetime coordinates $X^\mu$. We can interpret the Lagrangian as a difference of kinetic and potential energy, where the kinetic energy includes derivatives with respect to the time coordinate $X^0$ and the potential the rest. For a time independent solution, the kinetic energy disappears and the energy of the solution is therefore given essentially just by minus the action. Including the mass of the initial D-brane system $M$, we get
\begin{equation}\label{Energy def1}
M-S(\Psi)=M+\frac{1}{g_o^2}\left(\frac{1}{2}\int \Psi \ast Q \Psi+\frac{1}{3}\int \Psi\ast\Psi\ast\Psi \right).
\end{equation}
To make the definition precise, we first need to specify the coupling constant $g_0$ and normalization of the vertices.

The references \cite{SenUniversality}\cite{TaylorZwiebach}\cite{OhmoriReview} compute the coupling constant by identification of part of the action with kinetic energy of a moving D-brane. For a D$p$-brane, they find
\begin{equation}
\tt_p=\frac{1}{2\pi^2g_o^2},
\end{equation}
where $\tt_p$ is the tension of the D$p$-brane. However, this result is problematic. Consider OSFT formulated on a system of two distinct D-branes. Once we expand the trace in (\ref{SFT action Tr}), we find
\begin{equation}
S(\Psi)=-\frac{1}{g_o^2}\left(\frac{1}{2}\int \Psi_{11}\ast Q\Psi_{11}+\frac{1}{2}\int \Psi_{22}\ast Q\Psi_{22}+\int \Psi_{12}\ast Q\Psi_{21}+\dots\right).
\end{equation}
It is not clear whether $g_o$ should be related to the tension of the first or the second D-brane. Introducing more $g_o$s for different parts of the action is not an option because it would break the gauge symmetry. To solve this issue, we realize that the OSFT action is invariant under the transformation $g_o^2\rar \nnn g_o^2$, $\int\rar \nnn \int$. Therefore we can move the dependence on the D-brane dimension from $g_o$ to the normalization of the integration while preserving the expression for the energy from the aforementioned references.

We choose our coupling constant to be related to the D0-brane tension:
\begin{equation}
\frac{1}{g_o^2}=2\pi^2\tt_0.
\end{equation}
The vertices which appear in the action are given by correlators in the underlying BCFT. Let us consider the free boson theory in $D$ compact dimensions on a D$p$-brane background. The boundary state formalism, see (\ref{boundary state N}) and (\ref{boundary state D}), leads to a rather unpleasant expression
\begin{equation}
\la \Id\ra=\frac{V_p}{(2\pi)^{p-D/2} 2^{D/4}V_D^{1/2}},
\end{equation}
where $V_p$ is the volume of the D$p$-brane and $V_D$ is the volume of the compact space. This expression does not match the mass of the D$p$-brane, which should be directly proportional to its volume. Therefore we absorb the factor $\frac{(2\pi)^{D/2}}{2^{D/4}V_D^{1/2}}$ into definition of the integration operation\footnote{From the BCFT point of view, we can understand this change of normalization in the following way: Normalization of boundary states comes from the Cardy condition (\ref{Cardy condition}). The usual form of boundary states is derived assuming $\la i| i\ra$=1, which ensures that coefficients of boundary states are equal to one-point functions of bulk operators. However, the Cardy condition is invariant under a change of normalization \cite{Runkel1} given by $\la 0|0\ra\rar \mu^2 \la 0|0\ra$ and $B_\alpha^i\rar \mu^{-1}B_\alpha^i$, which leads to $\la \Id\ra^{(\alpha)}\rar \mu \la \Id\ra^{(\alpha)}$. By choosing $\mu=\frac{2^{D/4}V_D^{1/2}}{(2\pi)^{D/2}}$, we can set boundary correlators to the desired values.},
\begin{equation}
\int \Psi_1\ast\dots=\frac{2^{D/4}V_D^{1/2}}{(2\pi)^{D/2}}\la f_1\circ\Psi_1(0)\dots\ra=\la f_1\circ\Psi_1(0)\dots\ra^{eff},
\end{equation}
where we define an effective correlator as
\begin{equation}
\la \Id\ra^{eff}=\frac{V_p}{(2\pi)^p}.
\end{equation}

In most cases, we consider just a single D$p$-brane background. Then we can factorize the normalization factor and write the energy in terms of a dimensionless potential $\vv(\Psi)$:
\begin{equation}
M-S(\Psi)=\tt_p V_p +2\pi^2 \tt_0 \frac{V_p}{(2\pi)^p}\vv(\Psi).
\end{equation}
The dimensionless potential is computed using an effective matter correlator ${\la\Id\ra^{eff}=1}$. The universal part of the dimensionless potential is common to all open string field theories.

Finally, we divide the energy by the D0-brane tension $\tt_0$ to avoid its repetition in all results. Therefore we define the energy of a solution on the D$p$-brane background as
\begin{equation}\label{Energy def2}
E(\Psi)=\frac{1}{\tt_0}(\tt_p V_p-S(\Psi))=\frac{V_p}{(2\pi)^p} (1+2\pi^2\mathcal{V}(\Psi)),
\end{equation}
where we used the relation between D-brane tensions $(2\pi)^p\tt_p=\tt_0$ to remove $\tt_p$.
Specifically, in the free boson theory with Neumann boundary conditions on a circle and on a two dimensional torus, we find
\begin{eqnarray}
E^{circle}(\Psi)&=&R(1+2\pi^2\mathcal{V}(\Psi)),\\
E^{torus}(\Psi)&=&R_1 R_2 \sin\theta (1+2\pi^2\mathcal{V}(\Psi)).
\end{eqnarray}

These conventions allow a simple identification of most solutions. A D0-brane has energy 1 in any background, a D1-brane of length $2\pi R$ has energy $R$ and a generic D$p$-brane of volume $V_p$ has energy $\frac{V_p}{(2\pi)^p}$.

The volume factor is not well defined in minimal models and other CFTs without geometric interpretation. In such theories, we normalize the energy  using the one-point function $\la\Id\ra^\alpha$, which is equal to $B_\alpha^\Id$ in unitary models. In this thesis, we consider the Virasoro minimal models, where we define the energy as
\begin{equation}
E(\Psi)=\la\Id\ra^\alpha (1+2\pi^2\mathcal{V}(\Psi)).
\end{equation}

\subsection{Ellwood invariants}\label{sec:SFT:observables:Ellwood}
In many cases, just the knowledge of energy in not enough to identify a solution. Therefore we consider a different set of gauge invariant observables introduced in \cite{EllwoodHashimoto}\cite{GaiottoRastelliSU11}, which we call Ellwood invariants. They are given by
\begin{equation}
\la I | \vv(i,-i)| \Psi\ra= \la E[\vv]|\Psi\ra,
\end{equation}
where the state $\la E[\vv]|\equiv\la I | \vv(i,-i)$ is called the Ellwood state and $\la I |$ is the identity string field (\ref{identity definition}). The vertex operator $\vv$ must be an on-shell closed string primary operator of weight $(0,0)$. It has the form $\vv=c\bar c V^m$, where $V^m$ is a weight $(1,1)$ matter primary operator.

Gauge invariance of these observables can be shown following \cite{EllwoodInvariants}. We need to show that
\begin{equation}
\la I | \vv(i,-i)\left(Q |\Lambda\ra+|\left[\Psi,\Lambda\right]\ra\right)=0
\end{equation}
for any $\Psi$ and $\Lambda$. Since only the second term depends on $\Psi$, both must vanish separately. For the first term, we find
\begin{equation}
\la I | \vv(i,-i)Q=0
\end{equation}
independently on $\Lambda$. The vertex operator $\vv$ is on-shell and therefore it satisfies $\left[Q,\vv\right]=0$, $\la I|Q=0$ follows from (\ref{wedge and Q}).

The second term can be mapped on the upper half plane using the conformal transformations
\begin{equation}
f_1(z)=\frac{1+z}{1-z},\qquad f_2(z)=-\frac{1-z}{1+z}.
\end{equation}
The two parts of the commutator are given by
\begin{eqnarray}
\la I | \vv(i,-i)|\Psi \ast\Lambda\ra&=&\la \vv(i)f_1\circ\Psi(0) f_2\circ\Lambda(0)\ra, \\
\la I | \vv(i,-i)|\Lambda \ast\Psi\ra&=&\la \vv(i)f_1\circ\Lambda(0) f_2\circ\Psi(0)\ra.
\end{eqnarray}
The two maps $f_1$ and $f_2$ are related by the inverse transformation $-1/z$, which is part of the $SL(2,\mathbb{Z})$ group. Therefore both terms are equal and they cancel each other. Alternatively, we can map the correlators to a disk, where both terms are related just by a rotation.

The Ellwood's conjecture \cite{EllwoodInvariants} states that Ellwood invariants of a given solution $\Psi$ are related to the boundary state\footnote{A different prescription for the boundary state was introduced in \cite{KOZ}, but this construction is too complicated to be used for numerical solutions.} described by the solution as
\begin{equation}\label{Ellwood orig}
-4\pi i\la E[\vv]|\Psi\ra=\la\vv| c_0^- \ww B_\Psi\rra-\la \vv| c_0^-\ww B_0 \rra,
\end{equation}
where $\ww B_0\rra$ is the initial boundary state and $\ww B_\Psi\rra$ corresponds to the solution. Alternatively, we can make use of the fact that OSFT always admits the same tachyon vacuum solution $\Psi_{\rm{TV}}$, which has vanishing boundary state,
\begin{equation}
\la \vv | c_0^-\ww B_{\Psi_{\rm{TV}}}\rra=0.
\end{equation}
This implies
\begin{equation}
-4 \pi i \la I | \vv(i)|\Psi_{\rm{TV}}\ra=-\la \vv |c_0^-\ww B_0\rra
\end{equation}
and we can restate the Ellwood hypothesis as
\begin{equation}
\la \vv | c_0^-\ww B_\Psi\rra=-4 \pi i \la I | \vv(i)|\Psi-\Psi_{\rm{TV}}\ra.
\end{equation}
We do not know the tachyon vacuum solution exactly in the level truncation approximation, so this expression is only formal\footnote{However, this expression can be used to compute Ellwood invariants if we use a simple trick \cite{KMS}. The tachyon vacuum solution can be replaced by the string field $\frac{2}{\pi}c_1|0\ra$, which has the same Ellwood invariants as the tachyon vacuum solution.}.

The weakness of this construction is that it works only for weight $(1,1)$ matter operators and therefore it allows us to compute only a part of the boundary state. This problem was resolved in \cite{KMS} by definition of generalized Ellwood invariants. All known OSFT solutions include the universal subsector BCFT$'$, which is characterized only by its central charge. The field content of this theory is unspecified and therefore we can assume that for every weight $(h,h)$ matter primary operator $V^{(h,h)}$, there in an auxiliary primary operator $V^{aux}$ from BCFT$'$ with the following properties: It has weight $(1-h,1-h)$, it is orthogonal to $V^{(h,h)}$ and it satisfies $\la V^{aux}\ww B'\rra=1$. Then the composite matter operator $V^m=V^{(h,h)}V^{aux}$ has the correct weight $(1,1)$ and it can be used to construct an Ellwood invariant. Assuming that the BCFT$'$ boundary state remains unchanged, $\la V^{aux}\ww B'_0\rra=\la V^{aux}\ww B'_\Psi\rra=1$, the boundary state coefficient can be computed using (\ref{Ellwood orig}) with the generalized matter operator. For later purposes, let us define weight of an Ellwood invariant as the weight $h$ of the non-universal part of the matter vertex operator.

As an example, let's consider BCFT$'$ which includes a free scalar $Y$ with Dirichlet boundary conditions. Auxiliary vertex operators there are given by bulk operators $e^{2i\sqrt{1-h}Y}$, which satisfy the required properties. Notice that auxiliary operators have negative conformal weights when $h>1$. That means that BCFT$'$ must be a nonunitary theory. It does not lead to any direct technical problems in our algorithms, but it may be related to bad convergence of some Ellwood invariants, see section \ref{sec:FB circle:single:pade}.

In string field theories which have no universal subsector BCFT$'$, the construction of generalized Ellwood invariants requires tensoring the original BCFT with an auxiliary $c=0$ theory BCFT$^{aux}$ and lifting solutions to this new theory. See \cite{KMS} for more detailed discussion.

Full boundary states in string theories always factorize into the matter and ghost part \cite{KMS},
\begin{equation}
\ww B\rra=\ww B_m\rra \otimes \ww B_{gh}\rra.
\end{equation}
There is only one unique ghost boundary state, which is common to both $\ww B_\Psi\rra$ and $\ww B_0\rra$, see section \ref{sec:CFT:ghost}, and therefore we usually extract just the matter part of the boundary state. The ghost boundary state satisfies
\begin{equation}
\la 0|c_{-1} \bar c_{-1} c_0^-\ww B_{gh}\rra=-2,
\end{equation}
which leads to the definition
\begin{equation}\label{Ellwood definition}
E[\vv](\Psi)=2\pi i\la E[\vv]|\Psi-\Psi_{TV}\ra=2\pi i\la E[\vv]|\Psi\ra+\la V^m\ww B_0 \rra.
\end{equation}
These invariants should reproduce the matter boundary state described by the solution.

The definition above still has a freedom in normalization of the matter operator $V^m$. Boundary states are usually written in terms of operators with unit normalization,
\begin{equation}
\la V(z,\zb)V^\ast(w,\bar w)\ra = \frac{1}{|z-w|^{4h}},
\end{equation}
where $V^\ast$ is the operator conjugate to $V$. Therefore if we start with a generic operator, we usually first normalize it in this way. This fixes the normalization up to a sign, which can be further fixed if we know a nonzero one-point function $\la V(z,\zb)\ra$ for some reference boundary condition. Ellwood invariants with this normalization should reproduce the canonical boundary state.

However, in some cases, it is more convenient to choose a different normalization, which allows easier identification of solutions or to get rid of some unpleasant numerical factors. We will do so in the free boson theory, so we will discuss normalization of individual invariants in more detail later.

In the free boson theory (or in other theories where boundary states are written in terms of Ishibashi states with respect to some extended symmetry algebra), we find working directly with boundary states inconvenient. In order to get boundary state coefficients for non-fundamental primary operators, we would have to expand the generalized Ishibashi states and identify coefficients in front of Virasoro primaries. That would be laborious, so we instead use the fact that boundary state coefficients are defined by correlators
\begin{equation}
\la V\ww B_\alpha\rra=4^h\la V(i,-i)\ra^{UHP}_\alpha.
\end{equation}
This correlator can be computed using the doubling trick and standard CFT techniques. We use it to compute the BCFT$_0$ contributions to Ellwood invariants and to find their expected values.

At the end of this subsection, let us introduce the first Ellwood invariant, which is common to all OSFTs. It can be defined using any weight $(1,1)$ operator in the universal matter theory BCFT$'$. The definition usually uses the zero momentum graviton vertex operator:
\begin{equation}\label{E0 def}
E_0=-4\pi i \la E[c\bar c\del X^0\bar{\del}X^0]|\Psi-\Psi_{TV}\ra.
\end{equation}
This invariant provides an alternative way to compute the energy (see \cite{IshibashiEnergy} for a proof for a certain class of solutions) and, with the proper normalization, it should be equal to (\ref{Energy def2}). We choose its notation to match labeling of invariants in the given theory. The notation $E_0$ matches the series of $E_n$ invariants in the free boson theory on a circle, in minimal models, we change the notation to $E_\Id$, etc.

\subsection{Ellwood invariants in free boson theory}\label{sec:SFT:observables:FB}
In this subsection, we will introduce Ellwood invariants in the free boson theory. As we discussed before in section \ref{sec:SFT:observables:energy}, the free boson boundary state on a $D$-dimensional torus has an unpleasant normalization factor, which reduces to
\begin{equation}
\la \Id\ra=\frac{V_D^{1/2}}{(2\pi)^{D/2} 2^{D/4}}
\end{equation}
for Neumann boundary conditions. Since the $E_0$ invariant should match the energy, we define all invariants with an additional normalization
\begin{equation}
E[\vv](\Psi)=\frac{V_D^{1/2}2^{D/4}}{(2\pi)^{D/2}} 2\pi i\la E[\vv]|\Psi-\Psi_{TV}\ra=2\pi i\la E[\vv]|\Psi-\Psi_{TV}\ra^{eff},
\end{equation}
where the effective correlator is normalized as
\begin{equation}
\la \Id\ra^{eff}=\frac{V_D}{(2\pi)^{D}}.
\end{equation}

\subsubsection{Free boson on a circle}
First, Let us consider a single free boson compactified on a circle with radius $R$. The effective normalization of correlators in this theory is given by
\begin{equation}
\la \Id\ra^{eff}_{D1}=R.
\end{equation}

Following \cite{KMS}, we define a series of momentum invariants $E_n$ as
\begin{equation}
E_n=-4\pi i\la E[c\bar c\,\del X^0\bar\del X^0\cos \frac{n}{R}X\ V^{aux}]|\Psi-\Psi_{TV}\ra^{eff},
\end{equation}
where the auxiliary vertex operator has weight $-\frac{n^2}{R^2}$. This definition reflects Fourier modes of the $T_{00}$ component of the spacetime energy-momentum tensor and the definition of $E_0$ coincides with (\ref{E0 def}). Alternatively, we can define these invariants in a simpler way as
\begin{equation}
E_n=2\pi i\la E[c\bar c\,\cos \frac{n}{R}X\ V^{aux}]|\Psi-\Psi_{TV}\ra^{eff}.
\end{equation}
The auxiliary vertex operator in this definition has weight $1-\frac{n^2}{R^2}$.

If we consider a parity even string field, there are no other nontrivial momentum invariants and we can replace $\cos \frac{n}{R}X$ by $e^{\pm i\frac{n}{R}X}$ without changing the results. If we relax the parity even condition, there are twice as many momentum invariants because invariants defined using $\sin \frac{n}{R}X$ can be nonzero.

The initial values of $E_n$ invariants on a D1-brane are
\begin{equation}
E_n^{(D1)}=R\,\delta_{n0}
\end{equation}
and the expected values for a D0-brane solution are given by
\begin{equation}\label{En predicted}
E_n^{(D0)}=4^{\frac{n^2}{4R^2}}\la \cos\frac{n}{R}X(i,-i)\ra_D=\cos\frac{n}{R}x_0.
\end{equation}
A single D0-brane that respects the reflection symmetry can be located only at $x_0=0$ or $x_0=\pi R$, so the expected values are $E_n=\pm 1$. Nontrivial values appear only for two or more D0-brane solutions. We can find positions of the D0-branes by inverting the cosines, but we have to be careful about choosing the correct branches of arccosines for $n>1$.

Similarly to $E_n$, we define invariants $W_n$ which contain winding operators:
\begin{equation}
W_n=2\pi i\la E[c\bar c\,\cos n R \tilde X\ V^{aux}]|\Psi-\Psi_{TV}\ra.
\end{equation}
These invariants did not appear in \cite{KMS}, we mentioned them only in \cite{MarginalTachyonKM} without a closer description. The $W_0$ invariant coincides with $E_0$, so we can ignore it. The expected values of these invariants can be guessed using the T-duality. For a D1-brane, we have
\begin{equation}
W_n^{(D1)}=R\cos n R \tilde x_0,
\end{equation}
however, since we use a D1-brane as the initial boundary state, we will not encounter the T-dual position $\tilde x_0$ very often. The expected values for a D0-brane are
\begin{equation}
W_n^{(D0)}=\delta_{n0}.
\end{equation}

Apart from the momentum and winding primaries, the free boson theory contains the exceptional primaries mentioned in section \ref{sec:CFT:FB:bulk}. Most of them appear only at specific rational radii, so we focus on the zero momentum primaries, which are common to all radii. According to the SU(2) classification, these primaries are labeled as $(j,0,0)$, $j\in\mathbb{N}$, and they are given by products of left and right chiral primaries.

We will compute Ellwood invariants only for the first three primaries (\ref{primary states boson}). Their full vertex operators are given by
\begin{equation}
\vv_h=N_h c\bar cP_h \bar{P}_h V^{aux}.
\end{equation}
We choose the normalization $N_h$ so that all these invariants match $E_0$ for universal solutions. Using the one-point function
\begin{equation}
1=4^h N_h\la P_h(i)P_h(-i)\ra=\frac{4^h}{(2i)^{2h}}N_h\la P_h|P_h\ra,
\end{equation}
we find the normalizations in terms of the BPZ product of the corresponding states:
\begin{equation}
N_h=\frac{(-1)^h}{\la P_h |P_h\ra}.
\end{equation}
Now we can define a set of invariants $D_h$ as
\begin{equation}
D_h=2\pi i\frac{(-1)^h}{\la P_h |P_h\ra}\la E[c\bar cP_h\overline{P_h}V^{aux}]|\Psi-\Psi_{TV}\ra.
\end{equation}
If we choose $h=1$, this definition reproduces the $D$ invariant from \cite{KMS}.

The expected values of these invariants for a D1-brane are
\begin{equation}
D_h^{(D1)}=R
\end{equation}
and for a D0-brane we find
\begin{eqnarray}
D_1^{(D0)} &=& -1, \nn\\
D_4^{(D0)} &=& 1, \\
D_9^{(D0)} &=& -1. \nn
\end{eqnarray}

Finally, we introduce one more exceptional invariant, which appears only at the self-dual radius $R=1$. It is given by the first nontrivial primary with nonzero momentum, which has weight $1$ and spin $(1,1,0)$. We define it following \cite{MarginalTachyonKM} as
\begin{equation}
H=-4\pi\la E[c\bar c \del X \sin\bar X]|\Psi-\Psi_{TV}\ra.
\end{equation}
For a twist even string field, it does not matter whether $\del X$ appears in the left or the right part of the vertex operator. We have chosen its normalization to reproduce the expected result for the marginal solution (\ref{marginal pert expansion}) in the first order.

\subsubsection{Free boson on a 2D torus}
The spectrum of primary operators in the theory compactified on a 2D torus is much richer. There are far more momentum and winding primaries and, by decomposition of a U(1)$\otimes$U(1) Verma module into irreducible representations of the total Virasoro algebra, we find that there are partition $N$ primaries at level $N$. Computing all possible Ellwood invariants would be too complicated, so we introduce only few invariants that generalize the one-dimensional case.

We define momentum and winding invariants in a similar way as before
\begin{eqnarray}
E_{n_1,n_2}&=&2\pi i\la E[c\bar c\,\cos k(n_1,n_2).X\ V^{aux}]|\Psi-\Psi_{TV}\ra, \\
W_{n_1,n_2}&=&2\pi i\la E[c\bar c\,\cos \tilde k(n_1,n_2).\tilde X\ V^{aux}]|\Psi-\Psi_{TV}\ra,
\end{eqnarray}
where $ k(n_1,n_2)$ and $\tilde k(n_1,n_2)$ are the momentum and the T-dual momentum with quantum numbers $n_1$ and $n_2$, which are given by (\ref{FB momentum torus}) and (\ref{FB winding torus}) respectively. We have chosen not to compute any invariants with mixed momentum and winding.

The expected values of these invariants are as follows. In analogue with one dimension, for a D2-brane, we find
\begin{eqnarray}\label{En inv expected D2}
E_{n_1,n_2}^{(D2)}&=&R_1R_2\sin\theta\ \delta_{n_10}\delta_{n_20},\\
W_{n_1,n_2}^{(D2)}&=&R_1R_2\sin\theta\ \cos \tilde k(n_1,n_2).\tilde x_0.
\end{eqnarray}
Their behavior for a D0-brane is exchanged through the T-duality:
\begin{eqnarray}\label{En inv expected D0}
E_{n_1,n_2}^{(D0)}&=&\cos k(n_1,n_2).x_0,\\
W_{n_1,n_2}^{(D0)}&=&\delta_{n_10}\delta_{n_20}.
\end{eqnarray}
The expected values for a D1-brane are more complicated. A D1-brane on a torus is characterized by its winding numbers $m_1$ and $m_2$, which determine its length $L(m_1,m_2)$, the tangent vector $t(m_1,m_2)$ and the normal vector $n(m_1,m_2)$. We find that the $E_{n_1,n_2}$ invariants are nonzero only if the momentum is orthogonal to the D1-brane and the $W_{n_1,n_2}$ invariants are nonzero only if the T-dual momentum is parallel to the D1-brane,
\begin{equation}\label{En inv expected D1}
E_{n_1,n_2}^{(D1)}=\begin{cases}
L(m_1,m_2)\cos k(n_1,n_2).x_0 & {\rm{if}}\ k(n_1,n_2)\| n(m_1,m_2),\\
0  &  \rm{otherwise},
\end{cases}
\end{equation}
\begin{equation}
W_{n_1,n_2}^{(D1)}=\begin{cases}
L(m_1,m_2)\cos \tilde k(n_1,n_2).\tilde x_0 & {\rm{if}}\ \tilde k(n_1,n_2)\| t(m_1,m_2),\\
0  & \rm{otherwise}.
\end{cases}
\end{equation}

Next, we define generalizations of the $D_1$ invariant:
\begin{equation}
D_{1\mu\nu}=4\pi i\la E[c\bar c\,\del X^{\mu}\delb X^{\nu}]|\Psi-\Psi_{TV}\ra,
\end{equation}
where $\mu,\nu=X,Y$. There are 4 polarizations in general, but we get $D_{1XY}=D_{1YX}$ for twist even string field. We have also experimented with invariants defined using weight 2 primaries, but they are usually not well convergent and therefore we do not show these results.

The expected values of the $D_{1\mu\nu}$ invariants for a D2-brane are
\begin{eqnarray}\label{D1 inv expected D2}
D_{1XX}^{(D2)}=D_{1YY}^{(D2)}&=&R_1R_2\sin\theta,\\
D_{1XY}^{(D2)}=D_{1YX}^{(D2)}&=&0.
\end{eqnarray}
For a D1-brane, we find
\begin{eqnarray}\label{D1 inv expected D1}
D_{1XX}^{(D1)}=-D_{1YY}^{(D1)}&=&L(m_1,m_2)\cos 2\varphi(m_1,m_2),\\
D_{1XY}^{(D1)}=D_{1YX}^{(D1)}&=&L(m_1,m_2)\sin 2\varphi(m_1,m_2),
\end{eqnarray}
where $\varphi(m_1,m_2)$ is the angle between the D1-brane and the $X$ axis. Finally, for a D0-brane, we find
\begin{eqnarray}\label{D1 inv expected D0}
D_{1XX}^{(D0)}=D_{1YY}^{(D0)}&=&-1,\\
D_{1XY}^{(D0)}=D_{1YX}^{(D1)}&=&0.
\end{eqnarray}

\subsection{Ellwood invariants in minimal models}\label{sec:SFT:observables:MM}
If we consider a single $(p,q)$ minimal model, we can define only a finite set of Ellwood invariants. There are $(p-1)(q-1)/2$ bulk primaries, which are labeled by the Kac labels $(r,s)$. We define these invariants without any additional normalization as
\begin{equation}
E_{(r,s)}=2\pi i \la E[c\bar c\,\phi_{(r,s)} V^{aux}]|\Psi-\Psi_{TV}\ra.
\end{equation}
The invariant with the label $(1,1)\equiv \Id$ coincides with the previous definition of $E_0$. In unitary minimal models, these invariants should reproduce the Cardy boundary states (\ref{boundary state minmod}).

If we consider nonunitary models, there is a slight confusion thanks to negative signs of some two-point functions. We normalize the invariants following the one-point functions
\begin{equation}
\la \Id\ra^\alpha=\frac{S_\alpha^\Id}{\sqrt{|S_\Id^\Id|}}.
\end{equation}
That means that the expected values are also given by one-point functions instead of boundary state coefficients,
\begin{equation}
E_{(r,s)}^{(\alpha)}=\la \phi_{(r,s)}\ra^{\alpha}=\frac{S_\alpha^{(r,s)}}{\sqrt{|S_\Id^{(r,s)}|}}.
\end{equation}

String field theory with two minimal models is quite similar to the free boson theory on a torus. The state space spanned by the $L_{-n}^{(1,2)}$ generators decomposes into many irreducible representations of the $c^{(1)}+c^{(2)}$ Virasoro algebra, but we introduce Ellwood invariants only for the fundamental primaries,
\begin{equation}
E_{(r,s)(t,u)}=2\pi i \la E[c\bar c\,\phi_{(r,s)}^{(1)}\phi_{(t,u)}^{(2)} V^{aux}]|\Psi-\Psi_{TV}\ra.
\end{equation}

In principle, we could compute invariants for primaries that include some $L_{-n}^{(1,2)}$ descendants as well, but it would be quite technically quite challenging and most of these invariants would not converge well anyway. We would need to derive conservation laws for descendant states and write an algorithm similar to the one for the cubic vertex conservation laws (see the discussion at the end of section \ref{sec:SFT:Elw cons:general}). Therefore we leave this problem to future works.

\subsection{Out-of-Siegel equations}
When working in a given gauge, we solve only the projected equations of motion (\ref{equations projected}), while the remaining equations (\ref{equations remaining}) are not satisfied at finite level. However, they should be satisfied at least asymptotically, so we can use them as consistency checks.

The unsolved equations in Siegel gauge are given by
\begin{equation}
b_0 c_0 (Q\Psi+\Psi\ast\Psi)=0.
\end{equation}
Systematic study of these equations is not easy because it requires knowledge of vertices outside Siegel gauge, while we try to avoid evaluation of these vertices to save time and memory. Therefore we have decided to compute only the first nontrivial equation at level 2. Following \cite{KudrnaUniversal}, we define
\begin{equation}\label{Delta}
\Delta_S\equiv-\la 0|c_{-1}c_0b_2\left|Q\Psi+\Psi\ast\Psi\right\ra.
\end{equation}
The minus sign is just a convention added to make this quantity positive for the tachyon vacuum solution. We can also use this definition in Schnabl gauge, where (\ref{Delta}) is nonzero as well.

In this thesis, we mostly work with the SU(1,1) singlet basis. There we take advantage of the auxiliary vertices in the ghost sector, which contain some states outside Siegel gauge (see the algorithm in section \ref{sec:Numerics:V3:ghost SU11}). Therefore we are able to compute contraction of the full equations with the state $\la 0|c_{-1}j^{gh}_2$ and we can write $\Delta_S$ as
\begin{equation}\label{Delta2}
\Delta_S=-\la 0|c_{-1}j^{gh}_2\left|Q\Psi+\Psi\ast\Psi\right\ra.
\end{equation}
This expression equals to (\ref{Delta}) for solutions of the Siegel gauge equations because $\la 0|c_{-1}j^{gh}_2=\la 0|c_{-1}c_0b_2-\la 0|c_1$ and the equation corresponding to $\la 0|c_1=\la 0|c_{-1}L'^{gh}_2$ is one of the equations which we solve.

\section{Conservation laws for the cubic vertex}\label{sec:SFT:cubic cons}
One of the main technical challenges of OSFT is the evaluation of cubic vertices\footnote{To simplify the terminology, we will use the term cubic vertices for matrix elements of the cubic vertex.}. We have defined them using the CFT correlator (\ref{vertices CFT def}), but this formula is not well suited for practical calculations. Therefore we use a more efficient method for computing cubic vertices, which is based on conservation laws derived by Rastelli and Zwiebach \cite{RastelliZwiebach}\footnote{A slightly different derivation of conservation laws is described in \cite{BeccariaConservation}\cite{LumpsBeccaria}. Interestingly, conservation laws can be also derived in the sliver frame \cite{ConservationlawsArroyo}}. In this section, we develop a general procedure for derivation of these conservation laws and we show examples for the most important fields that we encounter in OSFT. A technical implementation of the conservation laws will be described in the next chapter.

These conservation laws are based on the usual CFT contour deformations techniques (see section \ref{sec:CFT:Bulk:correlators}). We take one creation operator from one of the string fields contracted with the cubic vertex, say $L_{-m}^{(2)}$, and express it as a contour integral similarly to (\ref{contour manipulation}). However, in this case, contour manipulations do not lead to simple expressions like (\ref{Ward identity}), partially because of the geometry of the cubic vertex and partially because the string field is usually not a primary field. Therefore we do not attempt to compute the OPE between the energy-momentum tensor and the insertions and, instead, we convert contour integrals into sums of annihilation operators acting on all three insertions. These expressions are much easier to evaluate using the Virasoro algebra. Since the conservation laws always remove at least one creation operator, we can reduce any vertex to (\ref{vertex primaries}) by using them repeatedly.

\subsection{General form of conservation laws}\label{sec:SFT:cubic cons:general}
First, we show how to derive conservation laws for a holomorphic field $\phi(z)$ of integer conformal weight $h$. For now, we assume that the field $\phi(z)$ is primary. We want to derive a conservation law in the form
\begin{equation} \label{conservation law gen}
\la V_3|\phi_{-m}^{(2)}=\la V_3| \left( \sum_{n} \cc(\phi_{-m})_n^{(1)}\phi_n^{(1)}+\sum_{n} \cc(\phi_{-m})_n^{(2)}\phi_{n}^{(2)}+\sum_{n} \cc(\phi_{-m})_n^{(3)}\phi_n^{(3)}\right),
\end{equation}
where $\cc(\phi_{-m})_n^{(i)}$ are constants which fully describe the conservation law. We use this type of notation to indicate which operator is the conservation law associated with because introducing a new symbol for every field would be too complicated. The sum over $n$ runs over mode numbers greater of equal to some $n_0$, where $n_0$ is usually the mode number of the first annihilation operator or the zero mode. We usually choose $n_0=0$, but sometimes it is more convenient (or necessary) to choose a different creation-annihilation operator splitting\footnote{In principle, the splitting can be even different for each entry of the vertex. Take for example the ghost theory, where we can work with three ground states of different ghost numbers, which are annihilated by different modes $b_n$ and $c_n$. Another issue is that it is not possible to remove all creation operators for fields of weight higher than 2 because the function (\ref{conservation law vm final}) violates the required asymptotic condition for small $m$. In such cases, one can decide to derive asymmetric conservation laws, which move the remaining creation operators to a single entry of the vertex.}. Some conservation laws also contain a constant term. For that purpose, we formally introduce the operator $\phi_{\Id}\equiv\Id$ and we let the sum over $n$ run over $\Id$ as well.

As we mentioned before, conservation laws are based on the idea of contour deformations in the geometry of the cubic vertex. For a given mode number $m$, we introduce the integral expression
\begin{equation}
\oint \frac{dz}{2\pi i}v_m(z)\phi(z),
\end{equation}
where $v_m(z)$ is a function which will be determined later. In order to allow smooth contour deformations, the function must be holomorphic in the whole complex plane except the three insertions at $-\sqrt{3}$, $0$ and $\sqrt{3}$. Regularity of the contour integral at infinity requires that $v_m(z)=O((1/z)^{-2h+2})$ as $z\rar\infty$. We can see this using the coordinate transformation $w=-\frac{1}{z}$, under which we find
\begin{eqnarray}
\oint_\inf \frac{dz}{2\pi i}v_m(z)\phi(z)&=&\oint_0 \frac{dw}{2\pi i} w^{-2} v_m(-1/w)\phi(-1/w)\nn \\
&=&\oint_0 \frac{dw}{2\pi i} w^{-2+2h} v_m(-1/w)\tilde\phi(w).
\end{eqnarray}
This integral vanishes if $w^{-2+2h} v_m(-1/w)=O(1)$, which leads to the aforementioned condition.

Now consider a contour $\mathcal{C}$ that encircles all three punctures. Then the expression
\begin{equation}
\oint_\mathcal{C} \frac{dz}{2\pi i} v_m(z) \left\la  \phi(z)\ f_1\circ \Psi_1(0)\ f_2\circ \Psi_2(0)\ f_3\circ \Psi_3(0) \right\ra
\end{equation}
must be identically zero for arbitrary insertions $\Psi_i$ because the contour integral vanishes at infinity thanks to the properties of $v_m(z)$. Therefore we can write
\begin{equation}\label{conservation law full C}
\la V_3| \oint_\mathcal{C} \frac{dz}{2\pi i} v_m(z) \phi(z)=0.
\end{equation}
By deforming the contour $\mathcal{C}$ into a sum of three contours $\mathcal{C}_i$ around the three punctures, we find
\begin{eqnarray}\label{conservation law aux1}
0&=&\la V_3| \sum_{i=1}^3 \oint_{\mathcal{C}_i} \frac{dz}{2\pi i} v_m(z) \phi(z)\nn \\
&=&\la V_3| \sum_{i=1}^3 \oint_{\mathcal{C}_i} \frac{dz_i}{2\pi i}f'_i(z_i) v_m(f_i(z_i)) \phi(f_i(z_i))\nn \\
&=&\la V_3| \sum_{i=1}^3 \oint_{\mathcal{C}_i} \frac{dz_i}{2\pi i} v_m^{(i)}(z_i) \tilde\phi(z_i),
\end{eqnarray}
where we use the local coordinates around the punctures given by $z=f_i(z_i)$ and where we define
\begin{equation}\label{conservation law vm local}
v_m^{(i)}(z_i)=\left(f'_i(z_i)\right)^{1-h} v_m(f_i(z_i)).
\end{equation}
Now we can expand the functions $v_m^{(i)}(z_i)$ in the local coordinates as $\sum_{n} v^{(i)}_{m,n} z_i^n$ and then we convert (\ref{conservation law aux1}) to a sum of operators,
\begin{equation}\label{conservation law final}
0=\la V_3| \sum_{i=1}^3 \oint_{\mathcal{C}_i} \frac{dz_i}{2\pi i} \sum_{n} v^{(i)}_{m,n} z_i^n \tilde\phi(z_i)
=\la V_3| \sum_{i=1}^3 \sum_{n} v^{(i)}_{m,n} \phi_{n-h+1}.
\end{equation}
This expression has the same form as (\ref{conservation law gen}), so we move $\phi_{-m}$ to the left hand side and we find
\begin{equation}
\cc(\phi_{-m})_n^{(i)}=-v^{(i)}_{m,n+h-1},\quad n\geq n_0.
\end{equation}

All that is left it to choose the correct function $v_m(z)$. From its holomorphic properties, we can guess that it can be written as a sum of terms of the form
\begin{equation}\label{conservation law ansatz}
(z-\sqrt{3})^{\eta_1} z^{\eta_2} (z+\sqrt{3})^{\eta_3}.
\end{equation}
The contour integral around the second puncture must reproduce the mode $\phi_{-m}$, so using $f_2(z)=\frac{2}{3}z+O(z^3)$, we find that the leading term has $\eta_2=-m+h-1$. The other two constants $\eta_1$ and $\eta_3$ are usually chosen to be equal to each other and using $(f_1(z)-\sqrt{3})=\frac{8}{3}z+O(z^2)$ and $(f_3(z)+\sqrt{3})=\frac{8}{3}z+O(z^2)$, we find $\eta_1=\eta_3= n_0+h-1$.\footnote{Strictly speaking, we can choose any $\eta_1\geq n_0+h-1$ and $\eta_3\geq n_0+h-1$ as long as $v_m(z)$ satisfies the regularity condition at infinity represented by $\eta_1+\eta_2+\eta_3\leq 2h-2$. Therefore it seems like we can derive several different forms of one conservation law for large enough $m$, but, after we add the lower order terms, we find that they differ maximally by the zero mode conservation law.}
Therefore we can write $v_m(z)$ as
\begin{equation}\label{conservation law vm final}
v_m(z)=C_m z^{-m+h-1} (z^2-3)^{n_0+h-1}+O(z^{-m+h}).
\end{equation}
We choose the constant $C_m$ so that $v_m^{(2)}(z_2)=z_2^{-m+h-1}+O(z_2^{-m+h})$. Using this function, the conservation law correctly reproduces $\phi_{-m}$, but sometimes it still contains lower creation operators,
\begin{equation}
0=\la V_3|\left(\phi_{-m}^{(2)}+\sum_{l=1-n_0}^{m-1}C_m^{l}\phi_{-l}^{(2)}+\dots\right).
\end{equation}
We remove these modes by subtracting a sum of $v_l(z)$ with $l<m$ from $v_m(z)$,
\begin{equation}\label{conservation law vm final2}
v_m(z)=C_m z^{-m+h-1} (z^2-3)^{n_0+h-1}-\sum_{l=1-n_0}^{m-1} C_m^l v_l(z),
\end{equation}
so that $v_m(z)$ has the expansion $v_m^{(2)}(z_2)=z_2^{-m+h-1}+O(z_2^{n_0+h+1})$ in the local coordinate. Then it leads to the correct form of the conservation law.

So far, we have assumed that the field $\phi$ is primary. However, some important fields (for example the energy-momentum tensor) are not. Therefore we consider a more general field, which has the following anomalous transformation law:
\begin{equation}\label{anomalous transformation}
\tilde \phi (w)=\left(\frac{dz}{dw}\right)^h\phi(z)+\mathcal{A}(z,w).
\end{equation}
This affects the calculations in two ways. First, the contraction of the contour $\mathcal{C}$ around infinity produces an anomaly and (\ref{conservation law full C}) changes to
\begin{equation}\label{conservation law anomaly inf}
\la V_3|\left( \oint_\mathcal{C} \frac{dz}{2\pi i} v_m(z)\phi(z)-\oint_{0} \frac{dw}{2\pi i}w^{2h-2}v_m(-\frac{1}{w})\mathcal{A}(-\frac{1}{w},w)\right)=0,
\end{equation}
where $w=-1/z$. Similar anomalies appear during the transformations to the local coordinates and the full conservation law reads
\begin{eqnarray}\label{conservation law anomaly}
0&=&\la V_3|\left( \sum_{i=1}^3 \oint_{\mathcal{C}_i} \frac{dz}{2\pi i} v_m^{(i)}(z_i) \left(\tilde\phi(z_i) -\mathcal{A}(f_i(z_i),z_i)\right)\right. \nn \\
&-& \left. \oint_{0} \frac{dw}{2\pi i}w^{2h-2}v_m(-\frac{1}{w})\mathcal{A}(-\frac{1}{w},w)\right).
\end{eqnarray}
The anomalous terms are proportional to the identity operator and they are not tied to any entry of the vertex. Therefore we conventionally add them to $\cc(\phi_{-m})^{(2)}_\Id$.

\subsection{Virasoro conservation laws}\label{sec:SFT:cubic cons:Virasoro}
As the first example, we compute some conservation laws for the energy-momentum tensor $T(z)$.
It transforms following (\ref{T transformation}) and, because the transformation law is anomalous, we have to use the equation (\ref{conservation law anomaly}) to derive its conservation laws. However, one of the anomalous terms disappears because $S(-1/w,w)=0$. Furthermore, we find that all three functions $f_i$ have the same Schwarzian derivative
\begin{equation}
S(f_i,z_i)=-\frac{10}{9}\frac{1}{(1+z_i^2)^2}=-\frac{10}{9}+\frac{20}{9}z_i^2-\frac{10}{3}z_i^4-\frac{40}{9}z_i^6+\dots.
\end{equation}
This expression is regular at $z_i=0$ and therefore the anomaly is nonzero only when $v_m^{(i)}(z_i)$ has poles, which is only around the puncture number 2.

The right hand side of conservation laws should not contain any creation operators. Therefore we choose $n_0=0$ and the leading term of the function $v_m(z)$ is
\begin{equation}
v_m(z)\sim \frac{z^2-3}{z^{m-1}}.
\end{equation}

Now we can compute the first few conservation laws. According to the previous equation, the conservation law for $L_{-1}$ is generated by the function $v_1(z)\sim z^2-3$. Around the puncture number 2, we find $v_1^{(2)}(z_2)=-9/2+O(z_2^2)$. Therefore we have to multiply $v_1$ by $-2/9$:
\begin{equation}
v_1(z)=-\frac{2}{9}(z^2-3).
\end{equation}
In the local coordinates, we find
\begin{eqnarray}
v_1^{(1)}(z_1)&=& -\frac{4}{3\sqrt{3}}z_1+\frac{8}{27}z_1^2-\frac{40}{81\sqrt{3}}z_1^3+\frac{40}{729}z_1^4+\frac{104}{729\sqrt{3}}z_1^5+O(z_1^6), \nn \\
v_1^{(2)}(z_2)&=& 1+\frac{11}{27}z_2^2-\frac{104}{729}z_2^4+O(z_2^6), \\
v_1^{(3)}(z_3)&=& \frac{4}{3\sqrt{3}}z_3+\frac{8}{27}z_3^2+\frac{40}{81\sqrt{3}}z_3^3+\frac{40}{729}z_3^4-\frac{104}{729\sqrt{3}}z_3^5+O(z_3^6).\nn
\end{eqnarray}
The contour integrals effectively work as replacements $z_i^n\rar L_{n-1}^{(i)}$ and therefore we get
\begin{eqnarray}
\la V_3|L_{-1}^{(2)}
&=&\la V_3|\left( \frac{4}{3\sqrt{3}}L_0-\frac{8}{27}L_1+\frac{40}{81\sqrt{3}}L_2-\frac{40}{729}L_3-\frac{104}{729\sqrt{3}}L_4+\dots\right)^{(1)} \nn \\
&+&\la V_3|\left(-\frac{11}{27}L_1+\frac{80}{729}L_3+\dots \right)^{(2)} \\
&+&\la V_3|\left(-\frac{4}{3\sqrt{3}}L_0-\frac{8}{27}L_1-\frac{40}{81\sqrt{3}}L_2-\frac{40}{729}L_3+\frac{104}{729\sqrt{3}}L_4+\dots \right)^{(3)}.\nn
\end{eqnarray}
There is no central term because $v_1$ is regular around the origin.

The conservation law for $L_{-2}$ requires the function
\begin{equation}
v_2(z)=-\frac{4}{27}\frac{z^2-3}{z}
\end{equation}
and we obtain
\begin{eqnarray}
\la V_3|L_{-2}^{(2)}
&=&\la V_3|\left( \frac{8}{27}L_0-\frac{80}{81\sqrt{3}}L_1+\frac{112}{243}L_2-\frac{304}{729\sqrt{3}}L_3+\frac{400}{19683}L_4+\dots\right)^{(1)} \nn \\
&+&\la V_3|\left(-\frac{5}{54}c-\frac{16}{27}L_0+\frac{19}{243}L_2-\frac{800}{19683}L_4+\dots\right)^{(2)} \\
&+&\la V_3|\left( \frac{8}{27}L_0+\frac{80}{81\sqrt{3}}L_1+\frac{112}{243}L_2+\frac{304}{729\sqrt{3}}L_3+\frac{400}{19683}L_4+\dots\right)^{(3)}. \nn
\end{eqnarray}
To derive the third conservation law, we start with the function $v_3=-\frac{8}{81}\frac{z^2-3}{z^2}$. However, we find $v_3^{(2)}(z_2)=\frac{1}{z_2^2}+\frac{7}{9}+O(z_2^2)$ in the local coordinates, so we have to add a multiple of $v_1(z)$ to remove the constant term:
\begin{equation}
v_3=-\frac{8}{81}\frac{z^2-3}{z^2}-\frac{7}{9}v_1(z).
\end{equation}
Then the conservation law reads
\begin{eqnarray}
\la V_3|L_{-3}^{(2)}
&=&\la V_3|\left(-\frac{68}{81\sqrt{3}}L_0-\frac{40}{243}L_1+\frac{152}{243\sqrt{3}}L_2-\frac{8792}{19683}L_3
+\frac{3320}{6561\sqrt{3}}L_4+\dots\right)^{(1)} \nn \\
&+&\la V_3|\left( \frac{80}{243}L_1-\frac{2099}{19683}L_3+\dots \right)^{(2)} \\
&+&\la V_3|\left( \frac{68}{81\sqrt{3}}L_0-\frac{40}{243}L_1-\frac{152}{243\sqrt{3}}L_2-\frac{8792}{19683}L_3
-\frac{3320}{6561\sqrt{3}}L_4+\dots\right)^{(3)}. \nn
\end{eqnarray}
We notice that there is no central term because both $v_3^{(2)}(z_2)$ and $S(f_2,z_2)$ are even functions and their product does not have a single pole. This property holds for higher odd $m$ as well and therefore only even conservation laws have a central term.


\subsection{Current conservation laws}\label{sec:SFT:cubic cons:current}
Next, we derive conservation laws for a weight one current. We allow a non-primary current with an anomalous transformation law of the form
\begin{equation}
\tilde j(w)=\frac{dz}{dw}j(z)+q\frac{d^2z}{dw^2}\left(\frac{dz}{dw}\right)^{-1}
\end{equation}
because we want to cover the ghost current $j^{gh}$.

We use (\ref{conservation law anomaly}) to compute the conservation laws. This time, both anomalous terms are potentially nonzero. The anomaly at infinity is equal to
\begin{equation}
2q\oint_{0} \frac{dw}{2\pi i}\frac{v_m(-1/w)}{w}.
\end{equation}
This contour integral is nonzero only if the expansion of $v_m(-1/w)$ contains a constant term and it follows from (\ref{conservation law vm final2}) that it contributes only to the $j_0$ conservation law. The other anomaly, which comes from the transformation to the local coordinates, includes the following derivatives of the maps $f_i$:
\begin{eqnarray}
\frac{d^2 f_1}{dz_1^2}\left(\frac{df_1}{dz_1}\right)^{-1}&=& \frac{4}{\sqrt{3}}+\frac{14}{9}z_1+\frac{28}{9\sqrt{3}}z_1^2+\frac{614}{243}z_1^3+O(z_1^4),\nn \\
\frac{d^2 f_2}{dz_2^2}\left(\frac{df_1}{dz_2}\right)^{-1}&=&-\frac{10}{9}z_2+\frac{230}{243}z_2^3+O(z_2^5),  \\
\frac{d^2 f_3}{dz_3^2}\left(\frac{df_3}{dz_3}\right)^{-1}&=&-\frac{4}{\sqrt{3}}+\frac{14}{9}z_3-\frac{28}{9\sqrt{3}}z_3^2+\frac{614}{243}z_3^3+O(z_1^4).\nn
\end{eqnarray}
We integrate these functions multiplied by $v_m(f_i(z_i))$ and, since all three are regular, the contour integrals are nonzero only if $v_m(f_i(z_i))$ have poles, which happens only around the puncture number 2.

The first conservation law for $j_0$ can be obtained using the constant function ${v_0(z)=1}$. It reads
\begin{equation} \label{conservation law current 0}
0=\la V_3| (j_0^{(1)}+j_0^{(2)}+j_0^{(3)}+2q).
\end{equation}
For the free boson current $\del X$, which has $q=0$, this conservation law represents momentum conservation and for the ghost current $j^{gh}$, which has $q=-3/2$, it represents the ghost number anomaly.

For higher modes, we follow the same procedure as for the energy-momentum tensor. Using functions
\begin{eqnarray}
v_1(z) &=& \frac{2}{3}\frac{1}{z}, \nn \\
v_2(z) &=& \frac{4}{9}\frac{1}{z^2}, \nn \\
v_3(z) &=& \frac{8}{27}\frac{1}{z^3}-\frac{5}{9}v_1(z), 
\end{eqnarray}
we derive the following conservation laws
\begin{eqnarray}
\la V_3|j_{-1}^{(2)}
&=&\la V_3|\left(-\frac{2}{3\sqrt{3}}j_0+\frac{16}{27}j_1-\frac{32}{81\sqrt{3}}j_2-\frac{16}{729}j_3+\frac{64}{729\sqrt{3}}j_4+\dots\right)^{(1)} \nn \\
&+&\la V_3|\left(-\frac{5}{27}j_1+\frac{32}{729}j_3+\dots \right)^{(2)} \\
&+&\la V_3|\left( \frac{2}{3\sqrt{3}}j_0+\frac{16}{27}j_1+\frac{32}{81\sqrt{3}}j_2-\frac{16}{729}j_3-\frac{64}{729\sqrt{3}}j_4+\dots\right)^{(3)},\nn
\end{eqnarray}
\begin{eqnarray}
\la V_3|j_{-2}^{(2)}
&=&\la V_3|\left(-\frac{4}{27}j_0+\frac{64}{81\sqrt{3}}j_1-\frac{128}{243}j_2+\frac{320}{729\sqrt{3}}j_3+\frac{256}{19683}j_4+\dots\right)^{(1)} \nn \\
&+&\la V_3|\left(-\frac{10}{9}q -\frac{10}{27}j_0+\frac{19}{243}j_2-\frac{800}{19683}j_4+\dots\right)^{(2)} \\
&+&\la V_3|\left(-\frac{4}{27}j_0-\frac{64}{81\sqrt{3}}j_1-\frac{128}{243}j_2-\frac{320}{729\sqrt{3}}j_3+\frac{256}{19683}j_4+\dots\right)^{(3)}, \nn
\end{eqnarray}
\begin{eqnarray}
\la V_3|j_{-3}^{(2)}
&=&\la V_3|\left( \frac{22}{81\sqrt{3}}j_0-\frac{16}{243}j_1-\frac{160}{243\sqrt{3}}j_2+\frac{10288}{19683}j_3
-\frac{3136}{6561\sqrt{3}}j_4+\dots\right)^{(1)} \nn \\
&+&\la V_3|\left( \frac{32}{243}j_1-\frac{893}{19683}j_3+\dots \right)^{(2)} \\
&+&\la V_3|\left(-\frac{22}{81\sqrt{3}}j_0-\frac{16}{243}j_1+\frac{160}{243\sqrt{3}}j_2+\frac{10288}{19683}j_3
+\frac{3136}{6561\sqrt{3}}j_4+\dots\right)^{(3)}, \nn
\end{eqnarray}
%
Notice that these conservation laws are slightly different from \cite{RastelliZwiebach}. However, the differences are proportional only to the zero mode conservation law (\ref{conservation law current 0}).

\subsection{Ghost conservation laws}\label{sec:SFT:cubic cons:ghost}
Finally, we will discuss ghost conservation laws. The $b$ ghost is a weight 2 tensor field and therefore its conservation laws can be obtained from the energy-momentum tensor conservation laws by a simple replacement $L_n\rar b_n$ and by setting the central charge to zero. Therefore we have to compute only $c$ ghost conservation laws.

In the ghost theory, we usually work with the ghost number 1 ground state $c_1|0\ra$. With respect to this vacuum, modes $c_k$ with $k\geq 1$ are annihilation operators and modes with $k\leq 0$ are creation operators.

The first conservation law for $c_0$ uses the function
\begin{equation}
v_0(z) = -3 \frac{1}{z^2(z^2-3)}
\end{equation}
and we obtain
\begin{eqnarray}
\la V_3|c_{0}^{(2)}
&=&\la V_3|\left( \frac{4}{3\sqrt{3}}c_1-\frac{8}{27}c_2-\frac{68}{81\sqrt{3}}c_3+\frac{176}{729}c_4+\frac{508}{729\sqrt{3}}c_5+\dots\right)^{(1)} \nn \\
&+&\la V_3|\left( \frac{16}{27}c_2-\frac{352}{729}c_4+\dots \right)^{(2)} \\
&+&\la V_3|\left(-\frac{4}{3\sqrt{3}}c_1-\frac{8}{27}c_2+\frac{68}{81\sqrt{3}}c_3+\frac{176}{729}c_4-\frac{508}{729\sqrt{3}}c_5+\dots\right)^{(3)}.\nn
\end{eqnarray}
By summing three cyclic permutations of this conservation law, we find
\begin{equation}
0=\la V_3|(c_{0}^{(1)}+c_{0}^{(2)}+c_{0}^{(3)}),
\end{equation}
which means that $c_0$ is a derivative of the star product. This conservation law can be also derived directly using the function
\begin{equation}
v_0^{sym}(z) = \frac{9(z^2+1)}{z^2 (z^2-3)^2},
\end{equation}
which reduces to $v_0^{sym(i)}(z_i)=\frac{1}{z_i^2}$ in the local coordinates.

To compute higher conservation laws, we use functions
\begin{eqnarray}
v_1(z) &=& -2 \frac{1}{z^3(z^2-3)}, \nn \\
v_2(z) &=& -\frac{4}{3} \frac{1}{z^4(z^2-3)}+\frac{2}{9} v_1(z),  \\
v_3(z) &=& -\frac{8}{27} \frac{1}{z^5(z^2-3)}+\frac{1}{27} v_2(z) \nn
\end{eqnarray}
and we find
\begin{eqnarray}
\la V_3|c_{-1}^{(2)}
&=&\la V_3|\left( \frac{8}{27}c_1-\frac{80}{81\sqrt{3}}c_2+\frac{40}{243}c_3+\frac{416}{729\sqrt{3}}c_4-\frac{2840}{19683}c_5+\dots\right)^{(1)} \nn \\
&+&\la V_3|\left( \frac{11}{27}c_1-\frac{80}{243}c_3+\frac{5680}{19683}c_5+\dots\right)^{(2)} \\
&+&\la V_3|\left( \frac{8}{27}c_1+\frac{80}{81\sqrt{3}}c_2+\frac{40}{243}c_3-\frac{416}{729\sqrt{3}}c_4-\frac{2840}{19683}c_5+\dots\right)^{(3)}, \nn
\end{eqnarray}
\begin{eqnarray}
\la V_3|c_{-3}^{(2)}
&=&\la V_3|\left( \frac{40}{81\sqrt{3}}c_1-\frac{112}{243}c_2+\frac{152}{243\sqrt{3}}j_2-\frac{800}{19683}j_3
-\frac{1720}{6561\sqrt{3}}j_4+\dots\right)^{(1)} \nn \\
&+&\la V_3|\left(-\frac{19}{243}c_2+\frac{1600}{19683}c_4+\dots \right)^{(2)} \\
&+&\la V_3|\left(-\frac{40}{81\sqrt{3}}c_1-\frac{112}{243}c_2-\frac{152}{243\sqrt{3}}j_2-\frac{800}{19683}j_3
+\frac{1720}{6561\sqrt{3}}j_4+\dots\right)^{(3)}, \nn
\end{eqnarray}
\begin{eqnarray}
\la V_3|c_{-3}^{(2)}
&=&\la V_3|\left( \frac{40}{729}c_1-\frac{304}{729\sqrt{3}}c_2+\frac{8792}{19683}c_3-\frac{13280}{19683\sqrt{3}}c_4
+\frac{8920}{177147}c_5+\dots\right)^{(1)} \nn \\
&+&\la V_3|\left(-\frac{80}{729}c_1+\frac{2099}{19683}c_3-\frac{17840}{177147}c_5+\dots\right)^{(2)} \\
&+&\la V_3|\left( \frac{40}{729}c_1+\frac{304}{729\sqrt{3}}c_2+\frac{8792}{19683}c_3+\frac{13280}{19683\sqrt{3}}c_4
+\frac{8920}{177147}c_5+\dots\right)^{(3)}. \nn
\end{eqnarray}

\subsection{Conservation laws and the twist symmetry}
Conservation laws can be used to prove the formula (\ref{twist vertex}) given the assumptions from section \ref{sec:SFT:basic:twist}.

Our proof is based on induction. The base case is the cubic vertex contracted with three primary operators. Assuming that the structure constants satisfy (\ref{twist 3-point}), it has the proper symmetry. In the inductive step, we have to show that a given vertex satisfies (\ref{twist vertex}) if all vertices at lower levels have the same symmetry. However, we can use the conservation laws (\ref{conservation law gen}) to compute vertices recursively, which means that it is enough to show that the conservation laws are compatible with the twist symmetry.
The sign in (\ref{twist vertex}) include three parts. $(-1)^{\phi_1+\phi_2+\phi_3}$ is related only to the primaries and $(-1)^{\Psi_1\Psi_2+1}$ follows from grassmannality of the string fields, so we are concerned with the part that includes the number operator, $(-1)^{N_1+N_2+N_3}$. Commutation of this operator through a mode $\phi_k$ produces a sign $(-1)^k$ and therefore the conservation laws have to be invariant under simultaneous replacement $\phi_{k}\rar (-1)^k\phi_k$ and exchange of vertex entries $(1)\leftrightarrow (3)$. Alternatively, we can formulate this condition in terms of the $\cc(\phi_{-m})_n^{(i)}$ constants as
\begin{equation}\label{conservation law C twist}
\cc(\phi_{-m})_n^{(i)}=(-1)^{m+n}\cc(\phi_{-m})_n^{(4-i)},\quad i=1,2,3.
\end{equation}
One can easily verify that the examples from subsections \ref{sec:SFT:cubic cons:Virasoro}, \ref{sec:SFT:cubic cons:current} and \ref{sec:SFT:cubic cons:ghost} satisfy this property. To prove this equation in full generality, we use the fact that the coefficients are given by the contour integrals  (\ref{conservation law aux1}). The condition (\ref{conservation law C twist}) is equivalent to the following symmetry of the local functions:
\begin{eqnarray}
v_m^{(i)}(-z)&=&(-1)^{m+1-h}v_m^{(4-i)}(z).
\end{eqnarray}
We can verify this equation directly. By substituting (\ref{conservation law vm final}) and (\ref{conservation law vm local}), we get the condition
\begin{eqnarray}
\left(\frac{f_i(-z)}{f_{4-i}(z)}\right)^{-m+h-1}\left(\frac{(f_i(-z))^2-3}{(f_{4-i}(z))^2-3}\right)^{n_0+h-1}
\left(\frac{f'_i(-z)}{f'_{4-i}(z)}\right)^{1-h}=(-1)^{m+1-h}.\qquad
\end{eqnarray}
The maps $f_i$, which are given by (\ref{vertex map3}), satisfy $f_i(-z)=-f_{4-i}(z)$ and $f'_i(-z)=f'_{4-i}(z)$. Therefore we find that the sign comes from the first term and that the remaining two terms are equal to 1, which concludes the proof.

\subsection{Further generalizations}
At the end of this section, we will briefly discuss the possibility of generalizing conservation laws to theories on orbifolds or on multiple D-branes backgrounds. The string field in such theories has matrix structure and, after expanding the trace in (\ref{SFT action Tr}), we observe that the action includes terms of the form
\begin{equation}
\int \Psi_{ij}\Psi_{jk}\Psi_{ki}.
\end{equation}
The off-diagonal part of the string field contains boundary condition changing operators. As an example, we can consider free boson theory with mixed Neumann-Dirichlet boundary conditions. The oscillators in this theory have half-integer labels, $\alpha_{n+\frac{1}{2}}$, where $n\in\mathbb{Z}$. They can be expressed in terms of contour integrals as
\begin{equation}
\alpha_{n+\frac{1}{2}} \leftrightarrow \oint \frac{dz}{2\pi i} z^{n+\frac{1}{2}} \del X(z),
\end{equation}
but they do not have a simple vertex operator representation in terms of the free field. For consistency, the field $\del X(z)$ must have a branch cut going between positions of the boundary condition changing operators.

Conservation laws in such theories can be computed using the formula (\ref{conservation law anomaly}), but we have to be very careful. The field $\phi$ now has branch cuts connecting the three insertions. Therefore, when we construct the function $v_m$, which now contains non-integer powers, we have to choose its branch cuts so that the phases coming from $\phi$ and $v_m$ cancel each other everywhere. Then we can use the usual contour manipulation techniques and derive the desired conservation laws.

Another issue in such setting is a lesser symmetry of the cubic vertex. We can use the cyclic symmetry only to bring the three string fields $\Psi_{ij}$, $\Psi_{jk}$, $\Psi_{ki}$ to some canonical order, but the symmetry between the three entries of the vertex is lost. Therefore we have to compute three independent sets of conservation laws for every combination of boundary condition changing operators.

\section{Conservation laws for Ellwood invariants}\label{sec:SFT:Elw cons}
In this section, we discuss our method of evaluation of Ellwood invariants using conservation laws, which were originally introduced in \cite{KMS}. These conservation laws are somewhat similar to the cubic vertex conservation laws from the previous section.

We begin with a closer look at the Ellwood state $\la E[\vv]|\equiv\la I|\vv(i,-i)$. We can write the identity string field as $\la I|=\la 0| U_{f_I}$, where
\begin{equation}
f_I(z)=\frac{2z}{1-z^2}.
\end{equation}
The vertex operator $\vv(i,-i)$ commutes with $U_{f_I}$ because $f_I(z)\circ \vv(i,-i)=\vv(i,-i)$. To see that, we observe that $f_I(\pm i)=\pm i$ and that the derivative $f'_I(\pm i)=0$ does not contribute because $\vv(i,-i)$ has conformal weight zero. Therefore we can write the Ellwood state as a wedge state with insertion
\begin{equation}\label{Ellwood state commuted}
\la E[\vv] |=\la 0| \vv(i,-i)U_{f_I}.
\end{equation}
This form allows us to extend the definition of Ellwood state even to vertex operators which are not primaries of weight $(0,0)$. Such states can no longer be interpreted as vertex operators acting on the identity string field, because the zero derivative $f'_I(\pm i)=0$ prevents us from moving $\vv$ through $U_{f_I}$, but they are well defined and they will appear in intermediate steps of our calculations.

By contracting the generalized Ellwood state (\ref{Ellwood state commuted}) with a string field $\Psi$, we get the correlation function
\begin{equation}\label{Ellwood correlator1}
\la E[\vv] |\Psi\ra=\la \vv(i,-i) f_I\circ\Psi(0)\ra.
\end{equation}
If $\Psi$ is a primary field of weight $h$, this expression simplifies to
\begin{equation}\label{Ellwood correlator2}
\la E[\vv] |\Psi\ra=(f_I'(0))^h\la \vv(i,-i) \Psi(0)\ra.
\end{equation}
This correlator can be computed using the standard CFT techniques and we will briefly discuss their implementation later. However, most components of the string field are not primary fields and direct evaluation of (\ref{Ellwood correlator1}) would not be practical. Therefore we choose a similar approach as for the cubic vertex in the previous section. We derive a set of conservation laws, which allow as to trade creation operators from the string field for annihilation operators. This leads to a very efficient recursive procedure, which decreases level of the string field in each step and eventually reduces all possible terms to (\ref{Ellwood correlator2}). However, these conservation laws sometimes affect the Ellwood state itself.

We start with a general procedure to derive these conservation laws and then we will show examples of conservation laws for the most common fields. We generalize the results from \cite{KMS}, but we choose a slightly different method of derivation of conservation laws. Similarly to section \ref{sec:SFT:cubic cons}, we primarily work in the global coordinates on the UHP, which makes the contour manipulations more clear than in \cite{KMS}.

\subsection{General form of conservation laws}\label{sec:SFT:Elw cons:general}
As in section \ref{sec:SFT:cubic cons:general}, we derive conservation laws for a conformal field $\phi$ of weight $h$ with the (possibly anomalous) transformation law (\ref{anomalous transformation}). We define a combination of modes
\begin{equation}
\Phi_m=\phi_m+(-1)^{m+h-1}\phi_{-m},
\end{equation}
which is motivated by conservation laws for the identity string field \cite{WedgeSchnabl}. In the local coordinates associated with the string field, we can write $\Phi_m$ as a contour integral,
\begin{equation}
\Phi_m=\oint_0 \frac{dw}{2\pi i} g_m(w) \phi(w),
\end{equation}
where
\begin{equation}\label{g_n def }
g_m(w)=w^{m+h-1}+(-1)^{m+h-1}w^{-m+h-1}.
\end{equation}

Now consider the following expression:
\begin{equation}
\oint_\mathcal{C} \frac{dz}{2\pi i}v_m(z)\la \vv(i,-i) \phi(z) f_I\circ \Psi(0)\ra,
\end{equation}
where $z$ is the global coordinate on the UHP, $v_m(z)$ is a function which we will determine later and the integration contour $\mathcal{C}$ encircles both $0$ and $\pm i$. The regularity at infinity requires that the function $v_m(z)$ behaves as $O\left((1/z)^{-2h+2}\right)$ for $z\rar\infty$. The contour can be contracted around $\inf$ and, assuming that $\phi$ is primary, we find that the integral vanishes. Otherwise, we find the same type of anomaly as in (\ref{conservation law anomaly inf}):
\begin{eqnarray}\label{Ellwood conservation anom}
0&=&\oint_\mathcal{C} \frac{dz}{2\pi i}v_m(z)\la \vv(i,-i) \phi(z) f_I\circ \Psi(0)\ra\\
&-&\oint_0 \frac{d\tilde z}{2\pi i} \tilde z^{2h-2}v_m(-\frac{1}{\tilde z})\aa(-\frac{1}{\tilde z},\tilde z)\la \vv(i,-i) \mathcal{I}\circ f_I\circ \Psi(0)\ra \nn,
\end{eqnarray}
where $\tilde z=-\frac{1}{z}$. The second contour integral can be evaluated directly because the correlation function does not depend on $\tilde z$ and the result is proportional to (\ref{Ellwood correlator1}).

The contour $\mathcal{C}$ from the first term of (\ref{Ellwood conservation anom}) can be split to contours around $\pm i$ and around 0. First, we will analyze the integral around zero. We evaluate this integral in the local coordinate $w$ given by $z=f_I(w)$,
\begin{eqnarray}
&&\oint_{0} \frac{dz}{2\pi i}v_m(z)\la \vv(i,-i) \phi(z) f_I\circ \Psi(0)\ra \nn\\
&=&\oint_{0} \frac{dw}{2\pi i}v_m(f_I(w))f'_I(w)\la \vv(i,-i) \phi(f_I(w)) \Psi(0)\ra \\
&=&\oint_{0} \frac{dw}{2\pi i}v_m(f_I(w))(f'_I(w))^{1-h}\la \vv(i,-i) \left(\tilde\phi(w)-\aa(f_I(w),w)\right) \Psi(0)\ra.\nn
\end{eqnarray}
The regular part of this integral must reproduce the combination of modes $\Phi_m$, which leads to the condition
\begin{equation}
v_m(f(w))(f'_I(w))^{1-h}=g_m(w).
\end{equation}
By solving this equation, we find
\begin{eqnarray}
v_m(z)&=&(-z)^{-m+h-1}(1+z^2)^{\frac{h-1}{2}}  \left(\left(1+\sqrt{1+z^2}\right)^m-(-1)^h \left(1-\sqrt{1+z^2}\right)^m\right)\nn\\
&=&2(-z)^{-m+h-1}\!\!\sum_{\substack{k=0\\k+h\ \rm{odd}}}^m \!\! \binom{m}{k}\, (1+z^2)^{\frac{k+h-1}{2}}.
\end{eqnarray}
The second line shows that the function $v_m(z)$ does not have any branch cuts and that it has the desired asymptotic behavior $O\left((1/z)^{-2h+2}\right)$. The integral around zero therefore reads
\begin{eqnarray}\label{Ellwood result1}
&&\oint_{0} \frac{dz}{2\pi i}v_m(z)\la \vv(i,-i) \phi(z) f_I\circ \Psi(0)\ra \nn\\
&=&\la E[\vv]|\Phi_m|\Psi\ra -\res_{w\rar 0}g_m(w)\aa(f_I(w),w) \la E[\vv]|\Psi\ra.
\end{eqnarray}

Next, we compute the integrals around $\pm i$, which get contributions from the OPE between $\phi$ and $\vv$. We parameterize the OPE as
\begin{eqnarray}\label{Ellwood OPE}
\phi(z) \mathcal{V}(i,-i)\sim\sum_k\frac{C_k\vv_k(i,-i)}{(z-i)^{h+h_\vv-h_k}}, \nn \\
\phi(z) \mathcal{V}(i,-i)\sim\sum_k\frac{\tilde C_k\vv_k(i,-i)}{(z+i)^{h+\bar h_\vv-\bar h_k}},
\end{eqnarray}
where the index $k$ labels a basis of bulk vertex operators. Using this OPE, we find
\begin{eqnarray}
&&\oint_{\pm i} \frac{dz}{2\pi i}v_m(z)\la \vv(i,-i) \phi(z) f_I\circ \Psi(0)\ra \nn\\
&=&\sum_k\oint_{i} \frac{dz}{2\pi i} \frac{C_k v_m(z)}{(z-i)^{h+h_\vv-h_k}}\la \vv_k(i,-i) f_I\circ \Psi(0)\ra \nn\\
&+&\sum_k\oint_{-i} \frac{dz}{2\pi i} \frac{\tilde C_k v_m(z)}{(z+i)^{h+\bar h_\vv-\bar h_k}}\la \vv_k(i,-i) f_I\circ \Psi(0)\ra \\
&=&\sum_k C_k \res_{z\rightarrow i}\frac{v_m(z)}{(z-i)^{h+h_\vv-h_k}}\la E[\vv_k] |\Psi\ra \nn\\
&+&\sum_k \tilde C_k \res_{z\rightarrow -i}\frac{v_m(z)}{(z+i)^{h+\bar h_\vv-\bar h_k}}\la E[\vv_k] |\Psi\ra. \label{Ellwood result2}
\end{eqnarray}
We emphasize that this part of the conservation law may change the Ellwood state. This happens when the OPE (\ref{Ellwood OPE}) contains vertex operators different from $\vv$ itself and the corresponding residua are nonzero. Since the conservation laws are part of a recursive algorithm, we need to compute conservation laws for all $\vv_k$ as well, although these operators are usually not primary. This is why we defined the generalized Ellwood state (\ref{Ellwood state commuted}) and why we derive the conservation laws in a way which does not rely on exact properties of $\vv$.

We have managed to find a closed formula for the residues in (\ref{Ellwood result2}), it involves a sum over products of binomials
\begin{equation}\label{residues Ellwood}
\res_{z\rightarrow i}\frac{v_m(z)}{(z-i)^n}=-2(-i)^{-m+h}\!\!\sum_{\substack{k=0\\k+h\ \rm{odd}}}^m \! \sum_{l=0}^{\frac{k+h-1}{2}}   i^{2l-n} \binom{m}{k}\binom{\frac{k+h-1}{2}}{l}\binom{2l\!-\!m\!+\!h\!-\!1}{n-1}.
\end{equation}
The residues at $-i$ are given by complex conjugation of this formula.

When we combine (\ref{Ellwood conservation anom}), (\ref{Ellwood result1}) and (\ref{Ellwood result2}) together, we get the full conservation law:
\begin{eqnarray}\label{Ellwood result final}
\la E[\vv]|\Phi_m&=&\res_{w\rar 0} w^{2h-2} v_m(-\frac{1}{w})\aa(-\frac{1}{w},w)\la E[\vv]|\nn\\
&+&\res_{w\rar 0}g_m(w)\aa(f_I(w),w) \la E[\vv]|\nn\\
&-&\sum_k C_k \res_{z\rightarrow i}\frac{v_m(z)}{(z-i)^{h+h_\vv-h_k}}\la E[\vv_k]| \\
&-&\sum_k \tilde C_k \res_{z\rightarrow -i}\frac{v_m(z)}{(z+i)^{h+\bar h_\vv-\bar h_k}}\la E[\vv_k]|. \nn
\end{eqnarray}
This formula may look complicated, but the results are quite simple in many cases of interest, see the examples in subsection \ref{sec:SFT:Elw cons:examples}.

Simple conservation laws usually appear when the vertex operator $\vv$ is primary with respect to the full symmetry algebra of the theory. However, the construction of Ellwood invariants requires that it must be primary only with respect to the full energy-momentum tensor and not with respect to some extended symmetry.
As an example, consider the operator $\del X$ in the free boson theory, which is not primary with respect to the U(1) current $\del X$, or the operator $c^{(2)} T^{(1)}-c^{(1)} T^{(2)}$ in a theory given by a product of two CFTs, which is not primary with respect to the constituent energy-momentum tensors $T^{(1,2)}$.

In the rest of this section, we will sketch how to deal with conservation laws for these Ellwood invariants. We have never implemented this algorithm, because we can still manage to evaluate (\ref{Ellwood result final}) in the free boson theory, but it would be necessary for non-fundamental primaries in any more complicated theory. The main issue here is the evaluation of the OPE (\ref{Ellwood OPE}). To avoid it, we make the conservation laws even more similar to those for the cubic vertex.

We take the problematic contour integrals around $\pm i$ and, instead of using the OPE, we can expand the function $v_m(z)$ around $\pm i$ as
\begin{equation}
v_m(z)=-\sum_{n}\ee_{n}^{\pm}(\phi_m)\ (z\mp i)^{n+h-1}
\end{equation}
and convert the contour integrals to operator modes acting on $\vv$,
\begin{equation}\label{Ellwood result operator}
\oint_{\pm i} \frac{dz}{2\pi i}v_m(z)\la \vv(i,-i) \phi(z) f_I\circ \Psi(0)\ra=-\sum_{n}\ee_{n}^{\pm}(\phi_m)\la (\phi_{n}^\pm \vv)(i,-i) f_I\circ \Psi(0)\ra,
\end{equation}
where we define $\phi_{n}^\pm\equiv \phi_{n}(\pm i)$ and the constants $\ee_{n}^{\pm}(\phi_m)$ are analogous to $\cc_{n}^{(i)}(\phi_m)$. Similarly to (\ref{conservation law gen}), we can now write the full conservation law as
\begin{equation}
\la E[\vv]|\Phi_m=\sum_{n} \ee_{n}^{+}(\phi_m) \la E[\phi_n^+ \vv] |+\sum_{n} \ee_{n}^{-}(\phi_m) \la E[\phi_n^- \vv] |,
\end{equation}
where the anomalies from the first two lines of (\ref{Ellwood result final}) are added to $\ee_\Id^{\pm}(\phi_m)$. In order to develop the full recursive algorithm, one can follow the steps from section \ref{sec:Numerics:V3}, namely to define matrix representations of $\phi_{n}^\pm$ acting on $\vv$ and combine them with $\ee_{n}^{\pm}(\phi_m)$ to form an analogue of $\kk_{n}^{(i)}(\phi_m)$.
\\

Finally, let us make a brief comment about evaluation of the correlator (\ref{Ellwood correlator2}). If the vertex operator $\vv$ is primary, then this correlator is given by (\ref{two-point bulk boundary}), but otherwise, we have to compute a relatively generic 3-point function. This can be done using the standard CFT techniques because the transformation to the local coordinates is no longer involved. In the free boson theory, we can use explicit formulas from section \ref{sec:CFT:FB:correlators}. However, for efficient evaluation of more complicated correlators, it is once again useful to develop a recursive algorithm with the same structure as for the cubic vertex. We would like to have a procedure which allows us to remove $\phi_{-m}$ from left or right part of $\vv$. Therefore we consider the contour integral
\begin{equation}
\oint_\mathcal{C} \frac{dz}{2\pi i}v_m(z)\la \vv(i,-i) \phi(z)\Psi(0)\ra,
\end{equation}
where $\Psi(0)$ is now just an ordinary primary field, and we convert it to operator expression similarly to (\ref{Ellwood result operator}). Then we can use matrix representations of $\phi_n$ acting on $\vv$ and $\Psi$ to complete the recursive algorithm similarly to section \ref{sec:Numerics:V3}.

The field $\phi$ is most likely going to be either a current or the energy-momentum tensor because the ghost part of $\vv$ is just $c\bar c$. The first case is simple, we can choose the usual function
\begin{equation}
v_m(z)=(z\mp i)^{-m}.
\end{equation}
When it comes to the energy-momentum tensor, it would be possible to use the function ${v_m(z)=(z\mp i)^{-m+1}}$, but it follows from (\ref{Ward identity}) that this contour manipulation would produce $L_{-1}$ operators. Fortunately, this can be fixed by a simple change of the function $v_m(z)$. We observe that the function
\begin{equation}
(z-z_1)^{-m+1}+\frac{(z-z_1)(z-z_2)}{(z_2-z_3)(z_3-z_1)^m}+\frac{(z-z_1) (z-z_3)}{(z_3-z_2) (z_2-z_1)^m}
\end{equation}
behaves as $(w-z_1)^{-m+1}+O(z-z_1)$ around the point $z_1$ and as $O(z-z_{2,3})$ around the points $z_{2,3}$, which means that the corresponding conservation law removes $L_{-m}$ without producing any $L_{-1}$ operators. Therefore the functions that allow us to remove $L_{-m}$ from the points $\pm i$ are
\begin{equation}
v_m(z)=(z\mp i)^{-m+1}+(\pm i)^{m+1}\left(1+z^2-2^{-m}z(z\mp i)\right).
\end{equation}

\subsection{Examples of conservation laws}\label{sec:SFT:Elw cons:examples}
After deriving the generic form of conservation laws, we will show some explicit examples, which we use to evaluate the Ellwood invariants defined in section \ref{sec:SFT:observables}. We assume that the vertex operator $\vv$ has a factorized form, $\vv=c\bar c V^{(1)}\dots V^{(N)}$, where $V^{(k)}$ belongs to the corresponding matter BCFT. To simplify the notation, we will show only the part of $\vv$ which interacts with the field $\phi$.

In order to show application of the formula (\ref{Ellwood result final}), we begin by re-deriving conservation laws for Virasoro operators from \cite{KMS}. The energy-momentum tensor has weight 2, so conservation laws include the combinations of modes
\begin{equation}
K_n\equiv L_m-(-1)^m L_{-m}.
\end{equation}
The relevant part of the vertex operator can be labeled only by its weights $(h,\bar h)$ and its OPE with $T$ is
\begin{equation}
T(z)V^{(h,\bar h)}(i,-i)\sim \frac{hV^{(h,\bar h)}(i,-i)}{(z-i)^2}+\frac{\del V^{(h,\bar h)}(i,-i)}{z-i}
\end{equation}
and similarly around $-i$. The functions $g_m(z)$ and $v_m(z)$ are given by
\begin{eqnarray}
g_m(z)&=&z^{n+1}-(-1)^{n}z^{-n+1},\\
v_m(z)&=&2(-z)^{-m+1}\!\!\!\!\sum_{\substack{k=0\\k=1\ \rm{mod}\ 2}}^m\!\!\! \binom{m}{k}\, (1+z^2)^{\frac{k+1}{2}}.
\end{eqnarray}

Now we can plug this input into (\ref{Ellwood result final}). The first anomalous term disappears because the energy-momentum tensor is a quasi-primary field. The Schwarzian derivative of $f_I(z)$ is $S(f_I(z),z)=\frac{6}{(1 + z^2)^2}$, so the second anomalous term gives us
\begin{equation}
\res_{z\rar 0}g_m(z)\frac{c}{2(1 + z^2)^2} \la E[V^{(h,\bar h)}]|=\frac{mc}{8}\left( i^m+(-i)^m\right)\la E[V^{(h,\bar h)}]|.
\end{equation}
The regular part of the conservation law comes from the OPE between $T$ and $V^{(h,\bar h)}$. We observe that the functions $v_m(z)$ behave as $O(z\mp i)$ around $\pm i$, so only the second pole from the OPE contributes to conservation laws and we get
\begin{equation}
-4m\left(h\, i^m+\bar h(-i)^m\right)\la E[ V^{(h,\bar h)}]|.
\end{equation}
By summing the two contributions, we obtain the final result
\begin{equation}\label{Ellwood cons Vir}
\la E[V^{(h,\bar h)}]K_m=m\left( i^m\left(\frac{c}{8}-4h\right)+(-i)^m\left(\frac{c}{8}-4\bar h \right) \right)\la E[V^{(h,\bar h)}]|.
\end{equation}
Notice that these conservation laws reproduce the original vertex operator $V^{(h,\bar h)}$.

Similarly, we can derive conservation laws for $b$ and $c$ ghosts, which are rather trivial:
\begin{equation}\label{Ellwood cons gh b}
\la E[ \vv]| (b_m-(-1)^m b_{-m})=0,
\end{equation}
\begin{equation}
\la E[ \vv]| (c_m+(-1)^m c_{-m})=0.
\end{equation}

Conservation laws for the ghost current $j^{gh}$ were derived in \cite{KudrnaUniversal}. The derivation includes both anomalous terms, but the result is quite simple:
\begin{equation}\label{Ellwood cons gh j}
\la E[ \vv]| (j^{gh}_m+(-1)^m j^{gh}_{-m})= \left(-\frac{1}{2} \left(i^m +(-i)^m \right) +3\delta_{m0}  \right)\la E[ \vv]|.
\end{equation}
Using this expression and (\ref{Ellwood cons Vir}) for $c=-26$ and $h=\bar h=-1$, we get conservation laws for the 'twisted' ghost Virasoros $L'^{gh}$:
\begin{equation}\label{Ellwood cons gh Lp}
\la E[ \vv]|K'^{gh}_m= \frac{1}{4} \left(i^m +(-i)^m \right) \la E[ \vv]|.
\end{equation}

Finally, we show conservation laws for $\alpha$ oscillators in case of a generic exponential primary operator, which generalize the results from \cite{KMS}:
\begin{equation}\label{Ellwood FB momentum}
\la E[ e^{i k_L X_L+i k_R X_R}]|(\alpha_m+(-1)^m \alpha_{-m})=-\sqrt{2} \left(i^m k_L+(-i)^m k_R\right)\la E[ e^{i k_L X_L+i k_R X_R}]|.
\end{equation}

\subsection{Free boson conservation laws}\label{sec:SFT:Elw cons:FB}
The free boson theory includes various operators which are primary with respect the full energy-momentum tensor, but not with respect to the U(1) current. In case of a single free boson, these operators are labeled according to SU(2) representations (see section \ref{sec:CFT:FB:bulk}), for two or more free bosons, they appear essentially at every level. We encounter these operators in the $D$-series of invariants and in the $H$ invariant.

A nontrivial primary operator can be always written as a sum of factorized operators, so it is enough to describe how the conservation laws work for a single boson. Any vertex operator consists of terms of the form
\begin{equation}\label{Ellwood FB gen op}
\del^{j_1}X\dots \del^{j_l}X\ \delb^{k_1}X\dots \delb^{k_r}X e^{i k_L X_L+i k_R X_R}(i,-i).
\end{equation}
The OPE between these operators and the $\del X$ current can explicitly written as
\begin{eqnarray}\label{Ellwood FB OPE}
&&\del X(w)\ \del^{j_1}\!X\dots \del^{j_l}\!X\ \delb^{k_1}\!X\dots \delb^{k_r}\!X  e^{i k_L X_L+i k_R X_R}(i,-i)\sim\\
&&-\frac{i}{2}\left(\frac{k_L}{w-i}+\frac{k_R}{w+i}\right) \del^{j_1}\!X\dots \del^{j_l}\!X\
   \delb^{k_1}\!X\dots \delb^{k_r}\!X e^{i k_L X_L+i k_R X_R}(i,-i)\nn \\
&&-\sum_{m=1}^l \frac{(j_m-1)!}{2(w-i)^{j_m+1}}\del^{j_1}\!X\dots\cancel{\del^{j_m}\!X} \dots\del^{j_l}\!X\
   \delb^{k_1}\!X\dots \delb^{k_r}\!X e^{i k_L X_L+i k_R X_R}(i,-i)\nn \\
&&-\sum_{m=1}^r \frac{(k_m-1)!}{2(w+i)^{k_m+1}}\del^{j_1}\!X\dots\del^{j_l}\!X\
   \delb^{k_1}\!X\dots\cancel{\delb^{k_m}\!X} \dots \delb^{k_r}\!X e^{i k_L X_L+i k_R X_R}(i,-i).\nn
\end{eqnarray}
Therefore we can compute the corresponding conservation laws directly using (\ref{Ellwood result final}) and (\ref{residues Ellwood}). The first term in the OPE leaves the vertex operator intact and it can be used to derive to (\ref{Ellwood FB momentum}). On the other hand, the remaining two terms reduce the weight of the vertex operator and lead to new Ellwood states. Therefore, in order to compute an Ellwood invariant of weight $h$, we usually need to find conservation laws for all vertex operators with lower weights. After we remove all $\alpha$ oscillators from the string field, we can compute the remaining correlator using the formulas from section \ref{sec:CFT:FB:correlators}.

\chapter{Numerical algorithms}\label{sec:Numerics}
In this chapter, we describe our computer algorithms which we used to obtain the results in this thesis and in some of our previous works \cite{MarginalKMOSY}\cite{KMS}\cite{Ising}\cite{MarginalTachyonKM}\cite{KudrnaUniversal}\cite{ArroyoKudrna}. Some parts of these algorithms have already been described in \cite{KudrnaUniversal}\cite{ArroyoKudrna} and some were inspired by \cite{GaiottoRastelli}. Readers who are not interested in details of numerical algorithms can skip most of this chapter, they should just check our extrapolation techniques in subsection \ref{sec:Numerics:observables:extrapolation}.

String field theory calculations can be divided into several different parts. The first (and technically the most complicated) is evaluation of matrix elements of the BRST charge and of the cubic vertex, which are needed for the action. This part includes finding convenient representations for the string field and the operator algebra and then implementation of the BPZ product and the conservation laws for the cubic vertex. Next, we need to solve the equations of motion, for which we use either Newton's method or the homotopy continuation method. Finally, to identify solutions, we need to compute observables, most notably the energy and Ellwood invariants. We also usually extrapolate them to infinite level for better precision.

We use a combination of Mathematica and C++ codes to execute our calculations. We have chosen this combination of programming languages to balance their respective strengths and weaknesses. Mathematica offers an environment for symbolic manipulations and tools for analysis and visualization of results, but it is not very efficient in manipulations with large quantities of real numbers. On the other hand, the C++ programming language is very fast when executing elementary operations and the OpenMP library \cite{OpenMP} offers tools for parallelization of calculations. However, manipulations with abstract objects are often difficult to encode in C++. The two programs communicate with each other using text files.

\section{Description of the string field}\label{sec:Numerics:string field}
We start with a description of the string field. The string field in the level truncation approximation is expanded into a basis of $L_0$ eigenstates as
\begin{equation}\label{string field basis}
|\Psi \ra=\sum_i t_i |i\ra,
\end{equation}
where $t_i$ are real or complex coefficients and $|i\ra$ are basis elements in the full BCFT. The coefficients mainly appear as variables in the equations of motion. We can encode them in a simple vector and they are easy to deal with, therefore we focus on description of the basis and later on matrix representations of various objects with respect to the basis.

The key feature of our numerical approach is factorization of calculations into the individual BCFTs whenever possible. Assuming that the full BCFT is given by BCFT$^{(1)}\otimes \ldots \otimes$BCFT$^{(N)}$, we can factorize a properly chosen basis as
\begin{equation}
|i\ra=|i_1\ra^{(1)}\otimes \ldots \otimes|i_N\ra^{(N)}.
\end{equation}
Therefore, before we deal with the full string field, we are first going to have a look at bases in the individual BCFTs.

The states and operators we work with should have an index which labels the theory they belong to, for example $L_{-n}^{(k)}$ belongs to BCFT$^{(k)}$. However, in most of this chapter, we usually deal with just one of the constituent BCFTs at a time. Therefore we will drop these upper indices to simplify the notation and restore them only when there is a risk of confusion.

\subsection{Universal matter sector}\label{sec:Numerics:string field:universal}
The universal matter BCFT is the simplest theory we encounter, so we will often use it to demonstrate our approach and to show examples.

The Hilbert space of this theory includes only a single vacuum Virasoro representation, so all states are of the form
\begin{equation}
L_{-i_1}\dots L_{-i_k}|0\ra,
\end{equation}
where $i_m\geq2$. If the central charge is greater than $1$, there are no null states. For illustration, we show the basis up to level 6 in the following table:
\begin{equation} \label{Virasoro basis}
\begin{array}{rcc}
\rm{state}              & \rm{level} & i \\
|0\ra                   & 0 & 1 \\
L_{-2}|0\ra             & 2 & 2 \\
L_{-3}|0\ra             & 3 & 3 \\
L_{-4}|0\ra             & 4 & 4 \\
L_{-2}L_{-2}|0\ra       & 4 & 5 \\
L_{-5}|0\ra             & 5 & 6 \\
L_{-3}L_{-2}|0\ra       & 5 & 7 \\
L_{-6}|0\ra             & 6 & 8 \\
L_{-4}L_{-2}|0\ra       & 6 & 9 \\
L_{-3}L_{-3}|0\ra       & 6 & 10 \\
L_{-2}L_{-2}L_{-2}|0\ra & 6 & 11
\end{array}
\end{equation}
We canonically order the basis states first by their level and then by mode numbers of Virasoro operators. The ordering of states at the same level is essentially irrelevant, but the ordering by level is crucial for some of our algorithms.

In a computer program, it is convenient to represent this basis as a vector made out of a structure that encodes all important information about the states. The operator content of a state can be described by a vector that contains the mode numbers. However, we have defined a more sophisticated environment in Mathematica, which we call OP, which allows us to compute commutators between operators, see later. A state of the form $L_{-i_1}\dots L_{-i_k}|0\ra$ is represented as ${\rm OP}[L[-i_1],\dots,L[-i_k], \rm vac]$. Apart from the explicit form of basis states, it is useful to store various "quantum numbers" (level, twist, etc.) which characterize their properties. These numbers follow from the operator content, but computing them repeatedly would be unnecessarily complicated and time consuming.

There are two ways how to generate this basis. The first possibility is to use a function that generates partitions (IntegerPartitions in Mathematica) to get all mode numbers that sum to a given level, remove all partitions that include 1 and then transform these lists of integers to actual states as $\{i_1,i_2,\dots,i_k\} \rar L_{-i_1}L_{-i_2}\dots L_{-i_k}|0\ra$.

The second possibility is to use a recursive procedure. We start with the ground state and then we repeat the following steps: Assuming we already have a basis up to level $L$, we act on all of its elements with a single creation operator so that we get states at level $L+1$. Then we canonically order the operators in the new states and remove all states that are equal to zero, duplicate or otherwise undesirable. Then we add the remaining states to the basis and repeat until we reach the desired level.

The first method is simpler, but the second method is more universal and it allows generalization to more complicated BCFTs where the first method does not work. This happens for example when mode numbers cannot be mapped to partitions.

\subsection{Ghost theory}\label{sec:Numerics:string field:ghost}
In the ghost theory, we can choose several different bases depending on the gauge or other conditions imposed on the string field. We represent them in a similar way as the basis of the universal matter theory, but some algorithms need auxiliary states for evaluation of vertices, so we will discuss the individual cases in more detail.

We will start with our most common setting, the SU(1,1) singlet string field in Siegel gauge. The basis of the SU(1,1) singlet part of the ghost state space has the same structure as in the universal matter theory, so we can get this basis using replacements $L'\rar L'^{gh}$ and $|0\ra\rar|\Omega\ra$\footnote{We always work with the ghost number 1 vacuum $|\Omega\ra\equiv c_1|0\ra$ in our algorithms. This way, we avoid the necessity of having an explicit representation for $c_1$ and, in case of the $bc$ basis at ghost number 1, we have the same number of $b$ and $c$ operators.}.

However, we also need auxiliary states of the form
\begin{equation}
j^{gh}_{-m}L'^{gh}_{-i_1}\dots L'^{gh}_{-i_k}|\Omega\ra
\end{equation}
for evaluation of the cubic vertex, see section \ref{sec:Numerics:V3:ghost SU11}. The basis of this auxiliary space can be constructed simply by multiplying elements of the singlet basis by one $j_{-m}^{gh}$ operator. The auxiliary states are required only up to one level less that the singlet states because the part of conservation laws which generates these states always reduces level at least by one.

Next, we take a look at the usual $bc$ basis. We order the operators in basis elements as
\begin{equation}\label{basis ghost bc}
b_{-i_1}\dots b_{-i_k}c_{-j_1}\dots c_{-j_k}|\Omega\ra.
\end{equation}
This ordering, where $b$ ghosts are always in from of $c$ ghosts, simplifies evaluation of cubic vertices because we can use just $b$ conservation laws.

In addition to physical states at ghost number 1, we also need auxiliary states with ghost numbers 0 and 2 for some intermediate calculations. Their bases have the same form, but they contain one additional $b$ or $c$ mode. Again, it is enough to consider auxiliary states only up to level $L-1$, similarly to the SU(1,1) singlet basis.

The Siegel gauge condition removes states with $c_0$ from the basis. We just keep the states $b_{-2}c_0|\Omega\ra$ and $c_0|\Omega\ra$, which are needed for the first out-of-Siegel equation, see subsection \ref{sec:Numerics:observables:Siegel}.

To generate this type of basis, it is convenient to start with independent construction of states with $b$ or $c$ operators only, which can have arbitrary ghost numbers. Then we select states with the desired ghost number out of the tensor product of these two auxiliary bases.

Finally, the basis in the ghost theory can be written using the ghost Virasoro or ghost current operators. These two bases require no auxiliary objects and they can be treated similarly as in the universal matter sector.

\subsection{Free boson theory and minimal models}\label{sec:Numerics:string field:FB MM}
Basis elements in the universal matter theory or the ghost theory are descendants of a single ground state. That changes when we consider a more typical BCFT, like the free boson theory or the Virasoro minimal models, which has multiple primary operators. The structure given by the fusion rules in such theory motivates us to treat the string field differently than in the universal sector.

Consider a theory which includes several different Verma modules. In principle, we could contain all states in a single vector, but it is more convenient keep the Verma modules in separate vectors. Therefore we split the index labeling basis elements into two,
\begin{equation}
i\rar (p,i_p).
\end{equation}
The first index now labels primary operators and the second index their descendants. The representation of such basis in a computer code changes from a vector to a two-dimensional object. As we are going to see later, this structure allows as a direct implementation of fusion rules.

In most aspects, the free boson theory and the Virasoro minimal models can be treated similarly because the differences between $\alpha$ oscillators and Virasoro generators and different labeling of primary operators do not play a significant role in our code. However, there are some differences that have to be handled separately.

In the free boson theory, we impose the parity even condition, which is satisfied by the states (\ref{states FB even}). However, these states are not eigenstates of the $\alpha_0$ operator. Therefore, when working with just with the free boson theory, we consider states of definite momentum (both positive and negative) and we impose the parity even condition only on the full basis. The transition between these two types of bases is implemented by addition of multiplicative factors to the full vertices, see later.

In the Virasoro minimal models, we encounter a different issue, null states. These must be removed and we select a new irreducible basis $|i^{IR}\ra$, which includes only representatives of nontrivial states. This basis is then used for construction of the full string field. We will discuss the process of elimination of null states later in subsection \ref{sec:Numerics:V2:MM}.

\subsection{Full string field}\label{sec:Numerics:string field:full}
As we mentioned before, the basis of the full string field factorizes as
\begin{equation}\label{basis full}
|i\ra=|i_1\ra^{(1)}\otimes \ldots \otimes|i_N\ra^{(N)},
\end{equation}
so we construct it simply by tensor product of bases of the constituent theories. Then we select states that pass through the level cut-off and satisfy other desired properties (most often, the twist even condition).

As an example, we show the first few basis states of the SU(1,1) singlet universal string field:
\begin{equation}
\begin{array}{rcccc}
state                              & level & i  & i_1 & i_2\\
|\Omega\ra                         & 0     & 1  & 1   & 1  \\
L_{-2}^m |\Omega\ra                & 2     & 2  & 2   & 1  \\
L'^{gh}_{-2} |\Omega\ra            & 2     & 3  & 1   & 2  \\
L_{-3}^m |\Omega\ra                & 3     & 4  & 3   & 1  \\
L'^{gh}_{-3} |\Omega\ra            & 3     & 5  & 1   & 3  \\
L_{-4}^m |\Omega\ra                & 4     & 6  & 4   & 1  \\
L_{-2}^m L_{-2}^m|\Omega\ra        & 4     & 7  & 5   & 1  \\
L_{-2}^m L'^{gh}_{-2}|\Omega\ra    & 4     & 8  & 2   & 2  \\
L'^{gh}_{-4}|\Omega\ra             & 4     & 9  & 1   & 4  \\
L'^{gh}_{-2}L'^{gh}_{-2}|\Omega\ra & 4     & 10 & 1   & 5
\end{array}
\end{equation}
The numbers $i_1$ and $i_2$ represent projections of the basis states to (\ref{Virasoro basis}) and the ghost counterpart respectively. We order the basis states first by their level and then by their operator contents. The ordering by level allows up to truncate the string field to a lower level just by restricting the range of $i$.

Interestingly, all we need to remember about the basis is the multiindex $(i_1,\dots,i_N)$. The explicit form of states is not necessary because all objects that we encounter in string field theory can be constructed from simpler objects in the constituent theories (see (\ref{kinetic term no Siegel}) for a nontrivial example). However, it useful to store some additional information about the basis states (level, twist, etc.) to avoid repeated evaluation of these numbers.

In theories where we label some of the constituent bases by doubled index, this structure is inherited by the full basis as well. Elements of the full basis are then labeled by $(p,i)$, where $p$ labels products of primary operators and $i$ their descendants. The multiindex which represents projections to the constituent bases changes to
\begin{equation}
(i_1\dots,i_M,(p_{M+1},i_{M+1}),\dots,(p_N,i_N)),
\end{equation}
where we have $M$ theories with simple bases and $N-M$ theories with composite bases. Coefficients of the string field should have the double index structure as well, but, for simplification of some matrix operations, it is easier to have the coefficients in a single vector. Therefore we keep two copies of the coefficients, $t_i$ and $t_{(p,i)}$, and use the one which is more convenient in the given context.

\section{Matrix representations of operators}\label{sec:Numerics:matrix rep}
For evaluation of vertices and some other OSFT calculations, it is essential to know how various (usually annihilation) operators act on the string field, respectively on its basis. For that purpose, it is useful to compute matrix representations of operators. In this section, we show how to compute matrix representations in the constituent theories. These can be then used to construct representations of more complicated operators (total Virasoros, the BRST charge $Q$, etc.).

We are going to illustrate our approach on the universal matter theory. There we need to compute expressions of the form
\begin{equation}
L_{m}|i\ra=L_{m}L_{-n_1}\dots L_{-n_k}|0\ra.
\end{equation}
Such expression can be easily evaluated using the Virasoro algebra (\ref{Virasoro algebra}), but the implementation of such algebra in a computer code can be a bit tricky. Fortunately, Mathematica offers a simple solution. We use the previously mentioned environments OP, which represents a chain of Virasoro operators $L_{n_1}L_{n_2}\dots L_{n_k}$ as ${\rm{OP}}[L[n_1],L[n_2],\dots,L[n_k]]$. The patter matching in Mathematica easily allows us to encodes basic properties of a product of operators (associativity, distributivity, etc.) and we implement the Virasoro algebra as
\begin{eqnarray}
&&{\rm{OP}}[A\_\_\_,L[m\_],L[n\_],B\_\_\_]:=  \nn\\
&&\qquad{\rm{OP}}[A,L[n],L[m],B]+{\rm{OP}}[A,(m-n)L[m+n]\\
&&\qquad+c/12(m^3-m){\rm{KroneckerDelta}}[m+n,0],B]/;m>n.\nn
\end{eqnarray}

This implementation of the Virasoro algebra is quite simple, but the actual calculations can be quite time consuming, especially for expressions like $L_n (L_{-2})^k |0\ra$. Therefore we want to avoid repeated evaluation of these expressions and we store their results for later use in form of a matrix representation. We denote matrix representations by the symbol $\mm$ with the relevant operator in brackets. The action of a Virasoro operator on a basis state in this notation reads
\begin{equation}\label{matrix representation}
L_m|i\ra=\sum_j \mm (L_m)_{ij} |j\ra.
\end{equation}

This representation is however still quite inefficient. To see why, let's have a look at the representation of the $L_{-2}$ operator on the basis (\ref{Virasoro basis}):
\begin{equation}
\mm (L_2)=\left(
\begin{array}{ccccccccccc}
0 & 0 & 0 & 0 & 0 & 0 & 0 & 0 & 0 & 0 & 0 \\
c/2 & 0 & 0 & 0 & 0 & 0 & 0 & 0 & 0 & 0 & 0 \\
0 & 0 & 0 & 0 & 0 & 0 & 0 & 0 & 0 & 0 & 0 \\
0 & 6 & 0 & 0 & 0 & 0 & 0 & 0 & 0 & 0 & 0 \\
0 & 8+c & 0 & 0 & 0 & 0 & 0 & 0 & 0 & 0 & 0 \\
0 & 0 & 7 & 0 & 0 & 0 & 0 & 0 & 0 & 0 & 0 \\
0 & 0 & 5+c/2 & 0 & 0 & 0 & 0 & 0 & 0 & 0 & 0 \\
0 & 0 & 0 & 8 & 0 & 0 & 0 & 0 & 0 & 0 & 0 \\
0 & 0 & 0 & c/2 & 6 & 0 & 0 & 0 & 0 & 0 & 0 \\
0 & 0 & 0 & 10 & 0 & 0 & 0 & 0 & 0 & 0 & 0 \\
0 & 0 & 0 & 0 & 24+3c/2 & 0 & 0 & 0 & 0 & 0 & 0 \\
\end{array}
\right)\nn
\end{equation}
The matrix is sparse and therefore we would spend most of the time during a matrix multiplication by summing over zero elements. To remove the zeros, we store only values and positions of nonzero elements. In Mathematica, we can represent each line of the matrix as a List with a pair of a real number and an integer in each element; in C++, we define an equivalent structure. Using the Mathematica notation, the matrix above reduces to
\begin{equation}\label{LxyL red}
\mm (L_2)=\left(
\begin{array}{l}
\{\}  \\
\{\{c/2,1\}\} \\
\{\} \\
\{\{6,2\}\} \\
\{\{8+c,2\}\} \\
\{\{7,3\}\} \\
\{\{5+c/2,3\}\} \\
\{\{8,4\}\} \\
\{\{c/2,4\},\{6,5\}\} \\
\{\{10,4\}\} \\
\{\{24+3c/2,5\}\}
\end{array}
\right)
\end{equation}
Now we do not have to sum over the whole range of the index $j$ in (\ref{matrix representation}), but only over the nonzero elements. For clarity of notation, we will continue to use the expression (\ref{matrix representation}), but from now on, this type of sum is to be understood as
\begin{equation}\label{reduced sum}
\sum_j \mm (L_m)_{ij} |j\ra \equiv \sum_{j=1}^{l_i} \mm (L_m)_{ij,1}|\mm (L_m)_{ij,2}\ra,
\end{equation}
where $l_i$ is the number of nonzero elements in the $i$-th row of the matrix. It is also convenient to introduce a formal operator $L_{\Id}\equiv \Id$ and its matrix representation.

Matrix representations in more complicated theories may receive another label that indicates which part of state space they act on. For example, the operator $b_n$ changes ghost number of a state from $g$ to $g-1$. Therefore we label these matrix representations as
\begin{equation}\label{matrix representation b}
b_m|i^{(g)}\ra=\sum_j\mm(b_m)_{ij}^{(g)}|j^{(g-1)}\ra
\end{equation}
and similarly for $c_n$.

The Hilbert space of the free boson theory or the minimal models consists of several Verma modules, so we define matrix representations for each Verma module separately,
\begin{eqnarray}
\alpha_m|(p,i)\ra=\sum_j\mm(\alpha_m)^p_{ij}|(p,j)\ra,\\
L_m|(p,i)\ra=\sum_j\mm(L_m)^p_{ij}|(p,j)\ra.
\end{eqnarray}

Finally, there is a specific case of $j^{gh}_m$ operators acting on SU(1,1) singlet states, which can produce states both from the physical and the auxiliary basis. Therefore we define two separate matrices for $j^{gh}_m$:
\begin{equation}\label{matrix representation j}
j^{gh}_m|i\ra^{phys}=\sum_j\mm(j^{gh}_m)^{phys}_{ij}|j\ra^{phys}+\sum_j\mm(j^{gh}_m)^{aux}_{ij}|j\ra^{aux}.
\end{equation}

\subsection{Recursive algorithm for matrix representations}
The direct method of computing matrix representations can be sometimes quite slow (although it is still much faster than evaluation of cubic vertices at high levels). Therefore we have developed a faster alternative, which is unfortunately limited only to some theories.

In the first step of this algorithm, we compute matrix representations of creation operators, say $\mm(L_{-n})$. The evaluation of these matrices is much faster because expressions like $L_{-n}L_{-i_1}\dots L_{-i_k}|0\ra$ require less operations to get the operators to the canonical ordering. The number of nontrivial matrix elements is also smaller because we do not need those that go over the maximal level.

Next, we introduce a notation that represents separation of the first operator from a given basis element,
\begin{equation}\label{operator separation}
|i\ra= L_{-m}|\hat i\ra.
\end{equation}
It is convenient to store the two numbers, $m$ and $\hat i$, together with other basis data because we use them frequently in this and in other algorithms. Of course, this operation is not defined if $|i\ra$ equals to the vacuum (or more generally, to a primary state), so these cases must be handled separately.

Now we can write
\begin{equation}\label{matrix representation recursive}
L_n|i\ra =L_n L_{-m} |\hat i\ra =L_{-m} L_n |\hat i\ra +(n+m)L_{n-m}|\hat i\ra+\frac{c}{12}(m^3-m)\delta_{nm} |\hat i\ra.
\end{equation}
By replacing the Virasoro operators by their matrix representations, we get a recursive formula
\begin{equation}
\mm (L_n)_{ij}=\sum_k \mm (L_n)_{\hat i k}\mm (L_{-m})_{kj}+(n+m)\mm (L_{n-m})_{\hat i j}+\frac{c}{12}(m^3-m)\delta_{nm}\delta_{\hat i j}.
\end{equation}
To be sure that the r.h.s. of this expression contains no unknown terms for all values of the indices, we compute the matrix representations in ascending order in $n$ and for each $n$, we start with $\mm (L_n)_1$ and then we go in ascending order in $i$. Since $\hat i <i$ and $n-m<m$, it is guaranteed that the terms on the r.h.s. are always known by the time they are needed.

This algorithm can be easily generalized to other theories, but it is guaranteed to work well only if the basis contains only a single type of operator. When there are more, complications may arise. As an example, let's consider matrix representations of $b_n$ on the ghost number 1 basis. In the expression analogous to (\ref{matrix representation recursive}), we encounter terms like $b_{-m}b_n|\hat i\ra$, where the state $|\hat i\ra$ has ghost number 2. In order to compute representations at ghost number 2, we need representations at ghost number 3, and so on. We also find that we need matrix representations of $c_n$. In the end, the algorithm gets so complicated that the simpler direct approach is more efficient.

A somewhat similar problem arises for matrix representations of $j^{gh}$ on the singlet basis. This algorithm mixes representations of $j^{gh}$ and $L'^{gh}$. However, in this case, the number of auxiliary objects is manageable and the algorithm works well.

\section{Kinetic term and Gram matrices}\label{sec:Numerics:V2}
Once we have matrix representations of all necessary operators, we can shift our attention towards the OSFT action. We start with the kinetic term, which is given by the 2-vertex with the insertion of the BRST operator. The 2-vertex can be expressed in terms of the ordinary BPZ product, so the kinetic term is fully determined by matrix elements of the BRST operator\footnote{In this section, we normalize the BPZ product so that it reproduces the kinetic term of the dimensionless potential (\ref{Energy def2}), that is $\la 0|c_{-1}c_0c_1|0\ra=1$. We return to the overall normalization of BCFT correlators only when evaluating observables.}
\begin{equation}
Q_{ij}=\la i|Q|j\ra.
\end{equation}

If we restrict our attention to Siegel gauge, this expression simplifies to
\begin{equation}
Q_{ij}=\la i|c_0 L^{tot}_0|j\ra.
\end{equation}
Therefore the matrix representation of $Q$ can be written in terms of products of Gram matrices from the constituent BCFTs as
\begin{equation}\label{kinetic term Siegel}
Q_{ij}= h_i\la i_{gh}|c_0|j_{gh}\ra^{gh}\prod_s \la i_s|j_s\ra^{(s)},
\end{equation}
where $h_i$ are eigenvalues of the $L^{tot}_0$ operator and $s$ labels matter theories. Gram matrices will be denoted as $G_{ij}^{(s)}\equiv \la i|j\ra^{(s)}$ in matter theories and as $G_{ij}^{gh}\equiv \la i|c_0|j\ra^{gh}$ in the ghost theory.

The kinetic term in other gauges gets more complicated. The BRST charge can be expressed in several different forms, for example
\begin{eqnarray}
Q&=&\sum_{n=-\infty}^\infty c_{-n} L_n^m+\frac{1}{2}\sum_{n=-\infty}^\infty \normc c_{-n} L_n^{gh}\normc-\frac{1}{2} c_0, \label{Q 1} \\
&=&\sum_{n=-\infty}^\infty c_{-n} L_n^m+\frac{1}{2}\sum_{m,n=-\infty}^\infty (m-n)\normc c_{m}c_n b_{-m-n}\normc-c_0, \label{Q 2} \\
&=&\sum_{n=-\infty}^\infty c_{-n} L_n^m-\sum_{n=-\infty}^\infty n\normc c_{-n} j_n^{gh}\normc- c_0. \label{Q 3}
\end{eqnarray}
The first form is usually the most useful one. By contracting it with the basis states (\ref{basis full}), we find quite a complicated expression,
\begin{eqnarray}\label{kinetic term no Siegel}
Q_{ij}&=&\sum_{n=-\infty}^\infty \la i_{gh}|c_{-n}|j_{gh}\ra^{gh} \sum_{s}\left(\la i_s|L_n^{(s)}|j_s\ra^{(s)}\prod_{r\neq s}G_{i_r j_r}^{(r)}\right)\\
&+&\frac{1}{2}\left(\sum_{n=-\infty}^\infty \la i_{gh}|\normc L_{-n}^{gh}c_{n}\normc|j_{gh}\ra-G_{i_{gh}j_{gh}}^{gh}\right)\prod_s G_{i_s j_s}^{(s)}. \nn
\end{eqnarray}
We need several different objects to construct this matrix. In matter theories, we need Gram matrices and $\la i_s|L_n^{(s)}|j_s\ra^{(s)}$, in the ghost theory $\la i_{gh}|c_n|j_{gh}\ra^{gh}$ and $\la i_{gh}|L_{-n}^{gh}c_{n}|j_{gh}\ra^{gh}$. The range of the summation over $n$ can be restricted to $|n|\leq L$ thanks to the level truncation and the symmetry of the BPZ product relates expressions with $n$ and $-n$, so we can restrict the range further to $n\geq 0$ (if we introduce an appropriate multiplicity factor).
In the ghost current basis, a similar formula can be derived using (\ref{Q 3}), but it needs matrices $\la i_{gh}|j_{-n}^{gh}c_{n}|j_{gh}\ra^{gh}$ instead of $\la i_{gh}|L_{-n}^{gh}c_{n}|j_{gh}\ra^{gh}$.

All the new matrices are closely related to Gram matrices and they can be computed in a similar way. Therefore, in rest of this section, we will focus on evaluation of Gram matrices in the constituent theories.

\subsection{Universal matter sector}\label{sec:Numerics:V2:universal}
As usual, we illustrate how our algorithm works in the universal matter theory and discuss its modifications to other theories later.

Elements of the Gram matrix are defined as
\begin{eqnarray}
G_{ij}=\la i|j\ra&=&{\rm{bpz}}(L_{-m_1}\dots L_{-m_k} |0\ra)L_{-n_1}\dots L_{-n_l} |0\ra \nn \\
 &=&(-1)^{\sum_i m_i}\la 0| L_{m_k}\dots L_{m_1} L_{-n_1}\dots L_{-n_l} |0\ra.
\end{eqnarray}
They can be computed by commuting all creation operators to the left and all annihilation operators to the right. However, direct evaluation of all commutators would be too time consuming. Therefore we introduce a recursive algorithm which uses the matrix representations of Virasoro operators from the previous section.

This algorithm also needs the information encoded in (\ref{operator separation}), that is the mode number of the first Virasoro operator $m$ and the number of state $\hat i$ that remains after separation of the first operator. In fact, once we have this information (and the matrix representations $\mm (L_n)$), the explicit form of the basis elements (\ref{Virasoro basis}) is no longer needed and all we have to remember about them is the following table of numbers:
\begin{equation}
\begin{array}{l|lllllllllcc}
i          & 1 & 2 & 3 & 4 & 5 & 6 & 7 & 8 & 9 & 10 & 11 \\\hline
\hat i     & 0 & 1 & 1 & 1 & 2 & 1 & 2 & 1 & 2 & 3  & 5  \\
m          & 0 & 2 & 3 & 4 & 2 & 5 & 3 & 6 & 4 & 3  & 2  \\
\rm{Level} & 0 & 2 & 3 & 4 & 4 & 5 & 5 & 6 & 6 & 6  & 6  \\
\end{array}
\end{equation}

Using the matrix representations, we can compute an element of the Gram matrix as
\begin{eqnarray}\label{Gram matrix recursive}
G_{ij}&=&\la i |j\ra=\la L_{-m} \hat i | j\ra=(-1)^{m}\la \hat i |L_m|j\ra\\
&=&(-1)^{m}\sum_k \mm (L_m)_{j k}\la \hat i |k\ra=(-1)^{m}\sum_k \mm (L_m)_{j k}G_{\hat i k}, \nn
\end{eqnarray}
where the sum over $k$ is defined using (\ref{reduced sum}) and it runs only over nonzero elements of $\mm (L_m)$. This formula expresses $G_{ij}$ as a sum of elements at lower levels. To compute the whole matrix, we start by setting $G_{11}=1$ and then we compute higher elements recursively in ascending order. Thanks to that, the algorithm never needs to access a yet unknown element. The algorithm can be sped up using the fact that the Gram matrix is block diagonal, so we need to compute only elements with matching levels.
The algorithm usually runs so fast that there is no need to parallelize it.

In order to compute (\ref{kinetic term no Siegel}), we also need the matrices $\la i|L_m|j\ra$. These matrices can be easily computed from the Gram matrix as
\begin{equation}
\la i|L_m|j\ra=\sum_k\mm(L_m)_{jk}G_{ik}.
\end{equation}

\subsection{Ghost theory}\label{sec:Numerics:V2:ghost}
The definition of Gram matrix in the ghost theory must be generalized by insertion of one $c$ ghost to saturate the ghost number. In Siegel gauge, it is enough to compute just the matrix $\la i|c_0|j\ra$, but in other gauges, we need a larger set of matrices $\la i|c_n|j\ra$. The algorithms to compute these matrices differ depending on the choice of basis.

\subsubsection{SU(1,1) singlet basis}
In our most common basis, the SU(1,1) singlet basis in Siegel gauge, the evaluation of the Gram matrix is actually very simple. We can follow the approach from the universal matter sector and use the equation (\ref{Gram matrix recursive}), where we just replace all matter objects by their ghost counterparts. The reader may wonder what happens to commutators between $L'^{gh}_m$ and $c_0$, which appear when we manipulate with the Virasoro operators. We will show that they do not contribute to the Gram matrix.

The commutator in question is given by
\begin{equation}
\left[L'^{gh}_m,c_n\right]=-(m+n)c_{m+n}
\end{equation}
and we can see that it produces operators $c_n$ with $n\neq 0$, which are not present in (\ref{Gram matrix recursive}). However, the recursive algorithm must eventually reduce all matrix elements to $\la \Omega |c_0|\Omega\ra$ and the last commutator, which changes some $c_n$ back to $c_0$, produces a zero coefficient. Therefore we can effectively set the commutator to zero and use (\ref{Gram matrix recursive}) to compute the Gram matrix.

\subsubsection{$bc$ basis}
The algorithm in the $bc$ basis is more complicated. We have to manipulate with modes of $b$ and $c$ ghosts, which change ghost numbers of states. Therefore we need some auxiliary objects to compute the required matrices.

We begin with the algorithm for the full basis, which includes $c_0$. In that case, we need to know the set of matrices $G^{(n)}_{ij}\equiv \la i|c_{n}|j\ra$. The simplest way to compute these matrices is to act with the $c_n$ operators on one of the states. In terms of matrix representations, we get
\begin{equation}
G^{(n)}_{ij}=\la i^{(1)}|c_{n}|j^{(1)}\ra=\sum_k \mm(c_n)_{jk}^{(1)}\la i^{(1)} |k^{(2)}\ra=\sum_k\mm(c_n)_{jk}^{(1)}G_{ik}^{aux},
\end{equation}
where we indicate ghost numbers of states by upper indices and we define an auxiliary matrix $G_{ij}^{aux}\equiv \la i^{(1)} |j^{(2)}\ra$. Therefore this algorithm requires an auxiliary basis of states of ghost number 2. The auxiliary matrix does not have any explicit insertions, so we compute it using simple manipulations with $b$ operators,
\begin{equation}
G_{ij}^{aux}=(-1)^m\la \hat i^{(2)}|b_{m}|j^{(2)}\ra=(-1)^m\sum_k \mm(b_m)_{jk}^{(2)} G_{k \hat i}^{aux}.
\end{equation}
In the second step, we changed the order of the BPZ product to match the definition of the auxiliary matrix. While doing so, we have to be careful about potential signs coming from anticommutators of ghost operators, but it turns out that the overall sign remains the same.

The additional matrices $\la i|L_{-n}c_{n}|j\ra$ are easiest to compute using matrix representations of Virasoro operators on the ghost basis,
\begin{equation}\label{Gram matrix L insertion}
\la i|L_{-n}c_{n}|j\ra=(-1)^n \sum_k\mm(L_n)_{ik}G^{(n)}_{kj}.
\end{equation}

\subsubsection{$bc$ basis in Siegel gauge}
In Siegel gauge, we need just the Gram matrix $G_{ij}=\la i^{(1)}|c_0|j^{(1)}\ra$. We could compute it using the algorithm above and remove matrix elements corresponding to states outside Siegel gauge, but we choose a different approach which requires just Siegel gauge states and which serves as a toy example for the cubic vertex algorithm.

This algorithm needs auxiliary bases at ghost numbers 0 and 2 and an auxiliary matrix $G^{aux}_{ij}=\la i^{(0)}|c_0|j^{(2)}\ra$, which represents BPZ products of the auxiliary states. Elements of the Gram matrix can be computed as
\begin{eqnarray}
G_{ij}&=&\la b_{-m} \hat i^{(2)} |c_0|j^{(1)}\ra=(-1)^{m+1}\la \hat i^{(2)} |c_0 b_m|j^{(1)}\ra\nn\\
&=&(-1)^{m}\sum_k\mm(b_m)^{(1)}_{jk} G^{aux}_{k\hat i}.
\end{eqnarray}
There is an additional sign in the first line for commuting $b_m$ through $c_0$ and another sign in the second line for reversal of the order of the BPZ product to match the definition of $G^{aux}$.
Elements of the auxiliary matrix can be expressed in a similar way:
\begin{eqnarray}
G^{aux}_{ij}&=&\la b_{-m} \hat i^{(1)} |c_0|j^{(2)}\ra=(-1)^{m+1}\la \hat i^{(1)} |c_0 b_m|j^{(2)}\ra\nn\\
&=&(-1)^{m}\sum_k\mm(b_m)^{(2)}_{jk} G_{\hat ik}.
\end{eqnarray}

The crucial property of this algorithm is that it regularly switches between the two matrices. In this aspect, it differs from the previous algorithm, where we  compute the auxiliary matrix first. Therefore we have to evaluate both matrices simultaneously. One possible way how to do it would be to split the loops over $i$ and $j$ according to levels. This recursive algorithm always reduces level of matrix elements, so computing all elements of both matrices at a given level before moving to the next one would be a safe solution. However, this approach would be cumbersome and it would not demonstrate the method we use for the cubic vertex later.

Instead, we actively compute only elements of the Gram matrix. However, every time the algorithm tries to access some matrix element, it first checks whether the given element has already been evaluated. If not, the algorithm calls itself recursively to compute the missing element and then it continues with the current element.

There are two ways how to indicate status of matrix elements. The first possibility is to use directly the values of the elements, known elements are represented by real numbers and unknown elements by some non-numerical constant. The second possibility is to use auxiliary matrices of boolean numbers of the same sizes as $G$ and $G^{aux}$. The first option is simpler, but the second is safer in case of parallelization. In this case, we can use both options, but we have to use the second one for the cubic vertex.

\subsubsection{Virasoro and ghost current basis}
Finally, we move to the Virasoro and the ghost current basis. These bases describe the full state space, so we need the matrices $G^{(n)}_{ij}$, which have one $c_n$ insertion. We can evaluate them following the approach from the universal matter sector, we just have to take into consideration commutators of $c_n$ with $L^{gh}_m$ or $j^{gh}_m$. In the Virasoro case, we find
\begin{eqnarray}
G^{(n)}_{ij}&=&\la L^{gh}_{-m} \hat i |c_n|j\ra \nn \\
&=&(-1)^{m}\la \hat i|c_n L^{gh}_m|j\ra-(2n+m)(-1)^{m}\la \hat i|c_{m+n}|j\ra\\
&=&(-1)^{m}\sum_k\mm(L^{gh}_m)_{jk} G^{(n)}_{\hat ik}-(2n+m)(-1)^{m}G^{(m+n)}_{\hat ij}.\nn
\end{eqnarray}
The algorithm mixes matrices with different $c_n$ insertions, so we have to compute the elements in a correct order to avoid problems. First, we compute $G^{(n)}_{1j}$ in ascending order in $n$ and then the remaining elements in descending order in $n$.

Modification to the ghost current basis is simple, we replace the commutator $[L^{gh}_m,c_n]$ by $[j^{gh}_m,c_n]$ and we get
\begin{eqnarray}
G^{(n)}_{ij}=(-1)^{m+1}\sum_k\mm(j^{gh}_m)_{jk} G^{(n)}_{\hat ik}+(-1)^{m+1}G^{(m+n)}_{\hat ij}.
\end{eqnarray}

For the BRST charge $Q$, we also need the matrices $\la i|L^{gh}_{-n}c_{n}|j\ra$ or $\la i|j^{gh}_{-n}c_{n}|j\ra$. These are related to $G^{(n)}_{ij}$ as
\begin{eqnarray}
\la i|L^{gh}_{-n}c_{n}|j\ra&=&(-1)^n \sum_k\mm(L^{gh}_n)_{ik}G^{(n)}_{kj}, \\
\la i|j^{gh}_{-n}c_{n}|j\ra&=&(-1)^{n+1} \sum_k\mm(j^{gh}_n)_{ik}G^{(n)}_{kj}.
\end{eqnarray}

\subsection{Free boson theory}\label{sec:Numerics:V2:FB}
Basis states in the free boson theory are labeled by doubled index $(k,i)$, so we divide the Gram matrix into blocks based on momentum. We define matrices
\begin{equation}
G^{(k)}_{ij}\equiv\la (-k,i)|(k,j)\ra=\la (k,i)|(-k,j)\ra.
\end{equation}
We can compute these matrices using a formula similar to the universal matter sector:
\begin{equation}
G_{ij}^{(k)}=(-1)^{m+1}\sum_l \mm (\alpha_m)_{jl}^k G^{(k)}_{\hat i l}.
\end{equation}
However, the algebra of $\alpha$ oscillators is so simple that these matrices are diagonal and they can be computed directly.
An element that corresponds to the state $(\alpha_{-m_1})^{n_1}\dots (\alpha_{-m_l})^{n_l}|k\ra$ is given by
\begin{equation}
(-1)^{\sum_i (m_i+1)n_i}\prod_i (m_i)^{n_i} (n_i!).
\end{equation}

This Gram matrix is defined using pure momentum states. Once we move to the full basis, we need parity even states, so we have to transform the matrices. Consider the following BPZ product of two primary states $|k\ra_\pm$:
\begin{equation}
\la k_\pm |k\ra_\pm=\pm\frac{1}{2} (1+\delta_{k0}).
\end{equation}
Therefore the transformation is realized by multiplication by this factor. If there is just one free boson, there is no problem and we can just multiply all matrix elements by these numbers. However, if there are more bosons, we realize that the BPZ product cannot be fully factorized because $|k_1,k_2\ra_\pm\neq |k_1\ra_\pm |k_2\ra_\pm$. Therefore we multiply elements of the Gram matrix only by the sign, which equals to $(-1)^{\# \alpha}$ and which is factorizable, and we add the other part of the factor, which equals either to 1 or to $\frac{1}{2}$, only to the full matrix $Q_{ij}$.

\subsection{Minimal models and treatment of null states}\label{sec:Numerics:V2:MM}
At the end of this section, we are going to discuss the Gram matrix in minimal models and how to use it to remove null states. The approach presented here is quite general and it should also work in any other CFT with null states.

Similarly to the free boson theory, the Gram matrix splits into blocks which correspond to primary operators, so we define $G^{(p)}_{ij}\equiv\la (p,i)|(p,j)\ra$. Evaluation of these matrices is quite simple, each can be computed using (\ref{Gram matrix recursive}), the only difference is that the first element $G_{11}^{(p)}$ now equals to the two-point structure constant for the given primary.

However, these Gram matrices are not invertible because the basis includes null states. That would cause problems later in Newton's method, so the null states need to be eliminated. The first step is to identify them. That is quite easy because standard linear algebra tells us that null states can be read off from solutions of homogenous linear equations given by the Gram matrix. We find convenient to execute this part of the algorithm in Mathematica because it has build-in function to identify null states and because it can give us exact results in terms of rational numbers, which means that we avoid various problems with finite numerical precision.

Once we know the null states in a given Verma module, we arrange their coefficients into a matrix and, using more linear algebra, we transform the matrix (respectively each independent block) to a triangular form as
\begin{equation}\label{null states triang}
\left(\begin{array}{ccccccc}
a_{11} & \dots  & a_{1k} & 0      & \dots   & 0      & 1      \\
a_{21} & \dots  & a_{2k} & 0      & \dots   & 1      & 0      \\
\vdots &        & \vdots & \vdots & \iddots & \vdots & \vdots \\
a_{n1} & \dots  & a_{nk} & 1      & \dots   & 0      & 0      \\
\end{array}\right)
\end{equation}

Now we can split the full basis into a basis of the irreducible part of the Verma module and into a complementary basis, which represents the null states. The irreducible basis corresponds to the first columns in (\ref{null states triang}) with elements $a_{ij}$ and the representatives of the null states to the remaining columns, which have a single unit element\footnote{The splitting of the basis can be done in many different ways, this particular option has the advantage that it eliminates the most complicated states with high powers of $L_{-1}$ and $L_{-2}$. Different ways of splitting the basis may change OSFT solutions by addition of null states, but gauge invariant observables are not affected.}. For simplicity, we assume that all null states are grouped together, but in some cases, the columns of (\ref{null states triang}) may be permuted.

We can eliminate null states simply by removing all states from the complementary basis. Consider a state that includes some representatives of null states. We can always add such combination of null states so that the coefficients in front of these states become zero. Therefore we can effectively forget about the states from the complementary basis and work just with the irreducible basis.

Next, we have to transform the matrix representations $\mm(L_m)^k_{ij}$ to be compatible with the irreducible basis. First, we select only the rows of the matrices that correspond to these basis elements. However, these rows may still contain representatives of null states, so when we encounter such element, say $\{x,j\}$, we replace it using (\ref{null states triang}) as
\begin{equation}
\{\dots,\{x,j\},\dots\}\rar \{\dots,\{-x a_{j1},1\},\{-x a_{j2},2\},\dots ,\{-x a_{jk},k\},\dots\}.
\end{equation}
Finally, we have to re-label indices with respect to the new irreducible basis.

We also have to keep in mind that separating the first Virasoro operator from a state from the irreducible basis, $|i\ra^{IR}\rar L_{-m}|\hat i\ra^{IR}$, can produce a representative of a null state, which has to be removed by addition of null states. That means that $\hat i$ changes from a simple number to a vector, $\hat i \rar \{\{\hat i_{1,1},\hat i_{1,2}\},\dots,\{\hat i_{k,1},\hat i_{k,2}\}\}$. If $\hat i$ belongs to the irreducible basis, we set $\hat i \rar \{\{1,\hat i\}\}$ for consistency.

To show how to work with the irreducible representations, we provide an algorithm to compute the Gram matrix directly in the irreducible basis (although it can be of course obtained just by restriction of the full matrix),
\begin{equation}\label{Gram matrix MM IR}
G_{ij}^{(p)IR}=(-1)^{m}\sum_{k,l}\mm (L_m)_{j k}^{p\ IR}\ \hat i_{l,1}G^{(p)IR}_{\hat i_{l,2} k}.
\end{equation}
The change of $\hat i$ to a vector introduces one additional sum compared to (\ref{Gram matrix recursive}). This notation is quite complicated, so we will keep the sum over elements of $\hat i$ implicit. From now on, we are going to also assume that all objects are already compatible with the irreducible basis, so we will drop the label $IR$.

\section{Evaluation of the cubic vertex}\label{sec:Numerics:V3}
In this section, we provide an algorithm to compute matrix representation of the cubic vertex, which determines the interaction term in the action. We define  matrix elements of the full vertex as\footnote{As for the kinetic term, we normalize the vertex to match the dimensionless potential.}
\begin{equation}
V_{ijk}=\la V_3|i\ra|j\ra|k\ra.
\end{equation}
As usual, the elements factorize into the constituent theories:
\begin{equation}\label{vertices factorization}
V_{ijk}=\prod_s V_{i_ij_sk_s}^{(s)}.
\end{equation}
We compute vertices in each sector independently and combine them together only when necessary. The reason is that we cannot store the full set of vertices at high levels because it would require far too much memory (if we consider a basis with tens of thousands of states, the memory would of order of petabytes).

We have already described the general strategy to compute the vertices in section \ref{sec:SFT:cubic cons}. In one step of the algorithm, we take a creation operator from one of the entries of the vertex and trade it for a sum of annihilation operators, which acts on the remaining states. This operation reduces the overall level of the vertex, so only three primary states remain after a finite number of steps and the corresponding vertex can be computed directly. In the following text, we are going to describe the implementation of this algorithm using our matrix representations.

\subsection{Universal matter theory}\label{sec:Numerics:V3:matter}
The conservation laws for Virasoro operators derived in section \ref{sec:SFT:cubic cons} take form
\begin{equation}
\la V_3| L_{-m}^{(2)}=\la V_3|\sum_{n}\sum_{a=1}^3 \cc(L_{-m})_n^{(a)}L_n^{(a)}.
\end{equation}
We remind the reader that the summation index $n$ runs over nonnegative integers and over the identity ($L_{\Id}\equiv\Id$). To convert this expression to a matrix form, we contract it with basis states and replace $L_n$ by their matrix representations,
\begin{eqnarray}
&&\la V_3|\sum_{n}\sum_{a=1}^3 \cc(L_{-m})_n^{(a)}L_n^{(a)}|i\ra|j\ra|k\ra \nn \\
&&=\la V_3|\left(\sum_{n,l}\cc(L_{-m})_n^{(1)}\mm(L_n)_{il}|l\ra|j\ra|k\ra+\sum_{n,l}\cc(L_{-m})_n^{(2)}\mm(L_n)_{jl}|i\ra|l\ra|k\ra\right. \nn\\
&&\quad \left. + \sum_{n,l}\cc(L_{-m})_n^{(3)}\mm(L_n)_{kl}|i\ra|j\ra|l\ra\right) \nn\\
&&=\la V_3|\left(\sum_{l}\kk(L_{-m})_{il}^{(1)}|l\ra|j\ra|k\ra + \sum_{l}\kk(L_{-m})_{jl}^{(2)}|i\ra|l\ra|k\ra\right. \nn\\
&&\quad \left.+\sum_{l}\kk(L_{-m})_{kl}^{(3)}|i\ra|j\ra|l\ra \right).
\end{eqnarray}
In the last expression, we have introduced a new set of matrices called $\kk$, which are given by
\begin{equation}\label{vertices K matrix}
\kk(L_{-m})_{ij}^{(a)}=\sum_n\cc(L_{-m})_n^{(a)}\mm(L_n)_{ij}.
\end{equation}
These matrices are the most efficient representation of conservation laws. Similarly to $\mm$, we remember only their nonzero elements. Now we can write a recursive formula for cubic vertices as
\begin{eqnarray}\label{vertices recursive}
V_{ijk}&=&\la V_3|i\ra|j\ra|k\ra=\la V_3|L_{-m}^{(2)}|i\ra|\hat j\ra|k\ra \\
&=&\sum_l \kk(L_{-m})_{il}^{(1)}V_{l\hat j k}+\sum_l \kk(L_{-m})_{\hat j l}^{(2)}V_{ilk}+\sum_l \kk(L_{-m})_{kl}^{(3)}V_{i\hat j l}. \nn
\end{eqnarray}
If the state $|j\ra$ equals to the vacuum, which obviously contains no $L_{-m}$, we use the cyclic symmetry of the vertex and apply the conservation law on a different entry. After a repeated use of this formula, we eventually get to $V_{111}=\la V_3 |0\ra |0\ra |0\ra=1$.

To improve efficiency of this algorithm, we take advantage of the cyclic and the twist symmetry of the cubic vertex. That allows us to compute and store only vertices which have indices ordered as $i\leq j\leq k$. Vertices with other ordering of indices are given by
\begin{equation}\label{vertices ordering}
V_{ijk}=\begin{cases}
V_{(ijk)}       & \quad \rm{for\ even\ permutation},\\
(-1)^{N_i+N_j+N_k}V_{(ijk)}  & \quad \rm{for\ odd\ permutation},
\end{cases}
\end{equation}
where $(i,j,k)$ is the ordered permutation of the three indices and $(-1)^{N_i}$ are eigenvalues of the twist operator (\ref{twist state}). If we use this symmetry, we have to compute only about one sixth of the total amount of vertices. Of course, this comes with a price. Every time the algorithm (\ref{vertices recursive}) needs to access an element $V_{ijk}$, it is necessary to reorder the indices and to add a sign following (\ref{vertices ordering}). This takes some time, but it is faster than computing six times more vertices.
Parallelization of this algorithm adds few more complications, which we discuss later in subsection \ref{sec:Numerics:V3:parallel}.

With just few minor modifications, this algorithm can be also used to compute vertices in the ghost theory with the ghost Virasoro or the ghost current basis. The conservation laws and matrix representations of matter Virasoro operators must be replaced by their ghost counterparts and the cubic vertex contracted with three ground states now equals to $V_{111}=\la V_3| \Omega\ra |\Omega\ra |\Omega\ra=K^3$.

The ordering of vertices using the cyclic and twist symmetry is useful during the initial evaluation of vertices, but it would significantly slow down Newton's method later. Therefore, after we compute vertices in all sectors and deallocate all auxiliary objects, we find convenient to restore the full set of vertices using (\ref{vertices ordering}). This usually takes less time than the initial evaluation of vertices, which is time consuming due to the complexity of conservation laws. The full set of vertices requires more memory, but this is usually not a big problem, see appendix \ref{sec:time}.

\subsection{Ghost vertices in the $bc$ basis}\label{sec:Numerics:V3:ghost bc}
Evaluation of ghost vertices in the $bc$ basis is more complicated. As we already discussed in section \ref{sec:Numerics:V2:ghost}, the problem is that manipulations with $b$ and $c$ operators change ghost numbers of the involved states. An algorithm which efficiently deals with this issue have been introduced in \cite{GaiottoRastelli}. It uses $b$ ghost conservation laws, but it could be easily reformulated using $c$ ghost conservation laws as well. $b$ ghost conservation laws are more convenient because they are very similar to Virasoro conservation laws. Unlike the algorithms in section \ref{sec:Numerics:V2:ghost}, this algorithm does not depend on whether we impose Siegel gauge or not.

We start with an overview of the algorithm. Our aim it to compute vertices with ghost numbers equal to $(1,1,1)$. We begin by separating one $b$ creation operator from the middle entry and we exchange it for a sum of annihilation operators. This operation changes ghost numbers of the three states to $(1,2,1)$. When we act with an annihilation operator on one of the states, the ghost numbers become either to  $(1,1,1)$ or to $(0,1,2)$ (after rearrangement). Vertices with ghost numbers $(0,1,2)$ are auxiliary objects similar to $G^{aux}$ from section \ref{sec:Numerics:V2:ghost}. These vertices can be also computed using the same $b$ ghost conservation laws. This time, we have to apply the conservation law on the ghost number 0 state because the other two options would produce states with undesired ghost numbers. The removal of one $b$ ghost changes the ghost numbers to $(1,1,2)$ and action of the conservation law again produces states with ghost numbers $(1,1,1)$ or $(0,1,2)$. This shows that the algorithm closes on these two types of vertices and no other auxiliary objects are needed.

To formalize the algorithm, we define matrices of physical and auxiliary vertices as
\begin{eqnarray}
V_{ijk}^{phys} &=& \la V_3|i^{(1)}\ra|j^{(1)}\ra|k^{(1)}\ra, \\
V_{ijk}^{aux} &=& \la V_3|i^{(0)}\ra|j^{(1)}\ra|k^{(2)}\ra.
\end{eqnarray}
The matrix representations (\ref{vertices K matrix}) for $b_{-m}$ need be modified to take into account the ghost number structure. For physical vertices, we define
\begin{eqnarray}
\kk(b_{-m})_{ij}^{(1)phys}&=&\sum_n\cc(b_{-m})_n^{(1)}\mm(b_n)_{ij}^{(1)}, \nn \\
\kk(b_{-m})_{ij}^{(2)phys}&=&\sum_n\cc(b_{-m})_n^{(2)}\mm(b_n)_{ij}^{(2)},  \\
\kk(b_{-m})_{ij}^{(3)phys}&=&\sum_n\cc(b_{-m})_n^{(3)}\mm(b_n)_{ij}^{(1)} \nn
\end{eqnarray}
and for auxiliary vertices
\begin{eqnarray}
\kk(b_{-m})_{ij}^{(1)aux}&=&\sum_n\cc(b_{-m})_n^{(2)}\mm(b_n)_{ij}^{(1)}, \nn \\
\kk(b_{-m})_{ij}^{(2)aux}&=&\sum_n\cc(b_{-m})_n^{(3)}\mm(b_n)_{ij}^{(1)},  \\
\kk(b_{-m})_{ij}^{(3)aux}&=&\sum_n\cc(b_{-m})_n^{(1)}\mm(b_n)_{ij}^{(2)}. \nn
\end{eqnarray}
To clarify the somewhat confusing notation, we remind the reader that the upper index of $\mm(b_n)$ denotes ghost number of the matrix representation (see (\ref{matrix representation b})) and upper indices of $\kk(b_{-m})$ or $\cc(b_{-m})$ are related to entry of the cubic vertex. Notice that we rotated the indices of $\cc(b_{-m})$ in the definition of $\kk(b_{-m})^{aux}$ because we apply the conservation law on the first entry of the vertex.

Furthermore, we have to be careful about anticommutators between ghost operators. Application of a $b$ conservation law includes movement of a $b_{-m}$ operator towards the 3-vertex, replacement of operators and movement of the new operators back to their proper place. Fortunately, the signs that appear during this process can be determined just from the ghost numbers. We find for example
\begin{equation}
\la V_3|i^{(1)}\ra b_{-m}^{(2)}|j^{(2)}\ra|k^{(1)}\ra=-\la V_3|b_{-m}^{(2)}|i^{(1)}\ra |j^{(2)}\ra|k^{(1)}\ra
\end{equation}
because a ghost number 1 state has odd number of anticommuting operators.

With this in mind, we can write recursive relations for cubic vertices as
\begin{eqnarray}
V_{ijk}^{phys}&=&-\sum_l (-1)^{N_{\hat j}+N_l+N_k} \kk(b_{-m})_{il}^{(1)phys}V_{l k \hat j}^{aux}\\
&+&\sum_l \kk(b_{-m})_{\hat j l}^{(2)phys}V_{ilk}^{phys}+\sum_l \kk(b_{-m})_{kl}^{(3)phys}V_{l i\hat j}^{aux} \nn
\end{eqnarray}
and
\begin{eqnarray}
V_{ijk}^{aux}&=&-\sum_l (-1)^{N_{\hat i}+N_l+N_k} \kk(b_{-m})_{j l}^{(2)aux}V_{l\hat i k}^{aux}\\
&+&\sum_l \kk(b_{-m})_{\hat il}^{(1)aux}V_{ljk}^{aux}+\sum_l \kk(b_{-m})_{kl}^{(3)aux}V_{\hat i j l}^{phys} \nn
\end{eqnarray}
We have explicitly ordered indices of auxiliary vertices according to their ghost number, but we have to additionally use (\ref{vertices ordering}) for physical vertices .

\subsection{SU(1,1) singlet ghost vertices}\label{sec:Numerics:V3:ghost SU11}
Next, we are going to describe the algorithm for the cubic vertex contracted with SU(1,1) singlet ghost states, which was developed in \cite{KudrnaUniversal}. We use this algorithm in most of our calculations. It it somewhat similar to the algorithm using the $bc$ basis, but this algorithm involves auxiliary vertices outside Siegel gauge instead of vertices with different ghost numbers.

In this basis, there are difficulties with $L'^{gh}$ conservation laws. The conservation laws for $L^{gh}$ and $j^{gh}$ derived in section \ref{sec:SFT:cubic cons} have the usual structure
\begin{equation}
\la V_3| L_{-m}^{gh(2)}=\la V_3|\sum_{n}\sum_{a=1}^3 \cc(L^{gh}_{-m})_n^{(a)}L_n^{gh(a)},
\end{equation}
\begin{equation}
\la V_3| j_{-m}^{gh(2)}=\la V_3|\sum_{n}\sum_{a=1}^3 \cc(j^{gh}_{-m})_n^{(a)}j_n^{gh(a)}.
\end{equation}
However, when we combine them together to match the structure of $L'^{gh}_n=L^{gh}_n+nj^{gh}_n+\delta_{n0}$, we find that coefficients in front of $j_n^{gh}$ do not cancel,
\begin{eqnarray}
\la V_3| L'^{gh(2)}_{-m}&=&\la V_3|\sum_{n}\sum_{a=1}^3 \cc(L^{gh}_{-m})_n^{(a)}L'^{gh(s)}_n-\la V_3|\sum_{a=1}^3 \cc(L^{gh}_{-m})_0^{(a)} \Id \nn  \\
&-&\la V_3|\sum_{n}\sum_{a=1}^3 \left(n\,\cc(L^{gh}_{-m})_n^{(a)}+m\,\cc(j^{gh}_{-m})_n^{(a)}\right) j_n^{gh(a)}\\
&=&\la V_3|\sum_{n}\sum_{a=1}^3 \left(\cc(L'^{gh}_{-m})_n^{(a)L}L'^{gh(a)}_n+\cc(L'^{gh}_{-m})_n^{(a)J}j^{gh(a)}_n\right), \nn
\end{eqnarray}
where we have introduced new symbols $\cc(L'^{gh}_{-m})^{L}$ and $\cc(L'^{gh}_{-m})^{J}$. The term with the identity operator can be merged with either of them, but it is more convenient to add it to $\cc(L'^{gh}_{-m})^{J}$, so that $\cc(L'^{gh}_{-m})^{L}=\cc(L^{gh}_{-m})$. The other symbols are therefore equal to
\begin{equation}
\cc(L'^{gh}_{-m})_n^{(a)J}=n\,\cc(L^{gh}_{-m})_n^{(a)}+m\,\cc(j^{gh}_{-m})_n^{(a)}-\delta_{n\Id}\,\cc(L^{gh}_{-m})_0^{(a)}.
\end{equation}

We observe that $L'^{gh}$ conservation laws do not close back on $L'^{gh}$ because they contain modes of the ghost current. When we act with $j^{gh}_n$ on a singlet state, the commutator (\ref{commutator Lp j}) can produce $j^{gh}_{-k}$ with $k>0$. Therefore we get in general a sum of singlet states and auxiliary states of the form
\begin{equation}
j^{gh}_{-k}L'^{gh}_{-n_1}\dots L'^{gh}_{-n_l}c_1|0\ra.
\end{equation}
We emphasize that these auxiliary states contain only one $j^{gh}_{-k}$ operator. They are not in Siegel gauge because the expansion of $j^{gh}_{-k}$ includes the $c_0$ operator.

Fortunately, cubic vertex conservation laws act only on one entry of the vertex at a time, so the vertex is never contracted with more than one auxiliary state. Therefore we define a matrix of auxiliary vertices as
\begin{equation}
V_{ijk}^{aux} = \la V_3|i\ra^{aux}|j\ra|k\ra.
\end{equation}
We always move the auxiliary state to the first position using the cyclic symmetry. Thanks to the twist symmetry, we can also rearrange the last two indices so that $k\geq j$.

We cannot use the $L'^{gh}$ conservation laws to compute these auxiliary vertices because they would produce more $j^{gh}$ operators and lead to more complicated auxiliary vertices. Instead, we use $j^{gh}$ conservation laws, which remove the only $j^{gh}$ creation operator and replace it with annihilation operators. This operation again produces a sum of both physical and auxiliary vertices, but it never increases the number of $j^{gh}$ operators. After a finite number of steps, the ghost current always disappears and the algorithms returns to physical vertices.

Matrix representations of $L'^{gh}$ and $j^{gh}$ conservation laws are more complicated than in other cases. First, we define
\begin{equation}
\kk(L'^{gh}_{-m})_{ij}^{(a)phys}=\sum_n\left(\cc(L'^{gh}_{-m})_n^{(a)L}\mm(L'^{gh}_{n})_{ij}+\cc(L'^{gh}_{-m})_n^{(a)J}\mm(j^{gh}_{n})_{ij}^{phys}\right),
\end{equation}
and
\begin{equation}
\kk(L'^{gh}_{-m})_{ij}^{(a)aux}=\sum_n\cc(L'^{gh}_{-m})_n^{(a)J}\mm(j^{gh}_{n})_{ij}^{aux}.
\end{equation}
These two objects represent $L'^{gh}$ conservation laws which lead to physical and auxiliary vertices respectively. The definition uses the matrix representations of $j^{gh}$ introduced in (\ref{matrix representation j}). Similarly, we define matrix representations of $j^{gh}$ conservation laws as
\begin{equation}
\kk(j^{gh}_{-m})_{ij}^{(a)phys}=\sum_n\cc(j^{gh}_{-m})_n^{(a)}\mm(j^{gh}_{n})_{ij}^{phys}
\end{equation}
and
\begin{equation}
\kk(j^{gh}_{-m})_{ij}^{(a)aux}=\sum_n\cc(j^{gh}_{-m})_n^{(a)}\mm(j^{gh}_{n})_{ij}^{aux}.
\end{equation}

Using these objects, we can finally write explicit recursive formulas for SU(1,1) singlet vertices:
\begin{eqnarray}
V_{ijk}^{phys}&=&\sum_l \kk(L'^{gh}_{-m})_{il}^{(1)phys}V_{l\hat j k}^{phys}+\sum_l \kk(L'^{gh}_{-m})_{il}^{(1)aux}V_{l\hat j k}^{aux} \nn\\
&+&\sum_l \kk(L'^{gh}_{-m})_{\hat jl}^{(2)phys}V_{ilk}^{phys}+\sum_l \kk(L'^{gh}_{-m})_{\hat jl}^{(2)aux}V_{lki}^{aux} \\
&+&\sum_l \kk(L'^{gh}_{-m})_{kl}^{(3)phys}V_{i\hat j l}^{phys}+\sum_l \kk(L'^{gh}_{-m})_{kl}^{(3)aux}V_{li\hat j}^{aux} \nn
\end{eqnarray}
and
\begin{eqnarray}
V_{ijk}^{aux}&=&\sum_l \kk(j^{gh}_{-m})_{\hat il}^{(2)phys}V_{ljk}^{phys}+\sum_l \kk(j^{gh}_{-m})_{\hat il}^{(2)aux}V_{ljk}^{aux} \nn\\
&+&\sum_l \kk(j^{gh}_{-m})_{jl}^{(3)phys}V_{\hat ilk}^{phys}+\sum_l \kk(j^{gh}_{-m})_{jl}^{(3)aux}V_{lk\hat i}^{aux} \\
&+&\sum_l \kk(j^{gh}_{-m})_{kl}^{(1)phys}V_{\hat ijl}^{phys}+\sum_l \kk(j^{gh}_{-m})_{kl}^{(1)aux}V_{l\hat ij}^{aux}. \nn
\end{eqnarray}
We rotated the upper indices in the second formula because we apply the $j^{gh}$ conservation law on the first entry of the vertex. To finish the algorithm, we have to order the indices of both types of vertices. We order all three indices of physical vertices using (\ref{vertices ordering}) and the last two indices of auxiliary vertices.

Unfortunately, we have not found any way to extend this algorithm to non-singlet states. Modes of the ghost current generally do not annihilate the SU(1,1) primary states (\ref{spin primary}),
\begin{equation}
j^{gh}_n |j,m\ra\neq 0.
\end{equation}
Therefore we cannot very well use $L'^{gh}$ conservation laws because the algorithm would not conserve SU(1,1) representations. That would lead to many complications with unclear solution. Therefore it seems that using the $bc$ basis is a better choice when working with non-singlet states in Siegel gauge.

\subsection{Free boson and minimal models}\label{sec:Numerics:V3:FB MM}
Finally, we are going to describe an algorithm to compute cubic vertices in theories with multiple primary operators. In our case, these are the free boson theory and the Virasoro minimal models, but this algorithm should work also in other CFTs with a similar structure of the state space.

A three-point function in a CFT can be nonzero only if the three involved primary operators obey fusion rules (\ref{fusion rules}) and this pattern also translates to the cubic vertex. Therefore we immediately know that many vertices are identically equal to zero. We do not wish to waste time and memory by computing these vertices, so we came up with an approach that automatically eliminates these trivial vertices.

We define a structure that records all triplets of primary operators that satisfy fusion rules. We denote its elements as $f\equiv(p,q,r)$. Furthermore, in order to avoid repetitions, we order the operators so that they obey some canonical ordering $p\leq q\leq r$. Then we define
\begin{equation}
V^{(f)}_{ijk}=\la V_3|(p,i)\ra|(q,j)\ra|(r,k)\ra,
\end{equation}
where we label basis elements by doubled indices. These matrices store only vertices that are potentially nonzero.

Matrix representations of conservation laws acting on these vertices gain one extra index. In the free boson theory, we define
\begin{equation}
\kk(\alpha_{-m})_{ij}^{(a)k}=\sum_n\cc(\alpha_{-m})_n^{(a)}\mm(\alpha_n)^k_{ij}
\end{equation}
and similarly in minimal models we define $\kk(L_{-m})_{ij}^{(a)k}$ using the replacement $\alpha_n\rar L_n$. A recursive algorithm for cubic vertices then reads
\begin{eqnarray}
V_{ijk}^{(f)}&=&\sum_l \kk(\alpha_{-m})_{il}^{(1)p}V_{l\hat j k}^{(f)} \\
&+&\sum_l \kk(\alpha_{-m})_{\hat jl}^{(2)q}V_{ilk}^{(f)}+\sum_l \kk(\alpha_{-m})_{kl}^{(3)r}V_{i\hat j k}^{(f)}.\nn
\end{eqnarray}
This equation assumes that $j>1$, i.e. $|(q,j)\ra$ is not a primary state. Otherwise we make a cyclic shift of indices and we apply the algorithm on $i$ or $k$. An analogous formula applies to the minimal models, but we have to keep in mind that $\hat j$ possibly represents a sum over several states. Therefore the whole algorithm gains one additional summation similarly to (\ref{Gram matrix MM IR}).

As usual, the algorithm eventually arrives to the cubic vertex contracted with three primaries, which is given by (\ref{vertex primaries}),
\begin{equation}
V_{111}^{(f)}=K^{-h_p-h_q-h_r}C_{prq}.
\end{equation}

This algorithm does not exploit the cyclic or the twist symmetry. Generically, when all three primaries are different, the fixed order of primaries forbids it anyway. In case that two or all three primaries coincide, we could use the symmetry to reduce the number of vertices, but the gain of efficiency is usually not worth complicating the algorithm even more.

The approach above is written for a single BCFT. If we have a product of two or more theories with nontrivial primaries, we have to make few small modifications. The full fusion rules in such theory have the following structure:
\begin{equation}
f=((p^{(1)},p^{(2)},\dots),(q^{(1)},q^{(2)},\dots),(r^{(1)},r^{(2)},\dots)).
\end{equation}
We choose some canonical ordering only for primaries in these full fusion rules. When we make projections to the individual sectors, $f\rar f^{(s)}=(p^{(s)},q^{(s)},r^{(s)})$, we take an union of the projected fusion rules, but we leave the primaries unordered.

Next, we will discuss the transition from momentum eigenstates to parity even states in the free boson theory. Consider parity even states with momenta $0<p_1\leq p_2<p_3$, for which we find
\begin{eqnarray}
&&\la V_3| \alpha_{-M_1}^{(1)}|p_1\ra_{\eps_1}\alpha_{-M_2}^{(2)}|p_2\ra_{\eps_2}\alpha_{-M_3}^{(3)}|p_3\ra_{\eps_3}\\
&&\ \ =\frac{1}{8}\la V_3| \alpha_{-M_1}^{(1)}\alpha_{-M_2}^{(2)}\alpha_{-M_3}^{(3)}\left(\eps_3|p_1\ra|p_2\ra|-p_3\ra+\eps_1\eps_2|-p_1\ra|-p_2\ra|p_3\ra\right),\nn
\end{eqnarray}
where $\eps_i=(-1)^{\sharp \alpha^{(i)}}$. The two terms in the second line are equal, which can be easily seen using the parity symmetry. If we take into account the possibility of zero momenta, we get
\begin{eqnarray}
&&\la V_3| \alpha_{-M_1}^{(1)}|p_1\ra_{\eps_1}\alpha_{-M_2}^{(2)}|p_2\ra_{\eps_2}\alpha_{-M_3}^{(3)}|p_3\ra_{\eps_3}\\
&&\ \ =\eps_1\eps_2\frac{1}{4} (1+\delta_{p_1 0}+\delta_{p_2 0}+\delta_{p_3 0})\la V_3| \alpha_{-M_1}^{(1)}|-p_1\ra\alpha_{-M_2}^{(2)}|-p_2\ra\alpha_{-M_3}^{(3)}|p_3\ra. \nn
\end{eqnarray}
Similarly to what we found in section \ref{sec:Numerics:V2:FB}, structure constants in two or more dimensions are not factorizable. Therefore, to get to the parity even basis, we multiply the vertices only by $\eps_1\eps_2$ and we add the remaining part of structure constants (which equals to $1$, $\frac{1}{2}$ or $\frac{1}{4}$) only to the full vertex $V_{ijk}$, see (\ref{Jacobian full}).

\subsection{Star product}\label{sec:Numerics:V3:star}
In order to solve the equations of motion, we need to know just cubic vertices, but for some purposes, for example for the perturbative marginal solution (\ref{marginal pert solution}), it is useful to have an explicit expression for the star product. To compute a star product of arbitrary string fields, we need to know star products of all basis elements, which we parameterize as
\begin{equation}
|i\ra\ast|j\ra=\tilde V_{ij}^{\ \ k}|k\ra.
\end{equation}
To express $\tilde V_{ij}^{\ \ k}$, we contract this expression with a basis state $\la l|$ and we find
\begin{equation}
V_{lij}=\tilde V_{ij}^{\ \ k}G_{lk}.
\end{equation}
We can see that the star product and the cubic vertex are related by the Gram matrix. Therefore we get
\begin{equation}
\tilde V_{ij}^{\ \ k}=G^{kl}V_{lij},
\end{equation}
where $G^{kl}$ is the inversion of the Gram matrix. The evaluation of the inverse matrix requires in principle $O(N^3)$ operations, but in practice, it can be done much faster because the Gram matrix is factorizable and block diagonal.

The evaluation of a star product of given string fields can be sped up using similar tricks as in section \ref{sec:Numerics:Newton:jac}.

\subsection{Parallelization}\label{sec:Numerics:V3:parallel}
At the end of this section, we will discuss parallelization of cubic vertex algorithms. We use the OpenMP library \cite{OpenMP} in C++ to parallelize the calculations, but our approach does rely on this specific library.

As we already discussed in section \ref{sec:Numerics:V2:ghost}, our recursive algorithms have an issue with the order of calculation of vertices. Generally, these algorithms express a vertex at a higher level as a sum of vertices at lower levels, so we have to assure that the lower level vertices are already known by the time they are needed. If we use just a single thread, we can start with $V_{111}$ and then compute all vertices in ascending order. However, when we use more threads to compute multiple vertices simultaneously, it may happen that one thread needs a vertex that is still being computed by another thread. To avoid this problem, we could parallelize only calculations of groups of vertices at the same level, but it turns out that parallelized loops in this approach are too short and the algorithm does not work very efficiently.

The situation is even more complicated in the ghost theory because our algorithm switches between physical and auxiliary vertices. If we tried to use some kind of level by level approach, it would lead to another setback. We would have to compute all auxiliary vertices, while only about half of them must be computed for evaluation of physical vertices, the rest is not needed due to the structure of the ghost conservation laws.

Therefore our algorithm works in a different way to avoid these problems. It a priori assumes that no vertices have been evaluated yet and it always asks about their status. If some vertex is missing, then the algorithm calls itself recursively to compute it (even if another thread is already working on it). This parallelization scheme potentially leads to conflicts between threads, so we have to be careful. In section \ref{sec:Numerics:V2:ghost}, we have suggested two possibilities of indicating status of vertices: directly by their values or by auxiliary boolean numbers. The first option does not work well in combination with parallelization. In some cases, for example when we use long double number format in C++, the code produces unpredictable and incorrect numbers, which are probably results of two threads trying to read and rewrite the same number. Therefore we use the second option and we define an auxiliary boolean field, which has the same shape as the vertices. There are no problems with simultaneous access to boolean numbers, so we just have to make sure that they are always updated after the corresponding vertices.

The parallelization is most efficient when the parallelized loops are as long as possible, so we use the collapse clause from the OpenMP library to merge the three loops over the indices $i$, $j$ and $k$ into one. The evaluation of individual vertices often requires significantly different amount of time, so we choose the dynamic schedule of parallelization, where vertices are assigned to threads one by one depending on which threads are free.

\section{Solving the equations of motion - Newton's method}\label{sec:Numerics:Newton}
In the previous sections, we introduced various auxiliary objects which are needed for evaluation of the OSFT action and now we are ready to start solving the equations of motion. For now, we assume that there are no gauge conditions imposed. We will discuss the implementation of gauge conditions later, but it turn out that this approach works in Siegel gauge as well.

Using the previously defined expansion of the string field $|\Psi\ra=\sum_i t_i|i\ra$ and the matrices
\begin{eqnarray}
Q_{ij} &=& \la i|Q_B|k\ra, \\
V_{ijk} &=& \la V_3|i\ra|j\ra|k\ra,
\end{eqnarray}
we can write the level-truncated action as\footnote{In the $(L,3L)$ truncation scheme, which we use in the whole thesis, all sums run from 1 to the number of states $N$. If one decides to use some other scheme, the range of the last summation in the interaction term depends on the other two indices. That means $k$ goes only up to some $N(i,j)$, which is determined by levels of the states $|i\ra$ and $|j\ra$.}
\begin{eqnarray}
S(t)=-\frac{1}{g_o^2}\left(\frac{1}{2}\sum_{i,j}Q_{ij}t_it_j+\frac{1}{3}\sum_{i,j,k}V_{ijk}t_it_jt_k\right).
\end{eqnarray}
The equations of motion are given by variation of this action with respect to the $t_i$ variables,
\begin{equation}\label{equations basis}
f_i(t)=\sum_{j}Q_{ij}t_j+\sum_{j,k}V_{ijk}t_jt_k=0.
\end{equation}

In this section, we show how to find a single solution of these equations. The algorithm to find all solutions will be described in the next section.

We solve the equations of motion using the well known Newton's method, which is an iterative algorithm that starts with an approximate solution $t^{(0)}$ and gradually improves precision of the solution. The algorithm works as follows: Suppose that we already have an approximate solution $t^{(n)}$ after the $n$th step of the algorithm. We write the next approximation as $t^{(n+1)}=t^{(n)}+\Delta t$, insert it into the equations of motion and neglect terms with $\Delta t^2$. Then we find
\begin{equation}\label{Newton delta t}
\sum_j M_{ij}(t^{(n)})\Delta t_j=-f_i(t^{(n)}),
\end{equation}
where $M_{ij}$ is the Jacobian matrix
\begin{equation}\label{Jacobian matrix}
M_{ij}(t)=\frac{\del f_i(t)}{\del t_j}=-g_o^2\frac{\del^2 S(t)}{\del t_i\del t_j}=Q_{ij}+\sum_k(V_{ijk}+V_{jik})t_k.
\end{equation}
One step of the Newton's method is therefore given by
\begin{eqnarray}\label{Newton iterations}
t^{(n+1)}_i=t^{(n)}_i+\Delta t_i=t^{(n)}_i-\sum_j M^{-1}_{ij}(t^{(n)})f_j(t^{(n)}).
\end{eqnarray}
However, in practice, it is better to find $\Delta t$ by solving (\ref{Newton delta t}) because it is faster than computing the full inverse matrix.

The algorithm needs some criteria to stop the iterations once the solution is good enough. We define precision of the solution after the $n$th step as
\begin{equation}\label{Newton precision}
p^{(n)}=\frac{\| t^{(n)}-t^{(n-1)}\|}{\|t^{(n-1)}\|}
\end{equation}
and we stop the iterations when $p^{(n)}<10^{-12}$. We use the Euclidean norm to define the precision, but, for consistency, we have checked that other $p$-norms lead to very similar results. Another alternative would be to define the precision using the action as $p^{(n)}=\frac{|S( t^{(n)})-S(t^{(n-1)})|}{|S(t^{(n-1)})|}$. We have checked that this option usually also leads to similar precision as (\ref{Newton precision}).

The target precision $10^{-12}$ is chosen somewhat arbitrary. It is high enough to get observables with good precision, but there is some reserve for cases when it is impossible reach the machine precision due to numerical errors. In some rare cases, for example in the Ising model, Newton's method has problems to reach this even precision. Therefore it is useful to set another criteria on the rate of improvement of the precision (for example $\frac{p^{(n)}}{p^{(n-1)}}>0.1$) after the precision reaches some reasonable value (say $p^{(n)}<10^{-6}$). Finally, we have to set a maximal number of iterations (we use 20) to stop the algorithm in cases when it does not converge to a fixed solution.

Newton's method has generically quadratic rate of convergence. For stable solutions, we usually reach the target precision within 4 or 5 iterations assuming that the starting point is close enough to the solution (the solution from the previous level is usually good enough). If we start from a less precise initial point, the number of iterations can be around 10. Sometimes we encounter unstable solutions, for which the number of iterations strongly varies with level, but these are usually nonphysical anyway.

We solve the linear equations (\ref{Newton delta t}) using a variant of the LU decomposition, which requires $O(N^3)$ operations\footnote{In principle, we could try to implement some algorithm with lesser complexity, but these algorithms are much more complicated and it would not reduce the overall complexity anyway because the evaluation of the Jacobian also requires $O(N^3)$ operations.}. This algorithm can be easily parallelized in C++.
Sometimes, for example when a solution does not excite states from some Verma module, the Jacobian has a block diagonal structure. In such cases, we can solve the equations (\ref{Newton delta t}) independently for each individual block, which speeds up the calculations. The analysis and rearrangement of the Jacobian matrix has negligible complexity $O(N^2)$.

The right hand side of the equations (\ref{Newton delta t}) is quite easy to evaluate once we realize that it can be expressed in terms of $M_{ij}$ as
\begin{equation}\label{Newton eom from M}
f_i(t)=\frac{1}{2}\sum_j M_{ij}(t)t_j+\frac{1}{2}\sum_j Q_{ij}t_j.
\end{equation}
This expression has complexity $O(N^2)$ and therefore its evaluation requires only a negligible amount of time.

\subsection{Evaluation of the Jacobian matrix}\label{sec:Numerics:Newton:jac}
The most difficult part of Newton's method is evaluation of the Jacobian matrix (\ref{Jacobian matrix}). It has same complexity $O(N^3)$ as solving the linear equations (\ref{Newton delta t}), but it is significantly more complicated because of factorization of the cubic vertex (\ref{vertices factorization}).

To compute the Jacobian efficiently, we first realize that cubic vertices contribute to the action only when the total twist of all three states is $+1$. To make use of that, we split the vector $t_i$ into its twist even part $t_i^+$ and twist odd part $t_i^-$. Then we can write
\begin{equation}
\sum_k(V_{ijk}+V_{jik})t_k=2\sum_k V_{ijk}t_k^\eps,
\end{equation}
where $\eps=\Omega_i \Omega_j$. This process can be also used for a different increase of efficiency: Some components of $t_i$ are quite often equal to zero. When we split $t_i$ into $t_i^\pm$, we can take advantage of that and discard all zero elements. In this way, we avoid a pointless summation over zero elements.

If the state space is build on a single primary (or we decide not to use the structure given by fusion rules), we can write the Jacobian as
\begin{equation}
M_{ij}=h_i\prod_s G_{i_s j_s}^{(s)}+2\sum_k \left(\prod_s V_{i_s j_s k_s}^{(s)}\right) t_k^\eps,
\end{equation}
where we explicitly factorize the vertices. This formula assumes that we are in Siegel gauge, otherwise, we replace the constant term by (\ref{kinetic term no Siegel}).
The evaluation of this expression in C++ can be significantly sped up if we temporarily remember pointers that are used repeatedly. Thanks to the symmetry of the Jacobian, we can also compute just the upper triangular part of the matrix and determine the other half by reflection.

In theories with more primary operators, where we label states by doubled indices, the Jacobian gets far more messy because we want to take advantage of the fusion rules. Previously, in section \ref{sec:Numerics:V3:FB MM}, we introduced triplets of integers $f\equiv(p,q,r)$, which encode the fusion rules. Now, we define additional objects denoted as $\mathfrak{F}_{pq}^{ij}$, where $i$ and $j$ are two different indices between 1 and 3 and the indices $p$, $q$ label primary operators in the given theory. $\mathfrak{F}_{pq}^{12}$ is defined as a list of fusion rules $f$ that have $p$ at the first position and $q$ at the second position or the other way. $\mathfrak{F}_{pq}^{13}$ and $\mathfrak{F}_{pq}^{23}$ are defined in the same way, but, to avoid overcounting, every fusion rule can appear only in one of these three objects for given $p$ and $q$. Now we can write the Jacobian as
\begin{eqnarray}
M_{(p,i)(q,j)}&=&C_{pq}h_{(p,i)}\prod_s G_{i_s j_s}^{(s,p^{(s)})} \nn\\
&+&2\sum_{f\in\, \mathfrak{F}_{pq}^{12}}C_f\sum_k \left(\prod_s V_{i_s j_s k_s}^{(s,f^{(s)})}\right) t_{(f_3,k)}^\eps \nn \\
&+&2\sum_{f\in\, \mathfrak{F}_{pq}^{13}}C_f\sum_k \left(\prod_s V_{i_s k_s j_s}^{(s,f^{(s)})}\right) t_{(f_2,k)}^\eps \label{Jacobian full} \\
&+&2\sum_{f\in\, \mathfrak{F}_{pq}^{23}}C_f\sum_k \left(\prod_s V_{k_s i_s j_s}^{(s,f^{(s)})}\right) t_{(f_1,k)}^\eps. \nn
\end{eqnarray}
The notation in this formula is quite complicated, so let us clarify it. $p^{(s)}$ is a projection of a total primary to a given sector and $G^{(s,p^{(s)})}$ is the Gram matrix build over it. Similarly, $f^{(s)}$ is a projected fusion rule and $V^{(s,f^{(s)})}$ are the corresponding vertices. The second upper index is trivial in the universal sectors. The index $k$ runs over descendants of the primary that complements the primaries $p$ and $q$ in the fusion rule $f$.
Finally, $C_{pq}$ and $C_f$ are non-factorizable parts of structure constants. In the free boson theory, these constants are
\begin{eqnarray}
C_{pq}&=&\delta_{pq}(1+\delta_{p0})/2, \nn \\
C_f&\equiv &C_{(p,q,r)}=(1+\delta_{p0}+\delta_{q0}+\delta_{r0})/4.
\end{eqnarray}
In the Virasoro minimal models, we set these constants to 1 because we absorb structure constants into $Q$ and $V$.

\subsection{Implementation of gauge fixing conditions}\label{sec:Numerics:Newton:gauge}
Next, we will discuss how to adapt Newton's method to incorporate nontrivial gauge fixing conditions. This algorithm is based on \cite{ArroyoKudrna}.

We consider gauge conditions given by a generic linear operator $\mathcal{G}$
\begin{equation}\label{gauge condition}
\mathcal{G} \Psi=0.
\end{equation}
By expanding the string field into a basis, this equation transforms into a matrix equation
\begin{equation}\label{gauge condition basis}
\sum_j \mathcal{G}_{ij}t_j=0.
\end{equation}
To find OSFT solutions in this gauge, we have to solve both these equations and (\ref{equations basis}). However, since the gauge symmetry is broken by the level truncation approximation, the whole system of equations is overdetermined. The usual method to deal with this problem is to solve only a subset of the full equations of motion, which we write as
\begin{equation}\label{equations projected}
P(Q\Psi+\Psi\ast \Psi)=0,
\end{equation}
where $P$ is some projector of the appropriate rank. The remaining equations,
\begin{equation}\label{equations remaining}
(\Id-P)(Q\Psi+\Psi\ast \Psi)=0,
\end{equation}
are left unsolved. However, for physical solutions, they must be satisfied in the limit $L\rar \inf$.

We begin by solving the gauge conditions (\ref{gauge condition basis}). Using standard linear algebra, we can transform the matrix $\mathcal{G}$ to the form\footnote{The matrix here is simplified for illustrative purposes. In the actual algorithm, we order states by their levels, therefore the dependent and independent variables are mixed together and we have to work with a matrix with permuted columns.}
\begin{equation}
\raisebox{-0.4cm}{$\mathcal{G}=$}\begin{array}{cc}
\rule{0.5cm}{0pt}\overbrace{\rule{2.2cm}{0pt}}^{t_i^{(D)}} & \hspace{-0.6cm}\overbrace{\rule{3.8cm}{0pt}}^{t_i^{(I)}}\\
\multicolumn{2}{c}{
\left(
\begin{array}{cccccccc}
1      & 0      & \dots  & 0      & a_{11}   & a_{12}   & \dots  & a_{1n_I} \\
0      & 1      & \dots  & 0      & a_{21}   & a_{22}   & \dots  & a_{2n_I}\\
\vdots & \vdots & \ddots & \vdots & \vdots   & \vdots   & \ddots & \vdots \\
0      & 0      & \dots  & 1      & a_{n_D1} & a_{n_D2} & \dots  & a_{n_Dn_I}
\end{array}
\right)}
\end{array}.
\end{equation}
This form of the matrix allows us to split the variables $t_i$ into $N_I$ independent variables $t_i^{(I)}$ and $N_D$ dependent variables $t_i^{(D)}$. It also immediately gives us a solution for the dependent variables:
\begin{equation}\label{dependent variables solution}
t_i^{(D)}=-\sum_{j=1}^{N_I}a_{ij} t_i^{(I)}.
\end{equation}
Next, we substitute the dependent variables into the equations of motion, which become equations of the independent variables only. Before we solve them, we have to first choose the projector $P$. In principle, many choices are possible as long as the projector has the correct rank $N_I$. For example, Kishimoto and Takahashi used the Siegel gauge projector in their calculations in $a$-gauge \cite{KishimotoTakahashi1}\cite{KishimotoTakahashi2}. However, there is a canonical choice of the projector.

The most natural way to obtain the projected equations is to substitute the dependent variables into the action, $S(t)\rar S(t^{(I)},t^{(D)}(t^{(I)}))$, and then take derivatives of this action with respect to the independent variables,
\begin{equation}
\frac{\del S(t^{(I)},t^{(D)}(t^{(I)}))}{\del t_i^{(I)}}=0.
\end{equation}
These equations lead to the canonical projector
\begin{equation}
P_C=\left(
\begin{array}{cccccccc}
1         & 0         & \dots  & 0      & -a_{11}    & -a_{21}    & \dots  & -a_{n_D 1}   \\
0         & 1         & \dots  & 0      & -a_{12}    & -a_{22}    & \dots  & -a_{n_D 2}   \\
\vdots    & \vdots    & \ddots & \vdots & \vdots     & \vdots     & \ddots & \vdots       \\
0         & 0         & \dots  & 1      & -a_{1 n_I} & -a_{2 n_I} & \dots  & -a_{n_I n_D} \\
\end{array}
\right),
\end{equation}
which is related to the transposition of the matrix $\mathcal{G}$.

Now we can formulate Newton's method for the independent variables. The change of these variables in every step of the algorithm is
\begin{equation}
\sum_j M_{ij}^{(P)}(t^{(n)})\Delta t_j^{(I)}=-f_i^{(P)}(t^{(n)}),
\end{equation}
where we defined projections of the full Jacobian and of the equations of motion:
\begin{eqnarray}
M_{ij}^{(P)}&=&\sum_{k,l}P_{ik}P_{Cjl}M_{kl},\\
f_i^{(P)}&=&\sum_{j}P_{ij} f_j.
\end{eqnarray}
By solving these equations, we find the independent variables, the remaining variables can be computed using (\ref{dependent variables solution}).
Notice that if we use a non-canonical projector, the Jacobian is multiplied by a different projector from each side.

This algorithm works in all linear gauges, but we can avoid explicit use of the projector in Siegel gauge (with the canonical projector). As we mentioned before in section \ref{sec:SFT:string field:ghost}, the Siegel gauge condition reduces to $t^{(D)}_i=0$ in a proper basis. Therefore we can effectively forget about the dependent variables. We formulate the action directly in terms of the independent variables and derive the equations of motion by variation of this action.

\section{Solving the equations of motion - Homotopy continuation method}\label{sec:Numerics:homotopy}
In order to find all solutions of the equations of motion, we cannot rely on Newton's method, which always searches only for one solution at a time. Therefore we use a different numerical algorithm called the homotopy continuation method (see for example \cite{HomotopyAllowager}\cite{Homotopy1}). It is a very efficient algorithm for finding all solutions of polynomial equations. This algorithm is implemented in the NSolve function in Mathematica and in several numerical packages (for example \cite{HomotopyPHC}), but we have programmed our own version, which is adapted specifically to the OSFT equations.

We consider a system of $N$ polynomial equations in $N$ variables,
\begin{equation}
f_i(t)=0,
\end{equation}
which is called the target system. In OSFT, this system is given by the quadratic equations (\ref{equations basis}) (or by the projected equations (\ref{equations projected})).

When we use this method, it is convenient to fully evaluate the matrices $Q$ and $V$ (instead of using various factorized objects) and use them as the input of the algorithm. The homotopy continuation method can solve only few equations, so there are no memory problems with storing these matrices. In this way, we can use just one algorithm for all OSFTs independently on their background.

In order to solve the target equations, we introduce a second system of equation $g_i(t)=0$, which is called the start system. This system must have at least as many solutions as the target system and all of its solutions must be explicitly known. In our case of quadratic equations, we can choose the start system to be
\begin{equation}
g_i(t)=t_i(A_i t_i+B_i)=0,
\end{equation}
where $A_i$ and $B_i$ are some nonzero numbers\footnote{It is convenient to choose $A_i$ and $B_i$ of the same order as coefficients in the target equations, which improves numerical stability of the algorithm, for example $A_i=\max\limits_{j,k}|V_{ijk}|$, $B_i=\max\limits_{j}|Q_{ij}|$}.

Next, we define a homotopy
\begin{equation}
H_i(t_j,\alpha)=(1-\alpha) g_i(t_j) \gamma + \alpha f_i(t_j)=0,
\end{equation}
where $\alpha$ is a real number from the interval $0\leq\alpha\leq 1$ and $\gamma$ is some complex constant. The homotopy is equal to the start system (multiplied by $\gamma$) for $\alpha=0$ and to the target system for $\alpha=1$. Therefore, by varying $\alpha$, we can continuously deform the equations from the start system to the target system and track the solutions of $H_i(t_j,\alpha)=0$ along the way. The constant $\gamma$ has been introduced so that it is possible to reach complex solutions of the target system even from real solutions of the start system. Conventionally, we choose $\gamma=e^{i\theta}$, where $0<\theta<2\pi$. For almost all values of $\theta$, there are no bifurcations or singularities along the paths.

The path tracking of solutions from $\alpha=0$ to $\alpha=1$ has to be done in finite steps. We use the predictor-corrector method with adaptable step size in $\alpha$. Each step of the algorithm proceeds as follows:
\begin{enumerate}
   \item After the $n$-th step of the algorithm, we start with a parameter $\alpha^{(n)}$ and a solution $t_i^{(n)}$.
   \item We increase the homotopy parameter by a given step size $\Delta\alpha$ so that $\alpha^{(n+1)}=\min(\alpha^{(n)}+\Delta\alpha,1)$.
   \item We estimate the solution of the homotopy equations at $\alpha^{(n+1)}$. The simplest possibility is $t_i^{(n+1)}=t_i^{(n)}$, but the method works better if we estimate the solution by extrapolation of the previous path. We use a second order extrapolation in our code.
   \item Then we correct the predicted solution by Newton's method so that it satisfies $H_i(t_j^{(n+1)},\alpha^{(n+1)})=0$. However, Newton's method is allowed only a given maximal number of iterations $M$ to prevent jumps between paths.
   \item If Newton's method converges within $M$ iterations and the solution satisfies $\|t_i^{(n+1)}-t_i^{(n)} \|<\epsilon \| t_i^{(n)} \|$, where $\epsilon$ is a small parameter, we accept the solution and move to the next step (if the solution satisfies the previous condition with $\epsilon/2$, we additionally double the step size $\Delta\alpha$). Otherwise, the solution is not accepted, we reduce the step size $\Delta\alpha$ by a factor of 2 and return to point 1.
\end{enumerate}
We repeat these steps until we reach $\alpha=1$, which indicates that we have found a regular solution of the target system, or until $\Delta\alpha<\Delta\alpha_{min}$ (which is typically accompanied by divergence of the solution), which means that this path does not lead to a solution of the target system. We usually choose $\Delta\alpha_{min}=10^{-15}$.

It there are multiple paths leading to a solution, it means that it has higher multiplicity. However, OSFT solutions usually have multiplicity one (the only known exception is the trivial solution $\Psi=0$ in OSFT without gauge fixing), so solutions with higher multiplicity usually appear thanks to jumps between different paths. To prevent these jumps, we have introduced the restrictions on the maximal number of iterations of Newton's method $M$ and the relative change of the solution $\epsilon$. We typically choose $M=5$ and $\epsilon\approx 0.1$. If we find solutions with higher multiplicity, we recompute them with smaller $\epsilon$ to verify their status.

The complexity of this algorithm is given primarily by the number of solutions, which means that it is roughly $O(2^N)$ for quadratic equations. Since the solutions are treated independently, this algorithm can be parallelized in a very straightforward way in C++. We have been able to fully solve at most 26 equations with the available computer resources. Considering that the number of states in OSFT theories grows exponentially with level and each additional equation doubles the computer requirements, we can use this method only at very low levels. In the best case scenario, which is the universal twist even string field in Siegel gauge, we have been able to go up to level 6. In a more typical case, for example the free boson theory, we usually use this method at levels between 1 to 3, depending on the number of states it the given theory.

\section{Observables}\label{sec:Numerics:observables}
In this section, we describe how we compute string field theory observables and how we extrapolate them to infinite level.

\subsection{Energy}\label{sec:Numerics:observables:Energy}
The evaluation of energy is quite simple because we have already prepared all the necessary elements before. We usually write the energy in terms of the effective potential as
\begin{equation}
E(\Psi)=\nnn(1+2\pi^2 \vv(\Psi)),
\end{equation}
where $\nnn$ is its normalization in the given background. It terms of string field coefficients, the effective potential reads
\begin{eqnarray}\label{Energy num1}
\vv(t)=\frac{1}{2}\sum_{i,j}Q_{ij}t_it_j+\frac{1}{3}\sum_{i,j,k}V_{ijk}t_it_jt_k.
\end{eqnarray}
Assuming that the string field coefficients solve the equations of motion (\ref{equations basis}), we can substitute them into the cubic term and we find
\begin{equation}\label{Energy num2}
\vv(t)=\frac{1}{6}\sum_{i,j} Q_{ij}t_it_j.
\end{equation}
This expression is proportional to the kinetic term in the action. This formula has complexity only $O(N^2)$, so the energy can evaluated very quickly.

If we impose some gauge conditions, the situation gets more complicated because we do not solve all equations of motion. However, if we choose the canonic projector, we can still use the formula (\ref{Energy num2}). The reason is that we can always change the basis so that the dependent variables are identically equal zero, $t^{(D)}_i=0$. Then the projected-out equations appear in the action schematically as $\sum_i t^{(D)}_i\frac{\del S}{\del t^{(D)}_i}$ and therefore they do not contribute to the energy.

If we decide to use a non-canonical projector $P\neq P_C$ or if we want compute the action for a string field which is not a solution, we have to use the original formula (\ref{Energy num1}). The evaluation of the cubic term has complexity $O(N^3)$, which is the same as Newton's method, so we have to parallelize it and make it as efficient as possible. We can take advantage of the symmetry of the cubic vertex and sum only over $k\geq j\geq i$:
\begin{equation}
\sum_i\sum_{j\geq i}\sum_{k\geq j} M(i,j,k) V_{ijk} t_i t_j t_k,
\end{equation}
where
\begin{equation}
M(i,j,k)=\begin{cases}
1\qquad {\rm if}\ i=j=k, \\
3\qquad {\rm if}\ i=j\ {\rm or}\ j=k\ {\rm or}\ i=k, \\
6\qquad {\rm if}\ i\neq j\neq k.
\end{cases}
\end{equation}
If the string field is twist non-even, we have to sum only over terms with even number of twist odd fields as in the Jacobian.

In theories with nontrivial primary fields, we get a sum over the fusion rules:
\begin{equation}
\sum_f \sum_{i,j,k} M(f,i,j,k) V_{ijk}^{(f)} t_{p,i} t_{q,j} t_{r,k}.
\end{equation}
The range of $i,j,k$ depends on the symmetry of the fusion rule $f$ and the multiplicative factor as well.

A curious case arises when only one of the equations of motion is violated (which happens for example for marginal deformations or when computing the tachyon potential). Let's say it is the equation number $m$. Then we do not have to compute the whole cubic term and we can just make a correction to (\ref{Energy num2}):
\begin{equation}\label{Energy num3}
\vv(t)=\frac{1}{6}\sum_{i,j} Q_{ij}t_it_j+\frac{1}{3}t_m\left(\sum_{j}Q_{mj}t_j+\sum_{j,k}V_{mjk}t_jt_k\right).
\end{equation}

\subsection{Ellwood invariants}\label{sec:Numerics:observables:Ellwood}
When we expand the string field into a basis, its contraction with the Ellwood state becomes
\begin{equation}
\la E[\vv]|\Psi\ra =\sum_i t_i \la E[\vv]|i\ra.
\end{equation}
Now consider a basis of bulk fields $\la i,j|\equiv\la 0|\Phi_i(i)\bar\Phi_j(-i)$, where $\Phi_i(z)$ form a basis of chiral vertex operators at the given point. The Ellwood state can be written in terms of this basis as
\begin{equation}
\la E[\vv]|=\la 0|\vv(i,-i)U_{f_I}=\sum_{i,j}W_{ij}[\vv]\la i,j|U_{f_I},
\end{equation}
where $W_{ij}[\vv]$ are coefficients which depend on the vertex operator $\vv$. They are usually quite simple. In most cases, they factorize into a product of holomorphic and antiholomorphic part and for fundamental primaries, there is just a single nonzero coefficient.

It is natural to define matrix elements for the basis of Ellwood states as
\begin{equation}
E_{ijk}=\la i,j|U_{f_I}|k\ra.
\end{equation}
Using these elements, we can rewrite the formula for Ellwood invariants (\ref{Ellwood definition}) as
\begin{equation}
E[\vv](t)=\la V^m| B_0 \ra+2\pi i\sum_{i,j,k} W_{ij}[\vv] E_{ijk} t_k.
\end{equation}
The evaluation of this formula is usually very fast because there are typically only few nonzero $W_{ij}[\vv]$ coefficients. Therefore the main challenge is to compute the matrix elements $E_{ijk}$. As usual, they factorize into the constituent theories as
\begin{equation}
E_{ijk}=\prod_s E_{ijk}^{(s)}=\prod_s \la i,j|^{(s)} U_{f_I}|k^{(s)}\ra.
\end{equation}
We compute $E_{ijk}^{(s)}$ using recursive algorithms based the conservation laws which we derived in section \ref{sec:SFT:Elw cons}. In most settings, we consider only fundamental primaries, so the evaluation of $E_{ijk}^{(s)}$ is relatively easy. The only exception is the free boson theory, where we compute some more complicated invariants. Next, we will show how to compute these matrix elements in the constituent BCFTs.

\subsubsection{Universal matter theory}
In the universal matter sector, Virasoro operators feel only conformal weight of the auxiliary primary operator $V_{aux}$. This primary has the same holomorphic and antiholomorphic weight $h=\bar h$, which is given by 1 minus the weight of nontrivial matter part of $\vv$. Therefore we can label bulk states just by one index and we define $E_{jk}=\la j,j|U_{f_I}|k\ra$, where $\la j,j|=\la 0| V_{aux}^{(h_j,h_j)}(i,-i)$ and the index $j$ labels all required conformal weights.

Using the conservation law (\ref{Ellwood cons Vir}), we find recursive relations
\begin{eqnarray} \label{Ellwood recursive}
E_{jk}&=&\la j,j|U_{f_I}L_{-m}|\hat k\ra\nn \\
&=&(-1)^m\la j,j|U_{f_I}L_{m}|\hat k\ra-m\left(i^{-m}+(-i)^{-m}\right)\left(\frac{c}{8}-4h_j\right)\la j,j|U_{f_I}|\hat k\ra \quad \nn \\
&=&(-1)^m \sum_l\mm(L_m)_{\hat k l} E_{jl}-m\left(i^{-m}+(-i)^{-m}\right)\left(\frac{c}{8}-4h_j\right)E_{j\hat k}.
\end{eqnarray}
This recursive algorithm is much simpler that the ones we derived in previous sections because the index $j$ plays just a passive role. The first element for every $j$ is given by
\begin{equation}\label{Ellwood first elem}
E_{j1}=\la j,j|U_{f_I}|0\ra=2^{-2h_j}.
\end{equation}

\subsubsection{Ghost theory}
The ghost part of every operator $\vv$ is always $c(i)c(-i)$. Therefore the corresponding matrix $E$ has only one nontrivial index, $E_{k}=\la 0|c(i)c(-i)U_{f_I}|k\ra$.
Its first element is given by
\begin{equation}
E_{1}=\la 0|c(i)c(-i) U_{f_I}c_1|0\ra=\frac{1}{2}\la c(i)c(-i)c(0)\ra=i.
\end{equation}
Recursive relations for higher elements depend on the choice of basis, but all of them are very simple. Using (\ref{Ellwood cons Vir}), (\ref{Ellwood cons gh j}), (\ref{Ellwood cons gh Lp}) or (\ref{Ellwood cons gh b}), we find
\begin{eqnarray}
E_k&=&(-1)^m \sum_l\mm(L^{gh}_m)_{\hat k l}E_{l}-m\frac{3}{4}\left(i^{-m}+(-i)^{-m}\right)E_{\hat k},\\
E_k&=&(-1)^{m+1} \sum_l\mm(j^{gh}_m)_{\hat k l}E_{l}-\frac{1}{2}\left(i^{-m}+(-i)^{-m}\right)E_{\hat k},\\
E_k&=&(-1)^m \sum_l\mm(L'^{gh}_m)_{\hat k l}E_{l}-m\frac{1}{4}\left(i^{-m}+(-i)^{-m}\right)E_{\hat k},\\
E_k&=&(-1)^m \sum_l\mm(b_m)_{\hat k l}^{(2)}E_{l}.
\end{eqnarray}

\subsubsection{Minimal models}
In Virasoro minimal models, we compute only invariants given by fundamental primaries, so we can proceed similarly as in the universal matter sector. We denote the basis of primary operators as $\la j,j|=\la 0|\phi_{j}(i)\bar \phi_{j}(-i)$, where $j$ goes over Kac labels in the given model. We label basis states in minimal models by the doubled index, so we define $E_{j(p,k)}=\la j,j|U_{f_I}|(p,k)\ra$. The recursive formula for these matrix elements is derived using (\ref{Ellwood cons Vir}),
\begin{equation}
E_{j(p,k)}=(-1)^m \sum_l\mm(L_m)_{\hat k l}^p E_{j(p,l)}-m\left(i^{-m}+(-i)^{-m}\right)\left(\frac{c}{8}-4h_j\right)E_{j(p,\hat k)}.
\end{equation}
As usual in minimal models, the index $\hat k$ hides another sum similarly to (\ref{Gram matrix MM IR}). The first matrix element for each primary operator is given by the corresponding bulk-boundary correlation function,
\begin{equation}
E_{j(p,1)}=\la j,j|U_{f_I}|p\ra=2^{-2h_j+2h_p}\ \CbbL \alpha jp,
\end{equation}
where we consider boundary condition $\alpha$. The factor $2^{-2h_j}$ mostly cancels with (\ref{Ellwood first elem}).

\subsubsection{Free boson theory}
The free boson theory allows us to define a large number of Ellwood invariants, especially in more than one dimension. Many of them are not primary with respect to the U(1) symmetry, so the evaluation of Ellwood invariants is more complicated than in the cases above.

The basis of bulk operators is given by (\ref{Ellwood FB gen op}). We represent these states by two copies of the usual Fock space (\ref{basis FB}) and label the basis elements by two double indices, so the matrix $E$ has in principle 6 indices,
\begin{equation}
E_{(p,i)(q,j)(r,k)}=\la (p,i),(q,j)|U_{f_I}|(r,k)\ra.
\end{equation}
However, the index $r$ is redundant because it is uniquely determined by momentum conservation. Alternatively, we can represent the momentum indices in terms of fusion rules as for the cubic vertex, but in this case, it does give us any significant advantage.

The recursive algorithm is based on the conservation law (\ref{Ellwood result final}) without the anomalous terms. First, we compute all required residues,
\begin{eqnarray}
R_{nk}^L &=& \res_{z\rar i}\frac{v_n(z)}{(z-i)^k}, \\
R_{nk}^R &=& \res_{z\rar -i}\frac{v_n(z)}{(z+i)^k},
\end{eqnarray}
which are given by (\ref{residues Ellwood}). Next, we have to deal with the OPE (\ref{Ellwood FB OPE}). In order to do that, we define a generalization of the operation (\ref{operator separation}). This time, we need to remove an arbitrary oscillator from a state. Given a state $|(p,i)\ra=\alpha_{-n_1}\dots\alpha_{-n_l}|p\ra$, we define
\begin{equation}
|(p,\hat i_k)\ra=\alpha_{-n_1}\dots\cancel{\alpha_{-n_k}}\dots\alpha_{-n_l}|p\ra,
\end{equation}
so that
\begin{equation}
|(p,i)\ra=\alpha_{-n_k}|(p,\hat i_k)\ra.
\end{equation}
Using this notation, the recursive algorithm reads
\begin{eqnarray}
E_{(p,i)(q,j)(r,k)}&=&(-1)^{m+1} \mm(\alpha_m)_{\hat k l}^r E_{(p,i)(q,j)(r,l)}\nn\\
&-&\sqrt{2}\left(i^{-m}p+(-i)^{-m}q\right) E_{(p,i)(q,j)(r,\hat k)}  \\
&-&\sum_l R^L_{m n_l} E_{(p,\hat i_l)(q,j)(r,\hat k)}-\sum_l R^R_{m n_l} E_{(p,i)(q,\hat j_l)(r,\hat k)}. \nn
\end{eqnarray}
The first line represents $\alpha_m$ acting on the string field, the second line contractions of the $\del X$ current with the momentum operators in the Ellwood state and the third line its contractions with $\del^n X$ operators in the Ellwood state.

After we remove all $\alpha$ oscillators from $|(r,k)\ra$, we have to compute $E_{(p,i)(q,j)(r,1)}$. These matrix elements can be computed by the usual free boson contractions, so in this case, we do not need any special conservation laws. For example, using the first operator from $\la (p,i)|$, we find the recursive relation
\begin{eqnarray}
E_{(p,i)(q,j)(r,1)}&=&\left(\frac{q}{\sqrt{2}}\frac{(-1)^{m_1-1}}{(2i)^{m_1}}
-\frac{p+q}{\sqrt{2}}\frac{(-1)^{m_1-1}}{i^{m_1}}\right)E_{(p,\hat i_1)(q,j)(r,1)}\nn\\
&+&\sum_l \frac{(m_1+n_l-1)!}{(m_1-1)!(n_l-1)!}\frac{(-1)^{m_1-1}}{(2i)^{m_1+n_l}}E_{(p,\hat i_1)(q,\hat j_l)(r,1)}.
\end{eqnarray}
The first line comes from contractions of $\del^{m_1} X$ with the momentum operators and the second line from contractions with other $\del^n X$ operators at the point $-i$. By repeating this step, we can reduce all matrix elements to the basic correlation function
\begin{equation}
E_{(p,1)(q,1)(r,1)}=\la e^{ipX}(i)\,e^{iqX}(-i)\,f_I \circ e^{irX}(0)\ra=2^{r^2+pq/2}\delta_{p+q,2r}.
\end{equation}

\subsection{Out-of-Siegel equations}\label{sec:Numerics:observables:Siegel}
In principle, the evaluation of out-of-Siegel equations is not difficult, we can simply use the formula (\ref{equations basis}). The problem is that, in order to use Siegel gauge effectively, we have to avoid computation of vertices involving states outside Siegel gauge. Because of that, we compute only the first out-of-Siegel equation, which is given by (\ref{Delta}).

This equation involves only few nonzero matrix elements of $Q$, so it easier to evaluate them by hand rather than to write a complicated algorithm for a generic matrix element. When it comes to the quadratic term, the method of its evaluation depends on the basis. In the SU(1,1) singlet basis, we express $\Delta_S$ as (\ref{Delta2}). The ghost vertices involving $j_{-2}^{gh}c_1|0\ra$ are computed as a byproduct of evaluation of singlet vertices, so we just need to keep this specific set when we deallocate auxiliary vertices and use it for computation of $\Delta_S$. In the $bc$ basis, auxiliary vertices do not involve states outside Siegel gauge, so we add the state $b_{-2}c_0c_1|0\ra$ to our basis by hand and compute the corresponding vertices. The cubic vertex algorithm does not have any problems with this exceptional state.

The evaluation of the projected-out equations in other gauges is much easier. We have to compute all ghost vertices anyway, so we can easily evaluate (\ref{equations basis}) for any state. We also notice that we evaluate the full equations of motion in every iteration of Newton's method, so we can simply save values of the equations we are interested in in the last iteration.

\subsection{Extrapolations to infinite level}\label{sec:Numerics:observables:extrapolation}
The final subsection of this chapter will not focus on evaluation of observables, but on their analysis. Numerical calculations in the level truncation scheme give us only finite level data, which often change quite a lot with level, and therefore we cannot expect that they would have very good precision. However, it has been established that one can significantly improve the precision of results by extrapolating them to infinite level (see for example \cite{TaylorPerturbativeTV}\cite{GaiottoRastelli}\cite{MarginalTachyonKM}\cite{KudrnaUniversal}\cite{ArroyoKudrna}).

In order to estimate the asymptotic value of some quantity, we fit the known data points with some function of level $f(L)$ and then we take the limit $L\rar \inf$. However, this procedure can be done in many different ways (see the aforementioned references for some examples) and none of them is going to work perfectly for all solutions and all possible quantities. It is not guaranteed that all solutions have the same type of asymptotic behavior in the first place. On the other hand, we would like to avoid case-by-case analysis because that would be too complicated and we would tend to choose a method that agrees with the expected results. Therefore, after many experiments, we came with a method that gives good results for a large number of solutions with clear interpretation.

Our method is essentially a generalization of the procedure from \cite{GaiottoRastelli}. We fit the known data points with a polynomial in $1/L$\footnote{We have experimented with various others functions (for example with functions including non-integers powers of $L$), but none of them worked for a larger set of quantities or solutions. The only exception are rational functions, see \cite{ArroyoSchablgauge}\cite{ArroyoKudrna}, which have the same asymptotic behavior as polynomials in $1/L$. It is also possible to replace $1/L$ by $1/(L+a)$ (\cite{GaiottoRastelli} uses $a=1$), but the results have only a small dependence on $a$, so we choose $a=0$ for simplicity.},
\begin{equation}\label{fit 1}
f^{(M)}(L)=\sum_{k=0}^M\frac{a_k}{L^k},
\end{equation}
where $M$ is the order of the polynomial, which must be lower than the number of used data points. We use the function NonlinearModelFit from Mathematica to do the fits.

Unfortunately, it turns out that data coming from OSFT usually do not follow a smooth curve, but they rather oscillate around it. The energy, $\Delta_S$ and string field coefficients have an oscillation period of two levels (because of that, this problem was not observed for the tachyon vacuum solution) and the differences between even and odd levels are usually quite small. Ellwood invariants have an oscillation period of 4 levels and their amplitude seems to be connected to weights of the invariants, see the discussion in sections \ref{sec:FB circle:MSZ} and \ref{sec:FB circle:single:pade}. If we wanted to extrapolate all data points using a single function, we would have to use a low order fit (linear or quadratic) to smoothen the oscillations. However, the results of such procedure are usually not very precise.

Instead, we divide the data points into groups based on the period of oscillations, which means that we have either two groups with levels $(2,4,6,8,\dots)$ and $(3,5,7,9,\dots)$ or four groups with levels $(2,6,10,\dots)$, $(3,7,11,\dots)$, $(4,8,12,\dots)$ and $(5,9,13,\dots)$. Then we extrapolate each group independently and we take the average of the level infinity estimates as the final result. This approach leads to good results for most quantities, even for some highly oscillating invariants.

By many experiments, we have found that we usually get the best results when we use the highest order possible, which means that the function $f^{(M)}(L)$ is actually a polynomial interpolation of the data points (respectively of the selected subset). Therefore we take $M$ to be equal to the number of data points minus one.

Since every extrapolation technique produces a different result, we would like to have an error estimate to see how trustworthy the results are. Unfortunately, making a good error estimate is very difficult.

Errors can be estimated by making variations of the extrapolation procedure. Since we independently extrapolate two or four groups of data, it is natural to take the standard deviation of these results as the error estimate. These error estimates will be used in the rest of this thesis unless stated otherwise.

However, when take a typical solution with a clear identification and compare error estimates with the actual differences between level infinity extrapolations with the expected results, we find discrepancies. Errors of well convergent quantities are usually underestimated (up to one order), errors of mildly oscillating invariants are more or less fine and errors of highly oscillating (but convergent) invariants are overestimated (up to one order). This problem probably arises because the independent extrapolations are not distributed randomly around the expected values, but they follow some patterns, so the standard deviation does not work as intended.

Unfortunately, we have not found any better alternative. We have experimented with variations of the input data by dropping the lowest/highest level data points and with variations of the order of the fit. For the energy and other quantities with oscillation period of 2 levels, these choices give us similar errors as the method above. When it comes to Ellwood invariants, these error estimates do not work properly because there are typically only 3-5 points in each group and therefore we cannot decrease the number of points or the fit order by much. And when we do, it almost always leads to strong decrease of precision, so taking standard deviation over several such extrapolations is not justified either. Therefore this approach is not more reliable than the one we use.

We conclude that the presented error estimates cannot be taken too strictly, they should be viewed only as order estimates of the actual errors. One should make corrections based on typical behavior of the extrapolated quantity (for example, increase the error for the energy, but decrease it for the $D_1$ invariant), but there is no precise prescription how to do it. We should also note that these errors can be viewed as 'statistical'. There can be also 'systematic errors', which could be caused for example by a wrongly guessed asymptotic behavior of a solution and which are pretty much impossible to predict.

We illustrate typical properties of extrapolations and their errors on the example of the MSZ lump solution in section \ref{sec:FB circle:MSZ:extrapolations}.

Finally, let us make a comment about extrapolations of complex solutions. The extrapolation procedure described above works consistently for real solutions, but that chances for some complex solutions. We encounter three main types of complex solutions. First, there are pseudo-real solutions. These mostly behave the same as real solutions. Next, there are solutions that start as complex, but become real at some finite level (a typical example is the Ising model solution in section \ref{sec:MM:Ising:Id}). The behavior of such solutions changes dramatically when the imaginary part disappears. That means that we can extrapolate only data from levels where these solutions are real and we often have only very few data points to work with, which makes the extrapolations inaccurate. There is essentially no point in extrapolating solutions which are complex at all available levels and expected to become real at some higher level beyond our reach.

Finally, there are solutions which are inherently complex even in the infinite level limit. We cannot make any general statement about them because they are quite diverse and their meaning is often unclear. The reliability of extrapolations for such solutions must be judged case by case.

\chapter{Results - Universal solutions}\label{sec:universal}
Now we will move from theory and description of numerical methods to analysis of OSFT solutions. In this chapter, we discuss universal solutions, i.e. solutions that do not depend on the D-brane background. Among these solutions, there is the most famous OSFT solution, the tachyon vacuum, which describes decay of the original D-brane system through tachyon condensation. This solution have been studied extensively both numerically \cite{KosteleckySamuel}\cite{SenZwiebachTV}\cite{HataTVSiegEq}\cite{SenUniversality}\cite{EllwoodTaylorGauge}\cite{GaiottoRastelli}\cite{Kishimoto26}\!\! \cite{KishimotoTakahashi1}\cite{KishimotoTakahashi2}\cite{KudrnaUniversal}\cite{ArroyoSchablgauge}\cite{ArroyoKudrna}
and analytically \cite{AnalyticSolutionSchnabl}\cite{AnalyticSolutionOkawa}\cite{ErlerKBC1}\cite{ErlerKBC2}\cite{EllwoodTVCohomology}\!\! \cite{TVAnalyticEquations}\cite{TVTakahashiNum}\cite{TVEllwoodTakahashi}\cite{TVArroyoPade}\cite{SimpleAnalyticSolutionErlerSchnabl}\cite{TVJokel}\cite{TVArroyo}. We do not have any new results concerning this solution and therefore we will just review its properties following \cite{KudrnaUniversal} and \cite{ArroyoKudrna}.

Apart from the tachyon vacuum, we cannot expect many other physically relevant solutions. Essentially all we can possibly describe in this setting are multi-branes, which describe several copies of the original D-brane system, and possibly a ghost brane, which has minus the energy of the original D-brane system. These solutions have been studied analytically in \cite{Multibrane1}\cite{Multibrane2}\cite{Multibrane3}\cite{Multibrane4}\cite{Multibrane5}\cite{Multibrane6}\cite{Multibrane7}\cite{Multibrane8}. They have the correct quantization of energy, but they usually satisfy the equations of motion only formally because there are anomalies when the equations are contracted with certain states. Recently, we have tried to find such solutions numerically in \cite{KudrnaUniversal}, but these numerical solutions also have various pathologies, see later. Multi-brane solutions can be also constructed using the intertwining solutions \cite{ErlerMaccaferri}\cite{ErlerMaccaferri2}\cite{ErlerMaccaferri3}. The solutions in the first two papers are not universal, but the newest reference finally describes a solution with the desired properties.

The main focus of this thesis are solutions in Siegel gauge, but in this chapter, we also touch other options. We discuss the tachyon vacuum solution in Schnabl gauge and solutions without any gauge fixing. However, it seems that Siegel gauge is the best choice for a generic OSFT solution. Without any gauge fixing, Newton's method, which we use to improve solutions to higher levels, does not work well and therefore this approach is pretty much useless. We can find a relatively well-behaved tachyon vacuum solution in Schnabl gauge, but the solution offers lesser precision than in Siegel gauge while requiring more computer resources. We have done only brief experiments with non-universal solutions outside Siegel gauge, but the results suggest these solutions behave similarly. Therefore using other gauges makes sense only if one wants to compare numerical and analytic results or in case of complex or otherwise problematic solutions, where one can hope to find a gauge where these solutions behave better than in Siegel gauge.

\section{Siegel gauge solutions}\label{sec:universal:Siegel}
There are two independent conditions we can impose on the string field in Siegel gauge: the twist even condition and the SU(1,1) singlet condition. By imposing these conditions, we significantly reduce the number of states at a given level (see appendix B.3 in \cite{KudrnaUniversal}), but we also risk loosing solutions which do not satisfy these conditions. Therefore we explore all possibilities to see which ansatz allows us to find some interesting solutions.

We start with the most restrictive case, which means imposing both conditions. First, we use the homotopy continuation method, which is described in section \ref{sec:Numerics:homotopy}, to find seeds for Newton's method. We have managed to solve the equations of motion at level 2, where we have 3 equations with 7 solutions, at level 4 with 8 equations and 250 solutions and finally at level 6 with 21 equations and approximately 2096000 solutions. Interestingly, the number of solutions is smaller than the generic amount $2^N$, but only a small percentage of solutions is missing.

\begin{figure}
\centering
\includegraphics[width=14cm]{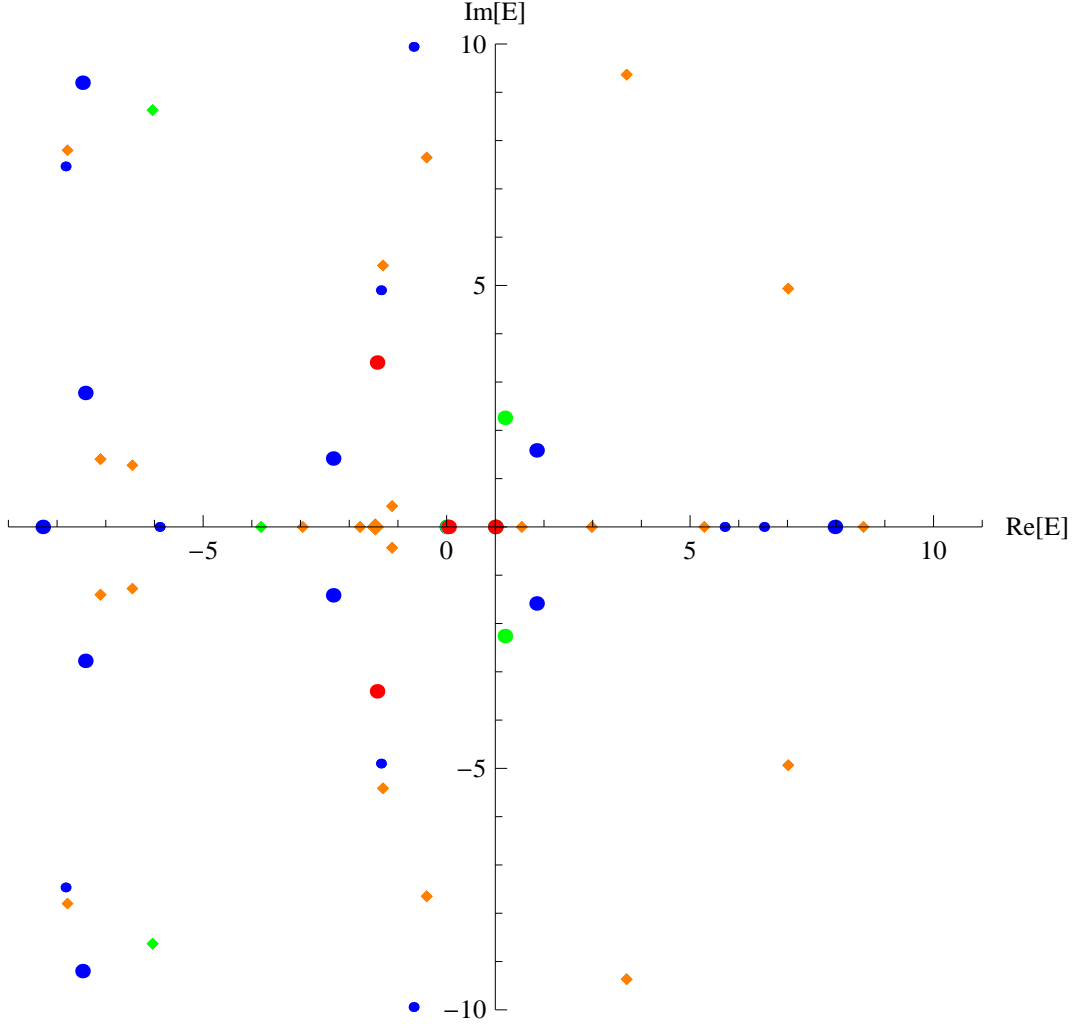}
\caption{Universal solutions in Siegel gauge at level 2 (red), level 4 (green), level 5 (orange) and level 6 (blue). Twist even solutions are denoted by circles, twist non-even solutions by diamonds. SU(1,1) singlet solutions have bigger marks that non-singlet solutions.}
\label{fig:universal sol}
\end{figure}

Although we have found a huge number of solutions, only few of them have energy of the same order as the initial D-brane. See figure \ref{fig:universal sol}, where we plot energies of solutions 'close' to the perturbative vacuum in the complex plane. In the next step, where we improve solutions to higher levels using Newton's method, even most of them must be discarded. In \cite{KudrnaUniversal}, we analyzed all solutions with $|E|<50$ and we have found only 17 solutions (up to complex conjugation) for which Newton's method converges within 20 iterations. Out of these, the tachyon vacuum is the only one which is well-behaved (not counting the trivial perturbative vacuum), plus there three more somewhat reasonable solutions, whose properties are summarized in table \ref{tab:Universal solutions}.

Next, we relax the twist even condition. The increased number states allows us to go only up to level 5, where we have found 65106 solutions of equations in 16 variables. However, only one of the new solutions appears in figure \ref{fig:universal sol} and none of them is stable when improved by Newton's method.

In the twist even non-singlet case, we can go once again up to level 6. This time, we have solved 26 equations with approximately 66 million solutions. Several of them appear in figure \ref{fig:universal sol}, but we have found no well-behaved solution when we improved the potentially interesting seeds using Newton's method.

Finally, we consider the twist non-even non-singlet case. We have solved 21 equations at level 5 with approximately 2 millions of solutions. Among these, we have found two stable solutions with energy of order 1, we show their properties in table \ref{tab:Universal solutions}. Interestingly, both solutions are even with respect to the combined symmetry $(-1)^J \Omega$ (see section \ref{sec:SFT:string field:ghost}).

\begin{table}
\centering\footnotesize
\begin{tabular}{|c|c|lll|c|}
\multicolumn{6}{c}{\normalsize Twist even SU(1,1) singlets}                                                    \\\hline
Solution \rowh{13pt}            & Level & \ps Energy$^{(\inf)}$ & $\ps E_0^{(\inf)}$ & $\ps \Delta_S^{(\inf)}$ & real   \\\hline
perturbative vacuum \rowh{10pt} & $-$   & $\ps 1              $ & $\ps 1           $ & $\ps 0                $ & yes    \\
tachyon vacuum                  & 30    & $   -9\dexp{-6}     $ & $\ps 0.00015     $ & $   -7\dexp{-6}       $ & yes    \\
"double brane"                  & 28    & $\ps 1.40+0.02 i    $ & $\ps 1.21-0.04 i $ & $\ps 0.23-0.12 i      $ & no     \\
"ghost brane"                   & 28    & $   -1.13+0.03 i    $ & $   -1.02+0.09 i $ & $\ps 0.08-0.04 i      $ & no     \\
copy of tachyon vacuum          & 28    & $   -0.06-0.27 i    $ & $   -0.15-0.04 i $ & $   -0.16-0.03 i      $ & no     \\\hline
\multicolumn{6}{l}{}                                                                                           \\
\multicolumn{6}{c}{\normalsize Non-even non-singlets}                                                          \\\hline
Solution \rowh{13pt}            & Level & \ps Energy$^{(\inf)}$ & $\ps E_0^{(\inf)}$ & $\ps \Delta_S^{(\inf)}$ & real   \\\hline
"half ghost brane" \rowh{10pt}  & 26    & $   -0.51           $ & $   -0.66        $ & $\ps 0.17             $ & pseudo \\
"half brane"                    & 24    & $\ps 0.67+0.002 i   $ & $\ps 0.51+0.09 i $ & $\ps 0.21+0.09 i      $ & no     \\\hline
\end{tabular}
\caption{Properties of selected solutions in Siegel gauge in the $(L,3L)$ truncation scheme. We show the maximal level we have reached, extrapolations of observables to the infinite level and reality of the solutions. Some of the numbers are slightly different from \cite{KudrnaUniversal} because we use a different type of extrapolations.}
\label{tab:Universal solutions}
\end{table}

Out of the stable solutions summarized in table \ref{tab:Universal solutions}, we are going to discuss in detail only the tachyon vacuum solution. The other solutions will mentioned only briefly, more about them can be found in \cite{KudrnaUniversal}. We cannot accept any of them as physical with the available data  because of problems like violation of the out-of-Siegel equation $\Delta_S$, discrepancy between the energy and the $E_0$ invariant, nonzero imaginary part, etc. On the other hand, we cannot entirely discard them because their behavior may improve at high enough level or in some other gauge\footnote{New results regarding the "double brane" and "ghost brane" solutions can be found in \cite{KishimotoTTnum}}.

The "double brane" is a complex solution whose real part of energy is close to 2 at low levels, which is something we would expect from the double brane. However, the energy seems to be asymptotically going toward approximately 1.4 and $E_0$ goes even farther away from 2, so its actual interpretation is probably different. What could save the solution is that the extrapolations suggest that it may become real at some high level (probably somewhere between levels 50 and 100) and therefore, after the transition, we can expect a change of behavior similar to the Ising model solution from section \ref{sec:MM:Ising:Id}.

The "ghost brane" solution behaves in a different way. It does not look very nice at finite levels, but both the energy and $E_0$ approach $-1$ asymptotically with relatively small errors. In comparison with the other non-standard solutions from table \ref{tab:Universal solutions}, it has the smallest value of $\Delta_S$. The solution is inherently complex, but the imaginary parts of its observables seem to go to zero asymptotically.

The energy and the $E_0$ invariant of the next solution suggest that it may be a Gribov copy of the tachyon vacuum, but it has a large imaginary part and its overall precision is not very good.

Out of the two twist non-even solutions, the "half ghost brane" is the more interesting one. It starts as complex, but its energy becomes real at level 22. We call it pseudo-real because the twist odd part of the solution is purely imaginary with respect to the complex conjugation (\ref{complex conjugation SFT}). Its energy is close to $-\frac{1}{2}$, but we have unfortunately access only to few real data points, which means that we cannot extrapolate the solution very reliably (see the discussion in section \ref{sec:MM:Ising:Id}) and its asymptotic behavior remains uncertain.

The final solution, the "half brane", has energy close to one half of the original D-brane energy. This solution is inherently complex and its precision is quite low, so it is the least likely one to be physical.

It is unlikely that it will be possible to significantly improve level of these solutions in near future, so exploring them outside Siegel gauge seems to be the best possibility to learn more about them. So far, we have only checked whether similar solutions can be found in Schnabl gauge \cite{ArroyoKudrna}, see the results in table \ref{tab:Schnabl solutions}. Properties of solutions in both gauges are mostly quite different. The best matching solutions are the possible copies of the tachyon vacuum, although one of them is complex and the other is real, so it is not clear whether they are related. The differences suggest that it is more likely that the additional solutions are just artifacts of the level truncation approach.

\subsection{Tachyon vacuum in Siegel gauge}\label{sec:universal:Siegel:TV}
The tachyon vacuum solution describes decay of the original D-brane system caused by tachyon condensation. The result of this process should be the closed string vacuum, which means that there should be no open string excitations around the solution and all of its gauge invariants should be equal to zero. In this thesis, we focus only on the gauge invariant observables. Showing that the cohomology of the BRST operator around the solution (\ref{Q Psi}) is trivial can be done for the analytic solution \cite{EllwoodTVCohomology}, but it is quite difficult task for the level truncated solution. The references \cite{EllwoodTaylorCohomology}\cite{GiustoCohomology}\cite{ImbimboCohomology} attempt to compute the cohomology, but the results are somewhat controversial because there seems to be a nontrivial cohomology at negative ghost numbers. These calculations clearly need to be repeated at higher levels and the method itself should be verified on other solutions.

\begin{table}[!t]
\centering
\begin{tabular}{|l|ll|l|}\hline
Level              &  \ps Energy        & $E_0$       & $\ps \Delta_S$    \\\hline
2                  & $\ps 0.040623400 $ & $0.1101382$ & $\ps 0.03332993 $ \\
4                  & $\ps 0.012178243 $ & $0.0680476$ & $\ps 0.01450131 $ \\
6                  & $\ps 0.004822879 $ & $0.0489211$ & $\ps 0.00841347 $ \\
8                  & $\ps 0.002069817 $ & $0.0388252$ & $\ps 0.00564143 $ \\
10                 & $\ps 0.000817542 $ & $0.0318852$ & $\ps 0.00412431 $ \\
12                 & $\ps 0.000177737 $ & $0.0274405$ & $\ps 0.00319231 $ \\
14                 & $   -0.000173730 $ & $0.0238285$ & $\ps 0.00257255 $ \\
16                 & $   -0.000375452 $ & $0.0213232$ & $\ps 0.00213597 $ \\
18                 & $   -0.000493711 $ & $0.0190955$ & $\ps 0.00181467 $ \\
20                 & $   -0.000562955 $ & $0.0174832$ & $\ps 0.00156995 $ \\
22                 & $   -0.000602262 $ & $0.0159666$ & $\ps 0.00137834 $ \\
24                 & $   -0.000622749 $ & $0.0148397$ & $\ps 0.00122487 $ \\
26                 & $   -0.000631156 $ & $0.0137381$ & $\ps 0.00109960 $ \\
28                 & $   -0.000631707 $ & $0.0129049$ & $\ps 0.00099569 $ \\
30                 & $   -0.000627118 $ & $0.0120671$ & $\ps 0.00090829 $ \\\hline
$\inf$ \rowh{12pt} & $   -9\dexp{-6}  $ & $0.00015  $ & $   -7\dexp{-6} $ \\
$\sigma$           & $\ps 2\dexp{-6}  $ & $0.00002  $ & $\ps 1\dexp{-6} $ \\\hline
\end{tabular}
\caption{Energy, Ellwood invariant and out-of-Siegel equation $\Delta_S$ for the tachyon vacuum solution in Siegel gauge up to level 30. In the last two lines, we show extrapolations of these quantities to level infinity and estimated errors of the extrapolations.}
\label{tab:TV}
\end{table}

\begin{figure}
\centering
\includegraphics[width=9cm]{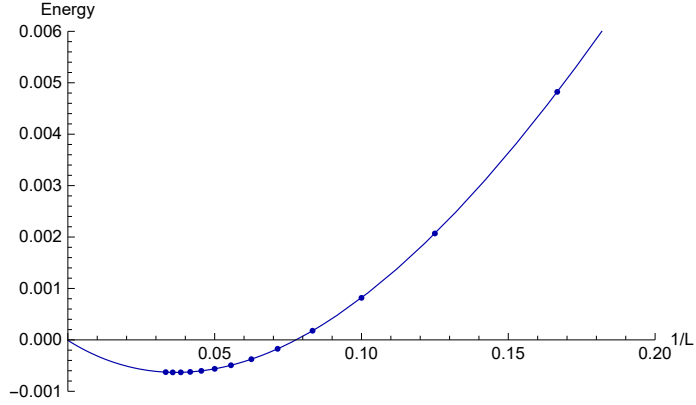}
\caption{Plot of the tachyon vacuum energy in Siegel gauge as a function of level. The solid line represents its infinite level extrapolation using a polynomial in $1/L$.}\label{fig:TV energy}
\end{figure}

\begin{figure}
\centering
\includegraphics[width=9cm]{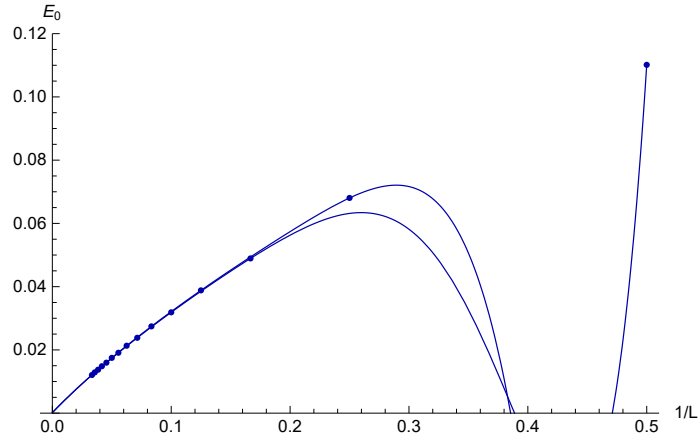}
\caption{Plot of the $E_0$ invariant of the tachyon vacuum solution in Siegel gauge. We show two independent extrapolations, which use levels of the form $4k$ and $4k+2$, $k\in \mathbb{N}$.}\label{fig:TV E0}
\end{figure}

\begin{figure}
\centering
\includegraphics[width=9cm]{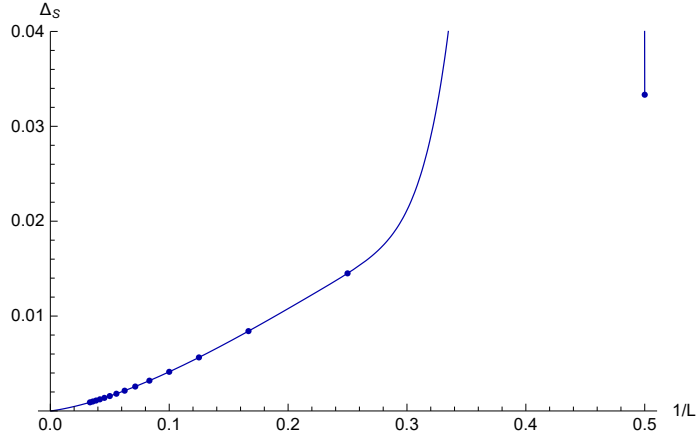}
\caption{The first out-of-Siegel equation $\Delta_S$ of the tachyon vacuum solution and its infinite level extrapolation.}\label{fig:TV Delta}
\end{figure}

We have managed to evaluate the tachyon vacuum solution up to an impressive level 30 \cite{KudrnaUniversal}. We show its gauge invariant observables and the first out-of-Siegel equation in table \ref{tab:TV} and we plot level dependence of these quantities in figures \ref{fig:TV energy}, \ref{fig:TV E0} and \ref{fig:TV Delta}.

The energy of the tachyon vacuum solution has a rather unusual feature. As a function of level, it first approaches zero, but then, at level 14, it overshoots the correct value and starts to move away from it. It was predicted \cite{TaylorPerturbativeTV}\cite{GaiottoRastelli} that the energy has a minimum at level 28. Using the level 30 data, it is finally possible to confirm this hypothesis directly, the data from table \ref{tab:TV} tell us that the energy at level 30 is higher than at level 28. The extrapolation shown in figure \ref{fig:TV energy} predicts that the energy at higher levels will go directly towards zero without any further oscillations. Interestingly, the energy of the Schnabl gauge solution (see figure \ref{fig:TV energy Schnabl}) has a minimum too, so this type of behavior is clearly common to a larger set of gauges. The $E_0$ invariant, which also measures the energy, does not suffer from such issues and it approaches zero monotonically, see figure \ref{fig:TV E0}.

In this thesis, we compute only one consistency check, the first out-of-Siegel equation $\Delta_S$. Figure \ref{fig:TV Delta} shows that it is satisfied with a very nice precision. Other out-of-Siegel equations were computed in \cite{RastelliZwiebach} and they are also satisfied quite well. In \cite{KudrnaUniversal}, we have also checked that the solution obeys the quadratic identities from \cite{SchnablQuadraticIdentities}.

The data from table \ref{tab:TV} are extrapolated using the methods described in subsection \ref{sec:Numerics:observables:extrapolation}. We observe that the asymptotical values are better than the data from level 30 approximately by two orders, for instance, the energy improves from $-6\dexp{-4}$ to $-9\dexp{-6}$. This clearly demonstrates that infinite level extrapolations are very useful tools for analyzing solutions and they can give us significantly more precise results than raw data from finite levels.

Notice that we get a better result for $E_0$ than in \cite{KudrnaUniversal}, even though we use the same data as input. That is because the extrapolation method which we use in this thesis is designed to deal with oscillations of Ellwood invariants and therefore it uses higher order polynomials in $1/L$. When extrapolating the energy and $\Delta_S$, we cannot use the error estimate described in subsection \ref{sec:Numerics:observables:extrapolation} because we have only data from even levels. So we have instead taken the standard deviation of extrapolations with the maximal level varying from 22 to 30. All three error estimates are several times smaller than the actual errors of extrapolations, which, as we are going to see later on many other examples, is unfortunately a typical property of extrapolations of these quantities.

\section{Tachyon vacuum in Schnabl gauge}\label{sec:universal:Schnabl}
When we search for solutions in Schnabl gauge, we can use the same techniques as in Siegel gauge. Using the homotopy method, we have been able to find all seeds at the same levels as in Siegel gauge, but the number of stable solutions seems to be smaller. Not counting the tachyon vacuum, we have found only two twist even solutions with reasonable properties \cite{ArroyoKudrna}, see table \ref{tab:Schnabl solutions}, and no stable twist non-even solutions.

The "half brane" solution is actually somewhat similar to the "double brane" solution in Siegel gauge at low levels, but asymptotically it behaves differently. It has large imaginary part, so it is not likely that it is a physical solution. The second solution is slightly better. It is real and the asymptotic values of its energy and $E_0$ are close to zero, so there is a chance that it is gauge equivalent to the tachyon vacuum. However, its precision is low and the first out-of-gauge equation is nonzero, so it is possible that this solution is nonphysical as well.

\begin{table}[!b]
\centering\footnotesize
\begin{tabular}{|c|c|lll|c|}\hline
Solution \rowh{13pt}            & Level & \ps Energy$^{(\inf)}$ & $\ps E_0^{(\inf)}$ & $\ps \Delta_S^{(\inf)}$ & real   \\\hline
perturbative vacuum \rowh{10pt} & $-$   & $\ps 1              $ & $\ps 1           $ & $\ps 0                $ & yes    \\
tachyon vacuum                  & 24    & $\ps 0.0005         $ & $   -0.0015      $ & $   -0.0003           $ & yes    \\
"half brane"                    & 24    & $\ps 0.63-0.32 i    $ & $\ps 0.46-0.45i  $ & $   -0.07-0.05 i      $ & no     \\
copy of tachyon vacuum          & 24    & $   -0.19           $ & $   -0.05        $ & $   -0.15             $ & yes    \\\hline
\end{tabular}
\caption{Properties of stable twist even solutions in Schnabl gauge. We have not found any stable twist non-even solutions.}
\label{tab:Schnabl solutions}
\end{table}

Therefore we focus only on the tachyon vacuum solution following \cite{ArroyoKudrna}. This solution is interesting for two reasons: It is a nontrivial test that the level truncation framework works outside Siegel gauge and it can be directly compared to the analytic tachyon vacuum solution \cite{AnalyticSolutionSchnabl}.

\begin{figure}
\centering
\includegraphics[width=10cm]{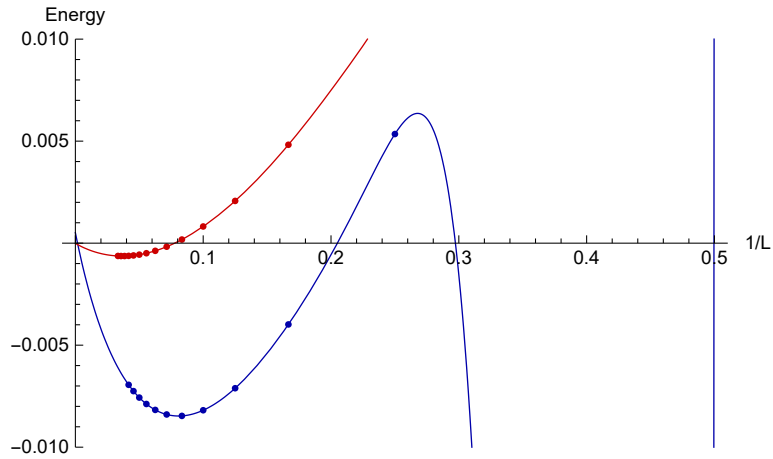}
\caption{Comparison of the tachyon vacuum energy in Schnabl gauge (blue) and in Siegel gauge (red). The solid lines represent infinite level extrapolations.}\label{fig:TV energy Schnabl}
\end{figure}

In table \ref{tab:TV Schnabl}, we show the gauge invariants of this solution and the first out-of-gauge equation $\Delta_S$. We have been able to compute the solution only up to level 24, which is 6 levels lower than in Siegel gauge, because we had to evaluate the full set of ghost vertices, which consume significantly more memory (see appendix \ref{sec:time}). When we compare this solution with the one in Siegel gauge, we find that both behave similarly. In particular, the energy also overshoots zero, it has a minimum (this time at level 12) and then it turn back towards the correct value. We plot the comparison of both energies in figure \ref{fig:TV energy Schnabl}. The extrapolation of energy in Schnabl gauge slightly overshoots 0 for a second time around level 500, but it is probably just an error of the extrapolation technique.

Surprisingly, extrapolations of the tachyon vacuum solution in Schnabl gauge are much less reliable than in Siegel gauge. We observe that the extrapolations improve precision of the solution compared to the level 24 data, but only by approximately one order, while we saw improvement by two orders in Siegel gauge. However, this issue mainly concerns coefficients of the solution.

\begin{figure}
\centering
\includegraphics[width=10cm]{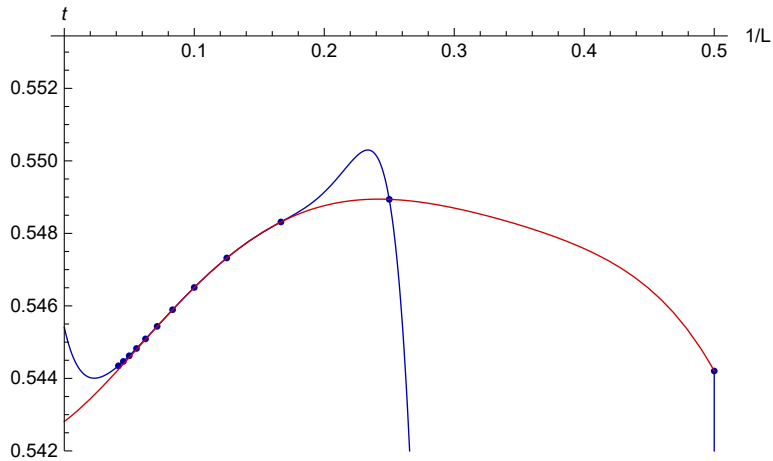}
\caption{Plot of the tachyon coefficient of the tachyon vacuum solution in Schnabl gauge. The blue line represents its extrapolation using all data points, the red line extrapolation using data only up to level 12. The horizontal axis is set to the analytic value $t_{analytic}=0.553466$.}\label{fig:TV tachyon Schnabl}
\end{figure}

Schnabl's analytic solution can be expanded into $L_0$ eigenstates (at least up to some finite level), so we can compare the numerical solution and the analytic one coefficient by coefficient and, ideally, we should observe convergence of the numerical coefficients towards the analytic ones. However, it is not entirely clear whether it really happens. We illustrate this problem on the tachyon coefficient $t$, which should be equal to $t_{analytic}=0.553466$. The values of the tachyon coefficient up to level 24 are given in table \ref{tab:TV tachyon Schnabl} and we observe that the extrapolated value $t^{(\inf)}=0.5454$ is quite far from analytic value, which is nicely illustrated in figure \ref{fig:TV tachyon Schnabl}. So the first conclusion seems to be that the two solutions are different, but we find the extrapolations have a strong level dependence, which may resolve the problem.

For comparison, have a look at the tachyon coefficient in Siegel gauge, which is also shown in table \ref{tab:TV tachyon Schnabl}. By extrapolating all data points up to level 30, we get $t^{(\inf,30)}=0.540493$. What happens if we use less data points to do the extrapolation? We find that even few data points give us quite accurate results, for example, we get $t^{(\inf,12)}=0.540520$ using only the data up to level 12.
On the other hand, the asymptotic values in Schnabl gauge change quickly as we add more levels. The level 12 extrapolation gives us $t^{(\inf,12)}=0.5428$, which differs from the maximal level extrapolation $t^{(\inf,24)}=0.5454$ already at the third digit. The general trend is that if we use more data points, we get closer to the analytic results. Other coefficients exhibit a similar type of behavior. In \cite{ArroyoKudrna}, we analyze the behavior of the tachyon coefficient in more detail and we have tried to deal with this instability by further extrapolation of the extrapolations. This method gives us a better agreement with the analytic solution, but the results depend more on details of the extrapolation technique, so their reliability is questionable. In the end, we are inclined to believe that the numerical solution converges towards the analytic one, but more data will be needed to prove it without a doubt.

We suspect that the source of these problems is that the gauge condition couples coefficients at different levels. We have made some experiments with other linear $b$ gauges and they suggest that if the coupling becomes too strong, the tachyon vacuum solution gains a stronger level dependence and it becomes difficult to extrapolate. However, more research will be needed to confirm this conjecture.

\begin{table}
\centering
\begin{tabular}{|l|ll|l|}\hline
Level    &  \ps Energy      & $\ps E_0$       & $\ps \Delta_S$   \\\hline
2        & $\ps 0.04062340$ & $\ps 0.1101382$ & $\ps 0.03332993$ \\
4        & $\ps 0.00534810$ & $\ps 0.0683413$ & $\ps 0.01648496$ \\
6        & $   -0.00398377$ & $\ps 0.0478479$ & $\ps 0.01026732$ \\
8        & $   -0.00711028$ & $\ps 0.0341758$ & $\ps 0.00719823$ \\
10       & $   -0.00818976$ & $\ps 0.0272661$ & $\ps 0.00541301$ \\
12       & $   -0.00846627$ & $\ps 0.0209872$ & $\ps 0.00425955$ \\
14       & $   -0.00839679$ & $\ps 0.0178483$ & $\ps 0.00345867$ \\
16       & $   -0.00817301$ & $\ps 0.0143618$ & $\ps 0.00287305$ \\
18       & $   -0.00788275$ & $\ps 0.0126889$ & $\ps 0.00242792$ \\
20       & $   -0.00756860$ & $\ps 0.0105109$ & $\ps 0.00207934$ \\
22       & $   -0.00725184$ & $\ps 0.0095251$ & $\ps 0.00179983$ \\
24       & $   -0.00694318$ & $\ps 0.0080520$ & $\ps 0.00157136$ \\\hline
$\infty$ & $\ps 0.0005    $ & $   -0.0015   $ & $   -0.0003    $ \\
$\sigma$ & $\ps 0.0002    $ & $\ps 0.0003   $ & $\ps 0.0001    $ \\\hline
\end{tabular}
\caption{Energy, Ellwood invariant and $\Delta_S$ of the tachyon vacuum solution in Schnabl gauge up to level 24. The last two lines show the infinite level extrapolations of there quantities and their errors. The errors of the energy and $\Delta_S$ are estimated as in Siegel gauge, that is by variation of the maximal level of data used for the extrapolations.}
\label{tab:TV Schnabl}
\end{table}

\begin{table}
\centering
\begin{tabular}{|l|l|l|}\hline
Level      & $t_{Schnabl}$ & $t_{Siegel}$ \\\hline
2          & 0.5442042     & 0.544204     \\
4          & 0.5489385     & 0.548399     \\
6          & 0.5483151     & 0.547932     \\
8          & 0.5473219     & 0.547052     \\
10         & 0.5465084     & 0.546261     \\
12         & 0.5458944     & 0.545608     \\
14         & 0.5454352     & 0.545075     \\
16         & 0.5450892     & 0.544637     \\
18         & 0.5448260     & 0.544272     \\
20         & 0.5446241     & 0.543964     \\
22         & 0.5444683     & 0.543702     \\
24         & 0.5443477     & 0.543476     \\
26         &               & 0.543280     \\
28         &               & 0.543107     \\
30         &               & 0.542955     \\\hline
$\inf$     & 0.5454        & 0.540493     \\\hline
Analytic   & 0.553466      &              \\\hline
\end{tabular}
\caption{Tachyon coefficient of the tachyon vacuum solution in Schnabl gauge and in Siegel gauge.}
\label{tab:TV tachyon Schnabl}
\end{table}


\section{Solutions without gauge fixing}\label{sec:universal:no gauge}
In the final section of this chapter, we discuss the possibility of solving the OSFT equations without any gauge fixing condition following \cite{KudrnaUniversal}. This is potentially an interesting approach, because we do not need to worry about the out-of-gauge equations (\ref{equations remaining}), but we will show that it is not very practical.

\begin{figure}
\centering
\includegraphics[width=14cm]{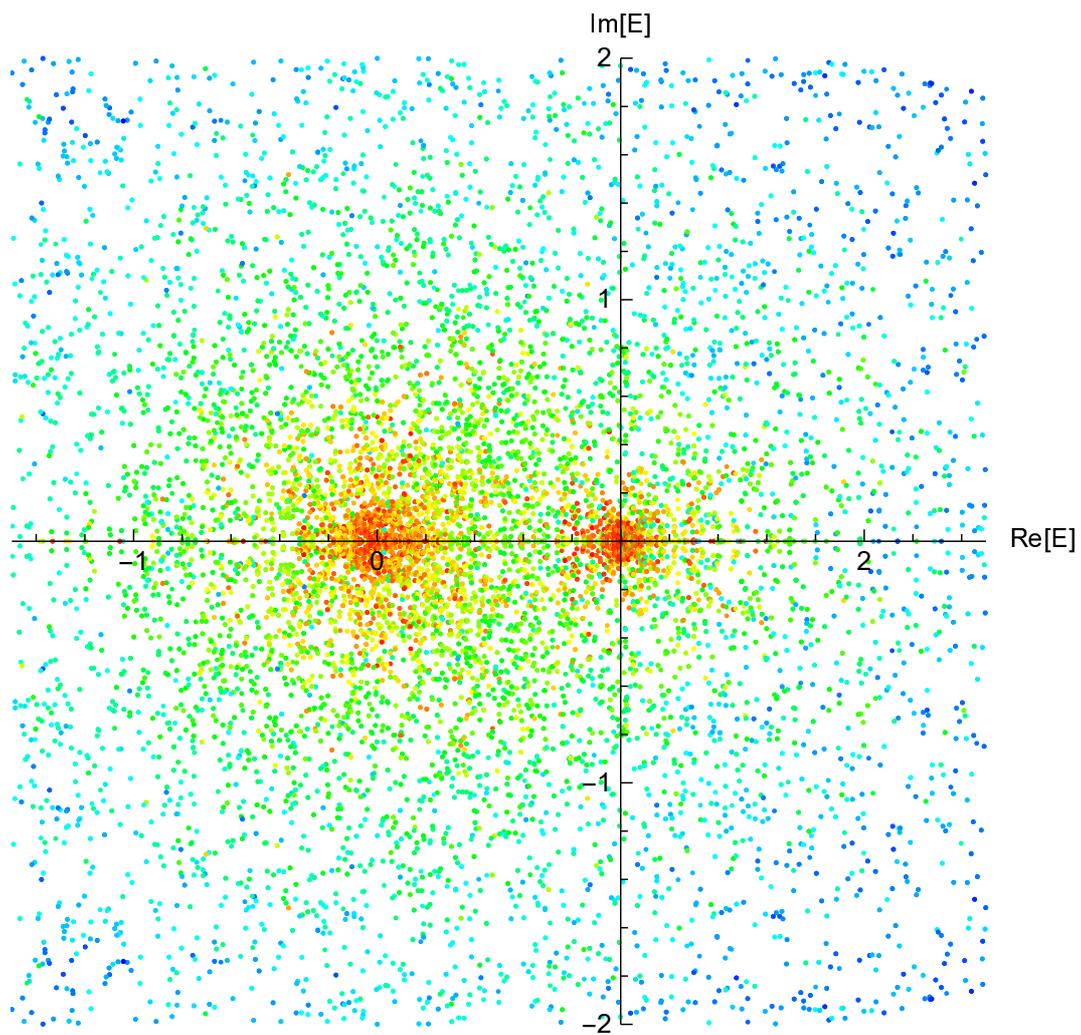}
\caption{Energy of twist non-even universal solutions without gauge fixing at level 4.}\label{fig:universal no gauge2}
\end{figure}

As we already mentioned in section \ref{sec:SFT:basic:eom}, the gauge symmetry (\ref{SFT symmetry}) is broken by the level truncation approximation. Therefore the truncated equations of motion have only a discreet set of solutions, which can be found numerically. Since there are no gauge fixing conditions, the state space is larger than in Siegel gauge. Therefore, using the homotopy continuation method, we have managed to solve the full equations of motion only up to level 4, both with and without imposing the twist even condition\footnote{Finding all solutions at level 6 is impossible with the computer resources we have access to. Even in the twist even case, there are approximately $2^{43}\cong 8.8\times 10^{12}$ solutions and we estimate the required CPU time to be around 500000 years.}. In figures \ref{fig:universal no gauge1} and \ref{fig:universal no gauge2}, we plot the energy of solutions at level 4 in complex plane. We observe that the number of solutions is much higher than in figure \ref{fig:universal sol}, even though we are at a lower level and the figures cover much smaller area. Since we have no better criteria, we distinguish the 'quality' of solutions using the absolute value of the difference between the two gauge invariant observables, the energy and the $E_0$ invariant. The best solutions are denoted by red color, the worst by blue.

There are far more twist non-even solutions than twist even solution (figure \ref{fig:universal no gauge1} contains approximately 750 points, while figure \ref{fig:universal no gauge2} contains approximately 17000 points), but the conclusion from both cases is the same. We observe a reddish cluster of solutions around $E=0$, which represents the tachyon vacuum, and another cluster around $E=1$, which represents the perturbative vacuum. The existence of these clusters is most likely caused by remnants of the gauge symmetry. Even though the gauge symmetry is broken, the OSFT action still contains approximately flat directions, along which we find many solutions with similar properties.

Interestingly, the exact perturbative vacuum solution $\Psi=0$ has a nontrivial multiplicity. In the twist even case at level 2, we have confirmed analytically that it has multiplicity 3. We estimate that the multiplicity at level 4 is over 200, but we are not sure about the exact number due to problems with numerical stability of the homotopy continuation method around the perturbative vacuum. As far as we know, this is the only case of an OSFT solution having a multiplicity higher than 1.

Unfortunately, these results do not allow us to make any decisive claims about existence of other solutions, like multi-branes or ghost branes. We do not observe any clusters that would correspond to such solutions, but it is likely that these solutions are not very precise at low levels (see for example the solutions from \cite{KudrnaUniversal} or positive energy lumps from section \ref{sec:FB circle:single:smallR}) and we may not be able to recognize them among many nonphysical solutions, which appear pretty much everywhere in the complex plane.

By now, the reader may wonder why we have not shown any solutions beyond level 4. The reason is that Newton's method does not work well in this setting. When we try to improve a solution from one level to the next one, we find that in most cases, Newton's method does not converge within a reasonable number of iterations, and when it does, the solution we find is usually very different from the initial one. We suspect that these problems are caused by the almost flat directions in the potential, but we have not been able to pinpoint their exact source in the numerical algorithm. In \cite{KudrnaUniversal}, we made some attempts to go around this problem using modifications of the homotopy continuation method, but we have never found a consistent way to improve solutions to higher levels.

The conclusion is that we can look for solutions without any gauge fixing condition only using the homotopy continuation method. That restricts the available level and it makes this approach much less effective that when we make a gauge choice.

\begin{figure}
\centering
\includegraphics[width=10cm]{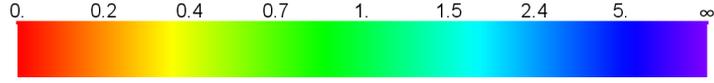}
\caption{Colors corresponding to the absolute value of the difference between the energy and the Ellwood invariant for solutions in figures \ref{fig:universal no gauge1} and \ref{fig:universal no gauge2}.}\label{fig:colors}
\end{figure}

\begin{figure}
\centering
\includegraphics[width=12cm]{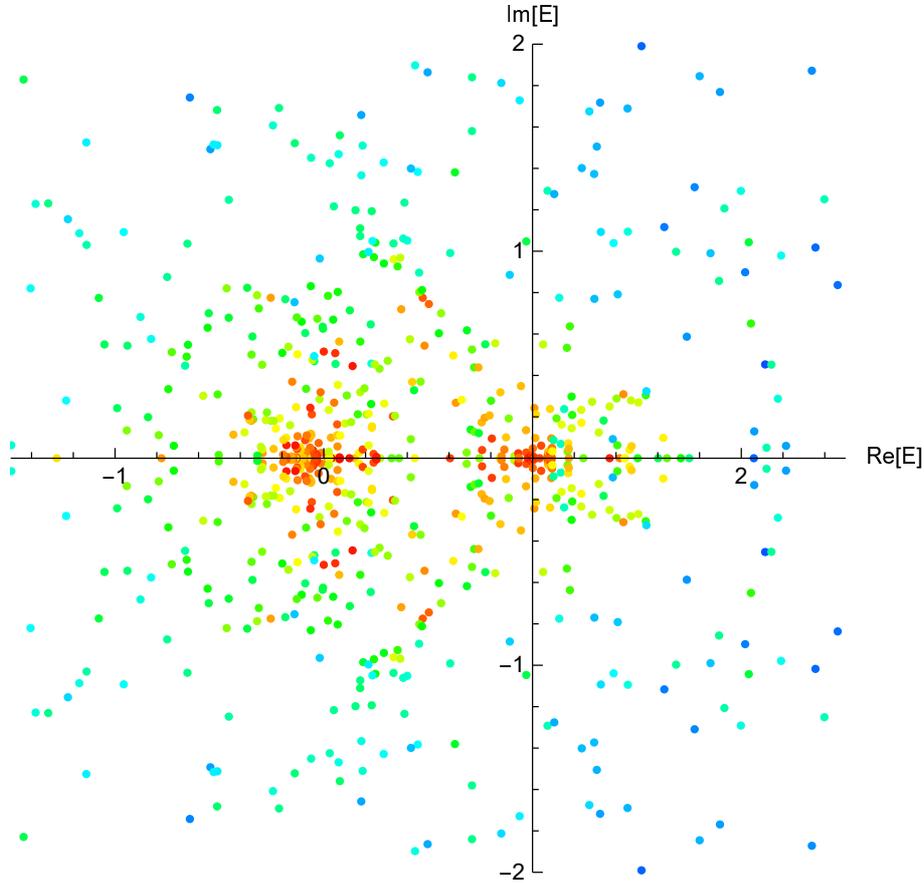}
\caption{Energy of twist even universal solutions without gauge fixing at level 4. The color of points depends on the difference between the energy and the Ellwood invariant, see figure \ref{fig:colors}. The best solutions are red, the worst are blue. }\label{fig:universal no gauge1}
\end{figure}

\chapter{Results - Free boson on a circle}\label{sec:FB circle}
In this chapter, we study string field theory describing free boson compactified on a circle. We choose the Neumann boundary conditions as the initial setting, which means that the open string background is a D1-brane wrapped around the circle. Thanks to the T-duality, we can make this choice without loss of generality. Most of known solutions in this theory describe D0-branes and they are known as lump solutions.

We have two main goals in this chapter. First, we use lump solutions to illustrate generic properties of OSFT solutions. Universal solutions, which were discussed in the previous chapter, are not convenient for this purpose because they have only a very limited set of gauge invariant observables and the tachyon vacuum is the only well-behaved solution among them. Our second goal is to make a systematic scan of solutions at different radii, to identify basic types of solutions and to discuss their properties and radius dependence.

Let us start by providing some references. First single lump solutions were found in \cite{LumpsNDHarvey}\cite{LumpsKoch} by solving differential equations for the tachyon field. First solutions in the usual level truncation scheme appeared in \cite{MSZ lump} at level (3,6). In \cite{LumpsBeccaria}, the level of calculations was improved to (8,16) by using the U(1) basis in the free boson theory instead of the Virasoro basis. This article also provides the first infinite level extrapolations. The reference \cite{KMS} made another improvement of level to 12. It also introduced new gauge invariants, which can be used for better identification of solutions, and discussed some double lump solutions. We should also mention \cite{LumpSiegEq}, where the validity of Siegel gauge is tested for the MSZ lump solution from \cite{MSZ lump}, and \cite{OhmoriLumps}, which provides lump solutions in several different gauges and attempts to find spectrum of fluctuations around these solutions.

Attempts to find analytic lumps solution were made in \cite{LumpAnalytic1}\cite{LumpAnalytic2}\cite{LumpAnalytic3}\cite{LumpAnalytic4}, but it seem that these solutions violate some parts of the equations of motion \cite{LumpAnalytic5}. Therefore the first analytic lump solution free of any problems is the Erler-Maccaferri solution \cite{ErlerMaccaferri}\cite{ErlerMaccaferri2}\cite{ErlerMaccaferri3}, which is constructed using boundary condition changing operators.

Free boson solutions have been discussed in many papers, but most of the references concentrate only on single lump solutions for $R>1$ and a systematic analysis is missing. We would like to change that, so in this thesis we study solutions at a much larger range of radii using seeds found by the homotopy continuation method. This allows us to make a classification of basis types of solutions in the free boson theory. The results can be summarized as follows:
The simplest non-universal solutions are single lump solutions, which describe a D0-brane positioned somewhere on the circle. Since we impose the parity even condition on the string field, the position of the D0-brane is restricted to $0$ or $\pi R$. So far, these solutions have been studied only for radii greater than 1, where they are real and well-behaved. We additionally investigate radii $R<1$ and we show that there are single lumps as well, but they are harder to find and they suffer from various problems, most notably they have nonzero imaginary part. Above $R=2$, we can find well-behaved double lump solutions, which describe two D0-branes positioned symmetrically around the origin. This configuration of two D0-branes has one free parameter, the distance between the D0-branes, so we inspect how is this modulus reproduced by OSFT solutions. If we consider even larger radii, we can find triple lumps, quadruple lumps, etc., but these solutions do not seem to have any new interesting properties, so we will not discuss them. In addition to lump solutions, we have found some exceptional solutions which can be identified as D1-branes with Wilson lines or as some specific marginal deformations. These solutions usually have some pathological properties and therefore we cannot say with certainty whether they are physical or not.

There is also a different class of free boson solutions that describes marginal deformations, but we postpone the discussion of this topic to chapter \ref{sec:marginal}.

\section{MSZ lump solution at $R=\sqrt{3}$}\label{sec:FB circle:MSZ}
We begin the presentation of results and detailed analysis of one single lump solution. We choose the lump solution at radius $R=\sqrt{3}$, which was originally found by Moeller, Sen and Zwiebach \cite{MSZ lump}. Their motivation for choosing an irrational radius was to avoid null states in Virasoro Verma modules over momentum primaries, but rationality of the radius does not matter in our basis. The original solution was constructed up to level $(3,6)$, later it was improved to level 12 using a Mathematica code \cite{KMS} and now we have managed to reach level 18 using a parallel C++ code. Compared to the older references, we add more gauge invariants, which describe (with limitations) the boundary state corresponding to the solution. We also compute the first out-of-Siegel equation as a consistency check.

\subsection{Observables}\label{sec:FB circle:MSZ:observables}

\begin{table}
\centering
\footnotesize
\hspace*{-0.25cm}
\vspace{0.2cm}
\begin{tabular}{|l|l|l|lll|ll|}\hline
Level    & Energy   & $\ps \Delta_S$  & $\ps D_1     $ & $\ps D_4    $ & $\ps D_9    $ & $\ps W_1     $ & $\ps W_2    $ \\\hline
2        & 1.094279 & $\ps 0.0294775$ & $   -1.013532$ & $   -7.32693$ & $   -17.8493$ & $   -0.487415$ & $   -5.22247$ \\
3        & 1.060535 & $\ps 0.0115445$ & $   -1.110775$ & $   -7.62362$ & $   -18.4784$ & $   -0.568038$ & $   -5.45267$ \\
4        & 1.035718 & $\ps 0.0067701$ & $   -0.752479$ & $\ps 9.29952$ & $\ps 80.5738$ & $\ps 0.370281$ & $\ps 14.9784$ \\
5        & 1.029357 & $\ps 0.0052315$ & $   -0.779229$ & $\ps 9.55604$ & $\ps 82.5396$ & $\ps 0.371251$ & $\ps 15.4396$ \\
6        & 1.020897 & $\ps 0.0036624$ & $   -0.945165$ & $   -29.0514$ & $   -471.005$ & $   -0.182729$ & $   -33.1531$ \\
7        & 1.018679 & $\ps 0.0031079$ & $   -0.959492$ & $   -30.2237$ & $   -488.474$ & $   -0.194398$ & $   -34.3871$ \\
8        & 1.014542 & $\ps 0.0024055$ & $   -0.909528$ & $\ps 68.8275$ & $\ps 2774.33$ & $\ps 0.136742$ & $\ps 57.2392$ \\
9        & 1.013510 & $\ps 0.0021156$ & $   -0.913363$ & $\ps 71.0525$ & $\ps 2859.65$ & $\ps 0.139243$ & $\ps 59.1281$ \\
10       & 1.011082 & $\ps 0.0017399$ & $   -0.963774$ & $   -141.185$ & $   -12871.4$ & $   -0.085347$ & $   -89.1121$ \\
11       & 1.010515 & $\ps 0.0015617$ & $   -0.966875$ & $   -144.871$ & $   -13210.2$ & $   -0.087330$ & $   -91.6589$ \\
12       & 1.008926 & $\ps 0.0013368$ & $   -0.945928$ & $\ps 265.863$ & $\ps 54043.9$ & $\ps 0.069163$ & $\ps 130.438$ \\
13       & 1.008578 & $\ps 0.0012164$ & $   -0.947155$ & $\ps 271.742$ & $\ps 55272.4$ & $\ps 0.070245$ & $\ps 133.646$ \\
14       & 1.007459 & $\ps 0.0010709$ & $   -0.972895$ & $   -461.998$ & $   -206883 $ & $   -0.048005$ & $   -180.059$ \\
15       & 1.007229 & $\ps 0.0009842$ & $   -0.973982$ & $   -470.859$ & $   -210984 $ & $   -0.048488$ & $   -183.914$ \\
16       & 1.006400 & $\ps 0.0008846$ & $   -0.961803$ & $\ps 764.747$ & $\ps 725405 $ & $\ps 0.041376$ & $\ps 236.018$ \\
17       & 1.006239 & $\ps 0.0008193$ & $   -0.962333$ & $\ps 777.639$ & $\ps 738123 $ & $\ps 0.041950$ & $\ps 240.492$ \\
18       & 1.005600 & $\ps 0.0007481$ & $   -0.978161$ & $   -1205.82$ & $   -2354459$ & $   -0.030433$ & $   -296.741$ \\\hline
$\inf$   & 0.999984 & $   -0.000010 $ & $   -1.0006  $ &               &               & $   -0.0007  $ &               \\
$\sigma$ & 0.000003 & $\ps 0.000003 $ & $\ps 0.0028  $ &               &               & $\ps 0.0045  $ &               \\\hline
Exp.     & 1        & $\ps 0        $ & $   -1       $ & $\ps 1      $ & $    -1     $ & $\ps 0       $ & $\ps 0      $ \\\hline
\end{tabular}
\hspace*{-0.3cm}
\begin{tabular}{|l|l|lllllll|}\hline
Level    & $E_0$   & $E_1$    & $E_2$    & $E_3$   & $\ps E_4    $ & $\ps E_5    $ & $\ps E_6    $ & $E_7$   \\\hline
2        & 1.09094 & 0.830804 & 1.03277  & 0       & $\ps 0      $ & $\ps 0      $ & $\ps 0      $ & 0       \\
3        & 1.06017 & 0.905713 & 1.08758  & 1.36793 & $\ps 0      $ & $\ps 0      $ & $\ps 0      $ & 0       \\
4        & 1.04623 & 0.918393 & 0.931471 & 1.41220 & $\ps 0      $ & $\ps 0      $ & $\ps 0      $ & 0       \\
5        & 1.03948 & 0.940750 & 0.933722 & 0.67817 & $\ps 0      $ & $\ps 0      $ & $\ps 0      $ & 0       \\
6        & 1.02921 & 0.946315 & 0.995166 & 0.67660 & $\ps 2.06251$ & $\ps 0      $ & $\ps 0      $ & 0       \\
7        & 1.02668 & 0.956761 & 0.996584 & 1.11184 & $\ps 2.11211$ & $\ps 0      $ & $\ps 0      $ & 0       \\
8        & 1.02301 & 0.959784 & 0.977037 & 1.12839 & $   -0.32773$ & $\ps 0      $ & $\ps 0      $ & 0       \\
9        & 1.02171 & 0.965702 & 0.976881 & 0.85903 & $   -0.35068$ & $\ps 3.66745$ & $\ps 0      $ & 0       \\
10       & 1.01787 & 0.967666 & 0.993958 & 0.86096 & $\ps 1.98806$ & $\ps 3.81063$ & $\ps 0      $ & 0       \\
11       & 1.01708 & 0.971569 & 0.993933 & 1.04829 & $\ps 2.01551$ & $   -4.09339$ & $\ps 0      $ & 0       \\
12       & 1.01549 & 0.972933 & 0.986990 & 1.05428 & $   -0.03537$ & $   -4.26896$ & $\ps 8.53484$ & 0       \\
13       & 1.01496 & 0.975609 & 0.986782 & 0.92037 & $   -0.04933$ & $\ps 7.35754$ & $\ps 8.69427$ & 0       \\
14       & 1.01297 & 0.976629 & 0.994649 & 0.92195 & $\ps 1.76507$ & $\ps 7.58467$ & $   -20.8014$ & 0       \\
15       & 1.01259 & 0.978671 & 0.994516 & 1.02444 & $\ps 1.78043$ & $   -6.94654$ & $   -21.1687$ & 0       \\
16       & 1.01172 & 0.979458 & 0.991005 & 1.02741 & $\ps 0.20548$ & $   -7.17135$ & $\ps 42.4557$ & 0       \\
17       & 1.01143 & 0.980980 & 0.990842 & 0.94802 & $\ps 0.19717$ & $\ps 9.67676$ & $\ps 43.1319$ & 27.0904 \\
18       & 1.01021 & 0.981611 & 0.995361 & 0.94919 & $\ps 1.59225$ & $\ps 9.91326$ & $   -67.1618$ & 27.5373 \\\hline
$\inf$   & 1.00020 & 0.9992   & 0.9995   & 0.998   & $\ps 0.99   $ & $\ps 1.9    $ &               &         \\
$\sigma$ & 0.00007 & 0.0002   & 0.0002   & 0.008   & $\ps 0.44   $ & $\ps 19     $ &               &         \\\hline
Exp.     & 1       & 1        & 1        & 1       & $\ps 1      $ & $\ps 1      $ & $\ps 1      $ & 1       \\\hline
\end{tabular}
\caption{Gauge invariant observables and the first out-of-Siegel equation $\Delta_S$ of the MSZ lump solution at radius $R=\sqrt 3$. The last three lines show extrapolations of observables to the infinite level (when possible), estimated errors of the extrapolations and the expected results.}
\label{tab:FB MSZ lump}
\end{table}

First, we should compare observables of the MSZ lump solution to the boundary state of a D0-brane. Table \ref{tab:FB MSZ lump} shows its observables up to level 18, their infinite level extrapolations and the predicted values. In this subsection, we focus only on results of the extrapolations, we will analyze how they work in more detail in the next subsection.

The energy of a D0-brane equals to 1 in our conventions. The energy of the solution at level 18 differs from the expected value approximately by $0.6\%$. That is not a bad result, but the infinite level extrapolation is far more accurate. It reproduces the expected value with an error lesser than $0.002\%$. As for the tachyon vacuum solution, the estimated error of the extrapolation is several times smaller than the actual difference. This seems to be a general feature of this type of error estimate, so we do not need to be concerned. An alternative way to measure the energy of the solution is the $E_0$ invariant. The asymptotic value of this invariant differs from 1 approximately by $0.02\%$, the precision is slightly worse than in case of the energy measured by the action, but it is still very good.

Next, we take a look at the $D_1$ invariant, which distinguishes whether the solution corresponds to Neumann or Dirichlet boundary conditions. For a D0-brane, which has Dirichlet boundary conditions, this invariant should be equal to minus the energy. The numerical values of the $D_1$ invariant approach $-1$, but, unlike for the $E_0$ invariant, the sequence is not monotonous and the precision at finite levels is only few percent. However, the infinite level extrapolation agrees with the expected value to three digits.

We have also computed two more invariants build using the exceptional primaries, $D_4$ and $D_9$, but these invariants oscillate with an increasing amplitude, which makes them essentially useless. We will discuss their behavior in more detail later in section \ref{sec:FB circle:single:pade}, where we will try to improve them using the \Pade approximant.

Another invariants that can distinguish between Neumann and Dirichlet boundary conditions are $W_n$ invariants. Unfortunately, only the $W_1$ invariant is convergent, the $W_2$ invariant oscillates wildly (similarly to $D_4$) and all higher invariants as well. These invariants should be equal to zero and the extrapolated value of $W_1$ indeed agrees with the expected value to three decimal places.

Finally, we get to $E_n$ invariants, which determine the position of the D0-brane. They should be equal to (\ref{En predicted}). From the data in table \ref{tab:FB MSZ lump}, we can see that their precision decreases with growing $n$, but all convergent invariants are close to 1. That means that the lump is located at the position $x_0=0$. Our ansatz admits one more lump solution, which differs from this one by signs of odd $E_n$ invariants and which corresponds to a D0-brane located at $x_0=\pi R$. If we want to describe a D0-brane at some other position, we can easily move the lump using the translation operator $e^{i p\Delta x}$.

We conclude that convergent observables agree very well with the expected boundary state. The out-of-Siegel equation $\Delta_S$ approaches zero, so the solution passes this consistency check as well and we have no doubts that it really describes a D0-brane.

Before me move on, we would like to point out two more properties of Ellwood invariants, which apply to essentially all OSFT solutions. First, we observe that values of invariants at any odd level are close to values on the previous even level\footnote{To a lesser extend, similar behavior also applies to other quantities, like the energy or coefficients of the string field. Larger changes of these quantities every two levels must have a different origin; we think that they are caused by appearance of new descendants of the zero momentum tachyon field.} (or on the next even level if the given invariant first appears at an odd level). This behavior is easy to understand. When we improve a solution by one level, its invariants are affected in two ways. First, new coefficients pass the level cut-off, and second, the presence of these coefficients changes coefficients at lower levels. The change of lower level coefficients is usually quite small and does not affect the invariants very much, so contributions from the new coefficients are dominant, see table \ref{tab:FB invariants z} for illustration. Since we usually consider twist even solutions, new descendants of a given primary appear every two levels. This explains the jumps of invariants every two levels. Furthermore, we find that the new contributions usually have alternating signs, which leads to oscillations with a period of 4 levels.

The second observation is that the behavior of Ellwood invariants, which varies from monotonous convergence to wild oscillations with exponentially increasing amplitude, is not random, but it is connected to properties of the corresponding Ellwood states and to some properties of solutions. We can deduce from table \ref{tab:FB MSZ lump} that the amplitude of oscillations grows with weights of invariants. We remind the reader that we define the weight of an invariant as weight of the corresponding vertex operator in the non-universal part of the matter theory. By studying lump solutions on a torus, we can make this conjecture more precise, the amplitude of oscillations of a given invariant depends on its weight in sectors where the solution is non-universal. We will discuss this behavior more in section \ref{sec:FB circle:single:pade}.

\subsection{Extrapolations to infinite level}\label{sec:FB circle:MSZ:extrapolations}
In this subsection, we use the MSZ lump to demonstrate how the extrapolation techniques described in section \ref{sec:Numerics:observables:extrapolation} work and to justify their choice.

\begin{figure}[!b]
   \centering
   \begin{subfigure}{0.47\textwidth}
      \includegraphics[width=\textwidth]{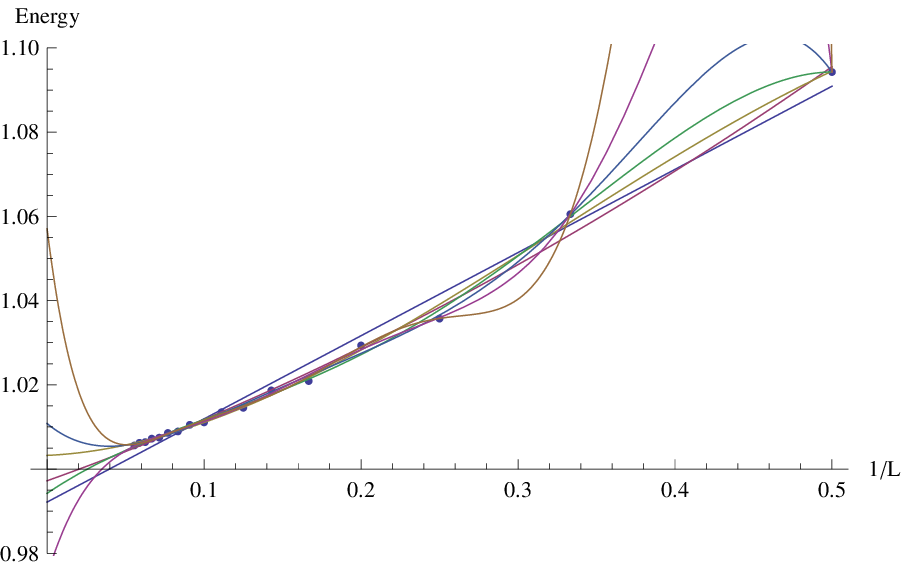}
   \end{subfigure}\qquad
   \begin{subfigure}{0.47\textwidth}
      \includegraphics[width=\textwidth]{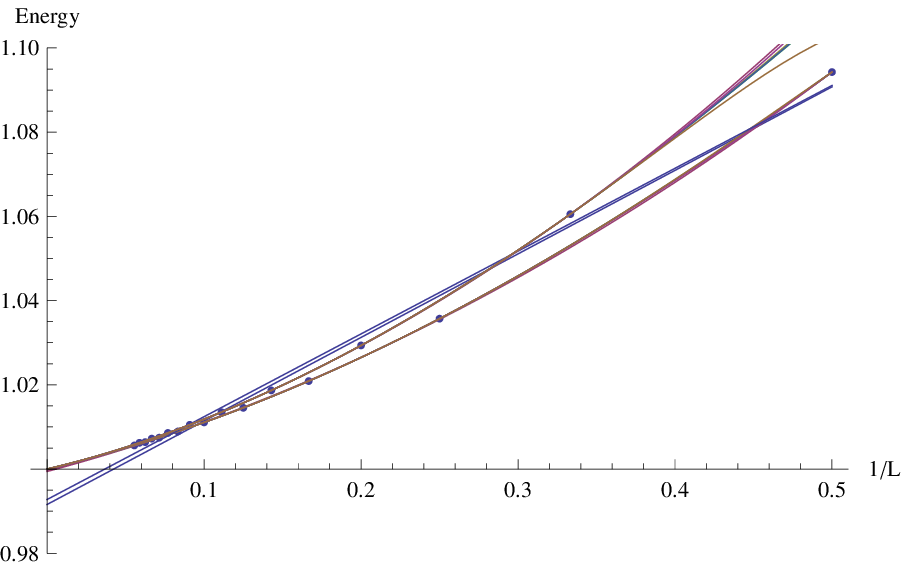}
   \end{subfigure}
\caption{Extrapolations of the energy of the MSZ lump solution using polynomials of different orders. In the left figure, we extrapolate all levels together, while we extrapolate even and odd levels separately in the right figure.}\label{fig:MSZ energy extrapolations}
\end{figure}

As usual, we start with the energy. It goes monotonically towards the correct value, so, in figure \ref{fig:MSZ energy extrapolations} on the left, we naively attempt to extrapolate all data points using polynomials in $1/L$ of different orders. It is immediately clear that this is not a viable option. The results are reasonable for low orders (but not very precise), but as we increase the order, the extrapolating functions begin to oscillate and they do not capture the actual trend at all. This problem was not noticed in case of the tachyon vacuum solution because universal solutions change only every two levels, but it concerns most non-universal solutions. The solution is simple, we must extrapolate even and odd levels independently. See figure \ref{fig:MSZ energy extrapolations} on the right, where we show that extrapolating functions in this scheme (from order 2) go very close to the correct value.

\begin{table}[!t]
\centering
\begin{tabular}{|l|l|l|}\hline
order & even levels & odd levels \\\hline
1     & 0.991590    & 0.992768   \\
2     & 0.999429    & 0.999610   \\
3     & 1.000051    & 1.000008   \\
4     & 0.999972    & 0.999969   \\
5     & 0.999960    & 0.999967   \\
6     & 0.999974    & 0.999975   \\
7     & 0.999982    & 0.999982   \\
8     & 0.999987    & -          \\\hline
\end{tabular}
\caption{Extrapolations of the energy of the MSZ lump solution using polynomials of different orders.}
\label{tab:MSZ energy extrapolations}
\end{table}

\begin{figure}[!]
   \begin{subfigure}{0.47\textwidth}
      \includegraphics[width=\textwidth]{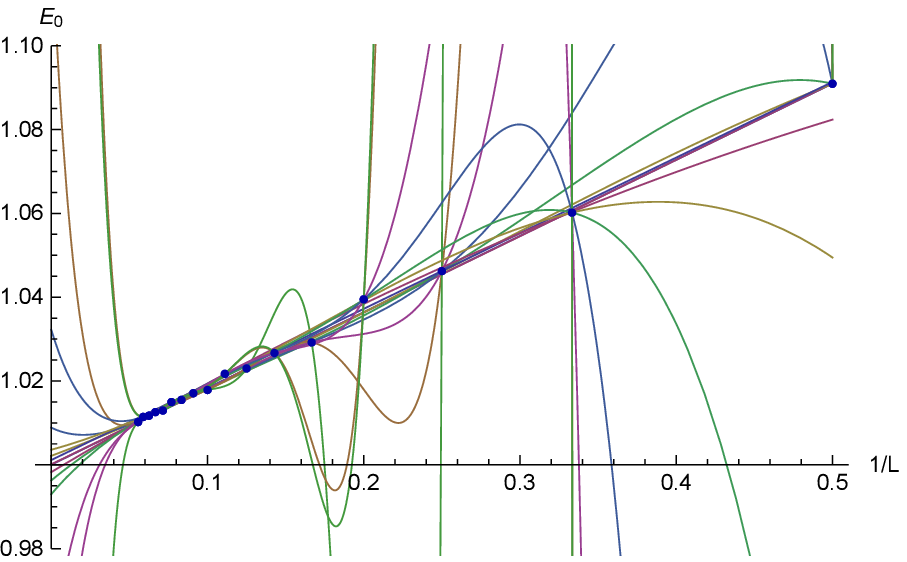}
   \end{subfigure}\qquad
   \begin{subfigure}{0.47\textwidth}
      \includegraphics[width=\textwidth]{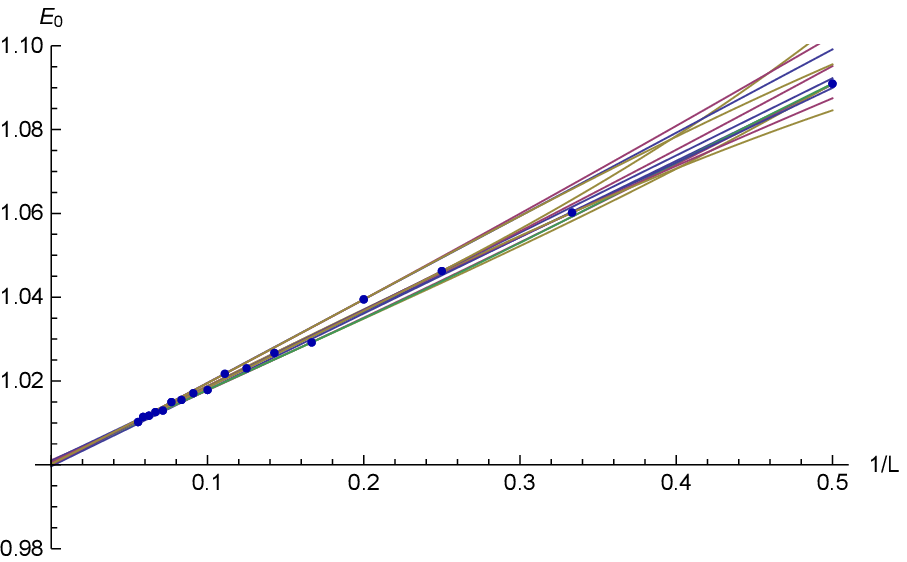}
   \end{subfigure}
\caption{Extrapolations of the $E_0$ invariant of the MSZ lump solution using polynomials of different orders. On the left, we show extrapolations using levels modulo 2, which behave very chaotically. On the right, we show extrapolations using levels modulo 4, all of which go close to a single point.}
\label{fig:MSZ E0 extrapolations}
\end{figure}

Once it is clear that we have to work with even and odd levels independently, we can investigate what are the best orders of extrapolating functions. Table $\ref{tab:MSZ energy extrapolations}$ shows how the extrapolated values of the energy depend on the order. There are some irregularities, but the general trend is obvious: The higher the order, the better results we get. This is why we use maximal order extrapolations, which directly interpolate all data points. The reader may notice that the polynomials in this case have different orders for even and odd levels, but this does not seem to cause any problems. The same extrapolation technique also works well also for $\Delta_S$ and for the coefficients in table \ref{tab:MSZ tachyon}.

\begin{figure}[!]
\centering
\includegraphics[width=12cm]{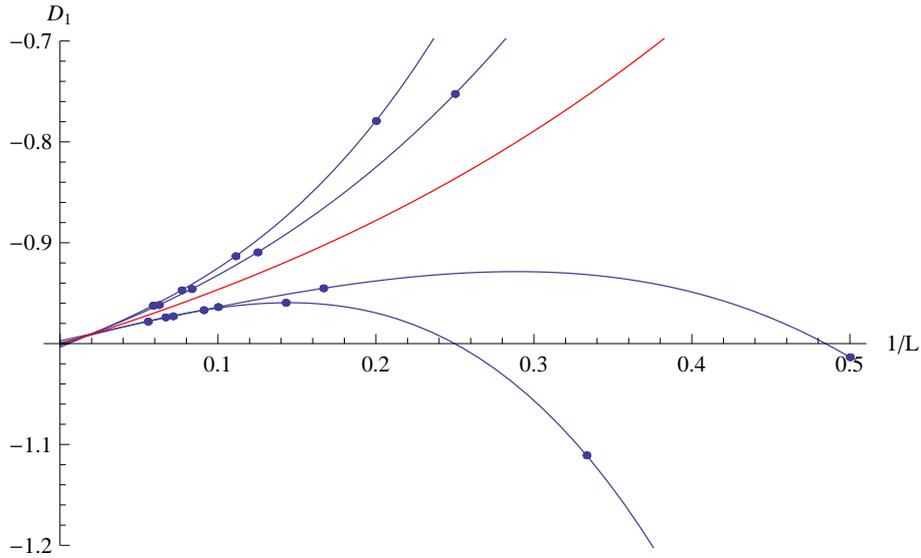}
\caption{Extrapolations of the $D_1$ invariant of the MSZ lump solution. The four blue lines are extrapolations of maximal order using levels modulo 4. The red line is the average of these four curves, which is probably the best approximation of the function around which the data points oscillate.}
\label{fig:MSZ D1 extrapolation}
\end{figure}

Next, we focus on extrapolations of Ellwood invariants, starting with $E_0$. Based on the experience with the energy, we do not even attempt to use all data points at once, but the left part of figure \ref{fig:MSZ E0 extrapolations} shows that even independent extrapolations of even and odd levels fail. In order to get consistent results, we have to increase the interval between levels to 4, see figure \ref{fig:MSZ E0 extrapolations} on the right. The reason is that Ellwood invariants oscillate with a period of 4 levels. The $E_0$ invariant is monotonously decreasing, which means that its oscillations are almost invisible, but from figure \ref{fig:MSZ D1 extrapolation}, where we show extrapolations of the $D_1$ invariant, it is obvious that the data points follow 4 different curves. When it comes to the order of extrapolating functions, we once again find that the highest possible order usually gives us the best results.

\begin{figure}[!]
\centering
\includegraphics[width=12cm]{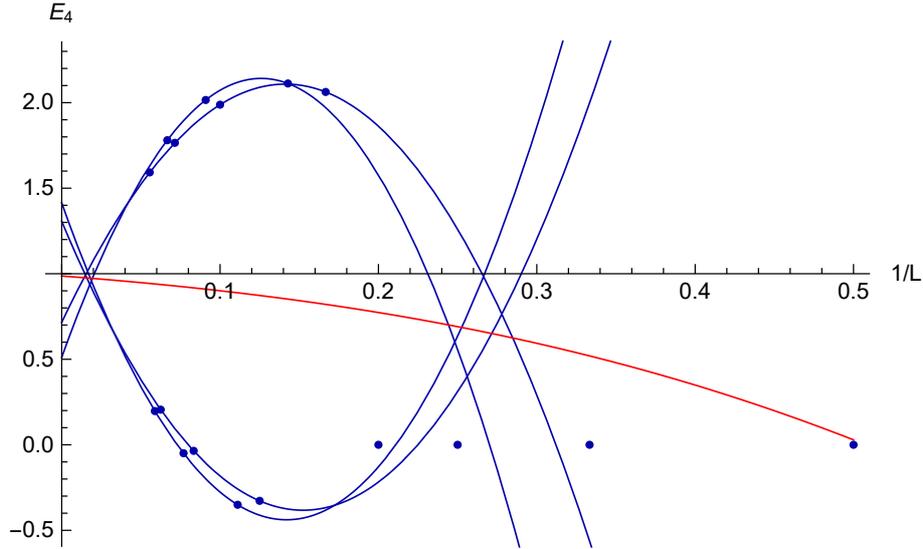}
\caption{Extrapolations of the $E_4$ invariant of the MSZ lump solution. The four blue lines represent extrapolations modulo 4 levels (we use only the nontrivial data points from level 6 above) and the red line is their average. Notice that the four level infinity values are positioned roughly symmetrically around the expected value. This leads to an unrealistically high error estimate.}
\label{fig:MSZ E4 extrapolation}
\end{figure}

Properties of error estimates of extrapolations highly depend on amplitude of the oscillations. Errors tend to be too small for mildly oscillating invariants because all independent extrapolations go above or below the correct value. On the contrary, errors are usually overestimated for highly oscillating invariants. See figure \ref{fig:MSZ E4 extrapolation}, where we extrapolate the $E_4$ invariant with the result $0.99\pm 0.44$. We can see that the estimated error is one order above the actual error. The reason is that the four independent extrapolations give us values which are positioned symmetrically around the expected result. Therefore the average of these values is quite precise, but the error given by the standard deviation is too big. Therefore the errors presented in this thesis cannot be taken too seriously and they should be understood only as order estimates. One could try to introduce some correction factor by hand, which would depend on weight of the invariant, but so far, we have not found a prescription that would work well enough.

Another generic problem of this extrapolation technique of Ellwood invariants is that the order of extrapolations is quite low, which limits the possible precision of results. Even for this solution, which we have computed to a very nice level 18, the maximal order is 4. In case of the free boson theory on a torus (or other theory with multiple symmetry generators), we cannot reach such high level and we often have to work with just linear or quadratic extrapolations, which leads to large errors. However, out of all extrapolation methods we have tested, this one gives us by far the best results, so we use it despite these issues.


\FloatBarrier
\subsection{Visualization of the lump}\label{sec:FB circle:MSZ:visualization}
Although it is possible to analyze lump solutions just by looking at their invariants, sometimes it is more instructive to plot how they depend on the coordinate $x$ along the circle. There are two quantities we can easily visualize.

\begin{figure}
\centering
\includegraphics[width=10cm]{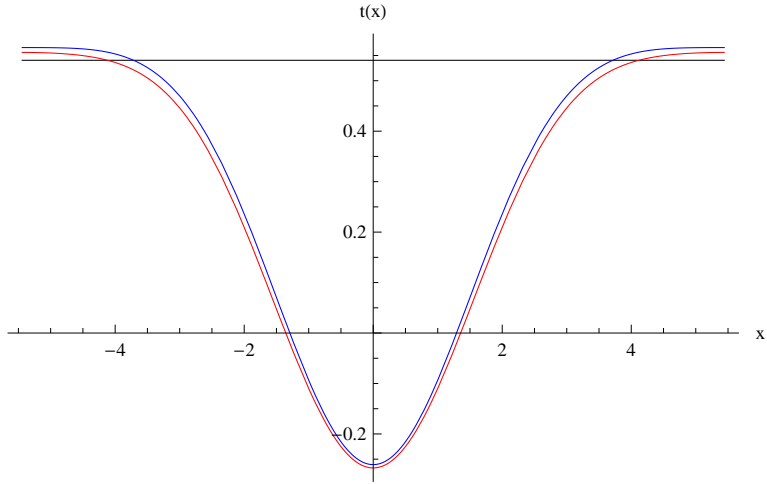}
\caption{Tachyon profile of the MSZ lump. The red curve denotes the profile at level 3, the blue curve the profile extrapolated to level infinity. The black line indicates the extrapolated value of the tachyon field of tachyon vacuum solution, $t_{TV}=0.54049$.}
\label{fig:MSZ tach profile}
\end{figure}

The first one is the tachyon profile, which was already used in the earliest references \cite{LumpsNDHarvey}\cite{LumpsKoch}. Before introduction of the $E_n$ invariants, it was the only option to determine the position of the D0-brane. To define it, we expand the string field as
\begin{equation}
|\Psi\ra=t_0 c_1 |0\ra+\sum_n t_n c_1\cos\frac {n X(0)}{R}|0\ra+\rm{descendants},
\end{equation}
where we singled out the Fourier modes of the tachyon field. Then we define the tachyon profile as
\begin{equation}\label{FB tachyon profile}
t(x)=\sum_n t_n \cos\frac {n x}{R}.
\end{equation}
The coefficients $t_n$ of course depend on level. For illustration, we show the first few coefficients of the MSZ lump solution in table $\ref{tab:MSZ tachyon}$. We observe that all coefficients are very stable and that higher coefficients are strongly suppressed. Therefore the whole profile does not change much with increasing level. In figure \ref{fig:MSZ tach profile}, we compare the level 3 profile $t^{(3)}(x)$ with the extrapolated profile $t^{(\inf)}(x)$ and we observe that the original picture from \cite{MSZ lump} was almost correct.

Interestingly, profiles of lump solutions can be very well approximated by gaussian curves of the form $a+b e^{\frac{-(x-x_0)^2}{2\sigma^2}}$ \cite{MSZ lump} or even better by periodic sums of gaussians $a+b \sum_{n\in \mathbb{Z}} e^{\frac{-(x-x_0-2\pi n R)^2}{2\sigma^2}}$, where the parameters $(a,b,\sigma)$ have only a small radius dependence. Concretely, for the MSZ lump we get
\begin{eqnarray}
a &=& 0.572091, \nn\\
b &=& -0.835713, \\
\sigma &=&1.47538. \nn
\end{eqnarray}
Notice that the parameter $a$ is close to value of the tachyon coefficient of the tachyon vacuum solution, $t_{TV}=0.54049$. Therefore lump solutions can be viewed as a sum of the tachyon vacuum solution and gaussian perturbations. This will be useful for construction of some double lump solutions in section \ref{sec:FB circle:double:superposition}.

\begin{table}
\centering
\footnotesize
\begin{tabular}{|l|llllll|}\hline
Level  & $t_0$    & $\ps t_1$      & $\ps t_2$      & $\ps t_3$       & $\ps t_4$        & $\ps t_5$          \\\hline
2      & 0.266124 & $   -0.384735$ & $   -0.119565$ & $\ps 0        $ & $\ps 0         $ & $\ps 0           $ \\
3      & 0.270292 & $   -0.396013$ & $   -0.125911$ & $   -0.0157120$ & $\ps 0         $ & $\ps 0           $ \\
4      & 0.278498 & $   -0.397604$ & $   -0.129218$ & $   -0.0162205$ & $\ps 0         $ & $\ps 0           $ \\
5      & 0.279641 & $   -0.398460$ & $   -0.130106$ & $   -0.0165435$ & $\ps 0         $ & $\ps 0           $ \\
6      & 0.282447 & $   -0.398276$ & $   -0.131421$ & $   -0.0169807$ & $   -0.00093274$ & $\ps 0           $ \\
7      & 0.282785 & $   -0.398375$ & $   -0.131657$ & $   -0.0170814$ & $   -0.00095517$ & $\ps 0           $ \\
8      & 0.284171 & $   -0.398036$ & $   -0.132328$ & $   -0.0173232$ & $   -0.00097584$ & $\ps 0           $ \\
9      & 0.284285 & $   -0.398020$ & $   -0.132420$ & $   -0.0173674$ & $   -0.00098820$ & $   -0.0000259147$ \\
10     & 0.285109 & $   -0.397713$ & $   -0.132823$ & $   -0.0175187$ & $   -0.00100186$ & $   -0.0000269265$ \\
11     & 0.285146 & $   -0.397677$ & $   -0.132867$ & $   -0.0175421$ & $   -0.00100957$ & $   -0.0000271398$ \\
12     & 0.285692 & $   -0.397422$ & $   -0.133136$ & $   -0.0176457$ & $   -0.00101919$ & $   -0.0000277935$ \\
13     & 0.285699 & $   -0.397385$ & $   -0.133159$ & $   -0.0176596$ & $   -0.00102442$ & $   -0.0000279542$ \\
14     & 0.286088 & $   -0.397175$ & $   -0.133351$ & $   -0.0177352$ & $   -0.00103156$ & $   -0.0000284167$ \\
15     & 0.286083 & $   -0.397143$ & $   -0.133365$ & $   -0.0177441$ & $   -0.00103533$ & $   -0.0000285401$ \\
16     & 0.286374 & $   -0.396968$ & $   -0.133509$ & $   -0.0178017$ & $   -0.00104085$ & $   -0.0000288867$ \\
17     & 0.286364 & $   -0.396940$ & $   -0.133518$ & $   -0.0178077$ & $   -0.00104369$ & $   -0.0000289838$ \\
18     & 0.286590 & $   -0.396794$ & $   -0.133630$ & $   -0.0178532$ & $   -0.00104809$ & $   -0.0000292541$ \\\hline
$\inf$ & 0.288142 & $   -0.394928$ & $   -0.134551$ & $   -0.0182668$ & $   -0.00110720$ & $   -0.0000322792$ \\\hline
\end{tabular}
\caption{First few tachyon coefficients of the MSZ lump solution with extrapolations. All estimated errors of the extrapolations are of order $10^{-7}$.}
\label{tab:MSZ tachyon}
\end{table}

There is one problem with the tachyon profile. It is not a gauge invariant quantity and therefore it does not have an actual physical meaning. There may be gauges in which it looks completely different. However, the profile of the Erler-Maccaferri analytic solution \cite{ErlerMaccaferri} has a quite similar shape, so it is likely that the profile preserves its basic properties in most gauges.

The second quantity we can plot is the energy density profile introduced in \cite{KMS}. It is made out of the $E_n$ invariants and therefore it is gauge invariant. We define it as the Fourier series
\begin{equation}\label{FB energy profile}
E^{(M)}(x)=\frac{1}{2\pi R}\left(E_0+\sum_{n=1}^M 2E_n \cos \frac{nx}{R}\right).
\end{equation}
We have seen that high $E_n$ invariants are not convergent, so we restrict the sum to $n\leq M$, where it is typically reasonable to take $M$ to be roughly $M\sim 2.5R$. The profile is of course level dependent, but for simplicity, we will show only extrapolated profiles.

For a D0-brane located at $x_0=0$, the exact profile should be equal to a delta function:
\begin{equation}
E^{(D0)}(x)=\delta(x)=\frac{1}{2\pi R}\left(1+\sum_{n=1}^\inf 2\cos \frac{nx}{R}\right).
\end{equation}
In order to compare it to profile of a solution, we have to truncate the sum over $n$ to the same number of harmonics as in (\ref{FB energy profile}). Therefore profiles of numerical solutions do not converge to delta functions, but to truncated Fourier series of delta functions. In practice, we observe finite size peaks at positions of D0-branes and oscillations around zero at places where the profile should vanish.

\begin{figure}
\centering
   \begin{subfigure}[t]{0.47\textwidth}
      \includegraphics[width=\textwidth]{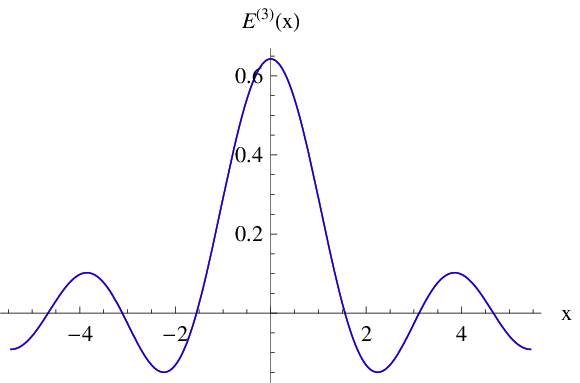}
   \end{subfigure}\qquad
   \begin{subfigure}[t]{0.47\textwidth}
      \includegraphics[width=\textwidth]{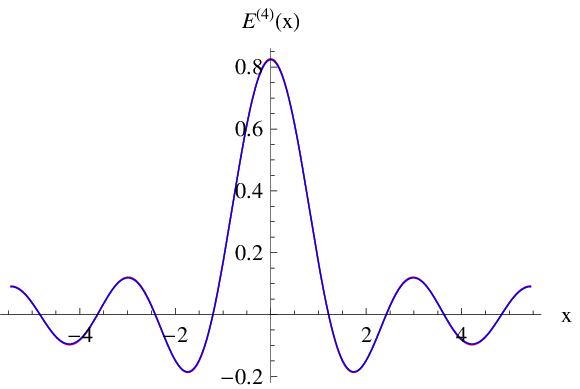}
   \end{subfigure}\vspace{5mm}
   \begin{subfigure}[t]{0.47\textwidth}
      \includegraphics[width=\textwidth]{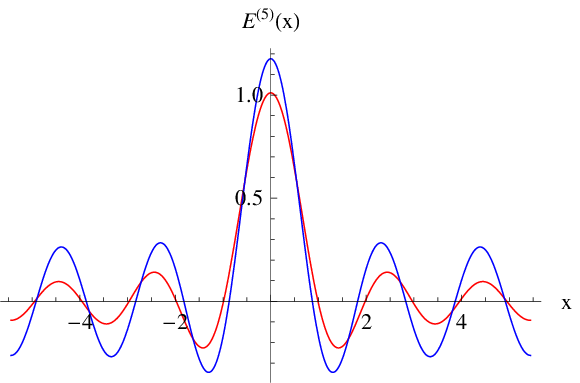}
   \end{subfigure}\qquad
   \begin{subfigure}[t]{0.47\textwidth}
      \includegraphics[width=\textwidth]{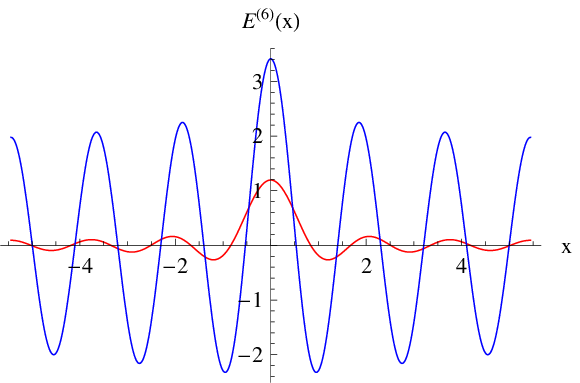}
   \end{subfigure}
\caption{Energy density profiles $E^{(M)}(x)$ of the MSZ lump solution for $M=3,4,5,6$. The blue lines denote the actual profiles extrapolated to level infinity and the red lines denote Fourier series of delta functions truncated to the same number of modes.}
\label{fig:MSZ energy profile}
\end{figure}

Several profiles of the MSZ lump solution of different orders are shown in figure \ref{fig:MSZ energy profile}. The profiles almost coincide with the predicted curves for $M=3,4$. For $M=5$, we can already see a difference because the extrapolation of $E_5$ is inaccurate and the profile for $M=6$ is completely dominated by the last nonconvergent invariant.

There is an interesting contrast between the tachyon profile and the energy density profile. The tachyon profile is not gauge invariant, but it is stable with respect to the level and to addition of high momentum states. On the contrary, the energy density profile is gauge invariant, but, like  $E_n$ invariants, it changes considerably with level and it gets ruined if we try to add high momentum invariants. Despite that, the energy density profile is a better tool to analyze solutions. Its definition does not depend on the initial boundary conditions and, as long as we include only convergent invariants, it gives us a good idea about the distribution of energy. On the other hand, the tachyon profile is defined only for the Neumann boundary conditions and it makes no sense if we consider some more complicated backgrounds (tilted D1-branes on a torus, D2-branes with flux \cite{KudrnaVosmera}).

\section{Single lump solutions at different radii}\label{sec:FB circle:single}
In the previous section, we analyzed one concrete lump solution, but now we will investigate how properties of single lump solutions depend on the compactification radius. We find that there are two types of single lump solutions with fundamentally different properties. The two types of solutions are separated by the self-dual radius $R=1$. Solutions above this radius are well-behaved and they have properties similar to the MSZ lump, solutions below $R=1$  are mostly complex and they have much worse properties.

\subsection{Lump solutions for $R>1$}\label{sec:FB circle:single:largeR}
First, let us focus on the case of $R>1$. Seeds for single lump solutions at these radii are easy to find, we just need to solve the equations of motion for relevant fields, which are few unless one picks a very high radius, and seeds at small radii can be even guessed.

\begin{figure}
\centering
\includegraphics[width=6cm]{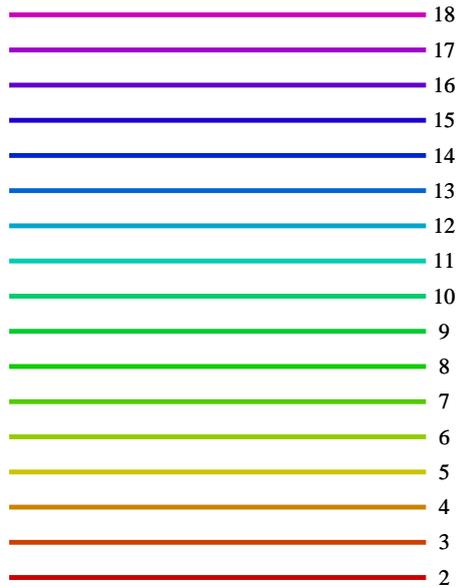}
\caption{Colors corresponding to different levels in figures in this section.}
\label{fig:FB colors}
\end{figure}

\begin{figure}
\centering
\includegraphics[width=12cm]{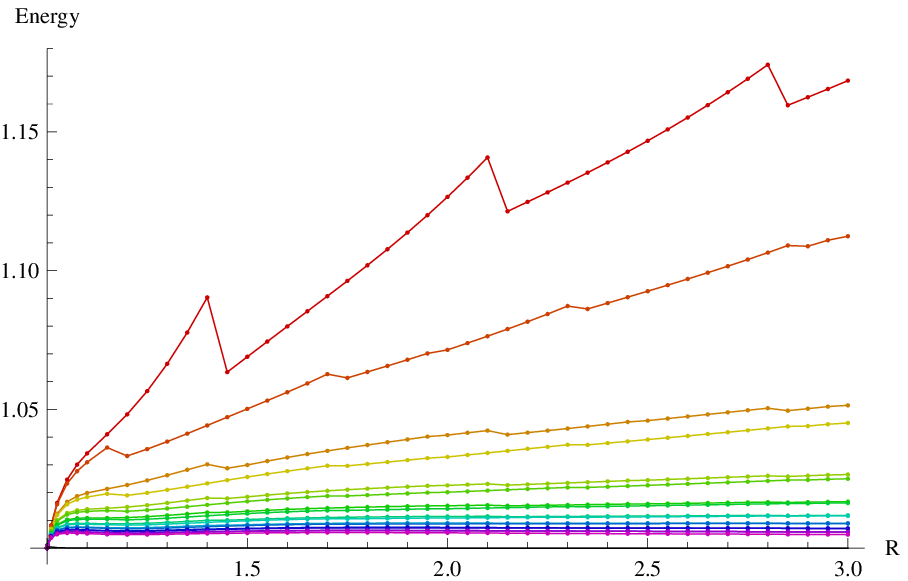}
\caption{Plot of the energy of single lump solutions between radii $1\leq R\leq 3$. Levels are denoted by colors following figure \ref{fig:FB colors}. The infinite level extrapolation is denoted by the black line, but it almost coincides with the horizontal axis. We use the same style for the remaining figures in this section.}\label{fig:FB Energy}\vspace{10mm}
\includegraphics[width=12cm]{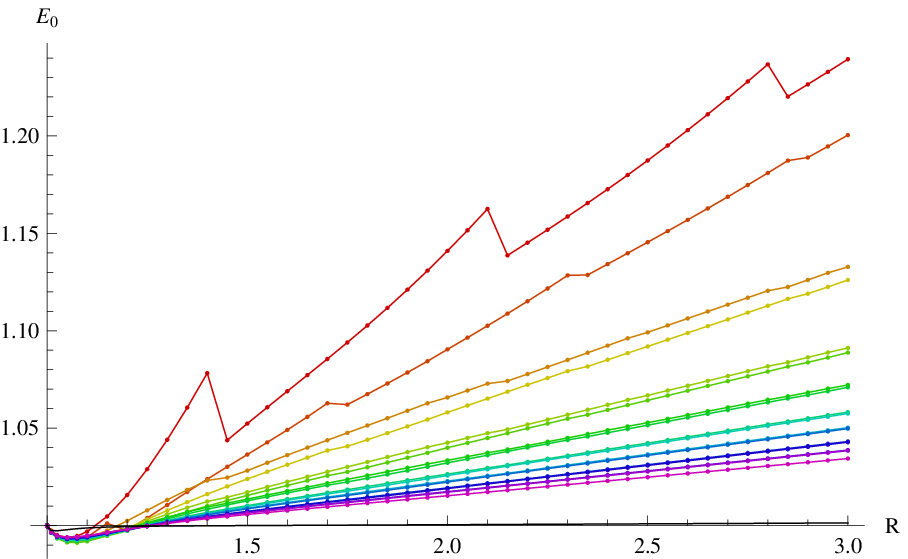}
\caption{Radius dependence of the $E_0$ invariant for single lump solutions.}
\label{fig:FB E0}
\end{figure}

\begin{figure}
\centering
\includegraphics[width=12cm]{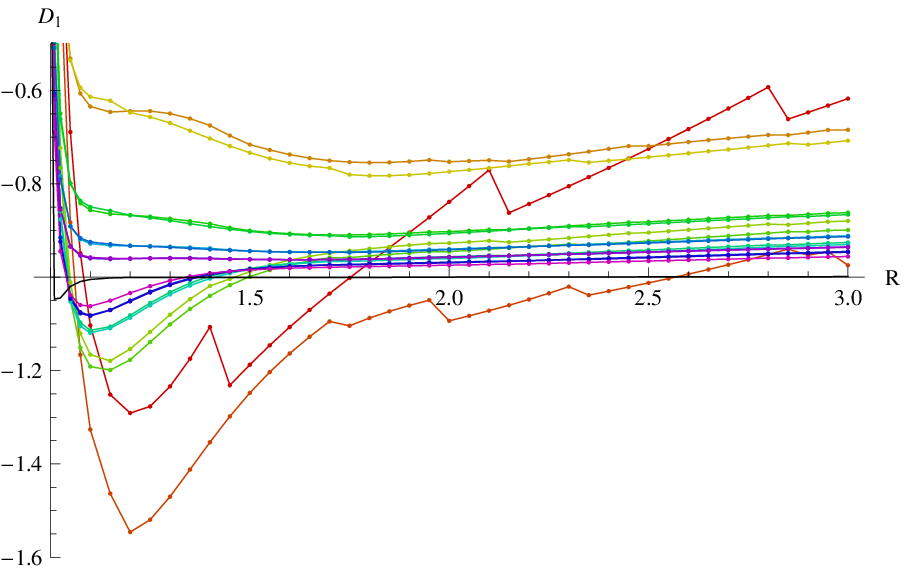}
\caption{Radius dependence of the $D_1$ invariant for single lump solutions.}
\label{fig:FB D1}\vspace{10mm}
\includegraphics[width=12cm]{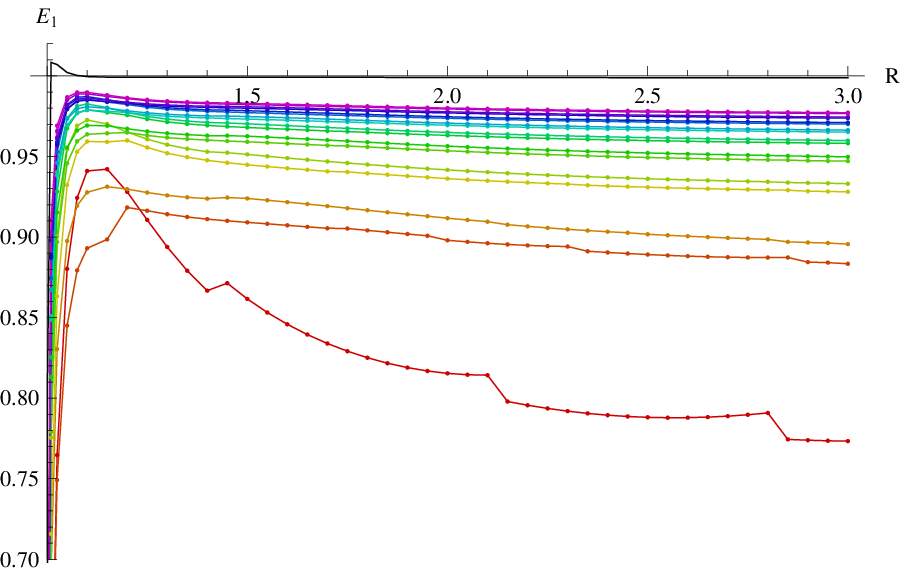}
\caption{Radius dependence of the $E_1$ invariant for single lump solutions.}
\label{fig:FB E1}
\end{figure}

\begin{figure}
\centering
\includegraphics[width=12cm]{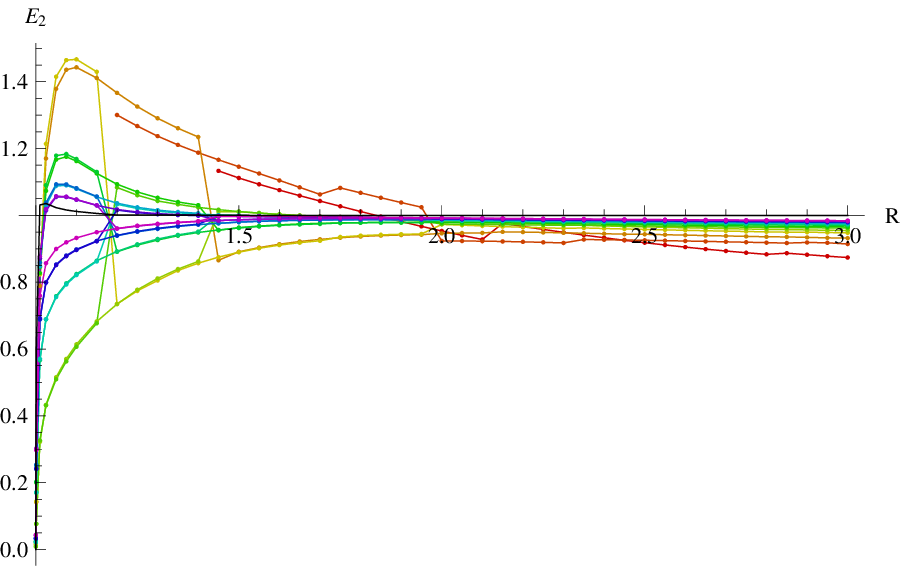}
\caption{Radius dependence of the $E_2$ invariant for single lump solutions.}
\label{fig:FB E2}\vspace{10mm}
\includegraphics[width=12cm]{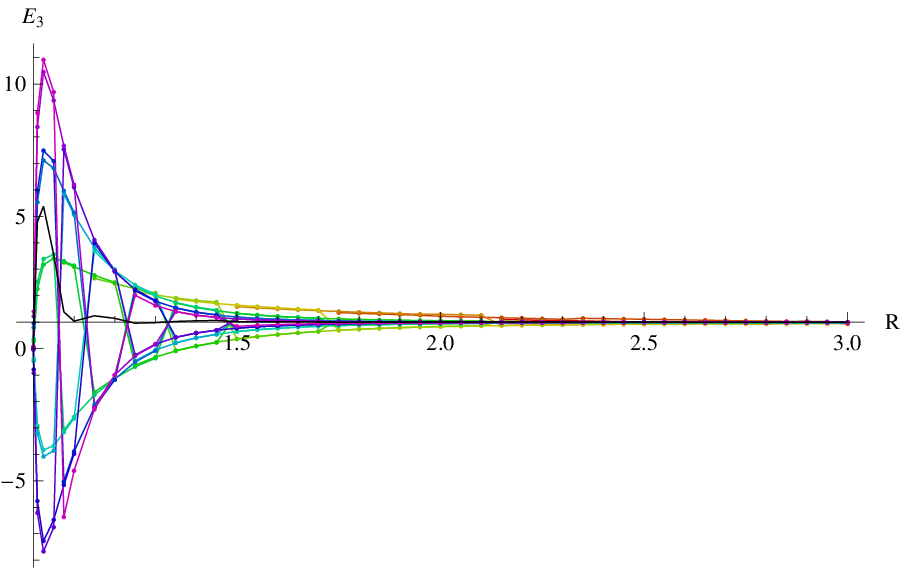}
\caption{Radius dependence of the $E_3$ invariant for single lump solutions.}
\label{fig:FB E3}
\end{figure}

\begin{figure}
\centering
\includegraphics[width=12cm]{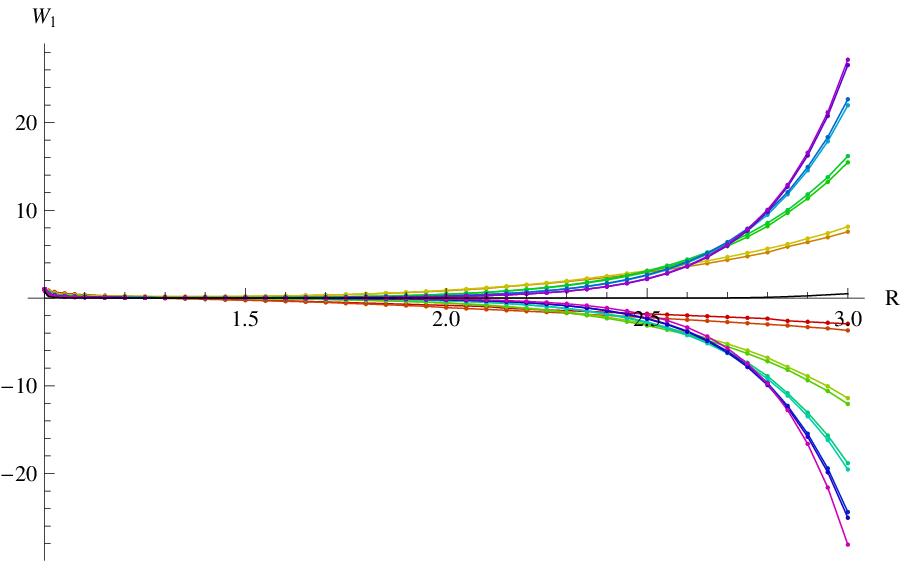}\vspace{10mm}
\includegraphics[width=12cm]{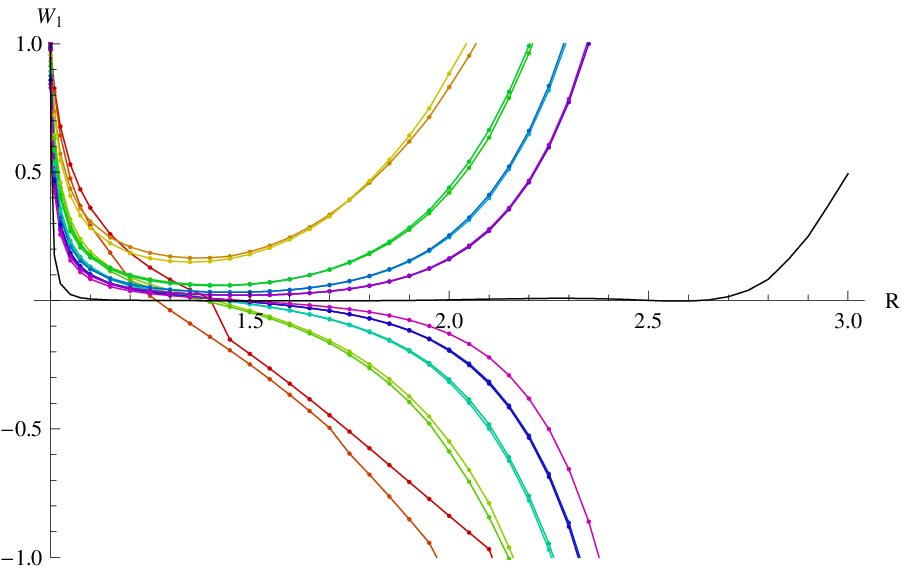}
\caption{Radius dependence of the $W_1$ invariant for single lump solutions. The first figure shows  the full range of data, in the second figure we zoom to data range $-1\leq W_1\leq 1$ to show its behavior at smaller radii.}
\label{fig:FB W1}
\end{figure}

\begin{figure}
\centering
\includegraphics[width=12cm]{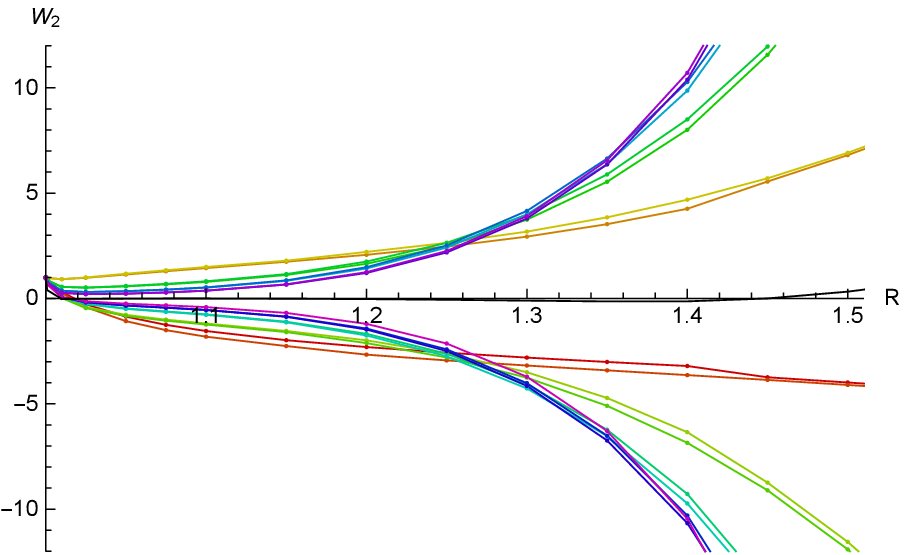}
\caption{Radius dependence of the $W_2$ invariant for single lump solutions. We show only data for $R\leq 1.5$ because it takes too high values at larger radii.}
\label{fig:FB W2}\vspace{10mm}
\includegraphics[width=12cm]{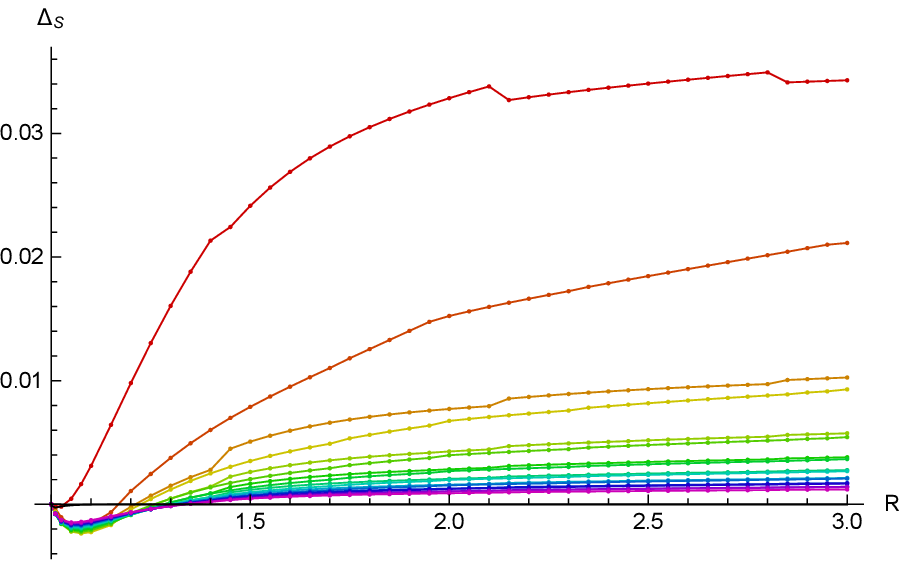}
\caption{Radius dependence of the out-of-Siegel equation $\Delta_S$ for single lump solutions}
\label{fig:FB Delta}
\end{figure}

\begin{figure}[!]
\centering
\includegraphics[width=12cm]{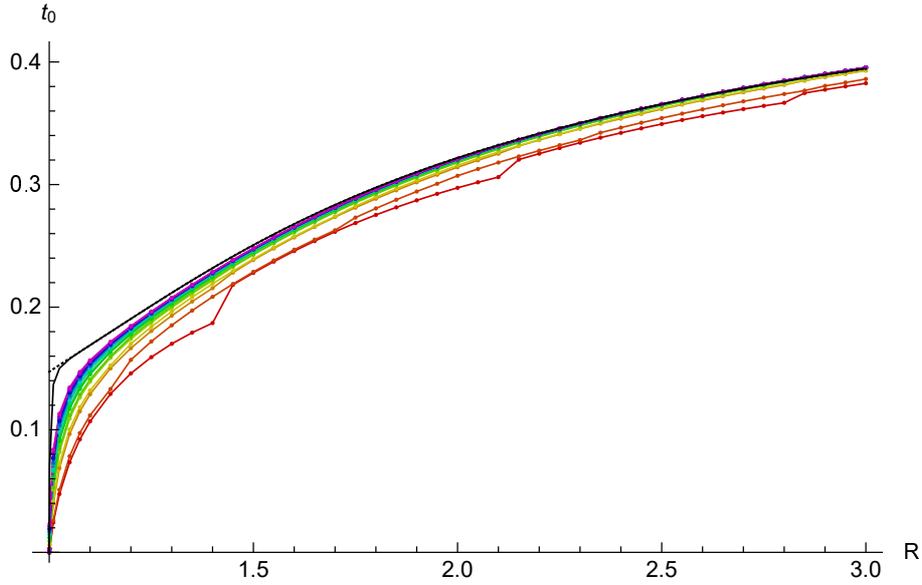}
\caption{Radius dependence of the zero momentum tachyon coefficient $t_0$ for single lump solutions. The black line denotes its infinite level extrapolation and the dotted black line, which almost coincides with the full black line, is a polynomial fit of the extrapolation, which we use to estimate the value of this coefficient at $R=1$.}\label{fig:FB t0}
\end{figure}

\begin{figure}[!]
\centering
\includegraphics[width=12cm]{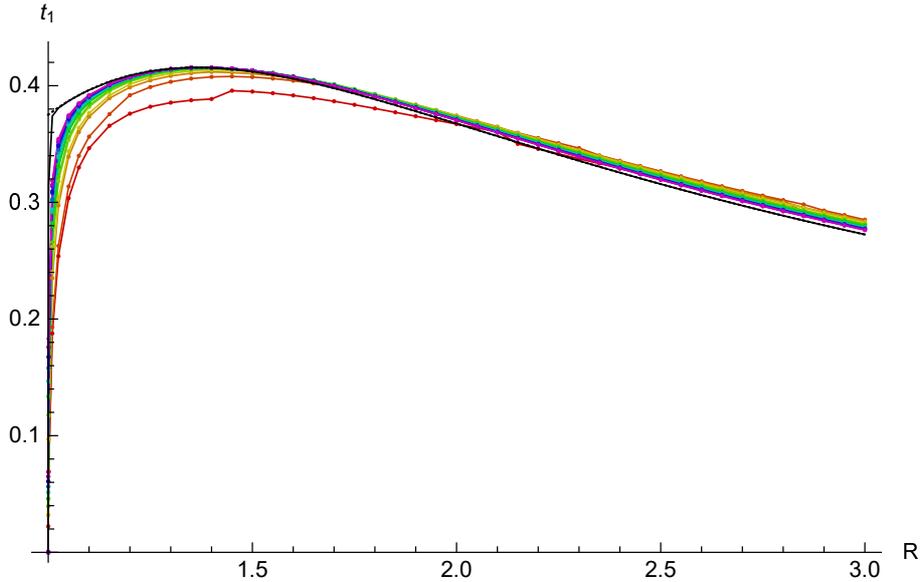}
\caption{Radius dependence of the first momentum coefficient $t_1$ for single lump solutions, which becomes a marginal field at $R=1$. The black line denotes the infinite level extrapolation and the dotted black line, which almost coincides with the full line, is a polynomial fit of the extrapolation.}
\label{fig:FB t1}
\end{figure}

We have been able to compute quite an extensive sample of solutions from $R=1$ to $R=3$ up to level 18. The amount of corresponding data it far too vast to be shown here, so instead we plot radius dependence of selected quantities in figures \ref{fig:FB Energy} to \ref{fig:FB t1}. We distinguish level of data in these figures by color, which follows the rainbow spectrum from red at level 2 to purple at level 18, see figure \ref{fig:FB colors}. We also plot infinite level extrapolations by black lines. The vertical axis is always set on the self-dual radius $R=1$ and the horizontal axis on the expected value of the given quantity.

The energy of single lump solutions is plotted in figure \ref{fig:FB Energy}. It is monotonously convergent in the whole region. At low levels, it changes significantly with the radius, but it is almost radius independent at high levels. The only exception is the area close to $R=1$, where it goes exactly to 1 at every level as the radius approaches 1. Unfortunately, this is not a sign of improved convergence, but, as we are going show, the solution itself approaches zero.

We also observe that the energy at a given level is not a continuous function of the radius. Discontinuities appear because there are radii where new states pass through the level cut-off, which causes sudden changes of the string field, and consequently of the energy and all other quantities. For example, at level 2, there are discontinuities at radii $R=n/\sqrt{2}$, $n\in \mathbb{N}$. These discontinuities have no physical meaning and, with the exception of some $E_n$ invariants, they get smaller with increasing level. At high levels, they are usually undistinguishable by naked eye.

After the energy, we look at the $E_0$ invariant, which is plotted in figure \ref{fig:FB E0}. This invariant shows a stronger radius dependence than the energy. It behaves approximately like a linear function, but it still converges nicely in the whole area and infinite level extrapolations are close to the expected value. Curiously, at low radii (roughly $R\lesssim 1.2$), it approaches 1 from below rather than from above. This is likely connected to the fact that the $E_0$ invariant of marginal solutions (see chapter \ref{sec:marginal}) is also lower than 1 at finite levels.

Next is the $D_1$ invariant, which is shown in figure \ref{fig:FB D1}. The oscillations, which we observed for the MSZ lump, are a generic feature, although they get slightly smaller with increasing radius. Infinite level extrapolations once again nicely reproduce the expected value. The behavior of the invariant near $R=1$ is more interesting. The curves at all levels abandon their current trend below approximately $R\lesssim 1.1$ and they quickly grow from approximately $-1$ to $+1$. This is a strong indication that the solution at $R=1$ becomes the perturbative vacuum. The same type of behavior repeats for all other invariants as well, they always approach the perturbative vacuum values near $R=1$.

The invariants $D_4$ and $D_9$ show the same kind of pathological behavior as for the MSZ lump at all radii, so we skip them and move to $E_n$ invariants.

We plot the first three $E_n$ invariants in figures \ref{fig:FB E1}, \ref{fig:FB E2} and \ref{fig:FB E3}. The first invariant approaches 1 in a very consistent way, but the other two invariants show a strong radius dependence. They converge well at large radii, but we observe large oscillations at low radii. The $E_3$ invariant does not even seem to be convergent below approximately $R\lesssim 1.2$. This supports our hypothesis that oscillations of Ellwood invariants depend on their conformal weights, which are $\frac{n^2}{4R^2}$ for the $E_n$ invariants. We also observe that the oscillations enhance discontinuities of these invariants. This is because $E_n$ invariants at low radii receive largest contributions from the highest available level, so addition of new states greatly changes these invariants.

Finally, we get to the $W_n$ invariants. Their conformal weights grow with the radius as $\frac{n^2R^2}{4}$, so only the first two invariants are convergent at least for some radii. We plot the $W_1$ invariant in figure \ref{fig:FB W1}. It converges quite well at low radii, but its oscillations grow uncontrollably with increasing radius. It is interesting that the extrapolated curve stays close to zero up to approximately $R\approx 2.7$, where its oscillations are already much bigger than 1 and the estimated errors of extrapolations are huge (for example, at $R=2.7$ we get $W_1^{(\inf)}=0.015\pm 3$). The second invariant $W_2$, which is shown in figure \ref{fig:FB W2}, converges only inside a very limited interval of radii because its conformal weight starts at 1. Unlike $E_n$ invariants, $W_n$ invariants do not exhibit large discontinuities and therefore we will use them later to investigate the dependence of oscillations on conformal weights.

The out-of-Siegel equation $\Delta_S$, which is plotted in figure \ref{fig:FB Delta}, is satisfied very well in the whole range of radii. We notice that in the region close to $R=1$, it switches sign and gets slightly worse, but elsewhere it behaves similarly as for the MSZ lump.

Next, we are going to discuss properties of single lump solutions around $R=1$. All observables suggest that the solution approaches the perturbative vacuum and we can easily confirm this conjecture directly by looking at its coefficients, all of which go to zero. Therefore we have
\begin{equation}
\lim_{L\rar\inf}\lim_{R\rar 1}\Psi=0.
\end{equation}
On the other hand, infinite level extrapolations behave very consistently even at quite small radii, which suggests that the solution at a fixed radius always describes a D0-brane, we just need to go to high enough level to see that at low radii. Therefore it is likely that when we exchange the limits, we find the marginal solution describing the Dirichlet boundary conditions,
\begin{equation}
\lim_{R\rar 1}\lim_{L\rar\inf}\Psi=\Psi_{\lB=\pi/2}^{mar}.
\end{equation}
Therefore we can use this approach to estimate some properties of the Dirichlet marginal solution. We are interested in the zero momentum tachyon coefficient $t_0$ and in the first momentum mode $t_1$, which becomes marginal at $R=1$. The radius dependence of these coefficients is plotted in figures \ref{fig:FB t0} and \ref{fig:FB t1} respectively. As we said before, both coefficients go to zero as $R\rar 1$, but infinite level extrapolations compensate this effect quite well until $R\approx 1.05$. Therefore we can drop the first few data points near $R=1$ and fit the rest of the extrapolated data with a polynomial, which can be then evaluated at $R=1$\footnote{A similar method was used in \cite{LumpsBeccaria} to estimate $t_1$. The result from this reference, $t_1^{(R=1)}=0.351$, is somewhat lower than our estimate, probably due to lower level of calculations in this reference.}. Using this method, we get
\begin{eqnarray}
t_0^{(R=1)} &=& 0.1473, \\
t_1^{(R=1)} &=& 0.3751.
\end{eqnarray}
The results do not depend much on orders of the extrapolating polynomials, we estimate that the errors are of order $10^{-4}$. These results nicely agree with the estimates from marginal deformations from chapter \ref{sec:marginal} and from the pseudo-real double lump solution from section \ref{sec:FB circle:double:R=2}.

\FloatBarrier
\subsection{Lump solutions for $R<1$}\label{sec:FB circle:single:smallR}
Single lump solutions below the self-dual radius are quite different from the ones described in the previous subsection. A D0-brane at radius $R<1$ has higher energy than a D1-brane and therefore lump solutions go against the RG frow from the BCFT point of view. There are no relevant operators apart from the usual zero momentum tachyon, so we must consider some irrelevant fields to see any nontrivial solutions. To find seeds for lump solutions, we have to choose a level which includes at least the first momentum state and the first descendants of the tachyon field, that means at least level 2. We find no real lump solutions at levels available for the homotopy continuation method, but there are some promising complex seeds. Next, we are going to discuss one solution of this type as an example.

We have chosen a special radius $R=\frac{1}{\sqrt{2}}$ and we show some properties of the lump solution at this radius in table \ref{tab:FB positive lump}. The energy of this solution is a familiar number and we find that it exactly matches the energy of the $\sigma$-brane solution in the Ising model (see \cite{Ising} or section \ref{sec:MM:Ising:Id}). We can also match $\Delta_S$, some Ellwood invariants and some of its coefficients, which means that this solution is dual to the Ising model solution, it is just written in a different basis. This duality can be explained as follows: In section \ref{sec:MM:Ising2}, we describe the duality between the double Ising model and the free boson theory on the orbifold $S^1/\mathbb{Z}_2$ with radius $\sqrt{2}$ (or with the T-dual radius $\frac{1}{\sqrt{2}}$). A similar duality can be written for the free boson theory on a normal circle, this theory is dual to the double Ising model with a non-diagonal bulk spectrum. The precise form of the duality is not important for us now because this particular Ising model solution needs only the Verma module of identity and the correct energy-momentum tensor. In the free boson theory at $R=\frac{1}{\sqrt{2}}$, we find that the operators
\begin{equation}
\frac{1}{2}(T(z)\pm \cos(X/\sqrt{2})(z))
\end{equation}
have the desired properties of energy-momentum tensors with $c=\frac{1}{2}$ and therefore they guarantee existence of the dual solution.

\begin{table}[!t]
\centering
\footnotesize
\begin{tabular}{|l|lll|}\hline
Level    & Energy                & $\ps \Delta_S$             & $\ps D_1$               \\\hline
2        & $1.59267-0.7268777i$  & $   -0.1551684+0.1181961i$ & $   -9.73471+5.239036i$ \\
4        & $1.41414-0.2015210i$  & $   -0.0881209+0.0512621i$ & $   -0.66854-1.991905i$ \\
6        & $1.28579-0.0766818i$  & $   -0.0659859+0.0295320i$ & $   -3.86207+0.373757i$ \\
8        & $1.21160-0.0305389i$  & $   -0.0544551+0.0187957i$ & $   -0.57514-0.822662i$ \\
10       & $1.16345-0.0100715i$  & $   -0.0472822+0.0117019i$ & $   -2.48552-0.002611i$ \\
12       & $1.12943-0.0012257i$  & $   -0.0423218+0.0053484i$ & $   -0.56951-0.245609i$ \\
14       & $1.10568           $  & $   -0.0330019           $ & $   -1.93951          $ \\
16       & $1.09045           $  & $   -0.0275429           $ & $   -0.95087          $ \\
18       & $1.07936           $  & $   -0.0239721           $ & $   -1.69824          $ \\
20       & $1.07084           $  & $   -0.0213384           $ & $   -1.04849          $ \\\hline
$\inf$   & $0.993             $  & $\ps 0.017               $ & $   -1.2              $ \\\hline
Exp.     & $1                 $  & $\ps 0                   $ & $   -1                $ \\\hline
\multicolumn{4}{l}{}\\[-5pt]\hline
Level    & $E_0$                & $E_1$                      & $\ps W_1$                \\\hline
2        & $1.060485+0.184547i$ & $5.39760-2.527244i$       & $   -0.288914+0.816358i$ \\
4        & $0.962899+0.142672i$ & $0.81572+1.067288i$       & $   -0.310926+0.474456i$ \\
6        & $0.922618+0.113783i$ & $2.39235-0.129987i$       & $   -0.327389+0.331617i$ \\
8        & $0.904803+0.086855i$ & $0.73997+0.454758i$       & $   -0.312953+0.237079i$ \\
10       & $0.892563+0.061735i$ & $1.68904+0.032173i$       & $   -0.316423+0.161724i$ \\
12       & $0.885097+0.031094i$ & $0.72730+0.138352i$       & $   -0.310218+0.079033i$ \\
14       & $0.914693          $ & $1.42710          $       & $   -0.223734          $ \\
16       & $0.930444          $ & $0.94066          $       & $   -0.173499          $ \\
18       & $0.939178          $ & $1.31871          $       & $   -0.147128          $ \\
20       & $0.945384          $ & $0.99694          $       & $   -0.126159          $ \\\hline
$\inf$   & $1.02              $ & $1.1              $       & $\ps 0.09              $ \\\hline
Exp.     & $1                 $ & $1                $       & $\ps 0                 $ \\\hline
\end{tabular}
\caption{Selected observables of the positive energy single lump solution at radius $R=\frac{1}{\sqrt{2}}$. The extrapolations use only the real data points from levels 14 to 20. We cannot very well estimate their errors.}
\label{tab:FB positive lump}
\end{table}

We have been able to evaluate this solution up to level 20. This is two levels higher than the MSZ lump solution, but 4 levels lower than the corresponding Ising model solution. Therefore we leave a more detailed discussion of its properties to section \ref{sec:MM:Ising:Id} and we review only its key features here.

The solution starts as complex, but its imaginary part steadily decreased and it finally becomes real at level 14. That allows us to accept the solution as physical, but the transition means that the real part of the solution dramatically changes its behavior when the imaginary part disappears (see figure \ref{fig:Ising Id} for plots of some of the invariants). For example, the real part of the $E_0$ invariant moves away from 1 below level 14, but it approaches the correct value above level 14. Other invariants behave similarly, although the trend is partially obscured by oscillations of some invariants.

This means that it is very difficult to extrapolate the solution. We have not found any consistent way to include the data from levels below 14, so there are only 4 real points for every observable (there are no new data at odd levels). In the end, we obtain only a rough approximation of the Dirichlet boundary state (Ellwood invariants have errors around 10-20\%, the energy slightly less than $1\%$), but the identification of the solution is unambiguous because there are no other boundary states with similar properties. Compared to the MSZ lump solution, the precision has dropped by 2 to 3 orders. The corresponding Ising model solution offers a slightly better precision, but we would probably need data from levels above 30 to get the same precision as for the usual lumps solutions.\\

Next, we will discuss radius dependence of positive energy lump solutions. We have computed these solutions up to level 20 at radii from $0.5$ to $0.99$. Since these solutions give us less information than those above the self-dual radius, we choose a less dense sampling.

\begin{figure}
\centering
\includegraphics[width=12cm]{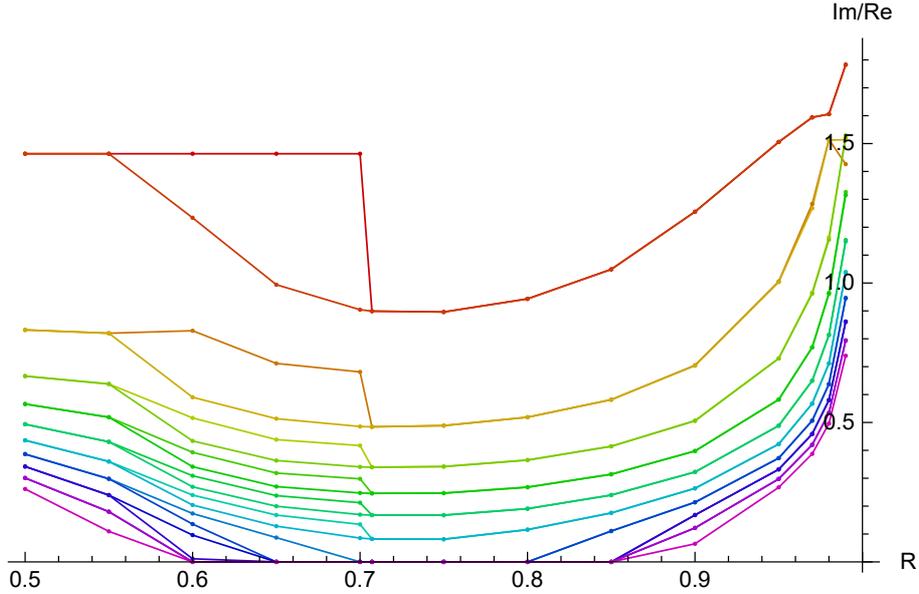}
\caption{Ratio between the imaginary part and the real part of single lump solutions at radii $0.5\leq R\leq 0.99$ up to level 20. Levels are denoted by colors following the rainbow spectrum similarly to figure \ref{fig:colors}, but the scale is now stretched to 20 levels. There are only 10 points at most radii because even and odd level data often coincide.}
\label{fig:FBP ImRe}
\end{figure}

We begin by investigating reality properties of these solutions. For that purpose we define a ratio of the imaginary part and the real part of the string field as $\sum_{i}{\rm Im}[t_i]/\sum_{i}{\rm Re}[t_i]$ and we plot it in figure \ref{fig:FBP ImRe}. This quantity is not invariant (it depends even on the choice of basis), but it is usually quite informative about how complex a solution is. We observe that solutions up to level 20 become real only in a limited interval of radii (approximately $0.6\lesssim R\lesssim 0.85$). Predictably, solutions get more complex further away from the self-dual radius because they need to go more against the RG flow. However, the imaginary part grows even more close to $R=1$. It is even possible that the imaginary part diverges in the $R\rar 1$ limit.

With the available data, we cannot say with certainty whether all lump solutions at all radii become real at sufficiently high level. Some extrapolations suggest that they might, but extrapolations of imaginary parts of different quantities often give us contradictory results, so it hard to make any definitive conclusion.

\begin{figure}
\centering
\includegraphics[width=12cm]{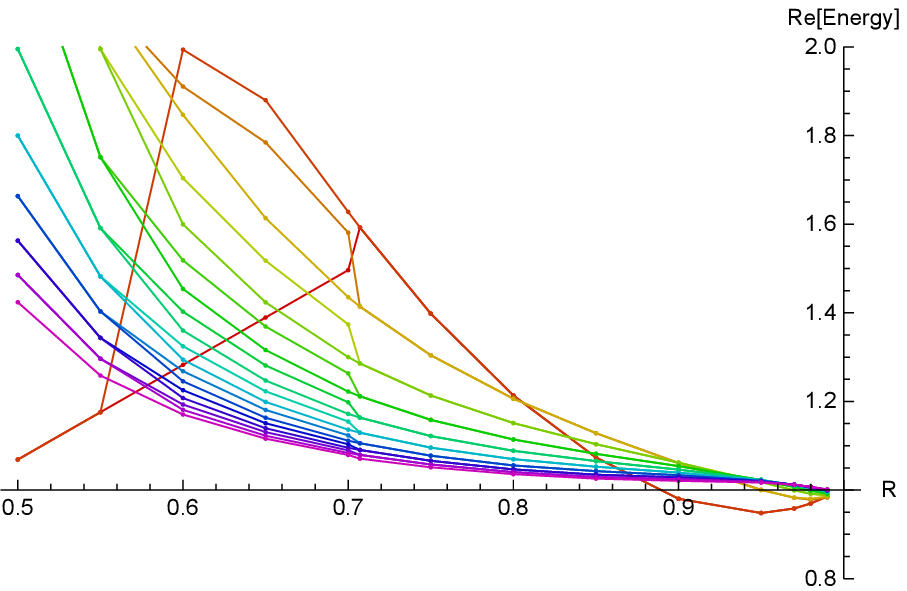}\vspace{10mm}
\includegraphics[width=12cm]{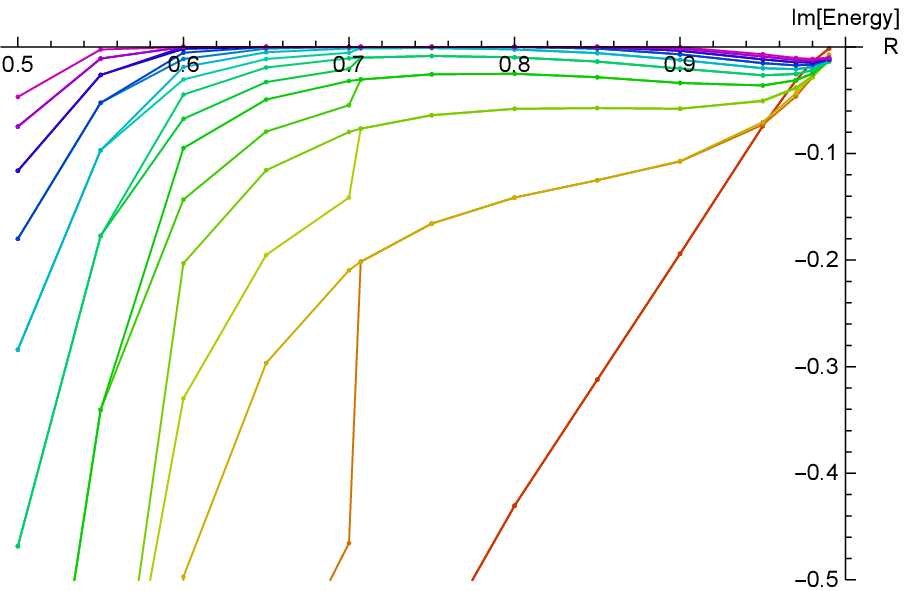}
\caption{Real and imaginary part of the energy of single lump solutions below the self-dual radius. We show no extrapolations because they are not reliable.}
\label{fig:FBP Energy}
\end{figure}

Figure \ref{fig:FBP Energy} shows the real and imaginary part of the energy. Surprisingly, it does not reflect the growth of imaginary coefficients near $R=1$. The real part of the energy actually behaves well close to the self-dual radius and even the imaginary part stays quite small. Unfortunately, we are not able to make any trustworthy extrapolations at these radii, so we have to make do with finite level data.

\begin{figure}
\centering
\includegraphics[width=12cm]{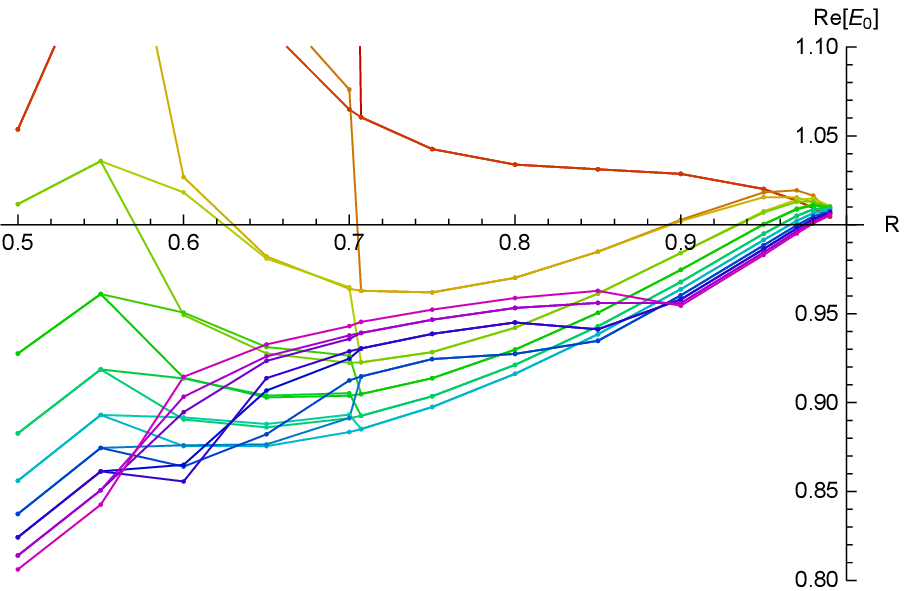}
\caption{Real part of the $E_0$ invariant of positive energy single lump solutions.}
\label{fig:FBP E0}\vspace{10mm}
\includegraphics[width=12cm]{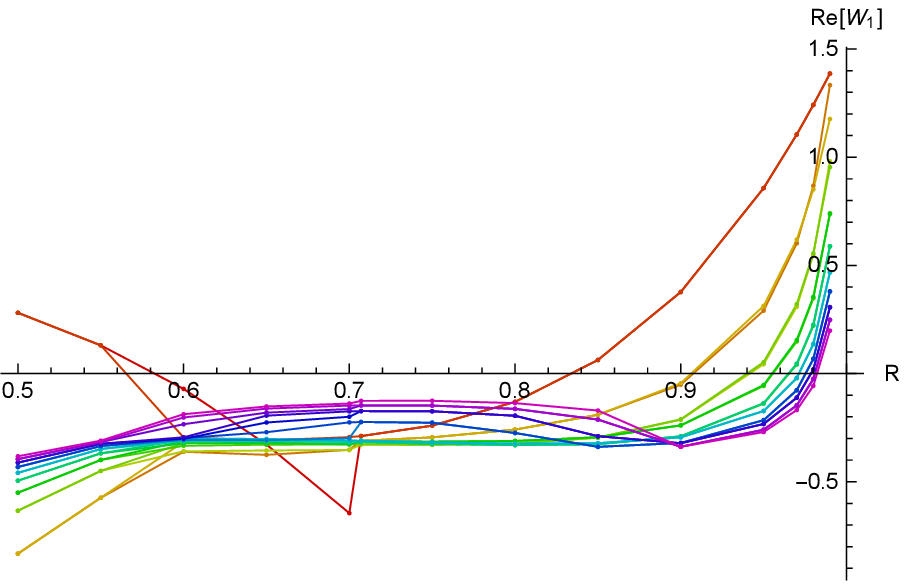}
\caption{Real part of the $W_1$ invariant of positive energy single lump solutions.}
\label{fig:FBP W1}
\end{figure}

Ellwood invariants agree with the expected boundary state only very roughly, the errors at level 20 are of order of tens of percents. Figures \ref{fig:FBP E0} and \ref{fig:FBP W1} show the real part of $E_0$ and $W_1$ respectively as examples. If we look at these invariants at high levels, we observe how the change of reality properties affects the invariants. The lines are not smooth functions of the radius and if we check the level dependence at fixed radius, we find that these two invariants approach the expected values only once they become real.

It is not clear what happens to these lump solutions in the $R\rar 1$ limit. Obviously, they do not approach the perturbative vacuum like those above $R=1$, and they do not seem to approach the marginal solution either. Some quantities change quickly as the radius approaches 1 and some may even diverge, so it is quite possible that there is no good $R\rar 1$ limit at finite level. Assuming that all solutions become real at high enough level, it may possible to first extrapolate them to level infinity and then take the limit like above $R=1$, but we cannot do so with the available data.

\FloatBarrier
\subsection{Convergence of Ellwood invariants and \Pade approximants}\label{sec:FB circle:single:pade}
We have seen that some Ellwood invariants of single lump solutions do not seem to be convergent. Moreover, it is clear that their oscillations are closely related to their conformal weights. This behavior does not apply only to lump solutions, but, as far as we can tell, to all non-universal OSFT solutions in any background. In this section, we investigate the oscillations and show our attempts to improve convergence of Ellwood invariants using \Pade approximants.

First, we would like to understand the origin of oscillations of Ellwood invariants. We have found that these oscillations are connected to the conservation laws derived in section \ref{sec:SFT:Elw cons}. Take for example the Virasoro conservation law (\ref{Ellwood cons Vir}) in the limit $|h|,|\bar h|\gg c$:
\begin{equation}
\la E[ \vv^{(h,\bar h)}]|L_{-n}=4n\left( i^{-n}h+(-i)^{-n}\bar h ) \right)\la E[ \vv^{(h,\bar h)}]|,
\end{equation}
where we neglected the central charge and the $L_n$ operator. Concretely, for $n=2$ and $h=\bar h$, we find
\begin{equation}
\la E[ \vv^{(h,h)}]|L_{-2}=-16 h\la E[ \vv^{(h,\bar h)}]|.
\end{equation}
Therefore the state $(L_{-2})^{L/2}|0\ra$ contributes at level $L$ to a given invariant as $(-16 h)^{L/2}$ plus subleading terms. Similarly, the state $(\alpha_{-1})^{L}|0\ra$ in the free boson theory contributes to $W_n$ invariants as $(-1)^{L/2}(2\sqrt{2} nR)^{L}$. These contributions grow exponentially with level. However, these high numbers can be in principle suppressed by string field coefficients, which decrease with level, or by mutual cancelation of terms. By investigating individual contributions to $D_4$, $D_9$ and $W_n$ for several lump solutions, we have found that both effects play a role and they actually significantly reduce the oscillations compared to these simple estimates, but not enough to fully suppress  them. Invariants with high conformal weights behave well only for universal solutions or for solutions which have universal dependence on some direction (for example, in free boson theory on a torus, we find lump solutions oriented in the $X$ direction, which have well convergent invariants constructed using the $Y$ field).

Unfortunately, we do not know whether this problem is just a 'low' level behavior and the invariants converge in the $L\rar \inf$ limit, or whether level truncated solutions lead to divergent series and some boundary state coefficients can be obtained only through some resummation process. To decide that, we would need an analytic solution with known $L_0$ expansion up to all levels, which have not been found so far. However, we at least try make some estimates about the convergence.

For that purpose, we make a simplified definition. We say that a given invariant is convergent if there is a level $L$ at which the value of the invariant is closer to the expected value than at level $L-4$. Now, consider the invariants $W_1$ and $W_2$, which have a relatively smooth radius dependence (see figures \ref{fig:FB W1} and \ref{fig:FB W2}). Using these invariants, we can try to determine critical radii at every level, below which the invariants converge. First, we fit the invariants with polynomials in $R$ to approximate them by smooth functions and to remove discontinuities. Then we compute where these polynomials cross each other to determine the critical radii and the associated conformal weighs. The results of this procedure are summarized in table \ref{tab:FB critical weight}. Although these results probably have large errors and there are some irregularities, the general trend is that the critical radius grows with level. This makes us slightly optimistic that Ellwood invariants might be convergent after all.

However, even if all invariants are convergent and if we could reach high enough level, some invariants would not be usable anyway due to loss of numerical precision. Consider the tachyon vacuum solution in the free boson framework. In our normalization conventions, all of its nonzero invariants should be the same. However, they differ due to numerical errors caused by cancelation of large numbers. For example, we find that the $D_4$ invariant at level 18 has an absolute error of order $10^{-7}$ (when using the usual double number format in C++) and the $D_9$ invariant has an error of order $10^{-3}$. Therefore invariants with even higher conformal weights or at higher levels would have numerical errors larger than the characteristic scale and there would be no point in computing such invariants.

\begin{table}
\centering
\begin{tabular}{|l|cc|cc|}\hline
      & \multicolumn{2}{c|}{$W_1$} & \multicolumn{2}{c|}{$W_2$} \\\hline
Level & $R$  & $h$  & $R$  & $h$  \\\hline
6     & 2.25 & 1.26 & 1.25 & 1.55 \\
7     & 2.32 & 1.35 & 1.26 & 1.59 \\
8     & 2.49 & 1.55 & 1.25 & 1.57 \\
9     & 2.51 & 1.57 & 1.26 & 1.58 \\
10    & 2.57 & 1.65 & 1.26 & 1.58 \\
11    & 2.59 & 1.68 & 1.27 & 1.60 \\
12    & 2.65 & 1.76 & 1.27 & 1.63 \\
13    & 2.67 & 1.78 & 1.28 & 1.64 \\
14    & 2.70 & 1.83 & 1.29 & 1.67 \\
15    & 2.71 & 1.84 & 1.30 & 1.70 \\
16    & 2.75 & 1.89 & 1.30 & 1.68 \\
17    & 2.76 & 1.91 & 1.31 & 1.73 \\
18    & 2.79 & 1.95 & 1.29 & 1.68 \\\hline
\end{tabular}
\caption{Critical radii computed from the invariants $W_1$ and $W_2$ at different levels and the associated conformal weights.}
\label{tab:FB critical weight}
\end{table}

Although raw data of some invariants are pretty much useless, there is still a possibility of improving their convergence using some resummation process. For that purpose we modify the definition (\ref{Ellwood definition}) by insertion of the operator $z^{L_0+1}$:
\begin{equation}\label{Ellwood modification}
E[\vv](z)=\la E[\vv]|z^{L_0+1}|\Psi-\Psi_{TV}\ra.
\end{equation}
This definitions turns Ellwood invariants into power series in $z$ and it allows us to distinguish contributions from different levels. The original values can be restored simply by setting $z=1$. For illustration, we show three modified invariants of the MSZ lump solution in table \ref{tab:FB invariants z}. These data support our previous claim that invariants change with increasing level mainly due to new states appearing at the highest level and that contributions from lower levels undergo only minor changes.

\begin{table}[t!]
\centering
\scriptsize
\begin{tabular}{|l|l|}\hline
Level \rowh{9pt} & $E_0(z)$                                                                                  \\\hline
2     \rowh{9pt} & $1.00801 +0.0829262 z^2                                                                $ \\
3                & $0.996668+0.0635063 z^2                                                                $ \\
4                & $0.974341+0.0596181 z^2+0.0122671  z^4                                                 $ \\
5                & $0.971233+0.057768  z^2+0.0104819  z^4                                                 $ \\
6                & $0.963599+0.0557043 z^2+0.00994823 z^4-0.000036609 z^6                                 $ \\
7                & $0.962678+0.0546188 z^2+0.00979042 z^4-0.000406542 z^6                                 $ \\
8                & $0.958906+0.053447  z^2+0.00947149 z^4-0.000565497 z^6+0.00174932 z^8                  $ \\
9                & $0.958596+0.0527582 z^2+0.00931698 z^4-0.000564807 z^6+0.00160117 z^8                  $ \\
10               & $0.956355+0.0520185 z^2+0.00911806 z^4-0.000669409 z^6+0.00152158 z^8-0.00047325 z^{10}$ \\\hline
\multicolumn{2}{l}{}\\[-4pt]\hline
Level \rowh{9pt} & $D_1(z)$                                                                                 \\\hline
2     \rowh{9pt} & $1.00801 -2.02154   z^2                                                                $ \\
3                & $0.996668-2.10744   z^2                                                                $ \\
4                & $0.974341-2.22358   z^2+0.496764   z^4                                                 $ \\
5                & $0.971233-2.25946   z^2+0.509003   z^4                                                 $ \\
6                & $0.963599-2.30147   z^2+0.543436   z^4-0.150734    z^6                                 $ \\
7                & $0.962678-2.31311   z^2+0.549798   z^4-0.158862    z^6                                 $ \\
8                & $0.958906-2.33471   z^2+0.56762    z^4-0.167302    z^6+0.0659523  z^8                  $ \\
9                & $0.958596-2.33992   z^2+0.57129    z^4-0.171262    z^6+0.06793    z^8                  $ \\
10               & $0.956355-2.35313   z^2+0.582412   z^4-0.176714    z^6+0.0712083  z^8-0.0439085  z^{10}$ \\\hline
\multicolumn{2}{l}{}\\[-4pt]\hline
Level \rowh{9pt} & $D_4(z)$                                                                                 \\\hline
2     \rowh{9pt} & $1.00801 -8.33494   z^2                                                                $ \\
3                & $0.996668-8.62029   z^2                                                                $ \\
4                & $0.974341-9.07319   z^2+17.3984    z^4                                                 $ \\
5                & $0.971233-9.21116   z^2+17.796     z^4                                                 $ \\
6                & $0.963599-9.37298   z^2+18.7402    z^4-39.3822     z^6                                 $ \\
7                & $0.962678-9.41628   z^2+18.9744    z^4-40.7445     z^6                                 $ \\
8                & $0.958906-9.49916   z^2+19.4825    z^4-42.3452     z^6+100.23     z^8                  $ \\
9                & $0.958596-9.51794   z^2+19.608     z^4-43.0705     z^6+103.074    z^8                  $ \\
10               & $0.956355-9.56856   z^2+19.9291    z^4-44.1197     z^6+106.475    z^8-214.857    z^{10}$ \\\hline
\end{tabular}
\caption{Modified invariants $E_0(z)$, $D_1(z)$ and $D_4(z)$ of the MSZ lump solution as polynomials in $z$ up to level 10.}
\label{tab:FB invariants z}
\end{table}

A common method to deal with a divergent series is the \Pade approximation. Given a series $f(z)=\sum_n a_n z^n$, the \Pade approximant is a rational function $f_p^q(z)$ of degree $(p,q)$, whose power series matches $f(z)$ up to order $p+q$. To improve convergence of an invariant, we replace it by its \Pade approximant and then we set $z=1$ to remove the artificial $z$ dependence. We usually take approximants of degree $(p,p)$ (if the polynomial has even degree) or $(p,p+1)$ (if it has odd degree).

\Pade approximations of Ellwood invariants sometimes do not work well (probably due to poles of $f_p^q(z)$ near $z=1$), so we introduce a second method called the \PadeBorel approximation. In this case, we first apply the Borel transformation to the function $f(z)$, $\mathcal{B} f(z)=\sum_n \frac{a_k z^k}{k!}$, then we replace this function by its \Pade approximant and finally we apply the inverse Borel transformation
\begin{equation}
\int_0^\inf e^{-t}\mathcal{B} f_p^q(t z)dt.
\end{equation}
These approximations have been successfully used to improve the convergence of the energy of some analytic solutions \cite{AnalyticSolutionSchnabl}\cite{TVArroyoPade}\cite{SimpleAnalyticSolutionErlerSchnabl} and they have been also applied to some numerical solutions \cite{MarginalKMOSY}\cite{Ising}\cite{MarginalTachyonKM}.

When we use these methods on Ellwood invariants, we encounter few issues. First, there is usually a common prefactor $z^h$, where $h$ is the weight of the fundamental primary that contributes to the given invariant. This term does not have any effect, so we drop it for simplicity. This is equivalent to inserting $z^N$ instead of $z^{L_0+1}$ into (\ref{Ellwood modification}). Second, modified invariants of twist even solutions include only even powers of $z$. The reader can easily check that there is no order (1,1) \Pade approximant for a function of the form $a_0+a_2 z^2$, so one has to use unusual orders of approximations (in this case $(0,2)$) when working with such functions. Therefore it is easier to make the replacement $z^2\rar \tilde z$. Then we take \Pade approximant of the polynomial in $\tilde z$. The \Pade approximation does not depend on whether we work with $z$ or $\tilde z$, but the \PadeBorel approximation does. We will denote the \PadeBorel approximation using $\tilde z$ as PB1 and using $z$ as PB2. The references \cite{MarginalKMOSY}\cite{Ising}\cite{MarginalTachyonKM} use the PB1 scheme.

These approximations usually reduce oscillations of invariants, but, unfortunately, it turn out that their use is often counterproductive. See table \ref{tab:FB Pade1}, where we apply the resummation techniques to the $D_1$ invariant of the MSZ lump solution. Starting with level 4, the approximations suppress the oscillations and give us almost monotonic convergence of the invariant. However, they also destroy the original pattern of level dependence and therefore we cannot use the usual extrapolation techniques. We are forced to fit the approximations just by linear or quadratic functions and the results are actually {\it less} precise than the extrapolation of the original data. This applies to other convergent invariants as well. The use of \PadeBorel approximations in \cite{MarginalTachyonKM} prevented us from making a good estimate of the $\lB(\lS)$ dependence, while we manage to do so in section \ref{sec:marginal:marginal} using the same data without \PadeBorel approximations.

\begin{table}
\centering
\begin{tabular}{|l|llll|}\hline
Level \rowh{12pt} & $\ps D_1$      & $\ps D_1^{P}$  & $\ps D_1^{PB1}$ & $\ps D_1^{PB2}$ \\\hline
2                 & $   -1.013532$ & $\ps 0.335390$ & $\ps 0.464621$  & $\ps 0.626043$  \\
4                 & $   -0.752479$ & $   -0.843193$ & $   -0.870075$  & $   -0.905075$  \\
6                 & $   -0.945165$ & $   -0.913092$ & $   -0.908299$  & $   -0.919990$  \\
8                 & $   -0.909528$ & $   -0.930971$ & $   -0.930396$  & $   -0.930085$  \\
10                & $   -0.963774$ & $   -0.944656$ & $   -0.943675$  & $   -0.943680$  \\
12                & $   -0.945928$ & $   -0.943369$ & $   -0.955824$  & $   -0.951481$  \\
14                & $   -0.972895$ & $   -0.975618$ & $   -0.960695$  & $   -0.958120$  \\
16                & $   -0.961803$ & $   -0.966478$ & $   -0.964733$  & $   -0.957612$  \\
18                & $   -0.978161$ & $   -0.974784$ & $   -0.970367$  & $   -0.966040$  \\\hline
$\inf$            & $   -1.0006  $ & $   -1.009   $ & $   -0.995   $  & $   -0.97    $  \\\hline
\end{tabular}
\caption{\Pade and \PadeBorel approximations of the $D_1$ invariant of the MSZ lump solution with extrapolations.}
\label{tab:FB Pade1}
\vspace{5mm}
\begin{tabular}{|l|llll|}\hline
Level \rowh{12pt} & $E_5^P$   & $\ps W_2^P$    & $\ps D_4^P$   & $\ps D_9^P$   \\\hline
2                 & $0      $ & $\ps 0.140372$ & $\ps 0.10875$ & $\ps 0.05115$ \\
4                 & $0      $ & $   -0.697061$ & $   -2.13552$ & $   -2.50729$ \\
6                 & $0      $ & $   -0.220903$ & $   -2.39626$ & $   -2.85202$ \\
8                 & $0      $ & $   -0.097232$ & $   -2.21191$ & $   -2.56816$ \\
10                & $3.81063$ & $   -0.181509$ & $   -2.95633$ & $   -3.13168$ \\
12                & $1.27485$ & $   -0.324430$ & $   -2.04616$ & $   -2.44598$ \\
14                & $0.52410$ & $   -0.255078$ & $   -1.09307$ & $   -0.42430$ \\
16                & $0.81695$ & $\ps 0.060441$ & $\ps 0.71694$ & $\ps 25.0938$ \\
18                & $0.85959$ & $   -0.076353$ & $\ps 0.79231$ & $\ps 4.06549$ \\\hline
Exp.              & $1      $ & $\ps 0       $ & $\ps 1      $ & $   -1      $ \\\hline
\end{tabular}
\caption{\Pade approximations of selected invariants with $h>1$ for the MSZ lump solution.}
\label{tab:FB Pade2}
\end{table}

Therefore it makes sense to use these approximations only for nonconvergent invariants. We show some examples in table \ref{tab:FB Pade2}. The \Pade approximations of the $E_5$ invariant with weight $\frac{25}{12}$ and of the $W_2$ invariant with weight $3$ give us roughly the expected results, but if we take invariants with even higher conformal weights, this method becomes unreliable as well. Take a look at \Pade approximations of the $D_4$ invariant, which should be equal to 1. The huge oscillations have disappeared, but the results below level 16 are far from correct. The approximation jumps to positive numbers at level 16, but we would need several more levels to confirm whether this is just a coincidence or not. The results for the $D_9$ invariant are even worse. Interestingly, \Pade approximations of both invariants change significantly at level 16, which could be caused by appearance of the $P_{16}$ Virasoro primary.

\FloatBarrier
\section{Double lump solutions}\label{sec:FB circle:double}
In this section, we analyze double lump solutions, which describe two D0-branes on a circle. These solutions have usually very similar properties as single lump solutions. Actually, some double lump solutions, concretely those describing D0-branes at opposite points of the circle, are double copies of single lump solutions from radii $R/2$. These solutions have the same string field and their observables are multiplied by two (with the exception of some necessary changes in $E_n$ and $W_n$ invariants). Similarly to single lumps solutions, double lump solutions are real and well-behaved above $R=2$ and complex below $R=2$.

The main focus of this section will not be agreement between observables and predicted boundary states, but distances between lumps. Generically, two D0-branes on a circle have two parametric modulus, but, because we impose the parity even condition, we can find only lumps positioned symmetrically around the origin, which can be described just by one parameter. However, the level truncation approximation breaks the exact symmetry and we find just a finite set of solutions, which describe lumps at discreet distances. We expect that the modulus should be restored in the infinite level limit, so the set of solutions should be getting more dense with increasing level.

Before we get to the actual results, we introduce a parametrization of position of the two lumps. The positions of two symmetrically placed D0-branes can be written as $x_1=\pi R s$ and $x_2=-\pi R s$ (or equivalently $x_2=\pi R(2-s)$), which gives them distance $2\pi R s$, where $s$ is a dimensionless parameter between 0 and 1\footnote{The relation between $s$ and the parametrization in \cite{KMS} is $s=1-a$.}. Furthermore, we can restrict the parameter to $s\leq\frac{1}{2}$ because the values $s$ and $1-s$ are related just by a translation by half of the circle length.

Given a double lump solution, we can determine the distance between lumps from the $E_n$ invariants. The formula (\ref{En predicted}) tells us that the expected values of these invariants are
\begin{equation}
E_n=2\cos(n\pi s).
\end{equation}
The best choice is to express the distance using the $E_1$ invariant as
\begin{equation}
s=\frac{1}{\pi}\arccos\left(\frac{E_1}{2}\right)
\end{equation}
because this invariant is usually the most precise one and determining $s$ from higher invariants is more complicated because we have to deal with the choice of a correct branch of the arccosine, see \cite{KMS} for more details.

\FloatBarrier
\subsection{Double lump solution at $R=3$}\label{sec:FB circle:double:example}
We begin with a closer look at one example of a double lump solution. The reference \cite{KMS} shows a double lump solution at $R=2\sqrt3$, but, since this solution is complex at some levels, we have chosen a different solution at radius $R=3$ here.

Solutions of level 2 equations of motion include two independent double lump solutions. One of them is just a double copy of the single lump solution from $R=\frac{3}{2}$. It describes D0-branes positioned directly opposite to each other, so we investigate the other solution, where the lumps have a more generic distance.

We show some observables of this solution in table \ref{tab:FB Double lump}. As for the MSZ lump solution, the energy, $E_0$, $D_1$ and $\Delta_S$ are in good agreement with the expected values, so we focus on $E_n$ invariants. The extrapolation of the $E_1$ invariant can be used to compute the relative distance between the two lumps, which is
\begin{equation}\label{FB DL distance}
s^{(1)}=0.3415\pm 0.0008.
\end{equation}
We use this distance to predict values of the other $E_n$ invariants in table \ref{tab:FB Double lump} and we can see that they agree with the predictions within the statistical errors. The actual distance between the two lumps in multiples of $2\pi$ is $s R=1.024\pm 0.003$. Since this number is higher than 1, there are no stretched tachyonic strings in the spectrum of excitations around this solution. We will discuss the importance of this issue in the next subsection.

\begin{table}[!]
\centering
\footnotesize
\begin{tabular}{|l|lllll|}\hline
Level    & \ps Energy    & $\ps \Delta_S $ & $\ps D_1    $  & $E_0$   & $E_1$    \\\hline
2        & \ps 2.17133   & $\ps 0.0275172$ & $   -1.83664$  & 2.13635 & 0.623568 \\
4        & \ps 2.06160   & $\ps 0.0042647$ & $   -1.41075$  & 2.05420 & 0.679687 \\
6        & \ps 2.03669   & $\ps 0.0018829$ & $   -2.12253$  & 2.03041 & 0.740993 \\
8        & \ps 2.02610   & $\ps 0.0010114$ & $   -1.73777$  & 2.02286 & 0.788949 \\
10       & \ps 2.02029   & $\ps 0.0005845$ & $   -2.08833$  & 2.01655 & 0.818299 \\
12       & \ps 2.01663   & $\ps 0.0003446$ & $   -1.83192$  & 2.01401 & 0.839221 \\
14       & \ps 2.01412   & $\ps 0.0001980$ & $   -2.06422$  & 2.01108 & 0.854017 \\
16       & \ps 2.01228   & $\ps 0.0001031$ & $   -1.87875$  & 2.00986 & 0.865900 \\
18       & \ps 2.01087   & $\ps 0.0000391$ & $   -2.04820$  & 2.00817 & 0.875055 \\\hline
$\inf$   & \ps 2.00013   & $   -0.000045 $ & $   -2.000  $  & 1.9986  & 0.955    \\
$\sigma$ & \ps 0.00001   & $\ps 0.000006 $ & $\ps 0.030  $  & 0.0002  & 0.004    \\\hline
Exp.     & \ps 2         & $\ps 0        $ & $   -2      $  & 2       & 0.955    \\\hline
\multicolumn{6}{l}{}\\[-5pt]\hline
Level    & $\ps E_2    $ & $\ps E_3    $   & $\ps E_4     $ & $E_5$   & $E_6$    \\\hline
2        & $   -1.12760$ & $   -1.87080$   & $   -0.260634$ & 0       & 0        \\
4        & $   -1.30866$ & $   -1.73432$   & $\ps 0.018296$ & 2.54195 & 2.24519  \\
6        & $   -1.29741$ & $   -1.85655$   & $   -0.163463$ & 1.31848 & 0.58457  \\
8        & $   -1.25599$ & $   -1.87273$   & $   -0.275079$ & 1.82313 & 2.24962  \\
10       & $   -1.23106$ & $   -1.91558$   & $   -0.399237$ & 1.45064 & 1.14686  \\
12       & $   -1.21002$ & $   -1.91998$   & $   -0.441497$ & 1.59977 & 2.15322  \\
14       & $   -1.19602$ & $   -1.93939$   & $   -0.508464$ & 1.43725 & 1.42902  \\
16       & $   -1.18330$ & $   -1.94115$   & $   -0.530770$ & 1.49886 & 2.08692  \\
18       & $   -1.17404$ & $   -1.95228$   & $   -0.573408$ & 1.40911 & 1.58523  \\\hline
$\inf$   & $   -1.083  $ & $   -1.994  $   & $   -0.83    $ & 1.21    & 2.01     \\
$\sigma$ & $\ps 0.006  $ & $\ps 0.004  $   & $\ps 0.02    $ & 0.04    & 0.19     \\\hline
Exp.     & $   -1.087  $ & $   -1.994  $   & $   -0.82    $ & 1.21    & 1.98     \\\hline
\end{tabular}
\caption{Selected observables of a double lump solution at $R=3$ up to level 18. We show only well convergent Ellwood invariants at even levels to reduce the amount of data. The expectation values of $E_n$ invariants are based on the lump distance computed from $E_1$.}
\label{tab:FB Double lump}
\end{table}

\begin{figure}[!]
   \centering
   \begin{subfigure}[t]{0.47\textwidth}
      \includegraphics[width=\textwidth]{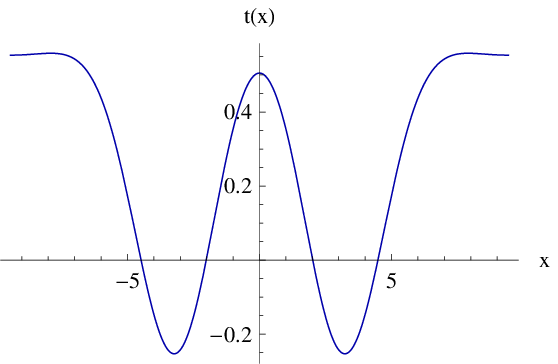}
   \end{subfigure}\qquad
   \begin{subfigure}[t]{0.47\textwidth}
      \includegraphics[width=\textwidth]{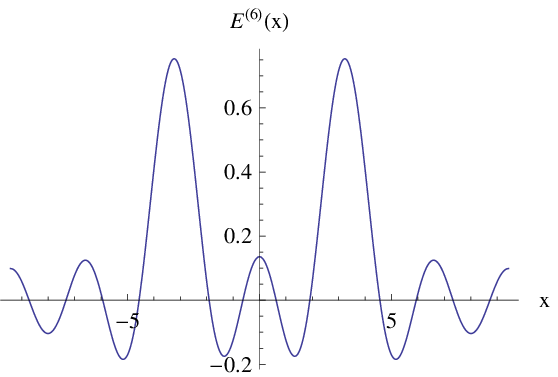}
   \end{subfigure}
\caption{Visualization of a double lump solution at $R=3$. On the left: Plot of the extrapolated tachyon profile $t(x)$. On the right: Plot of the energy density profile $E^{(6)}(x)$ using 6 momentum harmonics.}
\label{fig:Double lump profiles}
\end{figure}

Alternatively, we can try to determine the lump distance from positions of minima of the tachyon profile, which is shown in figure \ref{fig:Double lump profiles}. We obtain
\begin{equation}
s^{(tach)}=0.3430,
\end{equation}
which is quite close to (\ref{FB DL distance}) despite the fact that the tachyon profile is not gauge invariant. The maxima of the energy density profile, which is also shown in figure \ref{fig:Double lump profiles}, do not allow us to determine the distance with a good precision. By analyzing distances between maxima of truncated Fourier series of two delta functions, we find that they reproduce the correct distance only in the $n\rar \inf$ limit, which is not accessible for numerical solutions.

\FloatBarrier
\subsection{Superposition of lumps}\label{sec:FB circle:double:superposition}
When we want to find as many double lump solutions as possible, the traditional use of the level truncation approach is not the best choice. The homotopy continuation method, which can be used approximately up to level 3 for $R>2$, produces only a limited number of seeds. And as we improve them to higher levels, some of them often merge together, so the number of distinct solutions actually decreases with level. Therefore we introduce a new way to generate seeds for Newton's method, which is based on superposition of single lump solutions.

We take inspiration from tachyon profiles of lump solutions. Tachyon profiles of all known lump solutions on large enough radius follow a very simple pattern. Far enough from lump positions, the profile is almost flat and the tachyon field here has approximately the same value as for the tachyon vacuum solution. The lumps themselves are represented by gaussians subtracted from the constant part (see figures \ref{fig:MSZ tach profile} and \ref{fig:Double lump profiles}).

Therefore we can decompose a single lump solution located at the origin $|\Psi_{SL}(0)\ra$ into the tachyon vacuum solution and a lump 'perturbation' $|\Psi_{{lump}}(0)\ra$,
\begin{equation}
|\Psi_{SL}(0)\ra=|\Psi_{TV}\ra+|\Psi_{lump}(0)\ra.
\end{equation}
The tachyon vacuum is translation invariant, while the 'perturbation' can be moved to an arbitrary position. Using the parameter $s$, we define
\begin{equation}
|\Psi_{lump}(s)\ra=e^{i\pi R s p}|\Psi_{lump}(0)\ra.
\end{equation}
Now we can construct approximate double lump solutions $|\Psi_{DL}(s)\ra$ using superposition of single lumps as\footnote{This approximate solution is formally very similar to double lump solutions in \cite{ErlerMaccaferri}.}
\begin{eqnarray}\label{DL superposition}
|\Psi_{DL}(s)\ra&=&|\Psi_{TV}\ra+|\Psi_{lump}(s)\ra+|\Psi_{lump}(-s)\ra \nn \\
&=&|\Psi_{SL}(s)\ra+|\Psi_{SL}(-s)\ra-|\Psi_{TV}\ra.
\end{eqnarray}
When we translate this formula to coefficients of this string field, we find
\begin{equation}\label{DL seed}
t_i^{DL}(s)=2\cos(\pi R s k_i)t_i^{SL}-t_i^{TV},
\end{equation}
where $k_i$ is the momentum of the $i$th basis state. These approximate double lumps can be used as seeds for Newton's method.

In order to find how many double lump solutions exist for a given level and radius, we scan over a set of initial distances $s_{init}$ and then we make union of the double lump solutions we find. We have chosen to test the superposition method on three different radii, $R=3,4,5$. We have been able to go up to level 12 using 1000 of equally spaced initial points for each radii. The results not exactly as we expected.

\begin{figure}[t!]
\centering
   \includegraphics[width=10cm]{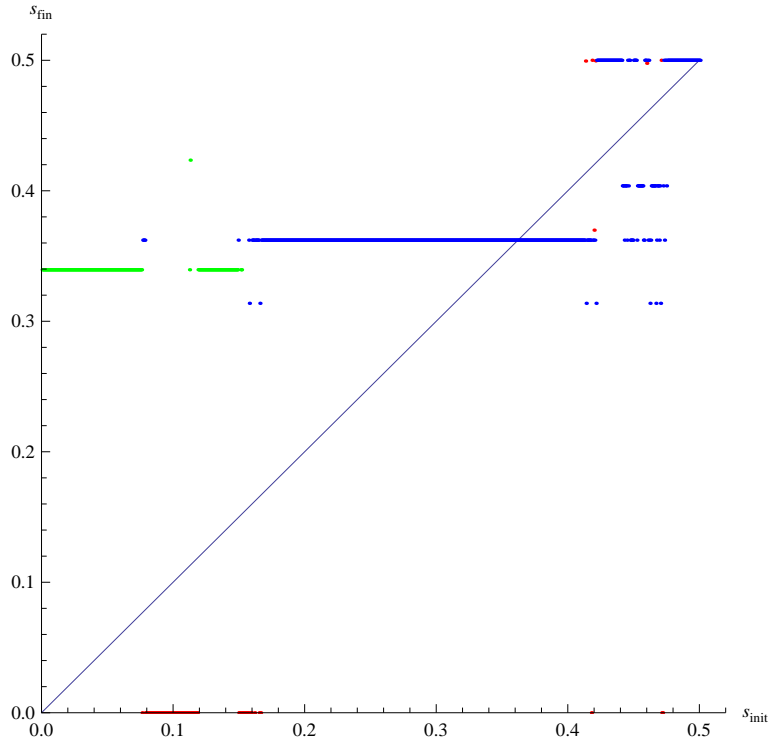}
\caption{Plot of final distances $s_{fin}$ against initial distances $s_{init}$ of double lump solutions obtained by the superposition method at level 12 and radius $R=3$. The sample contains 1000 equally spaced initial distances. Blue points represent double lumps solutions, green points other solutions and red points indicate  seeds that have not converged within 20 iterations. The "distances" for other solutions than double lumps do not have any physical meaning. The blue line indicated the ideal relation $s_{fin}=s_{init}$.}
\label{fig:Double lump sup R3 lev12}
\end{figure}

First, we are going to present few examples of results of this method. Figure \ref{fig:Double lump sup R3 lev12} shows 1000 relative distances that we have found at radius $R=3$ and level 12. Double lump solutions are represented by blue dots, green dots represent other solutions (usually single lumps) and red dots represent seeds that have not converged within 20 iterations of Newton's method. This figure illustrates a typical results of the superposition method at low radius. We find only few independent distances, which do not follow the ideal relation $s_{fin}=s_{init}$ at all. We observe that there are several larger areas where all seeds converge to the same solution. Solutions on borders of these areas alternate between several different distances. That is not surprising because domains of convergence of Newton's method often have fractal shapes.

\begin{figure}[t!]
\centering
   \includegraphics[width=10cm]{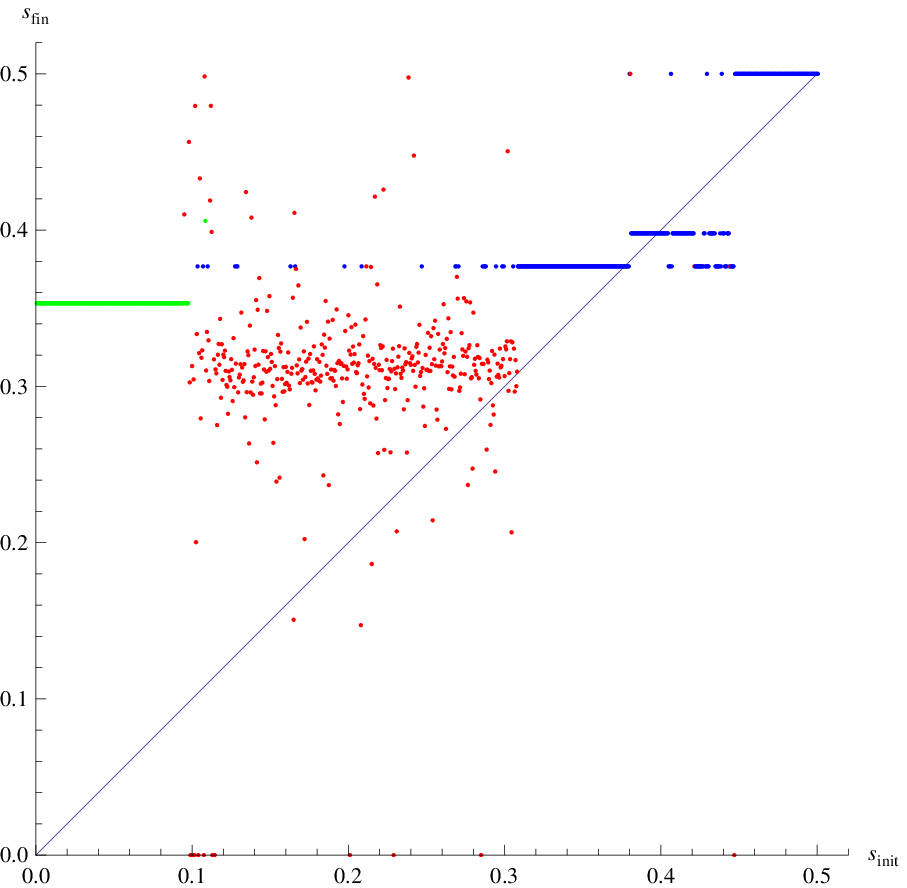}
\caption{Plot of final distances $s_{fin}$ against initial distances $s_{init}$ of double lump solutions obtained by the superposition method at level 4 and radius $R=4$. }
\label{fig:Double lump sup R4 lev4}
\end{figure}

\begin{figure}[t!]
\centering
   \includegraphics[width=10cm]{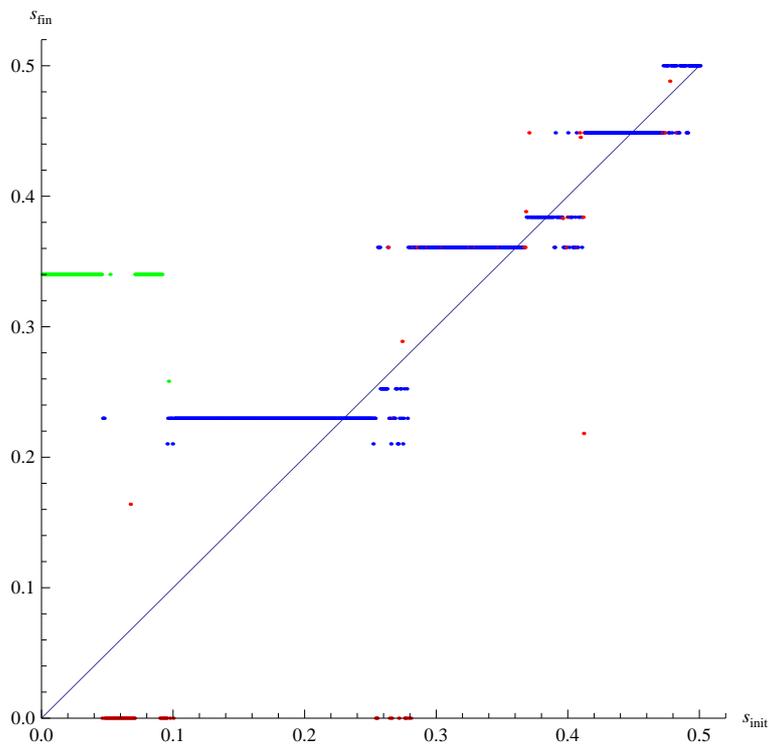}
\caption{Plot of final distances $s_{fin}$ against initial distances $s_{init}$ of double lump solutions obtained by the superposition method at level 11 and radius $R=5$.}
\label{fig:Double lump sup R5 lev11}
\end{figure}

Another example in figure \ref{fig:Double lump sup R4 lev4} shows the results from level 4 at radius $R=4$. This figure is atypical because there are many nonconvergent seeds and the "distances" computed from the approximate solutions concentrate around a certain point. This indicates existence of a complex double lump solution, which cannot be found from real seeds. In order to see it, we could add a small imaginary part to (\ref{DL seed}). Finally, in figure \ref{fig:Double lump sup R5 lev11}, we show the results from level 11 at radius $R=5$. The number of independent solutions has grown compared to the lower radii and there is finally a stronger correlation between initial and final values of the parameter $s$.

\begin{figure}
\centering
\includegraphics[width=10cm]{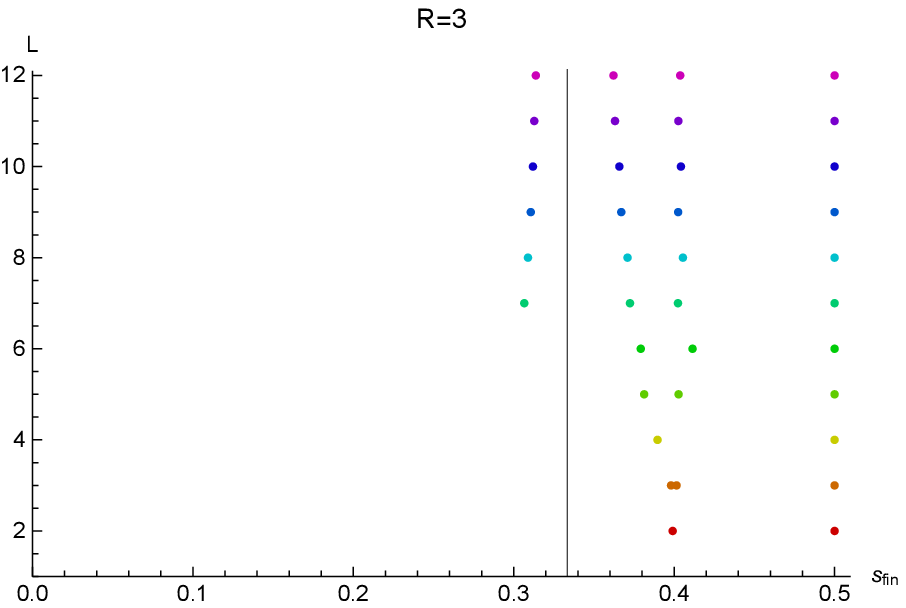}\vspace{5mm}
\includegraphics[width=10cm]{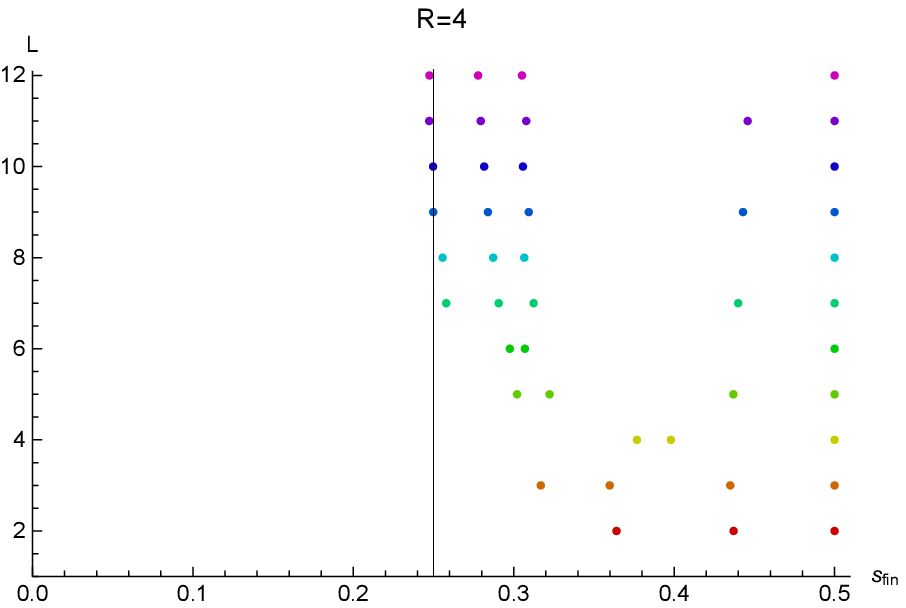}\vspace{5mm}
\includegraphics[width=10cm]{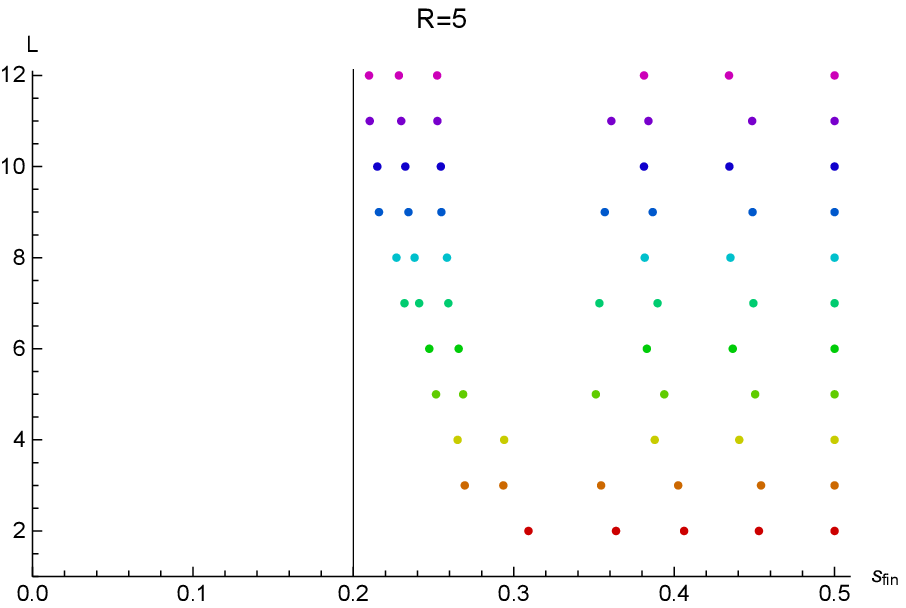}
\caption{Summary of relative distances of double lump solutions found by superpositions of single lumps at levels from 2 to 12. The top figure depicts $R=3$, the middle figure $R=4$ and the bottom figure $R=5$. The vertical lines correspond to $\Delta x=2\pi$.}
\label{fig:Double lump superposition}
\end{figure}

In the end, it is more important to know the number of independent double lump solutions than which seeds they come from. We summarize the relative distances we have found at different levels at the three radii in question in figure \ref{fig:Double lump superposition}. As we expected, the number of solutions grows with the radius, usually by one or two depending on the level. But their number does not significantly increase with the level. With the exception of some irregularities, there is only one new solution at level 3 at all three radii, another one at level 7, and that is all. That means either that it is necessary to go to very high levels to see a dense set of solutions or that the conjecture that the modulus should be restored in the $L\rar \inf$ limit is wrong. Of course, there is no guarantee that we have found all independent solutions, but we do not think that this is a likely possibility. At low levels, where we can find all solutions explicitly using the homotopy continuation method, the superposition methods gives us all double lump solutions. If we have missed some solutions, they must be quite different from traditional double lumps.

Next, we observe that the distribution of distances is not even. There are no solutions with low distances, see the discussion later, and there are several large gaps between the solutions. There are also sometimes big differences between distances at adjacent levels, which tells us that the way we truncate the string field has a large impact on which discreet distances we find.

\begin{figure}
   \centering
   \begin{subfigure}[t]{0.45\textwidth}
      \includegraphics[width=\textwidth]{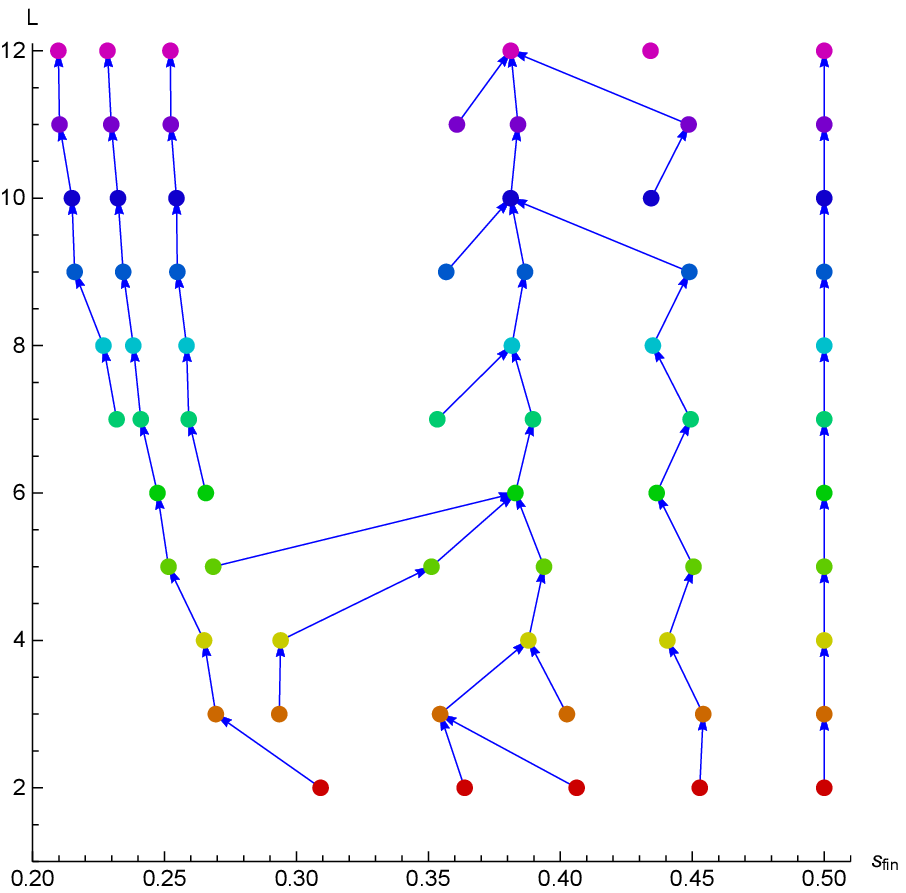}
   \end{subfigure}\qquad
   \begin{subfigure}[t]{0.45\textwidth}
      \includegraphics[width=\textwidth]{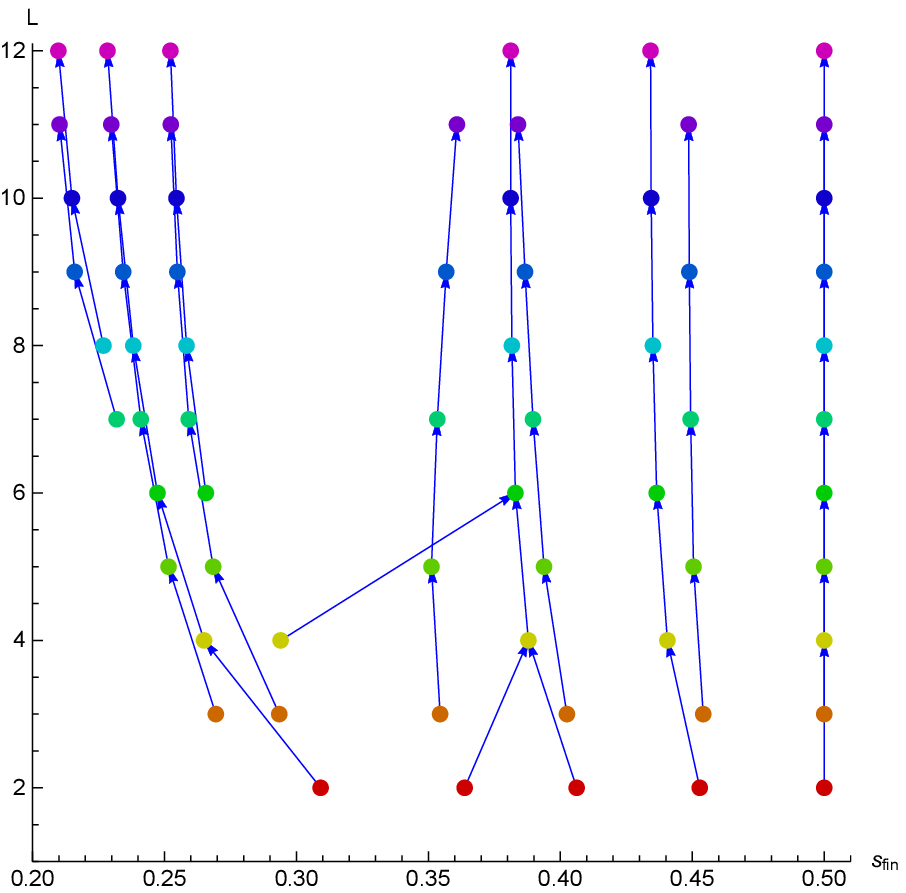}
   \end{subfigure}
\caption{Relations between double lump solutions at different levels at radius $R=5$. The arrows in the left figure represent improvement of solutions by one level using Newton's method, in the right figure improvement by two levels.}
\label{fig:Double lump sup R5 arrows}
\end{figure}

The differences between adjacent levels mean that there is no unique way to match solutions at different levels. For example, there is one solution  at radius $R=4$ that appears only at odd levels. To explore this issue, we have tried to find relations between all solutions at all levels using  Newton's method. This also gives us a partial check that there are no solutions missed by the superposition technique. Figure \ref{fig:Double lump sup R5 arrows} shows the most interesting part of the results, we compare improvement of solutions by one and two levels at radius $R=5$. We can see that some seeds can lead to different solutions depending on which pattern we choose for Newton's method. We also observe that solutions two levels apart are generally easier to match than solutions from adjacent levels. Sometimes we cannot be sure whether distances from even and odd levels converge to the same number and therefore we should be careful when extrapolating these solutions.

We notice that there is a relatively large minimal distance at all radii, below which we find no double lump solutions. This relative distance is approximately $s\approx 1/R$, which correspond to the absolute distance $\Delta x\!\approx\! 2\pi$.
The distance $2\pi$ has a well-known meaning. There are marginal modes in the spectrum of strings stretched between two D0-branes separated by this distance and if the separation decreases, the lightest modes become tachyonic. Therefore it seems that it is difficult to find double lump solutions with this type of tachyonic excitations. As we are going to see later in sections \ref{sec:FB D2:intersection} and \ref{sec:MM:Ising3:regular}, a similar rule applies to other models as well.

We have found one solution with $\Delta x<2\pi$, it is the solution with the smallest distance at $R=3$. The seed for this solution from level 7 can be improved to level 18 without any problems. Its invariants oscillate more than in case of other lump solutions, but otherwise it seems to be consistent. By extrapolating the $E_1$ invariant of this solution, we obtain
\begin{eqnarray}
s^{(\inf)}&=&0.3248\pm 0.0005, \\
\Delta x^{(\inf)}&=&2\pi (0.974\pm 0.002).
\end{eqnarray}
This confirms that the separation of the two D0-branes is lower than $2\pi$ and therefore there are slightly tachyonic stretched strings in their spectrum.

When we try to extend the solution below level 7, we find a problem. Newton's method either does not converge or converges to a clearly different solution. However, we can add a small imaginary part to the level 7 seed and then we find a complex double lump solution at levels 3 to 6. Therefore it seems that this solution is difficult to find because it is complex at low levels, similarly to the positive energy single lumps from section \ref{sec:FB circle:single:smallR}. This conclusion is in agreement with our findings in section \ref{sec:MM:Ising3:regular}.

\FloatBarrier
\subsection{Double lump solution at $R=2$}\label{sec:FB circle:double:R=2}
At the end of this section, we are going to describe two exceptional double lump solutions, which have some unusual properties.

We start with a double lump solution at radius $R=2$, which describes two D0-branes directly opposite to each other. Why do we find this solution, when we have not found any regular lump solutions at $R=1$? The reason lies in the structure of momentum fields. The marginal field at this radius is the second momentum field, the first momentum state is a relevant field and therefore it is actually possible to solve the equation for the marginal field by giving a purely imaginary value to the relevant field.

Observables of this solution are shown in table \ref{tab:FB DL R2}. This solution does not satisfy the usual reality condition, but has real energy, so we call is pseudo-real. The structure of the string field is as follows: Coefficients of even momentum states are real and coefficients of odd momentum states are purely imaginary, which can be deduced from the $E_n$ invariants. This structure is possible because odd momentum states contribute to the action only in pairs. Therefore the imaginary units cancel each other and the solution can have real energy. Curiously, the solution does not have the same symmetry as the boundary state it describes. The two D0-branes (and the real part of the string field) are invariant under translation by half of the circle length, but the imaginary part of the string field is not.

\begin{figure}
\centering
   \includegraphics[width=10cm]{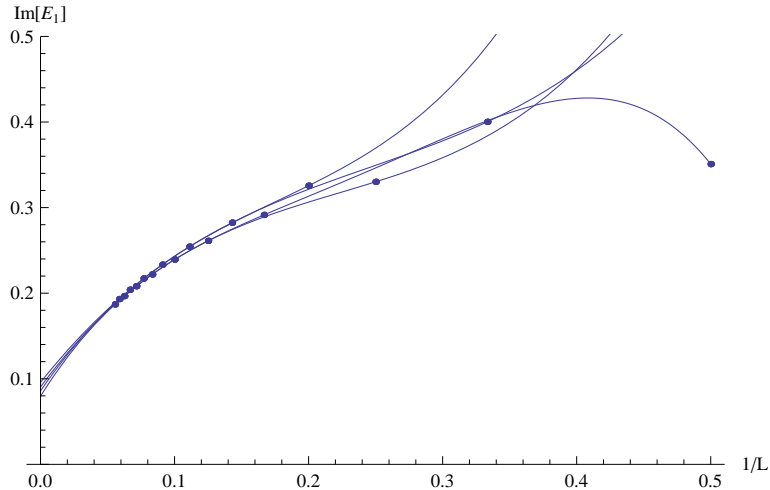}
\caption{Extrapolation of the imaginary part of the $E_1$ invariant of the pseudo-real double lump solution at $R=2$.}
\label{fig:FB DL Im E1}
\end{figure}

We are almost sure that this solution describes two D0-branes and not a generic $\cos X$ marginal deformation. The reason is that it can be smoothly deformed to nearby radii, where this type of marginal deformation is not allowed. The real invariants reproduce the expected boundary state with similar precision as other double lump solutions and $\Delta_S$ is close to zero, so the complex invariants $E_1$ and $E_3$ are the only issues. They should ideally disappear in the infinite level limit. They indeed decrease with level, but their extrapolations suggest that they do not go to zero and the same goes for imaginary coefficients of the string field. We show the extrapolation of the $E_1$ invariant in figure \ref{fig:FB DL Im E1} as an example. However, given the nice properties of the solution, we believe that its imaginary part will eventually disappear. We do not have much experience with solutions of this type and it is quite possible that its imaginary part has an unusual asymptotic behavior, which cannot be captured by our extrapolation techniques. It could decrease for example as $\frac{1}{\sqrt L}$.

This solution is useful because it allows us to extract some information about the marginal solution on the self-dual radius which corresponds to a D0-brane. We are interested in the zero momentum tachyon $t_0$ and in the marginal field, which is denoted as $t_2$ in this context. These values can be compared with the results from single lump solutions and from marginal deformations. The values of these two coefficients extrapolated to level infinity are
\begin{eqnarray}
t_0 &=&0.148207\pm 0.000008,  \\
t_2 &=&0.37534\pm 0.00002.
\end{eqnarray}

\begin{table}[h]
\centering
\begin{tabular}{|l|llll|}\hline
Level    & Energy  & $\ps \Delta_S$   & $\ps D_1$    & $\ps W_1$       \\\hline
2        & 2.05128 & $   -0.0130696$ & $   -2.67169$ & $   -2.67169$   \\
4        & 2.05729 & $   -0.0078289$ & $   -2.03525$ & $\ps 2.82656$   \\
6        & 2.04269 & $   -0.0064489$ & $   -3.23838$ & $   -2.17629$   \\
8        & 2.03348 & $   -0.0053473$ & $   -2.13530$ & $\ps 1.29709$   \\
10       & 2.02749 & $   -0.0045600$ & $   -2.79584$ & $   -1.18644$   \\
12       & 2.02332 & $   -0.0039786$ & $   -2.14875$ & $\ps 0.73298$   \\
14       & 2.02025 & $   -0.0035324$ & $   -2.57164$ & $   -0.76245$   \\
16       & 2.01791 & $   -0.0031791$ & $   -2.14580$ & $\ps 0.47297$   \\
18       & 2.01606 & $   -0.0028920$ & $   -2.44087$ & $   -0.53911$   \\\hline
$\inf  $ & 2.00006 & $   -0.0000340$ & $   -2.01   $ & $\ps 0.00   $   \\
$\sigma$ & 0.00001 & $\ps 0.000007 $ & $\ps 0.07   $ & $\ps 0.12   $   \\\hline
Exp.     & 2       & $\ps 0        $ & $   -2      $ & $\ps 0      $   \\\hline
\multicolumn {5}{l}{}\\[-5pt]\hline
Level    & $E_0$   & $\ps E_1$       & $\ps E_2$     & $\ps E_3$       \\\hline
2        & 1.92947 & $\ps 0.350852i$ & $   -2.29210$ & $\ps 0        $ \\
4        & 1.94841 & $\ps 0.330100i$ & $   -2.14142$ & $   -1.545560i$ \\
6        & 1.95570 & $\ps 0.291433i$ & $   -2.22384$ & $   -0.842663i$ \\
8        & 1.96266 & $\ps 0.261147i$ & $   -2.11874$ & $   -0.934474i$ \\
10       & 1.96659 & $\ps 0.239346i$ & $   -2.13678$ & $   -0.735506i$ \\
12       & 1.97047 & $\ps 0.221958i$ & $   -2.09152$ & $   -0.741258i$ \\
14       & 1.97295 & $\ps 0.208190i$ & $   -2.09884$ & $   -0.644309i$ \\
16       & 1.97545 & $\ps 0.196622i$ & $   -2.07378$ & $   -0.636889i$ \\
18       & 1.97716 & $\ps 0.186943i$ & $   -2.07759$ & $   -0.578248i$ \\\hline
$\inf  $ & 1.9963  & $\ps 0.088   i$ & $   -2.003  $ & $   -0.26    i$ \\
$\sigma$ & 0.0009  & $\ps 0.007   i$ & $\ps 0.001  $ & $\ps 0.02    i$ \\\hline
Exp.     & 2       & $\ps 0        $ & $   -2      $ & $\ps 0        $ \\\hline
\end{tabular}
\caption{Selected observables of the pseudo-real double lump solution at $R=2$.}
\label{tab:FB DL R2}
\end{table}

\FloatBarrier
\subsection{Twist non-even double lump solution at $R=\sqrt{3}$}\label{sec:FB circle:double:twist}
Finally, we present one twist non-even double lump solution. So far, this is the only interesting twist non-even solution we have found in the free boson theory. We have found it at the MSZ radius $R=\sqrt{3}$, but it can be smoothly deformed to nearby radii as well. The solution appears in two copies, which differ by sign of the twist odd part. Interestingly, the twist odd part does not contribute to any Ellwood invariant, so both solutions describe the same boundary state and they should be gauge equivalent. The twist odd part of the solution does not satisfy the usual reality condition, so this solution is pseudo-real similarly to the double lump in the previous subsection.

Some interesting observables of this solution are summarized in table \ref{tab:FB non-even double lump}. This solution describes the same boundary state as the twist even complex double lump which can be obtained by doubling the single lump solution from radius $R=\frac{\sqrt 3}{2}$. Therefore both solutions should be asymptotically the same, or at least gauge equivalent. The twist odd part of the solution decreases with level, but we cannot say for sure whether it disappears in the $L\rar \inf$ limit.

The solution behaves really badly at low levels, but it slowly improves. Asymptotic values of observables have a slightly better agreement with the expected boundary state than in case of the solution in table \ref{tab:FB positive lump}, but only because the observables are real and therefore we have more data points for extrapolations.

\begin{table}[!]
\centering
\footnotesize
\begin{tabular}{|l|llllll|}\hline
Level    & Energy  & $\ps \Delta_S$ & $\ps D_1$      & $E_0$   & $\ps E_1$ & $\ps W_1$     \\\hline
2        & 4.01352 & $   -0.753496$ & $   -65.8593$  & 1.89500 & 19.9072   & $   -48.9207$ \\
4        & 2.50133 & $   -0.463919$ & $   -21.7271$  & 1.69355 & 4.93716   & $\ps 3.70801$ \\
6        & 2.20056 & $   -0.120485$ & $   -11.0718$  & 1.88940 & 3.92636   & $   -7.21894$ \\
8        & 2.12761 & $   -0.070034$ & $   -4.40184$  & 1.92197 & 2.83124   & $\ps 0.64629$ \\
10       & 2.09388 & $   -0.049952$ & $   -5.97748$  & 1.93809 & 2.86987   & $   -3.02191$ \\
12       & 2.07437 & $   -0.039229$ & $   -3.54831$  & 1.94830 & 2.55657   & $\ps 0.12627$ \\
14       & 2.06163 & $   -0.032574$ & $   -4.58410$  & 1.95531 & 2.58583   & $   -1.86027$ \\
16       & 2.05264 & $   -0.028043$ & $   -3.23501$  & 1.96074 & 2.43516   & $   -0.06241$ \\
18       & 2.04596 & $   -0.024758$ & $   -3.93794$  & 1.96486 & 2.45429   & $   -1.32872$ \\\hline
$\inf  $ & 1.9986  & $   -0.00033 $ & $   -1.2    $  & 2.012   & 1.995     & $   -0.06   $ \\
$\sigma$ &         & $            $ & $\ps 0.9    $  & 0.009   & 0.044     & $\ps 0.48   $ \\\hline
Expected & 2       & $\ps 0       $ & $   -2      $  & 2       & 2         & $\ps 0      $ \\\hline
\end{tabular}
\caption{Selected observables of a twist non-even pseudo-real double lump solution at radius $R=\sqrt{3}$. We are not able to get a reasonable error estimate for the energy and $\Delta_S$.}
\label{tab:FB non-even double lump}
\end{table}

\FloatBarrier
\section{Wilson line solutions}\label{sec:FB circle:other}
In the final section of this chapter, we show few free boson solutions which are not lumps, but which describe D1-branes with nontrivial Wilson lines. These solutions suffer from various problems, so we are not sure whether they are really physical, but we think that they are worth further investigation.

\subsection{Radius independent Wilson line solutions}\label{sec:FB circle:other:WL}
First, we describe two U(1) universal solutions, by which we mean that the do not excite any of momentum states and therefore they do not depend on the radius. These solutions appear at level 2 and we have evaluated them up to level 20 at radius $R=0.1$. We have chosen such small radius in order to reduce the number of momentum states and because there is a large number of $W_n$ invariants with low conformal weights.

\begin{table}[!]
\centering
\scriptsize
\begin{tabular}{|l|llll|}
\multicolumn {5}{c}{\normalsize First solution}\\\hline
Level \rowh{8pt} &  \ps Energy              & $\ps \Delta_S$           & $E_0$                & $\ps D_1$                \\\hline
2                & $\ps 0.213763-0.259039i$ & $   -0.180560+0.290735i$ & $0.210728-0.003071i$ & $   -2.261770+1.86169i$  \\
4                & $\ps 0.261880-0.147938i$ & $   -0.119821+0.187750i$ & $0.206060-0.000874i$ & $\ps 0.821550-3.05230i$  \\
6                & $\ps 0.258631-0.099257i$ & $   -0.099764+0.136372i$ & $0.200969+0.004076i$ & $   -0.630079+1.25736i$  \\
8                & $\ps 0.252080-0.075209i$ & $   -0.085587+0.109749i$ & $0.199168+0.005189i$ & $\ps 0.912405-2.11730i$  \\
10               & $\ps 0.246542-0.061033i$ & $   -0.075606+0.093581i$ & $0.197581+0.006066i$ & $   -0.324981+1.10484i$  \\
12               & $\ps 0.242123-0.051668i$ & $   -0.068251+0.082608i$ & $0.196858+0.006354i$ & $\ps 0.799558-1.68726i$  \\
14               & $\ps 0.238569-0.044997i$ & $   -0.062585+0.074587i$ & $0.196179+0.006611i$ & $   -0.188202+0.97960i$  \\
16               & $\ps 0.235657-0.039987i$ & $   -0.058064+0.068416i$ & $0.195836+0.006681i$ & $\ps 0.709252-1.42756i$  \\
18               & $\ps 0.233227-0.036075i$ & $   -0.054353+0.063486i$ & $0.195492+0.006754i$ & $   -0.107857+0.87927i$  \\
20               & $\ps 0.231166-0.032928i$ & $   -0.051238+0.059434i$ & $0.195315+0.006745i$ & $\ps 0.641497-1.24802i$  \\\hline
$\inf$           & $\ps 0.2040  -0.0016  i$ & $   -0.01    +0.01    i$ & $0.1943  +0.0056  i$ & $\ps 0.226   +0.008  i$  \\\hline
Exp.             & $\ps 0.2               $ & $\ps 0                 $ & $0.2               $ & $\ps 0.2              $  \\\hline
\multicolumn {5}{c}{ }\\
\multicolumn {5}{c}{\normalsize Second solution}\\\hline
Level \rowh{8pt} & $\ps$ Energy             & $\ps \Delta_S$          & $E_0$                 & $\ps D_1$                \\\hline
2                & $   -0.191060-0.351735i$ & $\ps 1.55905-2.323930i$ & $0.237136-0.062213i$  & $   -0.705651+1.399001i$ \\
4                & $\ps 0.100912-0.279609i$ & $\ps 1.26697-0.799223i$ & $0.226314-0.031223i$  & $   -0.534327-0.904116i$ \\
6                & $\ps 0.188610-0.225048i$ & $\ps 1.05852-0.339393i$ & $0.224497-0.017720i$  & $   -0.438447+0.463401i$ \\
8                & $\ps 0.223805-0.183520i$ & $\ps 0.90560-0.135597i$ & $0.223062-0.010557i$  & $   -0.188985-0.962660i$ \\
10               & $\ps 0.239433-0.153592i$ & $\ps 0.79246-0.029478i$ & $0.221289-0.005634i$  & $   -0.212675+0.356629i$ \\
12               & $\ps 0.246770-0.131723i$ & $\ps 0.70731+0.031507i$ & $0.219964-0.002479i$  & $   -0.062573-0.858315i$ \\
14               & $\ps 0.250218-0.115252i$ & $\ps 0.64151+0.069038i$ & $0.218655-0.000107i$  & $   -0.111486+0.314412i$ \\
16               & $\ps 0.251700-0.102468i$ & $\ps 0.58932+0.093298i$ & $0.217664+0.001582i$  & $   -0.004990-0.767418i$ \\
18               & $\ps 0.252132-0.092278i$ & $\ps 0.54697+0.109542i$ & $0.216718+0.002929i$  & $   -0.055916+0.286049i$ \\
20               & $\ps 0.251983-0.083972i$ & $\ps 0.51192+0.120692i$ & $0.215976+0.003951i$  & $\ps 0.026323-0.696062i$ \\\hline
$\inf$           & $\ps 0.2206  -0.0002i  $ & $\ps 0.1216 +0.0988  i$ & $0.2063  +0.0123  i$  & $\ps 0.1186  -0.0579  i$ \\\hline
Exp.             & $\ps 0.2               $ & $\ps 0                $ & $0.2               $  & $\ps 0.2               $ \\\hline
\end{tabular}
\caption{Properties of two similar complex solutions, both of which probably describe two D1-branes separated by a Wilson line.}
\label{tab:FB double Wilson}
\end{table}

Table \ref{tab:FB double Wilson} shows some properties of the two solutions. The real part of the energy of both solutions is approximately equal to twice the energy of the original D1-brane. In this sense, these solutions are similar to the conjectured double brane, however, they do not describe two exact copies of the initial D1-brane, but rather two D1-branes separated by a Wilson line in the T-dual space.

Both solutions are complex and they have similar properties, the first one is slightly better because it has lower imaginary part and $\Delta_S$. The extrapolations in table \ref{tab:FB double Wilson} give us reasonably good results considering that the solutions are complex, but we should keep in mind that there may be large systematic errors. Extrapolations of imaginary parts of some observables are close to zero, but extrapolations of string field coefficients suggest that both solutions stay complex even asymptotically.

What is really interesting is the behavior of $W_n$ invariants. Positions of two D1-branes in the T-dual space can be parameterized in a similar way as for the usual double lumps. Due to the symmetry around origin, they must be located at $\tilde x_0$ and $-\tilde x_0$, where $\tilde x_0\in (0,\frac{\pi}{R})$. The expected values of the winding invariants are then
\begin{equation}
W_n=2R\cos(n R \tilde x_0)=2R\cos(\tilde k_n \tilde x_0),
\end{equation}
where $\tilde k_n=n R$ is the T-dual momentum. If we scale the invariants by $\frac{1}{R}$, we find that they depend only on $\tilde k$ and not on $n$ and $R$ separately\footnote{We can see that from the conservation law (\ref{Ellwood FB momentum}).}. In the limit $R\rar 0$, the ratio $W_n/R$ becomes a smooth function of $\tilde k$. Therefore it is better to characterize the D1-branes by the actual position $\tilde x_0$, which is radius independent, than by the relative distance like in case of double lumps.

\begin{figure}
\centering
   \includegraphics[width=10cm]{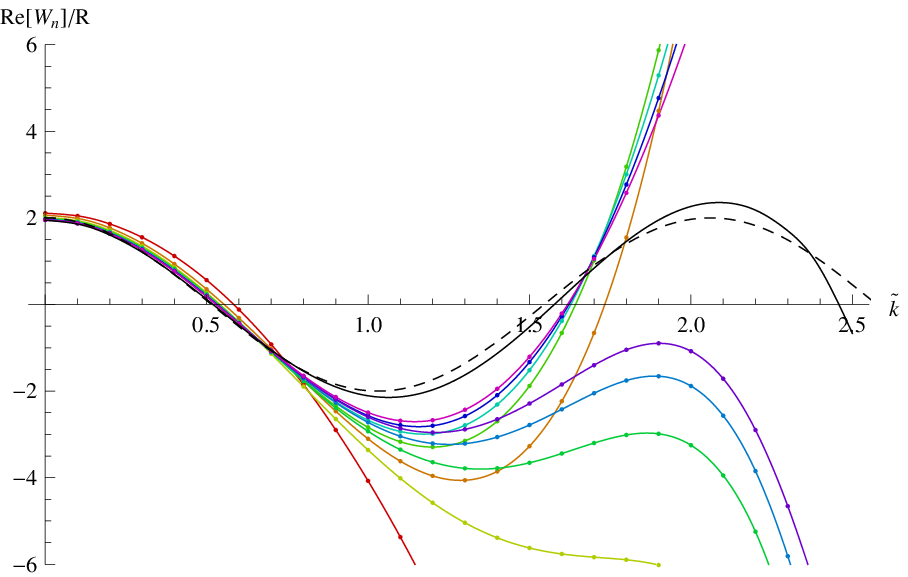}\vspace{10mm}
   \includegraphics[width=10cm]{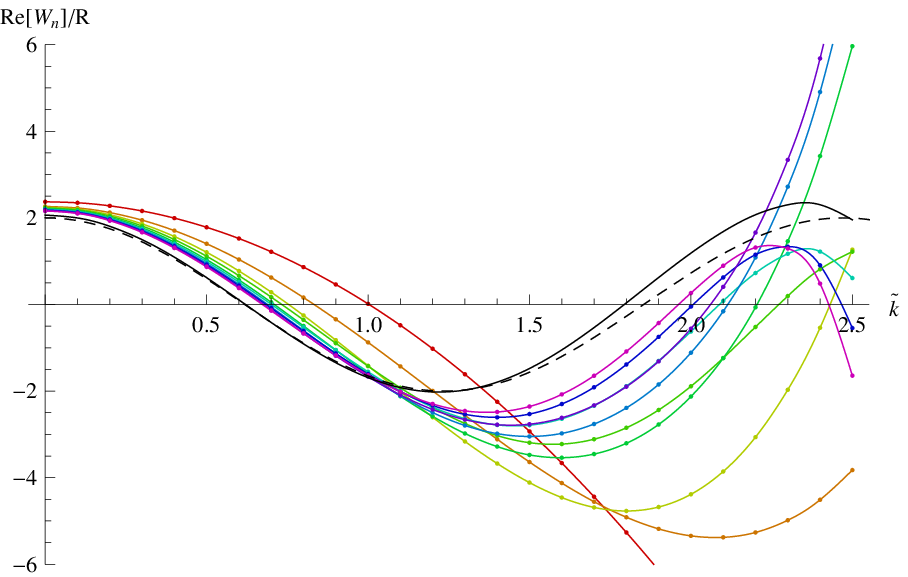}
\caption{Real part of scaled invariants $W_n/R$ as a function of the T-dual momentum $\tilde k=nR$ for the first (top figure) and second (bottom figure) Wilson line solution. The actual data from $R=0.1$ are represented by dots, the lines connecting them are interpolations representing smooth functions that will appear in the $R\rar 0$ limit. The solid black lines represent infinite level extrapolations and the dashed black lines fits of the extrapolations by the function $2\cos \tilde k\tilde x_0$.}
\label{fig:FB Wilson line}
\end{figure}

The real parts of scaled $W_n$ invariants of both solution are shown in figure \ref{fig:FB Wilson line}. Although the solutions are complex and the invariants oscillate, infinite level extrapolations are surprisingly well-behaved. In the region $\tilde k\lesssim 2$, they follow a cosine-like curve and stay more or less within the allowed range. We can estimate $\tilde x_0$ by fitting the extrapolated invariants by the function $2R\cos \tilde k\tilde x_0$. Using the first 10 invariants, we find
\begin{equation}
\tilde x_0=0.97\pi
\end{equation}
for the first solution and
\begin{equation}
\tilde x_0=0.81\pi.
\end{equation}
for the second solution. Since these numbers come from complex solutions, there may be relatively high errors. The first position could be equal to $\pi$, the second one seems to be quite generic. We are not sure why we have found these two particular positions out of all possible configurations. An interesting experiment would be to relax the twist condition and try to shift the D1-brane positions by giving a nonzero expectation value to the $\del X$ field.

Finally, let us mention that these solutions (assuming that they are physical) can be interpreted as double branes if the radius equals to $\frac{n\pi}{\tilde x_0}$, $n\in \mathbb{N}$. Then both D-branes are at the same position due to the identification of the T-dual coordinate and all $W_n$ invariants should be equal to $\pm 2R$.

\FloatBarrier
\subsection{Wilson line marginal deformation at $R=\sqrt{3}$}\label{sec:FB circle:other:WLmar}
In this subsection, we show a curious solution that we have found when we were investigation the MSZ radius $R=\sqrt{3}$. The seed for this solution appears at level 3. The solution starts as complex, but it quickly becomes real at level 8 (see table \ref{tab:FB Wilson sqrt3}, where we show some of its observables). Although it becomes real at a lower level than the positive energy lump from section \ref{sec:FB circle:single:smallR}, its extrapolations still have low precision because its invariants oscillate more than usually.

The extrapolated energy and $E_0$ are close to $\sqrt{3}$, which suggests that the solution still represents a D1-brane. Positive $D_1$ invariant and $E_1$ close to zero also support this identification. Finally, the $W_1$ invariant, which is negative, tells us that the solution describes a D1-brane with nonzero Wilson line. The T-dual position of the D1-brane is close to $\tilde x_0=\frac{\pi}{R}$, which is opposite to the initial D1-brane on the T-dual circle, but low precision of the extrapolation of $W_1$ does not allow us to determine the position with good accuracy.

D1-branes with Wilson lines are usually described by $\del X$ marginal deformations. The perturbative form of marginal solutions (\ref{marginal pert solution}) tells us that the solution should not excite any momentum states and that its coefficients should be radius independent. So how it is possible that we have found this solution? It does not even include the $\del X$ operator, which is forbidden by our ansatz, and, on the contrary, it excites momentum states, which it shouldn't. Our best guess is that this solution describes a large marginal deformation with $\lS=0$ and $\lB\neq 0$, which lies far away from the marginal branch of solutions. The role of momentum states is not clear. The asymptotic values of $E_n$ invariants are consistent with zero, but some momentum coefficients are clearly nonzero. Therefore it is possible that the momentum part of the string field is a pure gauge and that it can be removed by a gauge transformation.

\begin{table}[!]
\centering
\footnotesize
\begin{tabular}{|l|lll|}\hline
Level    & Energy              & $\ps \Delta_S$            & $\ps D_1$              \\\hline
4        & $2.31128-0.336101i$ & $   -0.0503426+0.060925i$ & $   -0.52878-17.4489i$ \\
6        & $2.18428-0.039012i$ & $   -0.0410234+0.014769i$ & $   -11.4734+4.63362i$ \\
8        & $2.04963          $ & $   -0.0307142          $ & $\ps 13.0762         $ \\
10       & $1.98344          $ & $   -0.0234379          $ & $   -11.3642         $ \\
12       & $1.94301          $ & $   -0.0198373          $ & $\ps 13.0737         $ \\
14       & $1.91497          $ & $   -0.0174573          $ & $   -10.4391         $ \\
16       & $1.89416          $ & $   -0.0157148          $ & $\ps 12.7869         $ \\
18       & $1.87800          $ & $   -0.0143630          $ & $   -9.60736         $ \\\hline
$\inf  $ & $1.739            $ & $   -0.0006             $ & $\ps 1.9             $ \\
$\sigma$ & $0.009            $ & $\ps 0.0048             $ & $\ps 9.6             $ \\\hline
Expected & $1.73205          $ & $\ps 0                  $ & $\ps 1.73205         $ \\\hline
\multicolumn {4}{l}{}\\[-5pt]\hline
Level    & $E_0$               & $\ps E_1$                 & $\ps W_1$              \\\hline
4        & $1.93968+0.357752i$ & $   -0.295764+0.694023i$  & $\ps 2.46710-2.34369i$ \\
6        & $1.70480+0.169881i$ & $   -0.415451+0.274095i$  & $   -4.57337+0.99667i$ \\
8        & $1.66896          $ & $   -0.341661          $  & $\ps 0.62677         $ \\
10       & $1.69052          $ & $   -0.217901          $  & $   -3.16189         $ \\
12       & $1.69357          $ & $   -0.176250          $  & $   -0.53983         $ \\
14       & $1.69351          $ & $   -0.141643          $  & $   -2.71935         $ \\
16       & $1.69396          $ & $   -0.124725          $  & $   -0.93343         $ \\
18       & $1.69386          $ & $   -0.106591          $  & $   -2.49373         $ \\\hline
$\inf  $ & $1.68             $ & $   -0.03              $  & $   -1.6             $ \\
$\sigma$ & $0.02             $ & $\ps 0.06              $  & $\ps 0.2             $ \\\hline
Expected & $1.73205          $ & $\ps 0                 $  & $   -1.73205         $ \\\hline
\end{tabular}
\caption{Selected observables of a solution which probably describes a Wilson line deformation of a D1-brane at $R=\sqrt{3}$. The extrapolations are taken from level 8, where the imaginary part disappears.}
\label{tab:FB Wilson sqrt3}
\end{table}

\chapter{Results - Marginal deformations}\label{sec:marginal}
In this chapter, we present our results concerning marginal deformations in Siegel gauge. Following \cite{MarginalKMOSY} and \cite{MarginalTachyonKM}, we choose  the free boson theory with Neumann boundary conditions on the self-dual radius $R=1$ as our initial setting. This allows us to make some comparisons with the results from chapter \ref{sec:FB circle} and we can use essentially the same computer code. This theory has three marginal operators ($\del X$, $\cos X$ and $\sin X$), but only $\cos X$ survives the conditions we impose on the string field.

The $\cos X$ marginal deformations interpolate between Neumann and Dirichlet boundary conditions. However, most of the results presented here should be valid in any model which has a current with the same OPE as $\cos X$. The solutions themselves and some of their observables (energy, $E_0$) are model independent. The full set of Ellwood invariants of course depends on the underlying theory because it is given by the bulk spectrum, but the relation $\lB(\lS)$ seems to be the same (within statistical errors) no matter which invariant we derive it from\footnote{The function $\lB(\lS)$ depends on the normalization of the OPE (\ref{current OPE}). Our results are valid for $k=\frac{1}{2}$.}. To check that, we have computed several marginal solutions in the double Ising model, which is described in section \ref{sec:MM:Ising2}, and we have found that they give us almost the same $\lB$ as the analogous free boson solutions.

We start with a brief review of the current understanding of marginal solutions in string field theory.
Numerical studies of marginal deformations in Siegel gauge were initiated in \cite{MarginalSen} and continued in \cite{MarginalKMOSY}\cite{MarginalTachyonKM}\cite{MarginalKishimotoNum}. The conventional numerical approach is to consider a fixed value of the marginal parameter $\lS$ and try to find marginal solutions using the usual level truncation scheme. However, the corresponding system of equations is overdetermined because $\lS$ is not a dynamical variable. The truncation to a finite level breaks the symmetry of the exact equations and therefore this system has solutions only for few values of $\lS$ (which are usually not real). We deal with this problem by leaving one equation unsolved, which equalizes the number of equations and variables. As a consequence, we find that the energy of marginal solutions evaluated from the kinetic term (\ref{Energy num2}), which we denote as $E^{kin}$, is not equal to the energy evaluated using the full action (\ref{Energy num1}), which we denote as $E^{tot}$. The difference between them is proportional to violation of the unsolved equation, see (\ref{Energy num3}). Therefore there is an additional consistency check that the two energies must converge to the same number.

In the original approach \cite{MarginalSen}, which we denote as the marginal approach, solutions are parameterized by $\lS$ and the corresponding equation is left unsolved. In this approach, we find a branch of solutions connected to the perturbative vacuum, which describes marginal deformations approximately up to $\lB\approx\frac{1}{2}$ (Dirichlet boundary conditions). The branch ends at a critical point $\lS^c\approx 0.46$, where it meets a different branch of solutions connected to the tachyon vacuum, but the second branch is clearly not physical. Surprisingly, the data suggest that there is a different critical point $\lS^\ast\approx 0.39$, where the marginal solution goes off-shell. The remaining part of the branch between $\lS^\ast\leq\lS\leq\lS^c$ probably does not satisfy the full equations of motion.

In a second approach, which was introduced in \cite{MarginalTachyonKM} and which we denote as the tachyon approach, we parameterize solutions by the tachyon field $t$ and we omit the corresponding equation. Solutions in this approach cover the moduli space up to roughly $\lB\approx 1$, which is approximately twice as much as in the marginal approach. However, some observables suggest that part of this branch may be off-shell as well.

The perturbative solution, which is given by (\ref{marginal pert expansion}) and (\ref{marginal pert solution}), has not been analyzed in the level truncation approach yet, so our results concerning this solution will be completely new.

There are also several papers describing analytic solutions for marginal deformations \cite{MarginalSchnablPert}\cite{MarginalKiermaier}\cite{MarginalFuchs}\cite{MarginalKiermaierOkawa}\!\!  \cite{MarginalInatomi}\cite{MarginalMaccaferri}\cite{MarginalMaccaferriSchnabl}\cite{MarginalLaroccaMaccaferri}. The solution from \cite{MarginalMaccaferri} is the most relevant for our purposes because it allows a direct evaluation of the relation between $\lS$ and $\lB$ \cite{MarginalMaccaferriSchnabl}. The function can be computed only numerically and it depends on parameters of the solution, but its overall properties remain the same. It has a similar behavior as the function $\lS=\frac{\lB}{1+\lB^2}$:
\begin{itemize}
  \item The function is linear for small $\lB$.
  \item It is odd.
  \item It has a maximum for a finite value of $\lB$.
  \item It goes to zero as $\lB\rar\inf$.
\end{itemize}
We will see that the Siegel gauge results are consistent with this behavior.
\\

In order to determine the function $\lB(\lS)$, we use various Ellwood invariants. The invariant given by the vertex operator $j\bar j$ should be equal to a constant independently on $\lB$, which can be easily checked for example for $j=\del X$, so all interesting invariants always depend on the chosen background. In our setting, there are six well convergent invariants with nontrivial $\lB$ dependence: $E_1$, $E_2$, $W_1$, $W_2$, $D_1$ and $H$. Their expected behavior can be derived using the results from \cite{RecknagelSchomerusMarginal}\cite{GaberdielBoson}\cite{GaberdielBosonSU2},
\begin{eqnarray}\label{Elw expected}
E_1(\lB) &=& -\sin \pi\lB, \nn \\
E_2(\lB) &=& \sin^2\pi\lB, \nn \\
W_1(\lB) &=& \cos\pi\lB,  \nn \\
W_2(\lB) &=& \cos^2\pi\lB, \\
D_1(\lB) &=& \cos 2\pi\lB, \nn \\
H  (\lB) &=&-\frac{1}{\sqrt{2}}\sin 2\pi\lB. \nn
\end{eqnarray}
$E_1$ and $W_1$ usually have the best precision because they have weight only $\frac{1}{4}$, the other invariants have weight 1 and therefore larger errors. These functions are easy to invert and we can compute $\lB$ from extrapolated values of these invariants, but for larger $\lB$, we have to be careful about which branch of arcsine or arccosine we use.

Since we able to find numerical solutions only for a limited range of $\lS$ (and therefore $\lB$), we will focus on computing $\lB(\lS)$ in the form of an expansion\footnote{It is also possible to invert the relation and write $\lS(\lB)=\sum_{i=1}^M \tilde a_i \lB^{2i-1}$. This form is better for global description of the relation between the two parameters because the function $\lB(\lS)$ has only a finite radius of convergence. However, when we look at the actual results from the marginal or perturbative approach, we find that the coefficients $\tilde a_i$ have much larger relative errors.}:
\begin{equation}\label{mar lambdaBCFT}
\lB(\lS)=\sum_{i=1}^M a_i \lS^{2i-1}.
\end{equation}
In order to determine the coefficients $a_i$, we expand both Ellwood invariants and their expected values as
\begin{equation}
E[\vv](\lS)=\sum_{i=1}^\inf E[\vv]^{(i)}\lS^i,
\end{equation}
\begin{equation}
f_\vv(\lB)=\sum_{i=1}^\inf f_\vv^{(i)}\lB^i.
\end{equation}
In the second equation, we replace $\lB$ by (\ref{mar lambdaBCFT}) and then we can compare both expressions order by order and solve for $a_i$.

As an example, consider the $E_1$ invariant. Up to order 5, it is given by
\begin{equation}
E_1(\lS)=E_1^{(1)}\lS+E_1^{(3)}\lS^3+E_1^{(5)}\lS^5+\dots
\end{equation}
and its expected behavior based on (\ref{Elw expected}) is
\begin{eqnarray}
f_{E_1}(\lS) &=&-\pi\lB(\lS)+\frac{\pi^3 }{6}\left(\lB(\lS)\right)^3-\frac{\pi^5 }{120}\left(\lB(\lS)\right)^5+\dots \\
   &=&-\pi a_1 \lS+\left(\frac{\pi^3}{6} a_1^3-\pi a_2\right) \lS^3+\left(-\frac{\pi^5}{120} a_1^5+\frac{\pi^3}{2} a_2 a_1^2-\pi a_3\right) \lS^5+\dots. \nn
\end{eqnarray}
By comparing these two equations, we find
\begin{eqnarray}
a_1&=& -\frac{E_1^{(1)}}{\pi }, \nn\\
a_2&=& \frac{-\left(E_1^{(1)}\right)^3-6 E_1^{(3)}}{6 \pi },\\
a_3&=& \frac{-3 \left(E_1^{(1)}\right)^5-20 E_1^{(3)} \left(E_1^{(1)}\right)^2-40 E_1^{(5)}}{40 \pi }.\nn
\end{eqnarray}

\section{Tachyon approach}\label{sec:marginal:tachyon}
We start by discussion of the tachyon approach following \cite{MarginalTachyonKM}. We use the same data for our analysis, but we abandon the \PadeBorel approximations. Instead, we extrapolate the data using the techniques described in section \ref{sec:Numerics:observables:extrapolation}, which allows us to get more reliable results.

Solutions in this approach are parameterized by the tachyon coefficient $t$ and we omit the corresponding equation, which means that the marginal parameter $\lS$ must be read off from the string field and it depends on level. Therefore this approach is not so good for determining the relation between $\lS$ and $\lB$, but it covers a larger part of the moduli space than the marginal approach, so we can obtain some data not available in the marginal approach.

Starting from level 2, we find that there is a pair of branches of marginal solutions connected to the perturbative vacuum, which differ by the sign of $\lS$. We pick the branch with positive $\lS$ for our analysis. The branch has finite length, which tends to decrease with level. The endpoint at level 18 is approximately $t^c\approx 0.632$ and we estimate that $t^c\approx 0.613$ asymptotically. See \cite{MarginalTachyonKM} for more details. At the endpoint, this branch meets with another branch of solutions, which however do not satisfy the full equations of motion. If we choose $t>t^c$, we find only complex solutions. The structure of branches is schematically similar to the left part of figure \ref{fig:mar branches 1} (if we replace $\lS$ by $t$).

We present our results in form of figures similarly to section \ref{sec:FB circle:single:largeR}. We use the same colors to distinguish levels (see figure \ref{fig:FB colors}) and black lines denote infinite level extrapolations. Dashed black lines (when present) denote the expected behavior of Ellwood invariants based on a fit of the $E_1$ invariant, see later. The data are shown only up to $t=0.625$ because there are problems with extrapolations close to the endpoint of the branch.

\begin{figure}
   \centering
   \includegraphics[width=14cm]{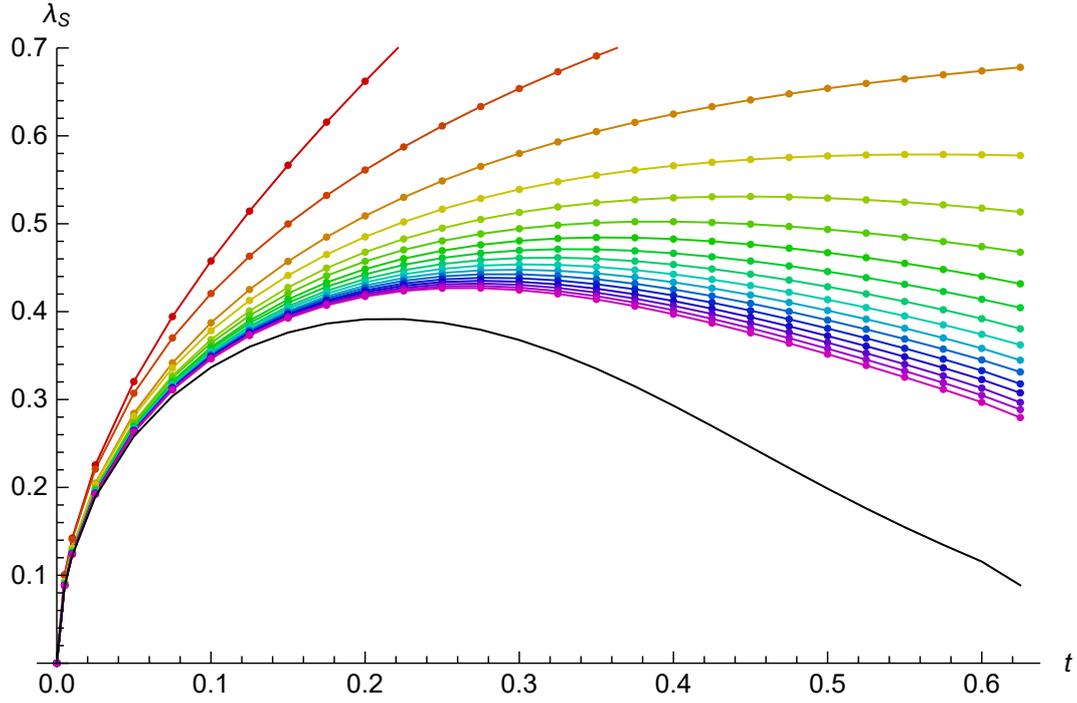}
   \caption{Marginal parameter $\lS$ as a function of the tachyon coefficient $t$ in the tachyon approach.}
   \label{fig:mar lambda}
\end{figure}

The relation between the tachyon coefficient $t$ and the marginal parameter $\lS$ is plotted in figure \ref{fig:mar lambda}. We observe that $\lS$ behaves as $\lS\sim\sqrt{t}$ for small $t$. At low levels, $\lS(t)$ is a monotonically increasing function, but from level 5, it has a maximum, after which it starts decreasing. This behavior is in agreement with \cite{MarginalMaccaferriSchnabl}. By extrapolating $\lS$ and fitting the results by a polynomial, we find an approximate position of the maximum in the infinite level limit:
\begin{eqnarray}
\lS^\ast&\approx &0.3920, \\
t^\ast&\approx &0.2138.
\end{eqnarray}
We notice that $\lS^\ast$ is surprisingly smaller than the length of the marginal branch in the marginal approach $\lS^c$ (see the next section), which suggests that part of the marginal branch is off-shell.

\begin{figure}
   \centering
   \begin{subfigure}[t]{0.47\textwidth}
      \includegraphics[width=\textwidth]{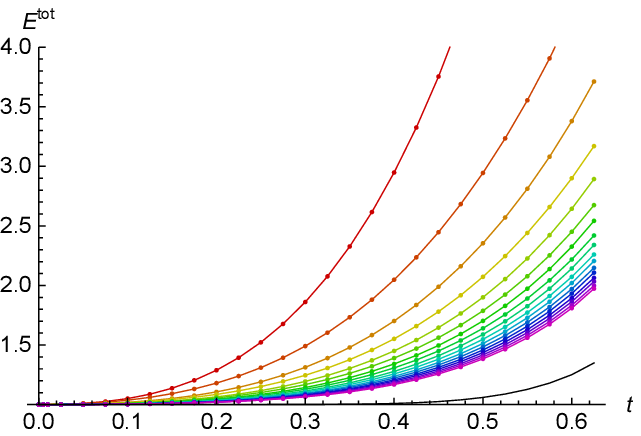}
   \end{subfigure}\qquad
   \begin{subfigure}[t]{0.47\textwidth}
      \includegraphics[width=\textwidth]{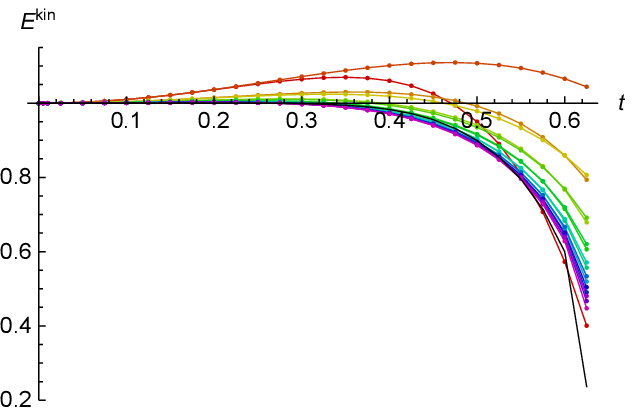}
   \end{subfigure}\vspace{5mm}
   \begin{subfigure}[t]{0.47\textwidth}
      \includegraphics[width=\textwidth]{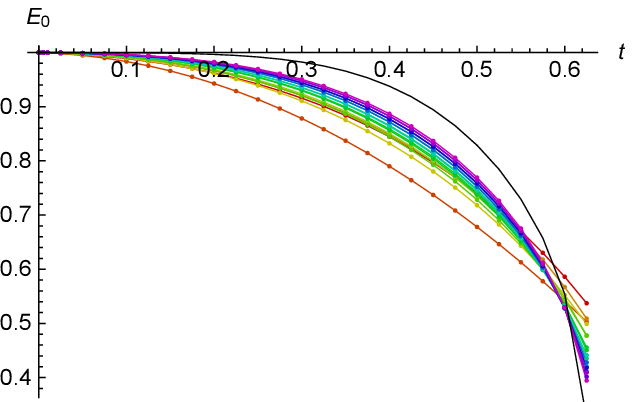}
   \end{subfigure}\qquad
   \begin{subfigure}[t]{0.47\textwidth}
      \includegraphics[width=\textwidth]{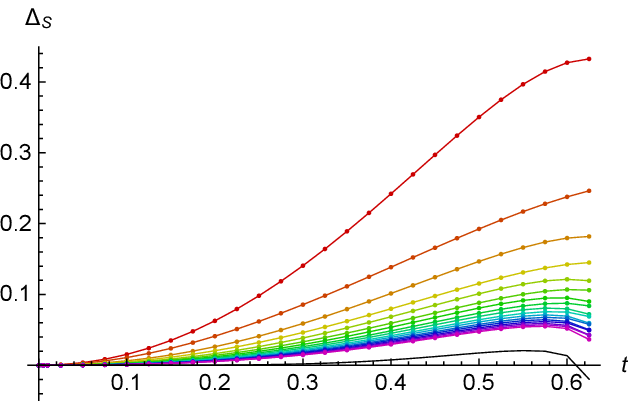}
   \end{subfigure}
   \caption{Energy of marginal solutions in the tachyon approach measured by the full action $E^{tot}$, by the kinetic term $E^{kin}$ and by the Ellwood invariant $E_0$. The last figure shows the out-of-Siegel equation $\Delta_S$. The extrapolations, which are denoted by black lines, are closer to the expected values than in \cite{MarginalTachyonKM}, but they are still quite far away.}
   \label{fig:mar energy tach}
\end{figure}

Next, figure \ref{fig:mar energy tach} shows the energy of marginal solutions measured in three different ways. The energy is close to 1 for small $t$, but it deviates quite a lot for large $t$ in all three figures. Near the end of the branch, the observables $E^{kin}$ and $E_0$ even move away from the correct value with increasing level. This suggests that part of the branch in the tachyon approach may be nonphysical as in the marginal approach. However, the solution does not exhibit any sudden changes, so it is difficult to pinpoint where exactly it goes off-shell.

\begin{figure}
   \centering
   \begin{subfigure}[t]{0.47\textwidth}
      \includegraphics[width=\textwidth]{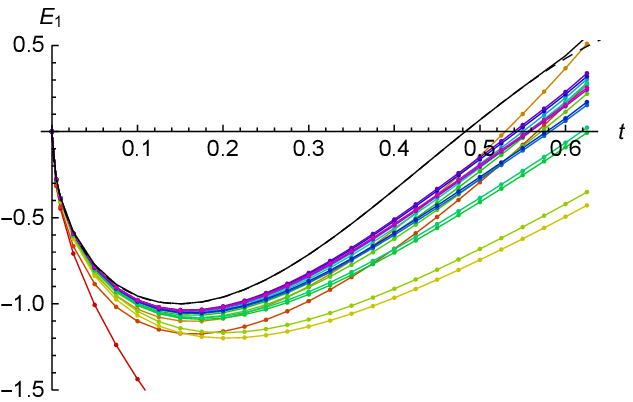}
   \end{subfigure}\qquad
   \begin{subfigure}[t]{0.47\textwidth}
      \includegraphics[width=\textwidth]{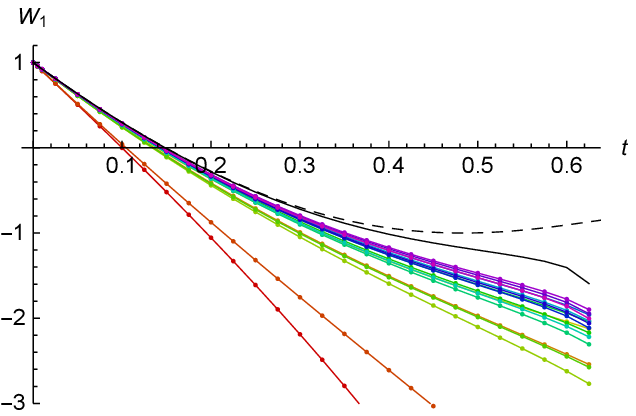}
   \end{subfigure}\vspace{5mm}
   \begin{subfigure}[t]{0.47\textwidth}
      \includegraphics[width=\textwidth]{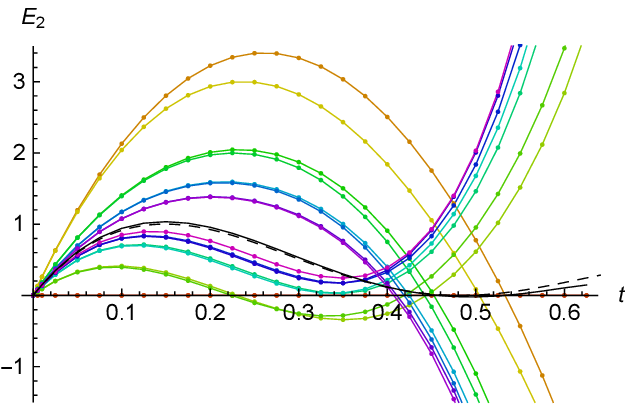}
   \end{subfigure}\qquad
   \begin{subfigure}[t]{0.47\textwidth}
      \includegraphics[width=\textwidth]{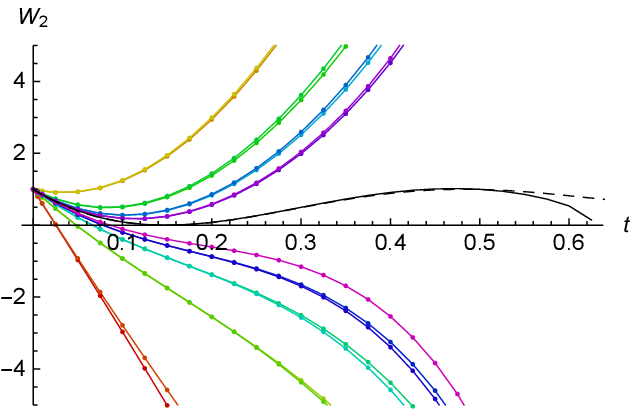}
   \end{subfigure}\vspace{5mm}
   \begin{subfigure}[t]{0.47\textwidth}
      \includegraphics[width=\textwidth]{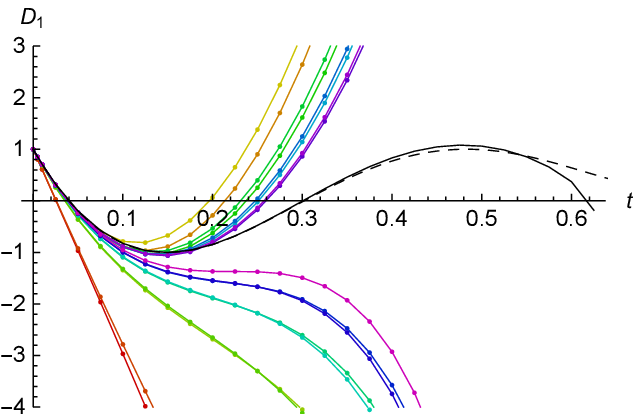}
   \end{subfigure}\qquad
   \begin{subfigure}[t]{0.47\textwidth}
      \includegraphics[width=\textwidth]{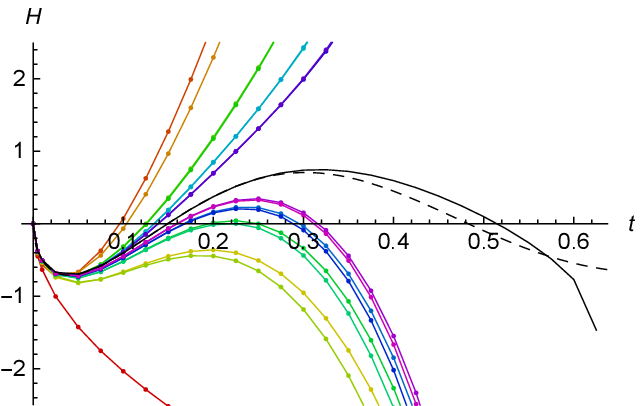}
   \end{subfigure}
   \caption{Six convergent Ellwood invariants in the tachyon approach. The solid black lines denote infinite level extrapolations, the dashed black lines the expected behavior based on a $\lB$ fit of the $E_1$ invariant.}
   \label{fig:mar invariants tach}
\end{figure}

Finally, we move to invariants which allow us to determine the relation between $t$ and $\lB$. We show the six well-behaved invariants in figure \ref{fig:mar invariants tach}. The figures are somewhat different from \cite{MarginalTachyonKM} because now we show raw data instead of \PadeBorel approximations. The invariants oscillate with significantly higher amplitudes, which are in some cases several times higher than the characteristic scale, but infinite level extrapolations are much smoother than in \cite{MarginalTachyonKM}. We can use them to fit the relation $\lB(t)$ by a function of the form
\begin{equation}
\lB^{(M)}(t)=\sqrt{t}\sum_{i=0}^M a_i t^i.
\end{equation}
The expected values of Ellwood invariants in figure \ref{fig:mar invariants tach} are based on $\lB$ obtained by $M=6$ fit of the $E_1$ invariant. The coefficients $a_i$ unfortunately exhibit a strong dependence on $M$ and on the number of data points used for the fit, so we will not discuss their exact values. The functions $\lB^{(M)}(t)$ are nevertheless always very similar in the region of interest.

Overall, the agreement between asymptotic behavior of invariants and the expected values based on the $\lB$ fit is better than in \cite{MarginalTachyonKM}. The only exception is the $W_1$ invariant, which goes outside the allowed range at large $t$. This behavior supports the possibility that the end of the branch is off-shell.

We observe that the branch of marginal solutions in this approach covers more than two fundamental domains of the moduli space, which means that $\lB$ goes above 1. This allows us estimate which $t$ and $\lS$ describe Dirichlet or Neumann boundary conditions. The $E_1$ invariant implies that Dirichlet boundary conditions are described by
\begin{eqnarray}
t^D &\approx & 0.1482, \\
\lS^D &\approx & 0.3752.
\end{eqnarray}
The other invariants give us similar results with differences of order $10^{-4}$. Recall that in section \ref{sec:FB circle:single:largeR}, we found
\begin{equation}
t^D \approx 0.1473,\qquad \lS^D \approx 0.3751
\end{equation}
by extrapolating single lump solutions to $R=1$ and that the pseudo-real double lump from section \ref{sec:FB circle:double:R=2} gives us
\begin{equation}
t^D \approx 0.1482,\qquad \lS^D \approx 0.3753.
\end{equation}
All three results are quite close to each other, which supports validity of these results and it tells us that the tachyon approach is reliable at least up to this point. Notice that $\lS^D<\lS^\ast$, which means that the marginal approach should cover the Dirichlet boundary conditions.

Using $E_1$, we find that Neumann boundary conditions correspond to
\begin{eqnarray}
t^N &\approx& 0.48, \\
\lS^N &\approx& 0.21.
\end{eqnarray}
The precision of these results is much smaller than in the Dirichlet case, the results from different invariants have dispersion of order $10^{-2}$. Moreover, it is not clear whether the branch is still on-shell at this point.

\begin{figure}
   \centering
   \includegraphics[width=10cm]{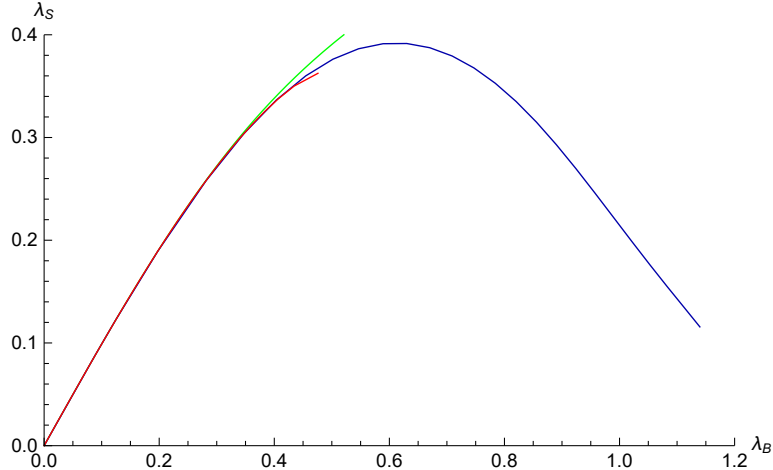}
   \caption{Comparison of $\lS$ as a function of $\lB$ in the tachyon approach (blue), the marginal approach (red) and perturbative approach (green). The plot is based on extrapolations of the $E_1$ invariant and $\lS$.}   \label{fig:mar lambda BS}
\end{figure}

Finally, we can use the results for $\lS$ and $\lB$ to reconstruct the function $\lS(\lB)$, which is captured in figure \ref{fig:mar lambda BS} by the blue line. The function has similar properties to those observed in \cite{MarginalMaccaferriSchnabl}, most notably, $\lS$ has a maximum at a finite value of $\lB$. Therefore it seems that the key properties of the $\lS(\lB)$ relation are gauge independent, although the precise form of the function changes. The figure also compares the results from the tachyon approach to the marginal and perturbative approaches, which are discussed in the following sections. We observe that all approaches lead to almost identical relation up to say $\lB\approx 0.4$, which is a nice consistency check.

\section{Marginal approach}\label{sec:marginal:marginal}
The so-called marginal approach is the original method to find non-perturbative marginal solutions in the level truncation approximation. In this approach, we keep the value of the marginal parameter $\lS$ fixed and we remove the related equation. We can restrict our analysis to $\lS\geq 0$ because observables of marginal solutions are (anti)symmetric with respect to change of the sign of $\lS$.

\begin{figure}
   \centering
   \begin{subfigure}{0.47\textwidth}
      \includegraphics[width=\textwidth]{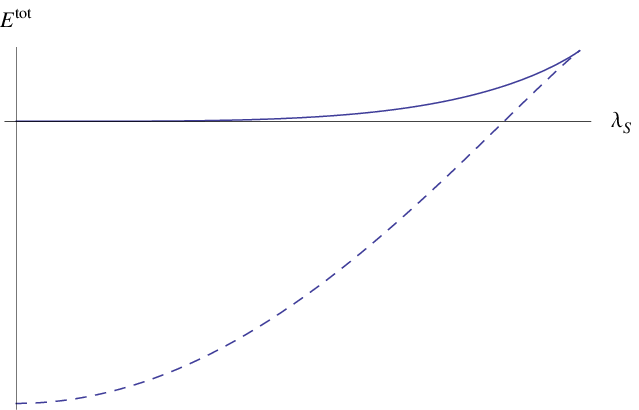}
   \end{subfigure}\qquad
   \begin{subfigure}{0.47\textwidth}
      \includegraphics[width=\textwidth]{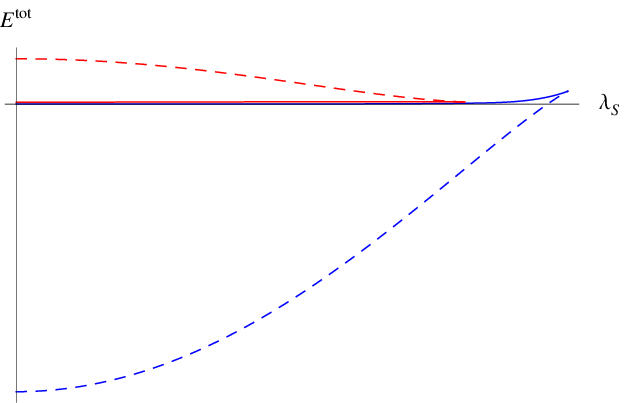}
   \end{subfigure}
   \caption{On the left: Schematic branch structure of OSFT solutions in the marginal approach. The solid line represents the total energy of the marginal branch, the dashed line the energy of the vacuum branch. On the right: Schematic branch structure of the toy model from \cite{MarginalTachyonKM}. The solid blue and red lines represent two marginal branches, the dashed lines two non-physical branches. Each of the marginal branches covers about half of the moduli space and the end on the blue branch is off-shell.}
   \label{fig:mar branches 1}
\end{figure}

There are two branches of real solutions: the marginal branch connected to the perturbative vacuum and the so-called vacuum branch connected to the tachyon vacuum. The two branches appear already at level 1 and they keep their properties to higher levels (see \cite{MarginalKMOSY}). As the parameter $\lS$ increases, the two branches approach each other and they meet at a critical point denoted as $\lS^c$, see the left part of figure \ref{fig:mar branches 1}, which schematically depicts the branch structure. For $\lS>\lS^c$, we find only some complex solutions, which seem to by nonphysical. The vacuum branch turns out to by off-shell as well because it violates the omitted equation of motion. The marginal branch behaves much better and it seems that it solves the full equations of motion at least within certain range of $\lS$.

As we are going to show, the marginal branch covers only a part of the moduli space (approximately up to $\lB\approx 0.5$). Therefore we expect that OSFT solutions cover the full moduli space in a similar way as in the toy model from \cite{MarginalTachyonKM}. There should be another branch (or possibly several branches) that describes rest of the moduli space (see figure \ref{fig:mar branches 1} on the right for the toy model example). However, finding the other branch turns out to be quite difficult. There are no other real solutions with reasonable properties (at least up to level 4). There is a branch of complex solutions with energy not far from 1, which may describe part of the missing moduli space (possibly $0.5\lesssim \lB \lesssim 1$ to match the tachyon approach), but these solutions are complex at least up to level 18. That means that we know their Ellwood invariants only with a very limited precision and we cannot determine whether this branch consistently describes marginal deformations or not.

We are therefore going to focus only on the marginal branch. The length of this branch varies with level as in the tachyon approach, see \cite{MarginalKMOSY}. It grows at first, it reaches its maximum at level 5 and then it slowly decreases. We estimate that its asymptotic length is around $\lS^c\approx 0.465$ and therefore we show our data only up to this value. We work with essentially the same level 18 data as in \cite{MarginalTachyonKM}. We have just added more data points for increased precision of some calculations and, as in the tachyon approach, we redo all infinite level extrapolations.

\begin{figure}
   \centering
   \begin{subfigure}[t]{0.47\textwidth}
      \includegraphics[width=\textwidth]{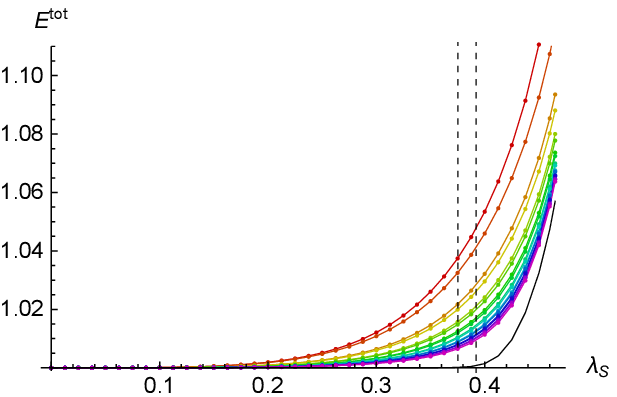}
   \end{subfigure}\qquad
   \begin{subfigure}[t]{0.47\textwidth}
      \includegraphics[width=\textwidth]{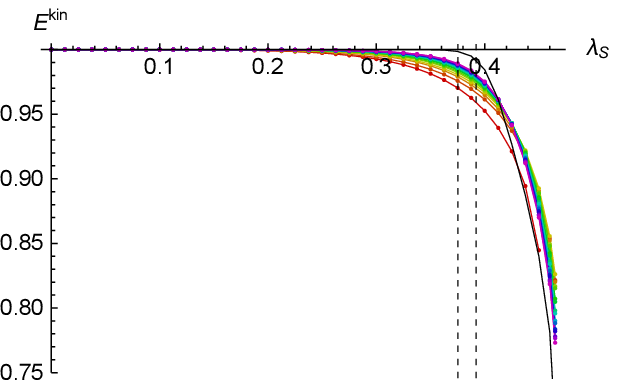}
   \end{subfigure}\vspace{5mm}
   \begin{subfigure}[t]{0.47\textwidth}
      \includegraphics[width=\textwidth]{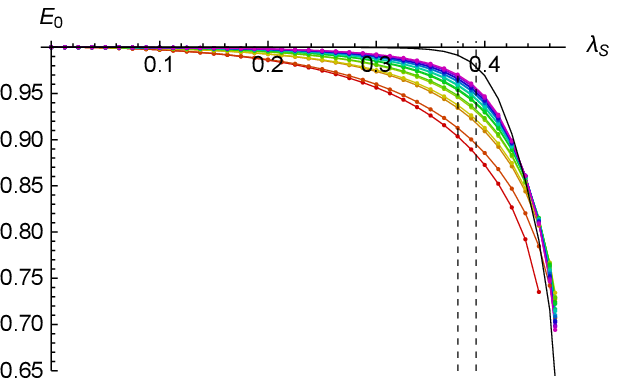}
   \end{subfigure}\qquad
   \begin{subfigure}[t]{0.47\textwidth}
      \includegraphics[width=\textwidth]{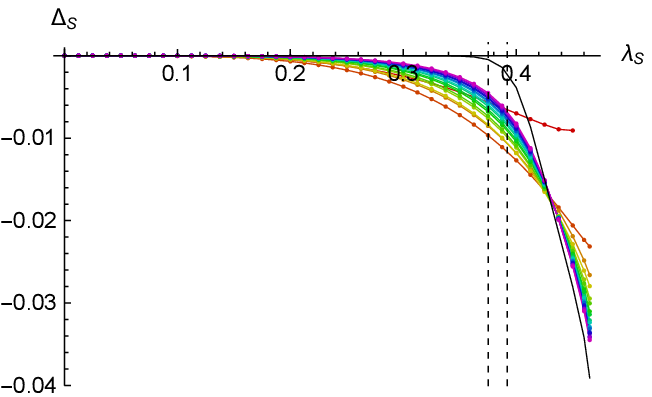}
   \end{subfigure}
   \caption{Energy of the marginal branch in the marginal approach measured by $E^{tot}$, $E^{kin}$ and $E_0$ and the first out-of-Siegel equation. }
   \label{fig:mar energy mar}
\end{figure}

\begin{figure}[!]
   \centering
   \begin{subfigure}[t]{0.47\textwidth}
      \includegraphics[width=\textwidth]{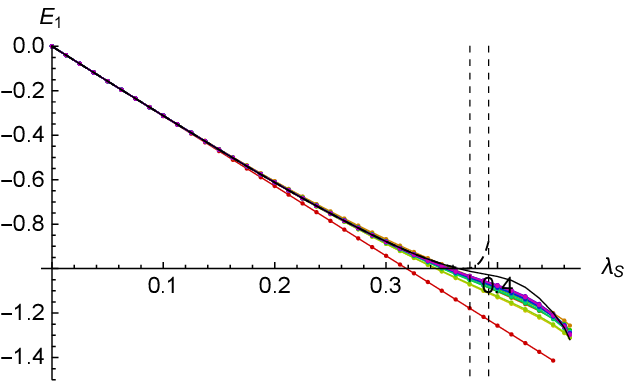}
   \end{subfigure}\qquad
   \begin{subfigure}[t]{0.47\textwidth}
      \includegraphics[width=\textwidth]{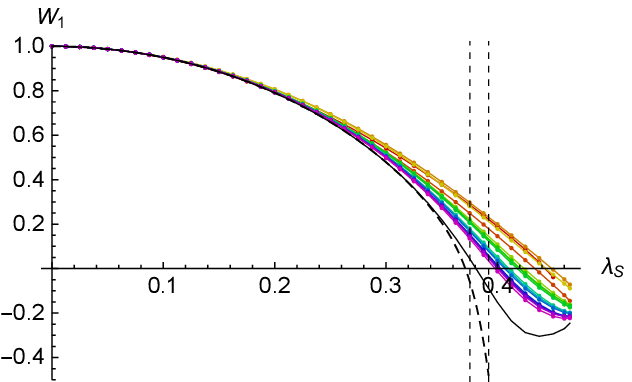}
   \end{subfigure}\vspace{5mm}
   \begin{subfigure}[t]{0.47\textwidth}
      \includegraphics[width=\textwidth]{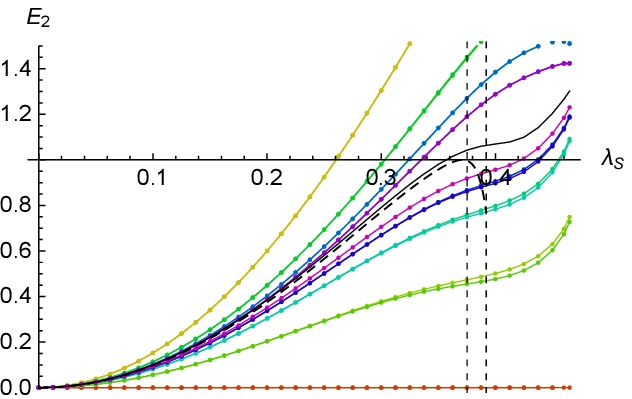}
   \end{subfigure}\qquad
   \begin{subfigure}[t]{0.47\textwidth}
      \includegraphics[width=\textwidth]{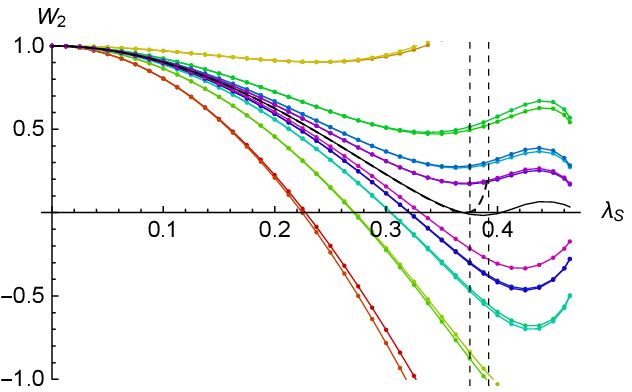}
   \end{subfigure}\vspace{5mm}
   \begin{subfigure}[t]{0.47\textwidth}
      \includegraphics[width=\textwidth]{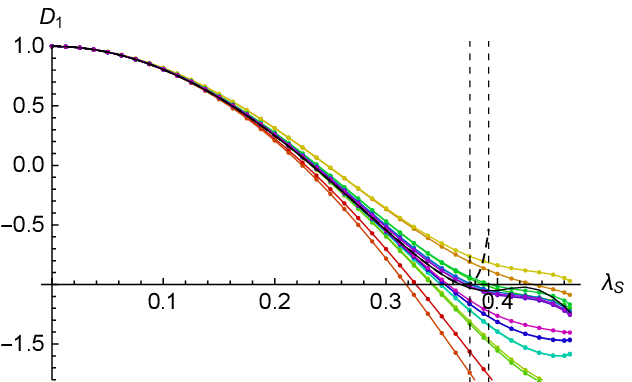}
   \end{subfigure}\qquad
   \begin{subfigure}[t]{0.47\textwidth}
      \includegraphics[width=\textwidth]{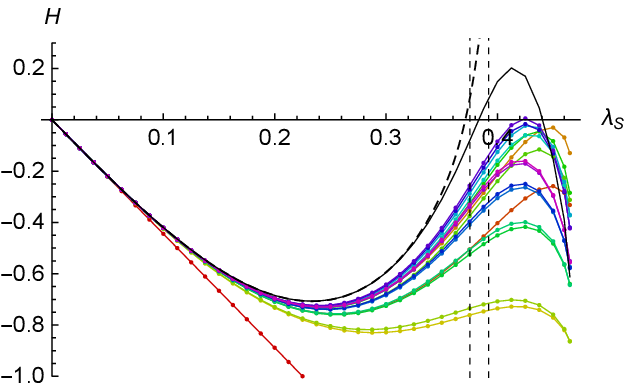}
   \end{subfigure}
   \caption{Six convergent Ellwood invariants in the marginal approach. The dashed lines denote the expected behavior based on a $\lB$ fit of the $E_1$ invariant. The fit is unstable in the region between the two vertical lines, so the expected behavior in this region should be viewed only as schematic.}
   \label{fig:mar invariants mar}
\end{figure}

The most important observables are shown in figures \ref{fig:mar energy mar} and \ref{fig:mar invariants mar}. We use the color scale from figure \ref{fig:FB colors} to distinguish levels and infinite level extrapolations are denoted by black lines as usual. Furthermore, we add two dashed vertical lines at $\lS=0.3752$ and $\lS=0.3920$, which mark the estimates for the Dirichlet point $\lS^D$ and the critical point $\lS^\ast$ from the tachyon approach respectively.

The first figure shows the energy measured in three different ways and the out-of-Siegel equation $\Delta_S$. The energy converges close to 1 for most of the length of the marginal branch (we are not even able to distinguish the extrapolated curve from the axis), but it clearly deviates from the expected value near the end. This is the first sign that part of the branch is off-shell similarly to the toy model from \cite{MarginalTachyonKM}. It detaches from the axis roughly at $\lS^\ast$ from the tachyon approach, but determining the exact point is difficult because the solutions get worse gradually and there is no sharp change in its behavior. The out-of-Siegel equation also deviates from 0 around the critical point $\lS^\ast$.

Well convergent invariants with a nontrivial $\lB$ dependence are shown in figure \ref{fig:mar invariants mar}. Compared to \cite{MarginalTachyonKM}, there are more oscillations, but infinite level extrapolations are quite similar. By inverting the functions (\ref{Elw expected}) and applying them to the invariants, we obtain the relation between $\lS$ and $\lB$ in the marginal approach. See figure \ref{fig:mar lambda BS}, which depicts the results from the $E_1$ invariant (by the red curve). By fitting this relation with the function (\ref{marginal pert expansion}), we can predict the behavior of other invariants, which is denoted by dashed black lines in figure \ref{fig:mar invariants mar}. The extrapolations agree quite well with these predictions up to $\lS\approx 0.37$, which shows that these solutions really describe marginal boundary states.

In the tachyon approach, we have found that Diriment boundary conditions should by described by $\lS^D\approx 0.3752$. The marginal approach is consistent with this result because the extrapolations of all invariants cross axes at similar values of $\lS$. However, some invariants should only touch the axis in a local minimum/maximum as indicated by the dashed lines. Instead, they go outside the allowed range, which suggests that solutions beyond $\lS^D$ are off-shell. This result is in conflict with the tachyon approach, which indicates that the branch should be on-shell up to the second vertical line, which denotes the estimated critical point $\lS^\ast$. The tachyon approach tells us with very high probability that $\lS^\ast>\lS^D$, so we think that the problem lies in the marginal approach. Invariants should change quite a lot between $\lS^D$ and $\lS^\ast$ and, based on properties of the function $\lS(\lB)$, they should in fact end with infinite derivative. This behavior is difficult to describe by finite level solutions. Therefore we think that Ellwood invariants near $\lS^\ast$ converge slower than elsewhere and we would need access to higher levels to see the desired behavior.

Next, we are going to have a closer look at the relation $\lB(\lS)$. In \cite{MarginalTachyonKM}, we failed in determining the coefficients of the perturbative expansion (\ref{mar lambdaBCFT}) because \PadeBorel approximations are difficult to extrapolate. However, our new extrapolations of Ellwood invariants can be consistently fitted with polynomials in $\lS$ and we can compute $a_i$ without any problems. The coefficients have quite small dependence on the number of used data points or on the order of the fit, so infinite level extrapolations are the main source of errors. The results for the first four coefficients are summarized in table \ref{tab:mar a fit mar}.

\begin{table}[!]
\centering
\begin{tabular}{|c|llll|}\hline
invariant & $a_1$              & $a_2$            & $a_3$            & $a_4$           \\\hline
$D_1$     & $1.0000\pm 0.0001$ & $1.077\pm 0.008$ & $3.16 \pm 0.04 $ & $12.2 \pm 1.1 $ \\
$E_1$     & $1               $ & $1.073\pm 0.002$ & $3.14 \pm 0.01 $ & $12.20\pm 0.07$ \\
$E_2$     & $1.01  \pm 0.03  $ & $1.15 \pm 0.15 $ & $3.5  \pm 1.1  $ & $14   \pm 7   $ \\
$W_1$     & $1.0001\pm 0.0002$ & $1.073\pm 0.001$ & $3.136\pm 0.005$ & $12.15\pm 0.06$ \\
$W_2$     & $1.00  \pm 0.04  $ & $1.07 \pm 0.05 $ & $3.2  \pm 0.8  $ & $12   \pm 7   $ \\
$H$       & $1               $ & $1.074\pm 0.007$ & $3.2  \pm 0.2  $ & $14   \pm 3  $ \\\hline
\end{tabular}
\caption{Extrapolations of the first four $a_i$ coefficients in the marginal approach. We use only data up to $\lS=0.2$ to do the fits to avoid the region where the $H$ invariant goes slightly out of the allowed range.}
\label{tab:mar a fit mar}
\end{table}

The $a_1$ coefficient must be exactly 1 because $E_1$ and $H$ get linear contributions only from the marginal field. The other invariants also give us 1 within the statistical error. Then we get approximately $a_2\approx 1.073$, $a_3\approx 3.136$ and $a_4\approx 12.15$. The most precise results come from the $W_1$ invariant, closely followed by the $E_1$ invariant. The coefficients from all invariants agree within the errors, which is a nice consistency check for the marginal solution. We observe that both the coefficients and their relative errors grow with increasing order, so we cannot compute much more than these four coefficients. Later, we will independently confirm these results in the perturbative approach.

\subsection{Marginal deformations in complex plane}\label{sec:marginal:marginal:complex}
A new interesting possibility is to study marginal deformations for complex $\lS$. This corresponds to extension of the underlying SU(2)$_1$ WZW model on the self-dual radius to SL(2,$\mathbb{C}$) theory. The boundary states described by these deformations are not physical (they describe for example D0-branes with complex $x_0$), but they can tell us more about analytic structure of marginal solutions.

First, we should check whether there are meaningful solutions for complex $\lS$. Table \ref{tab:mar complex} shows few observables of a sample solution with $\lS=0.2+0.2i$. Although the solution is necessarily complex, it behaves very well. It is even slightly more precise than the real solution with the same absolute value of $\lS$. Both the energy and $E_0$ go towards to 1 and their imaginary parts almost disappear. The first out-of-Siegel equation also approaches 0. The invariants with nontrivial $\lB$ dependence are complex and by inverting them, we find approximately $\lB\approx 0.180+0.212i$.

\begin{table}[!b]
\centering
\scriptsize
\begin{tabular}{|l|llll|}\hline
Level\rowh{8.5pt}  & $E^{tot}$                & $E_0$                 & $D_1$               & $\Delta_S$               \\\hline
2                  & $0.994016-0.0023239   i$ & $1.01175-0.0189320 i$ & $0.971164-1.58228i$ & $0.0021535-0.0001001  i$ \\
4                  & $0.997251-0.0012269   i$ & $1.00682-0.0084237 i$ & $0.805656-1.51345i$ & $0.0016287+0.0004286  i$ \\
6                  & $0.998263-0.0007942   i$ & $1.00484-0.0052388 i$ & $0.895782-1.55811i$ & $0.0010360+0.0003833  i$ \\
8                  & $0.998733-0.0005823   i$ & $1.00366-0.0036275 i$ & $0.851098-1.56613i$ & $0.0007439+0.0003139  i$ \\
10                 & $0.999003-0.0004592   i$ & $1.00298-0.0028502 i$ & $0.877603-1.57204i$ & $0.0005783+0.0002611  i$ \\
12                 & $0.999178-0.0003791   i$ & $1.00249-0.0022601 i$ & $0.860188-1.57834i$ & $0.0004728+0.0002224  i$ \\
14                 & $0.999300-0.0003229   i$ & $1.00216-0.0019345 i$ & $0.871621-1.57994i$ & $0.0003999+0.0001933  i$ \\
16                 & $0.999391-0.0002814   i$ & $1.00190-0.0016328 i$ & $0.862740-1.58394i$ & $0.0003466+0.0001708  i$ \\
18                 & $0.999460-0.0002493   i$ & $1.00170-0.0014577 i$ & $0.868886-1.58460i$ & $0.0003060+0.0001529  i$ \\\hline
$\inf$\rowh{9pt}   & $1.000000-4\dexp{-7}  i$ & $1.00005-0.00003   i$ & $0.862   -1.601  i$ & $6\dexp{-7}+5\dexp{-7}i$ \\
$\sigma$\rowh{8pt} & $8\dexp{-8}+3\dexp{-9}i$ & $7\dexp{-6}-0.00004i$ & $0.003   +0.001  i$ & $2\dexp{-7}+1\dexp{-9}i$ \\\hline
Exp.  \rowh{8.5pt} & $1                     $ & $1                  $ & $\cos(2\pi\lB)    $ & $0                     $ \\\hline
\end{tabular}
\caption{Selected observables of a complex marginal solution with $\lS=0.2+0.2i$.}
\label{tab:mar complex}
\end{table}

By analyzing more complex marginal solutions, we have arrived to a conjecture that the real part of the energy converges towards a real number in most of the complex plane with the exception of an area near the real axis for $|\re[\lS]|>\lS^c$. The exact shape of the area is difficult to determine because the imaginary part grows gradually as $\lS$ approaches the real axis. Marginal solutions elsewhere behave well even for $|\lS|>\lS^c$.

This leads us to believe that the exact solution in complex plane has a branch cut along the real axis starting at $\lS^c$. We have decided to test what happens to marginal solutions when they pass through the possible branch cut. Consider a circular path that crosses the real axis at $\lS<\lS^\ast$ and at $\lS>\lS^c$. When we start with a real marginal solution with $\lS<\lS^\ast$ and deform it in small steps along this path until it returns to the initial point, we can get two different results depending on the level and on the path. We obtain either a solution from the vacuum branch or a solution from the complex branch we briefly mentioned before. When we deform this solution along the path again, the vacuum branch solution returns to the marginal solution, while the complex solution goes to the vacuum branch solution. Therefore we need to go along the path two or four times to return to the original solution. This behavior is consistent with existence of a branch cut.

A special case of complex marginal deformations is a purely imaginary $\lS$. These solutions have real action, which means that they are pseudo-real, and their Ellwood invariants are described by hyperbolic functions of $\im[\lS]$. We have tracked these marginal solutions up to $\im[\lS]=2$ and we have not found any critical point like in the usual marginal approach. The branch clearly continues much further, possibly to infinity, but it gets more and more difficult to recognize marginal solutions from many other complex solutions. There are also large differences between even and odd levels, so Newton's method becomes unable to go between them after a certain point. In order to to track marginal solutions further, it would be necessary to modify the method to use solutions with lower $\lS$ as starting points, instead of solutions from lower levels.

\begin{figure}
   \centering
   \includegraphics[width=10cm]{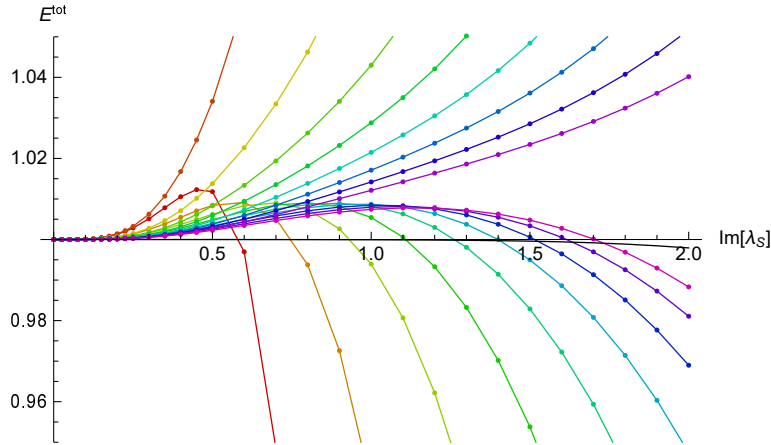}
   \caption{Full energy of complex marginal solutions for imaginary $\lS$.}
   \label{fig:mar im Etot}
\end{figure}

Figure \ref{fig:mar im Etot} shows the total energy of this branch of solutions to illustrate their behavior. We observe that data at even and odd levels follow different patterns, but infinite level extrapolations work better than in the marginal or the tachyon approach. The data from this branch can be also used to find the perturbative relation (\ref{mar lambdaBCFT}). The results are almost identical to table \ref{tab:mar a fit mar}.

\section{Perturbative approach}\label{sec:marginal:perturbative}
In the final section of this chapter, we discuss marginal solutions in the perturbative approach. We expand the string field as a power series in $\lS$,
\begin{equation}
\Psi(\lS)=\sum_{k=1}^\inf \lS^n \Psi_n,
\end{equation}
where the Siegel gauge solution for $\Psi_n$ is given by
\begin{equation}
\Psi_n=-\frac{b_0}{L_0^{tot}}\sum_{k=1}^{n-1}\Psi_k\ast \Psi_{n-k}.
\end{equation}
The solution must be computed recursively, starting with $\Psi_1=c_1 \cos X(0)|0\ra$. The star product can be evaluated numerically using the usual cubic vertices (see section \ref{sec:Numerics:V3:star}), the only problem is that $\Psi_n\ast \Psi_{k-n}$ sometimes has a nonzero coefficient in front of the marginal field and therefore $L_0^{tot}$ is not invertible. In that case, we set this coefficient to zero by hand.

When using this framework, we are able to reach higher level than in the other approaches because we need to evaluate the solution just once and we can use some of its symmetries to simplify calculations. Most notably, we utilize the fact that the even momentum part of the solution contains only even powers of $\lS$ and the odd momentum part only odd powers of $\lS$. We have computed order 6 solution up to level 22 and order 8 solution up to level 21. We have decided not show the full set of gauge invariant observables at every level, because it would require far too much space, so table \ref{tab:mar invariants pert} contains only extrapolations of the most important observables.

This solution is useful mainly for computing the $a_i$ coefficients in the perturbative expansion (\ref{mar lambdaBCFT}). Order 8 solution allows us to determine the first 4 coefficients. The results are summarized are in table \ref{tab:mar invariants pert} and we observe that they are in agreement with the results from the marginal approach (table \ref{tab:mar a fit mar}). The results from the perturbative approach are more precise because we have access to higher level data and there are no additional errors from polynomial fits in $\lS$. For illustration, the relation between $\lS$ and $\lB$ from this approach is plotted in figure \ref{fig:mar lambda BS} by green color. It agrees nicely with the other approaches at small $\lS$.

\begin{table}
\centering
\begin{tabular}{|c|l|}\hline
Invariant                             & Extrapolated value \\\hline
\multirow{2}{*}{\raisebox{-4pt}{$E^{kin}$}}\rowh{14pt} & $1-0.0000017\lS^4 -0.000465\lS^6 -0.006484\lS^8                        $ \\[4pt]
                                      & $\phantom{1}-0.121063 \lS^{10} -15.1874\lS^{12}-10.5166 \lS^{14}-271.659 \lS^{16}$ \\[2pt]\hline
$E_0$ \rowh{14pt}                     & $1-0.0000439\lS^2 -0.006225\lS^4 -0.046959\lS^6-0.35143 \lS^8          $ \\[2pt]\hline
$D_1$ \rowh{14pt}                     & $1-19.7391  \lS^2 +22.4594 \lS^4 +46.5328 \lS^6+164.963 \lS^8          $ \\[2pt]\hline
$E_1$ \rowh{14pt}                     & $\phantom{1} -3.14159 \lS   +1.79186 \lS^3 +4.23178 \lS^5+15.2851 \lS^7          $ \\[2pt]\hline
$E_2$ \rowh{14pt}                     & $\phantom{1} +9.79267  \lS^2 -10.8546 \lS^4 -22.5833 \lS^6-80.3573 \lS^8          $ \\[2pt]\hline
$W_1$ \rowh{14pt}                     & $1-4.93482 \lS^2 -6.54817 \lS^4 -20.6077 \lS^6-82.3808 \lS^8          $ \\[2pt]\hline
$W_2$ \rowh{14pt}                     & $1-9.9173  \lS^2 +11.3225 \lS^4 +23.6827 \lS^6+89.6873 \lS^8          $ \\[2pt]\hline
$H$   \rowh{14pt}                     & $\phantom{1} -4.44288 \lS   +24.4506 \lS^3 +22.6339 \lS^5+67.7032 \lS^7          $ \\[2pt]\hline
\end{tabular}
\caption{Extrapolations of gauge invariants of the marginal solution in the perturbative approach. We have computed only the quadratic energy $E^{kin}$ because the total energy $E^{tot}$ is a cubic function of the string field and its evaluation would require too much time.}
\label{tab:mar invariants pert}
\end{table}

\begin{table}
\centering
\begin{tabular}{|c|llll|}\hline
Invariant & $a_1$                  & $a_2$              & $a_3$            & $a_4$           \\\hline
$D_1$     & $0.999997\pm 0.000009$ & $1.076 \pm 0.001 $ & $3.16 \pm 0.03 $ & $12.2 \pm 0.2 $ \\
$E_1$     & $1                   $ & $1.0746\pm 0.0004$ & $3.144\pm 0.004$ & $12.19\pm 0.03$ \\
$E_2$     & $1.00    \pm 0.01    $ & $1.06  \pm 0.04  $ & $2.9  \pm 0.3  $ & $13   \pm 8   $ \\
$W_1$     & $1.000001\pm 0.000007$ & $1.0747\pm 0.0003$ & $3.143\pm 0.002$ & $12.14\pm 0.02$ \\
$W_2$     & $1.00    \pm 0.01    $ & $1.07  \pm 0.04  $ & $2.7  \pm 0.4  $ & $5    \pm 6   $ \\
$H$       & $1                   $ & $1.076 \pm 0.001 $ & $3.16 \pm 0.05 $ & $13   \pm 2   $ \\\hline
\end{tabular}
\caption{Extrapolations of the first four $a_i$ coefficients in the perturbative approach. The first three coefficients are derived using level 22 data, the last coefficient using level 21 data.}
\label{tab:mar a fit pert}
\end{table}

Next, we can ask a question whether the perturbative solution reproduces exactly the marginal branch from the marginal approach. In particular, we would like to know whether its radius of convergence is equal to $\lS^c$ or to $\lS^\ast$. To decide that, we do not need to know the solution to high levels, but rather to high orders. The following discussion is based on level 6 solution with order 400.

\begin{figure}
   \centering
   \includegraphics[width=12cm]{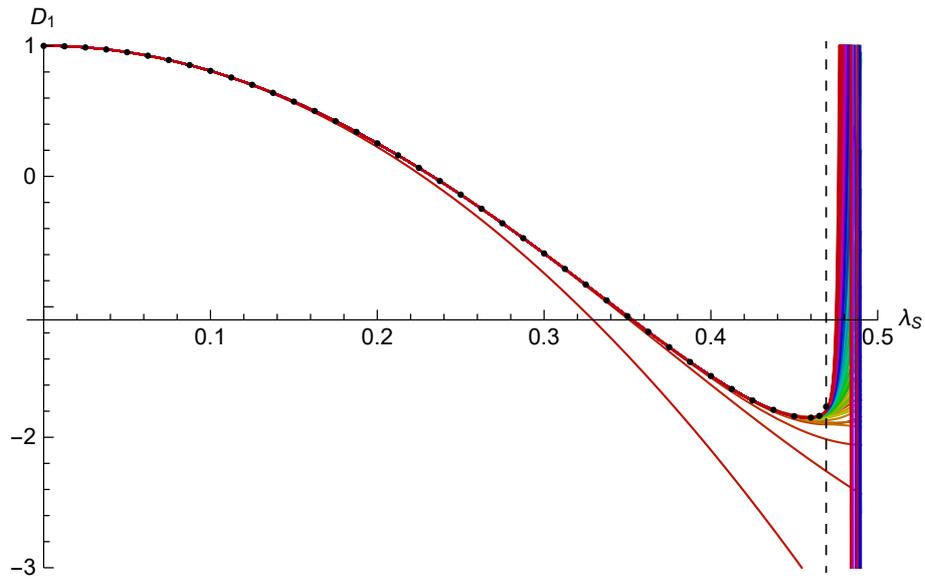}
   \caption{Comparison of the $D_1$ invariant at level 6 in the perturbative and in the marginal approach. Solid lines represent the perturbative solution at orders $5,10,15,\dots,400$ with color following the rainbow spectrum, black dots represent the numerical data from the marginal approach. We denote the critical radius at level 6, $\lS^c\approx0.4691$, by a dashed vertical line.}
   \label{fig:mar expansion}
\end{figure}

First, figure \ref{fig:mar expansion} shows comparison of the $D_1$ invariant from the marginal approach with its successive approximations in the perturbative approach. We observe that the perturbative solution approximates the data from the marginal approach quite well almost to the end of the marginal branch and then it starts to behave chaotically. This suggests that the radius of convergence is $\lS^c$ and therefore the perturbative solution covers the whole marginal branch.

To determine the radius of convergence more rigorously, recall that given a function $f(x)=\sum_n f_n x^n$, we can compute its radius of convergence as ${\lim\limits_{n\rar \inf} \left|\frac{f_n}{f_{n+1}}\right|}$. Components of the perturbative solution are either even or odd functions of $\lS$, so we have to modify this formula to $\lim\limits_{n\rar \inf}\sqrt{\left|\frac{f_{n}}{f_{n+2}}\right|}$ for the marginal solution.

\begin{figure}
   \centering
   \includegraphics[width=12cm]{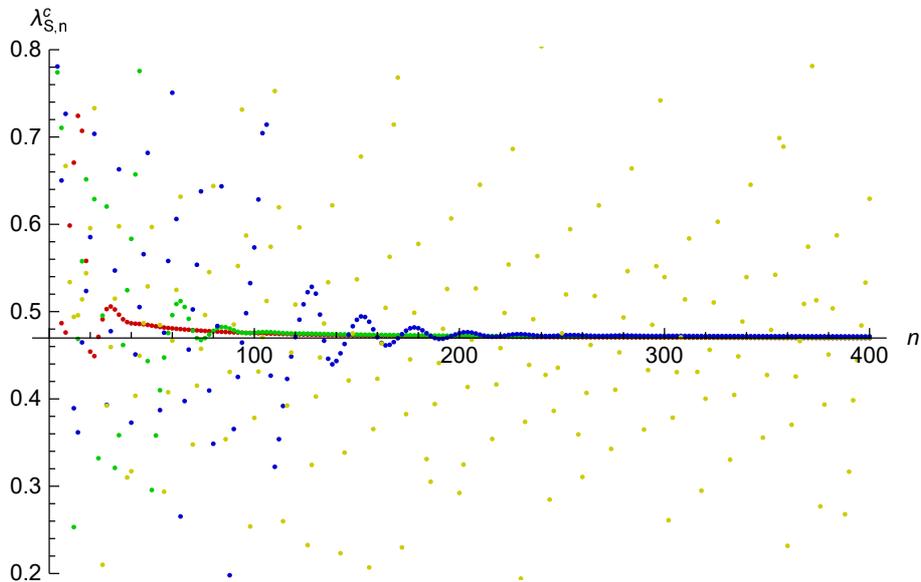}
   \caption{Estimated radius of convergence of the perturbative solution at orders up to 400. We estimate the radius using the $D_1$ invariant at levels 3 (red), 4 (green), 5 (blue) and 6 (yellow). The horizontal axis is set to the critical point at level 5, $\lS^c=0.469761$.}
   \label{fig:mar radius}
\end{figure}

We show estimates of the radius of convergence based on the $D_1$ invariant in figure \ref{fig:mar radius}. The estimates at levels 3, 4 and 5 are quite close to $\lS^c$. For example, a simple two parametric fit at level 3 gives us the asymptotic value $\lS^{c}(\inf)=0.46785$, while we the actual result from the marginal approach is $\lS^c= 0.46790$, which is a good agreement. However, we also observe that the data oscillate more and more with increasing level and we need to go to higher orders to get close to $\lS^c$. Finally, at level 6, we find just chaotic oscillations, which are only roughly centered around the expected value. Higher level data behave in a similar way. We have also checked that estimates of the radius of convergence from other invariants and from coefficients of the string field offer qualitatively the same results.

These properties of the solution lead us to an interesting question what happens in the limit
\begin{equation}
\lim_{\substack{n\rar\inf\\L\rar\inf}}\Psi(\lS).
\end{equation}
The data from the other approaches suggest that the two critical parameters ($\lS^\ast$ and $\lS^c$) are different. Therefore the exact solution should satisfy out-of-Siegel equations only for $\lS$ below $\lS^\ast$, but not above $\lS^\ast$, while this approach suggest that the radius of convergence is $\lS^c$.

There are several possible resolutions of this issue. $\lS^c$ could decrease to $\lS^\ast$ at high levels, but this does not seem to be very likely given the known data. There is a possibility that the solution is a nonanalytic function of $\lS$ with some kind of discontinuity at $\lS^\ast$. Different quantities evaluated on the solution can also have different radii of convergence, in particular, the radius of convergence of out-of-Siegel equations could by just $\lS^\ast$, while gauge invariants could converge up to $\lS^c$. Finally, there can be problems with the double limit itself. Estimates of the radius of convergence suggest that we need high orders to see regular behavior at high levels, so the results can differ depending on the order of the two limits.

\chapter{Results - Free boson on a two-dimensional torus}\label{sec:FB D2}
In this chapter, we investigate OSFT solutions of two-dimensional free boson theory with toroidal compactification. So far, this model has been neglected in the literature, there are only some very low level calculations in \cite{LumpsNDHarvey}\cite{LumpsNDKoch}\cite{Lumps2DMoeller}, which just show the existence of single lump solutions\footnote{There is also an analytic solution for D-branes with magnetic flux (based on the Erler-Maccaferri solution) \cite{FluxAnalytic}\cite{FluxAnalytic2}.}.

In the case of free boson theory on a circle, we have made a relatively exhaustive study of solutions at low radii, but we will not attempt to do the same here. It would be pretty much impossible anyway because a two-dimensional torus has three parameters, which means that we would need many samples to cover its moduli space, and the number of solutions is typically much larger than in one dimension. There are dozens of potentially interesting solutions already for $R_{1,2}\gtrsim 2$ and this number grows to hundreds and thousands with increasing radii. Our goals in this thesis are to provide a framework for future studies and to introduce so-called exotic solutions, which describe non-conventional boundary states.

\section{Solutions on torus with $R_1=2.4$, $R_2=2.2$, $\theta=0.4\pi$}\label{sec:FB D2:sample torus}
We are going to illustrate properties of solutions in this theory on a sample torus. We have picked a torus with radii $R_1=2.4$, $R_2=2.2$ and with angle $\theta=0.4\pi$. This is a relatively generic torus with high enough radii to see some interesting solutions. The energy of the initial D2-brane is $5.47809$, so we should see real solutions with energy up to this number. The number of states in this theory is significantly higher than in one dimension, so we can to reach only level 12 with similar computer resources.

We have used the homotopy continuation method to find seeds for Newton's method at level 1.9. After we sort out these seeds and remove those which differ only by translations, we are left with about 30 real and well-behaved solutions. We could probably find more if we considered seeds from other levels, but these are more than enough for illustrative purposes.

This model admits many Ellwood invariants, see section \ref{sec:SFT:observables:FB}, but the simplest way to identify solutions on a torus is to look at their tachyon and energy density profiles. Following (\ref{FB tachyon profile}) and (\ref{FB energy profile}), we define
\begin{equation}
t(x,y)=\sum_{n_1,n_2} t_{n_1,n_2} \cos k(n_1,n_2).(x,y)
\end{equation}
and
\begin{equation}
E(x,y)=\frac{1}{4\pi^2 R_1 R_2\sin\theta}\left(E_{0,0}+\sum_{n_1,n_2} 2E_{n_1,n_2} \cos k(n_1,n_2).(x,y)\right),
\end{equation}
where $k(n_1,n_2)$ is the momentum vector given by (\ref{FB momentum torus}). We restrict the sum in the energy density profile using weights of invariants as $\frac{|k(n_1,n_2)|^2}{4}<h$, where $h$ is chosen based on properties of $E_{n_1,n_2}$ invariants. We usually get the best profiles when $h$ is around 1.

\begin{figure}
   \centering
   \begin{subfigure}{0.45\textwidth}
   \includegraphics[width=\textwidth]{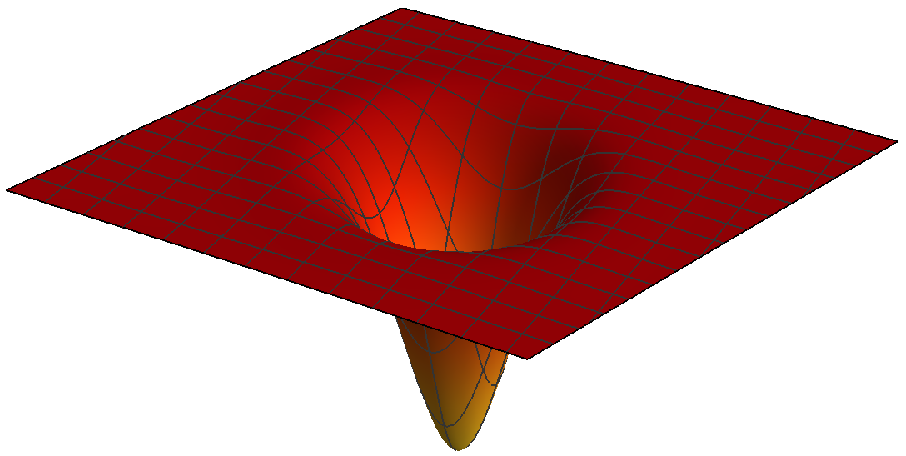}
   \end{subfigure}\qquad
   \begin{subfigure}{0.45\textwidth}
   \includegraphics[width=\textwidth]{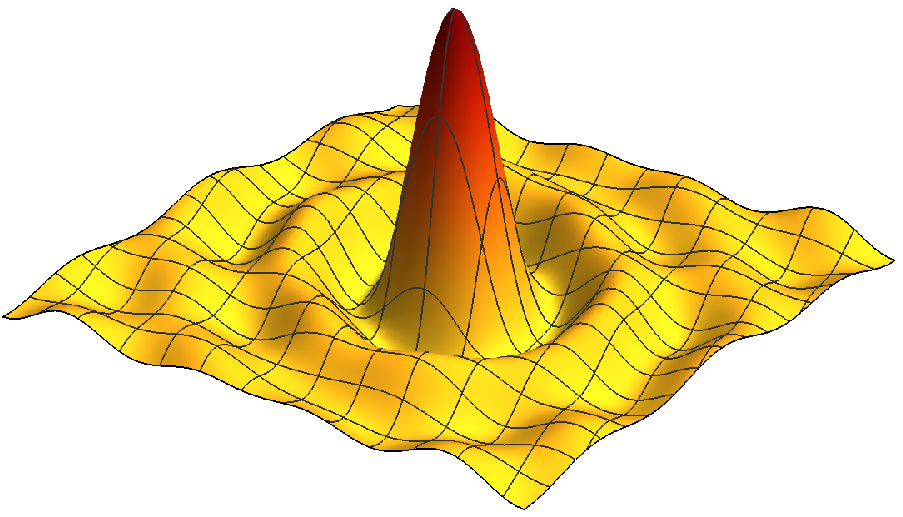}
   \end{subfigure}\vspace{10mm}

   \begin{subfigure}{0.45\textwidth}
   \includegraphics[width=\textwidth]{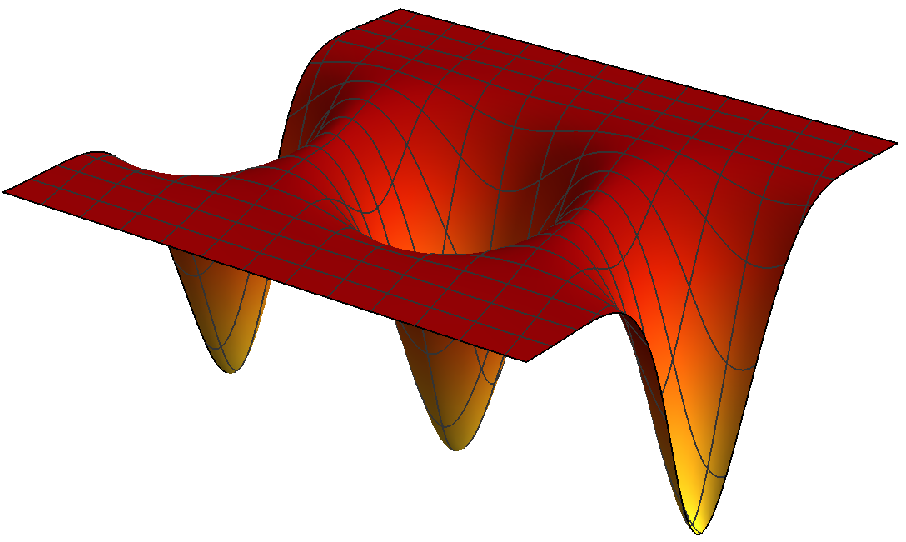}
   \end{subfigure}\qquad
   \begin{subfigure}{0.45\textwidth}
   \includegraphics[width=\textwidth]{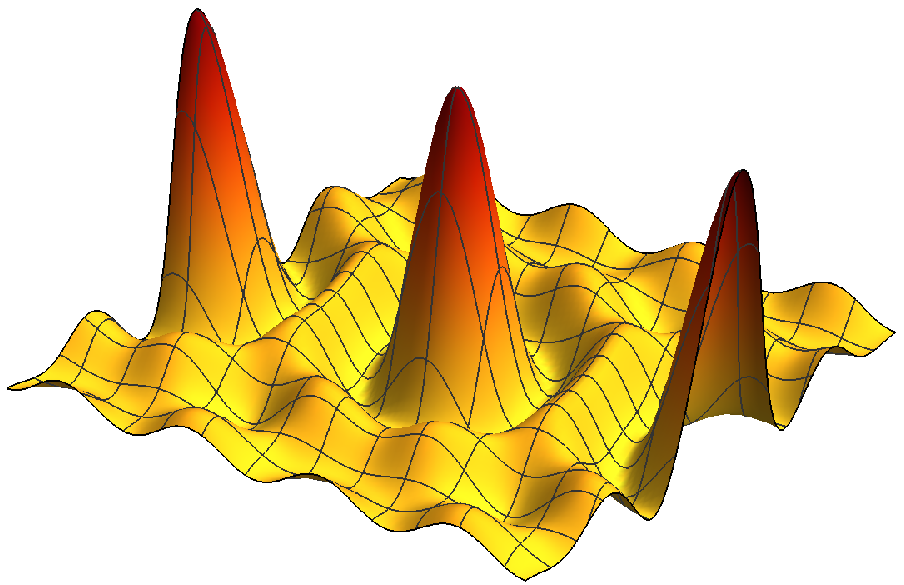}
   \end{subfigure}\vspace{10mm}

   \begin{subfigure}{0.45\textwidth}
   \includegraphics[width=\textwidth]{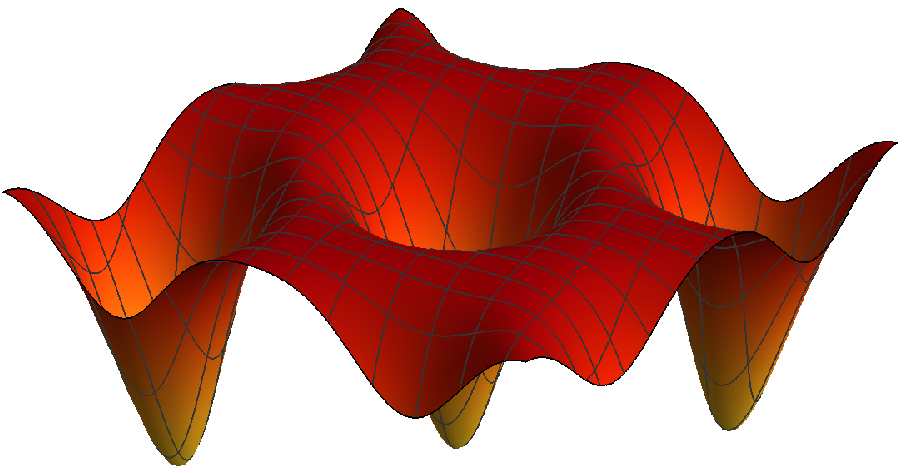}
   \end{subfigure}\qquad
   \begin{subfigure}{0.45\textwidth}
   \includegraphics[width=\textwidth]{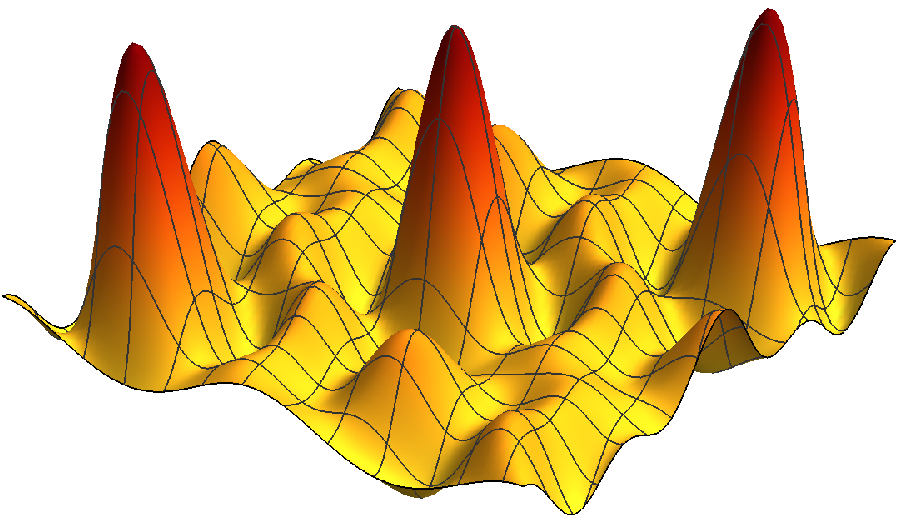}
   \end{subfigure}\vspace{10mm}

   \begin{subfigure}{0.45\textwidth}
   \includegraphics[width=\textwidth]{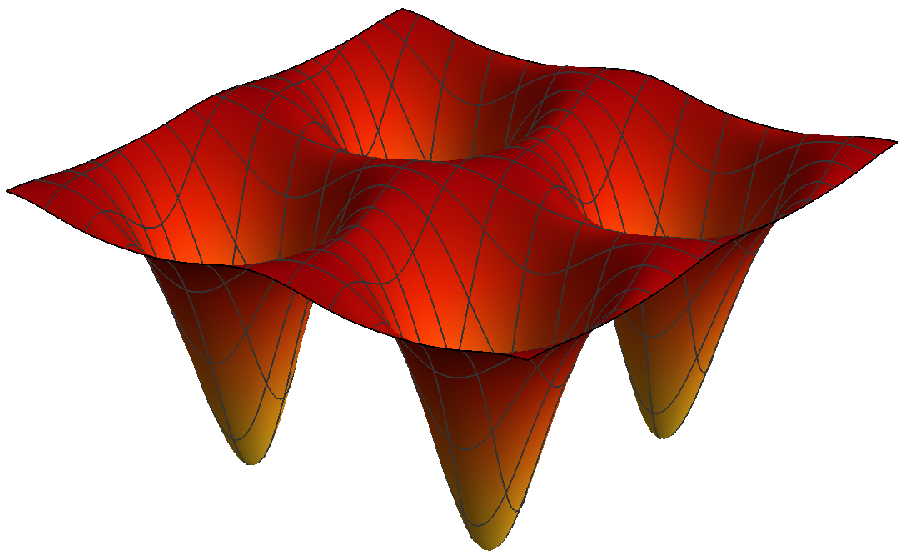}
   \end{subfigure}\qquad
   \begin{subfigure}{0.45\textwidth}
   \includegraphics[width=\textwidth]{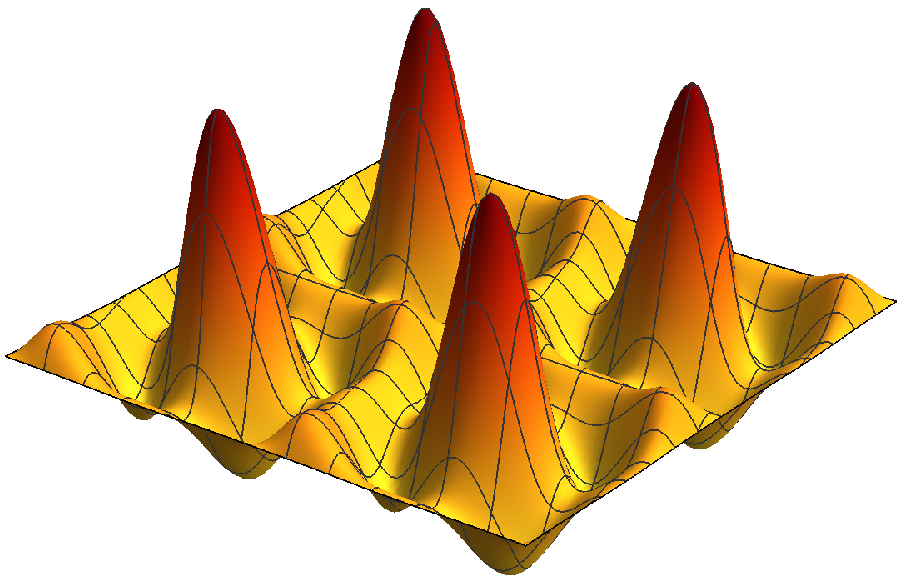}
   \end{subfigure}\qquad
   \caption{Examples of D0-brane solutions on the torus with parameters $R_1=2.4$, $R_2=2.2$ and $\theta=0.4\pi$. We show tachyon profiles of these solutions on the left and energy density profiles using the first 32 momentum invariants (which correspond to all relevant fields) on the right. All figures are based on extrapolations of level 12 data. From the top, the solutions represent one, two, three and four D0-branes.}
   \label{fig:torus examples 1}
\end{figure}

The solutions we have found describe many different boundary states. First, there are D0-brane solutions, which describe from one to four D0-branes in various geometrical configurations. Solutions describing five or more D0-branes would be more difficult to find because the D0-branes would be too close to each other, so the corresponding solutions are most likely complex. We show some typical examples of D0-brane solutions in figure \ref{fig:torus examples 1}. We observe that D0-branes are consistently represented by holes in tachyon profiles, which can be well approximated by two-dimensional gaussians, and by narrow peaks in energy density profiles. Unsurprisingly, properties of these solutions are very similar to properties of lump solutions on a circle.

\begin{figure}
   \centering
   \begin{subfigure}{0.45\textwidth}
   \includegraphics[width=\textwidth]{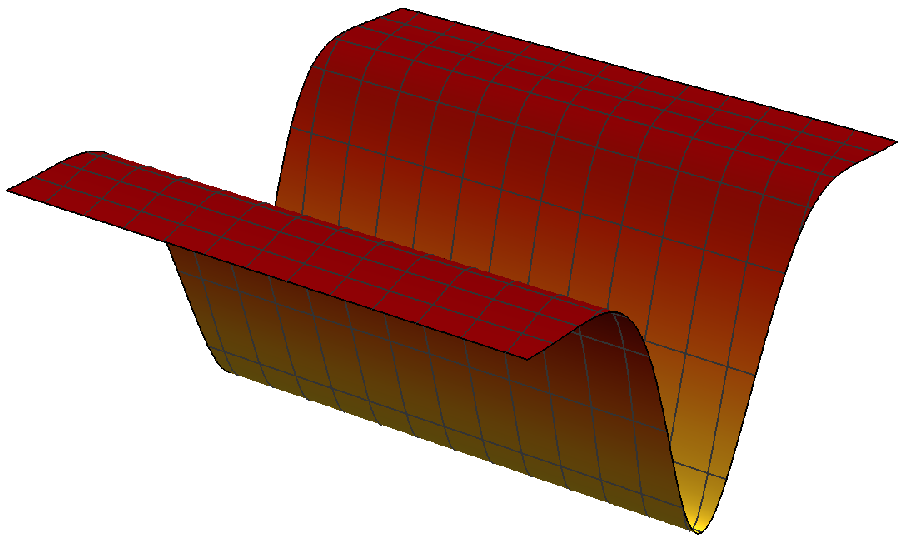}
   \end{subfigure}\qquad
   \begin{subfigure}{0.45\textwidth}
   \includegraphics[width=\textwidth]{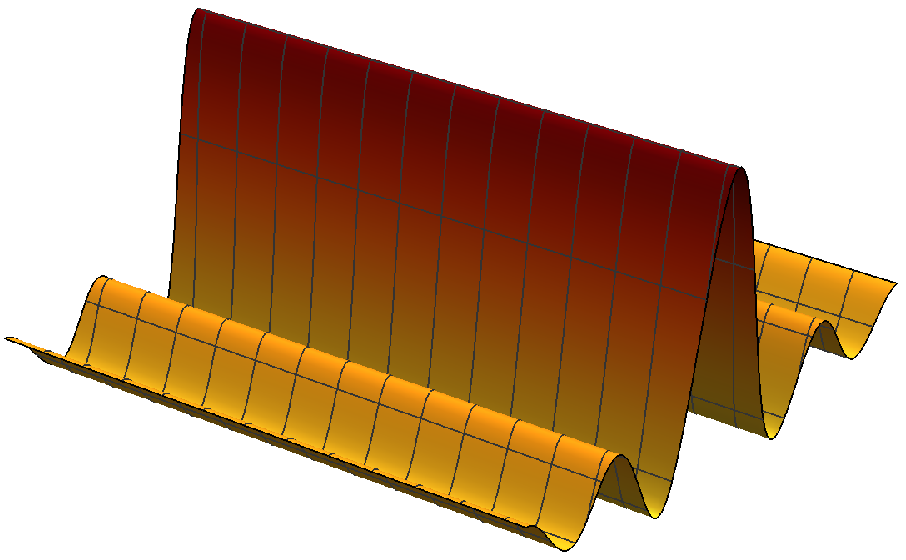}
   \end{subfigure}\vspace{10mm}

   \begin{subfigure}{0.45\textwidth}
   \includegraphics[width=\textwidth]{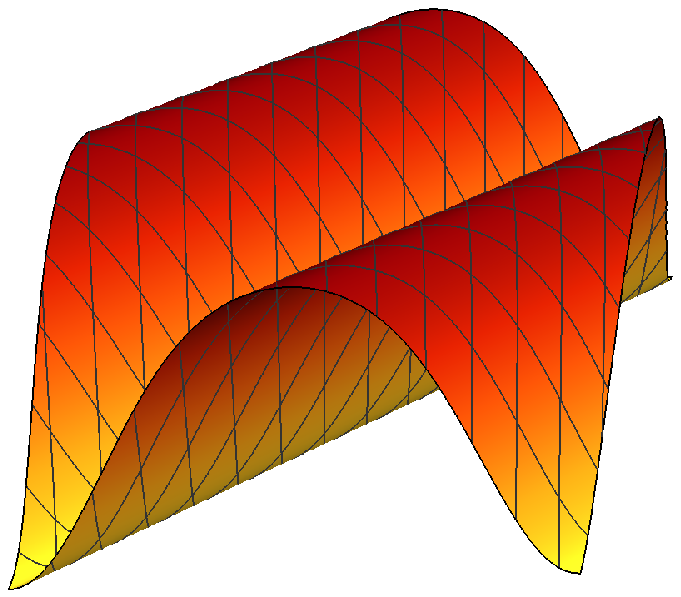}
   \end{subfigure}\qquad
   \begin{subfigure}{0.45\textwidth}
   \includegraphics[width=\textwidth]{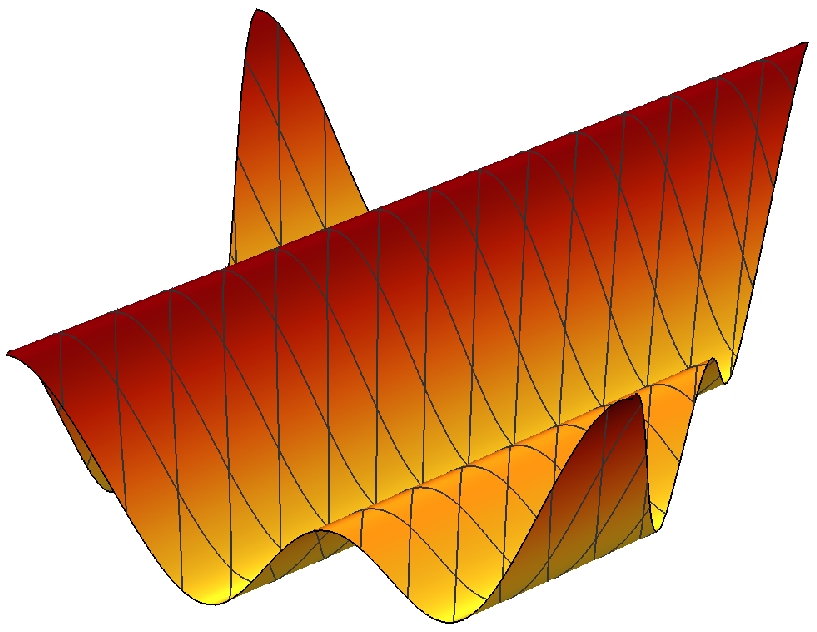}
   \end{subfigure}\vspace{10mm}

   \begin{subfigure}{0.45\textwidth}
   \includegraphics[width=\textwidth]{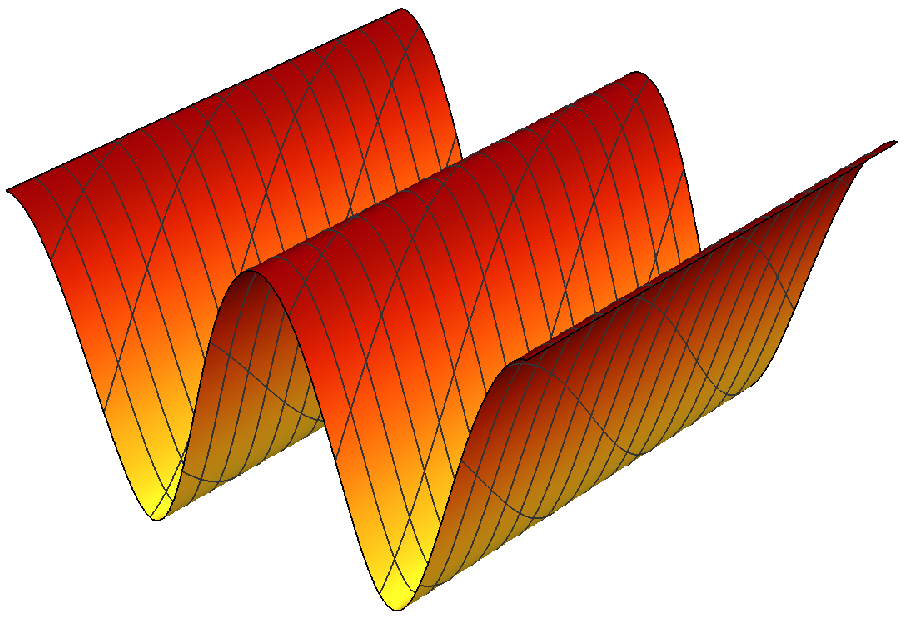}
   \end{subfigure}\qquad
   \begin{subfigure}{0.45\textwidth}
   \includegraphics[width=\textwidth]{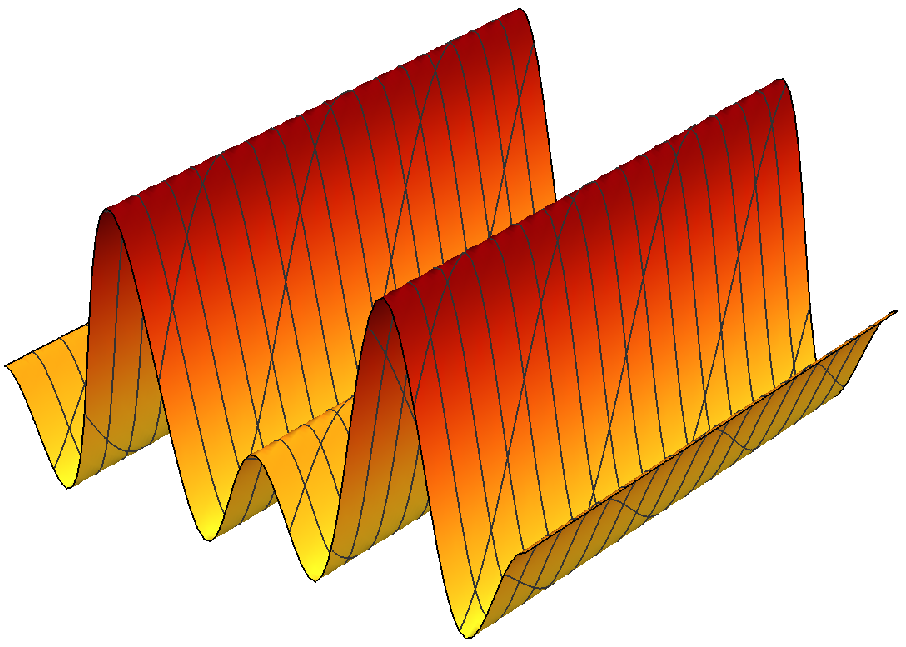}
   \end{subfigure}\vspace{10mm}

   \begin{subfigure}{0.45\textwidth}
   \includegraphics[width=\textwidth]{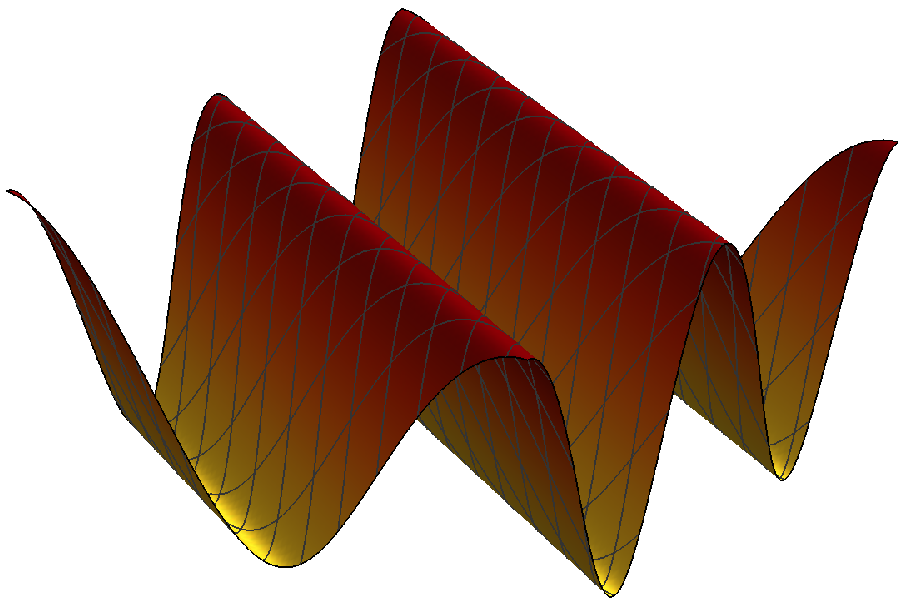}
   \end{subfigure}\qquad
   \begin{subfigure}{0.45\textwidth}
   \includegraphics[width=\textwidth]{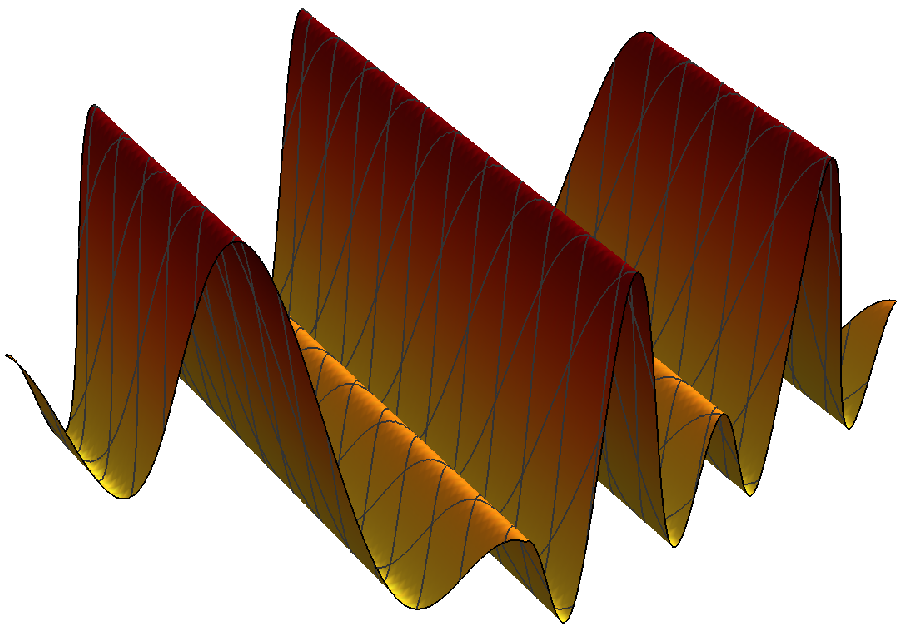}
   \end{subfigure}
   \caption{Examples of of D1-brane solutions on the torus with parameters $R_1=2.4$, $R_2=2.2$ and $\theta=0.4\pi$. From the top, the solutions represent a D1-brane with winding numbers (1,0), a D1-brane with winding numbers (1,1), two D1-branes with winding numbers (0,1) and a D1-brane with winding numbers (2,-1).}
   \label{fig:torus examples 2}
\end{figure}

Next, there are D1-brane solutions. Given the size of the torus, we find solutions describing one or two D1-branes, which can have various winding numbers. Figure \ref{fig:torus examples 2} shows several examples of these solutions. They are essentially just lump solutions from a circle which have been stretched in one direction and therefore they are simpler than D0-brane solutions.

\begin{figure}
   \centering
   \begin{subfigure}{0.45\textwidth}
   \includegraphics[width=\textwidth]{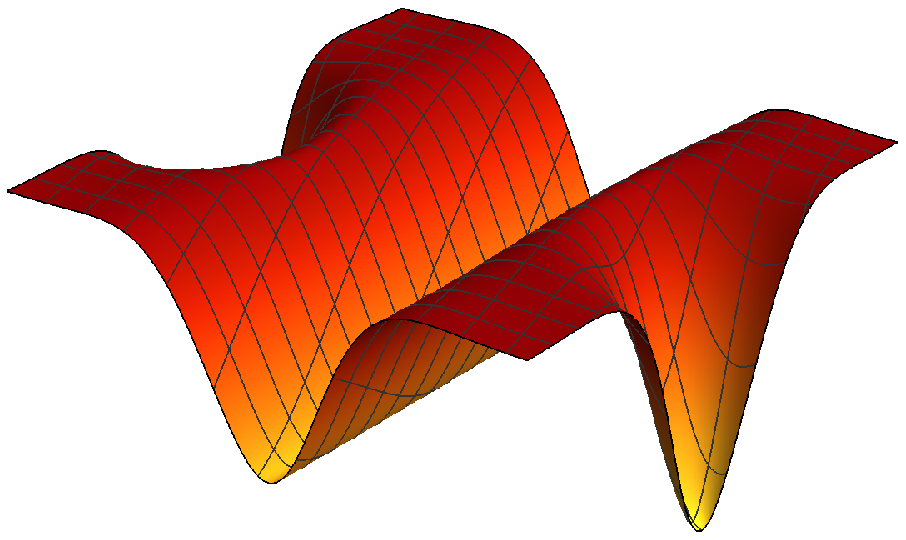}
   \end{subfigure}\qquad
   \begin{subfigure}{0.45\textwidth}
   \includegraphics[width=\textwidth]{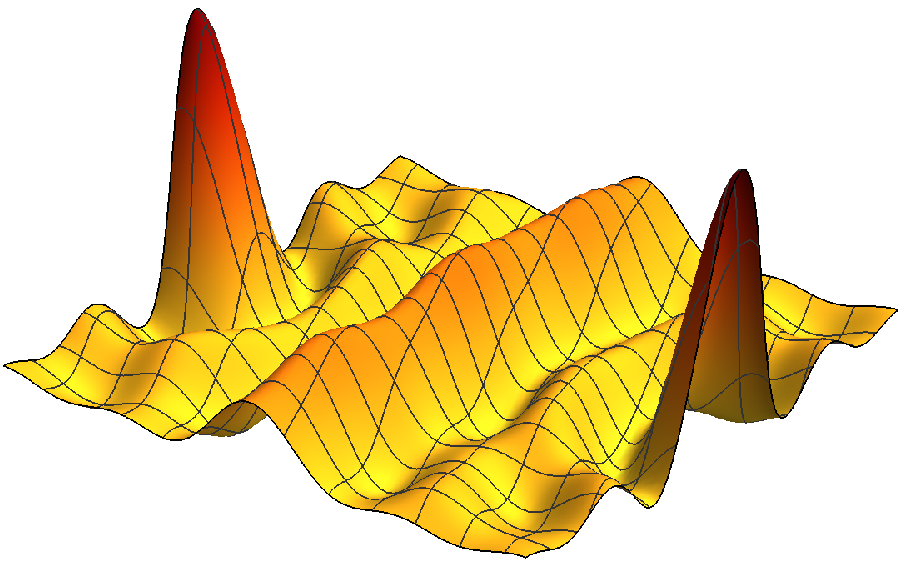}
   \end{subfigure}\vspace{10mm}

   \begin{subfigure}{0.45\textwidth}
   \includegraphics[width=\textwidth]{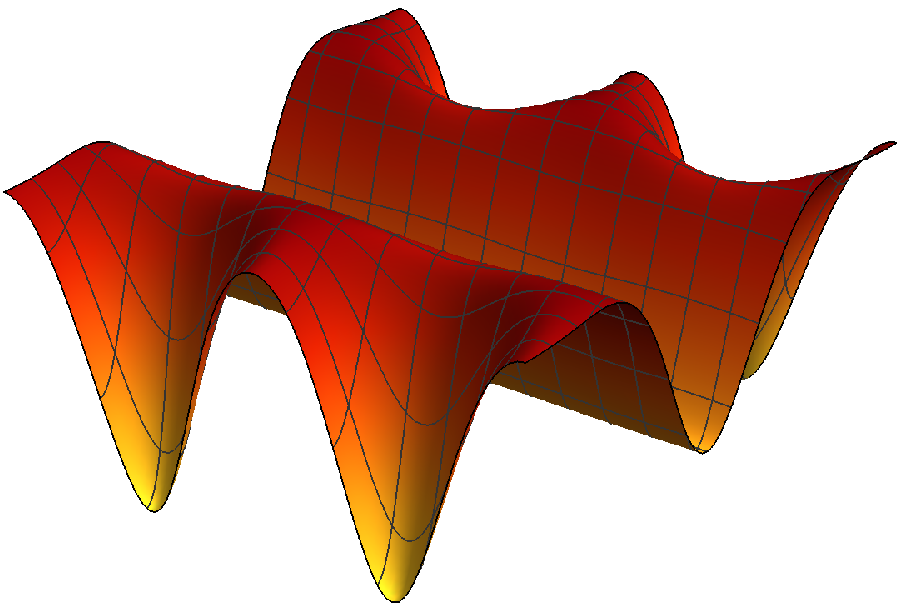}
   \end{subfigure}\qquad
   \begin{subfigure}{0.45\textwidth}
   \includegraphics[width=\textwidth]{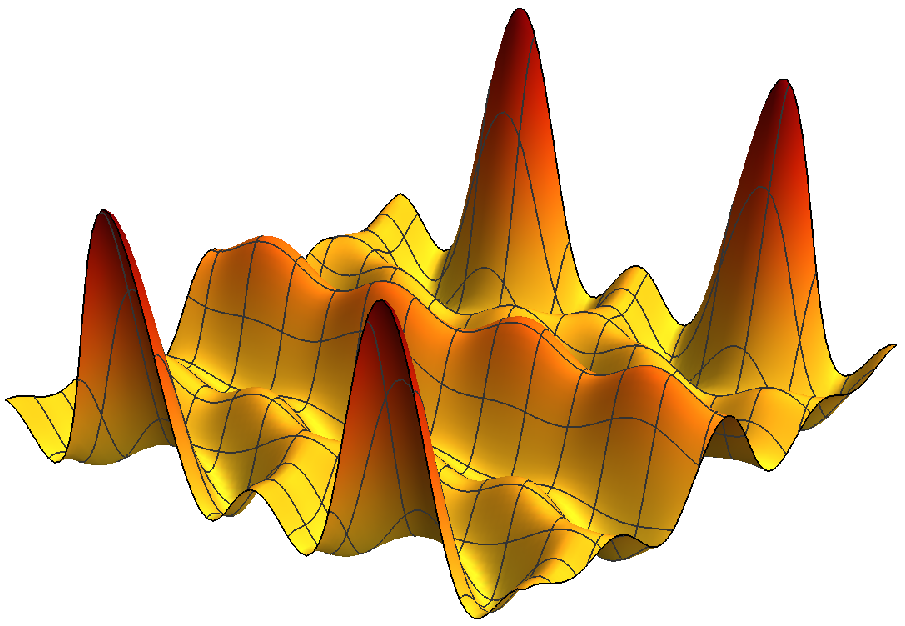}
   \end{subfigure}\vspace{10mm}

   \begin{subfigure}{0.45\textwidth}
   \includegraphics[width=\textwidth]{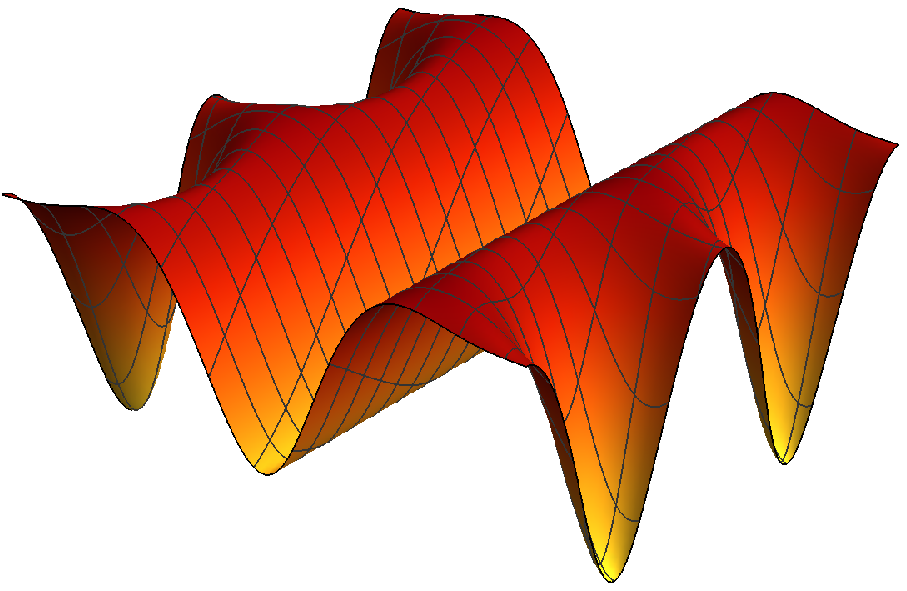}
   \end{subfigure}\qquad
   \begin{subfigure}{0.45\textwidth}
   \includegraphics[width=\textwidth]{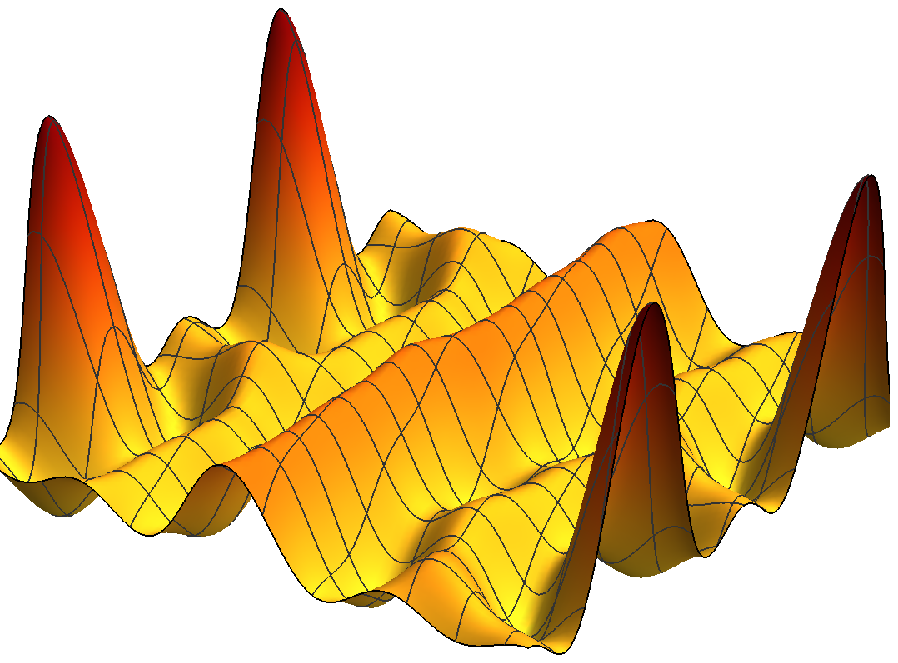}
   \end{subfigure}\vspace{10mm}
   \caption{Examples of solutions on the torus with parameters $R_1=2.4$, $R_2=2.2$ and $\theta=0.4\pi$. From the top, the solutions represent a D1-brane with winding numbers (0,1) and a D0-brane, a D1-brane with winding numbers (1,0) and two D0-branes and a D1-brane with winding numbers (0,1) and two D0-branes.}
   \label{fig:torus examples 3}
\end{figure}

Solutions can also describe both types of D-branes at once, figure \ref{fig:torus examples 3} shows solutions corresponding to one D1-brane and one or two D0-branes. Their tachyon profiles nicely combine both types of gaussians. The energy density profiles are slightly different, we observe that D0-brane peaks are more distinct than D1-brane ridges because their energy is concentrated into a smaller area, but both types of objects are easy to recognize.

We have not found any solutions describing D2-branes with magnetic flux. The reason is probably that these D-branes have much higher energy than the background, so we do not see them even among complex seeds\footnote{Some D2-branes with magnetic flux can be found if we consider more complicated backgrounds with higher energy \cite{KudrnaVosmera}.}.

Finally, there are some solutions that cannot be identified as conventional D-branes. We will discuss them in subsection \ref{sec:FB D2:sample torus:exotic}, but before that, we are going to analyze one example of a regular solution to show that there is also a quantitative agreement between gauge invariant observables and expected boundary states.

\FloatBarrier
\subsection{Solution for a D1-brane and two D0-branes}\label{sec:FB D2:sample torus:D1+2D0}
In this subsection, we analyze one solution as an example. We have picked the last solution from figure \ref{fig:torus examples 3}. The two profiles suggest that the solution describes one D1-brane with winding numbers (0,1) and two symmetrically positioned D0-branes. We can confirm this identification by looking at its observables, some of which are shown in table \ref{tab:torus D1+2D2}. The energy of this D-brane configuration should be equal to $1+1+2.2=4.2$ and the solution comes very close. On the other hand, the $E_{0,0}$ invariant overshoots the expected value and the extrapolation does not really help. We observe that the same problem repeats also for some other solutions on this torus, typically for those which describe D0-branes or D1-branes close to each other. This behavior is similar to the case of single lump solutions on a circle with radius close to 1 (see figure \ref{fig:FB E0}), therefore we think that the $E_{0,0}$ invariant will return to the correct value, we just do not know the solution to high enough level to see that.

Next, we consider the invariants $D_{1\mu\nu}$. For two D0-branes, they should be equal simply to
\begin{equation}
D_{1\mu\nu}^{(2D0)}=
\left(\begin{array}{cc}
-2 & 0 \\
0 & -2 \\
\end{array}\right).
\end{equation}
The invariants of an inclined D1-brane are a bit more complicated. Using (\ref{D1 inv expected D1}) for $\theta=0.4\pi$, we get
\begin{equation}
D_{1\mu\nu}^{(D1)}=
\left(\begin{array}{cc}
-1.77984 & 1.29313 \\
1.29313 & 1.77984 \\
\end{array}\right).
\end{equation}
Table \ref{tab:torus D1+2D2} shows that the extrapolated invariants are relatively close to sums of these numbers.

The invariants $E_{n_1,n_2}$ for two D0-branes depend on how the cosines of their positions from (\ref{En inv expected D0}) combine together. The expected values for our configuration are either 0 or $\pm 2$. For a D1-brane, these invariants should be equal to $2.2$ if the momentum is perpendicular to the D1-brane and to 0 otherwise. When combined, we find that about half of these invariants is identically equal to zero, but the rest can be compared with the solution. Table \ref{tab:torus D1+2D2} shows the first few invariants and we can see that they have approximately the same precision as $E_{0,0}$. The invariants $W_{n_1,n_2}$ are not very helpful because their weights start at $1.21$, which means that they exhibit large oscillations. Therefore we do not even show them.

Overall, the solution reproduces the expected boundary state with worse precision than other lump solutions from chapter \ref{sec:FB circle}. However, this can be expected because we are able to evaluate this solution only up to significantly lower level than in the theory on a circle. Another problem is that we can use only second order extrapolations for Ellwood invariants, which inevitably leads to larger errors. Nevertheless, we have many nontrivial Ellwood invariants which allow us to identify the solution without any doubts.

\begin{table}[!t]
\centering
\footnotesize
\begin{tabular}{|l|lllll|}\hline
Level    & Energy    & $\ps \Delta_S$  & $\ps D_{1XX}$ & $\ps D_{1YY}$ & $\ps D_{1XY}$ \\\hline
2        & 4.54498   & $\ps 0.0044617$ & $   -3.87181$ & $\ps 1.91503$ & $\ps 2.07439$ \\
3        & 4.48154   & $   -0.0029300$ & $   -4.62278$ & $\ps 0.65529$ & $\ps 1.89673$ \\
4        & 4.39872   & $   -0.0040695$ & $   -2.19747$ & $\ps 1.12165$ & $\ps 1.19696$ \\
5        & 4.37527   & $   -0.0042840$ & $   -2.19104$ & $\ps 0.95298$ & $\ps 1.13572$ \\
6        & 4.34246   & $   -0.0043615$ & $   -3.96225$ & $\ps 0.23927$ & $\ps 1.52174$ \\
7        & 4.3306    & $   -0.0042054$ & $   -4.05581$ & $\ps 0.09969$ & $\ps 1.50679$ \\
8        & 4.31265   & $   -0.0041583$ & $   -3.08741$ & $\ps 0.41405$ & $\ps 1.26954$ \\
9        & 4.30544   & $   -0.0039817$ & $   -3.08274$ & $\ps 0.37828$ & $\ps 1.25542$ \\
10       & 4.29395   & $   -0.0039039$ & $   -3.86062$ & $\ps 0.03092$ & $\ps 1.41193$ \\
11       & 4.28904   & $   -0.0037461$ & $   -3.88428$ & $   -0.01336$ & $\ps 1.40513$ \\
12       & 4.28098   & $   -0.0036626$ & $   -3.36758$ & $\ps 0.18634$ & $\ps 1.28986$ \\
13       & 4.27739   & $   -0.0035263$ & $   -3.36516$ & $\ps 0.17124$ & $\ps 1.28382$ \\\hline
$\inf$   & 4.2049    & $   -0.0009   $ & $   -3.74   $ & $   -0.22   $ & $\ps 1.28   $ \\
$\sigma$ & 0.0007    & $\ps 0.0001   $ & $\ps 0.17   $ & $\ps 0.04   $ & $\ps 0.05   $ \\\hline
Exp.     & 4.2       & $\ps 0        $ & $   -3.77984$ & $   -0.22016$ & $\ps 1.29313$ \\\hline
\multicolumn{6}{l}{}\\[-2mm]\hline
Level    & $E_{0,0}$ & $\ps E_{1,0}$   & $\ps E_{2,0}$ & $\ps E_{1,2}$ & $ E_{0,2}$    \\\hline
2        & 4.34810   & $   -0.131971$  & $\ps 3.90216$ & $   -1.04209$ & $   -1.47463$ \\
3        & 4.25774   & $   -0.163610$  & $\ps 3.76379$ & $   -1.64092$ & $   -1.60676$ \\
4        & 4.23520   & $   -0.195510$  & $\ps 3.87857$ & $   -1.65805$ & $   -1.71237$ \\
5        & 4.20944   & $   -0.196295$  & $\ps 3.97879$ & $   -1.78280$ & $   -1.76689$ \\
6        & 4.19803   & $   -0.210487$  & $\ps 4.01773$ & $   -1.79098$ & $   -1.80872$ \\
7        & 4.18718   & $   -0.208601$  & $\ps 4.00975$ & $   -1.83068$ & $   -1.81582$ \\
8        & 4.18875   & $   -0.216513$  & $\ps 4.02668$ & $   -1.83395$ & $   -1.83627$ \\
9        & 4.18304   & $   -0.214648$  & $\ps 4.05873$ & $   -1.86690$ & $   -1.85689$ \\
10       & 4.18100   & $   -0.219685$  & $\ps 4.06968$ & $   -1.86930$ & $   -1.86998$ \\
11       & 4.17757   & $   -0.217749$  & $\ps 4.06774$ & $   -1.88244$ & $   -1.87208$ \\
12       & 4.17955   & $   -0.221219$  & $\ps 4.07474$ & $   -1.88386$ & $   -1.88072$ \\
13       & 4.17732   & $   -0.219525$  & $\ps 4.09061$ & $   -1.89981$ & $   -1.89186$ \\\hline
$\inf$   & 4.169     & $   -0.228   $  & $\ps 4.165  $ & $   -1.96   $ & $   -1.96843$ \\
$\sigma$ & 0.004     & $\ps 0.002   $  & $\ps 0.004  $ & $\ps 0.02   $ & $\ps 0.007  $ \\\hline
Exp.     & 4.2       & $   -0.2     $  & $\ps 4.2    $ & $   -2      $ & $   -2      $ \\\hline
\end{tabular}
\caption{Selected observables of a solution describing one D1-brane and two D0-branes on a torus. About half of the $E_{n_1,n_2}$ invariants is identically zero (for example $E_{0,1}$, $E_{1,1}$, $E_{1,-1}$, $E_{2,1}$,$\dots$), so we show only the nontrivial invariants.}
\label{tab:torus D1+2D2}
\end{table}

\FloatBarrier
\subsection{Exotic solutions}\label{sec:FB D2:sample torus:exotic}
In the free boson theory on a circle, we have been able to uniquely identify all well-behaved solutions, but that is no longer true on a torus. There are so-called exotic solutions that clearly do not represent any combination of D0-branes, D1-branes or D2-branes. These solutions do not have any pathologies, they are real, stable in the level truncation scheme, stable with respect to deformations of the torus and they satisfy the out-of-Siegel equations quite well. Therefore it is not likely that they are artifacts of the level truncation approximation of the full theory. We interpret these solutions as non-conventional (symmetry-breaking) boundary states, which do not satisfy any current gluing conditions of the form
\begin{equation}
(\alpha^\mu_n +\Omega^\mu_{\ \nu}\bar \alpha_{-n}^\nu)\| B\rra=0.
\end{equation}
These boundary states have not been classified yet because this theory is irrational with respect to the $c=2$ Virasoro algebra and it is not known how to solve the Cardy conditions. Therefore OSFT solutions, which give us numerical approximations of the corresponding boundary states, can serve as hints for solving this problem analytically. In a project with Jakub Vo\v{s}mera \cite{KudrnaVosmera}, we analyze these solutions on a torus with $Z_6$ symmetry. We have managed to find analytic expressions for some exotic boundary states at one special radius, where the free boson theory is dual to a product of minimal models. A similar construction can be used in superstring theory \cite{VosmeraExotic}. In general, it seems that exotic solutions can be found on every torus where the initial D2-brane has high enough energy.

\begin{figure}
   \centering
   \begin{subfigure}{0.45\textwidth}
   \includegraphics[width=\textwidth]{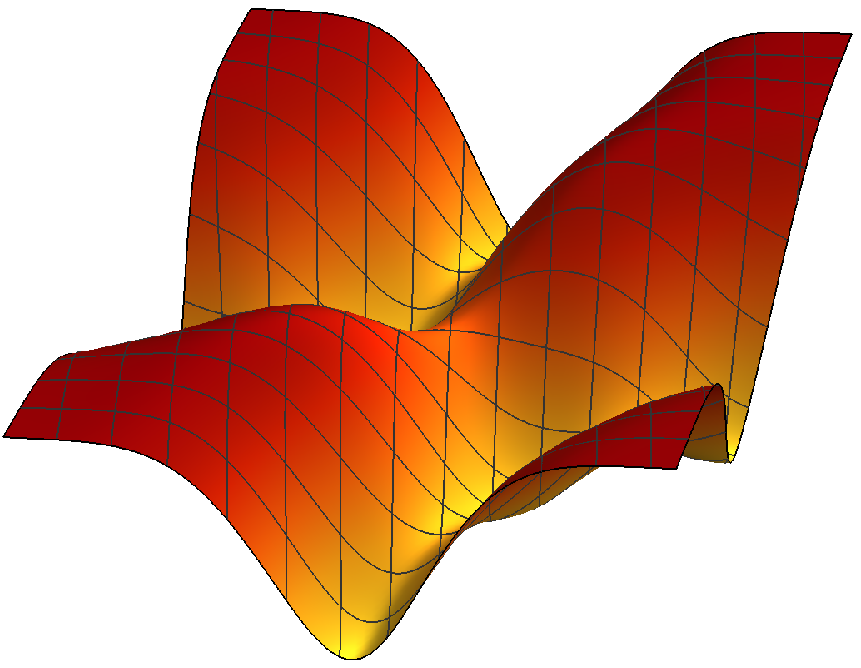}
   \end{subfigure}\qquad
   \begin{subfigure}{0.45\textwidth}
   \includegraphics[width=\textwidth]{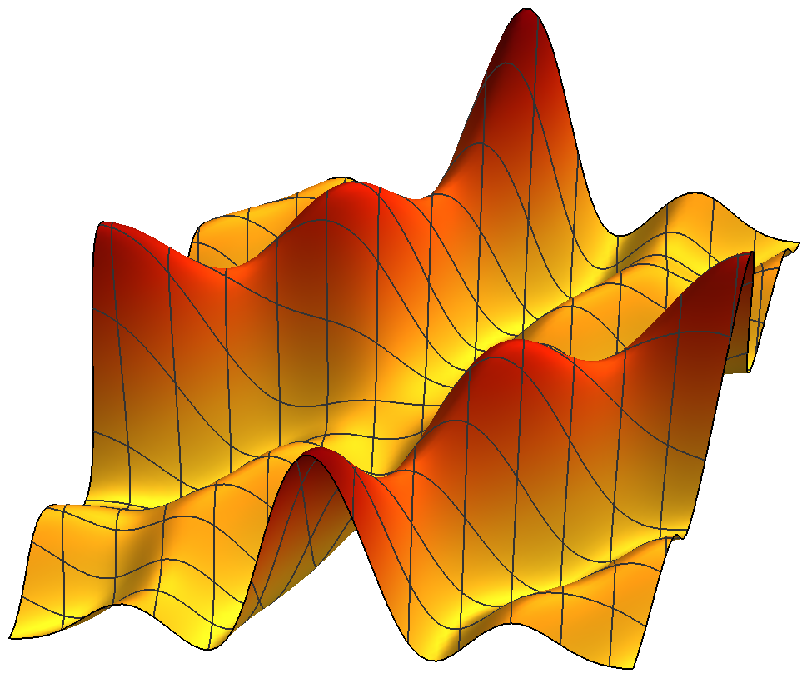}
   \end{subfigure}

   \begin{subfigure}{0.45\textwidth}
   \includegraphics[width=\textwidth]{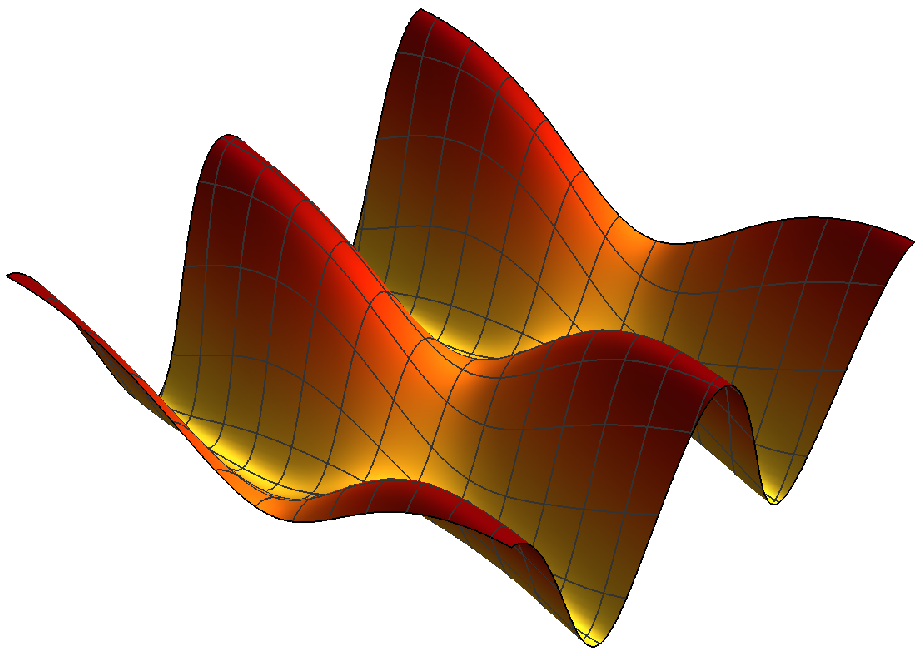}
   \end{subfigure}\qquad
   \begin{subfigure}{0.45\textwidth}
   \includegraphics[width=\textwidth]{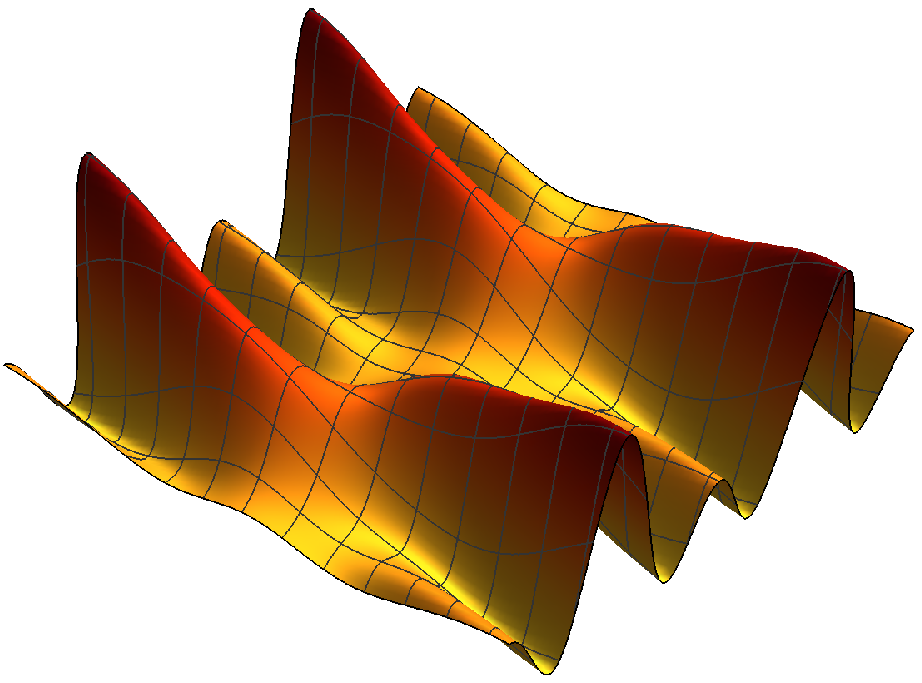}
   \end{subfigure}\vspace{-10mm}

   \begin{subfigure}{0.45\textwidth}
   \includegraphics[width=\textwidth]{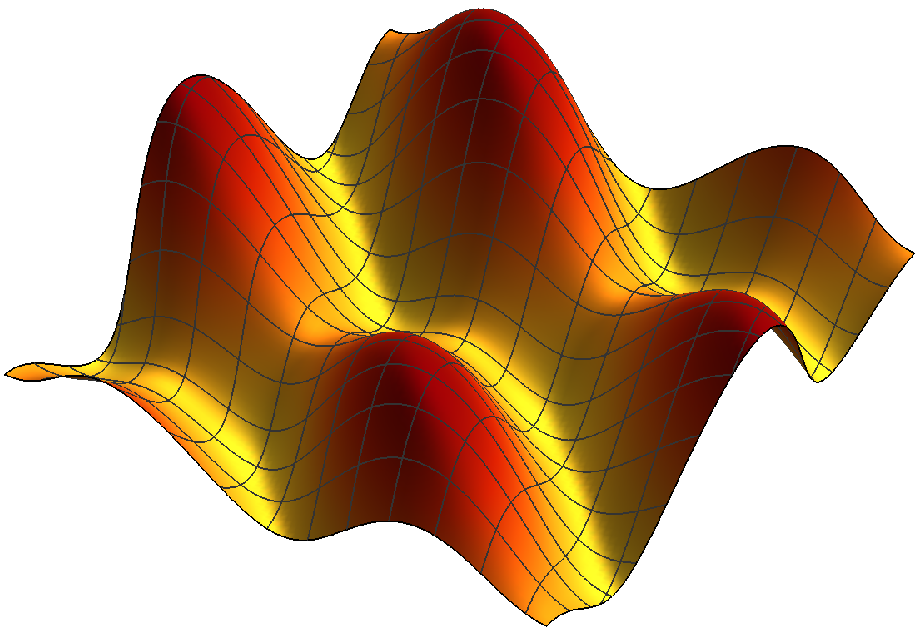}
   \end{subfigure}\qquad
   \begin{subfigure}{0.45\textwidth}
   \includegraphics[width=\textwidth]{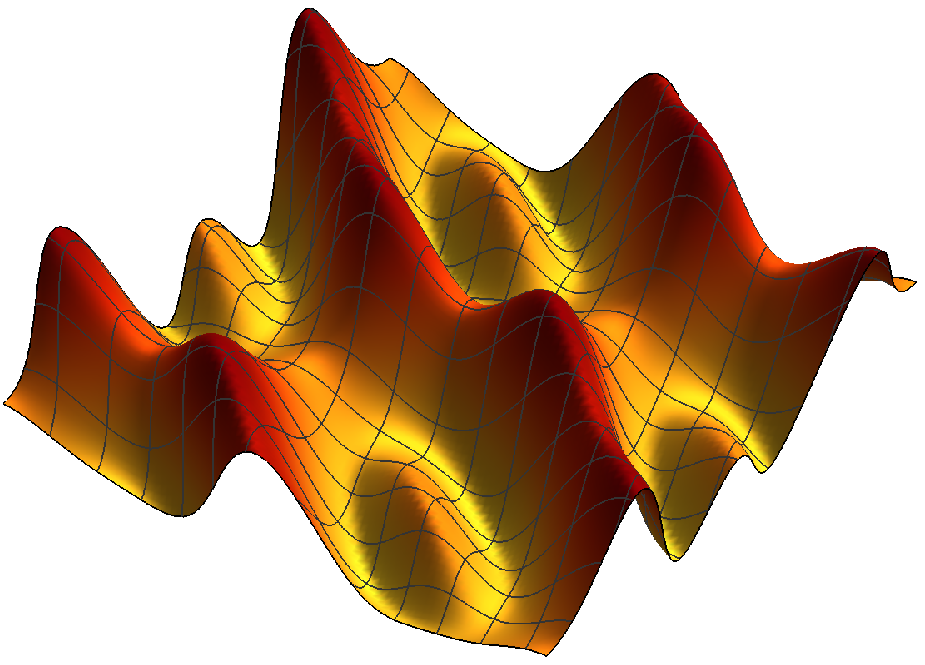}
   \end{subfigure}\vspace{-5mm}

   \begin{subfigure}{0.45\textwidth}
   \includegraphics[width=\textwidth]{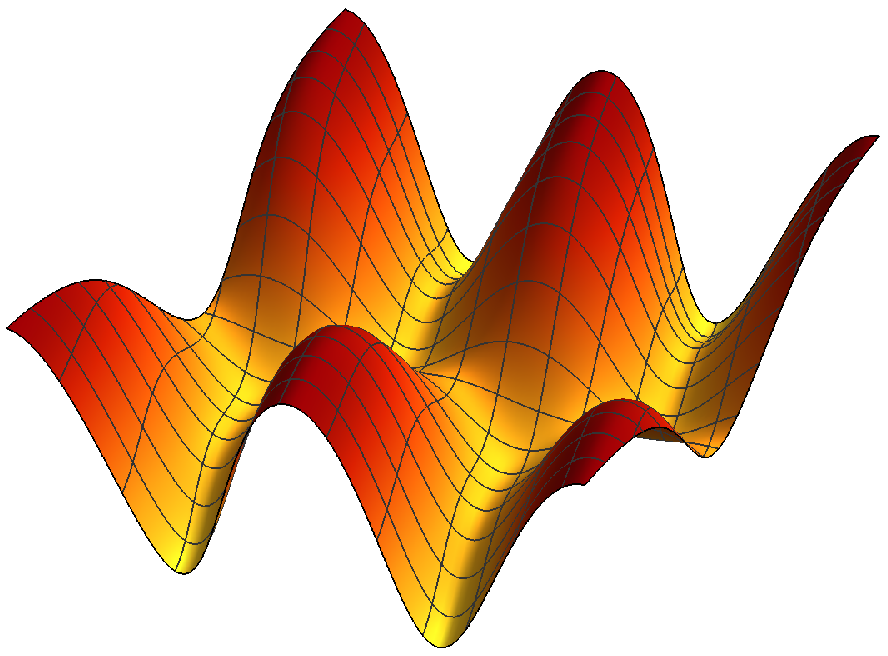}
   \end{subfigure}\qquad
   \begin{subfigure}{0.45\textwidth}
   \includegraphics[width=\textwidth]{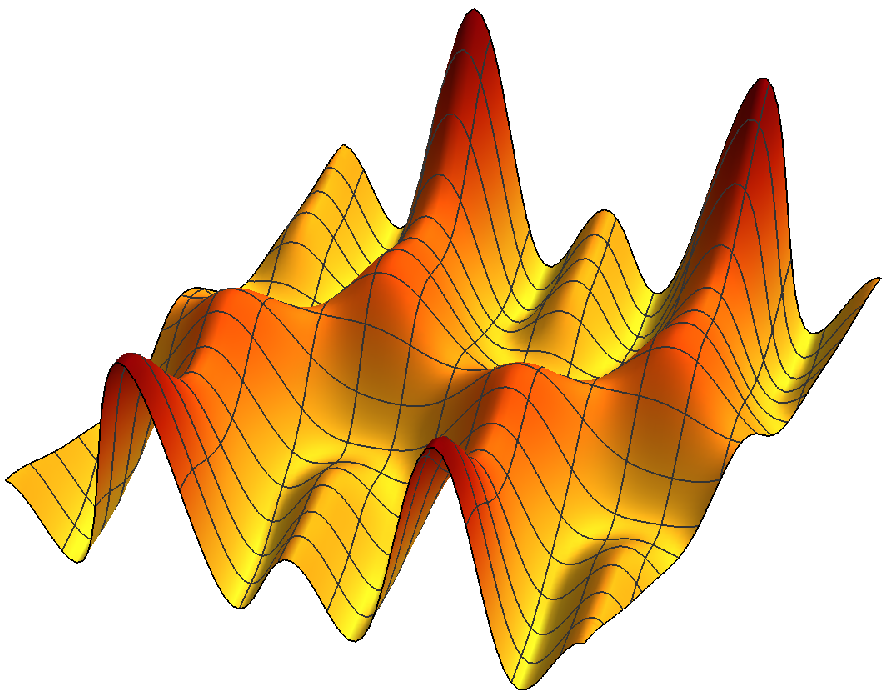}
   \end{subfigure}
   \caption{Examples of exotic solutions on the torus with parameters $R_1=2.4$, $R_2=2.2$ and $\theta=0.4\pi$, which correspond to non-conventional boundary states.}
   \label{fig:torus examples 4}
\end{figure}

In the case of the torus we have chosen in this section, we have found four types of exotic solutions. Their profiles are shown in figure \ref{fig:torus examples 4}. Take for example the first solution. Its profiles remind us of a D1-brane with winding numbers (1,1) (compare it with the second solution in figure \ref{fig:torus examples 2}), but it has been deformed from a straight line to a wave-like shape and it does not have constant energy density along the line. We considered the possibility that this solution is a gauge transformation of a D1-brane solution, but we have ruled it out by comparing its invariants with a D1-brane boundary state, see table \ref{tab:torus exotic 1}. The numbers are somewhat similar, but there is a clear difference. For example, its energy is $3.84$, while the D1-brane has energy $3.72$. We can also easily rule out all other possible D-brane configurations, which means that this solution represents a new boundary state.

The next two solutions in figure \ref{fig:torus examples 4} also look a bit like deformed D1-branes. The last solution reminds us a D1-brane net with hexagonal pattern, which somewhat similar to the honeycomb solution discussed in \cite{KudrnaVosmera}. A similar hexagonal pattern is relatively common among exotic solutions, but these solutions never have enough energy to describe a net of D1-branes.

Notice that all solutions in figure \ref{fig:torus examples 4} have the same type of correspondence between the tachyon and the energy density profile as regular solutions, valleys in one profile correspond to ridges in the other profile and the other way around. However, it seems that energy density of exotic solutions is spread unevenly across a larger area, while energy density of regular D-branes it is always localized to points and lines.

Unfortunately, all we can learn about these non-conventional D-branes right now is part of their boundary state. Therefore it would be very useful if we could compute spectra of excitations around these solutions, which would tell us more about the corresponding BCFTs.

\begin{table}[!]
\centering
\footnotesize
\begin{tabular}{|l|llllll|}\hline
Level     & Energy       & $E_{0,0}$ & $D_{1XX}$ & $\ps D_{1YY}$  & $D_{1XY}$ & $\ps \Delta_S$  \\\hline
2         & 4.07435      & 4.01834   & 1.61830   & $   -0.592648$ & 2.63862   & $\ps 0.0229351$ \\
3         & 3.96899      & 3.89948   & 1.37557   & $   -1.064910$ & 2.93799   & $\ps 0.0051528$ \\
4         & 3.92584      & 3.90130   & 1.81978   & $   -0.029252$ & 2.20095   & $\ps 0.0035096$ \\
5         & 3.90694      & 3.88053   & 1.77841   & $   -0.098403$ & 2.23391   & $\ps 0.0022101$ \\
6         & 3.89131      & 3.87178   & 1.49440   & $   -0.659290$ & 2.56811   & $\ps 0.0016847$ \\
7         & 3.88446      & 3.86438   & 1.46926   & $   -0.702204$ & 2.58998   & $\ps 0.0012443$ \\
8         & 3.87649      & 3.86429   & 1.55750   & $   -0.477520$ & 2.41466   & $\ps 0.0010255$ \\
9         & 3.87314      & 3.86061   & 1.54940   & $   -0.489404$ & 2.41926   & $\ps 0.0008036$ \\
10        & 3.86832      & 3.85711   & 1.44725   & $   -0.685991$ & 2.53359   & $\ps 0.0006965$ \\
11        & 3.86638      & 3.85494   & 1.44071   & $   -0.696202$ & 2.53814   & $\ps 0.0005633$ \\
12        & 3.86315      & 3.85499   & 1.47941   & $   -0.597697$ & 2.46093   & $\ps 0.0005055$ \\\hline
$\inf$    & 3.84030      & 3.840     & 1.32      & $   -0.82    $ & 2.53      & $   -0.00006  $ \\
$\sigma$  & $3\dexp{-7}$ & 0.004     & 0.04      & $\ps 0.10    $ & 0.08      & $\ps 0.00001  $ \\\hline\hline
Exp. (D1) & 3.72333      & 3.72333   & 1.37178   & $   -1.37178 $ & 3.46142   & $\ps 0        $ \\\hline
\end{tabular}
\caption{Selected observables of the first exotic solution from figure \ref{fig:torus examples 4}. The last line represents boundary state of a D1-brane with winding numbers (1,1), which we show for comparison.}
\label{tab:torus exotic 1}
\end{table}

\section{Search for intersecting D1-branes}\label{sec:FB D2:intersection}
In the previous section, we have shown examples of solutions describing various combinations of D0-branes and D1-branes, but one type of solution is missing: intersecting D1-branes. There is always a possibility that such solution does not appear at a given initial level or that it gets lost at some step of Newton's method, so we have tried to find intersecting D1-branes at several different tori. The ideal configuration for this task is most likely a square torus, which eliminates potential issues connected to asymmetry of the torus. However, we have never found any solution describing intersecting D1-branes. We have even tried a superposition method similar to the one described in section \ref{sec:FB circle:double:superposition}, but without success.

\begin{figure}
   \centering
   \includegraphics[width=7.5cm]{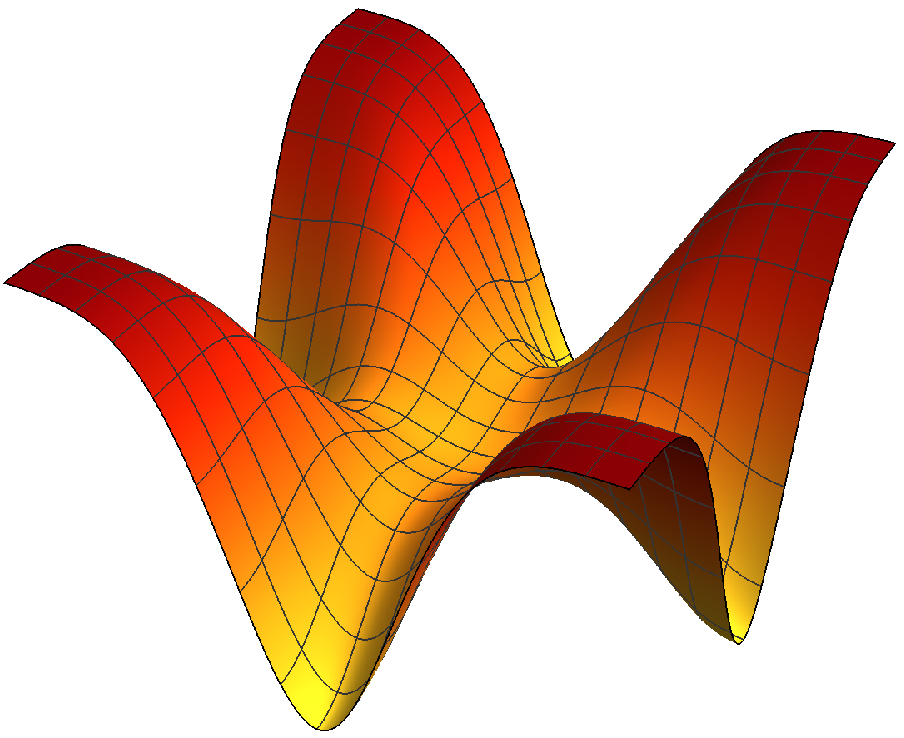}\vspace{10mm}
   \includegraphics[width=7.5cm]{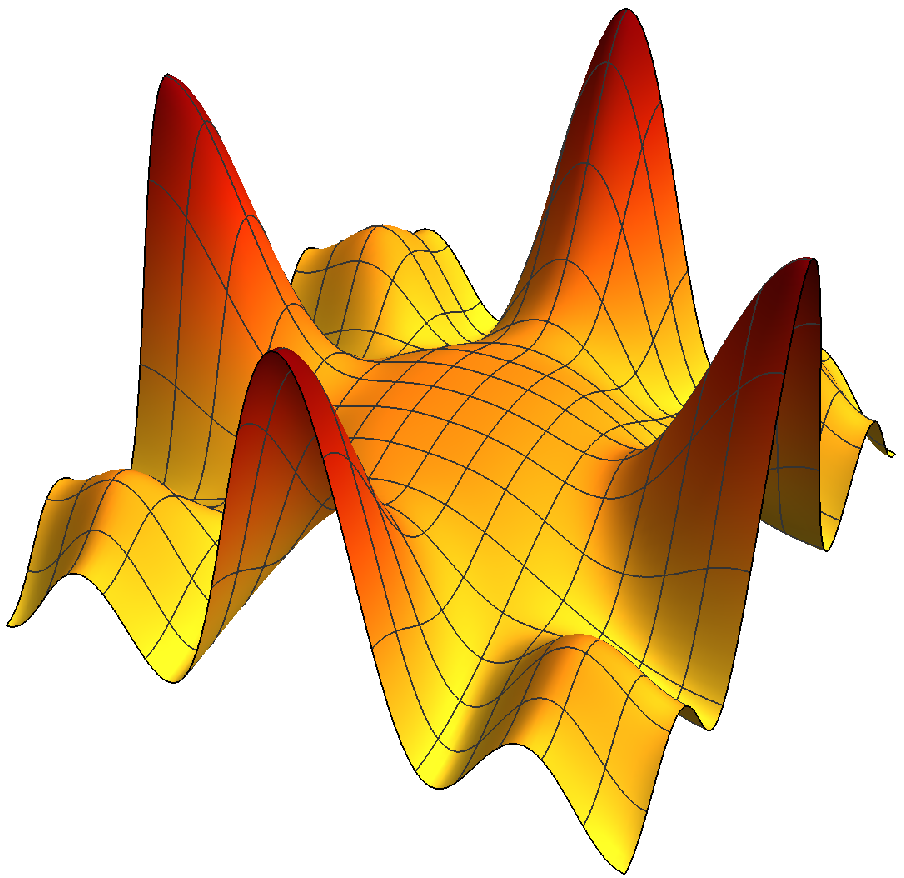}
   \caption{Tachyon profile (top figure) and energy density profile (bottom figure) of an exotic solution on a square torus with $R_1=R_2=1.8$. The solution possibly describes recombination of intersecting D1-branes.}
   \label{fig:torus square exotic}
\end{figure}

In the theory on a torus, there is one class of solutions with the correct $Z_4$ symmetry and energy concentrated along two intersecting lines. Figure \ref{fig:torus square exotic} and table \ref{tab:torus exotic 2} show some properties of one such solution at radius $R=1.8$. However, its energy is clearly not equal to $3.6$, which is the energy of D1-branes intersecting at right angle, and the energy density decreases at the intersection, while we would expect its increase. Therefore the solution does not represent intersecting D1-branes, but it is an exotic solution describing some non-conventional boundary conditions.

The absence of solutions for intersecting D1-branes supports our hypothesis from section \ref{sec:FB circle:double:superposition} (see also section \ref{sec:MM:Ising3:regular}) that it is difficult to find solutions describing multiple D-branes with stretched tachyonic strings in the spectrum. In this particular case of intersecting D1-branes, stretched strings have mixed Neumann and Dirichlet boundary conditions and the ground states in this sector, which are described by boundary condition changing operators, have weights $\frac{\theta}{\pi}(1-\frac{\theta}{\pi})/2$, which equals to $\frac{1}{8}$ for orthogonal branes. These modes are strongly tachyonic and the first descendants created using $\alpha_{-\frac{1}{2}}$ are also tachyonic. We can speculate that the aforementioned exotic solution describes a recombination of the intersecting D1-branes caused by condensation of these stretched tachyons, but more data would be needed to confirm this claim.

Finding solutions describing intersecting D1-branes is probably going to be very difficult. We expect that it will be necessary to solve at least level 2 equations on large enough torus ($R>2$), but the number of states in this theory grows quickly with level and solving the equations will require large amount of computer resources. The $Z_4$ symmetry on a square torus could help a bit because it could be used to reduce the number of independent variables for symmetric solutions. Furthermore, since the tachyons on intersecting D1-branes are very light, we expect that seeds for these solutions are going to be complex and it will be difficult to identify them among many other complex solutions.

\begin{table}
\centering
\begin{tabular}{|l|lllll|}\hline
Level    & Energy       & $E_{0,0}$ & $E_{0,1}$ & $D_{1XX}$ & $\ps \Delta_S$  \\\hline
2        & 2.83077      & 2.80245   & 0.664629  & 0.801166  & $\ps 0.0197752$ \\
3        & 2.77168      & 2.72540   & 0.712624  & 0.665611  & $\ps 0.0033006$ \\
4        & 2.75182      & 2.73322   & 0.734841  & 0.883228  & $\ps 0.0029023$ \\
5        & 2.74015      & 2.71925   & 0.746719  & 0.835579  & $\ps 0.0014871$ \\
6        & 2.73331      & 2.71829   & 0.754972  & 0.675610  & $\ps 0.0013920$ \\
7        & 2.72884      & 2.71306   & 0.763239  & 0.652171  & $\ps 0.0008809$ \\
8        & 2.72540      & 2.71512   & 0.767574  & 0.679369  & $\ps 0.0008568$ \\
9        & 2.72312      & 2.71244   & 0.770689  & 0.669485  & $\ps 0.0005938$ \\
10       & 2.72105      & 2.71192   & 0.773397  & 0.618965  & $\ps 0.0005918$ \\
11       & 2.71969      & 2.71030   & 0.776438  & 0.612229  & $\ps 0.0004323$ \\
12       & 2.71831      & 2.71125   & 0.778291  & 0.623014  & $\ps 0.0004381$ \\
13       & 2.71741      & 2.71016   & 0.779653  & 0.619087  & $\ps 0.0003314$ \\\hline
$\inf$   & 2.70623      & 2.707     & 0.800     & 0.52      & $   -0.00003  $ \\
$\sigma$ & $7\dexp{-6}$ & 0.001     & 0.001     & 0.01      & $\ps 0.00002  $ \\\hline
\end{tabular}
\caption{Selected observables of an exotic solution on a square torus with $R_1=R_2=1.8$. The solution has a $Z_4$ symmetry and therefore we find $D_{1YY}=D_{1XX}$, $D_{1XY}=0$ and $E_{1,0}=E_{0,1}$.}
\label{tab:torus exotic 2}
\end{table}

\chapter{Results - Minimal models}\label{sec:MM}
In this chapter, we discuss solutions in OSFT theories which include the Virasoro minimal models or their products. We start with the well-known Ising model, which is the simplest unitary minimal model. This model has already been investigated in \cite{Ising}, so we mostly review the results from this reference. Then we investigate the Lee-Young model, which is the simplest nonunitary minimal model, to see whether it behaves differently from unitary theories. Next, we discuss the product of two Ising models. This model has also been considered in \cite{Ising}, but we improve the results from this reference and add some new solutions. This model is special, because it is dual to the free boson theory on the orbifold $S^1/Z_2$ with $R=\sqrt{2}$, so at the end of this chapter  we consider the product of the Ising model and the tricritical Ising model to have an example of a more generic product of two minimal models.

\section{Ising model}\label{sec:MM:Ising}
The Ising model is the simplest unitary minimal model with label $m=3$ or alternatively $(p,q)=(3,4)$. Its central charge is $c=\frac{1}{2}$ and it has three primary operators, which are listed in table \ref{tab:Ising operators}. Nontrivial fusion rules of these operators are
\begin{equation}\label{fusion ising}
\eps \times \eps = \Id, \qquad \sigma \times \sigma = \Id+\eps, \qquad  \sigma \times \eps = \sigma.
\end{equation}

The Ising model has three boundary states, which are shown in table \ref{tab:Ising boudary states}. Since none of the primaries has $\sigma$ in the OPE with itself, the boundary spectra include only the operators $\Id$ and $\eps$ .

\begin{table}[b!]
\centering
\begin{tabular}{|c|c|c|}\hline
operator            & Kac label & $h$            \\\hline
$\Id$   \rowh{12pt} & (1,1)     & $0$            \\
$\sigma$\rowh{13pt} & (1,2)     & $\frac{1}{16}$ \\
$\eps$ \rowh{14pt}  & (1,3)     & $\frac{1}{2}$  \\\hline
\end{tabular}
\caption{Properties of primary operators in the Ising model.}
\label{tab:Ising operators}
\end{table}

\begin{table}[!t]
\centering
\begin{tabular}{|c|c|c|}\hline
Boundary condition  & Boundary state                                                                        & Boundary spectrum \\\hline
$\Id$  \rowh{13pt}  & $\frac{1}{\sqrt2}|\Id\rra+\frac{1}{\sqrt2}|\eps\rra+\frac{1}{\sqrt[4]{2}}|\sigma\rra$ & $\Id$             \\
$\eps$ \rowh{13pt}  & $\frac{1}{\sqrt2}|\Id\rra+\frac{1}{\sqrt2}|\eps\rra-\frac{1}{\sqrt[4]{2}}|\sigma\rra$ & $\Id$             \\
$\sigma$\rowh{13pt} & $|\Id\rra-|\eps\rra$                                                                  & $\Id$,  $\eps$    \\\hline
\end{tabular}
\caption{Boundary states in the Ising model and their spectra.}
\label{tab:Ising boudary states}
\end{table}

For our purposes, the $\Id$-brane and the $\eps$-brane have essentially the same properties. Both have only the identity operator in their spectra, there are no nontrivial boundary structure constants and the equations of motion, which depend only on the central charge, are identical. The only difference lies in the sign of one of the bulk-boundary structure constants, which changes sign of the $E_\sigma$ invariant. Therefore it is enough to investigate OSFT formulated on the $\Id$-brane and on the $\sigma$-brane.

From the technical point of view, the Ising model has one surprising (and unpleasant) property: solutions in this model exhibit a numerical instability. Newton's method usually allows us to find solutions with precision comparable to the precision of the number format we use, but the maximal accessible precision in the Ising model decreases with level. We can partially reduce this problem by changing the C++ number format from double to long double, which increases the working precision from 15 to 18 digits, but we are still unable to reach the usual precision $10^{-12}$ at the highest available levels.

The source of the problem most likely lies in the Gram matrix in the Ising sector of the theory. After we choose an irreducible basis and remove all null states, we find that there are some very small eigenvalues of the restricted Gram matrix. For example, the smallest eigenvalue at level 24 is of order $10^{-9}$. These small eigenvalues propagate to the matrix representation of the kinetic term $Q_{ij}$ and to the Jacobian in Newton's method. The Jacobian is therefore a badly conditioned matrix and its inversion generates large numerical errors. It is not clear whether this problem can be dealt with by some special choice of the irreducible basis. Curiously, this problem does not seem to repeat in other minimal models.

\subsection{Solutions on the $\Id$-brane}\label{sec:MM:Ising:Id}
There are no boundary states with energy lower than the $\Id$-brane energy, so we have look for solutions with the same or higher energy. The spectrum, which includes only descendants of the ground state, offers only one relevant operator, $c_1|0\ra$. This makes the search for solutions more difficult and, in order to find some nontrivial solutions, we have to consider equations including some irrelevant fields, first of which are the Virasoro descendants at level 2.

We can easily solve the equations of motion at this level using the homotopy continuation method, but we find no interesting real solutions. However, there is a complex solution (plus its complex conjugate) with reasonable properties. This solution is stable in the level truncation scheme and its imaginary part quickly decreases as we improve it to higher levels. The imaginary part completely disappears at level 14 and, as we are going to show, we can identify it as the $\sigma$-brane. Historically, it is the first positive energy solution found (see \cite{Ising}) and it is the one known to the highest level, so we will use it as a representative of positive energy solutions and discuss it more detail. Solutions of this type appear in OSFT quite often, other examples are positive energy lumps from section \ref{sec:FB circle:single:smallR} or some the double Ising model solutions from section \ref{sec:MM:Ising2:other}.

\begin{figure}[!]
\centering
   \begin{subfigure}[t]{0.47\textwidth}
      \includegraphics[width=\textwidth]{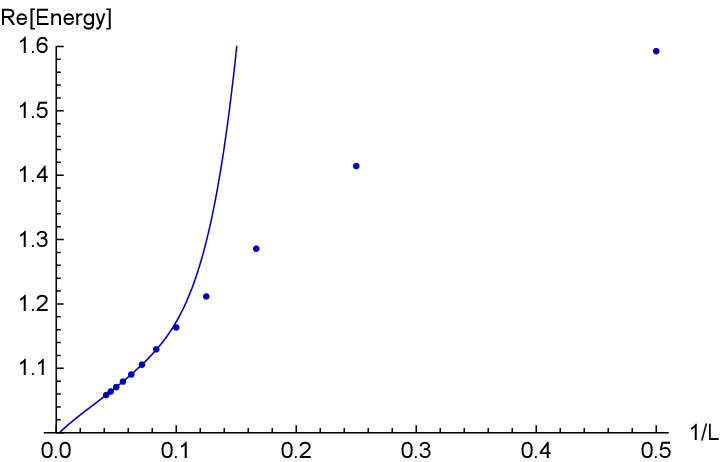}
   \end{subfigure}\qquad
   \begin{subfigure}[t]{0.47\textwidth}
      \includegraphics[width=\textwidth]{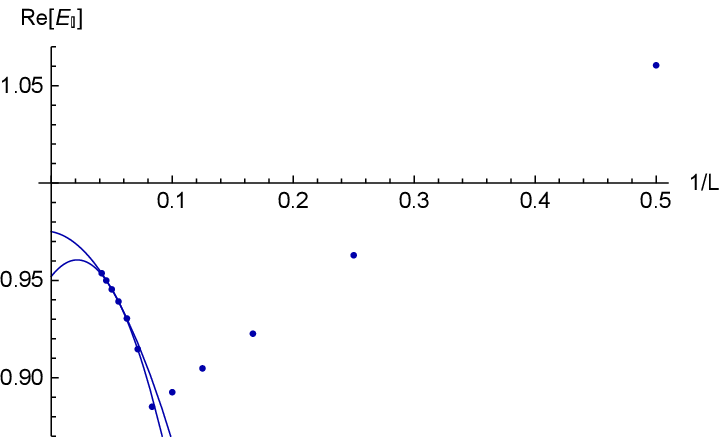}
   \end{subfigure}\vspace{10mm}
   \begin{subfigure}[t]{0.47\textwidth}
      \includegraphics[width=\textwidth]{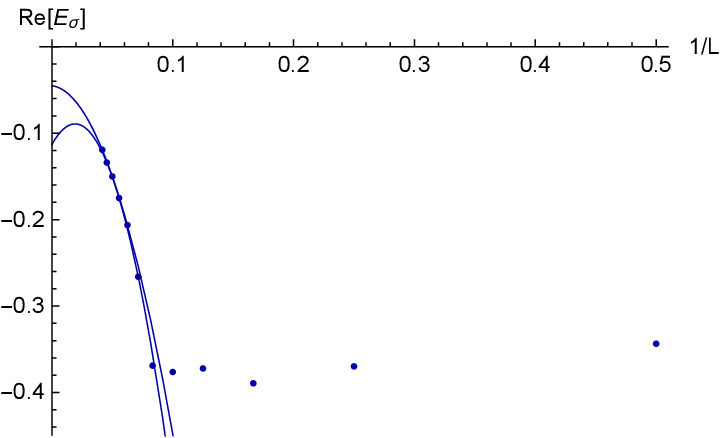}
   \end{subfigure}\qquad
   \begin{subfigure}[t]{0.47\textwidth}
      \includegraphics[width=\textwidth]{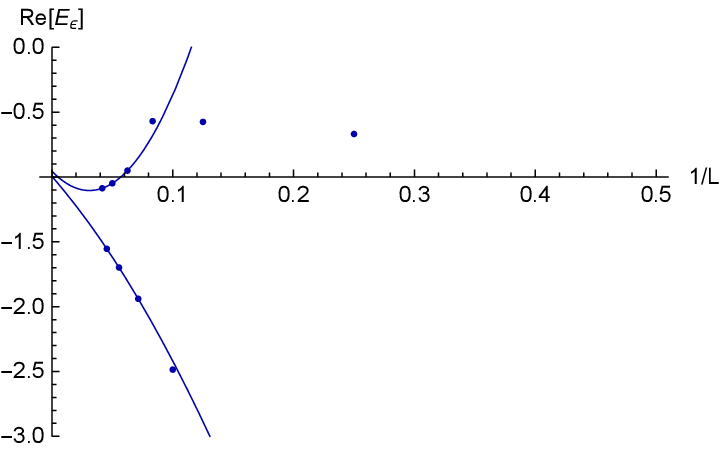}
   \end{subfigure}\vspace{10mm}
   \begin{subfigure}[t]{0.47\textwidth}
      \includegraphics[width=\textwidth]{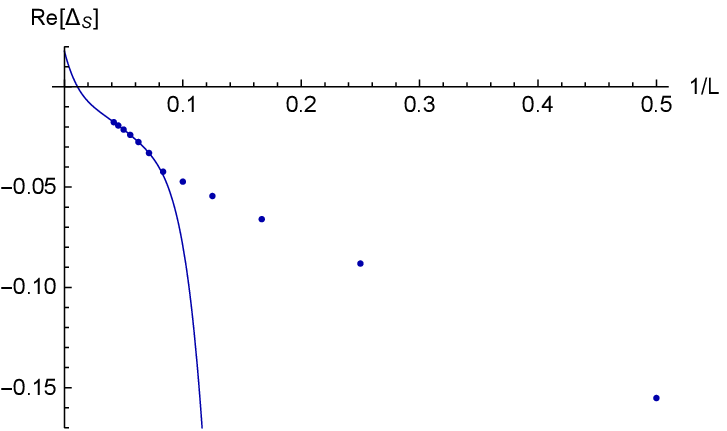}
   \end{subfigure}
   \caption{Real parts of observables of the $\sigma$-brane solution in the Ising model on the $\Id$-brane background. We show level dependence of the energy, the three Ellwood invariants and the out-of-Siegel equation $\Delta_S$. Solid lines represent extrapolations of these quantities using levels from 14 to 24. Horizontal axes in all figures are set to expected values of these observables.}
   \label{fig:Ising Id}
\end{figure}

We have been able to evaluate the solution up to a very nice level 24. The properties of the solution are summarized in table \ref{tab:Ising Id-sigma} and we plot level dependance of real parts of various quantities in figure \ref{fig:Ising Id}. The energy of the solution and the $E_\Id$ invariant are close to 1, so we can unambiguously identify the solution as the $\sigma$-brane because there are no other candidates with similar energy. However, convergence of the gauge invariants towards the expected values is not very good, especially when compared with lump solutions or with the solution in the next subsection. The problem is that the solution undergoes a dramatic change at level 14. That is most apparent on the real part of the $E_\Id$ invariant, see top right part of figure \ref{fig:Ising Id}. At low levels, it moves away from 1, but the trend reverses at level 14 and then it starts climbing back towards the correct value. Other quantities behave similarly, although the change at level 14 is less apparent.

This behavior cannot be described well by a polynomial function, which unfortunately means that we cannot use the data below level 14 for infinite level extrapolations\footnote{We have experimented with other types of extrapolating functions that could possibly describe disappearance of the imaginary part, for example with square roots of polynomials, but we have found none that would lead to consistent results for multiple quantities.}. Therefore we use the usual type of extrapolations only on data from level 14 above. The results are not much better than in \cite{Ising}, although we have more sophisticated extrapolation techniques now. We get the energy with a decent precision $0.5\%$, but $E_\Id$ and $E_\sigma$ show quite big differences between the two independent extrapolations (see figure \ref{fig:Ising Id}) and both results are below the correct values. Surprisingly, the extrapolation of the $E_\eps$ invariant, which has the highest conformal weigh, gives us the best result. The source of these problems does not seem to be just the low amount of data. For example, regular level 12 lump solutions, which offer the same number of data points, give us a significantly better agreement with the expected boundary state. The $\sigma$-brane solution has most likely some unusual level dependence even after becoming real, but we would need access to more levels to understand it.

In addition to the data from \cite{Ising}, we have also computed the first out-of-Siegel equation $\Delta_S$ to check consistency of the solution. It is predictably worse than for the tachyon vacuum or for lump solutions, but better than for the complex solutions in table \ref{tab:Universal solutions}. The infinite level extrapolation do not work very well for this quantity, but we are optimistic that it will eventually disappear.

We have also tried to search for other solutions, for example for the $\eps$-brane or for some multi-brane solutions. For this purpose, we have found all seeds at level 4 and we have improved them to higher levels, but none of the solutions is good enough to be considered physical. Some of these solutions can be found in appendix A of the reference \cite{Ising}, but all of them have high imaginary parts and their invariants cannot be matched with any combination of the Ising model boundary states.

\begin{table}[!]
\centering
\begin{tabular}{|l|lll|}\hline
Level    & Energy             & $\ps \Delta_S$           & $\ps $ Im/Re          \\\hline
2        & $1.59267+0.72688i$ & $   -0.155169+0.118196i$ & $\ps 0.78840        $ \\
4        & $1.41414+0.20152i$ & $   -0.088121+0.051262i$ & $\ps 0.43838        $ \\
6        & $1.28579+0.07668i$ & $   -0.065986+0.029532i$ & $\ps 0.30746        $ \\
8        & $1.21160+0.03054i$ & $   -0.054455+0.018796i$ & $\ps 0.22100        $ \\
10       & $1.16345+0.01007i$ & $   -0.047282+0.011702i$ & $\ps 0.15222        $ \\
12       & $1.12943+0.00123i$ & $   -0.042322+0.005348i$ & $\ps 0.07487        $ \\
14       & $1.10568         $ & $   -0.033002          $ & $\ps 0              $ \\
16       & $1.09045         $ & $   -0.027543          $ & $\ps 0              $ \\
18       & $1.07936         $ & $   -0.023972          $ & $\ps 0              $ \\
20       & $1.07084         $ & $   -0.021338          $ & $\ps 0              $ \\
22       & $1.06405         $ & $   -0.019285          $ & $\ps 0              $ \\
24       & $1.05850         $ & $   -0.017627          $ & $\ps 0              $ \\\hline
$\inf  $ & $0.995           $ & $\ps 0.017             $ & $                   $ \\\hline
Exp.     & $1               $ & $\ps 0                 $ & $                   $ \\\hline
\multicolumn{4}{l}{}\\[-5pt]\hline
Level    & $E_\Id$            & $\ps E_\sigma$         & $\ps E_\eps$            \\\hline
2        & $1.06048-0.18455i$ & $   -0.34358-0.97082i$ & $   -9.73471-5.23904i $ \\
4        & $0.96290-0.14267i$ & $   -0.36976-0.56423i$ & $   -0.66854+1.99191i $ \\
6        & $0.92262-0.11378i$ & $   -0.38933-0.39436i$ & $   -3.86207-0.37376i $ \\
8        & $0.90480-0.08685i$ & $   -0.37217-0.28194i$ & $   -0.57514+0.82266i $ \\
10       & $0.89256-0.06174i$ & $   -0.37629-0.19232i$ & $   -2.48552+0.00261i $ \\
12       & $0.88510-0.03109i$ & $   -0.36891-0.09399i$ & $   -0.56951+0.24561i $ \\
14       & $0.91469         $ & $   -0.26607         $ & $   -1.93951          $ \\
16       & $0.93044         $ & $   -0.20633         $ & $   -0.95087          $ \\
18       & $0.93918         $ & $   -0.17497         $ & $   -1.69824          $ \\
20       & $0.94538         $ & $   -0.15003         $ & $   -1.04849          $ \\
22       & $0.94994         $ & $   -0.13398         $ & $   -1.55407          $ \\
24       & $0.95367         $ & $   -0.11918         $ & $   -1.08669          $ \\\hline
$\inf  $ & $0.96            $ & $   -0.08            $ & $   -0.98             $ \\
$\sigma$ & $0.02            $ & $\ps 0.05            $ & $\ps 0.03             $ \\\hline
Exp.     & $1               $ & $\ps 0               $ & $   -1                $ \\\hline
\end{tabular}
\caption{Properties of the $\sigma$-brane solution in the Ising model on the $\Id$-brane background. The extrapolations use data only from levels 14 to 24, where the solution is real. Using our extrapolating techniques, we have not been able to make reliable error estimate for the energy and $\Delta_S$. The last column of the first part of the table shows the ratio between the real and the imaginary part of the solution (see section \ref{sec:FB circle:single:smallR}). It is not an invariant quantity, but it gives us a good idea how quickly the imaginary part decreases.}
\label{tab:Ising Id-sigma}
\end{table}

\FloatBarrier
\subsection{Solutions on the $\sigma$-brane}\label{sec:MM:Ising:sigma}
The theory on the $\sigma$-brane is more complicated than on the $\Id$-brane, but finding solutions in this theory is actually easier because the boundary spectrum contains the $\eps$ operator. Therefore the first two nontrivial solutions appear at level $1/2$. They differ only by the sign of the $E_\sigma$ invariant and they describe the $\Id$-brane and the $\eps$-brane. We have been able to improve these solutions two levels higher than in \cite{Ising}, that is to level 22. The results are presented in table \ref{tab:Ising sigma-Id}. The overall behavior of these solutions is quite similar to behavior of the MSZ lump solution from section \ref{sec:FB circle:MSZ}.

Extrapolations of the energy and Ellwood invariants give us a really good agreement with the expected boundary state (the errors are of order $10^{-6}$ to $10^{-4}$). Their precision is even better than for the MSZ lump. Notice that the precision has been improved quite significantly compared to \cite{Ising} thanks to better extrapolation techniques. We have also computed the first out-of-Siegel equation $\Delta_S$, which is satisfied with precision comparable to the tachyon vacuum solution.

\begin{table}[b!]
\centering
\begin{tabular}{|l|lllll|}\hline
Level    & Energy   & $E_{\Id}$ & $\ps E_\sigma$ & $E_\eps$ & $\ps \Delta_S $ \\\hline
2        & 0.749172 & 0.733703  & $\pm 0.739416$ & 0.893387 & $\ps 0.0210426$ \\
4        & 0.726558 & 0.722133  & $\pm 0.778236$ & 0.487621 & $\ps 0.0040501$ \\
6        & 0.719329 & 0.715848  & $\pm 0.801822$ & 0.721123 & $\ps 0.0019628$ \\
8        & 0.715961 & 0.714011  & $\pm 0.810113$ & 0.629844 & $\ps 0.0011938$ \\
10       & 0.714036 & 0.712159  & $\pm 0.816787$ & 0.704802 & $\ps 0.0008065$ \\
12       & 0.712795 & 0.711535  & $\pm 0.820182$ & 0.664923 & $\ps 0.0005809$ \\
14       & 0.711929 & 0.710651  & $\pm 0.823308$ & 0.701302 & $\ps 0.0004371$ \\
16       & 0.711292 & 0.710350  & $\pm 0.825168$ & 0.679188 & $\ps 0.0003397$ \\
18       & 0.710803 & 0.709832  & $\pm 0.826989$ & 0.700599 & $\ps 0.0002706$ \\
20       & 0.710416 & 0.709658  & $\pm 0.828170$ & 0.686603 & $\ps 0.0002198$ \\
22       & 0.710102 & 0.709318  & $\pm 0.829366$ & 0.700679 & $\ps 0.0001813$ \\\hline
$\inf  $ & 0.707104 & 0.70709   & $\pm 0.84059 $ & 0.7069   & $   -0.0000043$ \\
$\sigma$ & 0.000003 & 0.00001   & $\ps 0.00007 $ & 0.0004   & $\ps 0.0000003$ \\\hline
Exp.     & 0.707107 & 0.707107  & $\pm 0.840896$ & 0.707107 & $\ps 0        $ \\\hline
\end{tabular}
\caption{Energy, Ellwood invariants and the first out-of-Siegel equation for the $\Id$-brane and the $\eps$-brane solutions in the Ising model on the $\sigma$-brane background. The two solutions differ only by the sign of $E_\sigma$, the $\Id$-brane has positive sign and the $\eps$-brane negative sign. For simplicity, we show only data from even levels.}
\label{tab:Ising sigma-Id}
\end{table}

These two solutions seem to be the only new interesting solutions that can be found on the $\sigma$-brane. We have investigated many other seeds up to level 4, but we have found none worth a closer attention. However, the theory on the $\sigma$-brane contains all solutions from the $\Id$-brane. These solutions have exactly the same string field, but there are changes in some observables due to different bulk-boundary correlators. The existence of these solutions can be proved using the argument from section \ref{sec:SFT:basic:twist}. The part of the string field which includes the $\eps$ field enters the action only quadratically and therefore it can be consistently set to zero. The equations for the remaining variables exactly match those on the $\Id$-brane and therefore both theories share the same solutions.

When we take the solution from the previous subsection and evaluate its observables on this background, we find that the energy and $E_{\Id}$ get multiplied by $\sqrt{2}$, $E_\sigma$ becomes exactly 0 and $E_\eps$ gets multiplied by $\sqrt{2}$. Therefore the solution describes the $\ww\Id\rra \oplus \ww\eps\rra$ boundary state. A better understanding of the process of transferring solutions between various D-branes can be made using topological defects, see \cite{SchnablDefects}.

\FloatBarrier
\section{Lee-Yang model}\label{sec:MM:LeeYang}
The Lee-Yang model is the first nonunitary Virasoro minimal model. It has labels $(p,q)=(2,5)$ and its central charge is negative, $c=-\frac{22}{5}$. This model contains only one nontrivial primary called $\Phi$. This primary has negative conformal weight, $h_\Phi=-\frac{1}{5}$, and its fusion rule is
\begin{equation}
\Phi\times\Phi=\Id+\Phi.
\end{equation}

The Lee-Yang model has negative normalization of correlators in the bulk (see the discussion in section \ref{sec:CFT:MM:bulk boundary}), which means that there is a difference between boundary states and bulk-boundary one-point functions. We have decided to normalize Ellwood invariants to match one-point functions, which are
\begin{eqnarray}
\la\Id\ra^{(\Id)} =   -0.922307,& \quad &\la\Phi\ra^{(\Id)}=0.725073, \\
\la\Id\ra^{(\Phi)}=\ps 0.570017,& \quad &\la\Phi\ra^{(\Phi)}=1.17319.
\end{eqnarray}
Notice that $\la\Id\ra^{(\Id)}$ is negative, which means that there are problems with interpretation of the one-point function of the identity operator as boundary entropy, but that does not affect OSFT calculations.

The Ising model theory has a nice solution describing change of the $\sigma$-brane into the $\Id$-brane, but the solution describing the opposite process is complex at low levels and much less precise. In the Lee-Yang model, we surprisingly do not encounter such problems. We have easily found both a solution describing change of the $\Id$-brane into the $\Phi$-brane (see table \ref{tab:Lee-Yang 1}) and a solution describing the opposite process (see table \ref{tab:Lee-Yang 2}). Both solutions are real and their invariants match the predicted values very well. Apparently, solutions that change the sign of the energy behave similarly as solutions which just decrease it, so these solutions have nice properties even if the absolute value of the energy goes up.

\begin{table}[!]
\centering
\begin{tabular}{|l|llll|}\hline
Level    & Energy   & $E_{\Id}$ & $E_\Phi$ & $\ps \Delta_S$ \\\hline
2        & 0.949550 & 0.406086  & 1.34985  & $\ps 0.335531$ \\
4        & 0.673948 & 0.313684  & 1.23071  & $\ps 0.046525$ \\
6        & 0.635840 & 0.448849  & 1.21301  & $\ps 0.017921$ \\
8        & 0.621146 & 0.456154  & 1.19984  & $\ps 0.010891$ \\
10       & 0.612533 & 0.496203  & 1.19619  & $\ps 0.008129$ \\
12       & 0.606563 & 0.499312  & 1.19097  & $\ps 0.006669$ \\
14       & 0.602092 & 0.517315  & 1.18943  & $\ps 0.005739$ \\
16       & 0.598595 & 0.518922  & 1.18658  & $\ps 0.005074$ \\
18       & 0.595779 & 0.528969  & 1.18575  & $\ps 0.004564$ \\
20       & 0.593461 & 0.529963  & 1.18395  & $\ps 0.004155$ \\
22       & 0.591519 & 0.536334  & 1.18343  & $\ps 0.003818$ \\
24       & 0.589870 & 0.537024  & 1.18218  & $\ps 0.003532$ \\\hline
$\inf $  & 0.56996  & 0.5692    & 1.17320  & $   -0.00003 $ \\
$\sigma$ & 0.00005  & 0.0002    & 0.00002  & $\ps 0.00004 $ \\\hline
Exp.     & 0.570017 & 0.570017  & 1.17319  & $\ps 0       $ \\\hline
\end{tabular}
\caption{Properties of a Lee-Yang model solution describing the $\Phi$-brane on the $\Id$-brane background.}
\label{tab:Lee-Yang 1}
\end{table}

\begin{table}[!]
\centering
\begin{tabular}{|l|llll|}\hline
Level    & $\ps$ Energy   & $\ps E_{\Id}$  & $E_\Phi$ & $\Delta_S$ \\\hline
2        & $   -0.704011$ & $   -0.742719$ & 0.660445 & 0.138435   \\
4        & $   -0.793257$ & $   -0.775846$ & 0.693859 & 0.069899   \\
6        & $   -0.829450$ & $   -0.829777$ & 0.701306 & 0.049445   \\
8        & $   -0.849716$ & $   -0.840367$ & 0.708663 & 0.038162   \\
10       & $   -0.862757$ & $   -0.859573$ & 0.710889 & 0.030953   \\
12       & $   -0.871854$ & $   -0.865122$ & 0.713890 & 0.025977   \\
14       & $   -0.878557$ & $   -0.874918$ & 0.714912 & 0.022350   \\
16       & $   -0.883698$ & $   -0.878272$ & 0.716539 & 0.019596   \\
18       & $   -0.887764$ & $   -0.884212$ & 0.717128 & 0.017436   \\
20       & $   -0.891060$ & $   -0.886450$ & 0.718152 & 0.015699   \\
22       & $   -0.893784$ & $   -0.890440$ & 0.718537 & 0.014272   \\\hline
$\inf $  & $   -0.92230 $ & $   -0.9218  $ & 0.72501  & 0.00002    \\
$\sigma$ & $\ps 0.00003 $ & $\ps 0.0001  $ & 0.00006  & 0.00005    \\\hline
Exp.     & $   -0.922307$ & $   -0.922307$ & 0.725073 & 0          \\\hline
\end{tabular}
\caption{Properties of a Lee-Yang model solution describing the $\Id$-brane on the $\Phi$-brane background.}
\label{tab:Lee-Yang 2}
\end{table}

The solution on the $\Id$-brane, where the spectrum includes only descendants of the identity, can be also found on the $\Phi$-brane and there it describes sum of the $\Id$-brane and the $\Phi$-brane. We have not found any other solutions with clear interpretation, but there are several solutions that are most likely nonphysical, even though they are real and stable in the level truncation scheme. We show properties of one such solution on the $\Phi$-brane background in table \ref{tab:Lee-Yang 3}. The individual observables are well convergent, but there is a clear discrepancy between the energy and $E_\Id$ and $\Delta_S$ probably does not go to zero. The other solutions suffer from similar problems. The existence of such solutions shows the importance of out-of-Siegel equations and other consistency checks.

The most likely interpretation of these nonphysical solutions is that they are artifacts of the implementation of the Siegel gauge condition, they satisfy only the projected equations (\ref{equations projected}), but not the full equations of motion. For some reason, the Lee-Yang model contains more real and stable nonphysical solutions than the unitary models or free boson theories.

\begin{table}[!]
\centering
\begin{tabular}{|l|llll|}\hline
Level    & $\ps$Energy    & $\ps E_{\Id} $ & $E_\Phi$ & $\ps \Delta_S$ \\\hline
2        & $   -0.626305$ & $   -0.174675$ & 2.32972  & $\ps 0.027745$ \\
4        & $   -0.448673$ & $   -0.155463$ & 2.08181  & $   -0.188828$ \\
6        & $   -0.391326$ & $   -0.240781$ & 2.01388  & $   -0.153561$ \\
8        & $   -0.352190$ & $   -0.246105$ & 1.97870  & $   -0.124487$ \\
10       & $   -0.324215$ & $   -0.274888$ & 1.96092  & $   -0.106074$ \\
12       & $   -0.303447$ & $   -0.277211$ & 1.94860  & $   -0.094096$ \\
14       & $   -0.287478$ & $   -0.291869$ & 1.94119  & $   -0.085922$ \\
16       & $   -0.274831$ & $   -0.293345$ & 1.93532  & $   -0.080097$ \\
18       & $   -0.264567$ & $   -0.302388$ & 1.93149  & $   -0.075795$ \\
20       & $   -0.256067$ & $   -0.303504$ & 1.92821  & $   -0.072523$ \\\hline
$\inf $  & $   -0.163   $ & $   -0.350   $ & 1.908    & $   -0.055   $ \\
$\sigma$ & $\ps 0.002   $ & $\ps 0.002   $ & 0.001    & $\ps 0.002   $ \\\hline
\end{tabular}
\caption{An example of a nonphysical solution in the Lee-Yang model. Notice a big difference between the energy and $E_\Id$ and nonzero $\Delta_S$.}
\label{tab:Lee-Yang 3}
\end{table}

\section{Double Ising model}\label{sec:MM:Ising2}
In this section, we consider tensor product of two Ising models, which is the simplest example of a product of two minimal models. We follow the reference \cite{Ising}, but we update the analysis of the solution from this reference and we add few more interesting solutions.

We start with a brief description of this model. It has central charge $c=1$, which suggests that there may be some relation to the free boson theory. It was found that the double Ising model with diagonal bulk partition function\footnote{It is also possible to choose a non-diagonal bulk partition function, this double Ising model is dual to the free boson theory on ordinary circle.}, which is the theory we are interested in, is dual to the free boson theory on the orbifold $S^1 /\mathbb{Z}_2 $ with radius $R=\sqrt{2}$ \cite{OshikawaAffleckIsing}. Boundary states in this model can be also interpreted in terms of defects in the simple Ising model through a folding trick\footnote{An analysis of defects in the Ising model and other minimal models using OSFT can be found in \cite{DefectsOSFT}.}.

Special properties of this theory make it fully solvable. There is an infinite number of bulk primary operators with respect to the $c=1$ Virasoro algebra, but they have a simple classification. They belong to four infinite towers found in \cite{YangDoubleIsing}. We show conformal weights of these primaries, their multiplicities and some examples in table \ref{tab:Ising2 primaries}.

\begin{table}[!]
\centering
\begin{tabular}{|l|c|l|l|}\hline
$h=\bar h$ \rowh{14pt}                                                            & Multiplicity
     & (Ising)$^2$ Examples                                  & Orbifold Examples                      \\\hline
$\mr{2}{*}{$n^2=0,1,4,\ldots$} $                                                  & $\mr{2}{*}{1}$
     & $\Id\otimes\Id$                                     & $\Id$                                  \\
   & & $\eps\otimes\eps$                                   & $-2\partial X \bar\partial X$          \\
$\mr{2}{*}{$\frac{(n+1)^2}{2}=\frac{1}{2},2,\frac{9}{2},\ldots$}$                 & $\mr{2}{*}{2}$
     & $\frac{1}{\sqrt{2}}(\Id\otimes\eps+\eps\otimes\Id)$ & $\sqrt{2}\cos(\sqrt{2} X)$             \\
   & & $\frac{1}{\sqrt{2}}(\Id\otimes\eps-\eps\otimes\Id)$ & $\pm\sqrt{2}\cos(\sqrt{2} \tilde X)$   \\
$\frac{(2n+1)^2}{8}=\frac{1}{8},\frac{9}{8},\frac{25}{8},\ldots $                 & 1
     & $\sigma\otimes\sigma$                               & $\pm \sqrt{2}\cos(\frac{X}{\sqrt{2}})$ \\
$\mr{2}{*}{$\frac{(2n+1)^2}{16}=\frac{1}{16},\frac{9}{16},\frac{25}{16},\ldots$}$ & $\mr{2}{*}{2}$
     & $\Id\otimes\sigma , \sigma\otimes\Id$               & twist fields                           \\
   & & $\eps\otimes\sigma, \sigma\otimes\eps $             & excited twist fields                   \\\hline
\end{tabular}
\caption{Properties of spinless bulk primary fields in the double Ising model and some low level examples with their free boson duals.}
\label{tab:Ising2 primaries}
\end{table}

Boundary states in this model were classifies by Affleck and Oshikawa \cite{OshikawaAffleckIsing}. Among them, we find 9 factorized boundary states, which are given simply by products of the ordinary Ising model boundary states, and the remaining boundary states belong to two one-parametric families of boundary states denoted as $D_O(\phi)$ and $N_O(\tilde{\phi})$.

The interpretation of the double Ising model boundary states in terms of the free boson theory is as follows: Eight of the nine factorized boundary states are fractional D0-branes or fractional D1-branes. The last factorized boundary state, $\sigma\otimes\sigma$, is part of the continuous $D_O$ family and it corresponds to a D0-brane at the central position $\phi=\frac{\pi R}{2}$. The $D_O$ boundary states describe D0-branes at arbitrary positions. Finally, the $N_O$ boundary states correspond to D1-branes with a generic value of the Wilson line. A summary of properties of these boundary states can be found in table \ref{tab:Ising2 D-branes}.

\begin{table}[!]
\centering
\begin{tabular}{|l|c|c|c|c|c|}\hline
(Ising)$^2$      \rowh{13pt}      & Free boson       &  Energy              & Position          & T-dual position          & Twist charge \\\hline
$\Id\otimes\eps$ \rowh{13pt}      & fractional D0    & $\frac{1}{2}$        & $\pi R$           & -                        & $+1$         \\
$\eps\otimes\Id$ \rowh{13pt}      & fractional D0    & $\frac{1}{2}$        & $\pi R$           & -                        & $-1$         \\
$\Id\otimes\Id$  \rowh{13pt}      & fractional D0    & $\frac{1}{2}$        & 0                 & -                        & $+1$         \\
$\eps\otimes\eps$ \rowh{13pt}     & fractional D0    & $\frac{1}{2}$        & 0                 & -                        & $-1$         \\[2pt]\hline
$\Id\otimes\sigma$ \rowh{13pt}    & fractional D1    & $\frac{1}{\sqrt{2}}$ & -                 & $\frac{\pi}{R}$          & $+1$         \\
$\sigma\otimes\Id$ \rowh{13pt}    & fractional D1    & $\frac{1}{\sqrt{2}}$ & -                 & 0                        & $+1$         \\
$\eps\otimes\sigma$ \rowh{13pt}   & fractional D1    & $\frac{1}{\sqrt{2}}$ & -                 & $\frac{\pi}{R}$          & $-1$         \\
$\sigma\otimes\eps$ \rowh{13pt}   & fractional D1    & $\frac{1}{\sqrt{2}}$ & -                 & 0                        & $-1$         \\[2pt]\hline
$\sigma\otimes\sigma$ \rowh{13pt} & centered bulk D0 & 1                    & $\frac{\pi R}{2}$ & -                        & 0            \\[2pt]\hline
$D_O(\phi)$          \rowh{13pt}  & generic bulk D0  & 1                    & $\phi R $         & -                        & 0            \\[2pt]\hline
$N_O(\tilde\phi)$    \rowh{13pt}  & generic bulk D1  & $\sqrt{2}$           & -                 & $\frac{\tilde{\phi}}{R}$ & 0            \\[2pt]\hline
\end{tabular}
\caption{Properties of D-branes in the double Ising model and their free boson duals.}
\label{tab:Ising2 D-branes}
\end{table}

Boundary states from the two continuous families are given by
\begin{eqnarray} \label{Ising2 DO}
\ww D_O(\phi)\rra&=&\sum_{k=0}^{\inf}|k^2\ra+\sqrt{2}\sum_{k=1}^{\inf}\cos\left(\frac{k\phi}{R}\right)|\frac{k^2}{8},S\rra,\\
\ww N_O(\tilde{\phi})\rra&=&\sqrt{2}\sum_{k=0}^{\inf}|k^2\rra+2\sum_{k=1}^{\inf}\cos\left(Rk\tilde{\phi}\right)|\frac{k^2}{2},A\rra, \label{Ising2 NO}
\end{eqnarray}
where we defined symmetric and antisymmetric combinations of Ishibashi states
\begin{eqnarray}
|\frac{n^2}{8},S\rra&=&\left\{
  \begin{array}{ll}
    \frac{1}{\sqrt{2}}\left(|\frac{(2k)^2}{8},1\rra+|\frac{(2k)^2}{8},2\rra\right) & \quad \text{if $n=2k$}\\ \nn
    |\frac{(2k+1)^2}{8}\rra & \quad \text{if $n=2k+1$}
  \end{array} \right. ,\\
|\frac{n^2}{2},A\rra&=&\frac{1}{\sqrt{2}}\left(|\frac{n^2}{2},1\rra-|\frac{n^2}{2},2\rra\right).
\end{eqnarray}
The parameters $\phi\in(0,\pi R)$ and $\tilde\phi\in(0,\pi/R)$ can be interpreted as positions of D-branes on the orbifold and its T-dual respectively. In the Ising picture, they can be interpreted as defect strength \cite{OshikawaAffleckIsing}.

We consider only the $\sigma\otimes\sigma$ initial boundary conditions as the initial background. Using the same argument as for the ordinary Ising model, it is easy to show that this background admits all solutions from fractional D0-branes and fractional D1-branes and therefore there is no need to consider these initial configurations. Starting from a generic bulk D-brane could be interesting, but our computer code cannot handle that and we would have to develop a new one.

\FloatBarrier
\subsection{Solution for the $\Id\otimes\Id$-brane}\label{sec:MM:Ising2:IdxId}
The spectrum on the $\sigma\otimes\sigma$-brane includes three relevant operators, $\Id\otimes\Id$, $\Id\otimes\eps$ and $\eps\otimes\Id$, and one marginal operator $\eps\otimes\eps$. Therefore it is reasonable to start by investigating solutions at level 1. There we find two copies of the solutions from section \ref{sec:MM:Ising:sigma}, which describe the four fractional D1-branes, and four new solutions, which describe the four fractional D0-branes. The new solutions differ only by signs of some invariants, so we pick the $\Id\otimes\Id$-brane solution as a representative.

This solution was originally found in \cite{Ising}, where it was computed up to level 16. In this thesis, we have managed to improve it up to level 18\footnote{We have found that level 16 data in \cite{Ising} contain some errors, which manifest as irregularities in the level dependence of the solution. The new data in table \ref{tab:Ising2 1} correct these errors.} and, as in the simple Ising model, we redo all infinite level extrapolations to improve their precision.

Properties of the solution are shown in table \ref{tab:Ising2 1}. Thanks to some additional data and better extrapolation techniques, we find that the observables have a better agreement with the $\ww \Id\otimes\Id\rra$ boundary state than in \cite{Ising}, the precision is improved approximately by one order. However, the precision of this solution is somewhat lower than precision of the $\Id$-brane solution from the ordinary Ising model, probably because we cannot reach as high level in this model. Finally, we check the out-of-Siegel equation $\Delta_S$, which is also satisfied quite well.

\begin{table}[!b]
\centering
\scriptsize
\begin{tabular}{|l|lllllll|l| }\hline
Level & Energy & $E_{\Id\Id}$ & $E_{\Id\sigma}$ & $E_{\sigma\sigma}$ & $E_{\Id\eps}$ & \ps $E_{\eps\sigma}$& \ps $E_{\eps\eps}$
& $\ps \Delta_S$ \\\hline
2        & 0.603170 & 0.580240 & 0.487853 & 0.615929 & 0.288578 & $   -0.487853 $ & $   -1.15740 $ & $   -0.016883$ \\
4        & 0.544466 & 0.533444 & 0.536014 & 0.632425 & 0.262306 & $\ps 0.818510 $ & $\ps 1.54237 $ & $\ps 0.000318$ \\
6        & 0.528701 & 0.518794 & 0.551676 & 0.661580 & 0.417921 & $\ps 0.299819 $ & $   -0.75631 $ & $\ps 0.002027$ \\
8        & 0.521383 & 0.513329 & 0.559485 & 0.662792 & 0.378672 & $\ps 0.627675 $ & $\ps 1.25240 $ & $\ps 0.002505$ \\
10       & 0.517135 & 0.509311 & 0.564906 & 0.672906 & 0.438685 & $\ps 0.442614 $ & $   -0.38662 $ & $\ps 0.002632$ \\
12       & 0.514345 & 0.507405 & 0.568323 & 0.673837 & 0.416738 & $\ps 0.591368 $ & $\ps 1.05267 $ & $\ps 0.002625$ \\
14       & 0.512365 & 0.505590 & 0.571196 & 0.679168 & 0.449660 & $\ps 0.493583 $ & $   -0.16680 $ & $\ps 0.002564$ \\
16       & 0.510883 & 0.504683 & 0.573202 & 0.679905 & 0.435826 & $\ps 0.580394 $ & $\ps 0.92258 $ & $\ps 0.002480$ \\
18       & 0.509729 & 0.503677 & 0.575029 & 0.683272 & 0.456921 & $\ps 0.519003 $ & $   -0.02450 $ & $\ps 0.002389$ \\\hline
$\inf $  & 0.50009  & 0.4979   & 0.5906   & 0.702    & 0.492    & $\ps 0.585    $ & $\ps 0.50    $ & $\ps 0.00021 $ \\
$\sigma$ & 0.00002  & 0.0003   & 0.0007   & 0.001    & 0.002    & $\ps 0.008    $ & $\ps 0.13    $ & $\ps 0.00004 $ \\\hline
Exp.     & 0.5      & 0.5      & 0.59460  & 0.70711  & 0.5      & $\ps 0.59460  $ & $\ps 0.5     $ & $\ps 0       $ \\\hline
\end{tabular}
\caption{Observables of the $\Id\otimes\Id$-brane solution found on the $\sigma\otimes\sigma$-brane background in the double Ising model. Ellwood invariants of this solution are symmetric with respect to exchange of their indices, $E_{ij}=E_{ji}$}
\label{tab:Ising2 1}
\end{table}

\FloatBarrier
\subsection{Marginal solutions}\label{sec:MM:Ising2:marginal}
The double Ising model includes a marginal operator, which corresponds to $\del X$ in the free boson picture, and therefore we can construct the same marginal solutions as in chapter \ref{sec:marginal}. Marginal deformations of the $\sigma\otimes\sigma$-brane correspond to change of position of the D0-brane on the orbifold in the free boson picture and the parameter $\phi$ is proportional to $\lB$.

Marginal solutions in the double Ising model have the same energy as marginal solutions on the self-dual radius, but Ellwood invariants are not identical because of different bulk theory. There are two convergent invariants that tell us position of the D0-brane on the orbifold, $E_{\sigma\sigma}$ and $E_{\Id\eps}=E_{\eps\Id}$. From (\ref{Ising2 DO}), we find
\begin{eqnarray}\label{Ising2 phi}
\phi_{(\sigma\sigma)} &=& \sqrt{2}\arccos \frac{E_{\sigma\sigma}}{\sqrt{2}}, \\
\phi_{(\Id\eps)} &=& \frac{1}{\sqrt{2}}\arccos E_{\Id\eps}.\nn
\end{eqnarray}
We have checked on several examples that the relation between $\lS$ and $\lB$ is almost the same is in chapter \ref{sec:marginal} (with proper normalization), so there is no need to repeat the whole analysis here. There is however one curious solution which does not require leaving the equation for the marginal field unsolved, so we are going to investigate it next.

The seed for this solution appears at level 2 and its properties are shown in table \ref{tab:Ising2 mar}. Energy close to 1 and unusual values of $E_{\sigma\sigma}$ and $E_{\Id\eps}$ clearly identify this solution as a member of the $D_O$ family of boundary states. We notice that the solution is pseudo-real because its energy is real, but some of its invariants are not. A closer inspection of the string field reveals that coefficients of descendants of $\Id\otimes\Id$ and $\eps\otimes\eps$ are real, while coefficients of descendants of $\Id\otimes\eps$ and $\eps\otimes\Id$ are purely imaginary. The imaginary coefficients allow this solution to satisfy the equation for the marginal field, which is violated by other marginal solutions. However, the solution can be considered physical only if its imaginary part disappears in the $L\rar \inf$ limit. Unfortunately, we cannot say with certainty whether it happens or not with the available data. Imaginary parts of Ellwood invariants and of the string field decrease with the level, but the extrapolated values are not zero within the estimated errors. The out-of-Siegel equation $\Delta_S$ is satisfied quite well, so we are more inclined to believe that the solution is physical.

\begin{table}
\centering
\scriptsize
\begin{tabular}{|l|lllllll|l|}\hline
Level & Energy & $E_{\Id\Id}$ & $E_{\Id\sigma}$ & $E_{\sigma\sigma}$ & \ps $E_{\Id\eps}$ & $E_{\eps\sigma}$ & $ \ps E_{\eps\eps}$ & $\ps \Delta_S$ \\\hline
2        & 1.00386      & 0.984280 & $0.102573 i$ & 0.745762 & $   -0.486664$ & $0.102573 i$ & $   -0.01095$ & $   -0.0022373$ \\
4        & 1.00311      & 0.989546 & $0.087641 i$ & 0.815353 & $   -0.397745$ & $0.317584 i$ & $\ps 1.97254$ & $   -0.0012044$ \\
6        & 1.00223      & 0.992165 & $0.074145 i$ & 0.836305 & $   -0.322005$ & $0.210310 i$ & $\ps 0.41588$ & $   -0.0008805$ \\
8        & 1.00171      & 0.993927 & $0.065231 i$ & 0.839813 & $   -0.324584$ & $0.204409 i$ & $\ps 1.38328$ & $   -0.0006851$ \\
10       & 1.00139      & 0.994903 & $0.058819 i$ & 0.844855 & $   -0.300697$ & $0.175139 i$ & $\ps 0.75156$ & $   -0.0005598$ \\
12       & 1.00117      & 0.995709 & $0.054028 i$ & 0.845967 & $   -0.303078$ & $0.165485 i$ & $\ps 1.18225$ & $   -0.0004733$ \\
14       & 1.00100      & 0.996209 & $0.050219 i$ & 0.848216 & $   -0.291521$ & $0.150744 i$ & $\ps 0.86989$ & $   -0.0004103$ \\
16       & 1.00088      & 0.996671 & $0.047143 i$ & 0.848774 & $   -0.293053$ & $0.143335 i$ & $\ps 1.10418$ & $   -0.0003622$ \\
18       & 1.00079      & 0.996975 & $0.044558 i$ & 0.850048 & $   -0.286272$ & $0.134027 i$ & $\ps 0.92113$ & $   -0.0003244$ \\\hline
$\inf$   & 1.0000009    & 0.99987  & $0.019    i$ & 0.85643  & $   -0.2663  $ & $0.058    i$ & $\ps 1.00   $ & $   -0.0000015$ \\
$\sigma$ & $1\dexp{-7}$ & 0.00003  & $0.001    i$ & 0.00003  & $\ps 0.0003  $ & $0.005    i$ & $\ps 0.05   $ & $\ps 0.0000002$ \\\hline
Exp. \rowh{10pt} & 1 & 1 & 0 & $\sqrt{2}\cos\frac{\phi}{\sqrt{2}}$ & $\ps \cos \sqrt{2}\phi$ & 0 & $\ps 1  $ & $\ps 0        $ \\\hline
\end{tabular}
\caption{Nontrivial observables of an exceptional $D_O(\phi)$ solution in the double Ising model.}
\label{tab:Ising2 mar}
\end{table}

The parameter $\phi$ can be determined from the equations (\ref{Ising2 phi}). From the extrapolated values of $E_{\sigma\sigma}$ and $E_{\Id\eps}$, we get
\begin{eqnarray}
\phi_{(\sigma\sigma)} &=& 1.30149\pm 0.00005, \\
\phi_{(\Id\eps)} &=& 1.3013 \pm 0.0003.
\end{eqnarray}
Both values nicely agree within the estimated errors. In terms of the orbifold radius, we find $\phi\doteq 0.2929\pi R$. This value seems to be quite generic and not a nice multiple of $\pi R$, so the solution does not seem to describe any special boundary state.

Properties of this solution are quite similar to properties of the pseudo-real double lump solution from section \ref{sec:FB circle:double:R=2}. This suggests that this type of solution may appear regularly in theories with a marginal field and at least one nontrivial relevant field.

\subsection{Other solutions}\label{sec:MM:Ising2:other}
Finally, we investigate other solutions that can be found from seeds at levels 2 and $2.5$. First, we find two copies of the positive energy solution from section \ref{sec:MM:Ising:Id}, these solutions change one of the $\sigma$ boundary conditions to $\Id\oplus\eps$.

Next, there are four solutions that describe the $\Id\otimes\Id$-brane plus the $\Id\otimes\eps$-brane and three other combinations of fractional D0-branes which share one of the two boundary conditions. Properties of the first solution are shown in table \ref{tab:Ising2 3}. We observe that the solution is complex even at level 18, but its imaginary part quickly decreases and we expect that it will disappear at a slightly higher level. Therefore we cannot very well extrapolate the solution, but, given the available boundary states, its identification is essentially unambiguous.

These solutions are unusual because they have the same energy as the initial D-brane, but they cannot be reached by marginal deformations. They actually resemble fusion of the two fundamental solutions from the ordinary Ising model: one of the $\sigma$ boundary conditions changes to $\Id$ or $\eps$ boundary condition and the other to $\Id\oplus\eps$ boundary conditions. Two other combinations of fractional D0-branes, $\Id\otimes\Id$-brane plus $\eps\otimes\eps$-brane and $\Id\otimes\eps$-brane plus $\eps\otimes\Id$-brane, are given by $D_O(\phi)$ for $\phi=0,\pi R$, so they can be reached using marginal deformations. We have not found any solutions describing two copies of the same fractional D0-brane.

\begin{table}[!]
\centering
\scriptsize
\begin{tabular}{|l|llll|}\hline
Level \rowh{8pt} & $\ps $Energy             & $\ps E_{\Id\Id}       $ & $\ps E_{\sigma\Id}$       & $\ps E_{\Id\eps}$  \\\hline
4                & $\ps 1.87748-0.504781 i$ & $\ps 1.13136+0.099003i$ & $\ps 0.975334+0.461219i$  & $\ps 2.48481+2.87412 i$  \\
6                & $\ps 1.54264-0.197854 i$ & $\ps 1.01525+0.124044i$ & $\ps 0.986160+0.235880i$  & $\ps 4.13545-0.398509i$  \\
8                & $\ps 1.39417-0.095249 i$ & $\ps 0.96636+0.110729i$ & $\ps 0.994346+0.176603i$  & $\ps 1.19361+1.08909 i$  \\
10               & $\ps 1.30747-0.048978 i$ & $\ps 0.93781+0.094606i$ & $\ps 0.995807+0.138475i$  & $\ps 2.46987+0.12748 i$  \\
12               & $\ps 1.24985-0.024704 i$ & $\ps 0.92090+0.078164i$ & $\ps 0.996415+0.110137i$  & $\ps 0.95986+0.60817 i$  \\
14               & $\ps 1.20835-0.011083 i$ & $\ps 0.90845+0.061419i$ & $\ps 0.996103+0.084122i$  & $\ps 1.83447+0.15992 i$  \\
16               & $\ps 1.17681-0.003504 i$ & $\ps 0.89990+0.042622i$ & $\ps 0.995992+0.057438i$  & $\ps 0.84641+0.29448 i$  \\
18               & $\ps 1.15187-0.000090 i$ & $\ps 0.89291+0.012763i$ & $\ps 0.995623+0.016960i$  & $\ps 1.49710+0.04071 i$  \\\hline
Exp.             & $\ps 1.                $ & $\ps 1.               $ & $\ps 1.18921           $  & $\ps 1.               $  \\\hline
\multicolumn{5}{l}{}\\[-5pt]\hline
Level \rowh{8pt} & $\ps E_{\eps\Id} $       & $\ps E_{\sigma\eps}$    & $\ps E_{\eps\eps}$        & $\ps \Delta_S$           \\\hline
4                & $   -0.094956+0.940152i$ & $\ps 11.8564-1.81555 i$ & $\ps 16.6023+0.64451 i$   & $   -0.108626+0.045886i$ \\
6                & $\ps 0.225972+0.401185i$ & $   -0.74198+2.71204 i$ & $   -14.8666+6.26311 i$   & $   -0.070187+0.026544i$ \\
8                & $\ps 0.408722+0.252949i$ & $\ps 5.24390-0.59979 i$ & $\ps 17.8804-5.03978 i$   & $   -0.055272+0.017220i$ \\
10               & $\ps 0.487133+0.193678i$ & $   -0.16765+1.44484 i$ & $   -16.1252+6.13839 i$   & $   -0.046814+0.012128i$ \\
12               & $\ps 0.539822+0.141016i$ & $\ps 3.55884-0.13435 i$ & $\ps 18.4175-4.60803 i$   & $   -0.041242+0.008783i$ \\
14               & $\ps 0.577196+0.107553i$ & $\ps 0.05565+0.78273 i$ & $   -16.7324+4.61361 i$   & $   -0.037234+0.006246i$ \\
16               & $\ps 0.601692+0.069458i$ & $\ps 2.76066+0.00123 i$ & $\ps 18.6088-2.90147 i$   & $   -0.034180+0.004007i$ \\
18               & $\ps 0.624495+0.020569i$ & $\ps 0.16691+0.14470 i$ & $   -17.0586+1.05563 i$   & $   -0.031754+0.001125i$ \\\hline
Exp.             & $\ps 1.                $ & $\ps 1.18921          $ & $\ps 1.               $   & $\ps 0                 $ \\\hline
\end{tabular}
\caption{Observables of a solution describing the $\Id\otimes\Id$-brane plus the $\Id\otimes\eps$-brane in the double Ising model. The invariants which are not shown ($E_{\Id\sigma}$, $E_{\sigma\sigma}$ and $E_{\eps\sigma}$) are identically zero. We omit level 2 because the seed at this level has different symmetries than the solution. We do not show any extrapolations because the imaginary part quickly decreases and we expect that the solution will go through a similar change like the $\sigma$-brane solution from section \ref{sec:MM:Ising:Id} once it becomes real.}
\label{tab:Ising2 3}
\end{table}

Next, there is a group of 8 solutions that come from seeds at level $2.5$. Properties of one of them are shown in table \ref{tab:Ising2 4}. Their energy probably converges to $\frac{1+\sqrt{2}}{2}$, which means that they describe various combinations of one fractional D0-brane and one fractional D1-brane. Similarly to the previous group, these solutions are complex with quickly decreasing imaginary parts, so there in no point in trying to do any extrapolations.

Finally, we have found one more complex solution, which probably describes the combination of all four fractional D0-branes. We do not show its invariants because it has large imaginary part and its agreement with the expected boundary state is not very good. We can identify this solution essentially only because of its symmetries, many of its invariants are exactly equal to zero and the sum of all four fractional D0-branes is the only boundary state with the same properties and somewhat similar energy.

\begin{table}[!]
\centering
\scriptsize \begin{tabular}{|l|llll|}\hline
Level & Energy               & $\ps E_{\Id\Id}$         & $\ps E_{\Id\sigma}$      & $\ps E_{\sigma\Id}$      \\\hline
4     & $1.88451-0.437943i$  & $\ps 1.22978+0.163223i$  & $   -0.700425-0.255489i$ & $\ps 0.184130+0.089962i$ \\
6     & $1.63788-0.174782i$  & $\ps 1.16719+0.132841i$  & $   -0.601278-0.187997i$ & $\ps 0.244663+0.017782i$ \\
8     & $1.52349-0.087907i$  & $\ps 1.14372+0.109771i$  & $   -0.556596-0.157770i$ & $\ps 0.260859+0.003974i$ \\
10    & $1.45607-0.047858i$  & $\ps 1.12892+0.092239i$  & $   -0.530230-0.131967i$ & $\ps 0.275136-0.002140i$ \\
12    & $1.41100-0.026247i$  & $\ps 1.12070+0.076684i$  & $   -0.511852-0.110828i$ & $\ps 0.280143-0.003754i$ \\
14    & $1.37842-0.013627i$  & $\ps 1.11419+0.062470i$  & $   -0.498444-0.090534i$ & $\ps 0.286102-0.004834i$ \\
16    & $1.35357-0.006065i$  & $\ps 1.11000+0.048108i$  & $   -0.487887-0.070242i$ & $\ps 0.288479-0.004390i$ \\
18    & $1.33387-0.001728i$  & $\ps 1.10630+0.031917i$  & $   -0.479480-0.046758i$ & $\ps 0.291736-0.003497i$ \\\hline
Exp.  & $1.20711          $  & $\ps 1.20711          $  & $   -0.594604          $ & $\ps 0.246293          $ \\\hline
\multicolumn{5}{l}{}\\[-5pt]\hline
Level & $E_{\sigma\sigma}$   & $\ps E_{\Id\eps}$        & $\ps E_{\eps\Id}$        & $\ps \Delta_S$           \\\hline
4     & $0.439257+0.198605i$ & $   -0.538939+0.348600i$ & $\ps 2.06158+2.89148i$   & $   -0.097059+0.056257i$ \\
6     & $0.543503+0.048769i$ & $   -0.380064+0.108448i$ & $\ps 4.48623-0.56085i$   & $   -0.060653+0.029968i$ \\
8     & $0.566331+0.024062i$ & $   -0.439752+0.095523i$ & $\ps 1.29902+1.14156i$   & $   -0.046574+0.019318i$ \\
10    & $0.593035+0.016159i$ & $   -0.314548+0.026343i$ & $\ps 2.81778+0.02888i$   & $   -0.038685+0.013497i$ \\
12    & $0.596983+0.012999i$ & $   -0.39584 +0.054834i$ & $\ps 1.15126+0.65277i$   & $   -0.033513+0.009746i$ \\
14    & $0.608076+0.010608i$ & $   -0.300341+0.013750i$ & $\ps 2.16614+0.12127i$   & $   -0.029802+0.007046i$ \\
16    & $0.608943+0.008589i$ & $   -0.374984+0.032383i$ & $\ps 1.06877+0.36820i$   & $   -0.026979+0.004907i$ \\
18    & $0.614963+0.005953i$ & $   -0.296819+0.007352i$ & $\ps 1.81532+0.08914i$   & $   -0.024745+0.002976i$ \\\hline
Exp.  & $0.707107          $ & $   -0.207107          $ & $\ps 1.20711         $   & $\ps 0                 $ \\\hline
\end{tabular}
\caption{Selected observables of a solution describing the $\Id\otimes\sigma$-brane plus the $\eps\otimes\eps$-brane in the double Ising model. We show no extrapolations because the solution is expected to become real in few more levels.}
\label{tab:Ising2 4}
\end{table}

\section{Ising$\otimes$tricritical Ising model}\label{sec:MM:Ising3}
At the end of this chapter, we will explore a more complicated theory, the product of the Ising model and the tricritical Ising model. This theory has more relevant operators and there is no symmetry between the two minimal models like in the double Ising model, so we will be able to investigate different phenomena.

We have already made a brief review of the Ising model in section \ref{sec:MM:Ising}, so now we will provide similar information about the tricritical Ising model. It is labeled by integers $(p,q)=(4,5)$ and the central charge is $c=\frac{7}{10}$. The model has 6 primary fields, which are listed in table \ref{tab:TriIsing operators}. Their fusion rules can be derived using (\ref{MM fusion rules}) and they can be found for example in \cite{DiFrancesco}. This model has 6 boundary states, which are associated with the six primaries, their coefficients are in table \ref{tab:TriIsing BS} in numerical form.

\begin{table}[h]
\centering
\begin{tabular}{|c|c|c|}\hline
operator  & Kac label & $h$            \\\hline
$\Id$    \rowh{13pt} & (1,1)     & $0$            \\
$\sigma$ \rowh{13pt} & (2,3)     & $\frac{3}{80}$ \\
$\eps$   \rowh{13pt} & (3,3)     & $\frac{1}{10}$ \\
$\sigma'$\rowh{13pt} & (2,1)     & $\frac{7}{16}$ \\
$\eps'$  \rowh{13pt} & (1,3)     & $\frac{3}{5}$  \\
$\eps''$ \rowh{13pt} & (3,1)     & $\frac{3}{2}$  \\[2pt]\hline
\end{tabular}
\caption{Primary operators in the tricritical Ising model.}
\label{tab:TriIsing operators}
\end{table}

The product of the two minimal models is an irrational CFT with respect to the full energy-momentum tensor and therefore the classification of all boundary states is unknown. We understand only the 18 factorized boundary states, which can be used as open string backgrounds. Coefficients of the factorized boundary states can be obtained by multiplying the coefficients from tables \ref{tab:Ising boudary states} and \ref{tab:TriIsing BS}. The whole table is too big to be shown here, so we just list the energies of these boundary states: 0.362537, 0.512704, 0.586597, 0.725073, 0.829573 and 1.17319. Most of the energies have multiplicity 4 and the boundary states in the multiplets differ only by signs of some components. Boundary spectra of these boundary states are given by products of usual boundary spectra of the two  models.

\begin{table}[!b]
\centering
\begin{tabular}{|c|llllll|}\hline
                 & $|\Id\rra$ & $\ps |\sigma\rra$ & $\ps |\eps\rra$ & $\ps |\sigma'\rra$ & $\ps |\eps'\rra$ & $\ps |\eps''\rra$\\\hline
$\ww\Id\rra    $ & 0.512704   & $\ps 0.775565   $ & $\ps 0.652170 $ & $\ps 0.609711    $ & $\ps 0.652170  $ & $\ps 0.512704   $\\
$\ww\sigma\rra $ & 1.173193   & $\ps 0.         $ & $\ps 0.570017 $ & $\ps 0.          $ & $   -0.570017  $ & $   -1.173193   $\\
$\ww\eps\rra   $ & 0.829573   & $\ps 0.479325   $ & $   -0.403063 $ & $   -0.986534    $ & $   -0.403063  $ & $\ps 0.829573   $\\
$\ww\sigma'\rra$ & 0.725073   & $\ps 0.         $ & $   -0.922307 $ & $\ps 0.          $ & $\ps 0.922307  $ & $   -0.725073   $\\
$\ww\eps'\rra  $ & 0.829573   & $   -0.479325   $ & $   -0.403063 $ & $\ps 0.986534    $ & $   -0.403063  $ & $\ps 0.829573   $\\
$\ww\eps''\rra $ & 0.512704   & $   -0.775565   $ & $\ps 0.652170 $ & $   -0.609711    $ & $\ps 0.652170  $ & $\ps 0.512704   $\\\hline
\end{tabular}
\caption{List of boundary state coefficients in the tricritical Ising model.}
\label{tab:TriIsing BS}
\end{table}

We choose the boundary state with the highest energy as the OSFT background, which gives us the highest chance to see interesting results. It is the $\ww \sigma\otimes\sigma \rra$ boundary state, which has energy $1.17319$ and its boundary spectrum includes operators $\Id$ and $\eps$ in the Ising model part and operators $\Id$, $\eps$, $\eps'$ and $\eps''$ in the tricritical Ising model part.

\subsection{Regular solutions}\label{sec:MM:Ising3:regular}
The number of solutions in this model is too high to show all of them in detail and they mostly do not have any new interesting properties anyway, so we will explore only some selected aspects of the solutions. We have used the homotopy continuation method at level 2 to find seeds for Newton's method and we have improved some interesting solutions up to level 14. We will focus mainly on real solutions. There are 17 solutions that can be uniquely identified as single factorized D-branes. This is what we hoped for because they describe exactly the 17 fundamental boundary states with energy lower that the initial one. Then, there are 13 solutions describing two factorized D-branes and 6 exotic solutions. We will postpone analysis of the exotic solutions to the next subsection and we will start with discussion of solutions corresponding to two D-branes.

\begin{table}[!]
\centering
\begin{tabular}{|c|ccc|}\hline
Boundary state                                   & Energy   & $h_{min}$                  & real seed \\\hline
$(\Id\otimes\Id)    \oplus(\Id\otimes\eps'')   $ & 0.725073 & $\frac{3}{2}  \rowh{14pt}$ & yes       \\
$(\eps\otimes\Id)   \oplus(\eps\otimes\eps'')  $ & 0.725073 & $\frac{3}{2}  \rowh{14pt}$ & yes       \\
$(\Id\otimes\eps'') \oplus(\eps\otimes\Id)     $ & 0.725073 & $2            \rowh{14pt}$ & yes       \\
$(\Id\otimes\Id)    \oplus(\eps\otimes\eps'')  $ & 0.725073 & $2            \rowh{14pt}$ & yes       \\
$(\Id\otimes\eps'') \oplus(\sigma\otimes\Id)   $ & 0.875241 & $\frac{25}{16}\rowh{14pt}$ & yes       \\
$(\eps\otimes\eps'')\oplus(\sigma\otimes\Id)   $ & 0.875241 & $\frac{25}{16}\rowh{14pt}$ & yes       \\
$(\Id\otimes\Id)    \oplus(\sigma\otimes\eps'')$ & 0.875241 & $\frac{25}{16}\rowh{14pt}$ & yes       \\
$(\eps\otimes\Id)   \oplus(\sigma\otimes\eps'')$ & 0.875241 & $\frac{25}{16}\rowh{14pt}$ & yes       \\
$(\Id\otimes\eps')  \oplus(\eps\otimes\Id)     $ & 0.949133 & $\frac{11}{10}\rowh{14pt}$ & yes       \\
$(\Id\otimes\eps'') \oplus(\eps\otimes\eps)    $ & 0.949133 & $\frac{11}{10}\rowh{14pt}$ & yes       \\
$(\Id\otimes\Id)    \oplus(\eps\otimes\eps')   $ & 0.949133 & $\frac{11}{10}\rowh{14pt}$ & yes       \\
$(\Id\otimes\eps)   \oplus(\eps\otimes\eps'')  $ & 0.949133 & $\frac{11}{10}\rowh{14pt}$ & yes       \\
$(\sigma\otimes\Id) \oplus(\sigma\otimes\eps'')$ & 1.025408 & $\frac{3}{2}  \rowh{14pt}$ & yes       \\[2pt]\hline
\multicolumn{4}{l}{}\\[-5pt]\hline
$(\Id\otimes\Id)    \oplus(\Id\otimes\eps')    $ & 0.949133 & $\frac{3}{5}  \rowh{14pt}$ & no        \\
$(\Id\otimes\eps)   \oplus(\Id\otimes\eps'')   $ & 0.949133 & $\frac{3}{5}  \rowh{14pt}$ & no        \\
$(\eps\otimes\Id)   \oplus(\eps\otimes\eps')   $ & 0.949133 & $\frac{3}{5}  \rowh{14pt}$ & no        \\
$(\eps\otimes\eps)  \oplus(\eps\otimes\eps'')  $ & 0.949133 & $\frac{3}{5}  \rowh{14pt}$ & no        \\
$(\Id\otimes\eps)   \oplus(\sigma\otimes\eps'')$ & 1.099300 & $\frac{53}{80}\rowh{14pt}$ & no        \\
$(\Id\otimes\eps')  \oplus(\sigma\otimes\Id)   $ & 1.099300 & $\frac{53}{80}\rowh{14pt}$ & no        \\
$(\eps\otimes\eps'')\oplus(\sigma\otimes\Id)   $ & 1.099300 & $\frac{53}{80}\rowh{14pt}$ & no        \\
$(\eps\otimes\eps'')\oplus(\sigma\otimes\eps'')$ & 1.099300 & $\frac{53}{80}\rowh{14pt}$ & no        \\[2pt]\hline
\end{tabular}
\caption{List of solutions in the Ising$\otimes$tricritical Ising model which describe two D-branes with energy lower than the initial D-brane energy $E_{init}=1.17319$. The first column shows identifications of solutions, the second column exact energies of these D-brane configurations, the third column minimal conformal weights of operators with the two mixed boundary conditions and the last column tell us whether the solutions have real seeds at level 2.}
\label{tab:TriIsing solutions}
\end{table}

The first part of table \ref{tab:TriIsing solutions} summarizes some basis properties of the solutions that we found. There are 72 configurations of two D-branes with energy lower than $1.17319$, while we have found only 13, which means that we have found only a fraction of the expected number of solutions. Why do we observe these solutions and not others? The results from section \ref{sec:FB circle:double:superposition} suggest that we should look at the spectrum of string stretched between the D-branes. The analysis of all possible D-brane configurations shows that there are 13 configurations with no tachyonic modes among the stretched strings and these are exactly the 13 configurations we have found. This result is in agreement with what we found for the double lumps in section \ref{sec:FB circle:double:superposition}, there are real solutions only for boundary states which do not support such tachyonic excitations.

Is it possible to find any other two D-brane solutions? Yes, but in order to do so, we have to start from complex seeds. There are hundreds of potentially interesting complex seeds at level 2 and there will be undoubtedly even more at higher initial levels, so we have not been able to analyze all possible solutions, but we have found several relatively good ones. There are two groups of solutions that behave similarly to the $\sigma$-brane solution in the Ising model, they have complex seeds and they become real at available levels (concretely at levels 8 and 9). These solutions are listed in the second part of table \ref{tab:TriIsing solutions}. Interestingly, energies of these solutions at low levels are higher than the energy of the initial D-brane, but they quickly drop down as we increase the level. We are not able to extrapolate observables of these solutions with a very good precision, but we are essentially sure about their identification because we have access to large number of Ellwood invariants. There are more complex solutions which are similar to other configurations of two D-branes, but we are not sure about their status because they have large imaginary parts.

\FloatBarrier
\subsection{Exotic solutions}\label{sec:MM:Ising3:exotic}
Apart from regular solutions describing factorized boundary states, this model also admits some exotic solutions, similarly to the free boson theory on a torus. We have not discussed such solutions in the double Ising model because non-conventional boundary states in that model (the two continuous families) can be understood due to the duality to the free boson theory. Such description is not available in this model and therefore we have been able to discover yet unknown boundary states.

There are six real exotic solutions, which belong to two groups. The solutions in each group have the same energy and they differ only by signs of some invariants. Properties of a representative of the first group are shown in table \ref{tab:TriIsing 1}. There are three solutions which share exactly the same energy. All of them have exactly $E_{\Id\sigma}=E_{\sigma\Id}=0$, but we can distinguish them by the invariants $E_{\Id\eps}$ and $E_{\sigma\sigma}$. However, only two of them are exotic, the third one describes two factorized D-branes. The extrapolated energy of the regular solution is 1.025403, which is very close to $2\times 0.512704=1.02541$, and, using more invariants, we can identify this solution as the $\sigma\otimes\Id$-brane plus the $\sigma\otimes\eps''$-brane, therefore it is one of the solutions that is mentioned in the previous subsection. On the other hand, there is no combination of factorized boundary states that can describe the other two solutions, which means that we have found a yet unknown type of boundary state. The solution in table \ref{tab:TriIsing 1} satisfies the out-of-Siegel equation $\Delta_S$ quite well, so it does not seem to be an artifact of the level truncation approach.

Because this solution has the same energy as one of the regular solutions, we are almost sure that its energy is exactly $1.02541=(2(1-5^{-1/2}))^{1/4}$. We can also make a guess about some other components of the corresponding boundary state by comparing the Ellwood invariants with numbers that appear in the factorized boundary states. The $E_{\Id\eps}$ invariant most likely converges to zero and some other invariants are zero trivially. The extrapolated value of $E_{\sigma\sigma}$ is very close to $-0.922307=-((1+5^{-1/2})/2)^{1/4}$ and $E_{\sigma\sigma'}$ is close to $0.725073=((1-5^{-1/2})/2)^{1/4}$. Therefore there is a good chance that one may be able find an analytic formula for this boundary state by some manipulations with the known boundary states, although we have not managed to do so.

Properties of a representative of the second group of exotic solutions, which includes 4 members, are shown in table \ref{tab:TriIsing 2}. It has energy approximately $1.015023$, this number is not very far from the energy of the previous solution, but it is clearly different considering high precision of the extrapolation. Nonzero Ellwood invariants of the solution seem to have quite generic values, only $E_{\sigma\Id}$ is close to $-0.5$. Finding an analytic formula for the corresponding boundary state is therefore going to be more difficult than for the previous solution.

\begin{table}
\centering
\begin{tabular}{|l|llllll|}\hline
Level    & Energy    & $\ps \Delta_S$   & $E_{\Id\Id}$ & $E_{\Id\eps}$ & $\ps E_{\sigma\sigma}$ & $E_{\sigma\sigma'}$ \\\hline
2        & 1.05144   & $\ps 0.01362309$ & 1.03650      & 0.0507342     & $   -0.813591$         & 0.756836            \\
4        & 1.03701   & $\ps 0.00247042$ & 1.02949      & 0.0339477     & $   -0.867658$         & 0.630909            \\
6        & 1.03271   & $\ps 0.00128945$ & 1.02694      & 0.0224416     & $   -0.886902$         & 0.704793            \\
8        & 1.03072   & $\ps 0.00085443$ & 1.02679      & 0.0175740     & $   -0.896038$         & 0.687469            \\
10       & 1.02958   & $\ps 0.00062781$ & 1.02616      & 0.0138610     & $   -0.901242$         & 0.709211            \\
12       & 1.02884   & $\ps 0.00049011$ & 1.02621      & 0.0118458     & $   -0.904841$         & 0.702521            \\
14       & 1.02832   & $\ps 0.00039848$ & 1.02591      & 0.0100506     & $   -0.907224$         & 0.712589            \\
16       & 1.02794   & $\ps 0.00033361$ & 1.02596      & 0.0089576     & $   -0.909153$         & 0.709097            \\\hline
$\inf$   & 1.0254052 & $   -0.0000052 $ & 1.02537      & 0.00028       & $   -0.9218  $         & 0.7249              \\
$\sigma$ & 0.0000005 & $\ps 0.0000001 $ & 0.00008      & 0.00005       & $\ps 0.0001  $         & 0.0014              \\\hline
Exp.     & 1.0254083 & $\ps 0         $ & 1.02541      & 0 ?           & $   -0.922307$ ?       & 0.725073 ?          \\\hline
\end{tabular}
\caption{Properties of a representative of the first group of exotic solutions in the Ising$\otimes$tricritical Ising model. We show only the first few nontrivial Ellwood invariants. Missing invariants with low conformal weights are identically zero, $E_{\Id\sigma}=E_{\sigma\Id}=E_{\sigma\eps}=E_{\Id\sigma'}=0$. The other solution differs by signs of $E_{\sigma\sigma}$ and $E_{\sigma\sigma'}$.}
\label{tab:TriIsing 1}
\end{table}

\begin{table}
\centering
\begin{tabular}{|l|llllll|}\hline
Level    & Energy       & $\ps \Delta_S$  & $E_{\Id\Id}$ & $E_{\Id\sigma}$ & $\ps E_{\sigma\Id}$ & $E_{\Id\eps}$ \\\hline
2        & 1.05350      & $\ps 0.0174586$ & 1.04422      & 0.230456        & $   -0.431675$      & 0.516233      \\
4        & 1.02882      & $\ps 0.0023479$ & 1.02259      & 0.240321        & $   -0.457615$      & 0.490381      \\
6        & 1.02324      & $\ps 0.0010973$ & 1.01827      & 0.244379        & $   -0.472315$      & 0.487585      \\
8        & 1.02086      & $\ps 0.0006633$ & 1.01742      & 0.248063        & $   -0.478752$      & 0.487383      \\
10       & 1.01954      & $\ps 0.0004513$ & 1.01651      & 0.249565        & $   -0.483137$      & 0.486865      \\
12       & 1.01871      & $\ps 0.0003297$ & 1.01635      & 0.251068        & $   -0.485770$      & 0.486933      \\
14       & 1.01813      & $\ps 0.0002527$ & 1.01596      & 0.251787        & $   -0.487845$      & 0.486705      \\
16       & 1.01771      & $\ps 0.0002005$ & 1.01592      & 0.252593        & $   -0.489277$      & 0.486771      \\\hline
$\inf$   & 1.0150227    & $   -0.0000071$ & 1.01491      & 0.25716         & $   -0.4997  $      & 0.48641       \\
$\sigma$ & $3\dexp{-7}$ & $\ps 0.0000002$ & 0.00006      & 0.00008         & $\ps 0.0002  $      & 0.00009       \\\hline
\end{tabular}
\caption{Selected observables of a representative of the second group of exotic solutions in the Ising$\otimes$tricritical Ising model. The other solutions differs by signs of $E_{\Id\sigma}$ and $E_{\sigma\Id}$. }
\label{tab:TriIsing 2}
\end{table}

\chapter{Summary and discussion}\label{sec:Discussion}
In this thesis, we have explored a large number of Siegel gauge solutions of bosonic open string field theory, which allows us to formulate some empirical rules governing these solutions. We have found that properties of OSFT solutions are closely related to the boundary states they describe and that they can be divided into several groups. We also notice that we do not see all solutions that are expected to exist and that there are some "selection rules", which tell us which solutions can be found, respectively, how difficult is to find them.

The first type of solutions describes a single D-brane with energy lower than the energy of the initial D-brane, which we denote as $E_{init}$. These solutions are easy to find, it is usually enough to solve level 1 equations to find them.
They can describe both conventional and non-conventional boundary states. These solutions are real, well-behaved and we can significantly improve their precision by employing infinite level extrapolations. Their energy can determined with precision to several decimal places. The precision of their Ellwood invariants, which describe the corresponding boundary state coefficients, depends on weights of the invariants. Invariants with low conformal weights reproduce the expected values quite well, but as the conformal weight increases, they begin to oscillate and the precision goes down. Oscillations of invariants with weights higher than approximately 1.5 are so big that we are not able to extract any useful information from these invariants. Therefore Ellwood's conjecture can be verified only for low level part of boundary states.

In many settings, solving level 1 equations in enough to find solutions describing all fundamental boundary states with lower energy (or at least the known ones). However, there are few exceptions. Consider for example $m=5$ minimal model with $(3,1)$ boundary conditions. The boundary spectrum includes the identity and the field $\phi_{(3,1)}$, which has weight $h=7/5$. Therefore there are no nontrivial relevant boundary operators and we do not find a solution for the $\Id$-brane, which has lower energy than the $(3,1)$-brane. Another setting with similar issue is a long and thin rectangular torus with $R_1>1$, $R_2<1$ and $R_1R_2>1$. D0-brane solutions are allowed by the energy, but find none because all momentum modes along the shorted direction are represented by irrelevant operators. These exceptional cases suggest a conjecture that real solutions of level 1 equations describe D-branes that are accessible from the reference background by RG flow. If true, this would mean that OSFT gives us a systematic approach to classification of low energy boundary states in a given CFT.

The next type of solutions describes a single D-brane with energy higher than $E_{init}$. These solutions go against the RG flow and therefore they are more difficult to find. To see these solutions, we need to solve equations that involve at least some irrelevant fields. That usually means we have to start at least with level 2, where we find the first descendants of the ground state. Seeds for these solutions are complex in the vast majority of cases. There are some solutions whose imaginary part disappears at high enough level and we can be reasonably sure that these solutions are physical. However, there are also solutions whose observables are complex even in the infinite level limit. Its unlikely that these solutions solve the full equations of motion. We think that they may be either approximations of solutions that cannot be brought to Siegel gauge or nonphysical artifacts of the level truncation approximation.

Another problem with this type of solutions is that it is difficult to extrapolate them. We observe that the behavior of the real part of a solution (and its observables) changes significantly when the imaginary part disappears. Therefore we can extrapolate only real data points (which are usually few), which leads to significantly lower precision of results than in case of purely real solutions, we often get only a qualitative agreement with the expected boundary state.
In general, it seems that the imaginary part of complex solutions grows with the ratio $\frac{E}{E_{init}}$. That means that solutions with higher energy are usually harder to find than solutions with lower energy. However, this rule is not absolute, for example, the imaginary part of single lump solutions from section \ref{sec:FB circle:single:smallR} grows not only for small $R$, but also close to $R=1$.

Finally, the energy of the final D-brane can be also equal to the energy of the reference D-brane. These D-branes often belong to continuous families, which are connected by marginal deformations. There are several approaches to marginal deformations in OSFT, but all describe only a part of the moduli space. Marginal solutions in the traditional marginal approach behave similarly as solutions with $E<E_{init}$ and they reliably describe approximately one fundamental domain ($|\lB|\lesssim \frac{1}{2}$) of the moduli space. There are also theories with multiple D-branes with the same energy, but without a marginal operator. The simplest example is the Ising model, where both the $\Id$-brane and the $\eps$-brane have energy $2^{-1/2}$. So far, we have not found any solution connecting such D-branes.

Another class of solutions describes two D-branes. Let us focus only on solutions with $E<E_{init}$, because solutions with $E\geq E_{init}$ suffer from the same problems as above. Unlike for single D-brane solutions, we are often unable to find solutions corresponding to all D-brane configurations allowed by the energy. The evidence presented in sections \ref{sec:FB circle:double:superposition} and \ref{sec:MM:Ising3:regular} strongly suggests that the key feature which decides properties of these solutions is the spectrum of strings stretched between the two D-branes.
If there are no stretched tachyonic modes, these solutions are real and well-behaved, otherwise they behave similarly as solutions with $E>E_{init}$ or they are missing entirely. Solutions supporting tachyonic excitations are complex (at least at low levels) and they give us only a rough approximation of the expected boundary state. It seems that properties of these solutions generally get worse as the weight of the lightest mode in the spectrum of stretched strings decreases. In particular, we note that we have never seen any solution describing two exact copies of the same D-brane. This tells us that finding a well-behaved double brane solution in the universal sector is going to be very difficult because it has both $E=2E_{init}$ and the lightest possible tachyons in the spectrum of excitations. Finding a solution for intersecting D1-branes could be slightly easier, but to have a chance to see such solution, one would have to solve at least level 2 equations in a theory with many primary operators and analyze a large number of complex seeds.

The reason why we have trouble seeing solutions that support stretched tachyon excitations around the final D-brane system is not clear. One possibility is that such D-brane configurations cannot be reached by RG flow, which would imply that massive fields play a crucial role in construction of these solutions\footnote{We have tried to compare coefficients of several massive fields of some double lump solutions from section \ref{sec:FB circle:double:superposition}. The results suggest that values of these coefficients grow with decreasing distance between the lumps, but they are not entirely conclusive. The reason is that the solutions with the shortest distances have strong level dependence and therefore asymptotic values of their coefficients are uncertain.}. That would explain why they behave similarly as solutions with positive energy.
Another possibility, suggested in \cite{ErlerMaccaferri3}, is that the problem lies in gauge fixing. Analytic double lumps solutions are given by superposition of single lumps (by an expression analogue to (\ref{DL superposition})). When one wants to describe coincident D-branes, it is necessary to use two different types of single lump solutions, which are in different gauges (one of them is written in terms excited boundary condition changing operators). It is possible that a similar construction is needed even in the numerical approach to see lumps close to each other. However, we have not found any gauge equivalent lump solutions in Siegel gauge and it would be difficult to combine two solutions in different gauges in the level truncation approach.

As far as we can tell, solutions describing three or more D-branes behave similarly as two D-brane solutions, but we have not yet studied them systematically, so we cannot confirm their properties with certainty.

To conclude, the numerical approach to OSFT provides good evidence for Sen's conjectures, the Ellwood conjecture and the background independence of OSFT in general. There are of course some limitations, but that can be expected when we truncate the full theory to a finite number of degrees of freedom.

Next, we would like to propose some possible future directions of the numerical approach to OSFT.

First of all, the level truncation approach has two unsolved technical issues. The first one is convergence of Ellwood invariant with high conformal weights. These invariants behave badly at low levels, but that does not necessarily imply that they are divergent. So it would be useful to show how they behave asymptotically (perhaps by level expansion of some analytic solution) and to find a way how to reliably extract these boundary state coefficients, either by some resummation technique or by finding an alternative prescription for Ellwood invariants.

The second problem is a missing numerical algorithm that would systematically allow us to compute spectra of excitations around solutions. It will be necessary to check whether the techniques from \cite{EllwoodTaylorCohomology}\cite{GiustoCohomology}\cite{OhmoriLumps} can be extended to higher levels and other backgrounds, whether they reproduce correct spectra and, if necessary, to invent new algorithms. Knowing the spectra would be especially useful for exotic solutions, which do not have an interpretation in terms of conventional D-branes.

Next, one could try to find applications for gauges different from Siegel gauge. Most of numerical calculations so far have been done in this gauge, but that may not always be the best choice. Our experiments suggest that other gauges are not beneficial for solutions that behave well in Siegel gauge, but they could help us with analysis of complex solutions, which can have better properties in some gauges, or to find solutions that are missing in Siegel gauge.

Finally, we should exploit the newly discovered possibility of investigating non-conventional D-branes in irrational CFTs. In the free boson theory on a torus, we can find huge number of such solutions, which could help us with classification of boundary states in this theory. It would be also interesting to investigate more complicated theories, for example, we can look for solutions which break a part of the current symmetry in WZW models.

\chapter*{Acknowledgments}
\noindent
I would like to thank my supervisor Martin Schnabl for guiding me through my doctoral studies.
I would like to thank my coworkers and fellow students at the Institute of Physics of the Czech Academy of Sciences for useful discussions, namely: Ted Erler, Ond\v{r}ej Hul\'{i}k, Renann Lipinski Jusinskas, Carlo Maccaferri, Toru Masuda, Masaki Murata, Tom\'{a}\v{s} Proch\'{a}zka, Joris Raeymaekers, Miroslav Rap\v{c}\'{a}k, Jakub Vo\v{s}mera. Additionally, I would like to thank Ted Erler for helping me with revision of the thesis for publication on arXiv.

This research has been supported in part by the Czech Science Foundation (GACR) grant 17-22899S and by the European Regional Development Fund and the Czech Ministry of Education, Youth and Sports (MSMT), project No. CZ.02.1.01/0.0/0.0/15 003/0000437,  by Grant Agency of the Czech Republic, under the grant 14-31689S and by the EURYI grant GACR EYI/07/E010 from EUROHORC and ESF.

Computational resources were provided by the CESNET LM2015042 and the CERIT Scientific Cloud LM2015085, provided under the programme "Projects of Large Research, Development, and Innovations Infrastructures.

\begin{appendix}

\chapter{F-matrices in minimal models}\label{sec:MM Fmatrix}
In this appendix, we show explicit form of F-matrices in the Virasoro minimal models and some of their identities following \cite{Runkel3}.

\section{Explicit formula for F-matrices}
We consider a minimal model with labels $(p,q)$, for which we define $t=\frac{p}{q}$. For Kac labels of primary operators, which we denote as $I=(r_I,s_I)$, we define $d_I=r_I- t s_I$ and $\tilde d_I=s_I- t r_I$. One usually has the freedom to replace $(r,s)$ by $(p-r,s-q)$, but the formulas from \cite{Runkel3} work only if we choose the following convention: If $p$ is odd, then we must choose Kac labels with odd $r$, otherwise we choose Kac labels with odd $s$.

Using these conventions, the F-matrices are given by
\begin{eqnarray}
&&\Fmatrix PQIJKL=\\
&&\frac{j\left((r_L-r_I-1+r_Q)/2,(s_L-s_I-1+s_Q)/2,-d_I,d_L\right)}{j\left((r_J-r_I-1+r_P)/2,(s_J-s_I-1+s_P)/2,-d_I,d_J\right)}\nn\\
&&\times\frac{j\left((r_J+r_K-1-r_Q)/2,(s_J+s_K-1-s_Q)/2,d_J,d_K\right)}{j\left((r_K+r_L-1-r_P)/2,(s_K+s_L-1-s_P)/2,d_K,d_L\right)}\nn\\
&&\times \alpha\left(\frac{-r_I\!+\!r_J\!+\!r_K\!+\!r_L}{2},\frac{r_K\!+\!r_L\!+\!1\!-\!r_P}{2},\frac{r_J\!+\!r_K\!+\!1\!-\!r_Q}{2},\tilde d_I,\tilde d_J,\tilde d_K,\tilde d_L,\frac{1}{t}\right)\nn \\
&&\times \alpha\left(\frac{-s_I\!+\!s_J\!+\!s_K\!+\!s_L}{2},\frac{s_K\!+\!s_L\!+\!1\!-\!s_P}{2},\frac{s_J\!+\!s_K\!+\!1\!-\!s_Q}{2},d_I,d_J,d_K,d_L,t\right), \nn
\end{eqnarray}
where we define the following expressions
\begin{eqnarray}
S_\rho(a,b)&=&\sin\pi(a+b\rho),\nn\\
m_{xy}(a,b)&=&t^{2 x y}\prod_{g=1}^x\prod_{h=1}^y\left[(ht\!-\!g)(a\!+\!ht\!-\!g)(b\!+\!ht\!-\!g)(a\!+\!b\!+\!(y\!+\!h)t\!-\!x\!-\!g)\right]^{-1},\nn\\
j(x,y,a,b)&=&m_{xy}(a,b)\beta_{xy}\left(-\frac{a}{t},-\frac{b}{t},\frac{1}{t}\right)\beta_{xy}(a,b,t).\\
\beta_{xy}(a,b,\rho)&=&\prod_{g=1}^y\frac{\Gamma(g\rho)\Gamma(a+g\rho)\Gamma(b+g\rho)}{\Gamma(\rho)\Gamma(a+b-2x+(y+g)\rho)},\nn
\end{eqnarray}

\begin{eqnarray}
\alpha(s,x,y,a,b,c,d,\rho)&=&\sum_{h=\max(x,y)}^{\min(s,x+y-1)}\frac{\prod\limits_{g=1}^{s-h}S_\rho(d,x-1+g)
\prod\limits_{g=1}^{h-y}S_\rho(-a,s-x+g)}{\prod\limits_{g=1}^{s-y}S_\rho(-a+d,s-y+g)}\nn\\
&\times& \frac{\prod\limits_{g=1}^{y-1-h+x}S_\rho(b,s-x+g)\prod\limits_{g=1}^{h-x}S_\rho(c,x-1+g)}
{\prod\limits_{g=1}^{y-1}S_\rho(b+c,y-1+g)}\\
&\times& \prod\limits_{g=1}^{h-x}\frac{S_\rho(0,x+y-h-1+g)}{S_\rho(0,g)}\prod\limits_{g=1}^{s-h}\frac{S_\rho(0,h-y+g)}{S_\rho(0,g)},\nn
\end{eqnarray}

\section{Some F-matrix identities}
Next, we mention few F-matrix identities, which are useful for manipulations with sewing relations and structure constants:

\begin{equation}
\Fmatrix pqijkl=\Fmatrix pqjilk=\Fmatrix pqklij,
\end{equation}

\begin{equation}
\sum_r \Fmatrix prijkl \Fmatrix rqilkj=\delta_{pq},
\end{equation}

\begin{equation}
\Fmatrix \Id\Id iiii=\frac{S_\Id^{\ \Id}}{S_\Id^{\ i}},
\end{equation}

\begin{equation}
\Fmatrix k\Id jiij=\frac{S_\Id^{\ k}}{S_\Id^{\ j}}\Fmatrix j\Id ikki,
\end{equation}

\begin{equation}
\Fmatrix \Id kiijj \Fmatrix k\Id jiij=\frac{S_\Id^{\ \Id}S_\Id^{\ k}}{S_\Id^{\ i}S_\Id^{\ j}},
\end{equation}

\begin{equation}
\Fmatrix pinjkl \Fmatrix n\Id liil=\Fmatrix nklijp \Fmatrix p\Id lkkl,
\end{equation}

\begin{equation}
\sum_s \Fmatrix qspjkb \Fmatrix plaisb \Fmatrix srlijk= \Fmatrix praijq \Fmatrix qlarkb,
\end{equation}

\begin{equation}
\BRmatrix pqijkl=e^{i\pi (h_i+h_l-h_p-h_q)} \Fmatrix pqijlk.
\end{equation}

\chapter{Characters}\label{sec:characters}
In this appendix, we show characters describing the main state spaces that appear in this thesis. We use them for two main purposes: to show equivalence between some Hilbert spaces and for state counting, which allows us to make estimates of computer requirements for our calculations, see appendix \ref{sec:time}.

The character of a Hilbert space $\hh$ is usually defined as
\begin{equation}
\chi(q)={\rm{Tr}}_{\hh} q^{L_0-c/24},
\end{equation}
although the central charge is essentially irrelevant for our purposes. It disappears in the full theory anyway because it must have $c^{tot}=0$. In the ghost theory, we often want to keep track of ghost numbers of states, so for that purpose, we introduce a generalized character
\begin{equation}
\chi(q,y)={\rm{Tr}}_{\hh} q^{L_0-c/24}y^{j_0^{gh}},
\end{equation}
where the variable $y$ counts the ghost number. The decomposition of this character with respect to the ghost number $g$ is
\begin{equation}
\chi(q,y)=\sum_{g=-\infty}^\infty \chi^{(g)}(q)y^g.
\end{equation}

Next, we introduce a notation for the character of a generic non-degenerate Virasoro Verma module, which appears repeatedly in this appendix,
\begin{equation}
\chi^{Vir}(q)\equiv\prod_{m=1}^\infty\frac{1}{1-q^m}=\frac{q^{1/24}}{\eta(\tau)},
\end{equation}
where $\eta(\tau)$ is the Dedekind eta function and $q= e^{2\pi i \tau}$. In addition to this character, we define the character of the vacuum Verma module, which does not contain any $L_{-1}$ operators,
\begin{equation}
\chi'(q)\equiv\prod_{m=2}^\infty\frac{1}{1-q^m}=(1-q)\chi^{Vir}(q).
\end{equation}
This character describes the Hilbert space of the universal matter theory BCFT'.

In the following text, we will show characters which describe twist non-even string fields. To obtain characters of twist even spaces, we need to project out twist odd states. The projection to twist even subspaces can be done as
\begin{equation}\label{character even proj}
\chi_{even}(q)=\frac{1}{2}\left(\chi(q)+\Omega(\chi(q))\right),
\end{equation}
where the character $\Omega(\chi(q))$ has switched signs in front of terms that correspond to twist odd states. We define such character as
\begin{equation}
\Omega(\chi(q))={\rm{Tr}}_{\hh} q^{L_0-c/24} \Omega.
\end{equation}
For character of a single Verma module of weight $h$, we find
\begin{equation}
\Omega(\chi(q))=(-1)^{-h+\phi+c/24}\chi(-q),
\end{equation}
where the phase compensates the factor $(-q)^{h-c/24}$. The Hilbert space of the full theory usually includes several primary operators, so we apply the projection independently on each part of space.

\section{Ghost theory}\label{sec:characters:ghost}
The generalized character of the ghost theory is \cite{RastelliZwiebach}
\begin{eqnarray}\label{character ghost 1}
\chi^{gh}(q,y) &=&q^{-1-\frac{c^{gh}}{24}}y(1+y)\prod_{m=1}^\infty (1+q^m y)\left(1+\frac{q^m}{y}\right)\\
&=&q^{-\frac{c^{gh}}{24}}\prod_{m=1}^\infty \frac{1}{1-q^m}\sum_{g=-\infty}^\infty y^g q^{\frac{g^2-3g}{2}}
 =q^{-\frac{c^{gh}}{24}}\chi^{Vir}(q) \sum_{g=-\infty}^\infty y^g q^{\frac{g^2-3g}{2}}.\nn
\end{eqnarray}
The first line reflects the $bc$ basis of the state space and the second line the Virasoro or the ghost current basis at fixed ghost numbers.

The Siegel gauge character can be obtained from (\ref{character ghost 1}) by removing the contribution from the $c_0$ mode:
\begin{eqnarray}\label{character ghost 2}
\chi^{gh}_{Sieg}(q,y)&=&\frac{1}{1+y}\chi^{gh}(q,y) \\
&=&q^{-1-\frac{c^{gh}}{24}}\chi^{Vir}(q) \sum_{g=-\infty}^\infty \sum_{s=|g-1|}^\infty (-1)^{s+g-1} y^g q^{\frac{s^2+s}{2}}. \nn
\end{eqnarray}
This character does not reflect decomposition of the state space with respect to the SU(1,1) symmetry. To se that, we need to construct characters of irreducible SU(1,1) representations. The weights of the primaries (\ref{spin primary}) with respect to $L'^{gh}_0$ are $h'_j=j(2j+1)$. By investigating the Kac determinant (\ref{Kac determinat}) for central charge $-2$, we find that all corresponding Verma modules are reducible. The first null state of a spin $j$ representation  appears at level $h'_{j+1/2}$. By removing all null states, we obtain characters of irreducible representations:
\begin{equation}\label{character ghost SU11}
\chi^{gh}_j(q)=q^{-1-c^{gh}/24}\chi^{Vir}(q)\left(q^{h'_j}-q^{h'_{j+1/2}}\right).
\end{equation}
Notice that the character does not depend on $m$. By summing over all representations, we get the character
\begin{eqnarray}\label{character ghost 3}
\chi^{gh}_{Sieg}(q,y)&=&\sum_{j=0\; {\rm mod}\; 1/2\phantom{|}}^\inf \sum_{m=-j}^j y^{2m+1} \chi^{gh}_j(q)\nn \\
&=&\sum_{g=-\inf\phantom{|}}^\infty \sum_{j=\frac{|g-1|}{2}\; {\rm mod}\; 1}^\infty y^g \chi^{gh}_j(q).
\end{eqnarray}
After substitution $s=2j$ and some simple manipulations, we find that this character agrees with the second line of (\ref{character ghost 2}), which shows the equivalence of the $bc$ basis and the SU(1,1) basis of the Siegel gauge state space.

The character of the SU(1,1) singlet subspace, which we use in most of our calculations, reads
\begin{equation}
\chi_{singlet}^{gh}=\chi_{j=0}^{gh}=q^{-1-c^{gh}/24}\chi^{Vir}(q)(1-q)=q^{-1-c^{gh}/24}\chi'(q).
\end{equation}
This character is very similar to the character of universal matter theory, it differs only by the prefactor.

Finally, we show the character of the auxiliary space defined in section \ref{sec:Numerics:V3:ghost SU11}, where we describe an algorithm to compute SU(1,1) singlet cubic vertices. These auxiliary states have one additional mode of the ghost current compared to SU(1,1) singlets, $j_{-k}^{gh} L'^{gh}_{-M} c_1 |0\ra$, and therefore the corresponding character is given by
\begin{eqnarray}
\chi_{aux}^{gh}(q)&=&(q+q^2+q^3+\dots)\ \chi_{singlet}^{gh}(q) \nn\\
&=& \frac{q}{1-q}\ \chi_{singlet}^{gh}(q)=q^{-c^{gh}/24}\chi^{Vir}(q).
\end{eqnarray}

\section{Universal sector}\label{sec:characters:univ}
The character of the universal string field theory is given simply by the product of the matter and the ghost part,
\begin{equation}
\chi_{univ}(q)=\chi'(q)\chi^{gh}(q),
\end{equation}
where $\chi^{gh}(q)$ depends on conditions imposed on the ghost sector. In the most common case, which means imposing Siegel gauge and the SU(1,1) singlet condition, the character reads
\begin{equation}
\chi_{univ}(q)=q^{-1}\left(\chi'(q)\right)^2.
\end{equation}
Without any gauge conditions or in a generic gauge before implementation of gauge fixing conditions, we find
\begin{equation}
\chi_{univ}(q)=q^{-1}\chi'(q)\chi^{Vir}(q).
\end{equation}

These characters count both twist even and twist odd states. To get the character of the twist even subspace, we need to apply the projection (\ref{character even proj}).

\section{Free boson theory}\label{sec:characters:FB}
First, we consider the character of a U(1) Verma module of momentum $k$,
\begin{equation}\label{character FB 1}
\chi^{(k)}(q)=q^{k^2-1/24}\prod_{m=1}^\inf\frac{1}{1-q^m}=q^{k^2-1/24}\chi^{Vir}(q),
\end{equation}
which is the same result as for the character of a non-degenerate Virasoro representation. However, in this thesis, we work with the parity even subspace, so we need to remove contributions from parity odd states. The characters of subspaces with positive and negative momentum are the same for $k\neq 0$, $\chi^{(k)}(q)=\chi^{(-k)}(q)$, so the projection to parity even states gives us simply
\begin{equation}
\chi^{(k)}_{even}(q)=\frac{1}{2}(\chi^{(k)}(q)+\chi^{(-k)}(q))=\chi^{(k)}(q).
\end{equation}
The projection for $k=0$ is a bit more complicated and we find
\begin{eqnarray}
\chi^{(0)}_{even}(q)&=&q^{-1/24}\frac{1}{2}\left(\prod_{m=1}^\inf\frac{1}{1-q^m}+\prod_{m=1}^\inf\frac{1}{1+q^m}\right) \nn \\
&=&q^{-1/24}\frac{1}{2}\left(\chi^{Vir}(q)+\prod_{m=1}^\inf\frac{1}{1+q^m}\right).
\end{eqnarray}
To get the character of the full theory, we need to sum over all momenta and add the universal matter and ghost characters:
\begin{equation}
\chi_{even}^{(D1)}(q)=q^{-1}\left(\chi'(q)\right)^2\left( \chi^{Vir}(q)\left(\frac{1}{2}+\sum_{k>0} q^{k^2}\right) + \frac{1}{2}\prod_{m=1}^\inf\frac{1}{1+q^m} \right).
\end{equation}

Characters in two dimensions are very similar, the only significant difference is the projection to parity even states at zero momentum, which gives us
\begin{equation}
\chi^{(0,0)}_{evev}(q)=q^{-2/24}\frac{1}{2}\left((\chi^{Vir}(q))^2+\prod_{m=1}^\inf\frac{1}{(1+q^m)^2}\right).
\end{equation}
Therefore the character of the full theory is
\begin{equation}
\chi_{even}^{(D2)}(q)=q^{-1}\left(\chi'(q)\right)^2\left( (\chi^{Vir}(q))^2\left(\frac{1}{2}+\sum_{k>0} q^{k^2}\right) + \frac{1}{2}\prod_{m=1}^\inf\frac{1}{(1+q^m)^2} \right).
\end{equation}
The momentum in two dimensions is a two component vector, $k=(k_1,k_2)$, so we have some freedom in splitting the momenta into positive and negative, which however does not affect the character.

\section{Minimal models}\label{sec:characters:MM}
All Verma modules in the Virasoro minimal models are reducible, so characters of irreducible representations are more complicated then in the free boson theory. Following \cite{DiFrancesco}, we define functions
\begin{equation}
K_{(r,s)}^{(p,q)}=q^{-1/24}\chi^{Vir}(q)\sum_{n\in\mathbb{Z}}q^{\frac{(2pqn+ps-qr)^2}{4pq}},
\end{equation}
where $(p,q)$ are labels of the given minimal model. Using these functions, we can write the character of an irreducible representation over a primary labeled by $(r,s)$ as
\begin{equation}
\chi_{(r,s)}^{(p,q)}=q^{c/24}\left(K_{(r,s)}^{(p,q)}-K_{(r,-s)}^{(p,q)}\right).
\end{equation}

The character of the full theory therefore reads
\begin{equation}
\chi^{(p,q)}=q^{-1}\left(\chi'(q)\right)^2\sum_{(r,s)}\chi_{(r,s)}^{(p,q)},
\end{equation}
where $(r,s)$ runs over the spectrum of primaries allowed by the given boundary conditions.

Generalization of this character to a theory given by a product of two minimal models is straightforward
\begin{equation}
\chi^{(p_1,q_1)(p_2,q_2)}=q^{-1}\left(\chi'(q)\right)^2\sum_{(r_1,s_1)}\sum_{(r_2,s_2)}\chi_{(r_1,s_1)}^{(p_1,q_1)}\chi_{(r_2,s_2)}^{(p_2,q_2)},
\end{equation}
where we assume that the boundary theory has factorized boundary conditions.

\chapter{State counting and computer requirements}\label{sec:time}
In this appendix, we discuss how to estimate time and memory requirements of OSFT calculations in the level truncation scheme. First, we make some general asymptotic estimates and then we show finite level examples that are related to some of our calculations.

The results in this thesis were obtained using computer clusters at MetaCentrum VO, which is an organization that operates and manages computing infrastructure  which belong to CESNET, CERIT-SC and co-operative academic centers within the Czech Republic. The best machine we have access to has approximately 6 TB of operational memory and more than 300 CPUs, but we typically use much less computer resources. Most of the calculations in this thesis required tens or few hundreds of gigabytes of computer memory and they were executed using 20-60 CPUs. Using these computer resources, it is possible to find one solution of OSFT equations involving tens of thousands of variables within few days, the time varies depending on the OSFT background, properties of the solution and computer specifications. The most complicated system of equations we have solved has slightly over 100000 equations.

Among the algorithms presented in chapter \ref{sec:Numerics}, there are three that potentially consume large amount of time or memory: The evaluation of cubic vertices, Newton's method and the homotopy continuation method. The estimate of computer requirements for the homotopy continuation method is very straightforward, so we will focus on the other two. Observables can be usually computed very quickly, the only exception is the case when one has to compute the full energy (\ref{Energy num1}), which has the same complexity as Newton's method.

To find the number of states in a given state space at a given level, we expand its character as
\begin{equation}
\chi(q)=q^{-c/24-\delta_{gh}}\sum_{k}N_k\, q^k,
\end{equation}
where the sum goes over all weights in the state space and $\delta_{gh}$ equals 1 if the space includes ghosts and zero otherwise. The prefactor cancels the difference between the leading power in the character and the definition of level in OSFT. The number of states up to level $L$ is therefore given simply by
\begin{equation}
N(L)=\sum_{k\leq L} N_k.
\end{equation}
We put a tilde over $N$ in constituent theories to distinguish it from the number of states in the full theory.

First, we are going to make some estimates for cubic vertices. The number of cubic vertices is $N^3$, but we never compute the full non-factorized vertices (at least not at high levels) and we evaluate only vertices in constituent theories. Generally, the memory required for cubic vertices in one constituent theory is $8\tilde N^3$~B\footnote{We consider 8 B for one real number when using the usual double number format in C++. If we decide to use the long double type, which usually requires 16 B, the memory requirements double.}. In principle, we need just approximately one sixth of this number because many vertices are related by the twist or cyclic symmetry. However, using the reduced set of vertices would cause problems later, so we restore the full set of vertices before we start with Newton's method. As we are going to show, it does not usually lead to significant increase of memory requirements anyway.

Memory requirements for cubic vertices are higher in the ghost theory because of auxiliary vertices. For the singlet vertices, we need additional memory $8\tilde N_{singl}(\tilde N_{singl}+1)\tilde N_{aux}/2$ B, the one half factor comes from the remaining symmetry between the two singlet states. The number of auxiliary vertices in the $bc$ basis is $\tilde N_0 \tilde N_1 \tilde N_2$, where the lower index denotes the ghost number.

On the other hand, the number of nontrivial vertices is usually reduced in non-universal theories. If a theory contains multiple primaries with nontrivial fusion rules, we split the state space as described in section \ref{sec:Numerics:string field:FB MM} and the number of nonzero vertices is
\begin{equation}
\sum_{f=(p,q,r)} \tilde N_p \tilde N_q \tilde N_r,
\end{equation}
where $\tilde N_p$ denotes the number of states in the space over a primary $p$. This reduction is especially effective in the free boson theory, where fusion rules are sparse.

The time needed to compute cubic vertices primarily scales with the number of vertices, that is as $\tilde N^3$. However, time required to compute one vertex is not constant. The average time is proportional to the average number of terms in the $\mathcal{K}(\phi_{-n})$ representation of conservation laws. Theoretically, it can be as high as $\tilde N$, but it is typically much smaller. We can estimate it as the level times the average number of terms in $\mathcal{M}(\phi_{k})$. This number has only a small level dependence, but it depends a lot on the operator algebra. For example, $\mathcal{M}(b_k)$ in the $bc$ basis has at most one term, while $\mathcal{M}(L_k)$ can have large number of terms because all commutators between Virasoro operators are nonzero.
In practice, the time needed to evaluate cubic vertices is also strongly affected by the dynamic parallelization. Repeated calculations of the same set of vertices often take significantly different time and these fluctuations overshadow the subleading corrections, so $\tilde N^3$ is usually good enough estimate of the time scaling.

When it comes to Newton's method, we need memory mostly for storing the Jacobian matrix. The required memory is $16 N^2$ B because we generically work with complex numbers. The time needed for Newton's method is proportional to $N^3$. Both evaluation of the Jacobian matrix and solving the subsequent linear equations have the same complexity, but the Jacobian generally takes longer because we need to multiply factorized parts of vertices. However, the evaluation of the Jacobian can be sometimes sped up by omitting zero coefficients of the string field and by utilizing the twist symmetry and fusion rules, see section \ref{sec:Numerics:Newton:jac}. The time needed to solve the linear equations is less affected by OSFT background or by properties of solutions. Only in case that the Jacobian is a block diagonal matrix, the time scaling reduces to $\sum_i N_i^3$, where $N_i$ are sizes of the individual blocks.

\section{Asymptotic behavior}\label{sec:time:asymptotic}
Consider a generic Virasoro Verma module. The number of states at level $L$ is given by the number of integer partitions of the level $P(L)$, which has a well-known asymptotic behavior for large $L$:
\begin{equation}
P(L)\sim \frac{1}{4L\sqrt{3}}\ e^{\pi\sqrt{2L/3}}.
\end{equation}
All characters in appendix \ref{sec:characters} are based on the generic Virasoro character, so we will assume that, in the roughest approximation, the number of states in all theories behaves as
\begin{equation}
N(L)\sim e^{\alpha\sqrt{L}},
\end{equation}
where the constant $\alpha$ depends on the particular theory.

In this approximation, we do not have to distinguish between number of states {\it at} level $L$ and {\it up to} level $L$ because the majority of states appears at the few highest levels. The number of states up to level $L$ is approximately
\begin{equation}
N(L)\sim\int_0^L e^{\alpha \sqrt{x}} dx\sim \frac{2}{\alpha}\sqrt{L}\ e^{\alpha\sqrt{L}},
\end{equation}
which is roughly $\frac{\sqrt{L}}{\alpha}$ times higher than the number of states at level $L$.

Now consider a tensor product of two theories characterized by $\alpha_1$ and $\alpha_2$. The number of states in this theory at level $L$ can estimated using the saddle point approximation,
\begin{equation}
N(L)\sim\int_0^L e^{\alpha_1\sqrt{x}+\alpha_2\sqrt{L-x}}dx\sim \frac{\sqrt{8 \pi } \alpha_1 \alpha_2 }{(\alpha_1^2+\alpha_2^2)^{5/4}}L^{3/4} e^{ \sqrt{\alpha_1^2+\alpha_2^2}\sqrt{L}}.
\end{equation}
We observe that the full theory has the same type of asymptotic behavior as the constituent theories, the parameter $\alpha$ just changes to $\sqrt{\alpha_1^2+\alpha_2^2}$. Similarly, for a product of $M$ theories with $\alpha_1,\dots,\alpha_M$, the parameters combine together as $\alpha^{tot}=\sqrt{\alpha_1^2+\dots+\alpha_M^2}$.

These formulas allow us to estimate which part of the computational process requires the most of time or memory. We assume that the full theory is given by a product of $M\geq 2$ constituent theories with roughly the same number of states,
\begin{equation}
\tilde N_1(L)\cong \dots \cong \tilde N_M(L) \sim e^{\alpha\sqrt{L}}.
\end{equation}
The number of states in the full theory is therefore approximately
\begin{equation}
N(L)\sim e^{\sqrt{M}\alpha\sqrt{L}}\sim \tilde N^{\sqrt{M}}.
\end{equation}

First, we consider the time requirements. The time needed to evaluate cubic vertices scales as $\tilde N^{3}$ and the time for Newton's method as $\tilde N^{3\sqrt{M}}$. Obviously, $\tilde N^{3\sqrt{M}}\gg \tilde N^{3}$ for all $M$, so the time needed for Newton's method will be always dominant at high enough level.

When it comes to memory, it scales as $\tilde N^{3}$ for cubic vertices and as $\tilde N^{2\sqrt{M}}$ for Newton's method. For $M=2$, we find $\tilde N^{3}>\tilde N^{2\sqrt{2}}$, which means that the memory needed for cubic vertices is dominant, while for $M=3$, we have $\tilde N^{3}<\tilde N^{2\sqrt{3}}$, so the memory required for the Jacobian matrix is higher. The Jacobian matrix dominates even more for $M>3$.

These estimates tell us that the main restrictions for $M\geq 3$ come from Newton's method. It requires both more time and more memory than cubic vertices. The time needed to execute calculations is usually the bigger issue because it scales as $N^3$, while the memory only as $N^2$.

In practice, constituent theories usually do not have exactly the same amount of states and there are various subleading contributions that we neglected. Despite that, the estimates above are mostly correct. We have seen just 2 exceptions. For $M=2$, which means the universal theory, the time requirements for cubic vertices and for Newton's method are comparable. The reason is the large number of auxiliary ghost vertices and large number of operations needed to computer one vertex. For $M=3$, we sometimes find that we need more memory for cubic vertices, which is caused by asymmetric number of states in constituent theories. See the examples later.

\section{Universal sector}\label{sec:time:univ}
First, we will discuss time and memory requirements for cubic vertices in the universal part of string field theory following \cite{KudrnaUniversal}.

Table \ref{tab:states ghost} shows numbers of states up to level $L$ in the universal matter theory and in various subspaces of the ghost theory. We also add numbers of states in the auxiliary ghost spaces which are needed for evaluation of cubic vertices.

Using this data, we can compute the number of cubic vertices in the universal sector and the corresponding memory requirements, see table \ref{tab:memory univ}. We denote memory required for a set of cubic vertices by the letter $\mathfrak{V}$. When counting physical vertices, we consider only those with ordered indices, this reduces the memory requirements approximately by a factor of 6. Once we have evaluated all required vertices and deallocated all auxiliary objects, we restore the full set of physical vertices. This slightly increases the memory requirements, but not much when compared with the memory needed for auxiliary ghost vertices. The number of vertices in the universal matter theory is much smaller than in the ghost theory (including the auxiliary vertices), which means that calculations in the ghost theory take most of the time.

We observe that imposing the Siegel gauge condition and the SU(1,1) singlet condition leads to a huge reduction of memory requirements for cubic vertices. However, imposing the SU(1,1) singlet condition does not save us much time. Computing all vertices in Siegel gauge using the $bc$ basis takes about the same amount of time as computing the SU(1,1) singlet vertices. The reason is that the algebra of Virasoro and ghost current operators is much more complicated than the algebra of $b$ and $c$ ghosts. Therefore evaluation of one vertex in the $bc$ basis is much faster than in other bases.

\begin{table}[]
\centering
 \begin{tabular}{|c|rr|rr|rr|}\hline
Level \rowh{14pt} & $\tilde N_{singl}$ & $\tilde N_{aux}$ & $\tilde N_{Sieg}^{(1)}$ & $\tilde N_{Sieg}^{(0,2)}$ & $\tilde N_{gen}^{(1)}$ & $\tilde N_{gen}^{(0,2)}$ \\\hline
2     & 2     & 1     & 2     & 1     & 4     & 2     \\
4     & 5     & 4     & 7     & 3     & 12    & 7     \\
6     & 11    & 12    & 17    & 8     & 30    & 19    \\
8     & 22    & 30    & 37    & 20    & 67    & 45    \\
10    & 42    & 67    & 76    & 44    & 139   & 97    \\
12    & 77    & 139   & 148   & 89    & 272   & 195   \\
14    & 135   & 272   & 275   & 171   & 508   & 373   \\
16    & 231   & 508   & 493   & 315   & 915   & 684   \\
18    & 385   & 915   & 857   & 561   & 1597  & 1212  \\
20    & 627   & 1597  & 1451  & 970   & 2714  & 2087  \\
22    & 1002  & 2714  & 2403  & 1635  & 4508  & 3506  \\
24    & 1575  & 4508  & 3902  & 2696  & 7338  & 5763  \\
26    & 2436  & 7338  & 6224  & 4360  & 11732 & 9296  \\
28    & 3718  & 11732 & 9774  & 6930  & 18460 & 14742 \\
30    & 5604  & 18460 & 15131 & 10847 & 28629 & 23025 \\
32    & 8349  & 28629 & 23119 & 16742 & 43820 & 35471 \\
34    & 12310 & 43820 & 34907 & 25511 & 66273 & 53963 \\\hline
\end{tabular}
\caption{Number of states in various parts of the universal sector. Starting from the left, we show the number of states in the SU(1,1) singlet subspace, which also equals to the number of states in the universal matter theory, $\tilde N'=\tilde N_{singl}$, and in the associated auxiliary space with a single $j_{-k}^{gh}$ operator. Then we count all states in Siegel gauge at ghost number one, and at ghost numbers 0 and 2. Finally, we consider the ghost theory without any gauge conditions imposed. The number of states at ghost number one $\tilde N_{gen}^{(1)}$ also equals to the number of states in a generic non-degenerate Virasoro Verma module. In the auxiliary spaces, we count only states that enter the cubic vertex algorithm, which means only states up to one level less than in the physical space.}
\label{tab:states ghost}
\end{table}

Next, we focus on the full universal theory. We show the number of twist even states and the memory requirements in table \ref{tab:states univ}. We consider two cases, first we impose the SU(1,1) singlet condition in Siegel gauge, then we keep the full state space without any gauge fixing. At available levels, the memory requirements are a bigger issue than the time requirements, so we will discuss them in more detail. However, we find that every two levels increase the memory requirements 3-4 times around level 30 (the number grows smaller with increasing level), while the time needed to find one solution increases 5-6 times. Therefore the time requirements for  Newton's method will become the main restriction at sufficiently high level.

In the SU(1,1) singlet case, the number of matter and ghost states is the same, so the memory required for cubic vertices equals approximately to $12\,\mathfrak{V}_{singl}$. The memory required for the Jacobian matrix, which is denoted by the letter $\mathfrak{J}$, is significantly lower, which agrees with the estimate from section \ref{sec:time:asymptotic}. At level 30, which is the highest level we have been able to reach, we need approximately 2.7 TB of memory. For the ghost vertices, we need $9/8(\mathfrak{V}^{singl}+\mathfrak{V}^{singl}_{aux})\cong 2.6$~TB of memory, which shows that using the unordered set of cubic vertices does not have a significant impact on the memory requirements. If we consider other gauge conditions, we find that we can go only up to level 24 with the same amount of computer resources as in the singlet case. We need approximately 2.9 TB of memory, which is mostly consumed by the ghost vertices.

Time needed for evaluation of cubic vertices and time needed to find one solution using Newton's method are comparable in the universal theory. Which takes longer depends of the level, properties of the solution and conditions imposed on the string field. When it comes to Newton's method, we find that evaluation of the Jacobian takes 3-4 times more time than solving the linear equations because we need to construct full vertices out of the factorized ones, which requires several elemental operations.

\begin{table}[]
\centering
\begin{tabular}{|c|rr|rr|rr|}\hline
Level  & $\mathfrak{V}_{singl}$ & $\mathfrak{V}_{singl}^{aux}$ & $\mathfrak{V}_{bc}$ & $\mathfrak{V}_{bc}^{aux}$ & $\mathfrak{V}_{gen}$ & $\mathfrak{V}_{gen}^{aux}$ \\\hline
2  & 32    B  & 24    B  & 32    B  & 16    B  & 160   B  & 128   B  \\
4  & 280   B  & 480   B  & 672   B  & 504   B  & 2.844 kB & 4.594 kB \\
6  & 2.234 kB & 6.188 kB & 7.570 kB & 8.500 kB & 38.75 kB & 84.61 kB \\
8  & 15.81 kB & 59.30 kB & 71.40 kB & 115.6 kB & 409.3 kB & 1.035 MB \\
10 & 103.5 kB & 472.7 kB & 594.3 kB & 1.123 MB & 25.87 MB & 9.978 MB \\
12 & 617.8 kB & 3.185 MB & 4.206 MB & 8.944 MB & 29.11 MB & 78.91 MB \\
14 & 3.198 MB & 19.05 MB & 26.73 MB & 61.35 MB & 167.7 MB & 539.2 MB \\
16 & 15.88 MB & 103.9 MB & 153.3 MB & 373.2 MB & 977.3 MB & 3.19  GB \\
18 & 73.13 MB & 518.7 MB & 803.2 MB & 2.010 GB & 5.067 GB & 17.48 GB \\
20 & 314.9 MB & 2.343 GB & 3.801 GB & 10.17 GB & 24.85 GB & 88.07 GB \\
22 & 1.253 GB & 10.16 GB & 17.25 GB & 47.86 GB & 113.8 GB & 412.9 GB \\
24 & 4.861 GB & 41.69 GB & 73.83 GB & 211.3 GB & 490.9 GB & 1.773 TB \\
26 & 17.97 GB & 162.3 GB & 299.5 GB & 881.5 GB & 1.959 TB & 7.377 TB \\
28 & 63.87 GB & 604.3 GB & 1.133 TB & 3.415 TB & 7.630 TB & 29.19 TB \\
30 & 218.7 GB & 2.109 TB & 4.202 TB & 12.95 TB & 28.46 TB & 110.4 TB \\
32 & 722.9 GB & 7.261 TB & 14.99 TB & 47.15 TB & 102.0 TB & 401.2 TB \\
34 & 2.263 TB & 24.16 TB & 51.58 TB & 165.3 TB & 353.0 TB & 1404. TB \\\hline
\end{tabular}
\caption{Minimum memory requirements for storing various sets of ghost vertices. From the left: Memory requirements for SU(1,1) singlet vertices (this number also applies to the matter vertices), for vertices in Siegel gauge in the $bc$ basis and finally for vertices not restricted by any gauge condition. The memory requirements for physical vertices are always accompanied by memory requirements for the corresponding auxiliary vertices. We consider only physical vertices which are maximally reduced by the cyclic and twist symmetry, the memory needed for all vertices is approximately six times higher. We assume 8 B of memory for one number in C++ double format, but we also temporarily need additional 1 B for auxiliary boolean field during evaluation of vertices. }
\label{tab:memory univ}
\end{table}

\begin{table}[!]
\centering
\begin{tabular}{|c|rrr|rrr|}\hline
Level \rowh{14pt} & $N^{singl}_{Univ}$ & $\mathfrak{V}^{singl}_{Univ}$ & $\mathfrak{J}^{singl}_{Univ}$ & $N^{gen}_{Univ}$ & $\mathfrak{V}^{gen}_{Univ}$ & $\mathfrak{J}^{gen}_{Univ}$ \\\hline
2     & 3      & 128   B  & 144   B  & 4       & 576   B  & 256   B  \\
4     & 8      & 1.953 kB & 1.000 kB & 14      & 14.48 kB & 3.062 kB \\
6     & 21     & 20.80 kB & 6.891 kB & 43      & 221.3 kB & 28.89 kB \\
8     & 51     & 166.4 kB & 40.64 kB & 118     & 2.376 MB & 217.6 kB \\
10    & 117    & 1.130 MB & 213.9 kB & 299     & 21.05 MB & 1.364 MB \\
12    & 259    & 6.966 MB & 1.024 MB & 712     & 157.0 MB & 7.735 MB \\
14    & 549    & 37.54 MB & 4.599 MB & 1607    & 1019. MB & 39.41 MB \\
16    & 1124   & 188.1 MB & 19.28 MB & 3473    & 5.799 GB & 184.0 MB \\
18    & 2236   & 870.8 MB & 76.29 MB & 7233    & 30.77 GB & 798.3 MB \\
20    & 4328   & 3.673 GB & 285.8 MB & 14585   & 150.8 GB & 3.170 GB \\
22    & 8176   & 14.99 GB & 1020. MB & 28593   & 690.1 GB & 12.18 GB \\
24    & 15121  & 58.22 GB & 3.407 GB & 54678   & 2.903 TB & 44.55 GB \\
26    & 27419  & 215.4 GB & 11.20 GB & 102253  & 11.85 TB & 155.8 GB \\
28    & 48841  & 765.9 GB & 35.55 GB & 187428  & 46.14 TB & 523.5 GB \\
30    & 85604  & 2.561 TB & 109.2 GB & 337366  & 172.0 TB & 1.656 TB \\
32    & 147809 & 8.469 TB & 325.6 GB & 597257  & 616.5 TB & 5.191 TB \\
34    & 251719 & 27.15 TB & 944.2 GB & 1041392 & 2131. TB & 15.78 TB \\\hline
\end{tabular}
\caption{Number of twist even states and memory requirements in the universal string field theory. $\mathfrak{V}$ denotes memory requirements for cubic vertices and $\mathfrak{J}$ memory requirements for the Jacobian matrix. The left part of the table corresponds to Siegel gauge with the SU(1,1) singlet condition imposed, the right part to the theory without any gauge condition.}
\label{tab:states univ}
\end{table}

\section{Free boson theory}
Properties of a free boson theory depend on its compactification, so we are going to show only few examples to illustrate scaling of the computer requirements. We always impose Siegel gauge and the SU(1,1) singlet condition.

First, we consider the theory on a circle. We have picked two radii $R=1$ and $R=3$ as examples and the corresponding data are presented in table \ref{tab:states D1}. We can see that the Hilbert space is much larger than in the universal theory. Level 30 in the universal theory, where we find approximately 85000 states, roughly corresponds to level 20 at $R=1$ and to level 18 at $R=3$. However, the memory requirements for the same amount of states are significantly reduced, approximately by one order. We find that cubic vertices require more memory than the Jacobian at available levels, which contradicts the asymptotic estimate for a theory composed of three BCFTs. The reason is the dominance of the free boson sector. There are $\tilde N_{singl}$ states in the matter and ghost sector, while the number of states in the free boson sector is proportional to $\tilde N_{gen}$. The memory needed for the Jacobian matrix becomes dominant at high enough level, in accordance with the asymptotic estimate, but such levels are above our reach, it is for example level 36 at $R=1$.

\begin{table}[]
\centering
\begin{tabular}{|c|rrr|rrr|}\hline
Level \rowh{14pt} & $N^{R=1}_{D1}$ & $\mathfrak{V}^{R=1}_{D1}$ & $\mathfrak{J}^{R=1}_{D1}$ &
$N^{R=3}_{D1}$ & $\mathfrak{V}^{R=3}_{D1}$ & $\mathfrak{J}^{R=3}_{D1}$ \\\hline
2     & 5      & 768   B  & 400   B  & 8       & 1.219 kB & 1.000 kB \\
4     & 20     & 20.52 kB & 6.250 kB & 36      & 41.54 kB & 20.25 kB \\
6     & 72     & 331.4 kB & 81.00 kB & 141     & 767.8 kB & 310.6 kB \\
8     & 238    & 3.751 MB & 885.1 kB & 496     & 9.586 MB & 3.754 MB \\
10    & 727    & 34.75 MB & 8.065 MB & 1580    & 95.01 MB & 38.09 MB \\
12    & 2074   & 269.0 MB & 65.64 MB & 4649    & 774.2 MB & 329.8 MB \\
14    & 5577   & 1.766 GB & 474.6 MB & 12808   & 5.310 GB & 2.444 GB \\
16    & 14252  & 10.61 GB & 3.027 GB & 33357   & 33.03 GB & 16.58 GB \\
18    & 34842  & 57.87 GB & 18.09 GB & 82823   & 185.6 GB & 102.2 GB \\
20    & 81945  & 291.0 GB & 100.1 GB & 197343  & 957.9 GB & 580.3 GB \\
22    & 186256 & 1.331 TB & 516.9 GB & 453541  & 4.484 TB & 2.993 TB \\
24    & 410687 & 5.863 TB & 2.454 TB & 1009673 & 20.15 TB & 14.83 TB \\\hline
\end{tabular}
\caption{Number of twist even states in the free boson theory on a circle with $R=1,3$ and the corresponding memory requirements.}
\label{tab:states D1}
\end{table}

The maximal level we can reach in this theory is decided by the time needed for Newton's method. It scales worse than in the universal case, increase of level by 2 means 10-20 times more time. At low radii, we are able to reach levels 18-20 in reasonable time. The number of states grows more or less linearly with the radius (for $R\gg 1$, the growth in table \ref{tab:states D1} is somewhat smaller due to the common zero momentum states), so the time requirements scale approximately as $R^3$ with the radius and we can accordingly reach only lower levels at large radii. Given the same number of variables, Newton's method runs somewhat faster than in the universal case because evaluation of the Jacobian can be sped up using the sparse fusion rules.

Next, we get to two dimensions, where we have picked two square tori with $R=1$ and $R=2$ as examples, see in table \ref{tab:states D2}. This theory involves 4 constituent BCFTs, so the number of states grows very quickly with level and the maximal available level drops to 12-14. We find that the number of states is roughly proportional to the area of the torus, which means that scaling of the torus has larger effect than scaling of the circle in one dimension, the time requirements on a torus grow roughly as $R^6$ with the radius.

We noticed that solving the equations of motion on the same background often takes significantly different amount of time. Many solutions do not excite states with certain momenta, which simplifies evaluation of the Jacobian matrix and makes it block diagonal. Therefore these solutions can be computed much faster than the estimate based on $N^3$.

Memory requirements in this theory are dominated by the Jacobian in accordance with the asymptotic estimate. Overall, they are quite low and they are usually not a significant restriction.

\begin{table}[]
\centering
\begin{tabular}{|c|rrr|rrr|}\hline
Level \rowh{14pt} & $N^{R=1}_{D2}$ & $\mathfrak{V}^{R=1}_{D2}$ & $\mathfrak{J}^{R=1}_{D2}$ &
$N^{R=2}_{D2}$ & $\mathfrak{V}^{R=2}_{D2}$ & $\mathfrak{J}^{R=2}_{D2}$ \\\hline
2     & 10     & 2.000 kB & 1.562 kB & 18     & 3.312 kB & 5.062 kB \\
4     & 63     & 64.45 kB & 62.02 kB & 137    & 129.4 kB & 293.3 kB \\
6     & 332    & 1.113 MB & 1.682 MB & 794    & 2.440 MB & 9.620 MB \\
8     & 1492   & 13.81 MB & 33.97 MB & 3800   & 32.34 MB & 220.3 MB \\
10    & 5906   & 134.3 MB & 532.2 MB & 15696  & 328.7 MB & 3.671 GB \\
12    & 21129  & 1.052 GB & 6.652 GB & 57943  & 2.662 GB & 50.03 GB \\
14    & 69628  & 7.302 GB & 72.24 GB & 195676 & 18.98 GB & 570.6 GB \\
16    & 214307 & 45.01 GB & 684.4 GB & 614317 & 119.6 GB & 5.492 TB \\\hline
\end{tabular}
\caption{Number of twist even states in the free boson theory on a square torus with $R=1,2$ and the corresponding memory requirements.}
\label{tab:states D2}
\end{table}

\section{Minimal models}
Computer requirements in Virasoro minimal models follow a similar pattern as in the case of the free boson theory. In general, we are able to go to slightly higher levels because the first minimal models have low number of primaries and the number of independent variables is reduced thanks to null states. However, fusion rules in minimal models are relatively dense, so they do not speed up the calculations by much.

As an example, we show some data concerning the Ising model theory in table \ref{tab:states MM1}. The memory requirements follow the prediction from section \ref{sec:time:asymptotic} because the Jacobian consumes more memory than the vertices. We notice that the memory for vertices $\mathfrak{V}$ is similar for both backgrounds, which means that most of the memory is needed for the universal sector and the Ising model contribution is small.

Table \ref{tab:states MM2} shows the number of twist even states and memory requirements in the double Ising model and in the Ising$\otimes$tricritical Ising model. In both cases, we consider the $\sigma\otimes\sigma$ background from chapter \ref{sec:MM}. Since these theories consist of 4 BCFTs, the number of states is significantly increased compared to the simple Ising model and we can reach only much lower level.

Memory requirements in minimal models are usually not very high, which means that we are again restricted mainly by the time requirements for Newton's method.

\begin{table}
\centering
\begin{tabular}{|c|rrr|rrr|}\hline
Level \rowh{14pt} & $N^{\Id}_{Ising}$ & $\mathfrak{V}^{\Id}_{Ising}$ & $\mathfrak{J}^{\Id}_{Ising}$ &
$N^{\sigma}_{Ising}$ & $\mathfrak{V}^{\sigma}_{Ising}$ & $\mathfrak{J}^{\sigma}_{Ising}$ \\\hline
2     & 4      & 192   B  & 256   B  & 5      & 256   B  & 400   B  \\
4     & 13     & 2.930 kB & 2.641 kB & 17     & 3.555 kB & 4.516 kB \\
6     & 40     & 28.61 kB & 25.00 kB & 55     & 33.61 kB & 47.27 kB \\
8     & 114    & 211.9 kB & 203.1 kB & 162    & 243.6 kB & 410.1 kB \\
10    & 302    & 1.336 MB & 1.392 MB & 440    & 1.491 MB & 2.954 MB \\
12    & 759    & 7.864 MB & 8.790 MB & 1128   & 8.555 MB & 19.42 MB \\
14    & 1815   & 41.03 MB & 50.27 MB & 2742   & 43.82 MB & 114.7 MB \\
16    & 4157   & 200.6 MB & 263.7 MB & 6364   & 210.7 MB & 618.0 MB \\
18    & 9178   & 913.1 MB & 1.255 GB & 14211  & 947.6 MB & 3.009 GB \\
20    & 19611  & 3.804 GB & 5.731 GB & 30663  & 3.912 GB & 14.01 GB \\
22    & 40699  & 15.38 GB & 24.68 GB & 64182  & 15.71 GB & 61.38 GB \\
24    & 82309  & 59.35 GB & 101.0 GB & 130789 & 60.30 GB & 254.9 GB \\
26    & 162621 & 218.5 GB & 394.1 GB & 260170 & 221.2 GB & 1009. GB \\\hline
\end{tabular}
\caption{Number of twist even states and memory requirements in the Ising model theory with $\Id$ and $\sigma$ boundary conditions.}
\label{tab:states MM1}
\end{table}

\begin{table}[]
\centering
\begin{tabular}{|c|rrr|rrr|}\hline
Level \rowh{14pt} & $N_{DI}^{\sigma\otimes\sigma}$ & $\mathfrak{V}_{DI}^{\sigma\otimes\sigma}$ & $\mathfrak{J}_{DI}^{\sigma\otimes\sigma}$ &
$N_{I\otimes TI}^{\sigma\otimes\sigma}$ & $\mathfrak{V}_{I\otimes TI}^{\sigma\otimes\sigma}$ & $\mathfrak{J}_{I\otimes TI}^{\sigma\otimes\sigma}$ \\\hline
2     & 8      & 448   B  & 1.000 kB & 12     & 912   B  & 2.250 kB \\
4     & 35     & 5.781 kB & 19.14 kB & 61     & 15.67 kB & 58.14 kB \\
6     & 139    & 51.42 kB & 301.9 kB & 267    & 190.5 kB & 1.088 MB \\
8     & 487    & 352.4 kB & 3.619 MB & 1010   & 1.610 MB & 15.57 MB \\
10    & 1543   & 2.007 MB & 36.33 MB & 3410   & 11.31 MB & 177.4 MB \\
12    & 4521   & 10.84 MB & 311.9 MB & 10540  & 67.84 MB & 1.655 GB \\
14    & 12404  & 52.90 MB & 2.293 GB & 30312  & 358.5 MB & 13.69 GB \\
16    & 32181  & 243.5 MB & 15.43 GB & 82066  & 1.663 GB & 100.4 GB \\
18    & 79647  & 1.034 GB & 94.53 GB & 211204 & 7.276 GB & 664.7 GB \\
20    & 189231 & 4.259 GB & 533.6 GB & 520373 & 29.57 GB & 3.940 TB \\\hline
\end{tabular}
\caption{Number of twist even states and memory requirements in the double Ising model and the Ising$\otimes$tricritical Ising model. We consider the $\sigma\otimes\sigma$-brane background in both cases.}
\label{tab:states MM2}
\end{table}

\section{Homotopy continuation method}
Computer requirements for this algorithm scale mainly with the number of solutions, which is $2^N$. Storing of (generally complex) solutions needs $16 N 2^N$ B of memory and the time requirements are proportional to the average number of steps needed to find one solution times $N^3 2^N$, where $N^3$ comes from Newton's method. The scaling of these numbers with level is asymptotically double exponential, $2^{e^{\alpha\sqrt{L}}}$, so we can use the homotopy continuation method only at very low levels. So far, the highest level where we have successfully used this algorithm is 6. Asymptotic estimates are not very precise at low levels and, since every state counts, so we have to read off the number of equations $N$ from a character in each case individually.

The main restriction for this algorithm is the overall CPU time. If one does not have enough operational memory, it is always possible to save solutions on a hard drive or to discard solutions which are too far away from the perturbative vacuum. We have managed to solve at most 26 equations, which took us about 4 days using 100 threads. Therefore, unless one has access to computer systems with significantly better performance, it is not possible to solve more than around 30 equations.

\end{appendix}


\end{document}